\documentclass[a4paper,11pt]{memoir} 
\pdfoutput=1
\title{New Developments in Cosmology}
\author{Stefano Gariazzo}

\def\printversion{No}

\setsecnumdepth{subsection}

\usepackage{cite}

\usepackage{notoccite}
\usepackage[latin1]{inputenc}
\usepackage{latexsym}
\usepackage{amssymb}
\usepackage{url}
\usepackage{emptypage}
\usepackage{etoolbox}

\usepackage{epsf}

\usepackage{amsmath}
\usepackage[pdftex]{graphicx}
\graphicspath{{immagini/}} % Location of the graphics files
\usepackage[final]{pdfpages}
\ifthenelse{\equal{\printversion}{No}}%
{
\usepackage[%
margin=2.cm
]{geometry}
}{
\usepackage[%
margin=2.cm,
inner=3.5cm
]{geometry}
}
\usepackage{array}
\usepackage{xspace}
\usepackage{bm}
\usepackage{slashed}
\usepackage{setspace}
\usepackage{varwidth}
\usepackage{multirow}
\usepackage{indentfirst}%forces indentation of the first paragraph
\usepackage{hyperref}
\ifthenelse{\equal{\printversion}{No}}%
{
\hypersetup{
    colorlinks=true,
    linkcolor=red,
    citecolor=blue,
    urlcolor=blue,
    linktoc=page,
    pdftitle={PhD Thesis},
    pdfauthor={Stefano Gariazzo},
}
}{
\hypersetup{
    linktoc=page,
    pdftitle={PhD Thesis},
    pdfauthor={Stefano Gariazzo},
}
}

\DeclareGraphicsExtensions{.pdf}

\newcolumntype{+}{>{\global \let \currentrowstyle \relax}}
\newcolumntype{^}{>{\currentrowstyle}}

\newcolumntype{A}{>{\bfseries}}

\DeclareGraphicsExtensions{.pdf} 

\definecolor{myblue}{rgb}{.8, .8, 1}
\definecolor{lightgreen}{rgb}{0.1,.5,0.1}
\definecolor{darkred}{rgb}{.7,0,0}
\definecolor{darkgreen}{rgb}{0,.6,0}
\definecolor{darkblue}{rgb}{0,0,.7}

\newcommand{\be}{\begin{equation}}
\newcommand{\ee}{\end{equation}}
\newcommand{\bea}{\begin{eqnarray}}
\newcommand{\eea}{\end{eqnarray}}

\newcommand{\citelesg}[1]{\cite{Lesgourgues-Mangano-Miele-Pastor-2013 #1}}

\newcommand{\lsn}{LS$\nu$}

\newcommand{\e}[1]{\ensuremath{\times 10^{#1}}}
\newcommand{\cosmomc}{\texttt{CosmoMC} \ignorespaces}
\newcommand{\camb}{\texttt{CAMB} \ignorespaces}
\newcommand{\cmbfast}{\texttt{CMBFAST} \ignorespaces}

\newcommand{\pchip}{\texttt{PCHIP}\xspace}
\newcommand{\lcdm}{$\Lambda$CDM}

\newcommand{\dmsij}[1]{\ensuremath{\Delta m^2_{#1}}}

\newcommand{\diag}{\ensuremath{\mathrm{diag}}}
\newcommand{\munu}{\ensuremath{_{\mu\nu}}}
\newcommand{\iju}{\ensuremath{_{ij}}}

\newcommand{\eff}{\ensuremath{\mathrm{eff}}}

\newcommand{\meff}[1]{\ensuremath{m^{\eff}_{#1}}}
\newcommand{\Neff}{\ensuremath{N_{\eff}}}
\newcommand{\neff}{\ensuremath{N_{\eff}}}
\newcommand{\DNeff}{\ensuremath{\Delta N_{\eff}}}

\newcommand{\fs}[1]{\ensuremath{_{\mathrm{FS}}^{#1}}}
\newcommand{\nr}[1]{\ensuremath{_{\mathrm{NR}}^{#1}}}
\newcommand{\eq}[1]{\ensuremath{_{\mathrm{eq}}^{#1}}}
\newcommand{\summnu}{\ensuremath{\Sigma m_\nu}}
\newcommand{\mnu}{\ensuremath{\Sigma m_\nu}}

\newcommand{\tot}{\ensuremath{_{\mathrm{tot}}}}
\newcommand{\inu}{\ensuremath{_{\mathrm{in}}}}

\newcommand{\mx}{\ensuremath{_{\mathrm{max}}}}
\newcommand{\Omtot}{\ensuremath{\Omega\tot}}

\newcommand{\zeroord}{\ensuremath{^{(0)}}}

\newcommand{\sigmav}{\ensuremath{\langle\sigma v\rangle}}
\newcommand{\alm}[1]{\ensuremath{a_{lm}^{#1}}}
\newcommand{\cl}[1]{\ensuremath{C_{l}^{#1}}}

\newcommand{\ec}[1]{\textbf{(C#1)}}

\newcommand{\psj}[1]{\ensuremath{P_{s,#1}}}
\newcommand{\logA}{\ensuremath{\log(10^{10}A_s)}}

\newcommand{\fnl}{\ensuremath{f_{\mathrm{NL}}}}

\allowdisplaybreaks	%split long equations in different pages

\newcommand{\hub}{\ensuremath{\mathcal{H}}}
\newcommand{\wla}{\ensuremath{w_\Lambda}}
\newcommand{\wlae}{\ensuremath{w_\Lambda^{\mathrm{eff}}}}
\newcommand{\rhodm}{\ensuremath{\rho_{dm}}}
\newcommand{\rhode}{\ensuremath{\rho_\Lambda}}

\newcommand{\nus}{\ensuremath{+\nu_s}}

\newcommand{\epssr}[1]{\ensuremath{\epsilon_{\mathrm{SR}}^{#1}}}
\newcommand{\etasr}[1]{\ensuremath{\eta_{\mathrm{SR}}^{#1}}}

\newcommand{\mpl}[1]{\ensuremath{m_{\mathrm{Pl}}^{#1}}}

\newcommand{\lsu}[1]{\ensuremath{_{\mathrm{LS}}^{#1}}}
\newcommand{\etals}[1]{\ensuremath{\eta\lsu{#1}}}

\newcommand{\mcc}{\ensuremath{\mathcal{C}}}
\newcommand{\mcf}{\ensuremath{\mathcal{F}}}
\newcommand{\mcm}{\ensuremath{\mathcal{M}}}
\newcommand{\mcn}{\ensuremath{\mathcal{N}}}
\newcommand{\mco}{\ensuremath{\mathcal{O}}}
\newcommand{\mcp}{\ensuremath{\mathcal{P}}}
\newcommand{\mcr}{\ensuremath{\mathcal{R}}}

\newcommand{\Mpc}{\ensuremath{\, \text{Mpc}}}
\newcommand{\mpcinv}{\ensuremath{\, \text{Mpc}^{-1}}}
\newcommand{\Hou}{\ensuremath{\,\, \text{Km s}^{-1}\text{ Mpc}^{-1}}}
\newcommand{\ev}{\ensuremath{\, \text{eV}}}

\newcommand{\Mev}{\ensuremath{\, \text{MeV}}}

\newcommand{\openone}{\ensuremath{\bm{1}}\xspace}
\newcommand{\nua}[1]{\ensuremath{\rlap{\kern-2.5pt\ensuremath{%
  \overset{\scriptscriptstyle(-)}{\phantom{\nu}}}}%
  {\ensuremath{{\nu}_{#1}}}}\xspace}

\newlength{\singlefigland}
\newlength{\singlefigbig}
\newlength{\singlefigsmall}
\newlength{\halfwidth}
\newlength{\onethirdwidth}
\begin{document}
\frontmatter
\hypersetup{pageanchor=false}
%!TeX root=main.tex
\begin{titlingpage}
\begin{center}

\includegraphics[height=3cm]{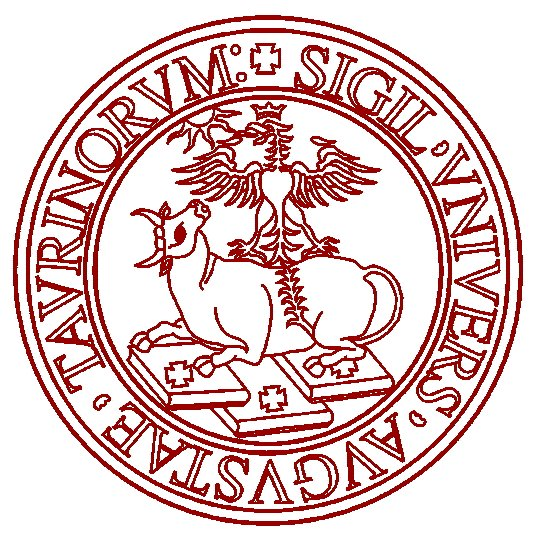}
\hspace{7cm}
\includegraphics[height=3cm]{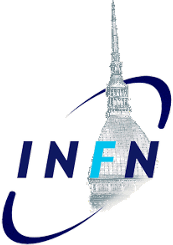}

\vspace{1.5cm}

% Upper part of the page
\textsc{\huge Universit\`a~di~Torino}\\[0.5cm]
\textsc{\LARGE Scuola~di~Dottorato~in~Scienza~ed~Alta~Tecnologia}\\[0.5cm]
\textsc{\LARGE Indirizzo di Fisica ed Astrofisica}\\[0.5cm]

\textsc{\large XXVIII ciclo}\\[1.5cm]
\textsc{\Large Tesi di Dottorato}\\[2cm]

% Title
\hrule \vspace{1cm}
{ \huge \bfseries New Developments in Cosmology}\\[1cm]

\hrule \vspace{2.0cm}
 
% Authors
\LARGE
Stefano Gariazzo
\vspace{2cm}

\Large
\begin{tabular}{rl}
% ~ \hspace{5cm} ~ &
\textsc{Relatore:} & Prof. Nicolao Fornengo\\
% &
\textsc{Co-relatore:}& Dott. Carlo Giunti\\
\end{tabular}

\vspace{2.0cm}
{\large Torino, 22 March 2016}

\vfill

\end{center}
\end{titlingpage}

\hypersetup{pageanchor=true}
% {
% \agg{cover_1stpage.tex}
% }

\cleardoublepage
% \phantomsection 
\chapter{Abstract (English)}
%!TeX root=main.tex 
% First author:
% \cite{Gariazzo:2013gua,Gariazzo:2014pja,Gariazzo:2014dla,Gariazzo:2015qea,Gariazzo:2015rra}
% 
% Co-author:
% \cite{Archidiacono:2014apa,DiValentino:2015zta,DiValentino:2016ikp,Murgia:2016ccp}
%
% Proceedings:
% \cite{Gariazzo:2015apa,Gariazzo:2016ehl,Gariazzo:2016mnp}
The evolution of the Universe is well described by the 
Standard Model of Cosmology,
parameterized through the so-called \lcdm\ 
($\Lambda$ + Cold Dark Matter) model,
based on the theory of General Relativity.
The \lcdm\ model has been widely studied in the past, and the
fundamental parameters that describe it have been constrained
using several different experimental measurements.
In the last years, the accurate observations
of the Cosmic Microwave Background (CMB) anisotropies
allowed to improve considerably the constraining power
of the cosmological analyses,
opening the way to precision cosmology.
Cosmology can help in studying constraints on the content of the
Universe at all times, and precision measurements of the cosmological
observables can improve even our knowledge on particle physics.
For example, constraints on the absolute neutrino mass scale
or on the presence of additional neutrinos beyond the three standard
ones can be derived using cosmological data.

The last results released by the Planck collaboration
are in strong agreement with the \lcdm\ model and
there is no strong evidence that the \lcdm\ model may be incomplete.
Despite the overall robustness, however,
some small inconsistencies appear.
For example, the local determinations of the Hubble parameter $H_0$
and of the matter fluctuations at small scales $\sigma_8$
are in tension with the estimates obtained
from the analyses of CMB data in the context of the \lcdm\ model.
% These two tensions can be solved if a light massive particle
% influences the evolution of the Universe.
We show that the presence of
a light sterile neutrino or a thermal axion
may reduce these tensions, suppressing
the matter fluctuations at small scales and
increasing the Hubble parameter.
These two light particles are motivated by the phenomenology of
short-baseline neutrino oscillations and by the strong CP problem
in Quantum ChromoDynamics, respectively.
We present also the most recent constraints
on the sterile neutrino and
on the thermal axion properties.

Another indication that not all the predictions of the standard
cosmological model are complete is related to
the possible presence of features
in the Primordial Power Spectrum (PPS) of curvature perturbations.
The initial fluctuations were generated
during the early inflationary phase of the Universe evolution
and they are the initial conditions for the subsequent evolution.
As a consequence, features in the PPS can be reconstructed
observing the WMAP and Planck spectra of CMB temperature anisotropies
at large scales.
The assumptions on the PPS shape, however, are crucial for all the
cosmological analyses.
If inflation is realized in a non-standard scenario,
the PPS may have a non-standard shape and if
it does not have the standard power-law shape,
the cosmological constraints can be (strongly) biased.
We study how the constraints
on the properties of massless and massive neutrinos and
of thermal axions change when a free PPS shape is considered instead
of the usual power-law one.
In addition, we study also how the constraints on primordial
non-Gaussianities change in the context
of a scenario involving ``inflationary freedom''.

We also show that a possible solution to the small 
$H_0$ and $\sigma_8$ tensions
may come from an additional non-gravitational interaction
between dark matter and dark energy,
if dark energy decays into dark matter.
This is not forbidden by any current observation, and this possibility
opens a new window to study the dark sector of our Universe.

\chapter{Abstract (Italiano)}
%!TeX root=main.tex 
% First author:
% \cite{Gariazzo:2013gua,Gariazzo:2014pja,Gariazzo:2014dla,Gariazzo:2015qea,Gariazzo:2015rra}
% 
% Co-author:
% \cite{Archidiacono:2014apa,DiValentino:2015zta,DiValentino:2016ikp,Murgia:2016ccp}
%
% Proceedings:
% \cite{Gariazzo:2015apa,Gariazzo:2016ehl}
L'evoluzione dell'Universo \`e ben descritta dal
modello cosmologico standard,
parametrizzato attraverso
il cosiddetto \mbox{modello} \lcdm\ ($\Lambda$ +
Cold Dark Matter - materia oscura fredda) e
basato sulla teoria della Relativit\`a Generale.
Il modello \lcdm\ \`e stato ampiamente studiato in passato,
e i parametri che lo descrivono sono strettamente vincolati dalle 
numerose osservazioni sperimentali.
Negli ultimi anni le accurate misure della
radiazione cosmica di fondo (CMB, da Cosmic Microwave Background)
hanno incrementato notevolmente la precisione delle determinazioni
nelle analisi,
aprendo la strada alla cosmologia di precisione.
Oggi i dati cosmologici ci permettono di studiare il contenuto
dell'Universo a tutte le epoche e le misure di precisione
delle osservabili cosmologiche permettono di migliorare anche
la nostra conoscenza della fisica delle particelle.
Per esempio, dai dati cosmologici si possono ottenere vincoli
sulla scala di massa dei neutrini e sulla eventuale presenza
di neutrini aggiuntivi in aggiunta ai tre neutrini standard.

Gli ultimi risultati pubblicati da parte della collaborazione Planck 
sono in notevole accordo con le predizioni del modello \lcdm\ 
e non compare nessuna evidenza significativa che il modello
\lcdm\ possa essere incompleto.
Al di l\`a della robustezza generale, comunque,
ci sono alcune piccole discrepanze.
Per esempio, le misure locali del parametro di Hubble $H_0$
e delle fluttuazioni di materia a piccola scala $\sigma_8$
sono in tensione con le stime ottenute dalle analisi dei dati
della CMB nel contesto del modello \lcdm.
Mostreremo che la presenza di un neutrino sterile leggero o
di un assione termico pu\`o ridurre tali tensioni,
sopprimendo le fluttuazioni di materia a piccola scala e
incrementando il parametro di Hubble.
Queste due particelle leggere emergono rispettivamente
come soluzione alle anomalie nei dati delle oscillazioni dei neutrini
a corto raggio
o al problema della CP forte nella cromodinamica quantistica.
Presenteremo quindi i vincoli pi\`u recenti
sulle propriet\`a di tali particelle.

Un'altra indicazione che non tutte le predizioni del modello
cosmologico standard sono complete \`e collegata alla forma dello
spettro di potenza iniziale (PPS, da Primordial Power Spectrum)
delle fluttuazioni di curvatura.
Queste condizioni iniziali sono state generate durante
il periodo di inflazione all'inizio dell'Universo
e ne determinano l'evoluzione successiva.
Le assunzioni sulla forma del PPS sono cruciali
per tutte le analisi cosmologiche.
Le osservazioni delle anisotropie di temperatura della CMB
osservate dagli esperimenti WMAP e Planck
suggeriscono la presenza di una forma anomala del PPS.
Se l'inflazione non pu\`o essere descritta nella maniera
pi\`u semplice e
il corrispondente PPS pu\`o deviare dalla legge di potenza standard,
i risultati delle analisi cosmologiche possono esserne influenzati.
Studieremo come i vincoli sulle propriet\`a dei neutrini
(massivi o privi di massa) e sugli assioni termici sono influenzate
dalla libert\`a nella forma del PPS, se questo pu\`o differire
dalla normale legge di potenza.
In aggiunta valuteremo anche come i vincoli sulle non-Gaussianit\`a
primordiali possono cambiare nel contesto di questa
``libert\`a inflazionaria''.

Una diversa possibilit\`a per risolvere le tensioni riguardanti
$H_0$ e $\sigma_8$ \`e collegata alla possibile \mbox{esistenza}
di una nuova interazione, di tipo non-gravitazionale,
fra materia oscura ed energia oscura,
in particolare se coinvolge
energia oscura che decade in materia oscura.
Questa interazione non \`e proibita da nessuna osservazione corrente
e rappresenta una possibilt\`a di aprire
una nuova finestra sullo studio
delle componenti \mbox{oscure} del nostro Universo.

\cleardoublepage
\setcounter{tocdepth}{2}
\tableofcontents

\mainmatter

% \chapter*{Introduzione}
% \addcontentsline{toc}{chapter}{Introduzione}
% \input{introITA.tex}

\chapter*{Introduction}
\addcontentsline{toc}{chapter}{Introduction}
%!TeX root=main.tex 
Recently the Nobel Prize in Physics was awarded to
Takaaki Kajita and Arthur B.~McDonald 
``for the discovery of neutrino oscillations,
which shows that neutrinos have mass'', a
result that confirms the hypothesis proposed almost sixty years ago by
B.~Pontecorvo~\cite{Pontecorvo:1957cp}. 
Pontecorvo was the first to suggest that
neutrinos may exist in different flavors and that they can oscillate.
Since the proposal of Pontecorvo, many years were needed
to measure neutrino oscillations, 
but finally this achievement opened a new window on physics,
imposing the existence of the mass of at least two neutrinos.
Neutrino oscillations, indeed,
require that the three neutrino mass eigenstates
have different masses $m_i$ (with $i=1,2,3$).
These masses can be measured in neutrino oscillation experiments,
as we will discuss in details in Chapter~\ref{ch:nu}.
The quantities that allow to describe the oscillations between
the three different flavor neutrinos are the squared-mass differences%
\footnote{We use the convention $\dmsij{ij}=m^2_i-m^2_j$.}
$\dmsij{21},\,\dmsij{31},\,\dmsij{32}$ and
the elements of the so-called PMNS mixing matrix,
originally proposed by Z.~Maki, M.~Nakagawa, S.~Sakata
\cite{Maki:1962mu} to describe the neutrino oscillation proposed by
B.~Pontecorvo \cite{Pontecorvo:1957cp}.
One standard possibility to write the unitary mixing matrix
for the three neutrino mixing 
paradigm is written in Eq.~\eqref{eq:3numix}.
Nowadays, most of the elements of the PMNS matrix are well determined
(see e.g.\ Ref.~\cite{Agashe:2014kda})
by the numerous experiments that probe neutrino oscillations
at different energies and distances.

Not all the quantities required to describe the neutrino physics, however,
are well known at present times.
The mixing matrix in the standard parameterization
of Eq.~\eqref{eq:3numix}
is described using 6 parameters:
three mixing angles
$\vartheta_{12}$, $\vartheta_{23}$ and $\vartheta_{13}$,
one Dirac phase $\eta_{13}$
and two Majorana phases $\lambda_{21}$ and $\lambda_{31}$,
that are physical only if neutrinos are Majorana particles.
The mixing angles are known with good precision,
apart for $\vartheta_{23}$, that is nearly maximal and we do not know if
it is larger or smaller than $45^\circ$.
The present knowledge about the phases, instead, is rather poor.
We have small indications that the favored value for the Dirac phase,
that may provide CP violation in the lepton sector,
is close to $3\pi/2$~\cite{Abe:2015awa},
but the statistical significance is small.

One of the interesting open questions concerns the nature of neutrinos.
All the known particles in the Standard Model (SM) of Particle Physics
are Dirac particles, but
neutrino is actually the only candidate for being a Majorana particle.
If they are Majorana particles, neutrinos coincide with their own antiparticles
and processes that violate the conservation of the
lepton number are possible.
The most studied process of this kind is the neutrinoless double $\beta$-decay,
that however has never been observed \cite{Bilenky:2014uka}.
Double-$\beta$ decay processes are possible for particular unstable atoms,
that may decay simultaneously through the emission of two electrons,
normally accompanied by the emission of two electron antineutrinos.
For these atoms, the observation of the double $\beta$-decay
is possible only because the single $\beta$-decay is forbidden by the
kinematics.
If the neutrino is Majorana, however,
in a small fraction of the cases the neutrino
is emitted and immediately absorbed inside the decaying nucleus,
that undergoes a double-$\beta$ decay emitting only two electrons,
with a violation of the lepton number.
Neutrinoless double $\beta$-decay is nowadays the only process that could
allow to measure the Majorana phases that appear in the mixing matrix,
since they are relevant only for processes that distinguish the
Majorana nature of the neutrinos \cite{Bilenky:2014uka}.

Another crucial unknown point is the absolute scale of the
neutrino masses.
Measurements of the neutrino mixing give information on the mass differences,
but we cannot learn from neutrino oscillation experiments what is the mass
of the lightest neutrino, that is $m_1$ in the normal ordering
and $m_3$ in the inverted ordering.
The absolute neutrino mass scale can be directly
determined measuring the endpoint of the
spectrum of the released electron in $\beta$-decay processes 
(see e.g.\ Ref.~\cite{Dragoun:2015oja})
or through the kinematics of neutrinoless double $\beta$-decay processes
\cite{Bilenky:2014uka},
if neutrinos are Majorana particles.
Currently, the direct measurements of the neutrino masses through
$\beta$-decay experiments provide an upper limit on the neutrino mass scale
of about $2.2\ev$ \cite{Weinheimer:2002rs}.
The future experiment KATRIN should reach a sensitivity of about $0.2\ev$
using the decay of tritium atoms \cite{Mertens:2015ila}.

Another unknown point pertains the squared mass differences.
% measured in neutrino oscillation experiments.
The squared-mass difference $\dmsij{21}$ is fully known
thanks to the matter effect in the oscillations inside the sun, 
also called the MSW effect after S.P.~Mikheev, A.Yu.~Smirnov and L.~Wolfenstein
\cite{Wolfenstein:1977ue,Mikheev:1986gs,Mikheev:1986wj}.
On the other hand, we know only the absolute values
of the squared-mass differences $\dmsij{31}$ and $\dmsij{32}$.
As a consequence, we know that the mass $m_2$ of the eigenstate $\nu_2$
is larger than the mass $m_1$ of the eigenstate $\nu_1$,
but we do not have information
on the ordering of the third mass eigenstate.
The neutrino mass ordering may be $m_1<m_2<m_3$ (normal ordering)
or $m_3<m_1<m_2$ (inverted ordering), depending on the sign of
$\dmsij{31}$ (or of $\dmsij{32}$).
Future experiments will investigate the neutrino mass ordering,
trying to measure
the matter effects on neutrino oscillations in the Earth
\cite{Aartsen:2014oha,Adrian-Martinez:2016fdl}
or using the phase difference in the oscillations 
of reactor electron antineutrinos, given by the different sign
of the squared-mass differences $\dmsij{31}$ and $\dmsij{32}$
in the oscillation probability formula
\cite{An:2015jdp}.

Short Baseline (SBL) neutrino oscillation experiments
suggest that the standard description of the three neutrino mixing
may be incomplete, since several anomalies appear (see Section~\ref{sec:sbl}
or Ref.~\cite{Gariazzo:2015rra}).
The global fit of SBL neutrino oscillation data improves if one assumes
an additional neutrino mass eigenstate $\nu_4$
with $\dmsij{41}\simeq1\ev^2$
(see Section \ref{sec:global}).
To the new neutrino mass eigenstate, a new flavor eigenstate should correspond.
This is called a ``sterile'' flavor state, since it is not coupled 
to the SM Lagrangian, but its interactions
with the SM particles and with the other neutrinos 
are possible only through neutrino oscillations.
The existence of the fourth neutrino state and the SBL anomalies
will be tested in future SBL neutrino oscillation experiments.

Direct mass detection and oscillation experiments, however,
are not the only way that we have to test the unknown neutrino properties,
although they represent the strongest tests that can be performed,
since their results are model independent.
Another exciting field of research, indeed, is cosmology.
From various cosmological measurements
it is possible to derive constraints on the absolute scale of neutrino masses
and
on the existence of additional particles.
In this case, however, the results are obtained in the context
of a specific cosmological model.

The standard description of our Universe is based on the theory of
General Relativity of A.~Einstein \cite{Einstein1916},
proposed one hundred years ago.
The Standard Model of Cosmology, also called the 
\emph{Hot Big Bang model} and described in Chapter~\ref{ch:cosmology},
predicts that the Universe started its evolution
in a very dense and hot configuration, that expanded for about 13 billions years
to become what we observe nowadays.
A crucial evidence in favor of the Big Bang model was
the detection of the Cosmic Microwave Background (CMB) radiation
\cite{Penzias:1965wn}, that is the thermal radiation left over
from the time of recombination.
It is the oldest light in the Universe,
originated when the photon energy decreased enough
to become smaller than the 
electron binding energy inside the hydrogen atoms.
Recombination indicates in fact the time at which the electrons and the protons
started to be bounded together in the hydrogen atoms.
Before recombination the Compton scattering of electrons and
the presence of high energy photons prevented those stable bounds
and the photons were continuously scattered.
After recombination, instead, the density of free electrons diminished
drastically, the photons started to propagate freely
and the CMB radiation was generated.
Further details are presented in the description of the CMB radiation 
and of its anisotropies developed in Chapter~\ref{ch:cmbr}.

The CMB radiation has become one of the pillars of the modern cosmology.
After the first detection by A.A.~Penzias and R.W.~Wilson \cite{Penzias:1965wn},
who were awarded the Nobel prize in 1978,
the discovery of the CMB anisotropies beyond the monopole and the dipole
by the COBE experiment in 1992 \cite{Smoot:1992td}
opened the window to a new way to test the evolution of the Universe.
With the precision measurements of the CMB spectrum
obtained by the WMAP \cite{Bennett:2012zja}
and Planck \cite{Ade:2013sjv,Adam:2015rua} experiments,
we have the possibility of testing the cosmological models with great accuracy
and to derive constraints on the cosmological parameters.

CMB observations, extensively discussed in Sec.~\ref{sec:cmb},
are not the only robust measurements that can be used to constrain the
cosmological models.
Baryon Acoustic Oscillations (see Section~\ref{sec:bao}), for example,
represent a robust tool that can give strong constraints
on the evolution using geometrical methods.
Other tests of the Universe evolution
at late times are the measurements
of the Hubble parameter, that gives the expansion rate today
(see Sec.~\ref{sec:h0}),
of the redshift-distance relation through the observations of SuperNovae
(see Sec.~\ref{sec:sn}),
and of the late time matter distribution
through the full power-spectrum of matter fluctuations (see Sec.~\ref{sec:mpk}),
the cluster counts (see Sec.~\ref{sec:cluster})
and the weak lensing detection through the observations of the cosmic shear
(see Sec.~\ref{sec:shear}).

In this dissertation we will use CMB data, together with the other 
observations of the Universe, to derive constraints on neutrino physics.
These constraints are model-dependent, in the sense that they depend
on the assumptions in the context of the Hot Big Bang model.
Additional mechanisms or phenomena that are not considered in the standard
description of the Universe evolution can dramatically change these results.
In our case, however, we will focus mainly
on the most simple parameterization of
the hot Big Bang model, that is the so-called \lcdm\ model
(see Section~\ref{sec:par_depend}), after the names
of the cosmological constant $\Lambda$ and of the cold dark matter (CDM),
that are the most abundant constituents of the Universe today.
We will detail extensively the properties of the cosmological constant
and of cold dark matter in the first two Chapters.

Cosmology cannot probe all the neutrino properties
that we listed above:
the cosmological evolution is basically insensitive to the mixing
of three neutrinos.
On the contrary, cosmological measurements provide strong constraints on
the neutrino masses and on the existence of additional particles that were
relativistic in the early Universe, as the 1~eV mass sterile neutrino
that we mentioned above.
These quantities can be constrained since the presence of massive neutrinos
has an impact on the CMB anisotropies and on the other cosmological quantities,
as we will describe in details in Section~\ref{sec:nucosmology}.
Part of the analyses presented in this Thesis have the aim of studying
the compatibility of the light sterile neutrino motivated by the SBL oscillations
with the most recent cosmological measurements, constraining the effects that
this additional neutrino has on the various observables.
These analyses will be presented in Chapters \ref{ch:lsn_cosmo} and
\ref{ch:pps_nu}, based on
Refs.~\cite{Gariazzo:2013gua,Archidiacono:2014apa,Gariazzo:2014pja}
and \cite{Gariazzo:2014dla,DiValentino:2016ikp},
respectively.

The presence of neutrinos in cosmology
may be particularly significant to solve the small
tensions that are present in the \lcdm\ model.
These regards the CMB estimates
and the determinations at small redshift 
of the Hubble parameter $H_0$ and
of the clustering parameter $\sigma_8$,
that measures the matter fluctuations inside a sphere of
8$h^{-1} \Mpc$ radius.
An additional light particle that is relativistic at the time of matter-radiation
equality and that becomes non-relativistic at late times
can reduce the amount of matter fluctuations at small scales thanks to its
free-streaming properties (see Section~\ref{sec:nuFS} for the neutrino case):
this goes in the required direction to reconcile local and cosmological estimates
of $\sigma_8$.
At the same time, the presence of additional ``dark radiation''
(i.e.\ relativistic particles, apart for photons) in the early Universe
requires an increase of the cold dark matter energy density and
of the cosmological constant
energy density at all times, in order to avoid a shift of the matter-radiation
equality epoch that would alter significantly the CMB spectrum.
This has the direct consequence of increasing the predictions of $H_0$,
reducing the difference between the local measurements
and the cosmological estimates for that parameter.

A crucial problem that appears when one tries to constrain the neutrino properties
from cosmology is that
from neutrino oscillations we expect that the sterile neutrino
is in full equilibrium with the active neutrinos in the early Universe:
the contribution to the radiation energy density of a sterile neutrino should 
be equal to the contribution of each active neutrino.
The expectation does not correspond to the results, however,
since the analyses of the most recent CMB data indicate with high precision
that there are approximately three neutrino-equivalent particles,
and the existence of a fourth one is strongly disfavored
(see Chapter~\ref{ch:lsn_cosmo}).
This is known as the thermalization problem of the sterile neutrino.
In the context of the standard cosmological model,
if there are four neutrinos, one of them cannot be in equilibrium with the others,
possibly as a consequence of some new physical mechanism in particle physics:
we will list some possibilities proposed in the literature
in Section~\ref{sec:nucosmo_conclusions}.

The thermalization problem can be solved in a different way
that does not involve new particle physics mechanisms.
If a new cosmological mechanism induces some effects in the evolution
that compensate the changes arising from the presence of an additional particle
(the sterile neutrino),
the tension may disappear.
One possibility is the scenario of ``inflationary freedom''.
Inflation is the initial phase of the Universe expansion, during which
the distances were stretched exponentially for a very short time.
Inflation is required to explain the ``horizon'' and the ``flatness'' problems,
that we will treat in Chapter~\ref{ch:cosmology},
as well as the extreme large scale homogeneity and isotropy of the Universe.
The simplest inflationary models predict an initial power spectrum of
curvature fluctuations that is a simple power-law.
Observations of the CMB spectrum suggests that there may be deviations from
such a featureless spectrum, especially at large scales.
If deviations from the power-law form exist also at small scales,
as a consequence of some freedom in the inflationary scenarios,
the effects of the additional dark radiation may be erased in the final results
by the shape of the
initial power spectrum of the Gaussian density fluctuations
and the final power spectrum of CMB anisotropies would be almost unchanged.
We study this possibility in Chapter~\ref{ch:pps_nu}, where we test
the degeneracies between the primordial power spectrum (PPS)
of scalar perturbations and the neutrino properties.
These degeneracies may give a partial solution to the thermalization problem,
that is still present when the recent CMB polarization data by Planck
are considered in the analyses.

The light sterile neutrino, however, is not the only candidate that could
help solving the $H_0$ and the $\sigma_8$ tensions.
Among the other possibilities,
we studied the thermal axion as a candidate of dark radiation.
Axions were proposed by R.D.~Peccei and
H.R.~Quinn~\cite{Peccei:1977hh,Peccei:1977ur} 
to solve the strong CP problem in Quantum Chromodynamics, as we will
explain in Section~\ref{sec:ax_intro}.
If one considers a thermal production mechanism
\cite{Turner:1986tb,Chang:1993gm,Masso:2002np},
it turns out that the axion can have a mass of the order of 1~eV,
it contributes to the radiation energy density in the early Universe
and it has free-streaming properties.
In brief, it behaves approximately as a massive neutrino and
therefore it can provide a solution to the $H_0$ and the $\sigma_8$ tensions.
In Chapter~\ref{ch:ther_ax},
based on Refs.~\cite{DiValentino:2015zta,DiValentino:2016ikp},
we will show the most recent constraints on the
thermal axion mass that arise from the cosmological analyses.
Also in this case we will study the degeneracies within the context of
inflationary freedom, as we did for the neutrino properties.

Another analysis that we will present concerns
the possible existence of non-Gaussianities,
i.e.\ deviations from the Gaussian distribution,
in the initial fluctuations that evolved to generate the CMB anisotropies
and the structures that we observe in our Universe.
Non-Gaussianities are expected to be generated during inflation, and the presence
of non-Gaussianities produces a distortion of the CMB (or matter) power spectrum.
Since non-Gaussianities and the initial power spectrum of scalar fluctuations
are both expected to be generated during inflation by the same mechanism,
there is the concrete possibility that they produce similar distortions
in the observed power spectrum of CMB (or matter) fluctuations.
In Chapter~\ref{ch:ng} we show that the distortions of the matter power
spectrum generated by non-Gaussianities may be mimicked by deviations of
the power spectrum of initial fluctuations from the simple power-law.
The immediate consequence is that the results obtained for the non-Gaussianities
may be significantly biased if some scenario 
involving ``inflationary freedom'' is assumed.
We devote Chapter~\ref{ch:ng} to test and discuss these degeneracies,
following the analyses published in Ref.~\cite{Gariazzo:2015qea}.

Up to now, we considered extensions of the \lcdm\ model including 
some new mechanism in the very beginning of the Universe life,
possibly connected with some particle physics model of inflation,
or
some new particles that arise from some model in particle physics
(sterile neutrinos, thermal axions).
These additional particles, however, are expected to give only a minor fraction
of the total energy density of the nowadays Universe.
The largest fraction of the Universe content today \cite{Adam:2015rua}
is provided by two fluids for which we do not have
a well assessed explanation in terms of particle physics:
the cold dark matter and the cosmological constant,
accounting for 26\% and 69\% of the total energy density today, respectively.
Cold dark matter indicates some massive component
that does not interact electromagnetically.
The cosmological constant, or in general the ``dark energy'',
is a diffuse fluid that is responsible
of the accelerated expansion of the Universe at late times.
These fluids are known only for their gravitational interaction and
nothing else is known about them.
In a minimal scenario, dark matter and dark energy do not have interactions
apart for gravity, but some non-gravitational coupling between them cannot
be excluded.
In Chapter~\ref{ch:cde}, based on Ref.~\cite{Murgia:2016ccp},
we will study exactly this case:
a phenomenological non-gravitational coupling between dark matter and dark energy,
and we will show how this coupling influences the Universe evolution.
We will explore two possibilities: dark matter decaying in dark energy
or dark energy decaying in dark matter.
Using cosmological data that probe different times,
we will study the compatibility
of the coupled scenario with the current cosmological measurements,
with a particular focus on the small tensions concerning
the Hubble parameter $H_0$
and the clustering parameter $\sigma_8$.

Chapter~\ref{ch:conclusions}, the last of this Thesis,
contains a resume and a brief discussion of our results.

% \mainmatter

\part{Overview of Standard Cosmology}

%!TeX root=main.tex 
\chapter{The Standard Model of Cosmology}
\label{ch:cosmology}

% \begin{abstract}
% In this Chapter we describe the Standard Model of Cosmology
% and the evolution of the Universe that it predicts.
% We introduce the most relevant cosmological quantities and
% we will write
% the evolution equations for the background and for the perturbations.
% These equations will be adopted in the following Chapter to derive
% the spectrum of CMB anisotropies and its dependencies on
% the various cosmological parameters.
% \end{abstract}

The evolution of our Universe is currently well described by the so-called
\emph{Standard Model of Cosmology}, or \emph{Hot Big Bang Model}.
This model is based on the renowned theory of General Relativity,
presented by A.~Einstein in 1915 and
published in 1916 \cite{Einstein1916}.
The fundamental elements of the cosmological model are
the \emph{Cosmological Principle}, 
which states that the Universe is homogeneous 
and isotropic on large scales,
and the \emph{Einstein Equations}, 
which describe the evolution of a physical system 
under the action of gravity.
In this first Chapter we will describe the 
Standard Model of Cosmology, 
particularly focusing on the equations that govern the 
thermal history of the Universe and the evolution of perturbations.
Since we will not develop the full calculations,
we suggest further readings for more details, 
e.g. Ref.~\cite{Dodelson-Cosmology-2003}.
We will work in natural units through all the text.

\section{Short Evolution History}\label{sec:evolution}
In the Big Bang model, the Universe started from a very hot and dense 
plasma, that cooled down during the expansion.
The initial phases of the Universe are not well known, 
since we do not have any confirmed theory to explain physics
at extremely large energies:
a complete theory of quantum gravity is required to fully describe
the initial phase of the Universe.

Possibly in the very early Universe an inflationary phase occurred.
Inflation is a theory that predicts an exponential expansion
during which the scale factor $a$ grows as $a(t)=\exp(Ht)$,
where $H$ is the Hubble factor (see Eq.~\eqref{eq:hubblepar}).
Inflation requires a constant energy density, with the consequence
that the first Friedmann equation (see Sec.~\ref{sec:feq})
becomes $H^2\simeq const$.
Using the cosmological constant notation, this becomes
$H\simeq\sqrt{\Lambda_I/3}$, where $\Lambda_I$
is the cosmological constant during inflation.

Inflation was proposed firstly in the eighties
\cite{Guth:1980zm,Linde:1981mu,Mukhanov:1981xt,Starobinsky:1982ee,
Hawking:1982cz,Albrecht:1982wi,
Lucchin:1984yf,Mukhanov:1990me}
to solve the \emph{horizon} and the \emph{flatness} problems.
The \emph{horizon} problem is connected to the fact that we observe
an extreme homogeneity between sky regions that are separated
by distances between them larger than the horizon radius.
These regions were not in causal contact in the past
if the standard evolution, without inflation, is assumed.
It appears unlikely that widely separated regions that could not be
in causal contact in the past can be so similar today.
This is not true if the Universe expanded exponentially
in the early phases of its history,
since regions that were in causal contact
before the end of inflation were stretched and widely separated.
Initial perturbations that were similar before the end of inflation
evolved independently after inflation,
possibly until today.

The \emph{flatness} problem indicates the fact that the curvature 
of the Universe is very close to 1 today:
the strongest constraints come from the Planck collaboration
\cite{Adam:2015rua},
which estimated that the curvature energy density is
$\Omega_k^0=0.000\pm0.005$ \cite{Ade:2015xua},
using the Planck full mission data on the CMB spectrum
(see Section~\ref{sec:feq}).
Going back in time, the bounds become very stringent,
since in a not flat and decelerating Universe
the curvature increases during the expansion (see Section~\ref{sec:feq}):
for example,
at the time of Big Bang Nucleosynthesis (BBN)
the total energy density $\Omtot=1-\Omega_k$
must fulfill the requirement $|\Omtot-1|\lesssim10^{-18}$,
in order to be compatible with the Planck bound today.
Since at earlier epochs the value would be even smaller,
this was considered as a fine-tuning problem.
In the context of inflation, this problem is solved by the 
exponential expansion which dilutes the curvature:
since the relation is 
$|\Omtot-1|\propto \exp(-\sqrt{4\Lambda_I/3}\; t)$ during inflation,
the longer was inflation, the closest $\Omtot$ was to 1 at its end.
To solve both the flatness and the horizon problems,
inflation should have lasted for at least 50 to 60
\emph{e-foldings},
a unit that measures the exponential variation of the scale factor:
$N$ e-foldings correspond to an increase in the scale factor
$a(t_{\mathrm{end}})= e^N a(t_{\mathrm{start}})$,
or equivalently $N=\ln(a(t_{\mathrm{end}})/a(t_{\mathrm{start}}))$.

As we will see in Section~\ref{sec:initial_conditions},
inflation is usually modeled with the introduction of
a scalar field $\phi$,
called \emph{inflaton}, that mimics the cosmological constant
behavior when rolling down a slowly varying potential $V(\phi)$.
Inflation ends when the scalar field decays into other particles,
with a consequent energy transfer to the plasma.
This phase takes the name of \emph{reheating}, since the temperature of
the plasma of coupled particles
is raised with the increase of its energy.

The Universe temperature continuously decreases.
As the temperature decreases, the kinematics of the processes
occurring in the plasma changes and some particles
that were abundant in the early Universe cannot be produced at later times:
for unstable particles, this means that they start to disappear,
being the production and decay processes out of equilibrium.
At the same time, some of the
symmetries that were perfect in the hot Universe
start to spontaneously break:
after the electroweak symmetry breaking the bosons of the weak 
interaction and most of the fermions start to have a mass.
Since they are still very energetic, each of them behave as relativistic 
particles until the temperature falls below its mass;
in other cases, such as for the $t$ quark, the mass is so high that they never
behave as relativistic particles.
The quarks still cannot be confined in hadrons since their kinetic
energy is too high.
As the temperature decreases, however, the kinetic energies decrease
and at a certain point the quarks can be confined: 
this is the transition to the hadron epoch.

Before this time, depending on its mass and its interaction rates,
DM can decouple.
When the DM particles can annihilate but they cannot be produced
because of the kinematics,
they stop interacting and they are \emph{freezed-out},
i.e.\ they stop interacting and their energy density is simply
diluted with the evolution.
The annihilation rate depends on the squared number density,
and consequently it decreases while the Universe expands.
At a temperature of around 1\Mev,
the equilibrium of neutrino-electron interactions is broken and also
the existing neutrinos decoupled from the rest of the plasma:
the relic neutrinos give origin to the 
\emph{Cosmic Neutrino Background} (CNB),
the neutrino analogous of the Cosmic Microwave Background (CMB)
radiation, composed by the cosmological photons.
The CNB today is very hard to detect directly, since these neutrinos
have an extremely low energy.
We have a number of
indirect signals that the number of relativistic species
at CMB decoupling is compatible with the presence of three relic
neutrinos, but we are still not sure that these additional particles
are truly the standard neutrinos.

Shortly after neutrino decoupling, the mean photon temperature
becomes too small to allow the production of electron-positrons pairs
and also the electrons start to decouple.
The energy density of electrons is transfered to photons
through the annihilation process $e^+e^-\rightarrow 2\gamma$.
In this phase the photons are reheated by this energy transfer, and
from now on the photon temperature is higher than the neutrino temperature.

During the hadron epoch, neutrinos play a role in the interactions
that bring protons and neutrons to equilibrium:
the number of neutrinos have an impact on the
relic neutron-to-proton ratio, that in turn influences
the relic abundances of light elements after 
the BBN.
As the photon energy diminishes below 0.1~MeV,
photons are no more able to break
the nuclear bounds and the light nuclei can be produced
in hadron scatterings.
Starting from protons and neutrons, the first element that is 
created is deuterium, $^2H$.
Inelastic scattering of deuterium and other nucleons originates
$^3He$, $^4He$, $^7Li$ and
some unstable elements such as $^3H$, $^7Be$,
that decay in $^3He$ and $^7Li$.

After the production of the light nuclei, photons have enough
energy to break electron-nucleus bounds and
matter is still ionized.
After matter-radiation equality,
that is the time at which the Universe evolution started
to be dominated by the matter energy density,
photons and relic neutrinos become
less and less important for the evolution of the Universe
and the matter perturbations
can start growing under the effect of gravity.
While the photons continue to cool down, their temperature
diminishes below $T\simeq0.1\ev$.
At this point
their energy becomes small enough to allow the creation of atoms:
photons are no more energetic enough to break the electron-nucleus
bounds and finally the Universe becomes transparent to photons,
that start to move freely.
This is the time of recombination, when the CMB was originated.
Since CMB photons interacted rarely in the following epochs,
the study of the CMB anisotropies gives us information
on the Universe at the time of recombination, that occurred about
380.000~years after Big Bang.
In the same way, the CNB anisotropies would give us
information on the Universe at the time of neutrino decoupling,
that occurred about 1~second after Big Bang.
The detection and the study of the CNB anisotropies
are far away from our current technological capabilities, however.

After CMB decoupling, the evolution of the matter perturbations
under the gravitational attraction
leads to the creation of the structures we observe today,
linearly at the beginning and
passing to a non-linear evolution after some time.
The last part of the Universe evolution, finally,
is no more dominated by matter at large scales:
an accelerated expansion of the largest scales was discovered in
the observation of far SuperNovae.
This cannot be the result of a matter dominated phase of the evolution,
but it can be explained assuming that
the Universe entered a Dark Energy (DE) dominated phase
that is responsible of the accelerated expansion.

After this qualitative introduction,
we are going to face in details
some of the calculations that must be deployed
in order to obtain the theoretical
predictions from the Standard Model of Cosmology.
In particular, we are interested in obtaining the predictions for
the power spectra of CMB anisotropies.
In the second part of this Thesis
these predictions will be compared with the various experimental results 
(presented in Chapter~\ref{ch:cosmomeasurements})
and we will
derive constraints on the quantities that describe the Universe.
The goal of this Chapter is to present all the necessary mathematical tools
and to obtain the evolution equation for the perturbations that describe
the Universe.
In Chapter~\ref{ch:cmbr} we will use these results to study in details the
spectrum of the CMB anisotropies and to show how they are influenced by the
various cosmological parameters.
Chapter~\ref{ch:nu}, finally, is devoted to introduce
the neutrinos and their properties, with a particular focus
on their impact in cosmology.

\section{The Expanding Universe}\label{sec:expuniv}
The expansion of the Universe is a very well assessed fact:
at earlier times the distances between us 
and distant galaxies were smaller than today.
The expanding behavior can be described using 
a scale factor $a=a(t)$, where today we have $a_0=a(t_0)=1$
\footnote{We will use the subscript 0 to refer to the today values
of the related quantities.}
and $a(t<t_0)<1$.
Using the scale factor we can define the \emph{comoving distance}
as the physical distance in units of the scale factor.
If two points are at rest in the expanding Universe,
the comoving distance between them 
is constant during the Universe evolution.
On the contrary, the physical distance evolves with time,
since it is proportional to the scale factor.
The comoving distance is used to measure the distances between two points in the \emph{comoving frame}, 
that is the reference frame
where the coordinates of an observer at rest 
do not change during the Universe evolution. 
An observer at rest has constant comoving coordinates and 
evolving physical coordinates, that scale with $a$.

We must also introduce the geometry of the space-time.
There are three possibilities: 
the Universe can be
\emph{flat}, \emph{open} or \emph{closed}.
The flat Universe is an Euclidean Universe, 
where if two particles start to move parallely, 
their motions will be parallel until they travel freely.
In an open (closed) Universe, instead, 
the particles will diverge (converge) during their motion even if they
move parallely at the beginning.
A flat, open or closed Universe has 
null, negative or positive curvature, respectively.
We will see that in General Relativity the geometrical properties 
of the space-time are related to energy: 
when the energy density is equal to the critical density, 
the Universe is flat and its curvature is null. 
Observations suggest that we live in a Universe 
that is flat (or very close to flat).

In the context of General Relativity, 
the expansion history of the Universe can be described
by the time evolution of the scale factor $a(t)$.
The \emph{Hubble factor} $H(t)$ is defined to
encode this time dependency:
\be
 \label{eq:hubblepar}
 H(t)\equiv\frac{\dot a}{a}\;,
\ee
where the dot indicates the derivative with respect to time, 
$\dot a=da/dt$.
It is interesting to measure the value of the Hubble factor today,
$H_0=H(t_0)$: 
this quantity is related to the critical energy density today, 
as we will discuss in Section~\ref{sec:feq}.

The Hubble factor today $H_0$, also called Hubble constant, 
is interesting also for another reason.
Consider two observers that are at rest 
in the comoving frame: 
they are moving away from each other with a velocity that depends 
on the evolution of the scale factor.
At low redshifts, the relative recessional velocity 
of two observers $v$ and their distance $d$ are related 
by the \emph{Hubble law}:
\be
 \label{eq:hubblelaw}
 v=H_0\, d\,,
\ee
where $H_0$ is measured to be about 70\Hou\ (see Section~\ref{sec:h0}), 
or equivalently the dimensionless Hubble constant is $h\simeq 0.7$, 
where $h$ is defined as $h\equiv H_0/(100\Hou)$.
We will discuss in more detail the Hubble law 
in Section \ref{sec:hubble}.

\section{Friedmann-Lema\^itre-Robertson-Walker Metric}\label{sec:flrw}
Under the assumption of the \emph{Cosmological Principle}, 
the most important properties of the Universe are 
homogeneity and isotropy. 
The observations of the galaxy distribution in the Universe 
and of the Cosmic Microwave Background (CMB) radiation
are in strong agreement with the hypothesis of the 
Cosmological Principle at scales larger than 100\Mpc: 
the Universe looks statistically
the same from all the possible points of view, 
in all the possible directions in which it is observed.
These properties corresponds to homogeneity, 
that is invariance under translations, 
and isotropy, that is invariance under rotations.
If we can state that at large scales
there are no privileged positions and directions, 
this is not true at small scales, 
at which the Universe is highly inhomogeneous:
we will need to introduce some perturbations 
to the homogeneous background and study them separately.
The background evolution is important since it gives 
the general behavior of the Universe, 
while all the structures of the visible Universe can be generated 
only by the small perturbations that we will introduce
in Section~\ref{sec:pertUniv}.

Homogeneity and isotropy of the Universe can be encoded 
into a coordinate system where the metric of the space-time 
does not depend on the position (in cartesian coordinates).
In the space-time reference frame described by the coordinates
$x^\mu=(x^0,x^i)$
\footnote{We use the convention that greek letter 
indices span the space-time coordinates ($0,\ldots,3$) and
latin letter indices span the space coordinates ($1,\ldots,3$).},
where 
$x^0=t$ is the time component and $x^i$ are 
the three space components, 
one can write the distance between two points:
\be\label{eq:ds2}
d s^2 \equiv g\munu dx^\mu dx^\nu\;,
\ee
where $d s^2$ is the squared distance between the points 
separated by $dx^\mu$
and $g\munu$ is the metric that describes
the geometrical properties of the space-time.
We use the convention that repeated indices are summed over.

The metric $g\munu$ must be a symmetric $4\times4$ tensor, 
with 4 diagonal and 6 off-diagonal independent components.
The metric for a homogeneous and isotropic Universe is called
Friedmann-Lema\^itre-Robertson-Walker (FLRW) metric. 
If one considers a local observer, general relativity 
can be approximated with the theory of special relativity,
described in the Minkowsky space-time with metric 
$\eta\munu = \diag(-1,+1,+1,+1)$.
The FLRW metric $g\munu$ can be approximated by
$g\munu\simeq\eta\munu$ only locally.
From the isotropy of the Universe we can infer that 
the off-diagonal terms, $g\munu$ with $\mu\neq\nu$, must vanish, 
since there are no privileged directions.
From the property of homogeneity we infer 
that $g\munu$ (in cartesian coordinates)
must be independent on the spatial coordinates, 
since there are no privileged observers.
For a flat Universe, the metric can then be written in the form
\be\label{eq:flrwmetric}
g\munu=\left(%
\begin{array}{cccc}%
-1 & 0 & 0 & 0\\%
 0 & a^2(t) & 0 & 0\\%
 0 & 0 & a^2(t) & 0\\%
 0 & 0 & 0 & a^2(t)
\end{array}
\right) 
\ee
and Eq.~\eqref{eq:ds2} becomes:
\be\label{eq:flrwds2}
ds^2=-dt^2+a^2(t) \delta\iju dx^i dx^j\;,
\ee
where we $\delta\iju=\diag(+1,+1,+1)$ is the Kronecker delta
in an Euclidean space.

To describe a closed or an open Universe, 
it is convenient to use spherical coordinates in the space
and introduce a new parameter: the curvature of the space-time, $k$.
The distance $ds^2$ can be written as
\be\label{eq:flrwkds2}
ds^2=
  -dt^2+a^2(t) 
  \left\{
    \frac{d r^2}{1-k r^2}+r^2
    (d\theta^2+\sin^2\theta d\phi^2 )
  \right\}\,,
\ee
where $(r,\theta,\phi)$ are the usual spherical coordinates.
The curvature is $k=0$ for a flat Universe, $k=+1$ for a closed 
Universe or $k=-1$ for an open Universe.
We will consider now the case of a flat Universe.

Given the metric $g\munu$, it is possible to study the free motion 
of a particle in the space-time.
It is necessary to obtain the \emph{Christoffel symbols}
$\Gamma^\rho\munu$, 
by definition symmetric in the $\mu$ and $\nu$ indices:
\be\label{eq:christoffel}
\Gamma^\rho\munu\equiv
  \frac{g^{\rho\tau}}{2}
  \left(
    \partial_\mu  g_{\nu\tau}+
    \partial_\nu  g_{\mu\tau}-
    \partial_\tau g_{\mu\nu} 
  \right)\;,
\ee
where we introduced the notation 
$\partial_\mu g_{\nu\tau}= \partial g_{\nu\tau}/\partial x^\mu$.
It is worth noting that the Christoffel symbols
are not tensors, since they do not transform in the correct way under
changes in the coordinate system.

The \emph{geodesic} is the trajectory of a particle
in the space-time, in absence of any forces: 
it is the generalized concept of straight line 
in presence of a non-trivial metric.
The Christoffel symbols appear in the \emph{geodesic equation}:
\be\label{eq:geodesic}
\frac{d^2 x^\mu}{d\lambda^2}=
  -\Gamma^\mu_{\alpha\beta}
  \frac{dx^\alpha}{d\lambda}
  \frac{dx^\beta}{d\lambda}\;,
\ee
where $\lambda$ can be any scalar monotonic parameter 
that describes the position on the geodesic, 
for example the conformal time $\eta$ that
we will introduce in Sec.~\ref{sec:hubble}.
To compute the geodesics, one should calculate the components 
of the Christoffel symbols from the metric $g\munu$, 
using the definition in Eq.~\eqref{eq:christoffel}, 
and insert them in Eq.~\eqref{eq:geodesic}.
For a flat Universe with the FLRW metric written 
in cartesian coordinates in Eq.~\eqref{eq:flrwmetric}, 
most of the derivatives of $g\munu$ vanish and most of the components
$\Gamma^\rho_{\mu\nu}$ vanish.
% Recalling that Roman letter indices run from 1 to 3 
% and Greek letter indices from 0 to 3, w
We have:
\begin{align}
\Gamma^0_{0\mu}=\Gamma^0_{\mu0} 
  & =  0\,,
  \label{eq:chr_flrw_1}\\
\Gamma^0\iju 
  & =  \delta\iju\,\dot a\, a\,,
  \label{eq:chr_flrw_2}\\
\Gamma^i_{0j}=\Gamma^i_{j0} 
  & =  \delta\iju\,\frac{\dot a}{a}\,,
  \label{eq:chr_flrw_3}\\
\Gamma^i_{\alpha\beta} 
  & =  0 \hspace{0.5cm}\text{otherwise}.
  \label{eq:chr_flrw_4}
\end{align}

The Christoffel symbols are necessary to define
the \emph{Ricci tensor}, 
symmetric in the indices $\mu$ and $\nu$, 
that we will use to write the Einstein equations:
\be\label{eq:riccitensor}
R\munu\equiv
\partial_\alpha \Gamma^\alpha\munu-
\partial_\nu \Gamma^\alpha_{\mu\alpha}+
\Gamma^\alpha_{\beta\alpha}\Gamma^\beta_{\mu\nu}-
\Gamma^\alpha_{\beta\nu}\Gamma^\beta_{\mu\alpha}\,.
\ee

The trace of the Ricci tensor is named \emph{Ricci scalar}: 
\be\label{eq:ricciscalar}
\mcr\equiv R^\mu_{\,\,\,\mu}=g^{\mu\nu} R\munu,
\ee
where $g^{\mu\nu}=\diag(-1,a^{-1},a^{-1},a^{-1})$
is the inverse of $g\munu$.

In a FLRW Universe the Ricci tensor 
and the Ricci scalar can be easily calculated.
The Ricci tensor is diagonal and its components are
\begin{align}
\label{eq:riccitensFLRW00} R_{00}	
  & = -3\frac{\ddot a}{a}				\,,\\
\label{eq:riccitensFLRWij} R\iju	
  & = \delta\iju \left(2\dot a^2+ a \ddot a\right)	\,,
\end{align}
while the Ricci scalar is simply the trace of the Ricci tensor:
\be\label{eq:ricciscalFLRW}
\mcr=6\left(\frac{\ddot a}{a}+\frac{\dot a^2}{a^2}\right).
\ee
These are the quantities to be used in the Einstein equations,
that we will discuss in the following Section.
After the introduction of the perturbations to the homogeneous and
isotropic Universe, the metric will become more complicate. 
We will discuss the perturbed Universe
in Section~\ref{sec:pertUniv}.

\section{Einstein Equations}\label{sec:eeq}
The evolution with time of the Universe can be derived
from the \emph{Einstein equations}:
\be\label{eq:eeq}
G\munu=8\pi G\, T\munu\,,
\ee
where $G\munu\equiv R\munu -1/2\,\mcr\, g\munu$
is the Einstein tensor and
$G=6.67\e{-11} \mathrm{m}^3\, \mathrm{s}^{-2}\, \mathrm{Kg}^{-1}$
is the Newton constant. 

The symmetric tensor $T\munu$ is the stress-energy tensor, 
that contains all the information about the energy content 
of the Universe.
For a perfect, isotropic and homogeneous fluid, 
it can be written as
\be\label{eq:tmunu_iso}
T\munu=\diag(\rho,p,p,p)\,,
\ee
where $\rho$ and $p$ are the energy density 
and the pressure of the fluid, respectively.
The definitions of $\rho$ and $p$ involve the momentum distribution
function $f$.
Using here the capital letter to denote the momentum $P$,
density and pressure are defined as:
\begin{align}
\rho=
g \int\frac{d^3P}{(2\pi)^3}\, f(P)\, E(P)
\,,\label{eq:rho_singlefluid}\\
p=
g \int\frac{d^3P}{(2\pi)^3}\, f(P)\, \frac{P^2}{3E(P)}
\,,\label{eq:p_singlefluid}
\end{align}
where $g$ is the degeneracy of the species.

Due to conservation laws, the covariant derivatives
of the stress-energy tensor must vanish:
\be\label{eq:covder}
D_\mu T^\mu_{\,\,\nu}\equiv 
\partial_\mu T^\mu_\nu+
\Gamma^\mu_{\alpha\mu}T^\alpha_\nu-
\Gamma^\alpha\munu T_\alpha^\mu=0\,.
\ee
This is the General Relativity equivalent
of the continuity equation %$\partial\rho/\partial t=0$
and
of the Euler equations %$\partial p/\partial x^i=0$ 
in the classical theory.
For the perfect fluid with stress-energy tensor
in Eq.~\eqref{eq:tmunu_iso},
the $\nu=0$ component of Eq.~\eqref{eq:covder} is
\be\label{eq:continuity_singlefluid}
\dot \rho+3\,\frac{\dot a}{a}(\rho+p)=
\dot \rho+3\,\frac{\dot a}{a}(1+w)\rho=
0\,,
\ee
where we used the equation of state $\rho=w p$
for the fluid we are considering.
This equation can be rearranged to obtain the relation
between $\rho$ and $a$ for different fluids:
\be\label{eq:continuity_rho_a}
\dot \rho+3\,\frac{\dot a}{a}(\rho+P)=
a^{-3}\,\frac{\partial(\rho\, a^{3(1+w)})}{\partial t}=
0\,,
\ee
which in turn gives that $\rho\, a^{3(1+w)}$ is constant over time.
Since different fluids have a different equation of state, 
the scaling of the energy density is different during the expansion: 
for radiation, the name used to indicate any relativistic fluid, 
$w=1/3$ and $\rho_r\propto a^{-4}$, while
for non-relativistic matter $w=0$ and $\rho_m\propto a^{-3}$.

Since our Universe is not made of a single perfect fluid, but rather
it is a mixture of different components with different properties,
the fact that the energy densities of different fluids evolve
differently imply the possibility of having different phases 
in the Universe history.
The Big Bang model predicts an initial radiation
dominated phase, when all species were relativistic, 
followed by a matter dominated phase, when
most of the species become non-relativistic 
and their total energy density diminishes more slowly
than the radiation energy density.
Moreover, observations show that in the recent history
the Universe expansion is accelerated, 
thus suggesting a new phase in the evolution. 
The current phase cannot be a radiation dominated
or a matter dominated phase, 
since these components do not give an accelerated expansion: 
it is necessary to introduce then something like
a cosmological constant $\Lambda$, 
which has a negative pressure:
the corresponding equation of state parameter is $w=-1$
and $\rho_\Lambda$ is constant over time
(see Eq.~\eqref{eq:continuity_rho_a}).
It is possible to include the cosmological constant
in the stress-energy tensor and consider it as a new fluid.
If today the Universe is in a $\Lambda$-dominated phase, 
the expansion is accelerated: 
this can be seen from the solutions of the Einstein Equations, 
in particular from the solution of the time-time component,
that we are going to treat.

\section{Friedmann Equations}\label{sec:feq}
If we insert the Eq.~\eqref{eq:tmunu_iso}
into Eq.~\eqref{eq:eeq}, for a FLRW Universe 
where the Ricci tensor and the Ricci scalar are those written in 
Eqs.~\eqref{eq:riccitensFLRW00},~\eqref{eq:riccitensFLRWij}
and \eqref{eq:ricciscalFLRW},
we obtain two different independent differential equations,
corresponding to the 00 and the $ii$ component of the tensor equation.
They are the so-called Friedmann Equations:
\begin{align}
\label{eq:freq1} H^2=
  \left(\frac{\dot a}{a}\right)^2	
  &=	\frac{8\pi G}{3}\,\rho		,\\
\label{eq:freq2} \dot H+H^2 =
  \frac{\ddot a}{a}	
  &=	-\frac{4\pi G}{3}\,(\rho+3p)	,
\end{align}
where $\rho=\sum \rho_i$ and $p=\sum p_i$ are
the total energy density and pressure of the Universe, respectively.
The total density and pressure
include the contributions from all the existing species: 
photons, 
baryons, 
dark matter (DM), 
cosmological constant,
neutrinos.
At different times, some of these species contribute as
relativistic components, being referred to as radiation,
(baryons and neutrinos before the non-relativistic transition, photons),
or as non-relativistic components,
falling into the category of matter (baryons
and neutrinos after the non-relativistic transition, DM%
\footnote{In the very early Universe, also
DM may have been relativistic,
thus accounting as radiation, but this depends on the specific model.}).
The cosmological constant component ($w=-1$) can be described
by some unknown species that contributes with a negative pressure.
It is also possible that there is some fluid that 
contributes with a negative pressure
but does not have a constant equation of state $w=-1$.
In this case the component that substitutes the cosmological constant
is usually referred to as Dark Energy (DE) and it
can have a generic $w<-1/3$, required to have an accelerated expansion,
with a possible dependence $w(t)$.
Moreover, if the Universe is not flat, the curvature $k$
can be described as an additional functional 
fluid in the Friedmann Equations:
one can compute the Ricci tensor for a curved FLRW Universe,
obtaining an additional term in Eq.~\eqref{eq:freq1}.
This can be considered as the contribution of the curvature fluid,
described by an energy density $\rho_k=-3k/(8\pi G a^2)$
and an equation of state $w_k=-1/3$.

From Eq.~\eqref{eq:freq1} one can define the
\emph{critical energy density}:
\be\label{eq:criticalrho}
\rho_c(t)\equiv\frac{3 H(t)^2}{8\pi G},
\ee
which is the total energy density of a flat Universe at a given time. 
Its value today, $\rho_c^0$,
depends only on the current value of the Hubble parameter $H_0$.
Using the critical density we can define
the \emph{density parameter} as 
the ratio between the absolute energy density $\rho$
and the critical density $\rho_c$,
for each different species $i$:
\be\label{eq:densityparam}
\Omega_i\equiv\frac{\rho_i}{\rho_c}\,,
\ee
where, for example, $i=\Lambda,k,m,r$
for cosmological constant, curvature, matter and radiation.
In term of the density parameters of the different species, 
the first Friedmann Equation becomes:
\be\label{eq:friedmann_eq_densparam}
H^2=
  H_0^2 (
    \Omega_\Lambda^0+
    \Omega_k^0 a^{-2}+
    \Omega_m^0 a^{-3}+
    \Omega_r^0 a^{-4}),
\ee
where we used the results of Eq.~\eqref{eq:continuity_rho_a}
for the different fluids.

As an example,
the matter contribution at the time of matter-radiation equality
takes into account baryons and charged leptons %
% \footnote{In Cosmology, it is common to denote with the name ``baryons''
% all the species that are electromagnetically coupled to the photons:
% hadrons and charged leptons.
% Even if they are not baryons, hence, the name includes also electrons.}
plus
the DM component that was non-relativistic at decoupling,
% \footnote{Decoupling for each species is when the interaction rate
% of the given species drops below the expansion rate of the Universe
% and corresponds to the fact that the species stop interacting with the rest
% of the plasma.},
named Cold Dark Matter (CDM),
and eventually other non-relativistic species,
such as massive neutrinos after their non-relativistic transition.
At least two neutrinos, in fact, must have small but non-zero masses,
whose values are currently unknown.
The neutrino mass is requested to explain the flavor oscillations,
that we will discuss in Chapter~\ref{ch:nu}.
The consequence is that at different times each neutrino
can contribute to $\Omega_r$ or to $\Omega_m$, depending on its mass:
a relativistic neutrino is considered radiation, 
while a non-relativistic neutrino accounts as matter.
Each massive neutrino, hence,
can account as radiation in the early Universe
and as matter when it becomes non-relativistic in the late Universe.
Eventually, if there are very light massive neutrinos
($m_\nu\lesssim T_\nu^0$), some of them can
be still relativistic today.
The correct behavior at all times must be evaluated numerically
and the non-relativistic transition of each neutrino can leave an imprint
on the cosmological observables.
We will discuss the neutrino effects in cosmology
in Section~\ref{sec:nucosmology}.

From Eq.~\eqref{eq:friedmann_eq_densparam},
the most important lesson we can learn is
that the evolution of the Universe depends
on the relative amounts of energy density 
corresponding to each fluid. 
At different times, one of the contributions is usually dominant
and the evolution rate $H=\dot a/a$ has a different behavior.
Recently the Planck collaboration determined the density parameters
for the different fluids, using the CMB measurements
of the Planck satellite \cite{Adam:2015rua,Ade:2015xua}:
these determinations tell us that we have approximately
$\Omega^0_\Lambda\simeq0.69$ for the cosmological constant, 
$\Omega^0_c\simeq0.26$ for the CDM, 
$\Omega^0_b\simeq0.05$ for the baryons and 
$\Omega^0_r\simeq10^{-5}$ for the relativistic components.
Thus
the cosmological constant gives 
the main contribution to the total energy density
and the Universe is in a $\Lambda$-dominated ($\Lambda$D) phase.
If we go back in time, however,
while $a$ decreases other contributions in
Eq.~\eqref{eq:friedmann_eq_densparam} start to dominate,
due to their different evolution with $a$:
before the $\Lambda$D phase there was a matter-dominated (MD) phase, 
while at the beginning of the evolution
the larger energy density was $\Omega_r$ and the Universe 
was in a radiation-dominated (RD) phase.
Even if from Eq.~\eqref{eq:friedmann_eq_densparam} we can expect also
a curvature-dominated phase, the current analyses show that
the Universe is almost flat, and we will neglect the possibility that
the space-time is open or closed.
The constraint of the Planck collaboration on the curvature 
is $\Omega^0_k=0.000\pm0.005$ \cite{Ade:2015xua}.

If we consider $a=1$ in Eq.~\eqref{eq:friedmann_eq_densparam}, finally, 
we obtain the following relation between all the density parameters:
\be\label{eq:sumdensityparam}
\Omega_\Lambda^0+\Omega_k^0 +\Omega_m^0 +\Omega_r^0 = 1.
\ee

We conclude defining the dimensionless quantity $\omega_i=\Omega_i h^2$,
where $h$ is the reduced Hubble parameter and $i$ indicates all the possible
fluids.
The dimensionless density parameter $\omega_i$ is proportional to the 
physical density of the component $i$ at present time and we will use it
in the following Chapters.

Coming back to the second Friedmann equation,
we can rewrite Eq.~\eqref{eq:freq2} evaluated today
in terms of the 
\emph{deceleration parameter}, named $q_0$:
\be\label{eq:deceleration}
q_0\equiv -\left(\frac{\ddot a}{a}\right)_{t=t_0} \frac{1}{H_0^2}\,,
\ee
that is positive for a decelerated expansion
and negative for an accelerated expansion of the Universe.
Using the equation of state of the different fluids
and the definition of $H_0$, 
it is possible to write:
\be\label{eq:decel_omega_w}
q_0=\frac{1}{2}\sum_i \Omega_i^0 \,(1+3w_i)\,.
\ee
If the cosmological constant $\Lambda$
or any other fluid with $w<-1/3$ dominates, 
$q_0$ can be negative, corresponding to an accelerated expansion.

\section{The Hubble Law and Distance Measurements}\label{sec:hubble}
One of the most difficult measurements
in the Universe are distance estimations.
A fundamental distance is the comoving distance,
that is the distance of two points
in the comoving frame and does not depend on the scale factor.
The physical distance, instead,
depends on the comoving distance and on the evolution history.

One important quantity is
the distance that light can have traveled since $t=0$.
Since in a time $dt$ light can travel a distance $dx=dt/a$,
the total comoving distance is
\be\label{eq:causalhorizonsize}
\eta=\int^t_0 \frac{dt}{a(t)}\,,
\ee
that is the maximum distance at which information
can be propagated in a time $t$, 
in the comoving frame: 
regions separated by distances greater
than $\eta$ are not causally connected.
We can think to $\eta$ as the size of the \emph{comoving horizon}.
As it is a monotonically increasing variable,
$\eta$ can be considered as a \emph{conformal time},
that describes the photon path
and can be used conveniently in place of the time $t$ 
in a number of calculations we will discuss in the following.
The corresponding physical distance,
that is the farthest distance we can observe today, 
is called the \emph{horizon distance}:
\be\label{eq:horizondistance}
d_H(t_0)=a(t_0)\int^t_0 \frac{dt}{a(t)}\,.
\ee
where $a(t_0)=1$ in the usual convention.
Points separated by a distance greater
than the horizon distance are not in causal contact.

Using the FLRW metric in polar coordinates in Eq.~\eqref{eq:flrwkds2}, 
the \emph{physical distance} among two objects
at a time $t$ can be written as
\be\label{eq:physdist}
d_p(t)=a(t) \int_0^r \frac{dr}{\sqrt{1-kr^2}}\,,
\ee
that for a flat Universe ($k=0$) becomes
\be\label{eq:d_ar}
d_p(t)=a(t)\, r\,.
\ee
In absence of peculiar motions in the comoving frame,
i.e. if $\dot r=0$,
the relative velocity between the considered objects
depends on their distance:
\be\label{eq:hubbleVelocity}
v\equiv\dot d_p=\dot a(t) \,r =H(t)\, d_p\,.
\ee
When $t=t_0$ we obtain the \emph{Hubble Law}:
\be\label{eq:hubbleLaw}
v=H_0\, d_p\,,
\ee
which tells us that the relative velocity
is higher for distant objects and
it is a strong probe of the expansion of the Universe.

To measure the Hubble parameter $H_0$,
one should obtain the distance and the velocity.
The latter one is straightforward
since it can be related to the redshift, $z$.
Due to cosmic expansion,
the light emitted by a distant observer is stretched while traveling
towards us, since the emitter is receding with respect to us.
It is convenient to define this stretching
of the wavelength of the emitted light in term
of the redshift $z$:
\be\label{eq:redshift}
1+z\equiv \frac{\lambda_{o}}{\lambda_{e}}=\frac{a(t_{o})}{a(t_{e})}\,,
\ee
that can be interpreted as a Doppler effect
between two objects with a relative velocity.
Subscripts $o$ and $e$ refer
to the observer and the emitter, respectively.
Usually the observer corresponds
to an experiment performed today on Earth 
and consequently the redshift is related
to the scale factor $a_e=a(t_e)$ at the emission time $t_e$, 
since $a(t_0)=1$:
\be\label{eq:redshift0}
1+z=a_e^{-1}\,.
\ee
In General Relativity, however,
the stretching of the wavelengths
does not arise only from something equivalent 
to the Doppler effect that occurs for the acoustic
and electromagnetic waves,
% that can be used only at small redshifts,
but also from the Universe expansion, that dilutes the photon energy
in a larger portion of space.
Moreover, the photons may be redshifted (or blueshifted)
by changes in the space-time properties
or in the gravitational potential
along the photon path: 
a photon is redshifted when exiting
a region with large gravitational potential
and it is blueshifted when leaving
a region with small gravitational potential.
% We will discuss this effect in more detail
% when treating the CMB anisotropies.

The most difficult part of the process to determine $H_0$
is the determination of the distance $d_p$.
The redshift can be used to connect the physical distance $d_p(t_0)$
and the \emph{luminous distance} $d_L$ of an object.
The luminous distance $d_L$ is defined as the distance
at which an observer $P_0$ at $t=t_0$ measures a flux $f$ 
from a source $P$, emitting a power $L$ in light:
\be\label{eq:lumindist}
d_L=\left(\frac{L}{4\pi f}\right)^{1/2}\,.
\ee
The spherical surface centered in $P$ and passing through $P_0$
at a time $t_0$ has an area $4\pi a_0^2 r^2$.
Since the expansion causes the photon to be redshifted
by a factor $a_0/a_e$ during the travel,
we can derive the relation between the luminous
and the physical distance:
\be\label{eq:luminousDistance}
d_L=\frac{r}{a_e}=(1+z)\,d_p\,,
\ee
where we used Eq.~\eqref{eq:d_ar} at $t=t_0$.

Determinations of the luminous distance are complicated
by the fact that we usually do not know 
the magnitude of the power $L$ for a given astrophysical object.
This is not true for particular objects,
that are supposed to behave as \emph{standard candles}:
they have always the same luminosity and we can obtain
their luminous distance simply measuring at Earth
the flux they produce.
Commonly used standard candles are, for example,
the Cepheids variable stars, since
their intrinsic brightness is related to the period of variation.
Other standard candles are type Ia SuperNovae (SN Ia),
which have always the same emission power since they originate
in a standard way:
when one of the two elements in a binary system is a white dwarf,
it can gradually accrete mass from the binary companion.
If the mass of the white dwarf is sufficient, during the accretion
the core can reach the ignition temperature for the carbon fusion.
At this point, a large part of the matter in the white dwarf undergoes
a runaway reaction, releasing enough energy
to unbind the star in a supernova explosion.

Another method to determine distances is to consider
the angular size $\delta\theta$ of a given object of length $l$,
aligned perpendicularly to the line of sight.
Its \emph{angular diameter distance}, $d_A$, is
\be\label{eq:angdiamdist}
d_A=\frac{l}{\delta\theta}
\ee
and it can be related to the physical distance through:
\be\label{eq:da_dl}
d_A=\frac{d_L}{(1+z)^2}\,.
\ee
As for the luminous distance,
determinations of the angular diameter distance suffer
the fact that it is difficult to know the size $l$ of generic objects.
In the context of cosmological observations, 
the angular diameter distance is especially used to study
the separation distance of the galaxies.
In fact, due to Baryon Acoustic Oscillations (BAO),
generated by the balance of the gravitational potential and the
radiation pressure between photons and baryons in the early Universe, 
there is a preferred separation distance between galaxies.
Since this typical distance depends on the evolution properties,
it can be used to constrain the cosmological parameters.
We will discuss BAO results in detail
in Section~\ref{sec:bao}.

\section{Boltzmann Equation}
\label{sec:boltzeq}
In the hot and dense primordial Universe, 
the interactions among particles were much more frequent than today 
and the species were maintained in equilibrium in most of the cases.
During the cooldown of the Universe, 
due to a decrease of the particle number densities
caused by the expansion, 
at certain times interactions were not able 
to maintain the chemical and thermal equilibrium between the 
involved species: most of the species 
decoupled from the rest of the primordial plasma at the
corresponding decoupling time.
This is a result arising from non-equilibrium phenomena, 
encoded in the \emph{Boltzmann equation},
which formalizes the fact that the rate of variation
for a given species is 
the difference between the production and annihilation rates.

We want to describe now the Boltzmann equation in a simple case.
Suppose we are interested in the number density 
of a species 1, $n_1$.
Let us assume that the species 1 is non-relativistic.
Suppose also that the only process involving the species 1 
is its annihilation with another species 2, 
during which elements of the species 3 and 4 are produced.
The inverse process must be considered as well.
The interaction is then summarized by $1+2\leftrightarrows 3+4$.
Under these assumptions, 
the integrated Boltzmann equation in the expanding Universe is:
\begin{align}
a^{-3}\frac{d(n_1 a^3)}{dt} = &
  \int\frac{d^3 p_1}{2 E_1 (2\pi)^3}
  \int\frac{d^3 p_2}{2 E_2 (2\pi)^3}
  \int\frac{d^3 p_3}{2 E_3 (2\pi)^3}
  \int\frac{d^3 p_4}{2 E_4 (2\pi)^3}
\nonumber \\
& \times (2\pi)^4 
  \delta(E_1+E_2-E_3-E_4)
  \delta^3(p_1+p_2-p_3-p_4)
  |\mcm|^2
\nonumber \\
& \times
  \{
  f_3 f_4 (1\pm f_1) (1\pm f_2) -
  f_1 f_2 (1\pm f_3) (1\pm f_4)
 \}\,.
\label{eq:boltzmannequation}
\end{align}
In the previous equation $n_i$, $f_i$, $p_i$ and $E_i$ are 
the number density, the distribution function, 
the momentum and the energy of the species $i$.
In the last line, the plus sign is for bosons 
and the minus sign is for fermions:
the terms $(1\pm f_i)$ represent the phenomena of
Bose enhancement and Pauli blocking, respectively.
In the absence of interactions, Eq.~\eqref{eq:boltzmannequation}
says that the density times the scale factor to the third
is conserved:
this is a consequence of the expanding Universe, 
and number densities of the particles scale with $a^{-3}$.
The interaction is encoded in the matrix element $\mcm$
in the second line of the right-hand side
and the last line tells us that
the production rate of the particle 1 is proportional 
to the abundance of the particles 3 and 4, $f_3$ and $f_4$,
while the disappearance rate is proportional to the abundances
of the particles 1 and 2, $f_1$ and $f_2$.
The Dirac delta functions in the second line 
give the four-momentum conservation.
Finally, the integrals sum over all the possible momenta:
either the matrix element and the distribution functions,
even if not explicitly written, 
depend on the particle momenta.

Equation~\eqref{eq:boltzmannequation} refers to the particle 1,
but corresponding equations hold for the other particles.
In practice, 
the \emph{kinetic equilibrium} 
is typically enforced by the interactions, since
scattering processes are fast enough to make all the particles
have a distribution that is close to a Bose-Einstein or a Fermi-Dirac.
This simplifies a lot the calculation.
All the uncertainty
in the correct form of the distribution of each species is encoded in 
a single function of time $\mu$, that is the chemical potential
if annihilations process are also in equilibrium.
In this case we can write
\be\label{eq:be-fd}
  f_j=\frac{1}{e^{(E_j-\mu_j)/T_j}\pm1}\,,
\ee
where $-1$ is for bosons and $+1$ is for fermions.
Since we are interested in temperatures smaller than $E-\mu$,
the $\pm1$ term in the denominator
is much smaller than the exponential and
the distribution functions can be approximated with:
\be\label{eq:mb}
  f_j\simeq e^{\mu/T} e^{-E/T}\,.
\ee

With this approximation, we show now that
the last line in Eq.~\eqref{eq:boltzmannequation}
can be simplified.
The number density of a species is defined as
\be\label{eq:numdens}
  n_i=g_i e^{\mu_i/T}\int\frac{d^3p}{(2\pi)^3}e^{-E_i/T}\,,
\ee
where $g_i$ is the degeneracy of the species $i$.
The equilibrium number density can be written under the approximation
of $m_i\ll T$ (relativistic) or $m_i\gg T$ (non-relativistic):
\be\label{eq:numdens_eq}
  n_i\zeroord=\left\{
  \begin{array}{ll}
    g_i\frac{T^3}{\pi^2}
      & \text{ for }m_i\ll T \\
    g_i\left(\frac{m_i T}{2\pi}\right) e^{-m_i/T}
      & \text{ for }m_i\gg T 
  \end{array}
  \right.\,.
\ee
% where at equilibrium $\mu_i=0$.
The out of equilibrium expression is then $n_i=n_i\zeroord e^{\mu_i/T}$:
using Eq.~\eqref{eq:mb} and this last expression
we can rewrite the last line of Eq.~\eqref{eq:boltzmannequation} as
\be\label{eq:boltzeq_lastline}
  e^{-(E_1+E_2)/T}
  \left(
  \frac{n_3 n_4}{n_3\zeroord n_4\zeroord}-
  \frac{n_1 n_2}{n_1\zeroord n_2\zeroord}
  \right)\,,
\ee
where we also used the energy conservation condition.

We can define \sigmav, the thermally averaged cross section, as
\begin{align}\label{eq:sigmav}
\sigmav = &
  \frac{e^{-(E_1+E_2)/T}}{n_1\zeroord n_2\zeroord}
  \int\frac{d^3 p_1}{2 E_1 (2\pi)^3}
  \int\frac{d^3 p_2}{2 E_2 (2\pi)^3}
  \int\frac{d^3 p_3}{2 E_3 (2\pi)^3}
  \int\frac{d^3 p_4}{2 E_4 (2\pi)^3}
\nonumber \\
& \times (2\pi)^4 
  \delta(E_1+E_2-E_3-E_4)
  \delta^3(p_1+p_2-p_3-p_4)
  |\mcm|^2\,.
\end{align}
This definition allows us to rewrite
in a more compact way the Boltzmann equation:
\be\label{eq:boltzmannequation_short}
  a^{-3}\frac{d(n_1 a^3)}{dt} = 
  n_1\zeroord n_2\zeroord
  \sigmav
  \left(
  \frac{n_3 n_4}{n_3\zeroord n_4\zeroord}-
  \frac{n_1 n_2}{n_1\zeroord n_2\zeroord}
  \right)\,.
\ee
This last expression is an ordinary differential equation for $n_1$
that can be applied to study
the freeze-out of DM.
Similar calculations can be exploited to derive the Boltzmann equations
needed to solve different scenarios, such as
the BBN and 
the recombination, that corresponds to 
the electron-photon decoupling and gives rise
to the last scattering surface.
The calculations in these two cases would be slightly different,
since the approximations we adopted here for non-relativistic species
are not valid for all the species involved in the different processes.

Through the Boltzmann equation it is possible to write the equilibrium
distributions and track the evolution into the out-of-equilibrium
phases for each species.
For stable particles, the distribution function after decoupling
evolves simply following the expansion history.
Tracking the full evolution it is then possible to obtain
the relic DM density today
or, using the corresponding Boltzmann equations,
the abundances of the light nuclei produced in the early Universe
and the isotropic photon distribution at the last scattering,
that evolved into the isotropic part of the CMB 
that we can observe today.
We will not treat the applications of this unperturbed Boltzmann equation
in detail:
we suggest 
Refs.~\cite{Dodelson-Cosmology-2003, %sez 3.2bbn, 3.3cmb, 3.4DM
Kolb:1990vq}
to the interested reader.
In the next Section, however, we will present the perturbed treatment
that is used to obtain the Boltzmann equation for the photon perturbations,
necessary to calculate the expected spectrum of the CMB anisotropies.
Before this, however, we must update our treatment to include the perturbations
of the space-time metric and of the distribution functions for each species.

\section{The Perturbed Universe}\label{sec:pertUniv}
\subsection{Metric}\label{ssec:perturbUniv}
If we look at the Universe near us today 
we have the immediate impression that
the hypothesis of the \emph{Cosmological Principle} we introduced 
at the beginning of this Chapter cannot be valid
at small scales.
At short distances the Universe is not homogeneous and isotropic,
with the direct consequence that the results we presented 
up to now are 
just approximations of the full solutions for the evolution.
To describe the perturbed Universe,
it is possible to define a perturbed metric,
that is no more characterized by one single function of time ($a$),
but it depends on two additional functions $\Psi$ and $\Phi$,
both of which are functions of space and time.
The perturbations are described by $\Psi$, 
that corresponds to the Newtonian potential,
and by $\Phi$, that describes the perturbations to the spatial curvature.
We will treat them as small quantities,
using series expansions truncated at the first order
and neglecting second order terms.
To write the perturbed metric, we must choose a \emph{gauge}, because
there is some freedom in selecting the variables used to describe
the fluctuations.
The physical results are insensitive to the gauge choice, 
but the complexity of the calculation can vary from gauge to gauge.
In the \emph{conformal Newtonian} gauge, the perturbed metric is
\begin{align}
  g_{00}(\vec x, t) & = 
    -1-2\Psi(\vec x, t) \label{eq:pertMetric0}\\
  g_{0i}(\vec x, t) & = 
    0 \label{eq:pertMetric1}\\
  g_{ij}(\vec x, t) & = 
    a^2(t)\,(1+2\Phi(\vec x, t))\, \delta_{ij}
    \label{eq:pertMetric2}\,.
\end{align}
We adopt the sign convention that 
positive $\Psi$ and negative $\Phi$ correspond to underdense regions,
while
negative $\Psi$ and positive $\Phi$ correspond to overdense regions.

We limit ourselves to the
treatment of the scalar perturbations in the metric
and we neglect the other possibilities:
vector and tensor perturbations.
The former ones arise from the generalization of a rotational fluid,
producing vortex motions that rapidly decays.
They are not predicted by the standard cosmological model.
The latter ones, instead, 
describe the contribution of tensor components,
such as gravitational waves.
These additional components would require additional functions 
to be parameterized:
we will not discuss all the details and we will
only mention some of the main results.

We want now to derive the Einstein Equations 
in the perturbed Universe. To do this, we must first calculate 
the Christoffel symbols, to get the Ricci tensor and the Ricci scalar.
Let us look at the first order terms in the Christoffel symbols,
starting from $\Gamma^0\munu$:
\be\label{eq:cs_0mn0}
  \Gamma^0\munu=\frac{1}{2}g^{0\alpha}
  (
  \partial_\nu g_{\alpha\mu}+
  \partial_\mu g_{\alpha\nu}-
  \partial_\alpha g\munu
  )\,,
\ee
where the only nonzero component of $g^{0\alpha}$ is 
$g^{00}=-1+2\Psi$ and we can write
\be\label{eq:cs_0mn1}
  \Gamma^0\munu=\frac{-1+2\Psi}{2}
  (
  \partial_\nu g_{0\mu}+
  \partial_\mu g_{0\nu}-
  \partial_0 g\munu
  )\,.
\ee
Neglecting the second order terms, we get
\begin{align}\label{eq:pertChrisTime}
  \Gamma^0_{00} 
    &= \partial_0\Psi\\
  \Gamma^0_{0i}=\Gamma^0_{i0}
    &= \partial_i\Psi=i k_i\Psi\\
  \Gamma^0_{ij}
    &= \delta_{ij} a^2 [H+2H(\Phi-\Psi)+\partial_0\Phi]\,,
\end{align}
Using the metric in Eqs.~\eqref{eq:pertMetric0} to \eqref{eq:pertMetric2}
we can calculate also the other Christoffel symbols:
\begin{align}\label{eq:pertChrisOther}
  \Gamma^i_{00} 
    &= \frac{\partial_i\Psi}{a^2} \\
  \Gamma^i_{0j}=\Gamma^i_{j0}
    &= \delta_{ij}(H+\partial_0\Phi) \\
  \Gamma^i_{jk}
    &= [
      \delta_{ij}\partial_k+
      \delta_{ik}\partial_j+
      \delta_{jk}\partial_i
    ]\Phi \,,
\end{align}

We can also express all these equations in the Fourier space,
simply replacing $\partial_i$ with $i k_i$ and each quantity 
with its Fourier transformed, such as $\Psi$ with $\widetilde\Psi$.
The Fourier convention we adopt is:
\be\label{eq:FourierConv}
  A(\vec x)=
  \int
  \frac{d^3k}{(2\pi)^3}
  e^{i\vec k\cdot\vec x}\;
  \widetilde A(\vec k)\,.
\ee
We will mostly work in Fourier space from now on, and we will neglect
the $\sim$ notation for all the quantities when it will be clear that
the quantities will be in the Fourier space.

The calculation of the Ricci tensor is a mechanical process
that requires the Christoffel symbols and some algebra.
The results are:
\begin{align}\label{eq:ricciTensPert0}
  R_{00} 
    &=
      -3\frac{\ddot a}{a}
      -\frac{k^2}{a^2}\Psi
      -3\partial_0^2\Phi
      +3H\partial_0(\Psi-2\Phi)
    \\
  R_{ij}
    &=
    \delta_{ij}
    \left[
      \left(
	2a^2H^2+a \ddot a
      \right)
      (1+2\Phi-2\Psi) \right.\nonumber \\
    &  \left. 
      +a^2 H \partial_0(6\Phi-\Psi)
      +a^2 \partial_0^2\Phi
      +k^2 \Phi
    \right]
    +k_i k_j(\Psi+\Phi)
    \label{eq:ricciTensPert1}
    \,,
\end{align}
where we adopted $k^2=\delta_{ij}k^i k^j$.

The contraction of the Ricci tensor with the metric gives the 
perturbed Ricci scalar:
\begin{align}
  \mcr = &
  (-1+2\Psi)\left(
    -3\frac{\ddot a}{a}
    -\frac{k^2}{a^2}\Psi
    -3\partial_0^2\Phi
    +3H\partial_0(\Psi-2\Phi)
  \right)\nonumber \\
  & +\left(
    \frac{1-2\Phi}{a^2}
  \right)
  \left\{3\left[\left(
    2a^2H^2+a\ddot a
    \right)\left(
    1+2\Phi-2\Psi
  \right)\right.\right.\nonumber\\
  &\left.\left.
    +a^2H\partial_0(6\Phi-\Psi)
    +a^2\partial_0^2\Phi
    +k^2\Phi
  \right]
  +k^2(\Phi+\Psi)
  \right\}\,,
  \label{eq:pertRicciScal}
\end{align}
that becomes Eq.~\eqref{eq:ricciscalFLRW} at zero-order
when $\Psi$ and $\Phi$ vanish.
The first-order part is:
\begin{align}
  \delta\mcr = &
  -12\Psi\left(
    H^2+\frac{\ddot a}{a}
  \right)
  +2\frac{k^2}{a^2}\Psi
  +6\partial_0^2\Phi
  \nonumber \\
  & 
  -6H\partial_0(\Psi-4\Phi)
  +4\frac{k^2}{a^2}\Phi
  \,.
  \label{eq:pertRicciScalFO}
\end{align}

To write the Einstein Equations in terms of the perturbed quantities
we will start from Eq.~\eqref{eq:eeq},
but we have to deal with the perturbed stress-energy tensor, before.
To obtain the first-order part of the stress-energy tensor, however,
it is necessary to study the first-order terms
in the distribution function $f_i$ for the different species,
generalizing the treatment of the Boltzmann equation in
Section~\ref{sec:boltzeq} to the case of a non-homogeneous Universe.

\subsection{Boltzmann Equations}\label{ssec:pertBoltzEq}
We discussed in Section~\ref{sec:boltzeq} the integrated version
of the Boltzmann Equation in the context of the homogeneous and
isotropic Universe, but we are now interested in the 
anisotropies of the distribution of cosmic photons
for the CMB observations and
in the inhomogeneities of the matter distribution,
that originated the structures in the current Universe
through the gravitational evolution.
These perturbed distributions are difficult to calculate,
since in the hot plasma before CMB decoupling
photons interact mainly with electrons through the Compton scattering
and electrons are coupled to protons.
Moreover, all the mentioned species, plus neutrinos and DM, 
are coupled to gravity.
Therefore,
it is necessary to solve simultaneously the Boltzmann equation
for each component,
to obtain the distribution functions $f_i$
for all the species.
The Boltzmann equation in its differential form can be schematically
written as
\be\label{eq:boltzeq_diff}
  \frac{df}{dt}=\mcc[f]\,,
\ee
where $\mcc$ contains all the possible collision terms.
For a non-interacting species this equation reduces to $df/dt=0$,
that is nontrivial to solve since the phase space elements
change with time, as a consequence of the nontrivial metric.

We want now to write the Boltzmann equation for photons.
It is convenient to express the total derivative in
Eq.~\eqref{eq:boltzeq_diff} as a sum of partial derivatives.
The momentum vector is defined as
\be\label{eq:momentum}
  P^\mu=\frac{dx^\mu}{d\lambda}\,,
\ee
where $\lambda$ is a monotonic
parameter that describes the particle path.
Since the photon is massless, 
\be\label{eq:momentumMassless}
 P^2=g\munu P^\mu P^\nu=0
\ee
and
there are only three independent components of $P^\mu$.
Defining the generalized magnitude of the momentum 
$p^2=g_{ij}P^i P^j$%
\footnote{Since in this section we will not need to denote the pressure,
we use $p$ to indicate the generalized magnitude of the momentum.}
, we can eliminate the time component
of $P^\mu$, using the metric in 
Eqs.~\eqref{eq:pertMetric0}-\eqref{eq:pertMetric2}:
\be
 P^0=\frac{p}{\sqrt{q+2\Psi}}\simeq (1-\Psi)\,p\,,
\ee
that is the perturbed version of $E=pc$ and 
it can be used to eliminate $P_0$ in favor of $p$.
From this equation we learn also that photons lose energy 
when exiting an overdense region.
Now we can re-express the total derivative in
Eq.~\eqref{eq:boltzeq_diff}:
\be\label{eq:df_dt}
  \frac{df}{dt}=
    \frac{\partial f}{\partial t}+
    \frac{\partial f}{\partial x^i}\cdot\frac{dx^i}{dt}+
    \frac{\partial f}{\partial p}\cdot\frac{dp}{dt}+
    \frac{\partial f}{\partial \hat p^i}\cdot\frac{d\hat p^i}{dt}\,,
\ee
where $\hat p^i$ is the direction of $P^i$.
The last term of this expression is at second order 
in the perturbations, since $f$ does not depend on $\hat p^i$ at 
zero order and in absence of the potentials $\Psi$ and $\Phi$
the photon goes straight,
hence $d\hat p^i/dt$ is also a first order term.

We can rewrite the second term:
\be\label{eq:dxi_dt}
  \frac{dx^i}{dt}=\frac{dx^i}{d\lambda}\frac{d\lambda}{dt}
  =\frac{P^i}{P_0}\,,
\ee
where we used the definition of $P^\mu$.
Since $P^i=C \hat p^i$ and $p^2=a^2(1+2\Phi)C^2$
(from the definition of $p^2$), we can always write
\be
 P^i=p \,\hat p^i\frac{1-\Phi}{a}\,,
\ee
and from Eq.~\eqref{eq:dxi_dt} we obtain
\be\label{eq:dxi_dt_fin}
 \frac{dx^i}{dt}=\frac{\hat p^i}{a}(1+\Psi-\Phi)\,.
\ee
For an overdense region the term in parentheses is less than one,
meaning that photons slow down.
Anyhow, in Eq.~\eqref{eq:df_dt} $dx^i/dt$ multiplies 
a first order term, since the momentum distribution at zero order 
does not depend on the position,
and we can neglect the potentials in Eq.~\eqref{eq:dxi_dt_fin}.

The last term we have to deal with is $dp/dt$.
For sake of brevity we do not present the complete calculations,
that involves the Christoffel symbols of the perturbed metric.
It is fully deployed, for example, 
in Ref.~\cite{Dodelson-Cosmology-2003}.
Neglecting the second order terms in $\Psi$ and $\Phi$,
the result is 
\be\label{eq:dp_dt}
  \frac{1}{p}\frac{dp}{dt}=
    -H
    -\frac{\partial \Phi}{\partial t}
    -\frac{\hat p^i}{a}\frac{\partial \Psi}{\partial x^i}\,.
\ee
The change in the photon momentum is described by
a term accounting for the momentum loss due to Hubble expansion ($H$)
plus two terms that depend on the perturbations:
if a photon is traveling in a deepening gravitational well
from one side it loses energy since the curvature is increasing
($\partial \Phi/\partial t$),
but from the other side
it gains energy because it is pulled towards the center
($\hat p^i \cdot\partial \Psi/\partial x^i$).

We can finally write the left-hand term of Eq.~\eqref{eq:boltzeq_diff}:
\be\label{eq:df_dt_fin}
  \frac{df}{dt}=
    \frac{\partial f}{\partial t}+
    p\,\frac{\hat p^i}{a}\cdot\frac{\partial f}{\partial x^i}
    \frac{\partial f}{\partial p}\left(
    H
    +\frac{\partial \Phi}{\partial t}
    +\frac{\hat p^i}{a}\frac{\partial \Psi}{\partial x^i}
   \right)\,.
\ee

The next step requires to expand
the perturbed photon distribution function. 
Following Ref.~\cite{Dodelson-Cosmology-2003}, we define
\be\label{eq:photDistFuncPert}
  f(\vec x, p, \hat p, t)=
  \left[
  \exp\left(
  \frac{p}{T(t)(1+\Theta(\vec x, \hat p, t))}
  \right)
  -1
  \right]^{-1}\,,
\ee
where we expanded the temperature at zero-order
as a function of time only, for the properties of
homogeneity and isotropy of the Universe,
while the perturbations are included in
a small function of space and momentum ($\Theta$).
We can then expand $f$ in terms of the perturbation:
\be
  f=f\zeroord-p\frac{\partial f\zeroord}{\partial p}\Theta\,,
\ee
where $f\zeroord$ is the Bose-Einstein distribution with $\mu=0$
(Eq.~(\ref{eq:be-fd})).

If we set the collision term to zero,
the zero-order term of Eq.~\eqref{eq:df_dt_fin} becomes:
\be
\left.\frac{df}{dt}\right|_{\text{zero order}}=
\frac{\partial f\zeroord}{\partial t}-
Hp\frac{\partial f\zeroord}{\partial p}=0\,.
\ee

For the first order, we have to extract all the terms proportional
to $\Psi$, $\Phi$ or $\Theta$ in  Eq.~\eqref{eq:df_dt_fin},
using the perturbed version of $f$.
The result gives 
\be
\left.\frac{df}{dt}\right|_{\text{first order}}=
-p\frac{\partial f\zeroord}{\partial p}
\left(
  \frac{\partial \Theta}{\partial t}+
  \frac{\hat p^i}{a}\frac{\partial \Theta}{\partial x^i}+
  \frac{\partial \Phi}{\partial t}+
  \frac{\hat p^i}{a}\frac{\partial \Psi}{\partial x^i}
\right)\,.
\ee
We may note that only physical distances ($ax^i$) appear in the 
equation.
The first two terms in the parentheses account for free-streaming,
while the last two terms arise from gravity.

Now we should calculate the collision term
for the processes involving photons.
For the epoch we are interested in,
photons interact only with electrons through Compton scattering.
We skip all the calculations
and we go directly to the final result.
To write the collision term, 
we need to define the \emph{monopole} part of the perturbation
to the distribution function, that is independent
of the direction vector:
\be\label{eq:photonMonopole}
  \Theta_0(\vec x,t)=
  \frac{1}{4\pi}\int
  d\Omega\;\Theta(\hat p, \vec x, t)\,,
\ee
where $\Omega$ is the solid angle element spanned by $p$.
The collision term is then \cite{Dodelson-Cosmology-2003}
\be
  \mcc[f(\vec p)]=
  -p\frac{\partial f\zeroord}{\partial p}
  n_e \sigma_T
  (\Theta_0-\Theta(\hat p)+\hat p\cdot\vec v_b)\,,
\ee
where $n_e$ is the electron number density,
$\sigma_T$ is the Thomson cross section
and $\vec v_b=\vec v_e$ is the baryon
% \footnote{We remember that
% the name ``baryons'' usually refers to all the charged particles,
% including electrons.}
velocity, carried by electrons,
that is small.
In particular, if $\vec v_b$ is negligible the collision term
drives $\Theta$ to $\Theta_0$,
hence all the higher moments are damped
and only the monopole term survives;
if $\vec v_b$ is not negligible, instead, the last term produces
a dipole moment in addition to the monopole.

With these results we can finally write a linear equation for
the perturbations to the photon distribution:
\be\label{eq:photDistPertEvol}
  \frac{\partial \Theta}{\partial t}+
  \frac{\hat p^i}{a}\frac{\partial \Theta}{\partial x^i}+
  \frac{\partial \Phi}{\partial t}+
  \frac{\hat p^i}{a}\frac{\partial \Psi}{\partial x^i}
=
  n_e \sigma_T\;
  [\Theta_0-\Theta(\hat p)+\hat p\cdot\vec v_b]\,.
\ee

It is convenient to move to the Fourier space
and to switch to the conformal time $\eta$.
We can change each time derivative into a conformal time derivative
introducing a $a^{-1}$ factor:
from now on, the overdots will indicate
conformal time derivatives.
The advantage of the Fourier transform is that
all the $\partial/\partial x^i$ becomes $k_i$
and the Fourier amplitudes obey ordinary differential equations.
Moreover, if the background is smooth and
the perturbations are small,
the space dependence is only encoded in the perturbation variables:
the Fourier transform of Eq.~\eqref{eq:photDistPertEvol}
originates a set of uncoupled differential equations for each mode
and the Fourier modes can be evolved independently.
In the case of the CMB perturbations, their smallness persists also
today and the Fourier transforms are extremely useful.
On the contrary, for the matter perturbations the nonlinearities
occur at small scales after some time and the Fourier transforms
lose part of their appeal.

Before writing the Fourier transformed version of 
Eq.~\eqref{eq:photDistPertEvol}, we define some useful quantities.
The cosine of the angle between the photon direction $\hat p$
and the wavenumber $\vec k$ is
\be\label{eq:cosineMu}
  \mu\equiv
  \frac{\vec k\cdot \hat p}{k}\,.
\ee
A photon traveling along the gradient (parallel to $\vec k$) 
corresponds to $\mu=1$, while a photon moving in a direction
where the temperature does not change has $\mu=0$.

The optical depth $\tau$, defined as the integral of the scattering
rate along the line of sight and measuring the total
amount of interactions that a photon experienced between $\eta$
and $\eta_0$, is 
\be\label{eq:opticalDepth}
  \tau(\eta)\equiv
  \int_\eta^{\eta_0} d\eta'\; n_e\sigma_T a\,.
\ee

With these definitions, we have finally the equation for
the evolution of the perturbation
to the photon distribution function:
\be\label{eq:photDistPertFour}
  \dot{\widetilde\Theta}
  +ik\mu\widetilde\Theta
  +\dot{\widetilde\Phi}
  +ik\mu\widetilde\Psi
  =
  -\dot\tau\;
  (\widetilde\Theta_0-\widetilde\Theta+\mu\widetilde v_b)\,,
\ee
where
$\widetilde\Theta$ is defined through Eq.~\eqref{eq:FourierConv}.

With similar calculations, it is possible to derive the corresponding
equations for dark matter and baryon perturbations: 
in these cases we will find also an equation for
the evolution of the velocity.
Naming $f_{dm}$ the momentum distribution of DM, we can define
\begin{align}
  \delta_{dm}(\vec x,t)
  &\equiv\frac{n_{dm} - n_{dm}\zeroord}{n_{dm}\zeroord}
    =\frac{\rho_{dm} - \rho_{dm}\zeroord}{\rho_{dm}\zeroord}
  \,,
  \label{eq:defDelta}
  \\
  v^i_{dm} 
  &\equiv\frac{1}{n_{dm}}\int\frac{d^3p}{(2\pi)^3}f_{dm}
    \frac{p\hat p^i}{E}
  \,,
  \label{eq:defV}
\end{align}
where we used the definition
\be
  n_{dm}\equiv
  \int\frac{d^3p}{(2\pi)^3}f_{dm}\,.
\ee
Due to the tight Coulomb scattering,
overdensities of electrons and protons are forced to a common value.
The same happens for the electron and proton velocity anisotropies,
that are maintained in equilibrium by the interactions.
For the baryons, considering together protons and electrons%
\footnote{Electrons, having a smaller mass than protons,
contribute less to the energy density when they are non-relativistic.},
the same definitions adopted for the quantities $\delta_{dm}$ and $v_{dm}$
can be used to define $\delta_b$ and $v_b$,
just substituting $f_{dm}$ with $f_b$.

With these definitions, the perturbation equations become:
\begin{align}
  \dot{\widetilde\delta}_{dm}+ik\widetilde v_{dm}+3\dot{\widetilde\Phi}&=0\,,
\label{eq:pertDMFourDelta}\\
  \dot{\widetilde v}_{dm}+\frac{\dot a}{a}\widetilde v_{dm}+ik\widetilde\Psi&=0\,,
\label{eq:pertDMFourV}\\
  \dot{\widetilde{\delta}}_b+ik\widetilde v_b+3\dot{\widetilde\Phi}&=0\,,
\label{eq:pertBarFourDelta}\\
  \dot{\widetilde {v}}_b+\frac{\dot a}{a}\widetilde v_b+ik\widetilde\Psi&=
  \dot\tau\,
  \frac{1}{R}\,
  (3i\widetilde{\Theta}_1+\widetilde v_b)
  \,,\label{eq:pertBarFourV}
\end{align}
where we defined the ratio
\be\label{eq:baryonToPhotonsR}
  R=\frac{3\rho_b}{4\rho_\gamma}\,.
\ee
The difference between
Eq.~\eqref{eq:pertDMFourV} and Eq.~\eqref{eq:pertBarFourV}
is a consequence of
the electromagnetic interaction between baryons and photons.
Here we used the definition of the first moment of $\Theta$:
\be
{\widetilde\Theta}_1=i\int_{-1}^1\frac{d\mu}{2}\mu\widetilde\Theta(\mu)\,.
\ee

Forgetting all the $\sim$, the relevant quantities to describe
the perturbations for non-relativistic particles are 
$\delta_{dm}$, $\delta_b$ and $v$, $v_b$:
all of them are functions of $k,\,\eta$.
For relativistic particles, more information is needed:
they have a monopole and a dipole perturbation
(corresponding to $\delta_{dm}$ and $v_{dm}$ for non-relativistic
dark matter),
but all the higher moments as well.
In other words, the photon perturbation $\Theta(k,\mu,\eta)$
(the Fourier transform of $\delta T/T$)
and
its equivalent for neutrinos $\mcn(k,\mu,\eta)$ 
(defined in Chapter~\ref{ch:nu})
depend also on the propagation direction.
The general definition of the higher moments for the temperature 
perturbations is:
\be\label{eq:photonMomenta}
  \Theta_l(k,\eta)\equiv
  \frac{1}{(-i)^l}
  \int_{-1}^1\frac{d\mu}{2}\,
  \mcp_l(\mu)\,\Theta(k,\mu,\eta)\,,
\ee
where we used the Legendre polynomial of order $l$, $\mcp_l$.
The higher moments describe the 
perturbations of the temperature field 
at increasingly smaller scales.
A similar definition applies to the massless neutrino distribution,
while massive neutrinos require an additional treatment
(see Subsection~\ref{ssec:nuperturb}).

The inverse relation of Eq.~\eqref{eq:photonMomenta} reads:
\be
  \Theta(k,\mu,\eta)
  =
  \sum_l
  (-i)^l\,(2l+1)\,
  \Theta_l(k,\eta)\,
  \mcp_l(\mu)\,,
\ee
which can be inserted into Eq.~\eqref{eq:photDistPertFour}
to obtain an infinite hierarchy of coupled equations for the
multipole moments $\Theta_l$:
\begin{align}
  \dot\Theta_0
  &=-k\Theta_1-\dot\Phi
  \label{eq:photonTheta0} \\
  \dot\Theta_1
  &=
  \frac{k}{3}(\Theta_0-2\Theta_2+\Psi)
  +an_e\sigma_T
  \left(\frac{iv_b}{3}-\Theta_1\right)
  \label{eq:photonTheta1}\\
  \dot\Theta_l
  &=
  \frac{k}{2l+1}
  \left[l\,\Theta_{l-1}-(l+1)\,\Theta_{l+1}\right]
  -an_e\sigma_T\Theta_l\qquad\forall l\geq2
  \label{eq:photonThetal}\,.
\end{align}

\subsection{Einstein Equations}\label{ssec:pertEE}
With the definitions of the perturbations presented above,
we can finally derive
the Einstein Equations in the perturbed Universe.

The first order component of $G\munu$ can be calculated using
the Ricci tensor and the Ricci scalar written
in Eqs.~\eqref{eq:ricciTensPert0} to \eqref{eq:pertRicciScal},
together with the perturbed metric
in Eqs.~\eqref{eq:pertMetric0} to \eqref{eq:pertMetric2}.
We obtain for the time-time component:
\be\label{eq:pet00}
  \delta G^0{}_0
  = 
  -6H\partial_0\Phi
  +6H^2\Psi
  -2\frac{k^2}{a^2}\Phi\,.
\ee
This has to be used with the time-time component of the 
stress-energy tensor, that is the energy density of all the particles
in the Universe (see Eq.~\eqref{eq:tmunu_iso})
and that can be obtained as the sum of the integrals
over the distribution functions of each species:
\be\label{eq:pt00}
  T^0{}_0(\vec x,t) =
  -\sum_i
    g_i
    \int
    \frac{d^3p}{(2\pi)^3}
    E_i(p)
    f_i(\vec p, \vec x, t)
  \,,
\ee
where $g_i$ is the degeneracy of the states (spin),
$E_i=\sqrt{p^2+{m_i}^2}$ and $i$ represents all the species.
The result gives
\begin{align}
  T^0{}_0 = &
    -\rho_\gamma(1+4\Theta_0)	  &&\text{(photons)} \notag\\
    &-\rho_{dm}(1+\delta_{dm})	  &&\text{(DM)}	  \notag\\
    &-\rho_{b}(1+\delta_{b})	  &&\text{(baryons)} \notag\\
    &-\rho_\nu(1+4\mcn_0)  &&\text{(massless neutrinos)}
  \,,
  \label{eq:pt00_fin}
\end{align}
where we used the perturbation variables for each species.
The perturbation variables for the neutrinos, $\mcn_l$, will be discussed
in the Chapter~\ref{ch:nu}.
We can now write the time-time component of the Einstein
equations in the perturbed space,
that is the first of the two equations we are going to obtain.
Changing again to the conformal time, it is:
\be\label{eq:pertFriedEq0}
  k^2\Phi+3\frac{\dot a}{a}\left(\dot\Phi-\Psi\frac{\dot a}{a}\right)
  =
  4\pi G a^2
  (4\rho_\gamma\Theta_0
  +4\rho_\nu\mcn_0
  +\rho_{dm}\delta_{dm}
  +\rho_{b}\delta_{b})\,,
\ee
that is the first evolution equation for $\Psi$ and $\Phi$.

To obtain the second evolution equation,
we have to focus on the spatial part of the Einstein tensor.
It is convenient to introduce the projection operator
$(\hat k_i\hat k^j-\delta^j_i/3)$
and to consider only the longitudinal and traceless part of $G_j^i$:
\be
(\hat k_i\hat k^j-\delta^j_i/3)
G_j^i
=
\frac{2}{3a^2}k^2(\Phi+\Psi)\,,
\ee
where the terms proportional to $\delta_{ij}$ are killed by the 
projection operator.
In the same way we can obtain the projection of $T^i_j$:
\be
(\hat k_i\hat k^j-\delta^j_i/3)
T_j^i=
\sum_i g_i
\int\frac{d^3p}{(2\pi)^3}
\frac{p^2(\mu^2-1/3)}{E_i(p)}
f_i(\vec p)\,,
\ee
where $(\mu^2-1/3)$ is proportional to the second Legendre polynomial
$\mcp_2(\mu)$, hence it picks up the quadrupole part of the
distribution.
Baryons and DM do not have a quadrupole term, that exists only
for photons and neutrinos and is related to their
\emph{anisotropic stress}.
The second evolution equation becomes
\be\label{eq:pertFriedEq1}
  k^2(\Phi+\Psi)
  =
  -32\pi G a^2
  (\rho_\gamma\Theta_2
  +\rho_\nu\mcn_2
  )
\ee
and we learn that $\Phi$ and $\Psi$ have opposite sign if the
quadrupole moments $\Theta_2$ and $\mcn_2$ are null.
In the practice, the photon quadrupole is large only
when the photon density becomes small and the main contribution
to the sum comes from the collisionless neutrino quadrupole in the
early Universe, when radiation is dominant.

\section{Adiabatic Initial Conditions}
\label{sec:initial_conditions}
\subsection{Initial Conditions}
\label{ssec:adiabInitial}

The solution of the Boltzmann equations requires a set of
initial conditions that must be fixed.
We recall that, at first order,
we have two equations for baryons and CDM, plus an infinite set of
equations for photons.
All the multipoles above the first two, however,
are negligible in the tightly-coupled limit,
since the Thomson scattering term in Eq.~\eqref{eq:photDistPertFour}
forces $\Theta$ to be equal to $\Theta_0+\mu v_b$:
a monopole part plus a dipole term aligned with $v_b$,
while all the higher multipoles are suppressed.
To compute the power spectrum of the CMB anisotropies
or of the matter perturbations, it is 
convenient to choose the initial conditions in the tightly-coupled regime
and for scales larger than the Hubble horizon,
in order to apply this simplification.
In this case,
given $N$ species, we have to deal only
with $2N$ first-order equations,
one for the monopole and one for the bulk velocity of each species.
Half of the $2N$ corresponding initial conditions seed
decaying modes that we do not observe today.
The combination of $N$ non-decaying solutions must be identified
when studying the mechanisms of generating the initial conditions
(inflation or other scenarios).

One particular combination has a simple physical interpretation:
in a homogeneous Universe, 
the Friedmann equations,
together with the equations of particle physics and thermodynamics,
allow us to determine the evolution of
the background densities $\rho_i\zeroord$ and pressures $p_i\zeroord$
for each species $i$.
The simplest realization of an inhomogeneous Universe that we can think of
is the following:
assume that some physical mechanism introduces a local time-shift,
which accounts for the fluctuations during inflation%
\footnote{For example,
in the single-field inflation scenario, the only clock in the quasi-De Sitter
Universe is represented by the inflaton,
whose fluctuations can be seen as local shifts with respect
to the average time.}.
In this situation,
we have the simplest realization of an inhomogeneous Universe,
where we can write the inhomogeneous densities and pressures:
\begin{align}
  \rho_i(i,\vec x)&=
  \rho_i\zeroord(t+\delta t(\vec x))
  \,\simeq\,
  \rho_i\zeroord(t)+\dot{\rho}_i\zeroord(t)\delta t(\vec x)\,,
  \label{eq:rTimeShift}\\
  p_i(i,\vec x)&=
  p_i\zeroord(t+\delta t(\vec x))
  \,\simeq\,
  p_i\zeroord(t)+\dot{p}_i\zeroord(t)\delta t(\vec x)\,.
  \label{eq:pTimeShift}
\end{align}
We assume that the time-shift function $\delta t(\vec x)$
is the same for all the species and it is at first order
in the perturbation.
Using the two last equations and the 
conservation equation~\eqref{eq:continuity_singlefluid} we obtain
\be
  \frac{\delta\rho_i}{\rho_i\zeroord+p_i\zeroord}=
  \frac{\dot{\rho}_i\zeroord}{\rho_i\zeroord+ p_i\zeroord}
  \delta t(\vec x)=
  -3\frac{\dot a(t)}{a(t)}\delta t(\vec x)\,,
\ee
that is independent on the species $i$.

In this perturbed Universe, at least for wavelengths larger than
the Hubble horizon,
all the species have an adiabatic sound speed $c^{}_{a,i}$,
defined as the ratio $\delta p_i/\delta\rho_i$:
\be\label{eq:adiabSoundSpeed}
  \frac{\delta p_i(t,\vec x)}{\delta\rho_i(t,\vec x)}=
  \frac{\dot{p}_i\zeroord(t)}{\dot{\rho}_i\zeroord(t)}\equiv
  c^2_{a,i}(t)\,.
\ee
The total perturbations are also described by
an effective sound speed $c_s$:
\be
  c_s^2(t)
  \equiv
  \frac{\sum_i \dot{\rho}_i\zeroord(t)c^2_{a,i}(t)}
  {\sum_i \dot{\rho}_i\zeroord(t)}\,,
\ee
that we can use to write the total pressure perturbation as
\be
  \delta p(t,\vec x)=c_s^2(t)\delta\rho(t,\vec x)\,.
\ee
If we do not assume the conditions in
Eqs.~\eqref{eq:rTimeShift} and~\eqref{eq:pTimeShift},
instead, we can only write the total pressure perturbation
as a sum over $N$ independent functions of $\vec x$:
\be
  \delta p(t,\vec x)
  =
  \sum_i c_{s,i}^2(t)\delta\rho_i(t,\vec x)\,,
\ee
that can be eventually rearranged using the entropy perturbations.
For any set of perturbations satisfying Eqs.~\eqref{eq:rTimeShift}
and~\eqref{eq:pTimeShift},
hence,
the fluctuations of the total effective fluid
have adiabatic properties and
the solutions of the perturbation equations are
\emph{adiabatic} or \emph{isentropic},
while in the more general case the solutions 
involve entropy perturbations.
In the simplest case, one can use the set of
Equations~\eqref{eq:rTimeShift} and~\eqref{eq:pTimeShift}
plus the other $2N$ Boltzmann equations
to obtain a basis of two independent sets of initial conditions.
If the basis is chosen appropriately,
one of the solutions becomes rapidly negligible: this is called
the \emph{decaying} mode, while the other one is the
\emph{growing} mode.

The full calculation of the adiabatic initial conditions is 
performed, for example,
in Chapter~6 of Ref.~\cite{Dodelson-Cosmology-2003}.
The initial conditions for each variable can be calculated
as a function of the gravitational potential at early times,
from the Boltzmann equations:
the problem reduces then to calculate the initial conditions
for the gravitational potential $\Phi$ only.
For the photon and neutrino monopoles it is possible to find
\be\label{eq:adiab1}
\Theta_0(k,\eta_i)=\mcn_0(k,\eta_i)=\Phi(k,\eta_i)/2\,,
\ee
at the early time $\eta_i$.
For the baryon and DM perturbations, the adiabatic solution is
\be\label{eq:adiab2}
\delta_{dm}=\delta_b=3\Theta_0\,.
\ee
Finally,
the appropriate initial conditions 
for bulk velocities and dipole moments are:
\be\label{eq:adiab3}
\Theta_1=
\mcn_1=
\frac{i v_b}{3}=
\frac{i v_{dm}}{3}=
-\frac{k\Phi}{6aH}\,.
\ee

\subsection{Initial Curvature Perturbations}
\label{ssec:inCurvPert}

To calculate the initial conditions for the curvature perturbations,
we must make some assumptions for the
physical process that excites the growing mode
in the very early Universe.
Inflation can be the mechanism that does the job,
being responsible for the initial perturbations
in the early Universe.

Inflation was firstly proposed in 1981 
\cite{Guth:1980zm,Linde:1981mu,Mukhanov:1981xt,Starobinsky:1982ee,
Hawking:1982cz,Albrecht:1982wi,
Lucchin:1984yf,Mukhanov:1990me}
to explain two theoretical problems affecting the Big Bang model:
the horizon and the flatness problems, that we discussed in
Section~\ref{sec:evolution}.
One possible implementation of the inflationary mechanism
requires the existence of a generic scalar field $\phi(\vec x,t)$,
which we call the inflaton.
The inflaton is required to contribute with a negative 
$\rho_\phi+3p_\phi$,
that can be calculated from the stress energy tensor of $\phi$:
\be
T^\alpha{}_\beta
=
g^{\alpha\nu}\;
\partial_\nu \phi\;
\partial_\beta \phi
-
g^\alpha{}_\beta
\left(
  \frac{1}{2}
  g^{\mu\nu}\;
  \partial_\mu\phi\;
  \partial_\nu\phi
  +V(\phi)
\right)\,,
\ee
where $V(\phi)$ is the potential of $\phi$.
The homogeneous zero-order part of the field, $\phi\zeroord$,
gives the homogeneous density and pressure:
\begin{align}
  \rho_\phi
  &=
  \frac{1}{2}
  \left(
    \frac{d\phi\zeroord}{dt}
  \right)^2
  +V(\phi\zeroord)\,,\\
  p_\phi
  &=
  \frac{1}{2}
  \left(
    \frac{d\phi\zeroord}{dt}
  \right)^2
  -V(\phi\zeroord)\,,
\end{align}
having considered the diagonal components of $T^\alpha{}_\beta$
and Eq.~\eqref{eq:tmunu_iso}.
If the potential is larger than the kinetic energy,
the field gives a negative pressure.
This can happen, for example, if the scalar field is trapped
in a false vacuum,
where it has small or vanishing kinetic energy since it is at a minimum,
but not in the true minimum:
the consequence is that the pressure is negative, the density is
almost constant and the Universe is in a phase
of exponential expansion.
The scenario with a scalar field trapped in a false vacuum
is not viable,
since the inflaton cannot exit the false vacuum
unless it tunnels quantum mechanically.
Detailed calculations showed that the exponential expansion
of the regions in the false vacuum prevents 
the transition of the full Universe to the true vacuum state.
\cite{Guth:1982pn,Hawking:1982ga}.

To avoid the problem of the Universe never reaching the true vacuum,
mechanisms involving a scalar field slowly rolling down a potential 
energy hill were proposed
\cite{Linde:1981mu,Albrecht:1982wi}.
If the potential is not too steep,
the inflaton energy density remains almost
constant and after some time it comes to dominate,
providing the desired exponential expansion.
From the Friedmann equation~\eqref{eq:freq1} it is possible to
derive the second-order differential equation for $\phi$:
\be\label{eq:inflatonEq}
  \ddot\phi\zeroord+2aH\dot\phi\zeroord+a^2V'=0\,,
\ee
using the dots to indicate derivatives
with respect to the conformal time $\eta$
and the primes to indicate
the derivatives with respect to the inflaton $\phi$.

Slow roll is usually quantified through two small parameters,
$\epssr{}$ and $\etasr{}$%
\footnote{We use the subscript ``SR'' to distinguish the
slow roll parameter $\etasr{}$ from the conformal time $\eta$.}
that vanish when $\phi$ is constant, since $H^2\propto(\rho_\phi)$.
We define
\be\label{eq:epssr_def}
  \epssr{}\equiv
  \frac{d(H)^{-1}}{dt}=
  \frac{-\dot H}{a H^2}\,,
\ee
that is always positive since $H$ is decreasing.
In the inflationary era, $\epssr{}$ is typically small,
while it can be large during the radiation or matter era,
during which its definition is valid,
but it loses its original meaning.
The complementary parameter $\etasr{}$ is instead:
\be\label{eq:etasr_def}
  \etasr{}\equiv
  \frac{1}{H}
  \left(\frac{d^2\phi\zeroord}{dt^2}\right)
  \left(\frac{d\phi\zeroord}{dt}\right)^{-1}
  =
  \frac{-1}{aH\dot\phi\zeroord}
  \left(
    3aH\dot\phi\zeroord+a^2V'
  \right)\,,
\ee
where we used Eq.~\eqref{eq:inflatonEq} to eliminate
the second derivative of $\phi\zeroord$.

Our goal at this point
is to predict the statistical properties of the perturbations
at a time $\eta$,
given the initial conditions inferred from inflation.
One of the assumptions is that the perturbations have
a gaussian distribution at the beginning.
This is preserved until the evolution remains in the linear regime.
Under this assumption, the statistical properties of the fluctuations
can be entirely encoded in the two-point correlation function.
For a stochastic gaussian field, different wavevectors are
uncorrelated and the two point correlation function in the Fourier
space is
\be\label{eq:powerSpectrumDef}
\langle A^\dagger(\vec k,t)A(\vec k',t)\rangle\equiv
(2\pi)^3\delta^3(\vec k -\vec k')\,P_A(k) \,,
\ee
where the coefficient $P_A(k)$ is called the power spectrum
of the quantity $A$.
In a statistically isotropic Universe,
the power spectrum is a function of the wavenumber $k$ only,
not of its direction $\hat k$.

We want now to derive the Primordial Power Spectrum (PPS)
of the initial curvature fluctuations,
from which it is possible to derive
the power spectra for the other quantities using the relations
presented in the previous Subsection.
Inflation is expected to excite also tensor fluctuations, or
gravitational waves.
These are not coupled to the energy density and do not affect the
growth of large scale structures of the Universe,
but they induce fluctuations in the CMB.
A detection of gravitational waves from the primordial Universe
would be a strong evidence of inflation,
but so far they were not observed.
We will not treat tensor perturbations here, but the interested reader
can find details of the calculations in
Ref.~\cite{Dodelson-Cosmology-2003}.
We report here only the PPS
that can be obtained
for the initial tensor perturbations:
\be
  P_T(k)=\frac{8\pi GH^2}{k^3}\,,
\ee
under the assumption that $H$ is constant.
More generally, $H$ has to be evaluated at the time when each mode
leaves the horizon.
Since the expression of the PPS of tensor perturbations is 
remarkably simple, a detection of gravitational waves would
give us a measure of the Hubble rate during inflation.
Since the inflaton energy density is usually dominated by its
potential energy, $H^2\propto\rho/\mpl{2}$ is proportional to
the inflaton potential $V$.
The PPS $P_T$ is consequently proportional to $V(\phi)$.

% Dodelson 6.5.1 p 176(162)
The calculation of the initial scalar fluctuations is more complicated.
All the density and metric perturbations are generated by quantum fluctuations
in the values of the inflaton field.
While tensor perturbations are not coupled to any of
the other perturbation variables, however,
scalar perturbations couple to energy density fluctuations.
Firstly we decompose the inflaton field in a background and
a perturbed component:
\be
  \phi(\vec x,t)=\phi\zeroord(t)+\delta\phi(\vec x,t)\,.
\ee
If we completely neglect the metric perturbations, we can derive a spectrum for
$\delta\phi$ that is similar to $P_T$, since in this approximation
both the quantities are decoupled from the metric perturbations:
\be\label{eq:pDeltaPhi}
P_{\delta\phi}=\frac{H^2}{2k^3}\,.
\ee
It is possible to show that the approximation under which $\Psi$ and $\Phi$
are negligible works well in a particular gauge,
called \emph{spatially flat slicing}.
In this gauge the metric is simple in its spatial part:
\be
  ds^2=
  -(1+2A)\,dt^2
  -2a\,\partial_i B \,dx^i\, dt
  +\delta\iju\, a^2\, dx^i\, dx^j\,,
\ee
where the functions $A$ and $B$ characterize the perturbations.
Under this assumption, Eq.~\eqref{eq:pDeltaPhi} is exact,
since the inflaton perturbations are decoupled from the metric ones.
It is then necessary to find a way to convert back the quantities to
the conformal Newtonian gauge.
This is possible since there is a gauge-invariant variable that
is proportional to $\delta\phi$:
\be
  \zeta=
  -\Phi_H
  -\frac{iaH}{k}v\,,
\ee
where $\Phi_H$ is the Bardeen's potential and
$v$ is the Bardeen's velocity, that in the spatially flat slicing is
\be
v=ikB-\frac{ik\,\dot\phi\zeroord\,\delta\phi}{(\rho+p)\,a^2}\,.
\ee
In the spatially flat slicing the Bardeen's potential is
$\Phi_H=aHB$ and the gauge-invariant quantity $\zeta$ becomes:
\be
\zeta=-\frac{aH}{\dot\phi\zeroord}\delta\phi\,.
\ee
With this relation we can immediately obtain the PPS for $\zeta$,
from Eq.~\eqref{eq:pDeltaPhi}:
\be
P_\zeta=
\left(\frac{aH}{\dot\phi\zeroord}\right)^2 P_{\delta\phi}=
\left.
  \frac{2\pi GH^2}{\epssr{}k^3}
\right|_{aH=k}\,.
\ee
This is the power spectrum of a gauge-invariant quantity:
if we compute $\zeta$ in the conformal Newtonian gauge we can relate
$P_\zeta$ to $P_\Phi$, and then we can use the relations in
Eqs.~\eqref{eq:adiab1}--\eqref{eq:adiab3} to obtain the power spectra
for all the other quantities.

In the conformal Newtonian gauge the Bardeen's potential is
$\Phi_H=-\Phi$, so we have
\be
  \zeta=
  -\Psi
  -\frac{ik_i \,\delta T^0{}_i\,H}{k^2(\rho+p)}\,.
\ee
It is possible to demonstrate that $\zeta$ is conserved on
super-horizon scales:
we can then evaluate the last expression after inflation and
we obtain a general result.
If we calculated the stress-energy tensor
for the inflaton perturbations,
we would find out that in the conformal Newtonian gauge,
after inflation, $\zeta=3\Phi/2$.
Assuming that $\Phi=-\Psi$ in absence of anisotropic stress
(see Eq.~\eqref{eq:pertFriedEq1}), 
we can finally use the spectrum $P_\zeta$ to obtain:
\be
P_\Psi=
P_\Phi(k)=
\frac{8\pi G}{9k^3}
\left.
  \frac{H^2}{\epssr{}}
\right|_{aH=k}\,,
\ee
which tells us that the ratio of the scalar to the tensor modes is of order
$\epssr{-1}$, so that the scalar modes dominate over the tensor ones.
With this solution and the results we presented in
Eqs.~\eqref{eq:adiab1}--\eqref{eq:adiab3}
it is then possible to calculate
the spectra of the initial perturbations for the other quantities,
relating them to the initial power-spectrum $P_\Phi(k)$
through the definition in Eq.~\eqref{eq:powerSpectrumDef}.

A spectrum with constant $k^3 P(k)$ is called a
\emph{scale-invariant}
or \emph{scale-free} spectrum.
Both the tensor and the scalar perturbations have an almost
scale-free power spectrum, where the deviation from
scale-invariance is proportional to the slow roll parameters and it is
typically small.
The scale-invariant spectrum is also referred to as 
``Harrison-Zel'dovich-Peebles spectrum'',
from the names of the people
that proposed it well before that inflation was developed
\cite{Harrison:1969fb,Sunyaev:1970a,Peebles:1970ag}.
The observations nowadays point towards a scalar perturbation spectrum
that is slightly away from scale-invariance, while the tensor spectrum
has never been measured.
The deviation from scale invariance can be parameterized through
the spectral indices $n_s$ and $n_T$,
for scalar and tensor perturbations respectively.
The spectra indices and the amplitudes of the PPS are defined using:
\begin{align}\label{eq:pHK}
 P_T(k)
 &=
 \left.\frac{8\pi GH^2}{k^3}\right|_{aH=k}=
 C_T\, k^{n_T-3}\,,
 \\
 \label{eq:pPK}
 P_\Phi(k)
 &=
 \left.\frac{8\pi GH^2}{9\epssr{}k^3}\right|_{aH=k}=
 \delta_H^2 
 \left(
  \frac{k}{H_0}
 \right)^{n_s-1} 
 \frac{50\pi^2}{9k^3}
 \left(
  \frac{\Omega_m}{D_1(a=1)}
 \right)^2\,,
\end{align}
where in this convention
$\delta_H$ and $C_T$ are the amplitudes of the power spectra
of scalar and tensor modes,
$\Omega_m$ is the fraction of critical density provided by matter and
$D_1$ is the \emph{growth function} of matter perturbations.

It is possible to relate the spectral indices to the slow roll
parameters using the logarithmic derivatives with respect to $k$:
\be
  \frac{d\ln(P_T)}{d\ln k}=n_T-3\,,
\ee
from which we can obtain the relationship between the tensor spectral
index and $\epssr{}$, that is
\be\label{eq:nT_sr}
  n_T=-2\epssr{}\,.
\ee
A similar relation can be derived for the scalar spectral index,
depending on both \epssr{} and \etasr{}:
\be\label{eq:ns_sr}
  n_s=1-4\epssr{}-2\etasr{}\,.
\ee
Please note that in this convention the scale-invariant spectrum
correspond to $n_T=0$ and $n_s=1$.

We conclude mentioning that many authors use the notation
$\mcp_A$ for the rescaled power spectrum:
\be
\mcp_A(k)\equiv\frac{k^3}{2\pi^2}P_A(k)\,.
\ee
With this definition $\mcp_A$ represents the contribution 
of each logarithmic interval in the Fourier space to the two-point
correlation function in the real space.
For practical reasons, if one does not deal
with a specific inflationary model, but rather is interested
in studying the cosmological evolution, the simplest way to 
write the power spectra of scalar and tensor perturbations is:
\begin{align}
  \mcp_t(k)&=A_T\left(\frac{k}{k_*}\right)^{n_T^{}}\\
  \mcp_s(k)&=A_s\left(\frac{k}{k_*}\right)^{n_s^{}-1}\,,
  \label{eq:plPPS}
\end{align}
where $k_*$ is the pivot scale and
the spectral indices are the same for each
$P_x$ and $\mcp_x$.

It is convenient to parameterize the power spectrum of tensor 
fluctuations in terms of the amplitude of the spectrum scalar modes $A_s$
and of the tensor-to-scalar ratio $r_{k_\star}$,
defined at the scale $k_\star$:
\be
  r_{k_\star}\equiv\frac{\mcp_t(k_\star)}{\mcp_s(k_\star)}\,.
\ee
With this definition of $r_{k_\star}$ and assuming $r=r_{k_*}$, we have
\be
  \mcp_t(k)=r\cdot A_s\left(\frac{k}{k_*}\right)^{n_T^{}}\,.
\ee
From Eqs.~\eqref{eq:pHK} and \eqref{eq:pPK} we can see that the tensor-to-scalar
ratio is proportional to the slow roll parameter $\epssr{}$ and is is
typically small.
In particular, under the hypothesis of single-field slow-roll inflation,
the tensor-to-scalar ratio is $r=\epssr{}=-n_T/2$.

Finally, we can relate the slow-roll parameters to the inflaton potential
and to its derivatives:
\begin{align}
  \epssr{}&=
  \frac{1}{16\pi G}
  \left(\frac{V'}{V}\right)^2\,,
  \\
  \etasr{}&=
  \epssr{}
  -
  \frac{1}{8\pi G}
  \frac{V''}{V}\,,
  \\
\end{align}
where the primes denote derivatives with respect to the zero-order field
$\phi\zeroord$.
These relations allow to write the spectral indices and the tensor-to-scalar
ratio as functions of $V$ and its derivatives:
\begin{align}
  n_s-1&=
  2\frac{V''}{V}
  -
  3\left(\frac{V'}{V}\right)^2\,,
  \\
  n_T&=
  -4\left(\frac{V'}{V}\right)^2\,,
  \\
  r&=
  8\left(\frac{V'}{V}\right)^2\,.
\end{align}
Measurements of a scale dependence of the spectral indices and of the
tensor-to-scalar ratio, then, can give information on the shape of the
inflaton potential and consequently on the inflationary mechanism.

%!TeX root=main.tex 
\chapter{Cosmic Microwave Background Radiation}\label{ch:cmbr}

% \begin{abstract}
With the quantities and the definitions presented
in the previous Chapter,
we now study the solutions of the Einstein and Boltzmann Equations
for the perturbations to the photon distribution function.
We will show the main features of the power spectrum of the
CMB anisotropies and we will
describe how the theoretical predictions are influenced by
variations of the different cosmological parameters.
% \end{abstract}

\section{Power Spectrum}\label{sec:cmb_ps}
The goal of a stochastic theory is to predict the statistical
properties of some physical quantity at a time $t$, given the
initial conditions at a time $t\inu$.
In the case of the theory of cosmological perturbations,
we want to obtain the statistical properties of the 
perturbations for some cosmological quantity,
such as the cosmological photon distribution function,
to be tested against the experimental measurements.
% In the case of the perturbations to the homogeneous
% photon distribution, they can be tested against
% the measured CMB anisotropies.
Assuming that the initial fluctuations are Gaussian,
as the current observations suggest,
it is possible to convert all the information encoded in the CMB maps
in the power spectrum of a two-point correlation function,
at least until the perturbations remain
in the linear regime.

The temperature anisotropy in the direction $\hat n$ can be 
expanded in spherical harmonics using
\be\label{eq:deltaT_spHar}
  \frac{\delta T}{T}(\hat n)
  \equiv
  \sum_{lm}\alm{}Y_{lm}(\hat n)\,.
\ee
This is related to the photon perturbation $\Theta$
at the time $\eta_0$, in the direction $-\hat n$
and at the position of the observer.
The $\alm{}$ coefficients can be extracted from the sky map with
\be\label{eq:alm_thetagamma}
\alm{}
\equiv
(-1)^l
\int\frac{d^3k}{2\pi^2}
Y_{lm}(\hat k)
\Theta_l(\eta_0,k)\,,
\ee
where $\hat k$ is the direction of $\vec k$ and
$\Theta_l$ is the photon perturbation in the Fourier space,
defined in Eq.~\eqref{eq:photonMomenta}.
This equations tells us that there is a linear relation between
the Fourier modes $\Theta_l$ and 
the multipoles $\alm{}$:
to any set of Gaussian-distributed cosmological perturbations
it corresponds a set of Gaussian-distributed $\alm{}$.
This situation is particularly interesting since the statistics of
a set of Gaussian-distributed $\alm{}$ is fully described
by the two point correlation function,
$\langle \alm{}a_{l'm'}^*\rangle$.
Eq.~\eqref{eq:alm_thetagamma} also implies
that different (theoretical) multipoles are uncorrelated,
as they are different modes of a gaussian random field.
If the power spectrum in the Fourier space is isotropic,
depending on the modulus of $\vec k$ but not on its direction,
the harmonic power spectrum is also isotropic and does not depend on $m$:
\be
\cl{}
\equiv
\langle \alm{}\alm{*}\rangle\,,\qquad\forall m.
\ee

Under the assumption of ergodicity,
it is possible to build an estimator for the true power spectrum,
since at a given $l$ all the multipoles \alm{}
should have the same variance $\cl{}$.
In the ideal case the best estimator would be:
\be\label{eq:cl_obs}
\cl{\mathrm{obs}}
\equiv
\frac{1}{2l+1}\sum_{m=-l}^l
\left|\alm{\mathrm{obs}}\right|^2\,.
\ee
This is not a realistic way to calculate the spectrum,
since in the real case the sky coverage is not complete
and the observation is affected by
the instrumental noise and the contamination of the anisotropic emission
(galaxy, point sources):
in this situation, building the optimal estimator
is a complicated task that we will not discuss.

Since we can observe only one realization
of the theory that describes the evolution of the primordial perturbations,
we can expect that the statistical fluctuations of the observed
spectrum have an impact on our best estimator.
It is easy to compute the average deviation at a given $l$
using an ideal full-sky experiment.
Each \cl{\mathrm{obs}} as computed in Eq.~\eqref{eq:cl_obs}
is obtained as the mean of $(2l+1)$ independent numbers,
each of them with mean zero and variance \cl{},
so that the \cl{\mathrm{obs}}
obey a $\chi^2$ distribution with $(2l+1)$ degrees of freedom.
The mean and variance of this distribution are 
$\cl{}$ and $\sqrt{2/(2l+1)}\,\cl{}$, respectively.
The distribution is asymmetric around its peak, especially at low $l$,
where the variance is larger.
This is a consequence of the fact that we have
less independent realizations of the same cosmic evolution at low $l$ 
(large angles).
This variance plays the role of a theoretical error
on the best estimator and it is called \emph{cosmic variance}.
Independently of the experimental errors, the cosmic variance is the 
minimum error for the CMB power spectrum at the multipole $l$,
as a consequence of the fact that
we can observe one single realization of the evolution history.
As we will discuss in Sec.~\ref{sec:cmb}, the most recent measurements
of the CMB spectrum are limited by the cosmic variance in a very wide
range of multipoles.

\section{Power Spectrum and Transfer Functions}
\label{sec:ps_transFunc}
We mentioned that the power spectrum of a given quantity $A$,
in the statistically isotropic Universe we are studying,
does not depend on the wavevector direction $\hat k$.
In the same way, we can note that the differential equations
for the perturbations
we presented in the previous Chapter are also independent of $\hat k$.
As a consequence,
the system of linear equations must be solved only once 
for each wavenumber $k$, given an arbitrary set of initial conditions.
For example, we could assume that the solution is normalized
to $\Theta_0(\eta\inu,\vec k)=1$.
In this case, the power spectrum of $\Theta_l$ at a given time will be
the product of the power spectrum of $\Theta_0$ at $\eta\inu$ multiplied
by the square of the solution $\Theta_l(\eta,\vec k)$
(see the definition in Eq.~\eqref{eq:powerSpectrumDef}).

The initial normalization,
in an Universe with only adiabatic conditions,
often refers to a dimensionless quantity $\mcr$, called the comoving
curvature perturbation.
In the comoving gauge, $\mcr$ represents the local fluctuation
of the spatial curvature, in comoving units.
In the Newtonian gauge this is defined as
\be
  \mcr
  \equiv
  \Psi
  -\frac{1}{3}
  \frac{\delta\rho\tot}{\rho\zeroord\tot+p\zeroord\tot}\,.
\ee
With this assumption,
all the evolution equations of the perturbations
can be solved using the arbitrary condition
$\mcr(\eta\inu,\vec k)=1$ and the power spectrum
of a given quantity $f$ will be then the square of the solution
multiplied by the initial power spectrum of $\mcr$.
In other words, one should solve the evolution equations for
some renormalized variables
\be
  f(\eta,k)
  \equiv
  f(\eta,\vec k)/\mcr(\eta\inu,\vec k)\,,
\ee
where we adopted the notation used in
Ref.~\citelesg{} to distinguish
the \emph{transfer functions} $f(\eta,k)$, depending on $k$,
from the corresponding not normalized quantity $f(\eta,\vec k)$,
depending on $\vec k$.
Here $f$ indicates one of the perturbation functions we defined in the previous
Chapter: $\Theta_l$, $\delta$, $\delta_b$, and so on.
Once one has the solution for the transfer function
$f(\eta,k)$ at any time $\eta$,
the power spectrum of $f(\eta,\vec k)$ can be
obtained from the initial spectrum of $\mcr$ multiplying
by the square of the transfer function corresponding to $f$,
that is $f(\eta, k)$ (see the definition in Eq.~\eqref{eq:powerSpectrumDef}):
\be
  \mcp_f(\eta,k)
  =
  \mcp_\mcr(k)\,
  [f(\eta,k)]^2\,.
\ee

\section{Acoustic Oscillations}\label{sec:acouOscill}
While a precise solution of the system of differential equations that
describe the cosmological perturbations can be obtained numerically,
several analytical approximations were developed in the past,
see e.g.~Ref.~\cite{Hu:1995em}.
These approximations helped in understanding all the complex physical
phenomena that occurred during the evolution.
The full analytical treatment is beyond
the scope of this thesis and we will only give
a qualitative description of the CMB spectrum.

When photons and baryons can be considered as a single tightly-coupled
fluid, the sound speed of the perturbations in the fluid is
\be\label{eq:soundspeed}
  c_s^2
  \equiv
  \frac{1}{3(1+R)}\,,
\ee
where $R$ is the baryon to photon ratio
defined in Eq.~\eqref{eq:baryonToPhotonsR}.
The ratio $R$ increases with the scale factor,
as the photon and baryon
densities scale differently.
The sound speed is then $c_s=1/\sqrt{3}$ during radiation domination,
when $R$ is small,
and decreases slowly to zero.

When the sound speed is different from zero,
acoustic waves propagate in the fluid.
Since the primordial perturbations drive the system locally
out of equilibrium,
gravitational attraction and radiation pressure are not
exactly compensated at each point and
the acoustic waves propagate causally.
The maximal distance at which they propagate is the \emph{sound horizon}.
The comoving sound horizon, that is the comoving distance traveled
by the wavefront in a time $\eta-\eta\inu$,
is given by
\be\label{eq:comovSoundHorizon}
  r_s(\eta)
  \equiv
  \int^\eta_{\eta\inu}c_s(\eta')\, d\eta'\,.
\ee
If $\eta\inu\ll\eta$, this quantity does not depend
on the initial time.

Acoustic waves are density waves in the coupled fluid,
whose perturbations can be encoded in the variations of the
temperature $\Theta_0(\eta,k)$.
However, the system does not behave like a simple harmonic oscillator.
This is the consequence of several phenomena:
first of all, the ratio $R$ increases with time,
changing the sound speed
and other properties of the fluid, like its inertia.
Secondly, the gravitational forces are seeded by the overdensities
of the baryon-photon fluid, but also by those of
the other species, as CDM or neutrinos.
All these effects are taken into account in the second-order
differential equation for $\Theta_0$:
\be\label{eq:acOscEq}
\ddot\Theta_0+
\frac{\dot R}{1+R}\dot\Theta_0+
k^2 c_s^2\,\Theta_0
=
-\frac{k^2}{3}\Psi
-\frac{\dot R}{1+R}\dot\Phi
-\ddot\Phi\,.
\ee
We analyze now the different terms that appear in this equation.

\subsection{Diffusion Damping}
\label{ssec:diffusiondamp}
% We start mentioning the fact that the tight-coupling approximation
% breaks down close to recombination, when
% perturbations at distances below the photon diffusion length
% are erased by random scattering processes.
When all the electrons were ionized, before recombination,
the photons had a mean free path that was much smaller than the size
of the Universe.
As a consequence of Compton scattering, the electron-proton fluid was
tightly coupled with the photons.
In the tight-coupling approximation, the scattering rate of the photons
is much larger than the expansion rate
and their trajectory can be described as a random walk,
with photons taking a random direction after each interaction with
an electron.
Since the interaction rate of the photons $\Gamma_\gamma$
can be obtained from the Thomson scattering
($\Gamma_\gamma=an_e\sigma_T$)
and the comoving mean free path of the photons is
$r_\gamma=\delta\eta=(an_e\sigma_T)^{-1}$,
an approximated expression for
the comoving distance traveled by a photon between an early time
$\eta\inu$ and a time $\eta$ will be
\be\label{eq:comDiffDist}
  r^2_d(\eta)
  \simeq
  \int^\eta_{\eta\inu}
  d\eta\,\Gamma_\gamma\, r^2_\gamma
  \simeq
  \int^\eta_{\eta\inu}
  \frac{d\eta}{a n_e \sigma_T}
  \,.
\ee
If $\eta\inu\ll\eta$, $r_d$ does not depend on $\eta\inu$.
Photon diffusion erases all the perturbations
with a wavenumber greater than $k_d=2\pi/r_d$, 
corresponding to small distances.

The damping effect, together with the driving contribute given by the 
gravitational terms in the right hand side of Eq.~\eqref{eq:acOscEq},
leads to an interesting phenomenology for the acoustic oscillations,
that we will study in three different stages:
radiation domination, matter domination before photon decoupling
and evolution after the photon decoupling.
This discussion is essential for understanding how the different
cosmological parameters can affect the CMB spectrum.

\subsection{%
Constant Acoustic Oscillations during Radiation Domination}
During radiation domination it is easy to obtain approximated analytic
solutions, since one can work in the limit $R=0$ or $c_s=1/\sqrt{3}$,
valid when baryon and CDM perturbations are negligible with respect to 
photon perturbations.
From the Einstein equations it is possible to find a second order
differential equation only for the perturbations of the
fluid we consider.
The growing solution corresponds to constant transfer functions
outside the sound horizon:
in this regime, the propagation of acoustic waves is negligible since
the comoving wavelength is much larger than the comoving sound horizon
and the modes are frozen at their initial values.
Inside the sound horizon, instead, the photon density modes
oscillate with a constant amplitude and metric fluctuations
decay with time.
The effects driven by the metric terms in Eq.~\eqref{eq:acOscEq}
are negligible with respect to photon pressure forces and
if we use $k\eta\gg1$ the driving term on the r.h.s.~can be neglected.
We obtain hence the equation of a simple harmonic oscillator.

\subsection{Damped Acoustic Oscillations after Equality}
\label{ssec:dampedAcousticOsc}
After matter-radiation equality and before photon decoupling,
several phenomena modify the evolution.
As the baryon fraction $R$ starts to increase,
the sound speed decreases, affecting
the amplitude of the acoustic oscillations.
At the same time, the increase of the baryon fraction
forces an increase of the coupling between the fluid and gravity,
and the zero-point of the oscillations is shifted.
If we neglect the time variation of $\Phi$ in Eq.~\eqref{eq:acOscEq},
the zero point of temperature oscillations
corresponds to $k^2c_s^2\Theta_0=-k^2\Psi/3$,
that is $\Theta_0=-(1+R)\Psi$. 
For a gravitational potential well with $\Psi<0$ the value of
$\Theta_0$ that corresponds to the equilibrium of the oscillations
increases with $R$.

After matter-radiation equality,
non-relativistic matter components start to 
influence the metric perturbations so that
$\Phi$ and $\Psi$ do not decay as quickly
as during the radiation domination, inside the Hubble radius.
The gravitational driving terms of Eq.~\eqref{eq:acOscEq}
become then more important and alter
the behavior of the acoustic oscillations.

Finally, when the fluid exits the tight-coupling approximation regime,
oscillations are damped at wavelengths smaller than
the diffusion length of the photons,
as we mentioned earlier.

% We now briefly describe the leading effects and we refer to
% Ref.~\cite{Hu:1995em} for a detailed treatment.
At the equality, temperature oscillations are roughly symmetric around
the zero-point $\Theta_0=\Phi$ inside the sound horizon and
constant at larger scales, where metric fluctuations are negligible.
At decoupling, the amplitude of photon oscillations is reduced on 
all the sub-horizon scales:
with respect to the zero-point at equality,
the zero-point of the oscillations is shifted down,
with a consequent enhancement
of the amplitude of the odd peaks with respect to the even ones.
These effects, plus the damping at small scales, are essentially
controlled by the duration of the transition between equality and
decoupling, by the baryon fraction $R$ at decoupling and by the
value of the diffusion length $r_d$.

\subsection{Gravitational Clustering after Decoupling}
After decoupling, photons stop interacting with the rest of the plasma
and the calculation of the perturbations concerns the
self-gravitation of non-relativistic matter components.
The evolution of matter perturbations leads to structure formation.
In the real Universe we cannot use
the approximation that $\Phi=-\Psi$ is constant over time
at all scales, valid in the ideal matter dominated Universe,
since at the beginning of the matter dominated era a residual decay
of $\Phi$ and $\Psi$ perturbations occurs.
Moreover, during the DE dominated stage a similar decay occurs.
The presence of massive neutrinos, finally,
breaks the approximation that $\Phi=-\Psi$ is always valid at
small scales.

\section{Temperature Anisotropies}\label{sec:tempAniso}
\subsection{Numerical Calculation}
As we stated in Section~\ref{sec:cmb_ps}, the main goal of the 
cosmological evolution theory is to predict the CMB spectrum,
or the final spectrum of the perturbations as a function
of the cosmological parameters.
% Once the transfer functions has been calculated,
% they should be converted into a CMB spectrum.

One possible way to do this is to adopt a brute-force method
and to integrate all the equations \eqref{eq:photonTheta0} to
\eqref{eq:photonThetal}
with at least $l\mx$ multipoles
for the photon perturbations, between an initial time $\eta\inu$
and today.
The temperature anisotropy spectrum
up to $l\mx$ is then given by:
\be\label{eq:CMBcldef}
  \cl{}
  =
  \frac{1}{2\pi^2}\int
  \frac{dk}{k}[\Theta_l(\eta_0,k)]^2\mcp_\mcr(k)\,.
\ee
The hierarchy of coupled photon equations is infinite, but any
numerical algorithm can integrate only a finite number of multipoles.
A truncation of the multipole series is needed, but this
can cause a reflection of power-down at lower multipoles.
Even if it is possible to avoid such a power-down with an appropriate
choice of $k\mx\simeq l\mx/\eta_0$, that ensures that only
the photon transfer functions $\Theta_l(\eta,k)$ with $l\gtrsim l\mx$ vanish,
the brute-force approach is extremely time-consuming
from the computational point of view.

A much more convenient approach is the so called
\emph{line-of-sight} approach.
For convenience, in this Subsection we will return to the
description in the real space.
The same calculations can be transposed in the Fourier space using
the spherical Bessel functions
(see Ref.~\cite{Zaldarriaga:1995gi}).
Consider a photon traveling along a geodesic between the last
scattering and us:
we know that the geodesic is not a straight line,
since the gravitational lensing effects modify the photon path.
These, however, are second-order effects and we are considering only
the first-order perturbations:
the geodesic is then approximated as a straight line,
since we are neglecting the spatial curvature.
A photon reaching us from the direction $-\hat n$ traveled in the
direction $\hat n\equiv\hat p$ from the last scattering surface.
Its comoving coordinates at the time $\eta$
were $\vec x=-(\eta_0-\eta)\,\hat n$,
so that the variation in the radial
coordinate is $dr=-d\eta$.
A function $\mcf(\vec x,\hat n,\eta)$ evolves along
the trajectory according to the total derivative
\be
  \frac{d\mcf}{d\eta}=
  \dot\mcf+\hat p\cdot\vec\nabla\mcf\,,
\ee
using the straight line approximation $d\hat n/d\eta=0$.
It is convenient to consider the function
$\mcf=\Theta(\vec x,\hat n,\eta)+\Psi(\vec x,\eta)$ and
to integrate the Boltzmann equation over the photon trajectory.
We can use the optical depth $\tau(\eta)$ written in 
Eq.~\eqref{eq:opticalDepth} to define the visibility function
$g(\eta)=-\dot\tau e^{-\tau}$,
that represents the probability for a photon reaching us today
to have experienced its last scattering
at the time $\eta$.
The last scattering time $\etals{}$ can be defined as the time that
gives the maximum of $g$.
With these definitions,
using the Boltzmann equation \eqref{eq:photDistPertFour}
and multiplying by $e^{-\tau(\eta)}$,
we obtain:
\be
  \frac{d}{d\eta}[e^{-\tau(\eta)}(\Theta+\Psi)]
  =
  g(\eta)
  (\Theta_0+\Psi+\hat n\cdot\vec v_b)+
  e^{-\tau(\eta)}(\dot\Psi-\dot\Phi)\,.
\ee
Integrating along the line of sight between an early time
$\eta\inu\ll\etals{}$, when $e^{-\tau(\eta\inu)}\simeq0$,
and today, when $e^{-\tau(\eta_0)}=1$, we obtain the temperature
anisotropy as seen by the observer in the direction $\hat n$:
\be\label{eq:tempAniLOS}
  \Theta(\vec o,\hat n,\eta_0)
  =
  -\Psi(\eta_0,\vec o)+
  \int^{\eta_0}_{\eta\inu} d\eta\,
  [g(\eta)(\Theta_0+\Psi+\hat n\cdot\vec v_b)+
  e^{-\tau(\eta)}(\dot\Psi-\dot\Phi)]\,,
\ee
where $\vec o$ refers to the observer position,
fixed at the origin for simplicity.
The first term on the r.h.s.~gives
a local isotropic redshift or blueshift of incoming photons
due to the local metric fluctuation today at the observer position,
that is usually small and we will neglect it.
The integral shows us that the observed temperature anisotropy in a
given direction depends on two terms:
the sum $(\Theta_0+\Psi+\hat n\cdot\vec v_b)$ around the
time of decoupling (when $g$ is not negligible) and the sum
$(\dot\Psi-\dot\Phi)$ between decoupling and today
(when $e^{-\tau}$ is not negligible).

Interestingly, from Eq.~\eqref{eq:tempAniLOS} we learn that the
photon perturbations $\Theta_l$ for $l>1$
are not needed to compute CMB anisotropies,
meaning that
this method is then much more economic than the brute-force method.

Equation~\eqref{eq:tempAniLOS} shows that four quantities are required
to obtained the temperature fluctuations:
$\Phi$, $\Psi$, $\Theta_0$ and $v_b$.
Since these must be obtained from the Einstein equations, however,
also the density perturbations and the bulk velocities for the other
species must be calculated.
To obtain a good precision on the first multipoles of the temperature
anisotropy, though, also the modes with $l>2$ must be calculated.
An economic truncation scheme requires the calculation up to
$l\mx^\gamma\simeq\mco(10)$ to obtained a sufficient precision
on the \cl{} up to $l\mx\simeq\mco(10^3)$ \cite{Ma:1995ey}.
As for the brute-force approach, $l\mx$ determines
the maximum wavenumber $k\mx\simeq l\mx/\eta_0$ at which the source
function has to be evaluated, corresponding to the information
about perturbations on the last scattering surface as seen today
under an angle $\theta\simeq\pi/l\mx$.
The advantage of the line-of-sight approach
over the brute-force method
is then given only by the factor $l\mx^\gamma/l\mx$,
that allows to gain few orders of magnitude in computation time.
The line-of-sight approach, used by all the modern Boltzmann 
codes, was firstly implemented in \cmbfast \cite{Seljak:1996is}.

\subsection{Physics of the CMB Anisotropies}
We want now to look at Eq.~\eqref{eq:tempAniLOS}
to study how the different
terms contribute to the CMB spectrum.

The most obvious contribution to the observed temperature
fluctuations in one direction is given by the temperature fluctuations
at the last scattering in the same direction, corrected
by a gravitational shift \cite{Sachs:1967er}:
this contribution comes from the $g(\eta)(\Theta_0+\Psi)$ term
in Eq.~\eqref{eq:tempAniLOS}.
Ideally, in the instantaneous decoupling limit
the last scattering surface can be seen as a flat
surface rather than a thick shell,
corresponding to a rapid increase of the mean free path of the photons
from 0 to infinity at $\eta=\etals{}$.
In this limit, the visibility function can be replaced by a
Dirac delta $\delta(\eta-\etals{})$ and
the integral of $g(\eta)(\Theta_0+\Psi)$ gives the 
\emph{Sachs-Wolfe (SW) contribution}:
\be
  \Theta^{\mathrm{SW}}(\vec o,\hat n,\eta_0)
  \simeq
  \Theta_0(\vec x\lsu{},\hat n,\etals{})
  +\Psi(\etals{},\vec x\lsu{})
  \,,
\ee
where $\vec x\lsu{}=(\etals{}-\eta_0^{})\,\hat n$.

For super-horizon scales and during matter-domination
it is possible to derive the relation $\Theta_0=-2/3\Psi$.
In a CMB map smeared over small scale fluctuations,
the SW contribution becomes then
\be
  \Theta^{\mathrm{SW,smoothed}}(\vec o,\hat n,\eta_0)
  \simeq
  -\frac{1}{2}\Theta_0(\vec x\lsu{},\hat n,\etals{})
  \simeq
  \frac{1}{3}\Psi(\etals{},\vec x\lsu{})
  \,.
\ee
This Equation tells us that hot regions in the observed CMB map
correspond to cold regions at the last scattering:
the reason is that photons leaving an overdense region lose
part of their energy to exit the gravitational potential well.

The second contribution from the integral in Eq.~\eqref{eq:tempAniLOS}
comes from the term proportional to $\vec v_b$.
Photons are emitted from the coupled baryon-electron fluid with 
a peculiar velocity that is different from point to point.
When they are projected along the line-of-sight, this velocity
induces a Doppler shift in the photon wavelength.
In the instantaneous decoupling limit,
the \emph{Doppler contribution} is:
\be
  \Theta^{\mathrm{Doppler}}
  \simeq
  \hat n\cdot\vec v_b(\etals{},\vec x\lsu{})\,.
\ee

Photons traveling from the last scattering surface
to a today observer
encounter several metric fluctuations:
every time they enter or exit a gravitational potential well,
they are blueshifted or redshifted.
The term in Eq.~\eqref{eq:tempAniLOS} that encodes this phenomenon is
the one containing to $e^{-\tau}(\dot\Psi-\dot\Phi)$.
In a static Universe,
variations in $\Psi$ correspond to the presence of over- or 
under-dense regions, while variations in $\Phi$ encode a local
correction to the average time-dilation, responsible for the
gravitational redshift during the Universe expansion.
Since the Universe is not static, the photon does not encounter the
same gradient when entering or exiting a local metric fluctuation:
while traveling along the line-of-sight, photons take a cumulative
temperature shift, accounted by the integral of
$\dot\Psi$ and $\dot\Phi$.
The combination of these shifts is the 
\emph{Integrated Sachs-Wolfe (ISW) contribution} to the temperature
fluctuations.
In the instantaneous decoupling limit, $e^{-\tau}$ can be replaced
by the Heaviside function $\theta(\eta-\etals{})$ and the ISW
contribution becomes
\be
  \Theta^{\mathrm{ISW}}(\vec o,\hat n,\eta_0)
  \simeq
  \int^{\eta_0}_{\etals{}}
  d\eta\,
  (\dot\Psi-\dot\Phi)
  \,.
\ee

\subsection{Features of the CMB spectrum}
\begin{figure}[t]
  \centering
  \includegraphics[width=\singlefigland]{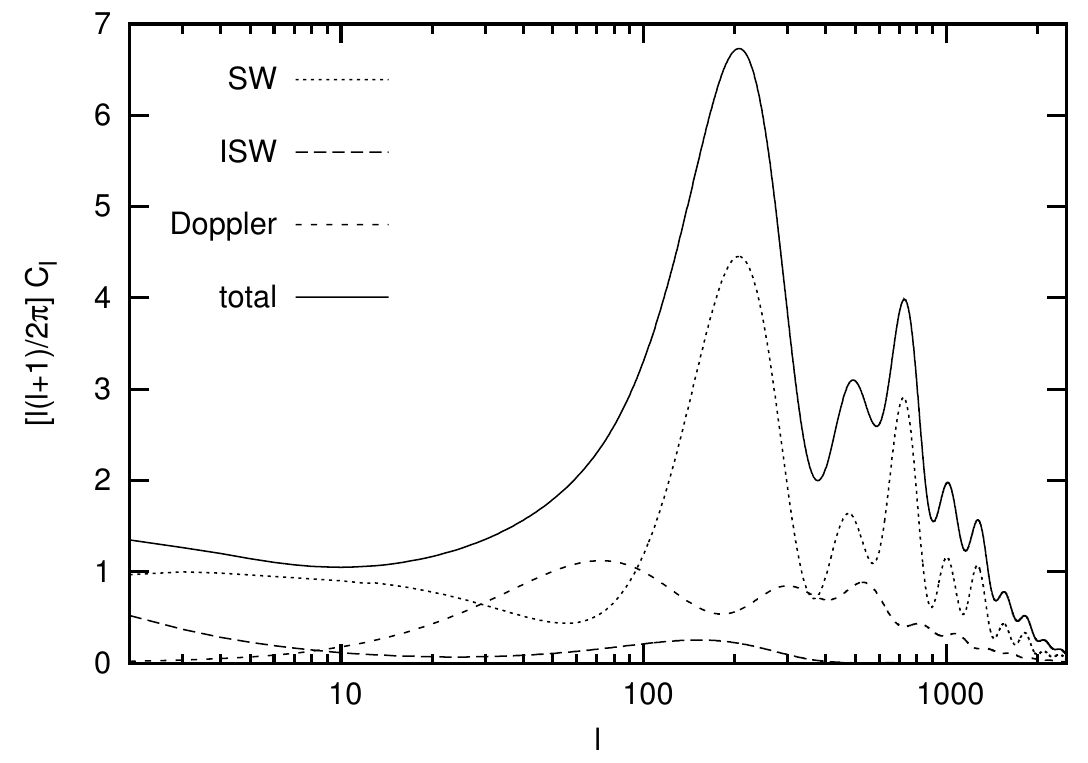}
  \caption[Spectrum of CMB anisotropies, with SW, Doppler and ISW contributions.]%
  {\label{fig:CMBcontribs}
  Full spectrum of CMB temperature anisotropies
  and individual contributions
  from the SW term, the Doppler term and the ISW term.
  The full spectrum is given by the sum of the above terms,
  plus the contributions coming from their correlations.
  The spectrum is obtained numerically 
  in a flat Universe without considering neutrino perturbations.
  From~\citelesg{}.}
\end{figure}

With the various contributions to the CMB spectrum we just
mentioned and the \cl{} formula in Eq.~\eqref{eq:CMBcldef}
it is possible to obtain the shape of the features
of the CMB spectrum.
An example of the full temperature power spectrum is plotted in 
Fig.~\ref{fig:CMBcontribs}.

In the Fourier space, the photon transfer function $\Theta(\eta,k)$
can be written
using the spherical Bessel functions $j_l(x)$,
that are peaked near $x\simeq l$.
Since the visibility function $g$ is peaked around recombination and
the PPS of curvature perturbations $\mcp_\mcr$
is nearly scale-independent,
we can derive mathematically a simple result:
in the Fourier space,
the SW contribution to the $\cl{}$
multiplied by $l^2$ is qualitatively similar
to the square of the SW contribution to the transfer function, namely
\be
l^2\cl{}\propto
[\Theta(\etals{},k)+\Psi(\etals{},k)]^2_{
k=l/(\eta_0-\etals{})}\,.
\ee
This comes from the fact that the
anisotropy multipoles at a given $l$ come mainly from the Fourier 
modes at $\lambda\simeq2\pi a(\etals{})/k$ on the last scattering
surface, that are seen today under an angle
$\theta=\lambda/d_A(\etals{})\simeq2\pi/l$.
In a flat space, $d_A(\etals{})=a(\etals{})/(\eta_0-\etals{})$,
that gives $2\pi/l\simeq2\pi/[k(\eta_0-\etals{})]$.
The full calculation is more complex, since a given wavenumber
contributes to an ensemble of multipoles and the relation we 
presented above gives only the value of $l$ corresponding to the
maximum contribution for a given $k$.
To develop a qualitative description of the CMB spectrum, however,
the approximation we adopted is sufficient.

We can look at the dotted line in Fig.~\ref{fig:CMBcontribs},
that represents the SW contribution to the CMB spectrum.
At large scales (small $l$), the nearly flat behavior gives the
so-called SW plateau, that corresponds to the modes that are
outside the sound horizon at decoupling, which are still frozen.
From $l\simeq100$ we can distinguish the acoustic peaks we discussed
in the previous Section,
which are modulated by the various effects already described.
We can see that the odd peaks are enhanced with respect to the
even ones, as a consequence of the high baryon fraction
(see Subsection~\ref{ssec:dampedAcousticOsc}).
The first peak is given by the correlation length on
the last scattering surface that corresponds to the sound horizon
at decoupling, while all the other peaks represent the
higher harmonics of the same feature.
These peaks are damped according to the diffusion damping
effect described in Subsection~\ref{ssec:diffusiondamp},
that gives a factor $e^{-(l/l_d)^2}$, with
$l_d\simeq k_d (\eta_0-\etals{})\simeq 2\pi(\eta_0-\etals{})/r_d$.
The damping effect is usually referred to as \emph{Silk damping}%,
% even if the work where it was firstly discussed refers to the
% damping of baryon density perturbations, that occurs at the same
% epoch 
\cite{Silk:1967kq}.

The second contribution comes from the Doppler term
(short-dashed line in Fig.~\ref{fig:CMBcontribs}).
On super-Hubble scales (at small $l$)
the contribution of the Doppler term is negligible,
since perturbations are frozen and the velocities in the
baryon-photon fluid are very small.
At smaller scales, instead, the contribution is sourced by
$\vec v_b$, that exhibits the same oscillatory pattern as $\Theta_0$,
but with a shift of $\pi/2$, as for any oscillator.

The last contribution comes from the ISW terms.
These would vanish if the evolution between decoupling and today
occurred in a perfectly matter-dominated Universe, since
in this case the metric perturbations would be static everywhere
and at any time.
Instead, the ISW term contributes in two different phases of the 
Universe history.
The first one is at the time of decoupling,
since the Universe is at the beginning of the matter-dominated
phase and the metric perturbations are still decaying together with
the photon perturbations: the residual time variation of $\Psi$
and $\Phi$ gives the Early ISW (EISW) effect.
Secondly, at late times the Universe enters a $\Lambda$-dominated
phase and the metric fluctuations start decaying again.
This Late ISW (LISW) effect can be considered as
a secondary anisotropy, since it comes from gravitational
interactions involving free-streaming photons that travel through
neighboring galaxy clusters.
We can identify the EISW and the LISW terms as two
separate contributions
to the long-dashed line in Fig.~\ref{fig:CMBcontribs}.
The EISW term cannot affect modes that were outside the sound horizon
at decoupling, so it is negligible at very large scales:
it gives the maximum contribution at $l\simeq200$ and it tends to 
decrease at larger $l$, as a consequence of the $k^{-2}$ coefficient
in the Doppler term.
The EISW then contributes with an enhancement of the first
acoustic peak.
The LISW contribution, instead, is present at any times, since it is
related to a decay of metric fluctuations at all scales caused
by the Universe entering a DE-dominated phase.
Since this effect decreases for the same reason of the EISW effect,
it is peaked at $l=2$ and it becomes sub-dominant for $l\gtrsim30$.

\section{Parameter Dependence}\label{sec:par_depend}
Up to now we presented how the CMB temperature spectrum looks like,
but we did not focus on how the different features are affected by
variations in the fundamental cosmological parameters
that we want to infer from the observations.
Before we describe how we can parameterize the standard
cosmological model and we
study how the parameters change the CMB spectrum, however,
we should discuss one last effect that is caused by
astrophysical phenomena after photon decoupling.

During the formation of the first stars, at redshift of order ten,
the Universe was partly reionized
by the light produced by the new stars.
A small fraction of CMB photons is then scattered by the free
electrons that are created in this context.
This effect is negligible for modes that entered the horizon well
after the reionization epoch ($l<l_\mathrm{step}$,
with $l_\mathrm{step}\simeq40$),
but it leads to a scale-independent
suppression of the CMB spectrum at smaller scales.
The effect is accounted by a factor $e^{-\tau_\mathrm{re}}$, where
$\tau_\mathrm{re}$ measures the optical depth to reionization:
this quantity is constrained to be $\tau_\mathrm{re}\simeq0.1$
by current observations.
The damping of the spectrum at $l\gg l_\mathrm{step}$ is completely
parameterized by $\tau_\mathrm{re}$, while around
$l\simeq l_\mathrm{step}$ the suppression depends on the details
of the reionization history,
which are not well constrained by the current data.

Now we have all the ingredients to describe how the cosmological 
quantities influence the CMB spectrum, restricting ourselves to
a flat Universe with three massless neutrinos:
we will describe the parameterization and the
neutrino effects in Section~\ref{sec:nucosmology}.
We emphasize that considering a Universe without neutrino perturbations
is not a realistic scenario,
but we want focus on the neutrino contribution separately.
We refer to the standard cosmological model as to the \lcdm~model,
from the names of the cosmological constant $\Lambda$ and of
CDM, that are two of the components of the Universe.
In the \lcdm{} model, we need six parameters to describe all the
phenomena we encountered:
\begin{itemize}
 \item to parameterize the PPS of scalar perturbations,
  we use its amplitude $A_s$ and its tilt $n_s$, see Eq.~\eqref{eq:plPPS};
 \item the baryon density fraction today is given by
  $\omega_b=\Omega_b h^2$;
 \item we can use either the CDM density fraction
  $\omega_c=\Omega_c h^2$
  or the total matter density fraction
  $\omega_m=\omega_b +\omega_c$.
  The former is more convenient if we consider additional massive
  components, for example massive neutrinos;
 \item the optical depth to reionization, $\tau_\mathrm{re}$;
 \item if we assume a spatially flat Universe, we can consider either
  the cosmological constant density fraction $\Omega_\Lambda$
  or the Hubble parameter today, $H_0$ or $h$,
  since for a fixed $\omega_m$
  they are related by 
  $h=\sqrt{\omega_m/(1-\Omega_\Lambda)}=H_0/(100\Hou)$.
  Since in the analyses reported in the second part
  of this Thesis we will use the public Boltzmann solver
  \camb \cite{Lewis:1999bs}, instead, we adopt a different parameterization
  for the \lcdm~model, that considers the characteristic angular size
  of the fluctuations in the CMB,
  also called the acoustic scale $\theta$,
  in place of the Hubble parameter.
  Since the acoustic scale is determined from the positions
  of the acoustic peaks, its measurement is quite robust and stable
  to changes in data combinations
  and in the assumed cosmological model.
  The situation is similar to that of the BAO feature in the context
  of the large scale structure surveys,
  with the advantage that the CMB acoustic peaks develop
  in a completely linear regime.
\end{itemize}
Since the CMB measurements give a very precise determination
of the photon temperature today, we consider
$\omega_\gamma=\Omega_\gamma h^2$ as a fixed parameter.
Since we fixed the amount of the other species (three massless neutrinos)
% \footnote{We neglect here the contribution of the neutrinos,
% but this leads to a significant change in the cosmological evolution.
% In a realistic case, neutrinos cannot be neglected.}
contributing to the radiation
energy density $\omega_R$ at the time of matter-radiation equality,
the redshift of equality depends only on $\omega_m$.
In the same way the redshift of coincidence, that occurs when 
the energy densities of matter and cosmological constant are equal,
is fixed by $\Omega_\Lambda$.

Given this set of parameters, we can list how they control
the features of the CMB temperature spectrum: we follow
the treatment of Ref.~\citelesg{}.
The shape of the CMB spectrum is controlled by:
\begin{itemize}
  \item[\ec1]
    the peaks location, depending on the angle
    $\theta=d_s(\etals{})/d_A(\etals{})$.
    The sound horizon at decoupling $d_s$ is controlled by
    the expansion history,
    controlled by $\omega_m$ through
    the redshift of matter-radiation equality,
    and by the sound speed at decoupling,
    affected by changes in $\omega_b$.
    The angular diameter distance, instead,
    depends on the expansion history after decoupling and
    is controlled by $\Omega_\Lambda$ or $h$, governing
    the coincidence redshift.
  \item[\ec2]
    the relative amplitude of odd to even peaks, that depends
    on the balance between gravity and pressure in the photon-baryon
    fluid through the ratio $\omega_b/\omega_\gamma$;
  \item[\ec3]
    the amplitude of all the peaks, depending on the expansion rate
    between equality and decoupling.
    Since decoupling is fixed by the interactions and by the evolution rate,
    the amplitude of the peaks
    is affected mainly by the redshift of equality
    ($\propto\omega_m/\omega_R$):
    for an earlier equality 
    (higher $\omega_m$) the peaks are smaller, because the damping
    of acoustic oscillations lasts longer.
    Moreover, if there is more time between equality and decoupling,
    the EISW effect is reduced and the first peak gets an even smaller
    contribution.
  \item[\ec4]
    the envelope of the secondary peaks, depending on the angle
    $\theta=\lambda_d(\etals{})/d_A(\etals{})$.
    The diffusion
    length $\lambda_d=a\, r_d$, controlled by the expansion
    history and recombination history before decoupling,
    depends essentially on the electron number $n_e$, that is
    the quantity in Eq.~\eqref{eq:comDiffDist} that changes more
    before recombination,
    and on the conformal time at decoupling, \etals{}.
    In the \lcdm{} model, $n_e$ is fixed and the integral in
    Eq.~\eqref{eq:comDiffDist} essentially does not depend on the
    expansion and on the electron fraction before equality.
    The angle $\theta$, then, depends essentially on $\omega_m$
    (entering $\lambda_d$) and on $\Omega_\Lambda$
    (entering $d_A$).
  \item[\ec5]
    the normalization of the power spectrum of initial fluctuations $A_s$,
    being the CMB spectrum proportional to $\mcp_\mcr$.
  \item[\ec6]
    the tilt $n_s$, for the same reason.
  \item[\ec7]
    the duration of the $\Lambda$-dominated phase.
    The part of the spectrum where the $n_s$ contribution is
    more evident is indeed the SW plateau.
    Here, however, a contribution from the LISW effect
    enhances the first multipoles.
    It depends on
    $\Omega_\Lambda/\Omega_m=\Omega_\Lambda/(1-\Omega_\Lambda)$ for
    a flat Universe:
    for a larger $\Omega_\Lambda$, the $\Lambda$-domination is
    longer and the LISW contribution is enhanced.
  \item[\ec8]
    the optical depth to reionization $\tau_\mathrm{re}$.
    Due to reionization, the behavior of the CMB spectrum
    at $l\gtrsim40$
    is different from that at $l\lesssim40$:
    the suppression at high $l$ depends on $\tau_\mathrm{re}$.
    This effect is not degenerate with the damping of acoustic
    oscillations, that affects only the multipoles starting
    from an higher $l$ and not in a constant way.
    If one considers the entire CMB spectrum, the step at $l\simeq40$
    breaks also the degeneracy with $A_s$.
\end{itemize}

The effects we listed do not take into account a number of other tiny
dependences that play a very small role in modeling the CMB
spectrum.
Some of these dependencies would concern the electron density $n_e$
and the redshift of recombination $\etals{}$,
that depends marginally on the baryon density and on
the primordial Helium fraction, usually denoted with $Y_p$.
These parameters affect the sound horizon at decoupling
(Eq.~\eqref{eq:comovSoundHorizon}),
the duration of the transition from equality to recombination and
the photon diffusion length (Eq.~\eqref{eq:comDiffDist}),
with also a small impact on the effects \ec1, \ec3 and \ec{4}.
The magnitude of these effects, however, is much smaller than the
magnitude of the primary effects \ec1--\ec8:
the baryon density impact through $z\lsu{}$ is much smaller than
its effect on the relative magnitude of the peaks,
and in the range currently allowed by the experimental
data the effect of $Y_p$ is negligible.
The approximation of considering a fixed recombination history,
therefore, is very strong for most of the purposes.

We listed eight different characteristics of the CMB spectrum
that can be controlled by only six parameters,
but until few years ago
the CMB measurements were not precise enough to strongly constrain
all the \lcdm~parameters, since
most of the effects listed above can be distinguished only with 
very precise measurements.
Cosmic variance at low-$l$ and instrumental noise at high-$l$ lead
to partial parameter degeneracies inside the experimental error.
The situation changed with the data release of the Planck experiment,
that measured the CMB spectrum in a wide range of multipoles,
obtaining an with unprecedented precision for the high-$l$ part of
the spectrum, up to $l\simeq2500$.
After having analyzed the full experimental data,
the Planck collaboration recently released the temperature and 
the polarization spectra, these latter ones measured
for the first time at high multipoles.
The joint analysis of temperature and polarization data allows
to reduce or break the degeneracies among the different parameters
and to improve the strength of the constraints on the cosmological parameters.
We will discuss in more detail the CMB experimental results in the
dedicated Section~\ref{sec:cmb}.

\section{Polarization spectra}
The CMB spectrum is not only characterized by temperature
fluctuations:
since Thomson scattering depends on the polarization of the photons,
when isotropy disappears at the time of recombination the quadrupole
momentum $\Theta_2(\eta,\vec x)$ of the growing anisotropies
is responsible for a net polarization of the scattered photons.
As a consequence, a polarization pattern appears on the
last scattering surface.
This is strongly correlated with the temperature pattern.

Photon polarization at last scattering can be detected as a vector
field on a sphere and can be decomposed in two modes:
% in analogy with electromagnetism,
% we have
an $E$-polarization (gradient field)
and a $B$-polarization (curl field) component.
As for temperature, it is possible to define an harmonic power
spectrum for the $E$ and $B$ modes auto-correlation and for the various
cross-correlation terms:
the different possibilities are given by
\be
  \cl{XY}
  =
  \langle \alm{X}\alm{*Y}\rangle\,,\qquad\forall m,
\ee
where $X,Y\in\{T,E,B\}$.

Polarization of the type $B$ is related to the gravitational waves
arising from inflation.
Gravitational waves are coupled only to species having
non-negligible tensor degrees of freedom, that are contained
in the non-diagonal part of the
spatial stress-energy tensor $\delta T\iju$.
These degrees of freedom vanish
for CDM, due to the smallness of the velocity dispersion,
and also for baryons and tightly-coupled photons,
due to the isotropic pressure enforced by interactions:
the only species coupled to gravitational waves are
photons, after decoupling,
and other collisionless species, before their
non-relativistic transition (neutrinos, for example).
The influence of neutrinos on tensor anisotropies was studied in
Ref.~\cite{Weinberg:2003ur} and implemented in
\camb~\cite{Lewis:1999bs}.
The neutrino contribution to CMB anisotropies, however,
can be only significant for modes crossing the horizon during
radiation domination or soon after matter-radiation equality,
i.e.~on small scales.

For parity invariance, the $TB$ and $EB$ cross-correlation spectra
are zero after the last scattering,
but they can be generated at the level of secondary anisotropies
through the weak lensing of last scattering photons.
Primary $B$ modes can be generated only if some tensor fluctuations
exist in the early Universe, and
they contribute to the CMB temperature spectrum only at small
multipoles (typically $l<150$).
Scalar fluctuations do not contaminate the primordial tensor
anisotropies, but
the main contribution to the $B$-modes auto-correlation
spectrum comes from a
leak from $E$- to $B$-type polarization driven by
gravitational lensing effects on small scales.
Consequently, the \cl{BB} spectrum is dominated by tensor
perturbations only at large scales.
Since the $B$-type polarization is subdominant with respect to
temperature and $E$-type polarization, the detection of the
contribution to the CMB spectra of primordial tensor perturbations
is a complicated experimental task.
We will discuss the current status of the experimental results
in Section~\ref{sec:cmb}.

The calculation of the spectra \cl{TE} and \cl{EE}, instead,
can be performed with the same procedure we presented for the
temperature anisotropies,
with the introduction of a new degree of freedom,
whose evolution can be described by a new Boltzmann equation.
The result of the calculation is a second hierarchy of
differential equations for polarization anisotropies,
coupled to the infinite set of equations describing
the temperature perturbations.
The contribution of polarization to the evolution of temperature
perturbations is small, so that our treatment of the
temperature perturbations is a very good approximation
of the full calculation.
We will not describe in details the calculation of the
polarization spectra, nor the different impact that
some physical effects, such as reionization, have on the $TE$ and
$EE$ spectra.
We conclude just remembering the importance of measuring and analyzing
the CMB polarization spectra to help removing parameter
degeneracies in the \lcdm~model.

% \input{structuresPK.tex}
%!TeX root=main.tex 
\chapter{Cosmological Measurements}\label{ch:cosmomeasurements}

% \begin{abstract}
This Chapter is devoted to describe all the cosmological measurements
that we will consider in our following analyses.
% \end{abstract}
We firstly review the status of CMB experiments
(Section~\ref{sec:cmb}), and then we present the other experimental data:
Baryon Acoustic Oscillations (BAO, Section~\ref{sec:bao}),
local measurements of the Hubble parameter $H_0$
(Section~\ref{sec:h0}),
distance calibration with the SuperNovae of type Ia
(Section~\ref{sec:sn}),
constraints on the matter power spectrum
(Section~\ref{sec:mpk}),
abundance of galaxy clusters
(Section~\ref{sec:cluster})
and cosmic shear observations
(Section~\ref{sec:shear}).

\section{Cosmic Microwave Background Radiation}\label{sec:cmb}
The CMB was discovered accidentally
by Penzias and Wilson in 1965~\cite{Penzias:1965wn},
who received the Nobel prize for their amazing discovery.
Since then the CMB science
had a terrific improvement.
The first detection of the CMB anisotropies above the dipole was
achieved by the COBE experiment in 1992 \cite{Smoot:1992td},
which stimulated a new generation of CMB detectors that
culminated with WMAP \cite{Bennett:2012zja} and Planck.
Most of the analyses we will present in the following chapters are based
on the measurements of the CMB anisotropies, mainly as detected by
the Planck satellite.
These results are described in Subsection~\ref{ssec:planck}.
We discuss also the results obtained by Earth-based high-precision
experiments such as ACT and SPT (Subsection~\ref{ssec:highl})
and the constraints from the $B$-mode polarization experiments,
such as the recent claims by BICEP2 and BICEP/Keck,
and the joint analysis presented by the BICEP/Keck and Planck collaborations
(Subsection~\ref{ssec:tensor}).

\subsection{Planck}
\label{ssec:planck}
Planck is a space-based mission designed to measure with
extreme accuracy the spectra of CMB anisotropies,
both in temperature and polarization.
Launched in 2009, Planck probes the microwave emission at nine
different frequencies, using two different instruments:
the Low Frequency Instrument (LFI) and
the High Frequency Instrument (HFI).
The different frequencies are used to separate
the foreground contributions, mainly coming from the Milky Way,
from the signal of the CMB.
HFI completed its survey in January 2012, while LFI collected data
until October 2013.
The data were analyzed and published in two branches.
The first release, in 2013, contained the data of the first 
15.5 months of operations \cite{Ade:2013sjv}.
With the second release in 2015 \cite{Adam:2015rua}
all the maps were published, but the analyses still requires further
studies of the polarization spectra
and a third version of the likelihood codes is expected.

\subsubsection{CMB temperature and polarization}
The Planck collaboration released the first public
data and codes in 2013 \cite{Ade:2013sjv}.
In this release, only the full temperature spectrum obtained
by the Planck data was presented \cite{Ade:2013kta}.
The CMB temperature auto-correlation spectrum is obtained from LFI
and HFI data using different methods for the low-$l$ and
for the high-$l$ part of the spectrum, that would require otherwise
an enormous computation time.
The spectrum at low multipoles, $2\leq l\leq49$, is obtained from the
maps between 30 and 353 GHz, using a fraction of sky equal to 91\%.
For the spectrum at high multipoles, $l\geq50$, the maps at 100, 143 and 217 GHz
were considered and a Gaussian approximation was adopted.
As the polarization data from the Planck satellite
were not satisfactory at the time
of the first release, the Planck collaboration decided to include
the WMAP polarization likelihood for the low multipoles
\cite{Bennett:2012zja, Hinshaw:2012aka} at $l\leq 23$ (denoted WP).

In the second data release \cite{Adam:2015rua},
the Planck collaboration presented the full mission data obtained
by the Planck satellite.
The analyses of the CMB maps to obtain the spectra and the likelihood
were also improved.
The second public likelihood code includes the $E$-mode polarization
through the $TE$ cross-correlation
and the $EE$ auto-correlation spectra.
The low-$l$ likelihood includes temperature and polarization up to
$l=29$, for a total sky fraction of 94\%,
and it is obtained from the 70 GHz (LFI) map,
cleaned with the measurements of the 30 GHz (LFI) and the 353 GHz (HFI)
for the polarized synchrotron and dust templates, respectively
\cite{Aghanim:2015xee}.
The high-$l$ part of the spectrum, instead, is obtained with the same
Gaussian approximation adopted in the first release, but for the
multipoles $30\leq l\leq2500$.
For the temperature spectrum,
the HFI maps at 100, 143 and 217 GHz were used with the
66\%, 57\% and 47\% of the sky retained, respectively.
For the polarization spectra, instead,
the same HFI maps were used with a fraction of sky of
70\%, 50\% and 41\%, respectively, to exclude the sky regions
where the dust signal is larger.

\subsubsection{CMB lensing}
The presence of large scale structures induces a dependency in the
CMB observables that is connected with gravitational lensing.
Late time geometry and clustering can then have an impact on CMB the
fluctuations, which in turn can be used to probe the strength
of the gravitational accretion after recombination.
Being originated at the last scattering, the CMB fluctuations are
more affected by the lensing due to structures at $z\simeq2$, that is
half-way to the last-scattering surface, while important effects
at low multipoles ($l\leq60$) are caused also 
by sources at smaller redshift.

Gravitational lensing in CMB maps is mainly observed as a smoothing
of the acoustic peaks and troughs in the
temperature and polarization maps,
a conversion from $E$- to $B$-mode polarization
and a production of late-time non-Gaussianities, that have the form of a
non-zero connected 4-point function.
The temperature and polarization likelihoods from Planck include 
the smoothing effect, that is then considered in all the analyses, but it
is possible to study separately the measurements of the
power spectrum $\cl{\phi\phi}$, where $\phi$ is the lensing potential.
This spectrum is extracted from the 4-point correlation functions involving both
temperature and polarization, as discussed in 
Refs.~\cite{Ade:2013tyw} and \cite{Ade:2015zua}
for the 2013 and 2015 releases, respectively.
The power of these lensing measurements is that they allow 
to constrain the late-time expansion, the geometry and 
the clustering of matter using CMB data alone.

The CMB lensing likelihood is constructed as
a simple Gaussian approximation of the estimated $\cl{\phi\phi}$,
covering the multipole range $40\leq l\leq400$.
The lower limit of this interval is conservatively chosen
in order to avoid problems in
the difficult reconstruction of the lensing potential at large scales,
that is the consequence of the large ``mean-field'' due to survey
anisotropies.
A less conservative choice could involve multipoles starting
from $l=8$.
The upper limit, instead, is fixed to exclude the multipoles
at which there is a marginal evidence of residual systematics
in the reconstruction of the lensing deflections from the temperature
maps only \cite{Ade:2015zua}.

\subsection{High-multipoles Experiments}
\label{ssec:highl}
The advantage of earth-based CMB missions is that they have
an higher angular resolution, but they are limited
in the sky coverage.
Detections from earth-based experiments can help to study the
high-$l$ tail of the CMB spectrum after appropriate calibrations
with the low-$l$ spectrum observed by space-based missions.
The study of the high-$l$ tail of the CMB spectrum allows to
constrain better the nuisance parameters used in the likelihood codes
to model some
unresolved foreground contributions, such as the kinetic SZ effect.
In part of the following analyses we will consider the results presented
by the two experiments ACT and SPT, that we introduce now.

The Atacama Cosmology Telescope (ACT) was settled in the Chilean
Andes and it mapped the sky in two distinct regions:
the equatorial stripe (ACTe) along the celestial equator
and a stripe along $-55^\circ$ of declination, that is
called the southern stripe (ACTs).
Observations lasted from 2007 to 2010 and covered approximately
600~deg$^2$ of sky.
The ACT survey covered the multipole ranges
$540\leq l\leq9440$ at the frequency of 148~GHz and
$1540\leq l\leq9440$ at the frequency of 218~GHz
\cite{Das:2013zf}.

The South Pole Telescope (SPT)
observed a different portion of sky of 2540~deg$^2$
at $2000\leq l\leq11\,000$ \cite{George:2014oba}.
In our analyses we used the incomplete results obtained for 
$650\leq l\leq3000$ at the frequency of 150~GHz
\cite{Keisler:2011aw, Story:2012wx}
and for
$2000\leq l\leq10\,000$ at the frequencies of 95, 150 and 210~GHz,
obtained observing a region of 800~deg$^2$
\cite{Reichardt:2011yv}.

For the ACT/SPT data we use the prescriptions
and the likelihood code described
in Ref.~\cite{Dunkley:2013vu}.

\subsection{Tensor Perturbations}
\label{ssec:tensor}
We stated that the search of primordial tensor perturbations
is of crucial importance for studying inflation.
The tensor fluctuations, however, have an amplitude that is suppressed
with respect to scalar fluctuations, and therefore it is much more difficult
to detect them experimentally.

In 2014, the BICEP2 experiment reported the first claim
\cite{Ade:2014xna} for the detection of a signal of $B$-mode polarization
anisotropies, that they associated to primordial tensor modes.
If the BICEP2 signal had been caused by the existence of primordial
gravitational waves, a preference for a tensor-to-scalar ratio
$r\simeq0.2$ would have been reported,
in apparent conflict with the Planck and WMAP
constraints $r\lesssim0.1$ \cite{Bennett:2012zja,Ade:2013zuv,Ade:2015xua},
that however are highly model dependent.
A subsequent study performed by the Planck collaboration showed
that the BICEP2 experiment observed a region were a non-negligible
contamination from dust emission was present \cite{Adam:2014bub}.
After some months the BICEP collaboration presented updated results
that consider in addition of the data taken by the Keck array
\cite{Ade:2015fwj,Array:2015xqh}. 
The signal of the existence
of $B$-modes was reported by the BICEP/Keck (BK) collaboration,
but not in association with their possible primordial origin.
In fact, a joint analysis of the BICEP/Keck and Planck collaborations
\cite{Ade:2015tva} finally demonstrated the dust origin of the
measured $B$-modes.
After removing the dust contribution, the signal of primordial tensor
modes disappears and the constraints on the tensor-to-scalar ratio 
are compatible with $r=0$.
As a consequence, we do not have
any evidence that inflationary tensor modes exist.

Several other experiments aim to measure the signals of primordial
gravitational waves.
The largest contribution to the $B$-mode polarization spectrum, however,
comes from the leak of $E$-mode polarization, that are partially
converted into $B$-modes through gravitational lensing.
The first detection of the lensing $B$-mode spectrum comes from the
SPTPol experiment \cite{Story:2014hni,Keisler:2015hfa}.
This detection is not important for constraining inflation, since
it does not concern primordial tensor modes,
but it confirms the predictions of General Relativity
about gravitational lensing.

\section{Baryon Acoustic Oscillations}\label{sec:bao}
BAO measurements and their implications in cosmology
has been reviewed in Refs.~\cite{Bassett:2009mm,Aubourg:2014yra}.
We present a less detailed treatment for length purposes.

Acoustic oscillations imprint a characteristic scale
in the clustering of matter,
providing a cosmological \emph{standard ruler}
that can be measured in the
power spectrum of CMB fluctuations and of large-scale structures,
at small redshift
\cite{Sakharov:1966aja,Peebles:1970ag,Sunyaev:1970a,
Blake:2003rh,Seo:2003pu}.
The BAO distance is computed from first principles, differently than
the distance measurements that involve SN Ia,
which are calibrated against
objects in the local Universe
\cite{Hamuy:1996su,Riess:1998cb,Perlmutter:1998np}.
The sharpening of the BAO precision at higher redshifts,
the difference between absolute and relative measurements and the
completely independent systematic uncertainties make
the BAO and SN Ia methods
highly complementary tools to measure the cosmic expansion history and
to test DE models.
Combining the SN Ia results for relative distances
and the BAO measurements,
it is possible to derive constraints on $H_0$ using an
\emph{inverse distance ladder}.
This essentially requires to use the SN Ia data
to transfer the information on the absolute calibration 
of the BAO scale
from the intermediate redshifts,
where it is measured with high precision,
to $z=0$.
The line-of-sight detection of BAO, indeed, allows
to obtain a direct determination of the
expansion rate $H(z)$ at the probed redshift,
in addition to the transverse
direction detection that allows to obtain
the angular diameter distance $d_A(z)$.

\subsection{BAO Physics}

\begin{figure}[p]
\includegraphics[width=\halfwidth]{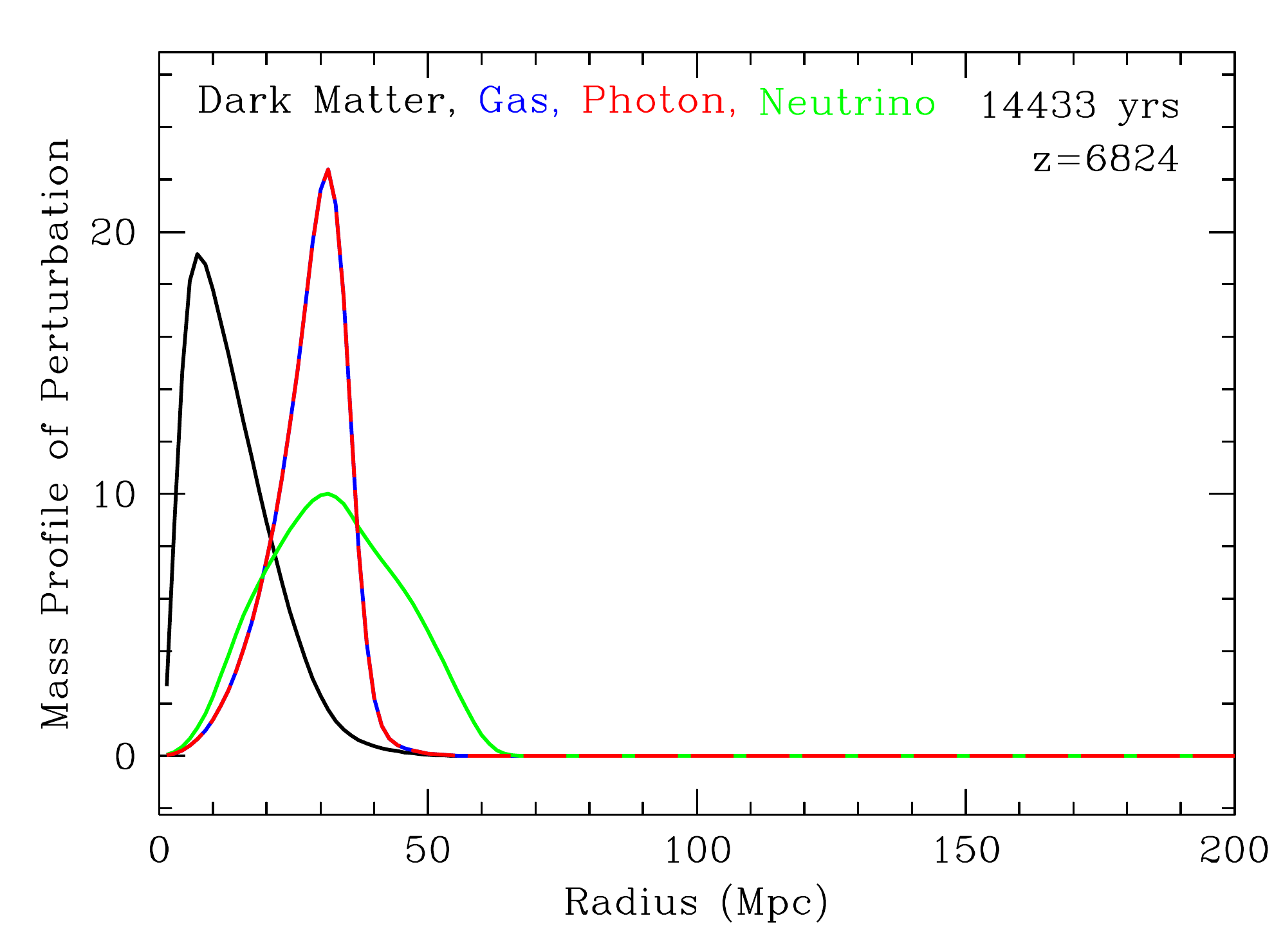}
\includegraphics[width=\halfwidth]{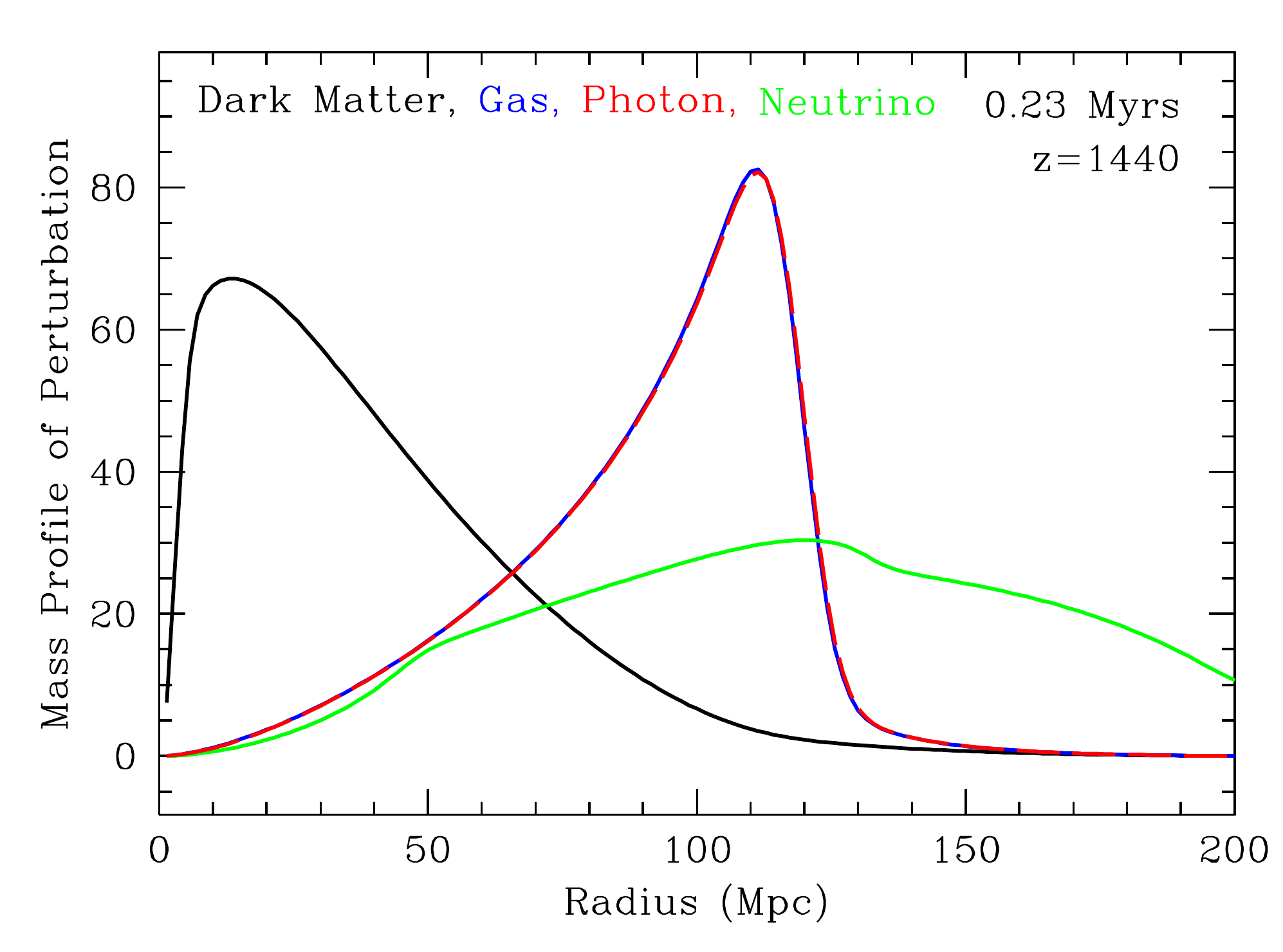}

\includegraphics[width=\halfwidth]{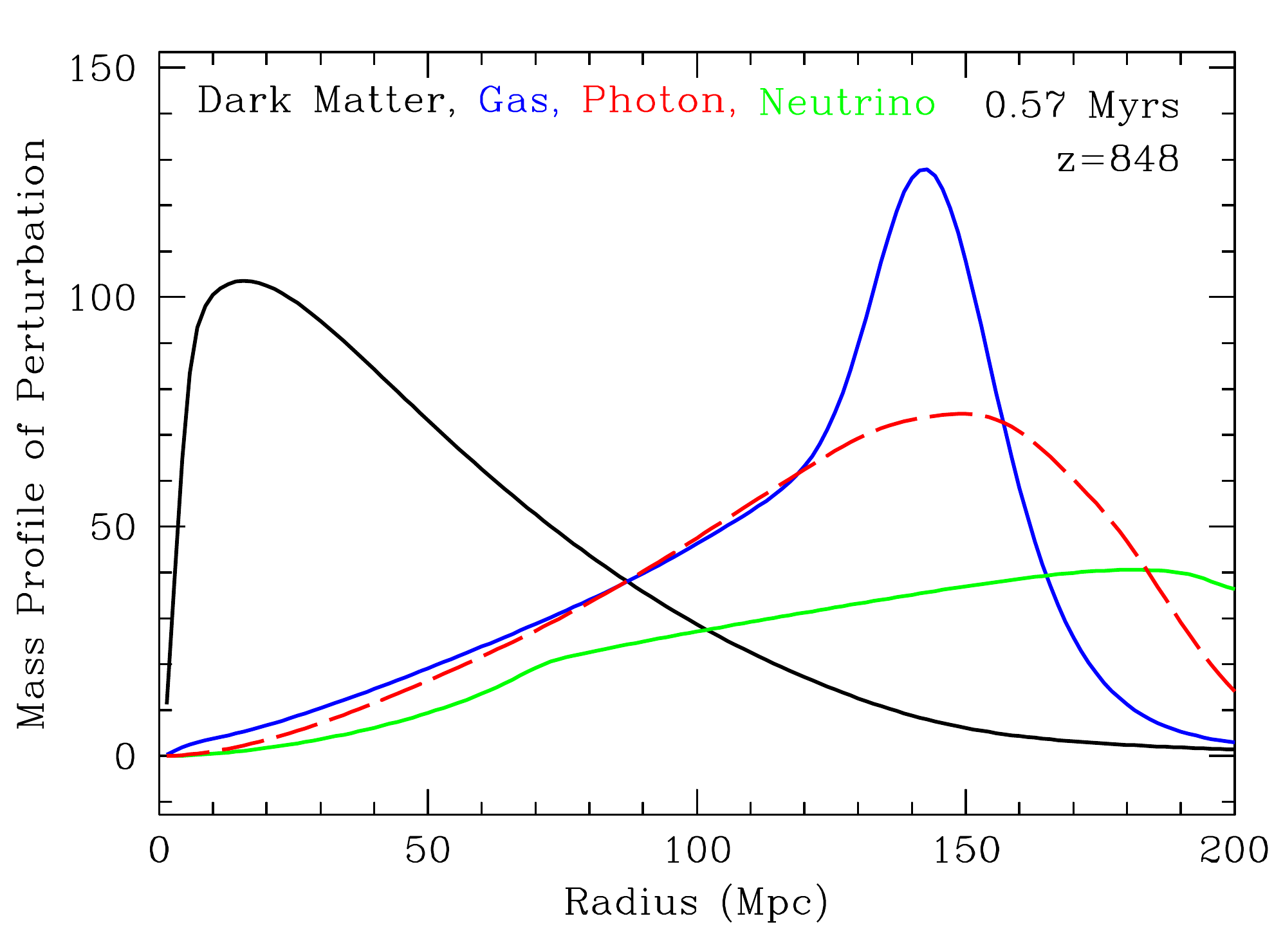}
\includegraphics[width=\halfwidth]{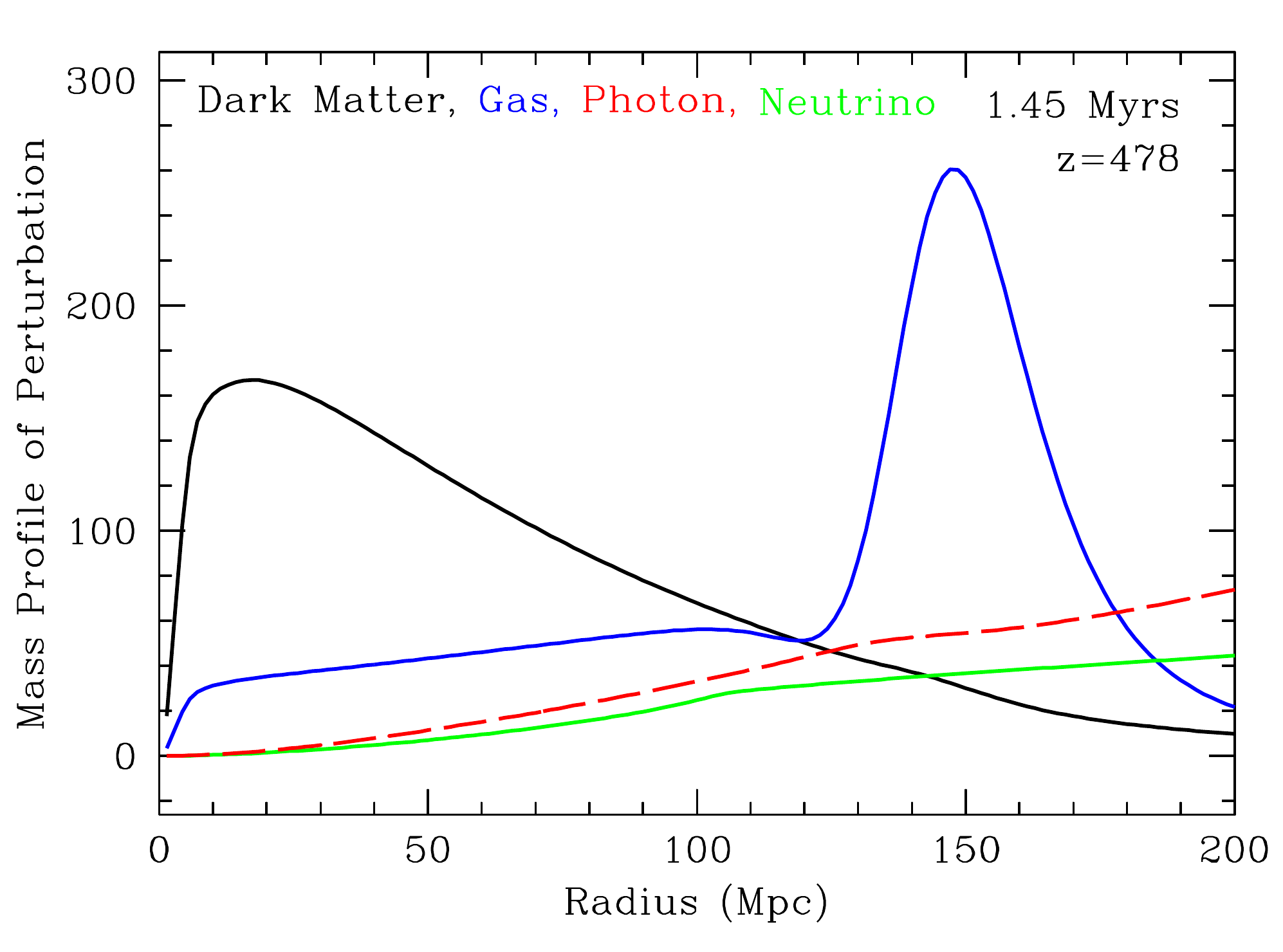}

\includegraphics[width=\halfwidth]{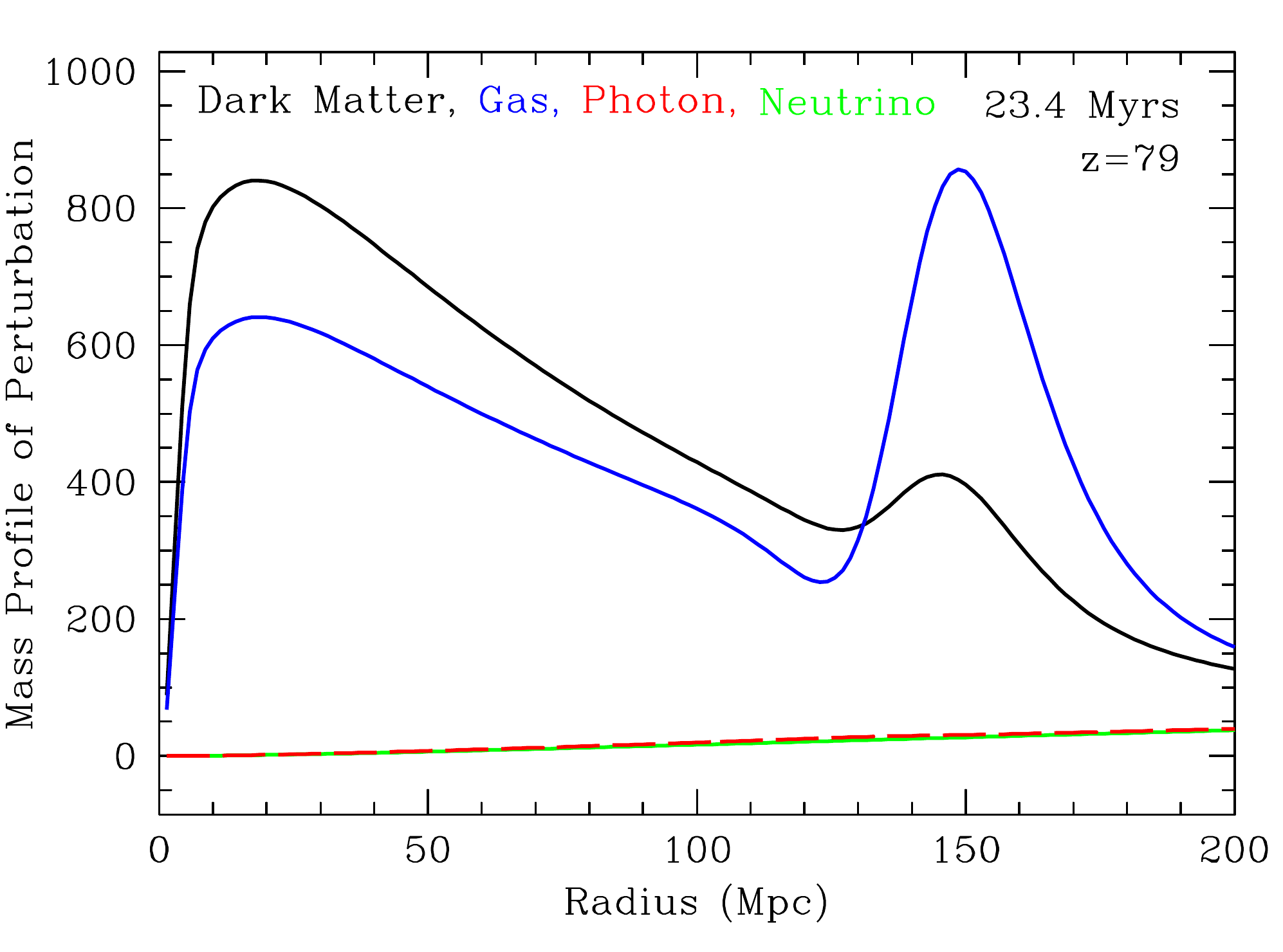}
\includegraphics[width=\halfwidth]{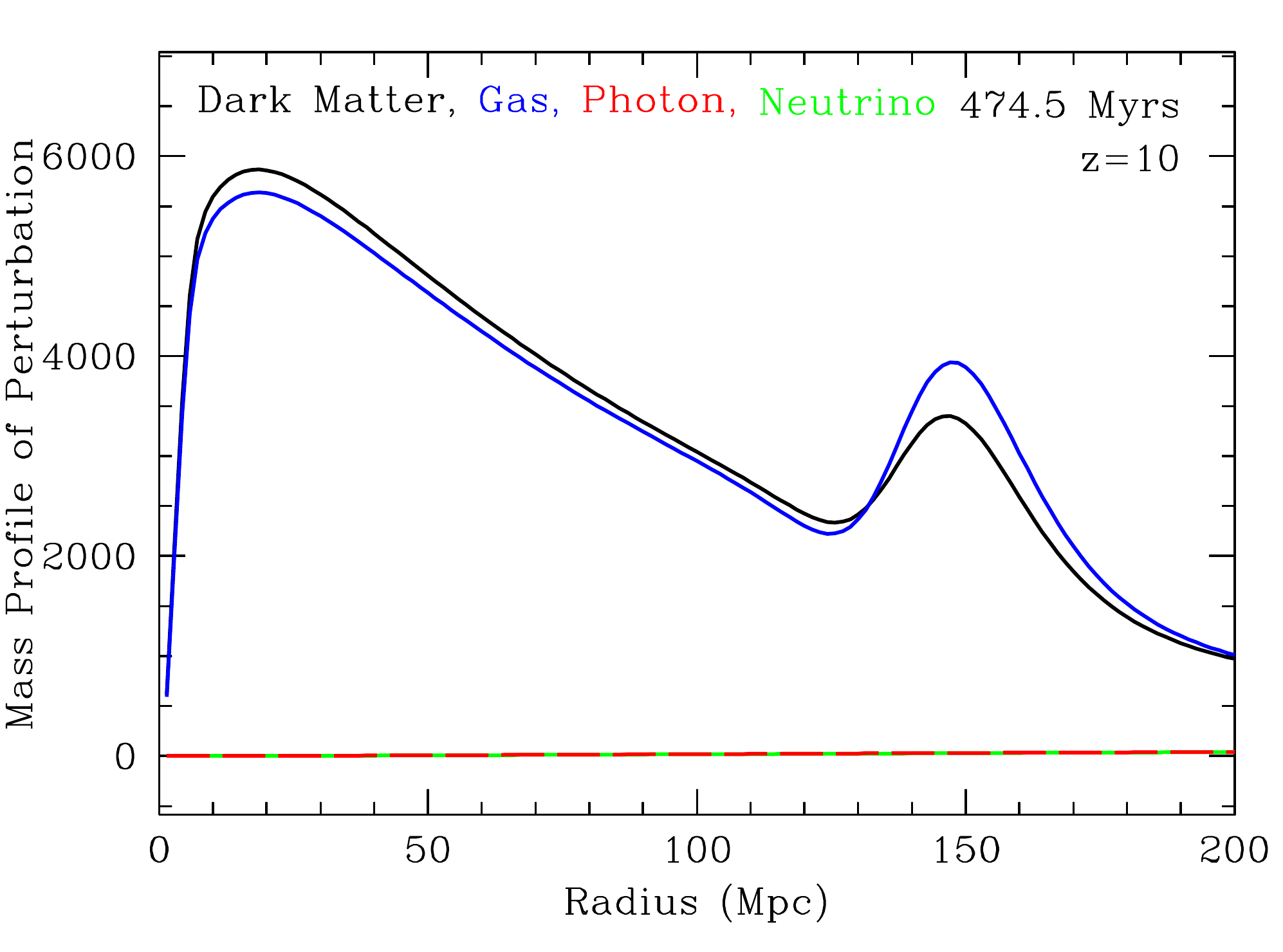}
\caption[Snapshots of evolution of the radial mass profile
versus comoving radius of an initially point-like overdensity
located at the origin]
{\label{fig:bao_mpk}
Evolution of the radial mass profile versus the comoving
radius of an initially point-like overdensity located at the origin.
The perturbations in
dark matter (black),
baryons (blue), 
photons (red) and 
neutrinos (green)
are plotted.
The perturbations are evolved 
from early times ($z=6824$, top left)
to $z=10$ (bottom right), long after the recombination.
At the initial time, photons and baryons travel outwards
as a pulse.
The drag of the coupled baryon-photon fluid
on dark matter is only gravitational and
it produces a delayed enhancement of the 
cold dark matter profile (top right).
The photon and the baryon profiles decouple at recombination
(middle left),
when the photons leak away from the baryons.
After recombination is complete (middle right),
the photons continue to free-stream away.
Gravitational instability now takes over,
and the overdensities start to grow (bottom left).
Dark matter pulls baryons towards the central overdensity,
while baryons drag dark matter towards the overdensity at 150~Mpc,
that is still visible in the mass profile at late times
(bottom right).
From Ref.~\cite{Eisenstein:2006nj}.
}
\end{figure}

The imprint of the BAO is visible in the matter power spectrum at
late times.
We discussed in the previous Chapter
the coupled baryon-photon oscillations,
consequence of the competing forces
of gravity and radiation pressure.
We want now to qualitatively describe how they influence the matter
distribution at late times.
For this purpose, we can consider a
single, spherical density perturbation
that propagates outwards in the tightly coupled fluid
as an acoustic wave
with a speed $c_s$, written in Eq.~\eqref{eq:soundspeed}.
Fig.~\ref{fig:bao_mpk}, from Ref.~\cite{Eisenstein:2006nj},
contains useful plots to help visualizing
the phenomena we are going to describe.
At the beginning, the photon (red line) and baryon (blue line) 
perturbations move together,
dragging the dark matter (black line) perturbations 
through gravity (top right panel).
Matter perturbations moves also outwards, but delayed (top left panel)
because the interaction is only gravitational, while
photons and baryons interact mainly electromagnetically.
At the time of photon decoupling, the radiation pressure on the baryons
disappears and the baryon wave stalls (middle left panel).
Neutrinos (green line),
that are already decoupled, and photons free stream away,
forming the Cosmic Neutrino Background (CNB) and
the CMB radiation (middle right panel).
The characteristic radius of the spherical shell formed
by the stalled baryon wave is imprinted in the baryon density
as a significant excess at this time.
From now on, the gravitational interaction is the only force
that drives the evolution, affecting dark matter and baryons.
Since the dark matter and the baryon profiles are peaked at different radii,
what happens is that the dark matter pulls baryons towards
the peak in the origin, 
while baryons continue to drag the dark matter towards the overdensity
at $\sim$150~Mpc (bottom left panel).
The final profiles have a significant overdensity near the center and
a smaller peak (bottom right panel)
at the scale $r_\mathrm{drag}$,
that is the sound horizon at the end of the baryon drag.

As it is impressed in the dark matter and baryon distributions,
the slight excess at $r_\mathrm{drag}$ appears
also in the distribution of galaxies
we can observe today.
As firstly suggested in Ref.~\cite{Eisenstein:1998tu}, this feature
can be used to constrain the cosmological parameters.
To do this, one has to reconstruct the typical distance that,
at each different redshift, has the role of a statistical standard ruler.
The easiest way to distinguish it is through the two-point correlation function
or through its Fourier counterpart, the power spectrum.
In the power spectrum, the excess
at the galaxy separation distance  $r_\mathrm{drag}$ appears
in the forms of oscillations, that are typically easy to recognize.
The line-of-sight and the tangential Fourier oscillating modes
can be measured separately,
so that both the Hubble parameter $H(z)$ and
the angular diameter distance $d_A(z)$ can be measured.
The experimentally observed modes, however,
contain components of both,
and consequently $H(z)$ and $d_A(z)$ are partially anti-correlated.

For the BAO analyis,
the situation is different from the analyses where the data points
are fixed and the different models must be fitted.
To compute the data points, indeed,
a redshift-distance relation must be assumed to convert
the points from the redshift space,
after which the power spectrum is constructed,
to the physical space, and vice versa.
The dependence on the fiducial model
is typically ignored in the analyses involving BAO data, since
the results are almost insensitive to this choice,
if one does not range far from the assumed fiducial model.
We will explore this point more in details in the next subsection,
where we will discuss the analyses of the BAO data.

\subsection{BAO Analysis}
In a flat Universe,
the comoving angular diameter distance $d_M(z)$ is
\be\label{eq:d_M}
  d_M(z)
  =
  \frac{c}{H_0}\int_0^z dz'
  \frac{H_0}{H(z')}\,.
\ee
The comoving angular diameter distance $d_M$
must not be confused with the
proper angular diameter distance $d_A$
we introduced in Section~\ref{sec:hubble}.
They are related by $d_M=d_A (1+z)$.
From Eq.~\eqref{eq:da_dl} we obtain that
the luminosity distance $d_L$, relevant to supernovae, is
$d_L=d_M (1+z)$.

The Hubble factor in Eq.~\eqref{eq:d_M}
can be calculated from the Friedmann
equation~\eqref{eq:freq1} with the contributions
of the energy densities of
all the species existing in the Universe.
The densities of CDM and baryons scale as $a^{-3}$,
while that of DE depends
on the equation of state, being proportional to $a^{-3(1+w)}$.
We will discuss in more details the aspects of neutrino cosmology
in the dedicated Chapter~\ref{ch:nu},
but we anticipate here that 
the most complicate energy density dependence to be obtained
is for the energy density of massive neutrinos,
since they behave differently when relativistic or non-relativistic.
The contribution of neutrinos,
photons and other relativistic particles can be written as
\cite{Aubourg:2014yra}
\be
  \Omega_{\nu+r}(a)=
  \frac{C}{H_0^2}
  \left[
    T_\gamma^4
    +
    T_\nu^4
    \sum_i
    I(m_i /k_B T_\nu)
\right]\,,
\ee
where $C$ is a normalization constant, obtained as a combination of
fundamental constants,
and $k_B$ is the Boltzmann constant.
The photon temperature $T_\gamma$ scales with $a^{-1}$, as the
neutrino temperature $T_\nu=T_\gamma(4/11)^{1/3} g_s$.
The factor $g_s=(3.046/3)^{1/4}$
encodes the small reheating of the
neutrinos at the electron decoupling.
The integral $I$ is defined as
\cite{Aubourg:2014yra}
\be
  I(r)=
  \frac{15}{\pi^4}
  \int_0^\infty dx
  \;
  x^2 \frac{\sqrt{x^2+r^2}}{e^x+1}
\ee
and must be evaluated separately for the different neutrino mass eigenstates.
For massless neutrinos $I(0)=7/8$, while for heavy neutrinos ($r\gg1$)
it tends to $I(r)\simeq 45\,\zeta(3)\,r/(2\pi^4)$,
where $\zeta$ is the Riemann function.
In the limit $r\gg1$, the integral $I(r)$ scales with $a$ and 
the energy density scales correctly with $a^{-3}$,
as for pressureless matter (CDM, baryons).

The BAO scale is set by the radius of the sound horizon
at the time of photon decoupling
(the end of the baryon drag), $r_\mathrm{drag}$%
% \footnote{This is not the same quantity we denoted with $r_d$
% in Chapter~\ref{ch:cmbr},
% where $r_d$ was the comoving diffusion distance
% of the photons.}%
,
that can be written as 
\be\label{eq:radSHdrag}
r_\mathrm{drag}=\int^\infty_{z_\mathrm{drag}}\frac{c_s(z)}{H(z)}dz\,,
\ee
where $z_\mathrm{drag}$ is the redshift of the drag epoch,
when photon-baryon decoupling occurred, and the sound speed $c_s$ is
defined in Eq.~\eqref{eq:soundspeed}.
The definition in Eq.~\eqref{eq:radSHdrag}
is sufficiently accurate for 
reasonable variations of the fiducial model, but it must be evaluated
numerically with a full Boltzmann code computation
to obtain very precise BAO measurements.

The robustness of BAO measurements comes from
the fact that a sharp feature
in the correlation function cannot be mimicked
by any kind of systematics.
The BAO scale is determined assuming a set of fiducial 
parameters in the cosmological model,
to define the redshift-distance relation.
In an isotropic fit, that does not distinguish the directions
parallel and perpendicular to the line-of-sight,
the measurement is encoded in the parameter $\alpha$,
that is the ratio of the measured BAO scale
divided by the one predicted by the fiducial model.
In an anisotropic fit, instead,
the ratios perpendicular and parallel to the
line of sight, $\alpha_\perp$ and $\alpha_\parallel$,
must be considered separately.
The errors on $\alpha_\perp$ and $\alpha_\parallel$
are usually correlated within
the same redshift slice in a real survey,
but they are uncorrelated across different redshift slices.
Even if the values of $\alpha$ are derived
within a fiducial model,
the BAO feature is independent of the choice of the fiducial model,
within a reasonable range.

While the various $\alpha$ are determined for a specific fiducial model,
the conversion to any other model is straightforward.
In an anisotropic fit,
a measurement at redshift $z$ of the parameter $\alpha_\perp$
constrains the ratio of the comoving angular diameter distance
to the sound horizon at the same redshift:
\be
  \frac{D_M(z)}{r_\mathrm{drag}}
  =
  \alpha_\perp
  \frac{D_{M,\mathrm{fid}}(z)}{r_{\mathrm{drag},\mathrm{fid}}}\,,
\ee
while a measurement of the parameter $\alpha_\parallel$ constrains
the Hubble parameter $H(z)$:
\be
  \frac{D_H(z)}{r_\mathrm{drag}}
  =
  \alpha_\parallel
  \frac{D_{H,\mathrm{fid}}(z)}{r_{\mathrm{drag},\mathrm{fid}}}\,,
\ee
having defined
\be
  D_H(z)=1/H(z)\,.
\ee

In the isotropic case, instead, the analysis measures a combination
of these distances.
If the redshift-space distortions are weak, 
the constrained quantity is the volume averaged distance $D_V$,
defined as
\be
  D_V(z)=[z D_M^2(z) D_H(z)]^{1/3}\,.
\ee
The constraint from the isotropic fit is then:
\be
  \frac{D_V(z)}{r_\mathrm{drag}}
  =
  \alpha
  \frac{D_{V,\mathrm{fid}}}{r_{\mathrm{drag},\mathrm{fid}}}\,.
\ee

The BAO measurement allows to constrain the cosmological parameters
through their impact on the sound horizon radius $r_\mathrm{drag}$
and on the distances $D_H$ and $D_M$.
For standard cosmological models, the error on $r_\mathrm{drag}$ as obtained
from the CMB analyses is small with respect to the errors on
the BAO measurements, so the constraints come mainly from the
distances $D_M$ and $D_H$ (or $D_V$ for the isotropic analyses).
We show in Table~\ref{tab:baodata} the results in terms of
$D_M/r_\mathrm{drag}$, $D_H/r_\mathrm{drag}$ or $D_V/r_\mathrm{drag}$ for the different experiments
we will consider in the cosmological analyses 
presented in the following Chapters.
The quoted redshift is usually an effective redshift,
determined using the statistical contributions of each sample
to the BAO measurement.
Since the anisotropic analyses yields to anti-correlated errors on
$D_M$ and $D_H$, the last column of Tab.~\ref{tab:baodata} contains
the correlation coefficient in the relevant case.

\begin{table*}
  \centering
  \begin{tabular}{c|c|cccc} 
   Name & Redshift 
      & $D_V/r_\mathrm{drag}$ & $D_M/r_\mathrm{drag}$ & $D_H/r_\mathrm{drag}$ & $r_{\mathrm off}$ \\
   \hline
   SDSS (DR7) & 0.35 & $8.88 \pm 0.17$ & & & \\
   6dFGS & 0.106 & $3.047 \pm 0.137$ & -- & -- & -- \\ 
   MGS   & 0.15  & $4.480 \pm 0.168$ & -- & -- & -- \\
   \hline
   BOSS DR9 & 0.57 & $13.67\pm0.22$ & & & \\
   BOSS DR11 \texttt{LOWZ}  
      & 0.32 & $8.467 \pm 0.167$ &      --            & --                &   --    \\
   BOSS DR11 \texttt{CMASS} 
      & 0.57 &   --              & $14.945 \pm 0.210$ & $20.75 \pm 0.73 $ & $-0.52$ \\
\hline
  \end{tabular} 
  \caption[BAO constraints used in the analyses.]%
  {\label{tab:baodata}
    BAO constraints used in the following Chapters.
    These values are taken from
    \cite{Padmanabhan:2012hf} (SDSS DR7),
    \cite{Beutler:2011hx} (6dFGS),
    \cite{Ross:2014qpa} (MGS),
    \cite{Anderson:2012sa} (BOSS DR9),
    \cite{Anderson:2013zyy} (BOSS DR11).
    }
\end{table*}

\subsection{BAO measurements}
The most precise BAO measurements
today come from the analysis of the
Baryon Oscillation Spectroscopic Survey (BOSS)
DR12 galaxy sample \cite{Cuesta:2015mqa,Gil-Marin:2015nqa},
that is the final BOSS release.
BOSS uses the same telescope of the original 
Sloan Digital Sky Survey (SDSS),
with improved spectrographs.
The total sample is composed of two distinct subsets of galaxies,
selected by different color cuts and luminosity fluxes:
the \texttt{CMASS} sample within $0.43<z<0.7$,
corresponding to an approximately constant
threshold for the galaxy stellar masses, and
the \texttt{LOWZ} sample, in the range $0.15<z<0.43$.
Both the samples are analyzed with reconstruction algorithms
in order to partly revert the non-linear effects and to improve the
measurement precision.
In part of the analyses presented in the next Chapters we will use the
former BAO data obtained from the BOSS samples as presented in the
DR9 \cite{Ahn:2012fh,Anderson:2012sa} and
DR11 \cite{Anderson:2013zyy} releases (Tab.~\ref{tab:baodata}).
In DR9, the \texttt{CMASS} statistics was not sufficient to perform
an anisotropic analysis, as it has been done in DR11 and DR12 instead.
Since the \texttt{LOWZ} sample is smaller,
only in the DR12 the anisotropic
analysis has been performed on it,
while in DR11 the results were firstly reported
only for the isotropic fit.

Part of the analyses presented in the next Chapters involves
other BAO measurements from the SDSS,
namely the SDSS DR7 isotropic results
\cite{Abazajian:2008wr,Percival:2009xn,Padmanabhan:2012hf}
and the recent re-analysis of the
SDSS main galaxy sample (MGS) data \cite{Ross:2014qpa},
that uses reconstruction to improve the former BAO measurement.
Further BAO measurements include the results from the 
Six Degree Field Galaxy Survey (6dFGS)
\cite{Jones:2009yz,Beutler:2011hx},
which carries small statistical weight
due to the less precise constraints,
and the results from the WiggleZ survey
\cite{Blake:2011en,Parkinson:2012vd}, which sample a fraction of sky
that partly overlaps with the BOSS volume.
Due to the overlap with the more precise BOSS data,
we do not consider the WiggleZ BAO measurements in our calculations.

We will not discuss, finally, the constraints on the BAO feature
at high redshifts, $z>2$, which can be obtained from the
auto-correlation of the Lyman-$\alpha$ forest fluctuations
in the spectra of high-redshift quasars.
The first detection of the BAO scale from the Lyman-$\alpha$ forest
was firstly obtained by BOSS DR9
\cite{Busca:2012bu,Slosar:2013fi,Kirkby:2013fh},
following the pioneering work \cite{Slosar:2011mq}.

\subsection{Redshift-Space Distortions}
\label{ssec:rsd}
\newcommand{\psgg}{\ensuremath{P^s_{gg}}}
\newcommand{\prgg}{\ensuremath{P^r_{gg}}}
\newcommand{\prmm}{\ensuremath{P^s_{mm}}}
\newcommand{\prgt}{\ensuremath{P^r_{g\theta}}}
\newcommand{\prtt}{\ensuremath{P^r_{\theta\theta}}}

% See \cite{Percival:2011abc}
The growth rate of the cosmic structures is a strong test
for discriminating between different cosmological models.
The evolution of these structures takes place in a Universe
where all the material moves within the comoving frame,
so that also the galaxies follow this peculiar velocity field.
The observed galaxy redshift depends both on the peculiar velocities
of the objects and on the global recessional velocity induced
by the Hubble flow.
If only the Hubble flow is considered when converting
from redshifts to distances, the local velocities cause
a distortion of the redshift reconstruction.
These distortions are referred to as
Redshift-Space Distortions (RSD,
see e.g.~Ref.~\cite{Percival:2011abc}).
RSD are more important for near objects,
since the velocity caused by the
Hubble flow is small and the peculiar motions can be relevant.
% Following Ref.~\cite{Percival:2011abc},
% we will adopt the ``plane-parallel'' approximation that
% the observed galaxy are sufficiently far away so that their
% separations are small when compared to the distances between them.

In the context of the standard General Relativity predictions
for the growth rate, it is possible to derive a relation at linear
order between the
redshift-space galaxy power spectrum $\psgg$ and
the real-space matter power spectrum $\prmm$.
This relation includes a dependence on the angle to the line of sight
\cite{Kaiser:1987a,Hamilton:1997zq}:
\be\label{eq:rsd1}
  \psgg(k,\mu)
  =
  \prmm(k)(b_\delta+b_v f \mu^2)\,,
\ee
where $b_\delta$ accounts for a linear deterministic bias
between galaxy and matter overdensity fields,
$b_v$ allows for a linear bias between galaxy
and matter velocity distributions, usually assumed to be one,
$f$ is the logarithmic derivative of the growth factor
with respect to the scale factor and
$\mu$ is the cosine of the angle to the line of sight.
We learn from Equation~\eqref{eq:rsd1} that
the component owing to RSD depends only on cosmological quantities: 
the growth rate, depending on the redshift,
and the amplitude of matter fluctuations at a given time.
It has been shown that the parameter combination $f(z)\sigma_8(z)$ 
is a good discriminant between models of modified gravity
that can be tested with RSD
\cite{Song:2008qt}
The parameter $\sigma_8$ is the root mean square of the amplitude of
matter fluctuations inside a sphere of 8$h^{-1}$~Mpc radius.

% Equation~\eqref{eq:rsd1} is not suitable to perform a non-linear evolution,
% since it is based on the assumption that the growth factor is scale-independent.
% Dropping this assumption, we can write a relation involving also
% the galaxy-galaxy spectrum \prgg,
% the galaxy-velocity divergence cross-correlation spectrum \prgt\
% and the velocity divergence auto-correlation spectrum \prtt,
% all in the real space:
% \be
%   \psgg(k,\mu)
%   =
%   \prgg(k)
%   -2\mu^2\prgt(k)
%   +\mu^4\prtt(k)\,,
% \ee
% where $\theta=\nabla\cdot u$ is the divergence of the 
% galaxy velocity field $u$.

The dominant non-linear contribution to the RSD signal,
at small scales,
is due to the peculiar motions of the galaxies inside the DM halos.
The peculiar velocities can be large enough that,
when misinterpreted as Hubble velocities,
lead to a stretching of the galaxy clusters reconstruction
in the real space along the line-of-sight.
The shape of the cluster in the real space
after the wrong reconstruction is
referred to as ``Fingers of God'' (FoG).
This effect can be approximated with an additional term
in Eq.~\eqref{eq:rsd1}
that reduces the power at small scales.
The approximations, however, are not very accurate and
a precise description still requires the higher-order solutions
in perturbation theory
\cite{Peacock:1996ci,Scoccimarro:2004tg,
Jennings:2010uv,Taruya:2010mx}.

RSD are related to distance measurements and not to angles,
but the distortions may affect also angles reconstruction.
This happens for example when determining
projected angular clustering of galaxies,
if the samples are selected using redshift-dependent quantities.
In general, clusters and voids within a sample tend to
``push-in'' and ``push-out'' the near galaxies, respectively,
so that both positive and negative overdensities are increased,
with a consequent distortion of the reconstructed power spectrum.

Currently, the most recent constraints on the RSD signal come
from the BOSS
experiment we mentioned in the discussion dedicated to BAO.
In particular, the last results come from
BOSS DR12 \cite{Gil-Marin:2015sqa},
but in our analyses we shall use the results given
by the analysis of the
BOSS DR11 data, presented in Ref.~\cite{Samushia:2013yga}.
Other experiments that presented results on the RSD are
6dFGS \cite{Beutler:2012px},
WiggleZ \cite{Blake:2011rj},
BOSS-\texttt{CMASS} with other different analysis methods
\cite{Beutler:2013yhm,Reid:2014iaa}
and the VIMOS Public Extragalactic Redshift Survey (VIPERS)
\cite{delaTorre:2013rpa}.

\section{Hubble parameter}\label{sec:h0}
We include in some of our analyses the constraints on the Hubble parameter $H_0$,
the expansion rate of the Universe today,
as determined in the local Universe.

The Hubble parameter can be constrained by
CMB observations in the context of the \lcdm~model.
The bounds on $H_0$ from CMB are typically
lower than the local measurements \cite{Ade:2013zuv, Ade:2015xua}.
One must remember that $H_0$ constraints from CMB are derived results and
they are considerably model dependent,
but they have the advantage of not suffering the existence of systematics
in the measurement. 
The most recent
Planck result in the \lcdm~model is $H_0=67.3\pm1.0\Hou$,
obtained using CMB temperature autocorrelation and polarization
on large scales only \cite{Ade:2015xua}.
Let us emphasize, however,
that the Planck value of $H_0$ reported above
has been obtained assuming the standard $\Lambda$CDM cosmological model.
If one extends the \lcdm\ model, the results for $H_0$ can change
significantly.
For example, if one considers as an additional parameter
the effective number of relativistic degrees of freedom $\Neff$
that we will introduce in Chapter~\ref{ch:nu},
the analysis of CMB data
lead to\footnote{
See page 185
of the tables with 68\% limits available at
\url{http://wiki.cosmos.esa.int/planckpla/index.php/File:Grid_limit68.pdf}.
}
$
H_0 = 68.0^{+2.6}_{-3.0} \, \text{km} \, \text{s}^{-1} \, \text{Mpc}^{-1}
$.

The cosmological constraint can be compared with the results obtained
by local determinations, 
that in turn can suffer the existence of unaccounted systematics,
but do not depend on a specific cosmological model.
Using the SN~Ia detected by HST, with Cepheid-calibrated distances,
the authors of Ref.~\cite{Riess:2011yx} found $H_0=73.8\pm2.4\Hou$. 
Using the same SN Ia set with different calibrations for the distance 
it is possible to derive some slightly different value:
for example,
when a new calibration of the NGC 4258 distance
is used to calibrate the HST Cepheid distances,
it is possible to obtain
$H_0=72.0\pm3.0\Hou$ \cite{Humphreys:2013eja}.
% 
% Riess recalibration with Humphreys:
% $H_0=73.0\pm2.4\Hou$ (unpublished!
% http://realserver4v.stsci.edu/t/data/2014/03/3951/AdamRiess033114.mp4)
% 
A different reanalysis of the HST SNe leads to $H_0=70.6\pm3.3\Hou$ 
(using NGC 4258 as a distance anchor)
and to $H_0=72.5\pm2.5\Hou$
(averaging over three different distance-calibration methods)
\cite{Efstathiou:2013via}.
Other calculations show that
$H_0 = 74.3 \pm 2.6 \, \text{km} \, \text{s}^{-1} \, \text{Mpc}^{-1}$,
obtained by the Carnegie Hubble Program
 \cite{Freedman:2012ny}
through a
recalibration of the secondary distance methods used in the HST Key Project,
or
$H_0 = 78.7 \pm 4.5 \, \text{km} \, \text{s}^{-1} \, \text{Mpc}^{-1}$,
from the strong gravitational lensing time delay measurements of
the system RXJ1131-1231,
observed as part of the
COSmological MOnitoring of GRAvitational Lenses (COSMOGRAIL) project
\cite{Suyu:2012aa}.

The significance of the tension between local and CMB results on $H_0$
depends hence on the calibrations of the SN Ia distances.
The result $H_0=70.6\pm3.3\Hou$ obtained in
Ref.~\cite{Efstathiou:2013via}
is consistent with the CMB result within 1$\sigma$, but typically
other determinations are in tension with the Planck result at the
level of 2 to 3$\sigma$.
If a reliable determination of $H_0$ from local measurements
will be confirmed in the future,
we will have a strong evidence that the \lcdm~model is not complete.

\section{Supernovae}
\label{sec:sn}
As we mentioned earlier,
Supernovae of the type Ia (SN~Ia) are believed to be
standard candles,
that means that a SN~Ia has always the same luminosity.
Under this hypothesis, SN~Ia are one of the best probes to verify
the redshift-distance relation,
since they provide a direct measurement
of the luminosity distance,
independently of the redshift determination.
For this reason,
SN~Ia can be used to constrain the Universe expansion history.
We shall include in the following analyses
the constraints obtained using the
Joint Lightcurve Analysis (JLA) compilation \cite{Betoule:2014frx},
which include the 
SN~Ia observations obtained by the SDSS-II and SNLS collaborations,
for a total of 740 SN~Ia.
The dataset includes several samples at low-redshift 
from different experiments ($z<0.1$),
the observations from all the seasons of the SDSS-II ($0.05<z<0.4$),
and those collected by SNLS in three years ($0.2<z<1$),
plus a number of SN~Ia at high redshift ($0.8<z<1.2$) from HST.

\section{Matter Power Spectrum}
\label{sec:mpk}
The gravitational collapse, that started to act in
the initial phases of the Universe evolution,
formed a number of structures that fill the Universe.
These structures are observed through the light they emit
when the gas is compressed and heated.
The analysis of this light permits to test our theoretical models
of structure formation, starting from the tiny density fluctuations
that were generated during inflation.
The increasing precision of the experiments requires a 
correspondingly good precision
in the predictions from theory,
from which we want to obtain the shape of the galaxy power spectrum 
(or the correlation function).
At the linear level, we can make predictions using
perturbation theory.
The problems appear when we want to go beyond the linear theory,
since the relationship between the observed
galaxy power spectrum and the prediction
for the matter power spectrum is complicated
by the existence of non-linear structure formation,
galaxy bias, and redshift space-distortions.
The non-linear structure formation occurs when the density perturbations
become large and the linear perturbation theory fails
to describe them.
While the linear theory is sufficient to describe CMB fluctuations,
at low-redshift the matter power spectrum is the consequence of
some non-linear evolution that can be estimated
by numerical simulations
\cite{Jennings:2010uv,Bird:2011rb,Marulli:2011he}
and then applied as a correction to the
linear prediction using an algorithm as
\texttt{Halofit} \cite{Smith:2002dz}.

Beside the non-linear evolution, there is the problem that
observations are affected by redshift-space distortions, as discussed
in the previous Section, and by the problem of the galaxy bias.
We observe the distribution of galaxies, but the theoretical 
predictions are obtained for the distribution
of the total matter fluctuations, that include also DM.
The complex phenomena that involve baryons in star and galaxy
formation cause a slight decoupling between galaxies and matter.
The simplest possibility is to assume the idea of
the linear bias \cite{Kaiser:1984sw}:
an overall, shape-independent amplitude
that scales from the matter power spectrum to the
galaxy power spectrum.
The bias parameter is directly related to the history of galaxy
formation of each population, and it is different for different
populations of galaxies.
For this reason, we expect that the bias parameter evolves
with redshift and with the environment of each population,
so that it is also scale-dependent.
Today, numerical simulations of galaxies allow
to predict the bias for each population of galaxies.

After considering all these effects that go beyond the linear
regime, the theory is quite robust
at low values of the comoving wavenumber $k$ (large scales),
where the large-scale clustering can be treated as linear.
The difference between the different models starts
to increase at smaller scales, approximately for
$k > 0.2 h$~Mpc${}^{-1}$.

The WiggleZ Dark Energy survey measured the matter power spectrum
in four redshift bins and seven regions on the sky,
giving 28 separate power spectra in total \cite{Parkinson:2012vd}.
All of these spectra are publicly available,
including the window functions and covariance matrices.
We use the measured matter power spectra in the four redshift bins
$0.1 < z < 0.3$,
$0.3 < z < 0.5$,
$0.5 < z < 0.7$ and
$0.7 < z < 0.9$.
Since the analysis of the matter power spectrum is limited by
the poor theoretical modeling of a number of effects,
such as non-linearities, galaxy bias and redshift-space distortions,
the WiggleZ collaboration presented several different methods
for modeling the theoretical power spectrum and tested them
against the N-body simulations named
``Gigaparsec WiggleZ'' (GiggleZ).
The WiggleZ likelihood and published results take into account
these analyses.

\section{Cluster Counts}\label{sec:cluster}
Another powerful probe to constrain the growth
of cosmic structures is the abundance of galaxy clusters.
The reliability of this probe is based on the calibration of the
mass-observable relation, which is currently the largest uncertainty.
It is the same problem we discussed for the RSD and the determination
of the matter power spectrum.
The cosmological information enclosed in the cluster abundance is
encoded in a constraint on the so-called cluster normalization condition
\cite{Allen:2011zs,Weinberg:2012es,Rozo:2013hha},
that is the combination
\be\label{eq:cluster1}
  \sigma_8
  \left(
    \frac{\Omega_m}{\alpha}
  \right)^\beta\,,
\ee
where $\alpha$ is a fiducial value adopted in each analysis and 
$\beta$ depends on the measured redshift.
The full calculation of the cluster counts requires a hard and
time consuming computation, that involves a geometrical determination
of the cosmological volume element and that considers
the number of halos for different redshift and mass bins.

We will use the measurements of the
Chandra Project \cite{Vikhlinin:2008ym,Burenin:2012uy}, 
that observes galaxy clusters in the X-rays constraining
$\sigma_8(\Omega_m/0.25)^{0.47}=0.813\pm0.013$
and from the 2013 and 2015 release of Planck
\cite{Ade:2013lmv,Ade:2015fva},
that counts the clusters through the Sunyaev-Zel'dovich effect.
The Sunyaev-Zel'dovich (SZ) effect 
\cite{Sunyaev:1970a,Sunyaev:1980vz}
is the result of
high energy electrons distorting the CMB spectrum
through inverse Compton scattering,
in which the low energy CMB photons receive an average energy boost
during collision with the high energy cluster electrons.
Observed distortions of the CMB spectrum are used to detect
the density perturbations of the Universe.
The Planck 2013 cluster count result can be written as
$\sigma_8(\Omega_m/0.27)^{0.3}=0.782\pm0.01$, 
obtained with a fixed mass bias,
or as
$\sigma_8(\Omega_m/0.27)^{0.3}=0.764\pm0.025$,
if the mass bias is free to vary.
The Planck collaboration improved the analyses
of the cluster counts in the 2015 release,
taking into account with increased accuracy
the possible dependence on the bias
between the galaxy and the matter distribution.
In this last case we do not write constraints in the form of
Eq.~\eqref{eq:cluster1},
since additional dependencies on the nuisance parameters
used to model the uncertainties have been introduced.

Some of the results from the cluster counts are in tension
with the CMB constraints on $\sigma_8$,
that is higher when obtained from the CMB than
when obtained from local measurements.
If more measurements of cluster counts are compared, however,
it seems that there is not a clear indication that the cluster count 
measurements are in tension with the CMB predictions.
A comparison between different methods is proposed for example
in Ref.~\cite{Mantz:2014paa},
where in Fig.~2 the constraints on $\sigma_8$ from the
\lcdm~predictions obtained from CMB analyses are compared with 
the results of
several experiment that probe the cluster counts
detected through X-ray, optical and SZ surveys.
The fact that some of the reported results are in good agreement with
the CMB predictions may indicate that the anomalous measurements
suffer the presence of unaccounted systematics, that possibly lead
to a wrong estimate of the mass calibration
(see also the discussion in Ref.~\cite{Ade:2015xua}).

The tension between local and cosmological estimations of $\sigma_8$
may be the indication that our comprehension of the
systematic effects that affect the experimental measurements
is rather limited,
but also that the \lcdm~model is incomplete and 
that some new physics is required.
For example, the free-streaming of a massive neutrino or
of a different light particle would reduce the value
of $\sigma_8$ on small scales and possibly reconcile local
and cosmological measurements 
(see e.g.~Refs.~\cite{Hamann:2013iba,Gariazzo:2013gua} and
the discussion in the following Chapters).

\section{Cosmic Shear}\label{sec:shear}
The presence of large scale structures along the line of sight
causes a distortion of the shape of distant galaxies,
that can be used to constrain the growth of fluctuations.

Today, the largest weak lensing (WL) survey is the 
Canada-France-Hawaii Telescope Lensing Survey (CFHTLenS)
\cite{Kilbinger:2012qz,Heymans:2013fya}.
This experiment provides results from 2 types of analysis:
from the analyses of 2D data
to estimate the shear correlation functions $\xi^\pm$
from 0.9 to 296.5~arcmin
\cite{Kilbinger:2012qz},
and from observations of the tomographic blue galaxy sample,
that allows to estimate the shear correlation functions
in six redshift bins, in the angular range $1.7<\theta<37.9$~arcmin.
\cite{Heymans:2013fya}.
These two determinations are not independent and we will use
only the results of the tomographic survey.

Since the non-linear scales contribute significantly to $\xi^\pm$,
it is important to have a good modeling of the non-linear evolution
to avoid the introduction of systematics in the analysis.
The analyses at the angular scales probed
by both the 2D and the tomographic data, however,
may be affected by the poor knowledge of the non-linear
evolution and by the consequent incomplete theoretical modeling.
To avoid the uncertainties related to the numerical calculations
in the non-linear regime, the CFHTLenS collaboration
proposed a set of ``conservative'' cuts on the observed data.
For the 2D analysis, the authors of Ref.~\cite{Kilbinger:2012qz}
propose to exclude angular scales $\theta<17'$ for $\xi^+$ and 
$\theta<54'$ for $\xi^-$.
For the tomographic analysis, instead,
different cuts are proposed for each redshift bin.
In the two lowest redshift bins, angular scales
$\theta<3'$ are excluded for $\xi^+$ and 
$\theta<30'$ are excluded for $\xi^-$.
In the two central redshift bins, the exclusions concern 
$\theta<30'$ only for $\xi^-$, while no cuts are applied for $\xi^+$.
Finally, in the highest redshift bins only a cut
$\theta<16'$ is applied to calculate $\xi^-$
\cite{Heymans:2013fya}.
The Planck collaboration argued that these ``conservative'' cuts
may be insufficient if one wants to investigate extensions
of the \lcdm~model \cite{Ade:2015xua,Ade:2015rim} and they proposed
a set of ``ultra-conservative'' cuts, that consists in completely
removing the $\xi^-$ analysis and restricting to angular scales
$\theta>17'$ for $\xi^+$, both in the 2D and the tomographic surveys.
At the small scales relevant for the CFHTLenS experiment 
the effects of baryonic feedback and intrinsic alignment
can also be important, 
but our knowledge and theoretical description
of these effects is quite limited nowadays.
More detailed discussions can be found in
Refs.~\cite{Kilbinger:2012qz,Heymans:2013fya,Ade:2015xua}.

Even if one applies the ultra-conservative cuts, however,
in the context of the \lcdm~model
the Planck results are in substantial tension
with the CFHTLenS results.
According to the author of Ref.~\cite{Raveri:2015maa},
this is a conclusion that cannot be obtained simply by studying
the marginalized posterior probabilities
for the cosmological parameters.
The tension can be explained invoking the presence
of some unaccounted systematics
in the analysis of the experimental data
or of an incomplete modeling of the theoretical predictions,
but can also be the result of the existence
of new physics beyond the standard model.
The importance of precise local measurements is therefore high,
since local measurements are not dependent
on a specific cosmological model
and they can help to explore cosmology in a model-independent way
\cite{Verde:2013rgh}.

A recent analysis \cite{Joudaki:2016mvz} of the CFHTLenS data
that takes into account several astrophysical systematics, however,
shows that the tension between Planck and the cosmic shear measurements
disappears when the systematics are considered jointly.
They find that the two data concordance tests are in agreement,
and that the level of concordance between the two datasets
depends on the exact details of the systematic uncertainties
included in the analysis.
The results of the concordance tests
based on the Bayesian evidence and on information theory
range from decisive discordance to substantial concordance
while the treatment of the systematic uncertainties becomes more conservative.
The least conservative scenario is the one most favored by the cosmic shear data,
but it is also the one that shows the greatest degree of discordance with Planck.
A future, robust result from local measurements that will 
take into account all the possible systematics
will either confirm
the tension with CMB estimates of the cosmological quantities,
probing that the \lcdm~model is incomplete and
possibly suggesting us where to look for new physics,
or confirm that the tension that we observe now is just due to an incomplete
knowledge of some astrophysical phenomenon.
These results are confirmed by an independent analyses by other authors
\cite{Kitching:2016hvn}.

%!TeX root=main.tex 
\chapter{Neutrino Physics}\label{ch:nu}
\chapterprecis{Part of this Chapter is based on
Ref.~\protect\cite{Gariazzo:2015rra}.}

% \begin{abstract}
% In this Chapter we review the theoretical description of neutrino oscillations,
% with a particular focus on the models of the oscillations between active
% and sterile neutrinos.
% We also describe
% the experimental data that lead to the so-called Short-Baseline Anomaly
% and the resulting constraints on the mass of the fourth neutrino mass eigenstate.
% Finally, we describe how neutrinos influence the evolution of the Universe,
% with a particular attention to the physics of CMB.
% \end{abstract}

After the proposal of Pauli in 1930, who conjectured the neutrino 
to explain the problem of the $\beta$ decay spectrum,
several years passed
before the neutrino was firstly observed in 1956
by Cowan~et~al.~\cite{Cowan:1992xc}.
B.~Pontecorvo was the first to guess that more than a single
flavor of neutrinos could exist, and also he proposed the possibility
that the neutrinos oscillate between the different flavors.
Only 30 years later neutrino oscillations were finally observed
in the SuperKamiokande and in the Sudbury Neutrino Observatory
experiments, which was recently awarded with the 2015 Nobel Prize in Physics.
The discovery of neutrino oscillations was the definitive confirmation
of the fact that neutrinos are massive particles, but their masses
are much smaller than the masses of all the other particles in the
Standard Model of electroweak interactions.

In this Chapter we will firstly introduce
and discuss the most important
aspects of the neutrino theory in particle physics, 
the short-baseline neutrino oscillation anomaly and its explanation
with a light sterile neutrino, and finally
we will show how cosmology can help to constrain
the neutrino absolute mass scale and other properties.

\section{Neutrino Masses and Oscillations}
\label{sec:mixing}
The electroweak interactions are described by the
Standard Model (SM) of particle physics
\cite{Glashow:1961tr,Weinberg:1967tq,Salam:1968rm},
a fantastic theory,
based on the $\text{SU}(2)_{L} \times \text{U}(1)_{Y}$ gauge symmetry,
which can explain the majority of terrestrial experimental observations.
The SM does not account for neutrino masses,
whose existence have been firmly verified
by the measurement of neutrino oscillations
in atmospheric, solar and long-baseline
neutrino oscillation experiments
(see e.g.\ 
Refs.~\cite{Giunti:2007ry,Bilenky:2010zza,Xing:2011zza,GonzalezGarcia:2012sz,
Bellini:2013wra,PDG-2012}).
The SM can be extended to include neutrino masses simply
through the introduction of
singlet fields
for the 
$\text{SU}(2)_{L} \times \text{U}(1)_{Y}$ gauge symmetry,
which are traditionally called \emph{right-handed neutrino} fields
or
\emph{sterile neutrino} fields.
They are \emph{right-handed} since they do not transform under the
$\text{SU}(2)_{L}$ transformations.
Assuming that they have zero hypercharge,
they can be called \emph{neutrino} fields since they are neutral.
Finally, they are \emph{sterile},
because they do not have SM electroweak interactions.
These right-handed sterile neutrino fields are included in
many models which extend the SM
(see e.g.\ Refs.~\cite{Volkas:2001zb,Mohapatra:2004,Mohapatra:2006gs,
Boyarsky:2009ix,Abazajian:2012ys,Drewes:2013gca}).
In the following we consider the general theory of neutrino mixing
that includes the three standard active left-handed flavor neutrino fields
$\nu_{e L}$,
$\nu_{\mu L}$,
$\nu_{\tau L}$
plus
$N_{s}$
sterile right-handed flavor neutrino fields
$\nu_{s_{1}R}$,
\ldots,
$\nu_{N_{s}R}$.
We can use these fields to write 
the most general Lagrangian mass term, that is%
\footnote{In the following we will adopt the convention that 
the superscript ``(F)'' indicates the flavor basis, while
the superscript ``(M)'' indicates the mass basis.}
\begin{equation}
\mathcal{L}_{\text{mass}}
=
\frac{1}{2} \, {\nu^{(\text{F})}_{L}}^{T} \, 
  \mathcal{C}^{\dagger} \, M \, \nu^{(\text{F})}_{L}
+
\text{h.c.}
\,,
\label{eq:201}
\end{equation}
where
$\mathcal{C}$
is the unitary charge-conjugation matrix\footnote{
We use the notations and conventions in Ref.~\cite{Giunti:2007ry}.
},
such that
$
\mathcal{C}
\,
\gamma_{\mu}^{T}
\,
\mathcal{C}^{-1}
=
- \gamma_{\mu}
$
and
$
\mathcal{C}^{T} = - \mathcal{C}
$, and 
\begin{equation}
\nu^{(\text{F})}_{L}
=
\begin{pmatrix}
\nu^{(\text{a})}_{L}
\\ \displaystyle
{\nu^{(\text{s})}_{R}}^{c}
\end{pmatrix}
,
\qquad
\nu^{(\text{a})}_{L}
=
\begin{pmatrix}
\nu_{e L}
\\ \displaystyle
\nu_{\mu L}
\\ \displaystyle
\nu_{\tau L}
\end{pmatrix}
,
\qquad
{\nu^{(\text{s})}_{R}}^{c}
=
\begin{pmatrix}
\nu_{s_{1}R}^{c}
\\ \displaystyle
\vdots
\\ \displaystyle
\nu_{s_{N_{s}}R}^{c}
\end{pmatrix}
\,.
\label{eq:202}
\end{equation}
Here we used the superscripts ``(a)'' and ``(s)'' to indicate
the column matrices of active and sterile neutrino fields, respectively.
For any field $\psi$ the charge-conjugated field
$\psi^{c}$
is given by
$\psi^{c} = \mathcal{C} \overline{\psi}^{T}$.
Charge conjugation transforms the chirality of a field:
for example, $\psi_{R}^{c}$ is left-handed.
In general, the mass matrix
$M$ is a complex symmetric matrix,
which can be diagonalized with the unitary transformation
\begin{equation}
\nu^{(\text{F})}_{L}
=
\mathcal{U} \, \nu^{(\text{M})}_{L}
,
\qquad
\text{with}
\qquad
\nu^{(\text{M})}_{L}
=
\begin{pmatrix}
\nu_{1L}
\\ \displaystyle
\vdots
\\ \displaystyle
\nu_{NL}
\end{pmatrix}
,
\label{eq:203}
\end{equation}
where
$N=3+N_{s}$
is the total number of neutrino fields.
The $N \times N$
unitary matrix $\mathcal{U}$ has the property that
\begin{equation}
\mathcal{U}^{T} M\, \mathcal{U}
=
\operatorname{diag}\!\left( m_{1}, \ldots, m_{N} \right)
\,,
\label{eq:204}
\end{equation}
where
$m_{1}, \ldots, m_{N}$ are real and positive masses
(see Refs.~\cite{Bilenky:1987ty,Giunti:2007ry}).
Using the definitions we just presented,
the Lagrangian mass term \eqref{eq:201} becomes
\begin{equation}
\mathcal{L}_{\text{mass}}
=
-
\frac{1}{2}
\sum_{k=1}^{N}
m_{k}
\overline{\nu_{k}} \nu_{k}
,
\label{eq:205}
\end{equation}
where $\nu_{k} = \nu_{kL} + \nu_{kL}^{c}$ are
massive Majorana neutrino fields,
since they satisfy the Majorana constraint
$\nu_{k} = \nu_{k}^{c}$.
This means that, in the general case of active-sterile neutrino mixing,
the massive neutrinos are Majorana particles%
\footnote{However, it is not excluded that the mixing is such that
there are pairs of Majorana neutrino fields with exactly the same mass
which form Dirac neutrino fields.}.

The unitary transformation \eqref{eq:203} has physical effects
connected with the non-invariance of the weak interaction Lagrangian
under a rephasing of the lepton fields.
We can write the leptonic charged-current weak interaction Lagrangian in
a matrix form, using the flavor basis where 
the mass matrix of the charged leptons,
$\ell_{e} \equiv e$,
$\ell_{\mu} \equiv \mu$,
$\ell_{\tau} \equiv \tau$,
is diagonal:
\begin{equation}
\mathcal{L}_{\text{CC}}
=
-
\frac{ g }{ \sqrt{2} }
\overline{\ell_{L}}
\gamma^{\rho} \nu^{(\text{a})}_{L} W_{\rho}^{\dagger}
+
\text{h.c.}
=
-
\frac{ g }{ \sqrt{2} }
\overline{\ell_{L}}
\gamma^{\rho} U \nu^{(\text{M})}_{L} W_{\rho}^{\dagger}
+
\text{h.c.}
\,,
\label{eq:206}
\end{equation}
where we used
\begin{equation}
\ell_{L}
=
\begin{pmatrix}
e
\\
\mu
\\
\tau
\end{pmatrix}
,
\qquad
\nu^{(\text{a})}_{L}
=
U \nu^{(\text{M})}_{L}
\qquad
\text{and}
\qquad
U
=
\left.
\mathcal{U}
\right|_{3 \times N}
.
\label{eq:207}
\end{equation}
The $3 \times N$ rectangular matrix $U$ is formed by the rows of $\mathcal{U}$
corresponding to the active neutrinos and it can be parameterized with a number
of mixing parameters smaller than those necessary for
the unitary matrix $\mathcal{U}$.
This is a consequence of the fact that weak interactions are not affected
by the arbitrariness of the mixing in the sterile sector.
It is possible to show \cite{Giunti:2007ry} that
the mixing matrix $U$ can be written in terms of
$3 + 3 N_{s}$ mixing angles
and
$3 + 3 N_{s}$ physical phases,
divided into
$1 + 2 N_{s}$ Dirac phases
and
$N - 1$ Majorana phases.
A convenient scheme for this parameterization is
\begin{equation}
U
=
\left[
\left(
\prod_{a=1}^{3}
\prod_{b=4}^{N}
W^{ab}
\right)
R^{23}
W^{13}
R^{12}
\right]_{3 \times N}
\operatorname{diag}\!\left(
1, e^{i\lambda_{21}}, \ldots, e^{i\lambda_{N1}}
\right)
.
\label{eq:f470}
\end{equation}
The unitary $N \times N$ matrix $W^{ab} = W^{ab}(\theta_{ab},\eta_{ab})$
represents a complex rotation in the $a$-$b$ plane,
described 
by a mixing angle $\theta_{ab}$ and a Dirac phase $\eta_{ab}$:
\begin{equation}
\left[
W^{ab}(\vartheta_{ab},\eta_{ab})
\right]_{rs}
=
\delta_{rs}
+
\left( c_{ab} - 1 \right)
\left(
\delta_{ra} \delta_{sa}
+
\delta_{rb} \delta_{sb}
\right)
+
s_{ab}
\left(
e^{i\eta_{ab}} \delta_{ra} \delta_{sb}
-
e^{-i\eta_{ab}} \delta_{rb} \delta_{sa}
\right)
,
\label{eq:d048}
\end{equation}
where
$c_{ab}\equiv\cos\vartheta_{ab}$
and
$s_{ab}\equiv\sin\vartheta_{ab}$.
The matrix $U$ in Eq.~\eqref{eq:f470} is insensitive
to the order of the product of the of $W^{ab}$ matrices.
The orthogonal matrix
$R^{ab}=W^{ab}(\theta_{ab},0)$
represents a real rotation in the $a$-$b$ plane.
We indicate with the square brackets with subscript $3 \times N$
the fact that the enclosed $N \times N$ matrix
is truncated to the first three rows.
The diagonal matrix on the right of Eq.~\eqref{eq:f470} collects the
Majorana phases
$\lambda_{21}, \ldots \lambda_{N1}$,
which are physical only if massive neutrinos are Majorana particles.
% \footnote{
% It is possible to choose any other diagonal matrix with $N-1$ phases,
% as for example
% $
% \operatorname{diag}\!\left(
% e^{i\lambda_{12}}, 1, e^{i\lambda_{32}}, \ldots, e^{i\lambda_{N2}}
% \right)
% $,
% etc.
% }.
% Moreover,
The product of $W^{ab}$ matrices in Eq.~\eqref{eq:f470}, finally, contains
a number of unphysical phases among the $\eta_{ab}$, which can be eliminated
for each value of the index $b=4,\ldots,N$
(see Ref.~\cite{Giunti:2007ry}).

In the limit of vanishing active-sterile mixing, 
the mixing matrix in the scheme \eqref{eq:f470} reduces to the 
three-neutrino ($3\nu$) mixing matrix in the standard parameterization
\begin{align}
\null & \null
U^{(3\nu)}
=
\left[
R^{23}
W^{13}
R^{12}
\right]_{3 \times 3}
\operatorname{diag}\!\left(
1, e^{i\lambda_{21}}, e^{i\lambda_{31}}
\right)
\nonumber
\\
=
\null & \null
\begin{pmatrix}
c_{12}
c_{13}
&
s_{12}
c_{13}
&
s_{13}
e^{-i\eta_{13}}
\\
-
s_{12}
c_{23}
-
c_{12}
s_{23}
s_{13}
e^{i\eta_{13}}
&
c_{12}
c_{23}
-
s_{12}
s_{23}
s_{13}
e^{i\eta_{13}}
&
s_{23}
c_{13}
\\
s_{12}
s_{23}
-
c_{12}
c_{23}
s_{13}
e^{i\eta_{13}}
&
-
c_{12}
s_{23}
-
s_{12}
c_{23}
s_{13}
e^{i\eta_{13}}
&
c_{23}
c_{13}
\end{pmatrix}
\begin{pmatrix}
1 & 0 & 0
\\
0 & e^{i\lambda_{21}} & 0
\\
0 & 0 & e^{i\lambda_{31}}
\end{pmatrix}
.
\label{eq:3numix}
\end{align}

% It is convenient to choose
% in Eq.~(\ref{f470})
% the order of the real or complex rotations for each index $b \geq 4$
% such that the rotations in the
% 3-$b$,
% 2-$b$ and
% 1-$b$ planes
% are ordered from left to right.
% In this way,
% the first two lines,
% which are relevant for the study of the oscillations of
% the experimentally more accessible flavor neutrinos $\nu_{e}$ and $\nu_{\mu}$,
% are independent of the mixing angles and Dirac phases
% corresponding to the rotations
% in all the 3-$b$ planes for $b \geq 4$.
% Moreover,
% the first line,
% which is relevant for the study of $\nu_{e}$ disappearance,
% is independent also of the mixing angles and Dirac phases
% corresponding to the rotations
% in the 2-$b$ planes for $b \geq 3$.
% For example,
% one can choose
% \begin{equation}
% U
% =
% \left[
% W^{3N}
% R^{2N}
% W^{1N}
% \cdots
% W^{34}
% R^{24}
% W^{14}
% R^{23}
% W^{13}
% R^{12}
% \right]_{3 \times N}
% \operatorname{diag}\!\left(
% 1, e^{i\lambda_{21}}, \ldots, e^{i\lambda_{N1}}
% \right)
% ,
% \label{examix1}
% \end{equation}
% or
% \begin{equation}
% U
% =
% \left[
% W^{3N}
% \cdots
% W^{34}
% W^{2N}
% \cdots
% W^{24}
% R^{1N}
% \cdots
% R^{14}
% R^{23}
% W^{13}
% R^{12}
% \right]_{3 \times N}
% \operatorname{diag}\!\left(
% 1, e^{i\lambda_{21}}, \ldots, e^{i\lambda_{N1}}
% \right)
% .
% \label{examix2}
% \end{equation}

We can now study the neutral-current Lagrangian:
\begin{equation}
\mathcal{L}_{\text{NC}}
=
-
\frac{ g }{ 2 \cos\vartheta_{\text{W}} }
\overline{\nu^{(\text{a})}_{L}} \gamma^{\rho} \nu^{(\text{a})}_{L} Z_{\rho}
=
-
\frac{ g }{ 2 \cos\vartheta_{\text{W}} }
\overline{\nu^{(\text{M})}_{L}} \gamma^{\rho} U^{\dagger} U \nu^{(\text{M})}_{L} Z_{\rho}
\,.
\label{eq:208}
\end{equation}
Given that the mixing matrix $U$ is a rectangular
$3 \times N$
matrix formed by the first three rows of the unitary matrix $\mathcal{U}$,
we have
\begin{equation}
U U^{\dagger} = \openone_{3\times3}
\,,
\qquad
\text{but}
\qquad
U^{\dagger} U \neq \openone_{N \times N}
\,.
\label{eq:209}
\end{equation}
Therefore, the GIM mechanism \cite{Glashow:1970gm}
is not operative in neutral-current weak interactions \cite{Schechter:1980gr}
and it is possible to have neutral-current transitions
among different massive neutrinos.

The effective number of active neutrinos
contributing to the decay of the $Z$-boson
is not affected, or is marginally affected
% \footnote{
% Sterile neutrinos at mass scales larger than the muon mass affect the determination of the Fermi constant
% $G_{\text{F}}$
% through muon decay
% \cite{Antusch:2006vwa}.
% Those at mass scales larger than $m_{Z}/2$
% can induce a kinematical suppression of $N_{\nu}^{(Z)}$
% \cite{Jarlskog:1990kt,Bilenky:1990tm}.
% }
by the introduction of sterile neutrinos.
This number has been determined with high precision by the LEP experiments
\cite{ALEPH:2005ab}:
\begin{equation}
N_{\nu}^{(Z)}
=
2.9840 \pm 0.0082
\,.
\label{eq:211}
\end{equation}
In the following we will consider sterile neutrinos with masses around 1~eV,
for which
$N_{\nu}^{(Z)}$
is given by
\cite{Jarlskog:1990kt,Bilenky:1990tm}
\begin{equation}
N_{\nu}^{(Z)}
=
\sum_{j,k=1}^{N}
\left|
\sum_{\alpha=e,\mu,\tau}
U_{\alpha j}^{*}
\,
U_{\alpha k}
\right|^2
=
3
\,.
\label{eq:212}
\end{equation}
For this reason the high-precision LEP measurement of $N_{\nu}^{(Z)}$
gives no constraint
on the number and mixing of these light sterile neutrinos.

If we want to study neutrino oscillations in vacuum,
we can conveniently use
the following general expression of the probability of
$\nua{\alpha}\to\nua{\beta}$
oscillations
\cite{Bilenky:2012zp,Bilenky:2015xwa}:
\begin{align}
P_{\nua{\alpha}\to\nua{\beta}}
=
\null & \null
\delta_{\alpha\beta}
-
4
\sum_{k \neq p}
|U_{\alpha k}|^2
\left( \delta_{\alpha\beta} -  |U_{\beta k}|^2 \right)
\sin^{2}\Delta_{kp}
\nonumber
\\
\null & \null
+
8
\sum_{\stackrel{\scriptstyle j>k}{\scriptstyle j,k \neq p}}
\left|
U_{\alpha j}
U_{\beta j}
U_{\alpha k}
U_{\beta k}
\right|
\sin\Delta_{kp}
\sin\Delta_{jp}
\cos(
\Delta_{jk}
\stackrel{(+)}{-}
\eta_{\alpha\beta jk}
)
\,,
\label{eq:221}
\end{align}
where
\begin{equation}
\Delta_{kp} = \frac{ \Delta{m}^{2}_{kp} L }{ 4 E }
\,,
\qquad
\Delta{m}^{2}_{jk} = m_{j}^2 - m_{k}^2
\,,
\qquad
\eta_{\alpha\beta jk}
=
\operatorname{arg}\!\left[
U_{\alpha j}^{*}
U_{\beta j}
U_{\alpha k}
U_{\beta k}^{*}
\right]
.
\label{eq:222}
\end{equation}
Here $p$ is an arbitrary fixed index,
which can be chosen in the most convenient way
depending on the case under consideration.
The choice of $p$ forces to have only one possibility for $j$ and $k$
such that $j>k$.
As a consequence,
in the case of three-neutrino mixing,
there is only one interference term in Eq.~\eqref{eq:221}.

The measurements of neutrino oscillations determined the existence of two
squared-mass differences, which guarantee that at least two neutrino mass
eigenstates are massive.
The analyses of the oscillations of neutrinos coming from the Sun
lead to the solar squared-mass difference
\begin{equation}
\Delta m^2_{\text{SOL}}
\simeq
7.5 \times 10^{-5} \, \text{eV}^2
\,,
\label{eq:213a}
\end{equation}
while from oscillations of neutrinos produced during the cosmic rays interactions
with the atmosphere it is possible to determine the 
atmospheric squared-mass difference
\begin{equation}
\Delta m^2_{\text{ATM}}
\simeq
2.4 \times 10^{-3} \, \text{eV}^2
\,.
\label{eq:213}
\end{equation}
We can conveniently label the masses of the three light neutrinos according
to the convention
\begin{equation}
\Delta{m}^{2}_{\text{SOL}}
=
\Delta{m}^{2}_{21}
\ll
\Delta{m}^{2}_{\text{ATM}}
=
\frac{1}{2}
\left|
\Delta{m}^{2}_{31}
+
\Delta{m}^{2}_{32}
\right|
,
\label{eq:B081}
\end{equation}
although different definitions has been adopted in the literature
(see e.g.\ Ref.~\cite{Bilenky:2015bvt}).
The sign of $\Delta{m}^{2}_{\text{SOL}}$ is determined thanks to the matter
effect in the neutrino oscillations in the Sun,
that give rise to the Mikheev-Smirnov-Wolfenstein (MSW) effect
\cite{Wolfenstein:1977ue,Mikheev:1986gs,Mikheev:1986wj}
(see also Ref.~\cite{Bethe:1986ej,Giunti:2007ry}).
On the contrary, we do not know the sign of $\Delta{m}^{2}_{\text{ATM}}$
and the absolute value in Eq.~\eqref{eq:B081}
is necessary.
As a consequence, there are two possible orderings of the neutrino masses:
the normal ordering
(NO)
with
$m_{1}<m_{2}<m_{3}$
and
$\Delta{m}^{2}_{31}, \, \Delta{m}^{2}_{32} > 0$,
and 
the inverted ordering
(IO)
with
$m_{3}<m_{1}<m_{2}$
and
$\Delta{m}^{2}_{31}, \, \Delta{m}^{2}_{32} < 0$.

According to Eq.~\eqref{eq:3numix}, the mixing in the $3\nu$ paradigm
can be described with 3 mixing angles, one Dirac phase and 2 Majorana phases
(given that the neutrinos are Majorana particles).
We report in Table~\ref{tab:global3nu}
the results of the determination of
the mixing angles and the squared-mass differences
as obtained in Ref.~\cite{Capozzi:2013csa}
from a global fit of neutrino oscillation data
(see also Refs.~\cite{Forero:2014bxa,Gonzalez-Garcia:2014bfa}).
The angle $\vartheta_{23}$ is the more uncertain, since its value
is known to be close to maximal ($\pi/4$),
but it can be smaller or larger than $\pi/4$.
For the Dirac CP-violating phase $\eta_{13}$
we have indications in favor of $\eta_{13} \approx 3\pi/2$ \cite{Abe:2015awa},
corresponding to maximal CP violation,
but at $3\sigma$ all the values of $\eta_{13}$ are allowed,
including the CP-conserving values $\eta_{13}=0,\pi$.

\begin{table}[t]
\begin{center}
%\resizebox{\textwidth}{!}{
\begin{tabular}{lccccc}
parameter
&
\begin{tabular}{c}
mass
\\[-0.1cm]
order
\end{tabular}
&
\begin{tabular}{c}
best
\\[-0.1cm]
fit
\end{tabular}
&
$1\sigma$ range
&
$2\sigma$ range
&
$3\sigma$ range
\\
\hline
$\Delta{m}^2_{\text{SOL}}/10^{-5}\,\text{eV}^2 $ & & 7.54 & 7.32 -- 7.80 & 7.15 -- 8.00 & 6.99 -- 8.18 \\
\hline
$\sin^2 \vartheta_{12}/10^{-1}$ & & 3.08 & 2.91 -- 3.25 & 2.75 -- 3.42 & 2.59 -- 3.59 \\
\hline
\multirow{2}{*}{$\Delta{m}^2_{\text{ATM}}/10^{-3}\,\text{eV}^2$}
& NO & 2.43 & 2.37 -- 2.49 & 2.30 -- 2.55 & 2.23 -- 2.61 \\
& IO & 2.38 & 2.32 -- 2.44 & 2.25 -- 2.50 & 2.19 -- 2.56 \\
\hline
\multirow{2}{*}{$\sin^2 \vartheta_{23}/10^{-1}$}
& NO & 4.37 & 4.14 -- 4.70 & 3.93 -- 5.52 & 3.74 -- 6.26 \\
& IO & 4.55 & 4.24 -- 5.94 & 4.00 -- 6.20 & 3.80 -- 6.41 \\
\hline
\multirow{2}{*}{$\sin^2 \vartheta_{13}/10^{-2}$}
& NO & 2.34 & 2.15 -- 2.54 & 1.95 -- 2.74 & 1.76 -- 2.95 \\
& IO & 2.40 & 2.18 -- 2.59 & 1.98 -- 2.79 & 1.78 -- 2.98 \\
\hline
\end{tabular}
%}
\end{center}
\caption[Neutrino mixing parameters obtained with a global analysis
of neutrino oscillation data.]%
{\label{tab:global3nu}
Values of the neutrino mixing parameters obtained in
Ref.~\cite{Capozzi:2013csa} with a
global analysis of neutrino oscillation data
in the framework of three-neutrino mixing
with the normal ordering (NO) and the inverted ordering (IO).
}
\end{table}

We can extend the framework of $3\nu$ mixing
with the introduction of non-standard massive neutrinos.
The requirement, however, is that mixing between active and non-standard 
neutrinos is small, since we do not want to spoil the 
successful $3\nu$ mixing explanation of
solar, atmospheric and long-baseline
neutrino oscillation measurements.
The non-standard massive neutrinos must be then mostly sterile
and in the following we will always assume the constraint
\begin{equation}
|U_{\alpha k}|^2 \ll 1
\qquad
(\alpha=e,\mu,\tau; \, k=4, \ldots, N)
\,.
\label{eq:smallmix}
\end{equation}

Even if more than one sterile neutrino has been considered in the literature,
we consider only the so-called 3+1 scheme,
where the ``+1'' refers to a non-standard massive neutrino, mostly sterile,
at the eV scale.
It generates a new squared-mass difference
\begin{equation}
\Delta m^2_{\text{SBL}}
\sim
1 \, \text{eV}^2
\,,
\label{dm2sbl}
\end{equation}
that allows to explain the 
anomalies found in some short-baseline (SBL) neutrino oscillation experiments
(see Section~\ref{sec:sbl}).
We assume that the three standard massive neutrinos
are much lighter than the eV scale.
A different possibility would concern an inverted sterile ordering,
where the additional neutrino has a mass much smaller than the active neutrinos,
which have then almost degenerate masses at the eV scale
in order to generate the same
$\Delta m^2_{\text{SBL}}\sim1 \, \text{eV}^2$.
This possibility is strongly disfavored by cosmological measurements
\cite{Ade:2015xua}
and by the experimental bounds on
neutrinoless double-$\beta$ decay, 
assuming that massive neutrinos are Majorana particles
(see Ref.~\cite{Bilenky:2014uka}).
In any case, the 3+1 scheme must be considered an effective mixing scheme,
since possible additional non-standard massive neutrinos beyond the first one
are allowed, 
if their mixing with the three active neutrinos is sufficiently small
to be negligible in the analysis of the data of current experiments.

We want now to consider Eq.~\eqref{eq:221} to obtain
the effective oscillation probabilities in short-baseline experiments,
for which
$\Delta_{21} \ll \Delta_{31} \ll 1$.
Consider the general
3+$N_{s}$
case in which
$\Delta{m}^2_{k1} \approx \Delta{m}^2_{\text{SBL}}$
and
$\Delta_{k1} \approx 1$
for $k\geq4$.
Choosing $p=1$ in Eq.~\eqref{eq:221},
we obtain
\begin{align}
P_{\nua{\alpha}\to\nua{\beta}}^{(\text{SBL})}
\simeq
\null & \null
\delta_{\alpha\beta}
-
4
\sum_{k=4}^{N}
|U_{\alpha k}|^2
\left( \delta_{\alpha\beta} -  |U_{\beta k}|^2 \right)
\sin^{2}\Delta_{k1}
\nonumber
\\
\null & \null
+
8
\sum_{k=4}^{N}
\sum_{j=k+1}^{N}
\left|
U_{\alpha j}
U_{\beta j}
U_{\alpha k}
U_{\beta k}
\right|
\sin\Delta_{k1}
\sin\Delta_{j1}
\cos(
\Delta_{jk}
\stackrel{(+)}{-}
\eta_{\alpha\beta jk}
)
\,.
\label{eq:223}
\end{align}

Let us consider the survival probabilities of active neutrinos:
we can define the effective amplitudes
\begin{equation}
\sin^2 2\vartheta_{\alpha\alpha}^{(k)}
=
4 |U_{\alpha k}|^2 \left( 1 - |U_{\alpha k}|^2 \right)
\simeq
4 |U_{\alpha k}|^2
\qquad
(\alpha=e,\mu,\tau; \, k\geq4)
\,,
\label{eq:asurv}
\end{equation}
where we have taken into account the constraint in Eq.~\eqref{eq:smallmix}.
The quadratically suppressed terms can be dropped
% also 
in the survival probabilities,
and we obtain
\begin{equation}
P_{\nua{\alpha}\to\nua{\alpha}}^{(\text{SBL})}
\simeq
1
-
\sum_{k=4}^{N}
\sin^2 2\vartheta_{\alpha\alpha}^{(k)}
\sin^{2}\Delta_{k1}
\qquad
(\alpha=e,\mu,\tau)
\,.
\label{eq:psurv}
\end{equation}
Each effective mixing angle
$\vartheta_{\alpha\alpha}^{(k)}$
parameterizes the disappearance of $\nua{\alpha}$
due to its mixing with $\nua{k}$.

We can now consider the probabilities of short-baseline
$\nua{\alpha}\to\nua{\beta}$
transitions between two different active neutrinos
or an active and a sterile neutrino.
The transition amplitudes are defined as
\begin{equation}
\sin^2 2\vartheta_{\alpha\beta}^{(k)}
=
4 |U_{\alpha k}|^2 |U_{\beta k}|^2
\qquad
(\alpha\neq\beta; \, k\geq4)
\,,
\label{eq:atran}
\end{equation}
which allow us to write the transition probabilities as
\begin{align}
P_{\nua{\alpha}\to\nua{\beta}}^{(\text{SBL})}
\simeq
\null & \null
\sum_{k=4}^{N}
\sin^2 2\vartheta_{\alpha\beta}^{(k)}
\sin^{2}\Delta_{k1}
\nonumber
\\
\null & \null
+
2
\sum_{k=4}^{N}
\sum_{j=k+1}^{N}
\sin 2\vartheta_{\alpha\beta}^{(k)}
\sin 2\vartheta_{\alpha\beta}^{(j)}
\sin\Delta_{k1}
\sin\Delta_{j1}
\cos(
\Delta_{jk}
\stackrel{(+)}{-}
\eta_{\alpha\beta jk}
)
\,.
\label{eq:ptran}
\end{align}
We can see from the first line that
each effective mixing angle
$\vartheta_{\alpha\beta}^{(k)}$
parameterizes the amount of
$\nua{\alpha}\to\nua{\beta}$
transitions
due to the mixing of $\nua{\alpha}$ and $\nua{\beta}$ with $\nua{k}$.
The second line in Eq.~\eqref{eq:ptran}, instead,
is the interference between the contributions of $\nua{k}$ and $\nua{j}$,
depending on the same effective mixing angles.

From Eqs.~\eqref{eq:asurv} and \eqref{eq:atran}
we can see that for each value of $k\geq4$
the transition amplitude
$\sin 2\vartheta_{\alpha\beta}^{(k)}$
and
the disappearance amplitudes
$\sin 2\vartheta_{\alpha\alpha}^{(k)}$
and
$\sin 2\vartheta_{\beta\beta}^{(k)}$
depend only on the elements in $k^{\text{th}}$ column of the mixing matrix
$\mathcal{U}$
and are related by\footnote{
This relation was derived in the case of 3+1 mixing
(see Eq.~\eqref{eq:appdis3p1})
in Refs.~\cite{Okada:1996kw,Bilenky:1996rw}.
}
\begin{equation}
\sin^2 2\vartheta_{\alpha\beta}^{(k)}
\simeq
\frac{1}{4}
\,
\sin^2 2\vartheta_{\alpha\alpha}^{(k)}
\,
\sin^2 2\vartheta_{\beta\beta}^{(k)}
\qquad
(\alpha=e,\mu,\tau)
\,.
\label{eq:appdis}
\end{equation}
The importance of this relation is crucial for the acceptance or rejection
of the 3+$N_{s}$ mixing schemes with sterile neutrinos through their test
against the experimental results,
because it constrains the oscillation signals that can be observed in
short-baseline experiments, both in the appearance and disappearance channels.
In particular,
the amplitudes of the short-baseline transition probabilities
between active neutrinos are quadratically suppressed
since both
$\sin^2 2\vartheta_{\alpha\alpha}^{(k)}$
and
$\sin^2 2\vartheta_{\beta\beta}^{(k)}$
are small for $\alpha,\beta=e,\mu,\tau$.

In the case of 3+1 neutrino mixing
\cite{Okada:1996kw,Bilenky:1996rw,Bilenky:1999ny,Maltoni:2004ei},
we have
$\Delta{m}^2_{41} = \Delta{m}^2_{\text{SBL}}$
and
$\Delta_{41} \sim 1$
in short-baseline experiments.
The transition and survival probabilities
become
\begin{equation}
P_{\nua{\alpha}\to\nua{\beta}}^{(\text{SBL})}
\simeq
\sin^2 2\vartheta_{\alpha\beta}
\sin^{2}\Delta_{41}
\quad
(\alpha\neq\beta)
,
\qquad
P_{\nua{\alpha}\to\nua{\alpha}}^{(\text{SBL})}
\simeq
1
-
\sin^2 2\vartheta_{\alpha\alpha}
\sin^{2}\Delta_{41}
\,,
\label{eq:pro3p1}
\end{equation}
where the transition and survival amplitudes are
\begin{equation}
\sin^2 2\vartheta_{\alpha\beta}
=
4 |U_{\alpha 4}|^2 |U_{\beta 4}|^2
\quad
(\alpha\neq\beta)
,
\qquad
\sin^2 2\vartheta_{\alpha\alpha}
=
4
|U_{\alpha 4}|^2
\left(1 -  |U_{\alpha 4}|^2 \right)
\,.
\label{eq:amp3p1}
\end{equation}
The appearance-disappearance constraint is
\cite{Okada:1996kw,Bilenky:1996rw}
\begin{equation}
\sin^2 2\vartheta_{\alpha\beta}
\simeq
\frac{1}{4}
\,
\sin^2 2\vartheta_{\alpha\alpha}
\,
\sin^2 2\vartheta_{\beta\beta}
\qquad
(\alpha=e,\mu,\tau)
\,.
\label{eq:appdis3p1}
\end{equation}

In Eq.~\eqref{eq:pro3p1}, the transition and survival probabilities 
depend only on the largest squared-mass difference, that in the 3+1 scheme is
$\Delta{m}^2_{41} = \Delta{m}^2_{\text{SBL}}$,
and on the absolute values of
the elements in the fourth column of the mixing matrix.
There is no difference between the transition probabilities of
neutrinos and antineutrinos,
since the absolute values of the elements $U_{\alpha4}$
do not depend on the CP-violating phases.
Even in the presence of CP-violating phases in the mixing matrix,
signals of CP violation cannot be measured in short-baseline experiments,
but it must be searched for in
experiments sensitive to the oscillations generated
by the smaller squared-mass differences
$\Delta{m}^2_{\text{ATM}}$
\cite{deGouvea:2014aoa,Klop:2014ima,Berryman:2015nua}
or
$\Delta{m}^2_{\text{SOL}}$
\cite{Long:2013hwa}.

\section{Short-baseline Anomalies and Constraints}
\label{sec:sbl}

The measurements obtained in short-baseline neutrino oscillation experiments
require the existence of at least
one additional squared-mass difference,
$\Delta{m}^2_{\text{SBL}}$,
which is much larger than
$\Delta{m}^2_{\text{SOL}}$
and
$\Delta{m}^2_{\text{ATM}}$.
Three types of experiments give indications
in favor of $\Delta{m}^2_{\text{SBL}}$:
the reactor antineutrino anomaly,
the Gallium neutrino anomaly,
and
the LSND anomaly.

\subsection{The reactor Antineutrino Anomaly}
\label{sub:reactor}

In the literature, one can find a discrepancy between
the rate of $\bar\nu_{e}$ observed in several
short-baseline reactor neutrino experiments
and the value expected from the calculation of
the reactor neutrino fluxes
\cite{Mueller:2011nm,Huber:2011wv,Abazajian:2012ys},
which predict more events than those observed.
Many authors studied this discrepancy
\cite{Mention:2011rk,Sinev:2011ra,Kopp:2011qd,
Giunti:2011gz,Giunti:2011hn,Giunti:2011cp,
Abazajian:2012ys,Ciuffoli:2012yd,Giunti:2012tn,
Giunti:2012bc,Zhang:2013ela,Kopp:2013vaa,Ivanov:2013cga,
Giunti:2013aea}, that is referred to as 
the \emph{reactor antineutrino anomaly}.

The significance of the reactor anomaly depends
on the uncertainties of the reactor antineutrino flux,
that is calculated
from the available database information on nuclear decays 
and from the electron spectra associated with the fission of
$^{235}\text{U}$,
$^{239}\text{Pu}$, and
$^{241}\text{Pu}$
measured at ILL in the 80's
\cite{VonFeilitzsch:1982jw,Schreckenbach:1985ep,Hahn:1989zr,
Haag:2014kia}.
These determinations of the
values and uncertainties of the reactor antineutrino fluxes
have been presented in
Refs.~\cite{Mueller:2011nm,Huber:2011wv,Abazajian:2012ys}.
There have been, however, some debate
\cite{Fallot:2012jv,Hayes:2013wra,Dwyer:2014eka,Ma:2014bpa,
Sonzogni:2015aoa,Fang:2015cma,Zakari-Issoufou:2015vvp,Hayes:2015yka},
especially after the discovery of an excess
at about 5 MeV in the reactor antineutrino spectrum measured by the
RENO \cite{Seo:2014xei},
Double Chooz \cite{Crespo-Anadon:2014dea}
and
Daya Bay \cite{Zhan:2015aha}
experiments.

The main process involved for
neutrino detection in reactor experiments
is the inverse neutron decay process
\begin{equation}
\bar\nu_{e} + p \to n + e^{+}
\label{eq:l011}
\end{equation}
that occurs in liquid-scintillator detectors.
This detection process has a cross section
$
\sigma_{\bar\nu_{e}p}(E_{e})
\propto
E_{e} p_{e}
$
(see Refs.~\cite{Bemporad:2001qy,Fukugita:2003en,Giunti:2007ry}),
where $E_{e}$ and $p_{e}$ indicate
the positron energy and momentum, respectively.
The recoil energy of the neutron is small and it can be neglected.
The neutrino energy $E$ can be calculated from
the kinetic energy $T_{e}$ of the positron,
that can be measured, through the relation
\begin{equation}
E
\simeq
T_{e} + m_{e} + m_{n} - m_{p}
\simeq
T_{e} + 1.8 \, \text{MeV}
\,,
\label{eq:l013}
\end{equation}
where $m_{p}$ and $m_{n}$ are
the proton and neutron masses, respectively.
As a consequence,
the threshold for the detection process
is about $1.8 \, \text{MeV}$ for the neutrino energy.

The anomaly is usually parameterized using the 
ratio $R\equiv N_{\text{exp}}/N_{\text{cal}}$ of the measured
($N_{\text{exp}}$)
and calculated
($N_{\text{cal}}$)
number of electron antineutrino events
in reactor experiments at different distances $L$.
The average ratio of the values $R$ obtained in several
different experiments
\cite{Declais:1994ma,Kuvshinnikov:1990ry,Declais:1995su,Zacek:1986cu,
Hoummada:1995zz,Vidyakin:1990iz,Afonin:1988gx,Greenwood:1996pb,
Apollonio:2002gd,Boehm:2001ik,Abe:2014bwa,Zhang:2015fya}
is
$\overline{R} = 0.933 \pm 0.021$,
indicating a deficit with a 
% nominal\footnote{
% We call ``nominal'' the statistical significances
% of the indications
% in favor of short-baseline neutrino oscillations
% in order to emphasize that they depend
% on the estimated standard uncertainties.
% }
statistical significance of about
$3.1\sigma$ (see also Ref.~\cite{Gariazzo:2015rra}).

One possible explanation of the
reactor antineutrino anomaly is the existence of
neutrino oscillations
with an oscillation length shorter than about 20 m.
From the relation between
the squared-mass difference $\Delta{m}^2$
and the corresponding oscillation length $L^{\text{osc}}$, that is
\begin{equation}
L^{\text{osc}}
=
\frac{ 4 \pi E }{ \Delta{m}^{2} }
\simeq
2.5 \,
\frac{E \, [\text{MeV}]}{\Delta{m}^{2} \, [\text{eV}^{2}]}
\,
\text{m}
\,,
\label{eq:Losc}
\end{equation}
given that the average energy of the antineutrinos detected 
in a reactor experiment is about 4 MeV,
these oscillations require a squared-mass difference
\begin{equation}
\Delta{m}^2_{\text{SBL}}
\gtrsim
0.5 \, \text{eV}^2
\,.
\label{eq:dm2rea}
\end{equation}

\subsection{The Gallium Neutrino Anomaly}
\label{sub:gallium}

The second anomaly we present is the Gallium neutrino anomaly
\cite{Abdurashitov:2005tb,Laveder:2007zz,Giunti:2006bj,Acero:2007su,
Giunti:2009zz,Giunti:2010zu,Giunti:2012tn},
a disappearance of $\nu_{e}$
measured in the short-baseline
Gallium radioactive source experiments
GALLEX
\cite{Anselmann:1994ar,Hampel:1997fc,Kaether:2010ag}
and
SAGE
\cite{Abdurashitov:1996dp,Abdurashitov:1998ne,Abdurashitov:2005tb,
Abdurashitov:2009tn}.
% 
% \begin{figure}
% \begin{center}
% \begin{minipage}[r]{0.49\linewidth}
% \begin{center}
% \subfigure[]{\label{fig:GaGe}
% \includegraphics*[width=\linewidth]{fig-04a.pdf}
% }
% \end{center}
% \end{minipage}
% \hfill
% \begin{minipage}[l]{0.49\linewidth}
% \begin{center}
% \subfigure[]{\label{fig:gal}
% \includegraphics*[width=0.8\linewidth]{fig-04b.pdf}
% }
% \end{center}
% \end{minipage}
% \end{center}
% \caption{ \label{fig:3.2}
% \subref{fig:GaGe}:
% ${}^{71}\text{Ga}\to{}^{71}\text{Ge}$
% transitions induced by
% ${}^{51}\text{Cr}$
% and
% ${}^{37}\text{Ar}$
% electron neutrinos.
% \subref{fig:gal}:
% Ratios $R$ of the measured
% ($N_{\text{exp}}$)
% and calculated
% ($N_{\text{cal}}$)
% number of electron neutrino events in
% the GALLEX and SAGE
% radioactive source experiments.
% The horizontal shadowed red band shows the average ratio $\overline{R}$ and its uncertainty.
% For each experiment
% the error bar shows the experimental uncertainty.
% }
% \end{figure}
The detectors of the GALLEX and SAGE
solar neutrino experiments have been tested with
intense artificial ${}^{51}\text{Cr}$ and ${}^{37}\text{Ar}$
radioactive sources,
which produce electron neutrinos through the electron captures
\begin{equation}
e^{-} + {}^{51}\text{Cr} \to {}^{51}\text{V} + \nu_{e}
,
\qquad
e^{-} + {}^{37}\text{Ar} \to {}^{37}\text{Cl} + \nu_{e}
\,.
\label{eq:EC}
\end{equation}
% The neutrino energies and branching ratios are given in
% Tab.~\ref{tab:src}.
The radioactive source was placed near the center of the detector
of each experiment, which detected
electron neutrinos with the reaction
\begin{equation}
\nu_{e} + {}^{71}\text{Ga} \to {}^{71}\text{Ge} + e^{-}
.
\label{eq:f31}
\end{equation}
% Figure~\ref{fig:GaGe}
% shows the transitions
% from the ground state of ${}^{71}\text{Ga}$ to the
% ground state or one of two excited states of ${}^{71}\text{Ge}$
% which are energetically allowed in the detection process.
The total detection cross section of this reaction is given by
\begin{equation}
\sigma
=
\sigma_{\text{gs}}
\left(
1
+
\xi_{175}
\frac{\text{BGT}_{175}}{\text{BGT}_{\text{gs}}}
+
\xi_{500}
\frac{\text{BGT}_{500}}{\text{BGT}_{\text{gs}}}
\right)
,
\label{eq:cs01}
\end{equation}
where
$\sigma_{\text{gs}}$
indicates the cross sections of the transitions
from the ground state of ${}^{71}\text{Ga}$
to the ground state of ${}^{71}\text{Ge}$,
$\text{BGT}_{\text{gs}}$
is the corresponding Gamow-Teller strength,
and
$\text{BGT}_{175}$
and
$\text{BGT}_{500}$
are the Gamow-Teller strengths of the transitions
from the ground state of ${}^{71}\text{Ga}$
to the two excited states of ${}^{71}\text{Ge}$
at about 175 keV and 500 keV
(see e.g.\ Ref.~\cite{Gariazzo:2015rra}).
The coefficients of
$\text{BGT}_{175}/\text{BGT}_{\text{gs}}$
and
$\text{BGT}_{500}/\text{BGT}_{\text{gs}}$
are determined by phase space:
$\xi_{175}({}^{51}\text{Cr}) = 0.669$,
$\xi_{500}({}^{51}\text{Cr}) = 0.220$,
$\xi_{175}({}^{37}\text{Ar}) = 0.695$,
$\xi_{500}({}^{37}\text{Ar}) = 0.263$
\cite{Bahcall:1997eg}.

Bahcall \cite{Bahcall:1997eg}
calculated accurately the 
cross sections of the transitions
from the ground state of ${}^{71}\text{Ga}$
to the ground state of ${}^{71}\text{Ge}$:
\begin{equation}
\sigma_{\text{gs}}({}^{51}\text{Cr})
=
55.3 \times 10^{-46} \, \text{cm}^2
,
\qquad
\sigma_{\text{gs}}({}^{37}\text{Ar})
=
66.2 \times 10^{-46} \, \text{cm}^2
,
\label{eq:sgs}
\end{equation}
and
\cite{Haxton:1998uc,Giunti:2012tn}
\begin{equation}
\text{BGT}_{\text{gs}}
=
0.0871 \pm 0.0004
\,.
\label{eq:f32}
\end{equation}
The Gamow-Teller strengths
$\text{BGT}_{175}$
and
$\text{BGT}_{500}$
have been measured in 1985 in the
$(p,n)$
experiment of Krofcheck et al.\ 
\cite{Krofcheck:1985fg,Krofcheck-PhD-1987}
and
in 2011 in the
$({}^{3}\text{He},{}^{3}\text{H})$
experiment of Frekers et al.\ 
\cite{Frekers:2011zz} with higher precision.

In analogy with the reactor anomaly,
the results for the Gallium anomaly are usually reported in terms
of the ratio $R\equiv N_{\text{exp}}/N_{\text{cal}}$
of the measured number of electron neutrino events
($N_{\text{exp}}$)
and the one calculated
($N_{\text{cal}}$)
with the Frekers et al.\ Gamow-Teller strengths.
The average ratio calculated with the results obtained in the
GALLEX and SAGE
radioactive source experiments
is
$\overline{R} = 0.84 \pm 0.05$,
indicating a deficit with a statistical significance of about
$2.9\sigma$.

The average neutrino travels distances
in the GALLEX and SAGE
radioactive source experiments
equal to
$\langle L \rangle_{\text{GALLEX}} = 1.9 \, \text{m}$
and
$\langle L \rangle_{\text{SAGE}} = 0.6 \, \text{m}$.
The produced neutrinos may have different energies,
depending on the electron-capture channel.
The largest branching ratios are for the 
$E=747\,\text{keV}$ neutrino for ${}^{51}\text{Cr}$
and for the
$E=811\,\text{keV}$ neutrino for ${}^{37}\text{Ar}$,
while the complete list of neutrino energies and the corresponding
branching ratios can be found for example in Tab.~2 of
Ref.~\cite{Gariazzo:2015rra}.
From Eq.~\eqref{eq:Losc}
we can estimate that
the Gallium neutrino anomaly
can be explained by neutrino oscillations if they are
generated by a squared-mass difference
\begin{equation}
\Delta{m}^2_{\text{SBL}}
\gtrsim
1 \, \text{eV}^2
\,.
\label{eq:dm2gal}
\end{equation}

\subsection{The LSND Anomaly}
\label{sub:LSND}

Finally, the LSND experiment
\cite{Athanassopoulos:1995iw,Aguilar:2001ty}
observed an excess of electron antineutrino events
in a beam of muon antineutrinos produced by $\mu^{+}$
decay at rest,
\begin{equation}
\mu^{+} \to e^{+} + \nu_{e} + \bar\nu_{\mu}
\,.
\label{eq:l008}
\end{equation}
The energy spectrum of the muon antineutrinos is
$
\phi_{\bar\nu_{\mu}}(E)
\propto
E^2 \left( 3 - 4 E / m_{\mu} \right)
$
(see Ref.~\cite{Fukugita:2003en})
for neutrino energies $E$ smaller than
$
E_{\text{max}} = (m_{\mu}-m_{e})/2 \simeq 52.6 \, \text{MeV}
$.
The experiment used a detector filled with liquid scintillator
to detect electron antineutrino events
at a distance $L \simeq 30 \, \text{m}$
through the inverse neutron decay process \eqref{eq:l011}.
The energy range is
$20 \lesssim E_{e} \lesssim 60 \, \text{MeV}$
for the energy $E_{e}$ of the detected positron.

From Eq.~\eqref{eq:Losc} and for the energy range of LSND,
we can estimate that the $\bar\nu_{e}$ appearance signal
can be explained by $\bar\nu_{\mu}\to\bar\nu_{e}$ oscillations
generated by a squared-mass difference
\begin{equation}
\Delta{m}^2_{\text{SBL}}
\gtrsim
0.1 \, \text{eV}^2
.
\label{eq:dm2lsnd}
\end{equation}

The statistical significance
of the electron antineutrino appearance signal
at LSND is of about $3.8\sigma$.
We must note, however, that the similar KARMEN experiment
\cite{Armbruster:2000uw,Armbruster:2002mp}
did not measure any excess of $\bar\nu_{e}$ events
over the background at a distance $L \simeq 18 \, \text{m}$.
Another experiment, MiniBooNE,
was designed to check the LSND signal
with about one order of magnitude larger distance and energy,
but with the same order of magnitude for the ratio $L/E$.
Unfortunately, the results of the MiniBooNE experiment are ambiguous,
since the LSND signal was not seen in the neutrino mode
($\nu_{\mu}\to\nu_{e}$)
\cite{AguilarArevalo:2008rc},
while the $\bar\nu_{\mu}\to\bar\nu_{e}$
signal observed in 2010 \cite{AguilarArevalo:2010wv}
with the first half of the antineutrino data
was not observed in the second half of the antineutrino data
\cite{Aguilar-Arevalo:2013pmq}.
Moreover, 
in the MiniBooNE data, both for the neutrino and antineutrino modes,
an excess in the low-energy bins appears.
This is widely considered an anomalous effects,
since it cannot be explained with neutrino oscillations
\cite{Giunti:2011hn,Giunti:2011cp}.

\section{Global Fits of short-baseline Data}
\label{sec:global}

\begin{table}[t]
\begin{center}
%\resizebox{\textwidth}{!}{
\begin{tabular}{c|cccc}
% 					 & 3+1         & 3+1         & 3+1         & 3+1         \\
					 & GLO         & PrGLO       & noMB        & noLSND      \\ \hline
$\chi^{2}_{\text{min}}$                  & 306.0       & 276.3       & 251.2       & 291.3       \\
NDF                                      & 268         & 262         & 230         & 264         \\
GoF                                      & 5\%         & 26\%        & 16\%        & 12\%        \\ \hline
$(\chi^{2}_{\text{min}})_{\text{APP}}$   & 98.9        & 77.0        & 50.9        & 91.8        \\
$(\chi^{2}_{\text{min}})_{\text{DIS}}$   & 194.4       & 194.4       & 194.4       & 194.4       \\
	$\Delta\chi^{2}_{\text{PG}}$     & 13.0        & 5.3         & 6.2         & 5.3         \\
	$\text{NDF}_{\text{PG}}$         & 2           & 2           & 2           & 2           \\
	$\text{GoF}_{\text{PG}}$         & 0.1\%       & 7\%         & 5\%         & 7\%         \\ \hline
$\Delta\chi^{2}_{\text{NO}}$             & $49.2$      & $47.7$      & $48.1$      & $11.4$      \\
	$\text{NDF}_{\text{NO}}$         & $3$         & $3$         & $3$         & $3$         \\
	$n\sigma_{\text{NO}}$            & $6.4\sigma$ & $6.3\sigma$ & $6.4\sigma$ & $2.6\sigma$ \\
\end{tabular}
%}
\end{center}
\caption[Global fit of short-baseline neutrino oscillation data
in the 3+1 scheme.]%
{ \label{tab:chi}
Results of the fit of short-baseline data
in the 3+1 scheme.
The four different possibilities take into account different
dataset combinations:
all MiniBooNE data (GLO),
only the MiniBooNE data above 475 MeV (PrGLO),
without MiniBooNE data (noMB)
and without LSND data (noLSND).
In the first three lines 
the minimum $\chi^{2}$ ($\chi^{2}_{\text{min}}$),
the number of degrees of freedom (NDF) and
the goodness-of-fit (GoF) are listed.
The five lines in the middle give the quantities
relevant for the appearance-disappearance (APP-DIS)
parameter goodness-of-fit (PG)
\protect\cite{Maltoni:2003cu}.
In the last three lines we list
the difference between the $\chi^{2}$
without short-baseline oscillations (NO)
and $\chi^{2}_{\text{min}}$
($\Delta\chi^{2}_{\text{NO}}$),
the corresponding difference of number of degrees of freedom
($\text{NDF}_{\text{NO}}$)
and the resulting
number of $\sigma$'s ($n\sigma_{\text{NO}}$)
for which the absence of oscillations is disfavored.
Adapted from \protect\cite{Gariazzo:2015rra}.
}
\end{table}

Since the discovery of the LSND anomaly, 
many analyses of short-baseline neutrino oscillation data
have been done 
\cite{GomezCadenas:1995sj,Goswami:1995yq,Okada:1996kw,
Bilenky:1996rw,Bilenky:1999ny,Giunti:2000wt,Barger:2000ch,
Peres:2000ic,Grimus:2001mn,Strumia:2002fw,Maltoni:2002xd,
Foot:2002tf,Giunti:2003cf,Sorel:2003hf,Barger:2003xm,Maltoni:2007zf,
Goswami:2007kv,Schwetz:2007cd,Karagiorgi:2009nb,Akhmedov:2010vy,
Nelson:2010hz}.
The interest for joint fits of neutrino oscillation data
increased after 
the discoveries of the Gallium neutrino anomaly
\cite{Giunti:2006bj,Acero:2007su,Giunti:2007xv,Giunti:2009zz,
Giunti:2010wz,Giunti:2010zu,Giunti:2012tn,Giunti:2010zs,
Giunti:2010jt,Giunti:2010uj}
and the reactor antineutrino anomaly
\cite{Mention:2011rk,Kopp:2011qd,Donini:2011jh,Conrad:2011ce,
Giunti:2011gz,Giunti:2011hn,Giunti:2011cp,Karagiorgi:2012kw,
Kuflik:2012sw,Donini:2012tt,Archidiacono:2012ri,Giunti:2012tn,
Giunti:2012bc,Conrad:2012qt}.
The most recent global fit of SBL neutrino oscillation data
was presented in Ref.~\cite{Gariazzo:2015rra} and
it is an update of the analysis of Ref.~\cite{Giunti:2013aea}.
These analyses include
\begin{itemize}
\item 
$\nua{\mu}\to\nua{e}$
appearance data from several experiments
\cite{Aguilar:2001ty,Aguilar-Arevalo:2013pmq,Borodovsky:1992pn,
Armbruster:2002mp,Astier:2003gs,Antonello:2013gut,Agafonova:2013xsk};

\item
$\nua{e}$ disappearance data from several
reactor neutrino experiments
\cite{Declais:1994ma,Kuvshinnikov:1990ry,Declais:1995su,Zacek:1986cu,
Hoummada:1995zz,Vidyakin:1990iz,Afonin:1988gx,Greenwood:1996pb,
Apollonio:2002gd,Boehm:2001ik,Abe:2014bwa,Zhang:2015fya}
(see Section~\ref{sub:reactor}),
from the Gallium radioactive source experiments
GALLEX
\cite{Anselmann:1994ar,Hampel:1997fc,Kaether:2010ag}
and SAGE
\cite{Abdurashitov:1996dp,Abdurashitov:1998ne,Abdurashitov:2005tb,
Abdurashitov:2009tn}
(see Section~\ref{sub:gallium}),
from the solar neutrino constraint on $\sin^{2}2\vartheta_{ee}$
\cite{Giunti:2009xz,Palazzo:2011rj,Palazzo:2012yf,Giunti:2012tn,
Palazzo:2013me}
and
from the
$\nu_{e} + {}^{12}\text{C} \to {}^{12}\text{N}_{\text{g.s.}} + e^{-}$
scattering data \cite{Conrad:2011ce}
of KARMEN 
\cite{Bodmann:1994py,Armbruster:1998uk}
and LSND \cite{Auerbach:2001hz},
with the method discussed in Ref.~\cite{Giunti:2011cp}.

\item
$\nua{\mu}$
disappearance obtained from
the data of the
CDHSW experiment \cite{Dydak:1983zq},
from the analysis \cite{Maltoni:2007zf} of
the data of
atmospheric neutrino oscillation experiments,
from the analysis \cite{Hernandez:2011rs,Giunti:2011hn} of the
MINOS neutral-current data \cite{Adamson:2011ku}
and from the analysis of the
SciBooNE-MiniBooNE
neutrino \cite{Mahn:2011ea} and
antineutrino \cite{Cheng:2012yy} data.
\end{itemize}

\begin{figure*}[t]
\null
\hfill
\includegraphics*[width=\halfwidth]{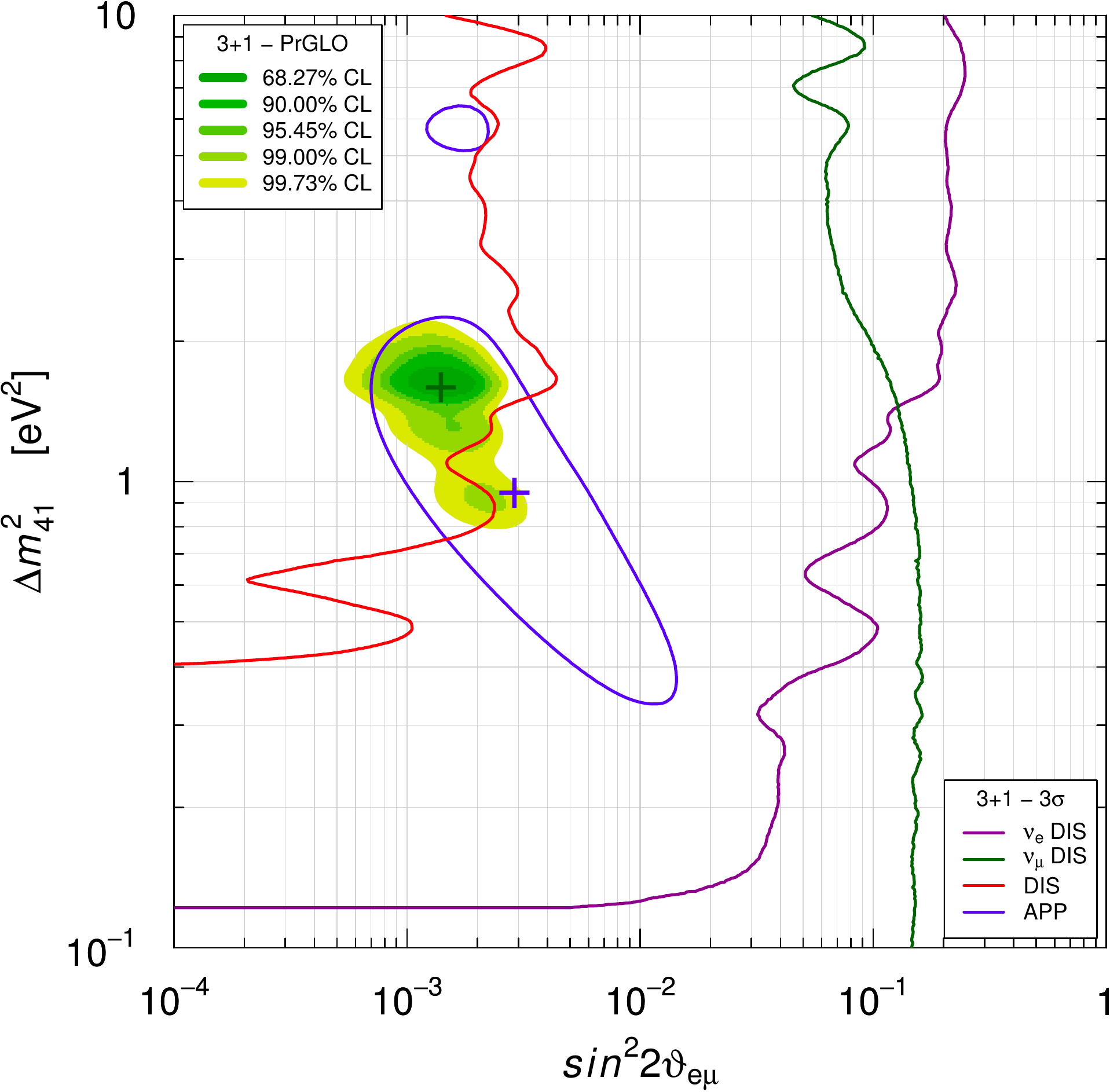}
\hfill
\includegraphics*[width=\halfwidth]{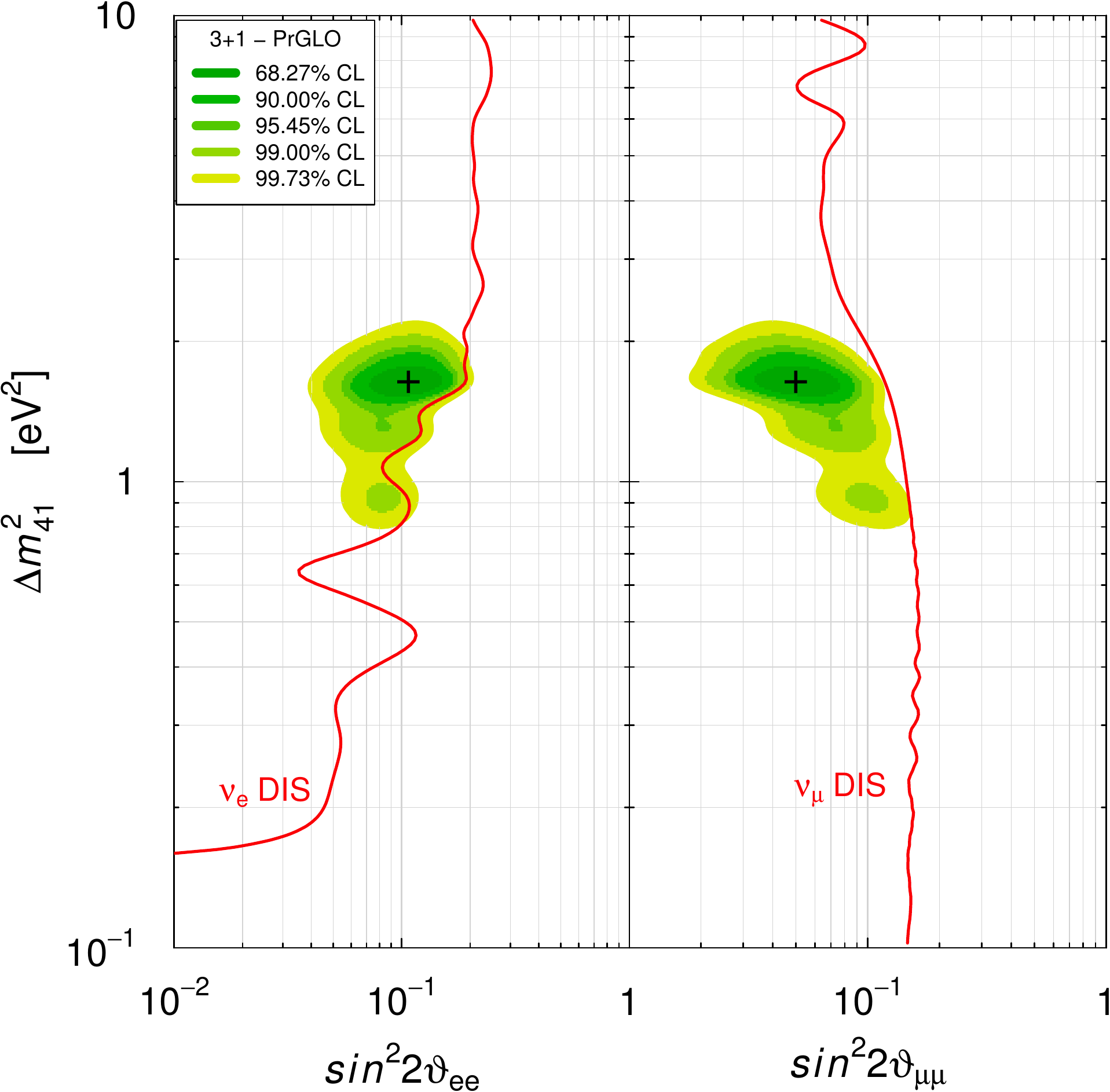}
\hfill
\null
\caption[Allowed regions for the mixing angle and
squared-mass difference in global fits of short-baseline
neutrino oscillation data.]{%
\label{fig:glo}
Allowed regions in the
$\sin^{2}2\vartheta_{e\mu}$--$\Delta{m}^{2}_{41}$,
$\sin^{2}2\vartheta_{ee}$--$\Delta{m}^{2}_{41}$
and
$\sin^{2}2\vartheta_{\mu\mu}$--$\Delta{m}^{2}_{41}$
planes
obtained in the pragmatic 3+1-PrGLO global fit
of short-baseline neutrino oscillation data.
These are compared with the $3\sigma$ allowed regions
obtained from
$\protect\nua{\mu}\to\protect\nua{e}$
short-baseline appearance data (APP),
the $3\sigma$ constraints obtained from
$\protect\nua{e}$
short-baseline disappearance data ($\nu_{e}$ DIS) and
$\protect\nua{\mu}$
short-baseline disappearance data ($\nu_{\mu}$ DIS),
and the
combined short-baseline disappearance data (DIS).
The best-fit points of the PrGLO and APP fits are
indicated by crosses.
From Ref.~\protect\cite{Gariazzo:2015rra}.
}
\end{figure*}

The statistical results obtained from the global fits of the data
listed above are summarized in
Table~\ref{tab:chi}.
The global (GLO) fit takes into account all the 
MiniBooNE data,
including the anomalous low-energy bins,
which are omitted in the pragmatic global (PrGLO) fit
\cite{Giunti:2013aea}.
The last two columns concern the results for
a fit without the MiniBooNE data (noMB)
and
one without the LSND data (noLSND).

\begin{table}[t]
\begin{center}
\begin{tabular}{c|cccc}
CL
&
$\Delta{m}^2_{41}[\text{eV}^2]$
&
$\sin^22\vartheta_{e\mu}$
&
$\sin^22\vartheta_{ee}$
&
$\sin^22\vartheta_{\mu\mu}$
\\
\hline
68.27\%
&
$ 1.57 - 1.72 $
&
$ 0.0011 - 0.0018 $
&
$ 0.085 - 0.13 $
&
$ 0.039 - 0.066 $
\\
\hline
90.00\%
&
$ 1.53 - 1.78 $
&
$ 0.00098 - 0.0020 $
&
$ 0.071 - 0.15 $
&
$ 0.032 - 0.078 $
\\
\hline
95.45\%
&
$ 1.50 - 1.84 $
&
$ 0.00089 - 0.0021 $
&
$ 0.063 - 0.16 $
&
$ 0.030 - 0.085 $
\\
\hline
99.00\%
&
$ 1.24 - 1.95 $
&
$ 0.00074 - 0.0023 $
&
$ 0.054 - 0.18 $
&
$ 0.025 - 0.095 $
\\
\hline
99.73\%
&
$ 0.87 - 2.04 $
&
$ 0.00065 - 0.0026 $
&
$ 0.046 - 0.19 $
&
$ 0.021 - 0.12 $
\end{tabular}
\end{center}
\caption[Marginal allowed intervals of the oscillation parameters
obtained in the global 3+1-PrGLO]
{ \label{tab:int}
Marginal allowed intervals of the oscillation parameters
obtained in the global 3+1-PrGLO fit
of short-baseline neutrino oscillation data.
From \protect\cite{Gariazzo:2015rra}.
}
\end{table}

From Tab.~\ref{tab:chi},
we can see that 
the absence of short-baseline oscillations
is nominally disfavored at about $6\sigma$
in all of the fits which include the LSND data,
because the improvement of the $\chi^2$ with
short-baseline oscillations
is much larger than the number of oscillation parameters.
On the other hand,
when the LSND data are not considered (noLSND fit),
the nominal exclusion of the case of no-oscillations drops
dramatically to $2.6\sigma$.
Therefore,
the LSND experiment
is clearly still crucial
for the indication in favor of short-baseline
$\bar\nu_{\mu}\to\bar\nu_{e}$.

In the GLO analysis, the goodness-of-fit is significantly 
worse than that in the PrGLO analysis
and the same applies for the appearance-disappearance
parameter goodness-of-fit.
This result confirms the fact that the MiniBooNE low-energy 
anomaly
is not compatible with neutrino oscillations,
requiring a small value of $\Delta{m}^2_{41}$
and a large value of $\sin^22\vartheta_{e\mu}$
\cite{Giunti:2011hn,Giunti:2011cp},
which are excluded by the oscillation data of other experiments
(further details are discussed in Ref.~\cite{Giunti:2013aea}).
Therefore,
it is very likely that the MiniBooNE
low-energy anomaly
must be explained with some mechanism
different from neutrino oscillations.
It is interesting to investigate what is the
impact of the MiniBooNE experiment
on the global analysis of short-baseline neutrino oscillation data.
With this aim,
we consider also the noMB fit without MiniBooNE data.
From Tab.~\ref{tab:chi}
we can see that the results of the
noMB fit are similar to those of the
PrGLO fit
and the nominal exclusion of the case of no-oscillations remains
at the level of $6\sigma$.
Therefore,
it is clear that the MiniBooNE experiment
has been rather inconclusive.
The MicroBooNE experiment at Fermilab
\cite{MicroBooNE-2007,Szelc:2015dga},
a large Liquid Argon Time Projection Chamber (LArTPC)
in which electrons and photons can be distinguished,
is going to investigate the cause of the MiniBooNE
low-energy excess of $\nu_{e}$-like events
and to check the LSND signal
(see the review in Ref.~\cite{Katori:2014qta}).
Since the low-energy anomaly of MiniBooNE is under discussion,
in the following we adopt the ``pragmatic approach''
advocated in Ref.~\cite{Giunti:2013aea}.
The PrGLO fit,
that does not take into account
the anomalous MiniBooNE low-energy bins,
is more reliable than the GLO fit,
which includes all the MiniBooNE data.

The allowed regions in the
$\sin^{2}2\vartheta_{e\mu}$--$\Delta{m}^{2}_{41}$,
$\sin^{2}2\vartheta_{ee}$--$\Delta{m}^{2}_{41}$ and
$\sin^{2}2\vartheta_{\mu\mu}$--$\Delta{m}^{2}_{41}$
planes as
obtained in the PrGLO fit
are shown in Fig.~\ref{fig:glo}.
These regions are relevant, respectively, for
$\nua{\mu}\to\nua{e}$ appearance,
$\nua{e}$ disappearance and
$\nua{\mu}$ disappearance
searches.
The corresponding marginal allowed intervals
of the oscillation parameters are given in Tab.~\ref{tab:int}.
Figure~\ref{fig:glo}
shows also the region allowed by
$\nua{\mu}\to\nua{e}$ appearance data
and
the constraints from
$\nua{e}$ disappearance and
$\nua{\mu}$ disappearance data.
We can see that the combined disappearance constraint
in the $\sin^{2}2\vartheta_{e\mu}$--$\Delta{m}^{2}_{41}$ plane
excludes a large part of the region allowed
by $\nua{\mu}\to\nua{e}$ appearance data,
leading to the well-known
appearance-disappearance tension
\cite{Kopp:2011qd,Giunti:2011gz,Giunti:2011hn,Giunti:2011cp,
Conrad:2012qt,Archidiacono:2012ri,Archidiacono:2013xxa,Kopp:2013vaa},
quantified by the parameter goodness-of-fit in Tab.~\ref{tab:chi}.

\section{Neutrino and Cosmology}
\label{sec:nucosmology}

This Section is devoted to extend the treatment presented
in the previous Chapters, where we ignored the
presence of the neutrino perturbations in the Universe evolution.
We will briefly review the impact of the neutrinos on the various
cosmological observables we mentioned earlier,
with a particular focus on the impact of a light sterile neutrino
with a mass at the eV scale.
A more detailed discussion is presented, for example,
in Ref.~\citelesg{}.

When considering the additional neutrino,
which is mostly sterile
as explained in Section~\ref{sec:mixing},
we will denote its mass with the symbol $m_{s}$.
In this section we use this notation,
keeping in mind that its real meaning in the 3+1 mixing scheme is
$m_{s}=m_{4}$.
Moreover,
in the discussion of the
combined analysis of cosmological data and short-baseline oscillation data
we consider
$m_{1},m_{2},m_{3} \ll m_{4}$,
so that
$m_{s} = m_{4} \simeq \sqrt{\Delta{m}^2_{41}}
= \sqrt{\Delta{m}^2_{\text{SBL}}}$.

This Section is organized as it follows:
in Subsection~\ref{sub:parameterization}
we introduce the parameterization of the neutrino energy density,
in Subsection~\ref{ssec:nuperturb} we discuss the definitions of the
neutrino perturbations,
in Subsection~\ref{sec:nuFS} we present the neutrino free-streaming,
in Subsection~\ref{sub:radiationeffects}
we briefly review the effects of neutrinos
which are relativistic in the early Universe
on observables such as
the Cosmic Microwave Background (CMB)
and
the nuclear abundances produced by Big Bang Nucleosynthesis (BBN).
In Subsection~\ref{sub:masseffects}
we discuss the effects of massive neutrinos,
% on the formation of Large Scale Structures (LSS),
which are important only after the sterile neutrinos became non-relativistic.
All these effects can be used to derive constraints
on the neutrino properties from the various cosmological data
we presented in the previous Chapter.
The constraints on the sterile neutrino properties will be
discussed in Chapter~\ref{ch:lsn_cosmo}.

\subsection{Neutrino Parameterization}
\label{sub:parameterization}

The neutrino contribution to
the radiation content in the early Universe 
can be conveniently parameterized in terms of
the effective number of degrees of freedom \Neff.
This is defined so
that the total energy density of relativistic species
$\rho_{r}$ is given by
\begin{equation}\label{eq:neff}
\rho_{r}
=
\left[1+\frac{7}{8}\left(\frac{4}{11}\right)^{4/3} \Neff\right] \rho_{\gamma}
=
\left[ 1 + 0.2271\, \Neff \right] \rho_{\gamma}
\,,
\end{equation}
where $\rho_{\gamma}$ is the energy density of photons.
$\Neff=1$ corresponds to the contribution of one single family
of active neutrinos which were in equilibrium in the early Universe
and
passed through an instantaneous decoupling at a temperature
of about 1 MeV.
The factor $7/8$ is for fermionic degrees of freedom,
while the factor $T_\nu^{id}/T_{\gamma}=(4/11)^{4/3}$
is the consequence
of the fact that after neutrino decoupling
there is an entropy transfer
between electrons and photons, caused by $e^\pm$ annihilations.
The superscript ``id'' indicates that this is the temperature
obtained in the instantaneous decoupling limit.
This entropy transfer enhances the photon temperature,
that becomes higher than the temperature of the decoupled neutrinos.
In the real history
the neutrinos did not decouple instantaneously
and part of them were not completely decoupled
from the electron-photon plasma
when the $e^\pm$ annihilation occurred.
For this reason, the effective number of active neutrinos
is slightly larger than three:
it is $\Neff^{\mathrm{SM}}=3.046$~\cite{Mangano:2001iu,Mangano:2005cc}.
Assuming that the active neutrino follows the usual thermal
history and that the non-standard contribution
to the effective number of relativistic species comes only
from additional sterile neutrinos,
the sterile neutrino contributes to the total radiation energy density with
$\DNeff=\Neff-3.046$.
This can be calculated as \cite{Acero:2008rh}
\begin{equation}
\label{eq:dneff}
\DNeff
\equiv
\frac{\rho_{s}^{\mathrm{rel}}}{\rho_\nu}
=
\left[\frac{7}{8}\frac{\pi^2}{15}{T_{\nu}^{id}}^4\right]^{-1}
\frac{1}{\pi^2}
\int dp \, p^3 f_{s}(p)
\,,
\end{equation}
where
$\rho_\nu$ is the energy density for one active neutrino species,
$\rho_{s}^{\mathrm{rel}}$ is the energy density
of the relativistic sterile neutrinos,
$p$ is the neutrino momentum and
$f_{s}(p)$ is the momentum distribution.
The same formula gives the corresponding contribution of
one single active neutrino if the momentum distribution function
$f_\nu(p)$ is used instead of $f_{s}(p)$.

After their non-relativistic transition,
neutrinos contribute to the matter energy density
of the Universe.
The contribution of one single neutrino with mass $m_\nu$
is given by \cite{Acero:2008rh}
\begin{equation}\label{eq:omeganu}
\omega_{\nu}
=
\Omega_{\nu} h^2
=
\frac{\rho_{\nu}}{\rho_{c}} \, h^2
=
\frac{h^2}{\rho_c}
\frac{m_{\nu}}{\pi^2}
\int dp \, p^2 f_{\nu}(p)
\,,
\end{equation}
where
$\rho_{\nu}$ is the energy density of a non-relativistic neutrino,
$f_{\nu}(p)$ is the momentum distribution,
$\rho_c$ is the critical density and
$h$ is the reduced Hubble parameter.
The sterile neutrino contribution can then be parameterized
in terms of the dimensionless number \cite{Acero:2008rh}
\begin{equation}\label{eq:omegas}
\omega_{s}
=
\Omega_{s} h^2
=
\frac{\rho_{s}}{\rho_{c}} \, h^2
=
\frac{h^2}{\rho_c}
\frac{m_{s}}{\pi^2}
\int dp \, p^2 f_{s}(p)
\,,
\end{equation}
where
$\rho_{s}$ is the energy density of
a non-relativistic sterile neutrino.
Alternatively, $\omega_{s}$ can be converted
in the effective sterile neutrino mass \cite{Ade:2013zuv}
\begin{equation}\label{eq:meffs}
\meff{s} \equiv 94.1 \, \omega_{s} \, \mathrm{eV}
\,.
\end{equation}

All the quantities that we introduced depend
on the neutrino momentum distribution $f_{\nu}(p)$ or $f_{s}(p)$.
We focus now on the sterile neutrino with mass of about 1~eV.
If the light sterile neutrino decouples
from the rest of the plasma when it is still relativistic,
$f_{s}(p)$ does not depend on $m_{s}$,
but it depends only on the production mechanism.
The simplest possibility is that one species of light sterile neutrinos
is generated by active-sterile oscillations in the early Universe
\cite{Dolgov:2003sg,Cirelli:2004cz,Melchiorri:2008gq,Hannestad:2012ky,
Mirizzi:2013gnd,Hannestad:2015tea}
and they share the same temperature of the active neutrinos.
In this case we have simply $\DNeff=1$
and
$\omega_{s} \simeq m_{s}/(94.1\,\mathrm{eV})$.

If for some reasons
the light sterile neutrino thermalizes at a temperature $T_{s}=\alpha T_\nu$,
its momentum distribution is given by the standard Fermi-Dirac distribution
\begin{equation}\label{eq:thermalnu}
f_{s}(p)=\frac{1}{e^{p/T_{s}}+1}
\,.
\end{equation}
We name this case the \emph{thermal scenario} (TH),
and
from Eqs.~\eqref{eq:dneff} and \eqref{eq:omegas} we obtain
\begin{equation}\label{eq:THquantities}
\DNeff = \alpha^4
\,,
\qquad
\omega_{s} = \alpha^3\,\frac{m_{s}}{94.1\,\mathrm{eV}}
\,,
\qquad
\meff{s} = \alpha^3 m_{s} = \DNeff^{3/4} m_{s}
\,.
\end{equation}

There are several possible mechanisms that give
a non-thermal sterile neutrino production.
A popular one is the non-resonant production scenario,
also called \emph{Dodelson-Widrow scenario} (DW) \cite{Dodelson:1993je},
which is motivated by early active-sterile neutrino oscillations
in the limit of zero lepton asymmetry and small mixing angle.
It is possible to calculate the neutrino momentum distribution 
for the DW scenario:
\begin{equation}\label{eq:DWnu}
f_{s}(p)=\frac{\beta}{e^{p/T_\nu}+1}
\,,
\end{equation}
where $\beta$ is a normalization factor.
This momentum distribution leads to
\begin{equation}\label{eq:DWquantities}
\DNeff=\beta
\,,
\qquad
\omega_{s} = \beta\,\frac{m_{s}}{94.1\,\mathrm{eV}}
\,,
\qquad
\meff{s} = \beta m_{s} = \DNeff m_{s}
\,.
\end{equation}

We can see from Eqs.~\eqref{eq:THquantities}
and \eqref{eq:DWquantities} that
the DW and the TH models have an exact degeneracy,
since they are related by
$\alpha=\beta^{1/4}$ and $m_{s}^{\mathrm{TH}}=m_{s}^{\mathrm{DW}}\beta^{1/4}$
\cite{Colombi:1995ze,Cuoco:2005qr}.

\subsection{Neutrino Perturbations}
\label{ssec:nuperturb}
We want now to extend the treatment of the perturbation theory presented
in Chapter~\ref{ch:cosmology} with the introduction of
the neutrino perturbations.
Neutrinos behave differently when relativistic or non-relativistic, and the full
treatment must take into account the two possibilities.
The treatment of the massless neutrino perturbations can be used to describe
any collisionless particle that is still relativistic today,
i.e.~any particle with mass $m\lesssim 10^{-4}$~eV.
Only one out of the three standard neutrinos can be still in this state,
given that its mass is sufficiently small.
The squared-mass differences
obtained from the analyses of the neutrino oscillation data, in fact,
tell us that the other two neutrino mass eigenstates are non-relativistic today.

In this Section we will show how it is possible to deal with neutrino
perturbations in the evolution equations of the Universe, but
we will not show how to find the solutions
in the numerical calculation.
The interested reader can see
Ref.~\citelesg{}
for a detailed treatment.

\subsubsection{Massless Neutrinos}
Details of neutrino decoupling would only impact perturbations that were
inside the Hubble horizon at the time of neutrino decoupling.
These scales are not observable today, since they are suppressed because of
diffusion damping, and anyway they are contaminated by foreground
emission in real dataset.
They are not observable neither in the spectrum of large scale structures,
since the non-linear evolution has strong effects that deleted the memory
of the previous linear evolution.

Neglecting the non-thermal distortions due to electron-positron annihilation,
that are very small,
we can consider the neutrino distribution function to be a
simple Fermi-Dirac distribution.
As a consequence, the neutrino perturbations can be calculated in the same
way of the photon perturbations,
apart for the sign in the
Fermi-Dirac distribution with respect to the Bose-Einstein one.
The main difference for the neutrinos, clearly,
is the absence of interaction terms with the baryons
in all the relevant differential equations.

Using $\mcn$ to denote the neutrino perturbations,
in analogy with $\Theta$ for the photons,
we can write the differential equation for the evolution of the neutrino
perturbations in the Fourier space:
\be\label{eq:BoltzEqMasslessNu}
\dot\mcn+i k \mu\mcn
=
-\dot\Phi -i k \mu\Psi\,.
\ee
The neutrino perturbation $\mcn$ can be treated as the photon perturbation
$\Theta$, being the only difference in the equations is that
for the neutrinos the limit $\sigma_T\rightarrow0$ applies.

This is not the most general treatment that can be developed.
To describe the neutrino perturbations when the distribution function 
is not of the standard Fermi-Dirac type
one should generalize the discussion
as shown for example in 
Ref.~\citelesg{}.
The extended treatment
can be used if the neutrino has a chemical potential or relevant non-thermal
distortions, but also for other decoupled relativistic relics.

\subsubsection{Massive Neutrinos}
To describe massive neutrinos we have to find a set of equations that interpolate
from the CDM equations (in the large mass limit) to the massless neutrinos
equations (in the small mass limit).
The simplest assumption is that
neutrinos are decoupled and still relativistic
at the time of imposing the initial conditions,
so that they have
a Fermi-Dirac momentum distribution $f_{\nu,0}$.
For the active neutrinos, this would be enough.
Since we want to deal with sterile neutrinos, we assume that $f_{\nu,0}$
has a generic form, but 
we require that it is time-independent after neutrino decoupling.

For massive neutrinos, the mass enters the expression for the energy and
some of the simplifications we assumed in Section~\ref{sec:pertUniv} are
no more valid.
The reason is that the gravitational interactions induce
a relative momentum shift that depends on the momentum itself.
We can still simplify the Boltzmann equations with the introduction of the
relative fluctuations of the phase-space distribution,
that we denote with $\Upsilon$:
\be
\Upsilon(\eta,\vec x, p, \hat n)
\equiv
\frac{f_\nu(\eta,\vec x, p, \hat n)}{f_\nu(\eta, p)}
-1\,,
\ee
at the first order in perturbations.
In the general case, in the relativistic limit we have:
\be\label{eq:UpsRel}
  \Upsilon(\eta,\vec x, p, \hat n)
  =
  -\frac{1}{4}
  \mcn(\eta,\vec x, p, \hat n)
  \frac{d\ln f_{\nu,0}(y)}{d\ln y}
  \,,
\ee
where $y=ap$.
This expression is valid only when neutrinos are relativistic,
since when each neutrino becomes non-relativistic
the non-thermal distortions induced by gravity
modify the distribution function.

If we write the Boltzmann equation for massive neutrinos and
we replicate the calculations developed
for photons and massless neutrinos,
we obtain the equation of motion for $\Upsilon$ in the real space:
\be\label{eq:BoltzEqMassiveNuReal}
\dot\Upsilon
+\frac{y}{\epsilon}\hat n\cdot\vec\nabla\Upsilon
=
\frac{d\ln f_{\nu,0}}{d\ln y}
\left(
  \dot\Phi
  +\frac{\epsilon}{y}\hat n\cdot\vec\nabla\Psi
\right)
\,,
\ee
where $\epsilon$ is the neutrino energy.
In the relativistic limit,
$y/\epsilon\rightarrow1$
and we can use Eq.~\eqref{eq:UpsRel} to recover
Eq.~\eqref{eq:BoltzEqMasslessNu}.

In analogy with the treatment of the photon perturbations,
it is possible to expand $\Upsilon$ in Legendre momenta to obtain
an infinite hierarchy of $\Upsilon_l$.
From Eq.~\eqref{eq:BoltzEqMassiveNuReal}, using the
axial symmetry around $\hat n$ and going to the Fourier space,
the equations for the $\Upsilon_l$ are
\begin{align}
 \dot\Upsilon_0
 &=
 -\frac{yk}{\epsilon}\Upsilon_1
 +\dot\Phi\frac{d\ln f_{\nu,0}}{d\ln y}
 \\
 \dot\Upsilon_1
 &=
 \frac{yk}{3\epsilon}(\Upsilon_0-2\Upsilon_2)
 -
 \frac{\epsilon k}{3y}\,
 \frac{d\ln f_{\nu,0}}{d\ln y}
 \,\Psi
 \\
 \dot\Upsilon_l
 &=
 \frac{yk}{(2l+1)\epsilon}\,
 [
  l\,\Upsilon_{l-1}
  -(l+1)\Upsilon_{l+1}
 ]
 \,,\qquad\forall l\geq 2\,.
\end{align}

Given that $y/\epsilon\rightarrow0$ in the deeply non-relativistic
limit,
one could show that
the neutrino perturbations evolve exactly as those of CDM
after each neutrino becomes non-relativistic.
This ca ne seen using the definitions in
Eqs.~\eqref{eq:rho_singlefluid}, \eqref{eq:defDelta}
and \eqref{eq:defV}
for massive neutrinos,
with the expression of $f_\nu$ at the first
order in perturbations.

\subsubsection{Adiabatic initial Conditions in Presence of Neutrinos}

In presence of neutrinos,
the relation $\Phi+\Psi=0$ is no more valid in the early Universe,
as a consequence of the neutrino anisotropic stress in
Eq.~\eqref{eq:pertFriedEq1}.
The presence of neutrinos induce a constant offset between the 
metric perturbations $\Phi$ and $\Psi$. 
The growing adiabatic solution becomes
\be\label{eq:adiabNu1}
  \Phi=-\Psi\left(1+\frac{2}{5}R_\nu\right)\,,
\ee
where we defined the neutrino ratio $R_\nu$ as
\be
R_\nu
\equiv
\frac{\rho_\nu\zeroord}{\rho_\gamma\zeroord+\rho_\nu\zeroord}
=
\frac{
  \frac{7}{8}\left(\frac{4}{11}\right)^{4/3}\Neff
}{
  1+\frac{7}{8}\left(\frac{4}{11}\right)^{4/3}\Neff
}
\,,
\ee
assuming that all the neutrinos are relativistic at the time
of applying the initial conditions.
It is possible to show that
the updated version of Eqs.~\eqref{eq:adiab1} and \eqref{eq:adiab2}
reads
\be\label{eq:adiabNu2}
\Theta_0(k,\eta_i)
=
\mcn_0(k,\eta_i)
=
\frac{\delta}{3}
=
\frac{\delta_b}{3}
=
-\frac{\Psi}{2}
\,,
\ee
where $\Psi$ can be substituted with $\Phi$ using the relation
\eqref{eq:adiabNu1}.

Equation \eqref{eq:adiabNu1} may give the wrong impression that
the initial conditions depend on the neutrino anisotropic stress.
The shift between $\Phi$ and $\Psi$
is explicit in the conformal Newtonian gauge, but
it disappears in other gauges, so that it is clear
that $R_\nu$ has no observable consequences.
Without changing gauge,
this can be seen from the fact that the equations contain the metric
perturbations in the $\dot\Phi$ and $k\Psi$ terms.
The contribution of $R_\nu$ does not affect $\dot\Phi$ and the
leading term of $k\Psi$ is of order $(k\eta)$.
As a consequence,
the equations for the evolution are affected by the presence of
neutrinos only at the next-to-leading order in a $(k\eta)$ expansion.

\subsection{Neutrino Free-streaming}
\label{sec:nuFS}
After decoupling, neutrinos evolve as freely falling particles.
Neutrino free-streaming does not affect all the scales,
since the Universe expands.
The characteristic quantities that describe the
distances related to neutrino free-streaming 
are the \emph{free-streaming scale} $\lambda\fs{}$, indicating the
scales at which free-streaming can be ignored,
and the \emph{free-streaming horizon} $d\fs{}$, corresponding to the
average distance traveled by neutrinos before a given time.

The free-streaming scale $\lambda\fs{}$,
or the corresponding wavenumber $k\fs{}$ in comoving Fourier space,
can be defined as the product of the
neutrino velocity $c_\nu$ by the Hubble time $t_H=1/H$,
normalized in analogy with the Jeans length:
\be
\lambda\fs{}(\eta)
=
a(\eta)\frac{2\pi}{k\fs{}}
\equiv
2\pi\sqrt{\frac{2}{3}}
\,\frac{c_\nu(\eta)}{H(\eta)}
\,.
\ee
This is the scale below which the free-streaming particle cannot
be confined inside a gravitational potential well.

The free-streaming horizon, instead, is defined as the integral
\be
d\fs{}(\eta)
= a(\eta)\, r\fs{}(\eta)
\equiv
a(\eta)
\int_{\eta\inu}^\eta c_\nu(\eta)d\eta
\,,
\ee
which is independent of $\eta\inu$ if $\eta\inu\ll\eta$ is 
chosen after the end of inflation.

While a neutrino is relativistic, its speed is $c_\nu=c=1$
and we have simply
\be
\lambda\fs{}=2\pi\sqrt{\frac{2}{3}}\frac{1}{H}
\,,
\qquad
d\fs{}=a\eta\,.
\ee
These quantities are very close to each other, since $a\eta=H^{-1}$
during radiation domination and $a\eta=2H^{-1}$
during matter domination.

The story is more complicated for neutrinos that become
non-relativistic during the Universe evolution.
Quantitatively, neutrinos become non-relativistic when their
mean momentum $\langle p\rangle$ becomes smaller
than their mass $m_\nu$.
If neutrinos follow a relativistic Fermi-Dirac distribution
with negligible chemical potential,
the average momentum can be calculated and it is
$\langle p\rangle=3.15\, T_\nu$.
Using the relation between the photon and neutrino temperatures,
it is possible to show that a neutrino becomes non-relativistic
during matter domination if its mass is
\be
5.28\e{-4}\ev\leq m_\nu\lesssim 1.5\ev
\,.
\ee
Since we are interested in neutrinos below 1.5~eV, we will firstly
discuss the free-streaming quantities during matter domination.
After the non-relativistic transition, the thermal velocity
$c_\nu$ scales with $a^{-1}$ or $\eta^{-2}$, since
\be
c_\nu
= \frac{\langle p\rangle}{m_\nu}
=
158
(1+z)
\left(
\frac{T_\nu}{T_\nu^{id}}
\right)
\left(
\frac{1\ev}{m_\nu}
\right)
\mathrm{Km}\,\mathrm{s}^{-1}
\,.
\ee
This means that the free-streaming length increases with $\eta$
and the comoving free-streaming length decreases with $\eta^{-1}$.
At the time of the non-relativistic transition, the comoving
free-streaming length passes through a maximum that corresponds to
the wavenumber $k\nr{}$. 
This can be approximated as
\be
k\nr{}
\equiv
k\fs{}(\eta\nr{})
\simeq
0.0178\,
\Omega_M^{1/2}
\left(
\frac{T_\nu^{id}}{T_\nu}
\right)^{1/2}
\left(
\frac{m_\nu}{1\ev}
\right)^{1/2}
h\,\mathrm{Mpc}^{-1}
\,,
\ee
valid only if the transition occurs during matter domination.
The comoving free-streaming horizon, instead,
becomes 
\be
r\fs{}(\eta>\eta\nr{})
\simeq
\sqrt{\frac{3}{2}}
\frac{4}{k\nr{}}
\left(
  1
  -
  \frac{1}{2}
  \left[
    \frac{1+z}{1+z\nr{}}
  \right]^{1/2}
\right)
\,.
\ee
Also in this case the expression is valid
during matter domination only.

For heavier neutrinos that becomes non-relativistic during
radiation domination, the things are slightly different.
The free-streaming length still increases as $\eta$, but the
comoving free-streaming length is constant,
since in this case the relation between $a$ and $\eta$ during
radiation domination must be considered.
Since the comoving free-streaming length starts to decrease after
matter-radiation equality, it encounters its maximum between
$\eta\nr{}$ and $\eta\eq{}$.
The minimum value of $k\fs{}$ is then
\be
k\nr{}
\equiv
k\fs{}(\eta\nr{})
\simeq
0.776\,
\Omega_R^{1/2}
\left(
\frac{T_\nu^{id}}{T_\nu}
\right)^{1/2}
\left(
\frac{m_\nu}{1\ev}
\right)^{1/2}
h\,\mathrm{Mpc}^{-1}
\,.
\ee
If we approximate 
\be
  c_\nu
  =
  \left\{
  \begin{array}{ll}
   1 & \text{ for }\eta\leq\eta\nr{}\\
   \eta\nr{}/\eta & \text{ for }\eta\nr{}<\eta\leq\eta\eq{}\\
   \eta\nr{}\eta\eq{}/\eta^2 & \text{ for }\eta>\eta\eq{}
  \end{array}
  \right.
  \,,
\ee
the comoving free-streaming horizon after matter-radiation equality
becomes
\be
r\fs{}(\eta>\eta\eq{})
\simeq
\sqrt{\frac{3}{2}}\,\frac{2}{\eta\nr{}}
\left[
  1
  +
  \frac{1}{2}\log
  \left(
  \frac{1+z\nr{}}{1+z\eq{}}
  \right)
  -
  \frac{1}{2}
  \left(
  \frac{1+z}{1+z\eq{}}
  \right)^2
\right]
\,.
\ee
The last term is usually negligible, but the logarithm may be large
for heavy particles becoming non-relativistic at high redshift.

In the next Subsections we will use all the defined quantities
to describe the neutrino effects on the main cosmological observables.
We will try to separate the background from the perturbation effects,
both for massless and massive neutrinos.

\subsection{Physical Effects as Radiation in the early Universe}
\label{sub:radiationeffects}

Before discussing the impact of neutrinos on the CMB spectrum,
we recall that it is complex to single out the effects of
a specific quantity, since it is connected with the other
quantities. 
It is often difficult (or impossible) to separate the
contributions of each parameter,
but we will do our best to isolate the effects of neutrinos from
those of all the others parameters.

% The presence of neutrinos impacts the primary CMB anisotropies in
% several ways, that can be classified as background or perturbation
% effects.
% Perturbation effects are more direct, since the impact of neutrino
% perturbations cannot be easily mimicked by variations 
% of other parameters.

The contribution of neutrinos as relativistic particles
can be described simply through the parameter $\Neff$
we have already defined. 
As relativistic components,
additional neutrino degrees of freedom change
the time of matter-radiation equality
(effect \ec3 in Section~\ref{sec:par_depend}),
whose redshift $z\eq{}$ is given by
\begin{equation}\label{eq:zeq}
1+z\eq{}
=
\frac{\rho_{m}}{\rho_{r}}
=
\frac{\omega_{m}}{\omega_{r}}
=
\frac{\omega_{m}}{\omega_{\gamma}}\frac{1}{1+0.2271\,\Neff}
\,,
\end{equation}
where we used Eq.~\eqref{eq:neff}.
To shorten the notation, we define conveniently
\be
\alpha\equiv1+0.2271\,\Neff\,.
\ee
A shift in the matter-radiation equality affects
the position (effect \ec1)
and the shape (effect \ec4) of the acoustic peaks of the CMB
(see Ref.~\cite{Archidiacono:2013fha}).
At recombination,
the extra radiation component enhances the expansion rate $H$.
This increase of $H$
generates a decrease of the comoving sound horizon
$r_{\text{s}} \propto H^{-1}$
\cite{Hou:2011ec}
and a reduction of the angular scale of
the acoustic peaks $\theta_{\text{s}}=r_{\text{s}}/D_A$,
leading to a shift of the CMB peaks towards higher multipoles
(see Fig.~2(a) of Ref.~\cite{Archidiacono:2013fha}).
In addition,
if matter-radiation equality is delayed,
the amplitude of the first CMB peak at $\ell\simeq200$
is increased by the early ISW effect,
since decoupling occurs when matter domination is
at an earlier stage
and the subdominant radiation component causes
a slow decrease of the gravitational potential
(see Figs.~2(a) and 2(b) of Ref.~\cite{Archidiacono:2013fha}).

These effects of additional relativistic neutrinos
can be partially compensated if other cosmological parameters
are simultaneously varied.
For example, if the total matter density
$\omega_{m}$ is also increased by a factor $\alpha$ without altering
the baryon density $\omega_b$,
so that the ratio between odd and even CMB peaks is not altered,
according to Eq.~\eqref{eq:zeq} $z\eq{}$ can be kept fixed
and the two effects discussed above do not appear.
After having restored the matter radiation equality, we should
consider the coincidence time (effect \ec7) that is altered
by the increase of $\omega_m$.
We can increase also the cosmological constant energy density 
$\Omega_\Lambda$ and all the important redshifts at
which the Universe change its evolution domination are preserved.
However,
we cannot obtain exactly the same CMB spectrum
as in the standard case,
because the additional relativistic neutrinos
increase the Silk damping effect at high multipoles
\cite{Bowen:2001in,Bashinsky:2003tk,Hou:2011ec}.
The damping depends on the ratio $r_{\text{d}} / r_{\text{s}}$,
where $r_{\text{d}} \propto H^{-1/2}$ is the photon diffusion length
at recombination \cite{Hou:2011ec}.
Since at fixed
$z\eq{}$
we have
$H^2 \propto \rho_{r}
= \alpha \rho_{\gamma}$,
an increase of $\Neff$
corresponds to an increase of $H$
and to an increase of $r_{\text{d}} / r_{\text{s}}\propto\alpha^{1/4}$,
which enhances the Silk damping at high-multipoles
\cite{Hou:2011ec}.

\begin{figure}[t]
  \centering
  \includegraphics[width=\singlefigland]{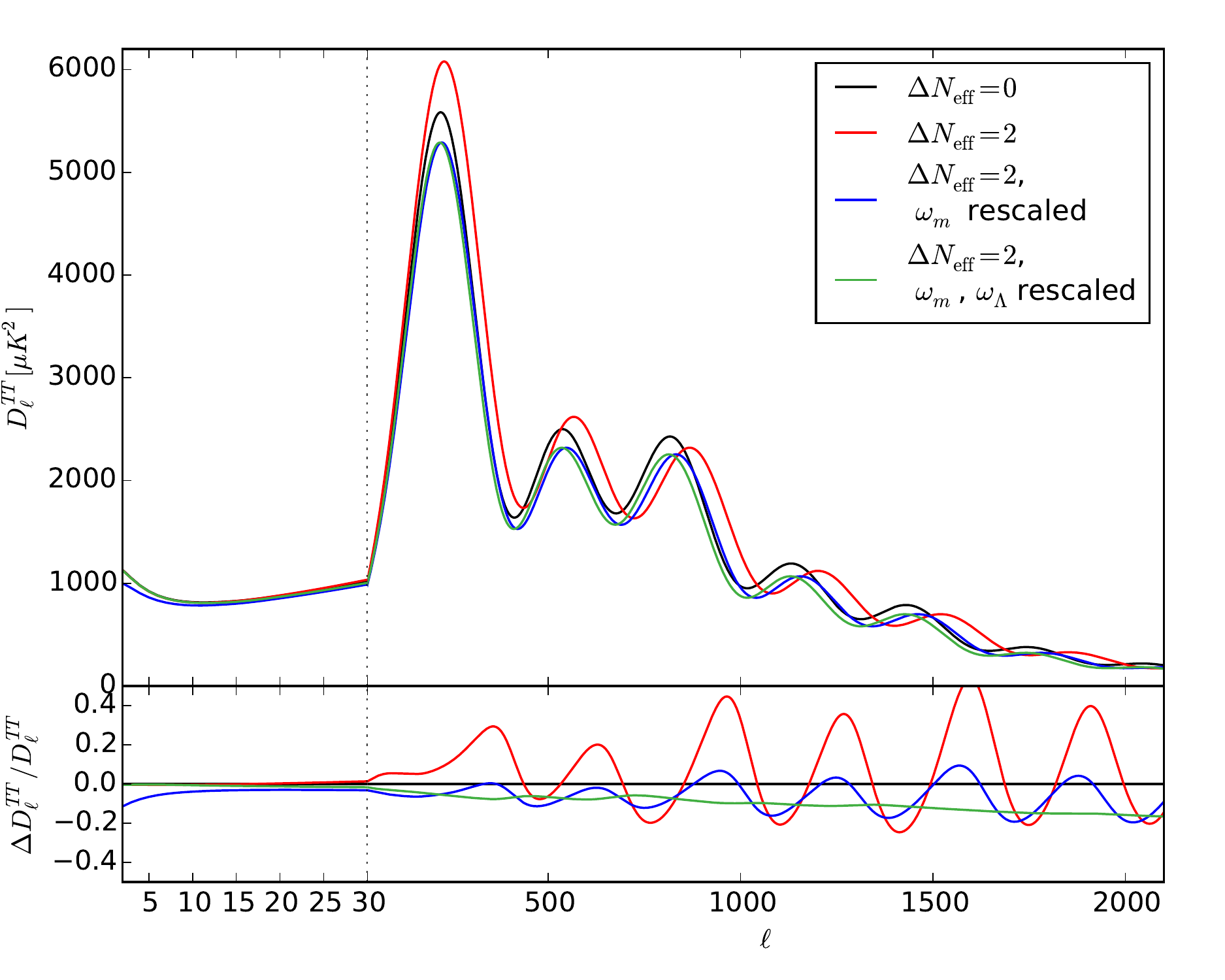}
  \caption[CMB spectrum for different values of \protect\Neff]{%
  \label{fig:neff_cl}
  Comparison of the CMB spectrum obtained for different
  values of the effective number of relativistic species
  $\protect\Neff$ and for different values of
  $\omega_m$ and $\omega_\Lambda$, rescaled to fix the
  matter-radiation equality and the coincidence times.
  The upper panel shows the spectrum $D_l=l(l+1)\,\cl{TT}/(2\pi)$,
  while the lower panel shows the relative difference with
  respect to the model with a standard neutrino content.
%   We can see how an increased amount of radiation in the early
%   Universe, if the redshifts of matter-radiation equality and
%   of coincidence do not change,
%   is to enhance the diffusion damping at small scales.
  From Ref.~\cite{Gariazzo:2016ehl}.
  }
\end{figure}

The effects of $\Neff$ are summarized in Fig.~\ref{fig:neff_cl}
from Ref.~\cite{Gariazzo:2016ehl},
where we compare the CMB spectrum predicted by a model
with $\DNeff=0$ (black line) with the spectrum obtained varying
$\Neff$ alone (red line with $\DNeff=2$),
the one with changed
$\Neff$ and $\omega_m$ (blue line, same $\DNeff$,
$\omega_m$ rescaled by $\alpha$)
and the last one with rescaled
$\Neff$, $\omega_m$ and $\Omega_\Lambda$
(green line, same $\DNeff$ and $\omega_m$,
$\Omega_\Lambda$ rescaled by $\alpha$).
It is easy to see that the change in the matter-radiation equality
has an effect on the amplitude, the position and the envelope
of the peaks,
that is partially restored changing the total matter density.
A residual effect is still present because of the different
time of matter-$\Lambda$ equality.
Once also the second equality is restored to the initial value,
the only remaining background effect is
the enhanced diffusion damping at high multipoles,
well visible comparing the black and green curves in the lower panel.

The effect of altering $\Neff$ is not limited to the background
evolution of the Universe, however.
At the level of perturbations, neutrino effects can be important when a mode
crosses the sound horizon and acoustic oscillations are driven
by metric fluctuations.
The presence of neutrinos varies the size of metric fluctuations inside the
free-streaming scale, below which neutrino cannot cluster.
During radiation domination, neutrinos are a large fraction of the total 
content of the Universe and they significantly reduce the metric fluctuations 
at distances smaller than their free-streaming scale.
We have seen that temperature fluctuations are boosted by time variations
of the metric fluctuations:
the presence of neutrinos has hence the result of reducing this boost
during the driven oscillation stage.
The temperature fluctuations for the modes that enter the sound horizon before
decoupling, especially during radiation domination, are then smaller.
An analytic approximation of the impact of neutrinos on the driven oscillations
has been derived in Ref.~\cite{Hu:1995en}.
The oscillation amplitude inside the sound horizon is reduced by a factor
$(1+4/15R_\nu)^{-1}$.
With respect to the neutrinoless model 
the CMB peaks are reduced by the square of:
\be\label{eq:peakSuppNeffPert}
\frac{\Delta C_l}{C_l}
=
\left(
  1+
  \frac{4}{15}
  \frac{\alpha-1}{\alpha}
\right)^{-2}
\,.
\ee
For small variations of $\Neff$ around three,
the above expression can be approximated with
\be
\frac{\Delta C_l}{C_l}
=
-0.072\,\DNeff
\,,
\ee
valid in the region of acoustic oscillations.
A more detailed calculation \cite{Bashinsky:2003tk}
reported a different formula:
\be
\frac{\Delta C_l}{C_l}
=
\left(
  1-
  0.2683\,R_\nu
  +\mco(R_\nu^2)
\right)^{2}
\,,
\ee
in good agreement with Eq.~\eqref{eq:peakSuppNeffPert}.
The authors of Ref.~\cite{Bashinsky:2003tk} report
also that relativistic neutrinos tend to pull temperature perturbations out
of gravitational potential wells, since neutrinos moves faster than the
temperature perturbations (traveling at a speed $c_s\simeq c/\sqrt{3}$).
This \emph{neutrino drag} effect causes a shift in the phase of the acoustic
oscillations, so that the peaks are shifted at smaller $l$.
The analytic approximation gives:
\be
\Delta l_\text{peak}
=
-
\frac{r_s(\etals{})}{r_A(\etals{})}
\left(
  0.1912\, R_\nu+\mco(R_\nu^2)
\right)
\,.
\ee

Finally, the effective number of relativistic species is
connected with BBN:
the number of relativistic degrees of freedom fixes
the expansion rate during BBN, that in turn fixes
the abundances of light elements.
BBN can thus give strong constraints on $\Neff$
through the observations of the primordial abundances of light elements
\cite{Steigman:2012ve,Iocco:2008va,Jacques:2013xr,Fields:2014uja}.
According to Ref.~\cite{Mangano:2011ar},
BBN limits the effective number of additional relativistic species
to $\DNeff<1$ at 95\% C.L., regardless of the inclusion
of CMB constraints on the baryon density $\Omega_b h^2$.
More recently, the authors of Ref.~\cite{Cyburt:2015mya} obtained
$\DNeff<0.2$ at 95\% C.L.
considering the BBN and CMB data.

\subsection{Physical Effects as massive Component}
\label{sub:masseffects}

The parameterization of massive particles as neutrinos is a not trivial
step of the description of the cosmological theory.
While a single parameter is enough to describe massless particles,
the energy density $\omega_\nu$ plays an important role in describing massive
neutrinos, but it may not catch all the effects that they induce on cosmology.
If one wants to go beyond the minimal picture, for example,
the mass of each neutrino mass eigenstate 
plays a role, as well any modification of the
phase-space distribution function due to the presence of chemical potentials
or non-thermal distortions.
We will assume for simplicity that the three active neutrinos
share the same mass, being the effects of the single neutrino masses 
extremely small to be detected.
The effects of different neutrino masses would be easier to detect in
the power spectrum of large scale structures, for which the suppression due
to neutrino free-streaming is larger.
Also the small non-thermal distortions due to 
electron-positron annihilation after a non-instantaneous neutrino
decoupling, that would alter differently the distribution functions of 
$\nu_e$, $\nu_\mu$ and $\nu_\tau$, have a very small imprint on the observable
quantities.

As we are assuming masses below 1.5~eV, neutrinos are still relativistic
at matter-radiation equality.
The redshift of equality, defined in Eq.~\eqref{eq:zeq},
must be calculated with the neutrinos as relativistic components.
The comparison between different masses can be performed at fixed
$\omega_m=\omega_b+\omega_c$, or better at fixed ratio
$\omega_b/\omega_c$.
The matter energy density today, however, is
$\omega_m=\omega_b+\omega_c+\omega_\nu$, since neutrinos became non-relativistic.
The difference between the model with $\sum m_\nu=0$ and the one with
$\sum m_\nu>0$ appears only after the neutrino non-relativistic transition,
that occurs at $z\nr{}\propto m_\nu$.
The neutrino mass has then an impact only at redshift
$z\lesssim z\nr{}$ on the comoving angular diameter distance to recombination
and on the redshift of the dark matter-dark energy equality.
For neutrinos that are non-relativistic at photon decoupling ($m_\nu\geq0.6\ev$),
there is an additional impact on the comoving sound horizon $r_s(\etals{})$
and on the damping scale $r_d(\etals{})$ at recombination.

One between $d_A(\etals{})$ and $z_\Lambda$ can be fixed changing $h$ or
$\Omega_\Lambda$, but not both simultaneously.
Since $d_A(\etals{})$ is related to the scale of the peaks, it is more 
interesting to fix it and to let $z_\Lambda$ change.
Most of the effects \ec1--\ec8 are unchanged for variations of these quantities,
with only two exceptions.
If $d_A(\etals{})$ is maintained fixed, the shift in $z_\Lambda$ induces a
change in the late ISW effect \ec7 that alters the spectrum of CMB
anisotropies at the largest scales.
Moreover, only for neutrinos heavier than 0.6~eV, the additional impact
on $r_d(\etals{})$ causes a variation in the diffusion damping \ec4.
This concludes what we can say about the modifications of the
background evolution,
but additional effects appear
at the perturbation level.

Neutrino masses can cause perturbation effects through the evolution of the
metric perturbations after decoupling (early ISW effect) or through the gravitational
driving of photon-baryon oscillations before decoupling,
as already discussed for the massless neutrinos.
The former effect gives the larger contribution in the multipoles range
$20\lesssim l\lesssim500$.
The depletion of the spectrum in this range can be roughly approximated with
\citelesg{,Lesgourgues:2014zoa}
\be
\frac{\Delta C_l}{C_l}
\simeq
-
\left(\frac{m_\nu}{10\ev}\right)\%
\,,
\ee
but the multipoles range that it affects depends on the neutrino free-streaming.
Since massive neutrinos can cluster at scales $k<k\nr{}$, while massless
neutrinos free-streams at the same scales, the metric perturbations experience
less decay in presence of neutrino masses.
For this reason, the early ISW effect is smaller for $k<k\nr{}$,
which is visible above a given angle on the CMB spectra.

\part{Beyond the Standard Model}

%!TeX root=main.tex 
\chapter{Light Sterile Neutrino in Cosmology}
\label{ch:lsn_cosmo}
\chapterprecis{This Chapter is based on
Refs.~\protect\cite{Gariazzo:2013gua,Archidiacono:2014apa,Gariazzo:2014pja}.}

In the previous Chapters we introduced the main ingredients of the analyses
we are going to present:
the Cosmic Microwave Background (CMB) radiation physics and observations,
the other cosmological observations,
the physics of neutrino and their effects in cosmology,
with a particular attention to the light sterile neutrino (\lsn)
with mass of around 1~eV
motivated by short-baseline (SBL) neutrino oscillations.
In this Chapter we study the constraints on the \lsn\ that can be obtained
from the analysis of CMB data and we show how other cosmological measurements
can influence these constraints.

%%%%%%%%%%%%%%%%%%%%%%%%%%
% JHEP 1311 (2013) 211   %
%%%%%%%%%%%%%%%%%%%%%%%%%%

\section{Light Sterile Neutrino Constraints with Planck 2013 Results}
\label{sec:jhep}
% \cite{Gariazzo:2013gua}

After the Planck collaboration published the 2013 release of data and codes
\cite{Ade:2013sjv,Ade:2013zuv},
a lively discussions started to grow
\cite{Jacques:2013xr,DiBari:2013dna,Boehm:2013jpa,Kelso:2013paa,
DiValentino:2013qma,Said:2013hta,Weinberg:2013kea,Verde:2013wza,
Verde:2013cqa,Wyman:2013lza,Hamann:2013iba,Battye:2013xqa}
on the value of the effective number of relativistic degrees of freedom
\Neff\
before photon decoupling
(see \cite{Giunti:2007ry,Lesgourgues:2012uu,
Lesgourgues-Mangano-Miele-Pastor-2013}),
which gives the energy density of radiation $\rho_r$ through the relation
presented in Eq.~\eqref{eq:neff}.
Since the value of $\Neff$ in the Standard Model (SM) is
$\Neff^{\text{SM}}=3.046$
\cite{Mangano:2001iu,Mangano:2005cc},
a positive measurement of
$\DNeff$
may be a signal that the radiation content of the Universe
was due not only to photons and SM neutrinos,
but also to some additional light particle called generically ``dark radiation''.

In this Chapter we consider the possibility that the dark radiation is made
of the light sterile neutrinos
(see Chapter~\ref{ch:nu})
whose existence is indicated by the results of
SBL neutrino oscillation experiments
(see Section~\ref{sec:sbl}).
Here we consider the simplest possibility of the 3+1 scheme presented in
Section~\ref{sec:mixing},
in which the three active flavor neutrinos
$\nu_{e}$,
$\nu_{\mu}$,
$\nu_{\tau}$,
are mainly composed of
three very light neutrinos
$\nu_{1}$,
$\nu_{2}$,
$\nu_{3}$,
with masses much smaller than 1 eV,
and there is a sterile neutrino
$\nu_{s}$
which is mainly composed of a new massive neutrino
$\nu_{4}$
with mass
$m_{4} \sim 1 \, \text{eV}$.

The problem of the determination of $\Neff$
from cosmological data is related to that of the Hubble constant $H_0$,
because these two quantities are positively correlated
in the analysis of the data
(see Subsection~\ref{sub:radiationeffects}
and Refs.~\cite{Hou:2011ec,Archidiacono:2013fha,Gariazzo:2013gua,
Archidiacono:2014apa,Gariazzo:2015rra}).
Since dedicated local astrophysical experiments
obtained values of $H_0$ which are larger than that obtained
by the Planck collaboration from the analysis of
cosmological data alone \cite{Ade:2013zuv}
(see Section~\ref{sec:h0}),
there is an indication that $\Neff$ may be larger than 3.046,
as a consequence of the correlation between $\Neff$ and $H_0$.
Here, we will consider the local measurement on $H_0$ from HST
\cite{Riess:2011yx} as a prior in the cosmological analyses.

Since the neutrino oscillation explanation of SBL data requires
the existence of a massive neutrino at the eV scale,
we discuss also the cosmological bounds on the effective sterile neutrino mass
$\meff{s}$
defined in Eq.~\eqref{eq:meffs}.
For the distribution function of the \lsn,
we consider the two cases discussed
in the previous Chapter and
by the Planck collaboration \cite{Ade:2013zuv}
(see also \cite{Acero:2008rh}):
the Thermal (TH) model, for which
$\meff{s}
=
(\DNeff)^{3/4}\,
m_{s}
$ (see Eq.~\eqref{eq:THquantities})
and the Dodelson-Widrow (DW) model \cite{Dodelson:1993je},
for which
$\meff{s}
=
\DNeff
\,
m_{s}
$ (see Eq.~\eqref{eq:DWquantities}).
The thermal and the Dodelson-Widrow models
are discussed in Subsection~\ref{sub:parameterization}.

A further important problem is the
compatibility of the cosmological bounds on
$\Neff$
and
$\meff{s}$
with the active-sterile neutrino mixing required to
fit SBL oscillation data.
The stringent bounds on
$\Neff$
and
$\meff{s}$
presented in Ref.~\cite{Ade:2013zuv,Ade:2015xua}
by the Planck collaboration imply
\cite{Mirizzi:2013gnd}
that
the production of sterile neutrinos in the early Universe,
that should occur given the mixing angles 
relevant for active-sterile oscillations,
is suppressed by some non-standard mechanism.
Here we adopt a phenomenological approach similar to the one in
Refs.~\cite{Archidiacono:2012ri,
Archidiacono:2013xxa,Kristiansen:2013mza}:
we use the results of the fit of SBL neutrino oscillation data
\cite{Giunti:2013aea}
as a prior for the analysis of cosmological data.
In this way,
in Subsection~\ref{ssec:jhep_SBL}
we derive the combined constraints on
$\Neff$
and
$\meff{s}$
and the related constraints on $H_0$ and $m_{s}$.

\begin{table}[t]
\begin{center}
\renewcommand{\arraystretch}{1.2}
\begin{tabular}{clccc}
\multicolumn{2}{c}{data} & $H_0^{\text{gbf}}$ & $H_0^{\text{mbf}}\pm1\sigma$ & $2\sigma$ \\
\hline
\multirow{3}{*}{
\begin{varwidth}{4em}
\center no\\SBL\\prior
\end{varwidth}}
	&CMB+$H_0$		&$73.6$	&$72.7^{+1.9}_{-1.7}$	&$69.0 \div 76.3$ \\
	&CMB+$H_0$+BAO		&$71.1$	&$71.5^{+1.4}_{-1.4}$	&$68.7 \div 74.4$ \\
	&CMB+$H_0$+BAO+LGC	&$71.1$	&$70.4^{+1.5}_{-1.3}$	&$68.1 \div 73.5$ \\
\hline
\multirow{4}{*}{
\begin{varwidth}{4em}
\center TH\\SBL\\prior
\end{varwidth}}
	&CMB			&$66.8$	&$66.6^{+1.1}_{-1.2}$	&$64.3 \div 68.9$ \\
	&CMB+$H_0$		&$68.7$	&$68.7^{+1.0}_{-1.1}$	&$66.5 \div 70.7$ \\
	&CMB+$H_0$+BAO		&$68.7$	&$68.8^{+0.8}_{-0.7}$	&$67.3 \div 70.4$ \\
	&CMB+$H_0$+BAO+LGC	&$69.1$	&$69.3^{+0.6}_{-0.6}$	&$68.1 \div 70.6$ \\
\hline
\multirow{4}{*}{
\begin{varwidth}{4em}
\center DW\\SBL\\prior
\end{varwidth}}
	&CMB			&$66.5$	&$66.9^{+1.2}_{-1.3}$	&$64.6 \div 69.4$ \\
	&CMB+$H_0$		&$68.1$	&$68.9^{+1.1}_{-1.0}$	&$66.9 \div 71.0$ \\
	&CMB+$H_0$+BAO		&$69.3$	&$69.1^{+0.8}_{-0.8}$	&$67.6 \div 70.6$ \\
	&CMB+$H_0$+BAO+LGC	&$69.5$	&$69.7^{+0.7}_{-0.5}$	&$68.6 \div 71.0$ \\
\hline
\end{tabular}
\end{center}
\caption[Results on $H_0$ from different dataset]{\label{tab:jhep_H0}
Global best-fit value $H_0^{\text{gbf}}$,
marginal best-fit $H_0^{\text{mbf}}\pm1\sigma$ (68.27\%) and
$2\sigma$ (95.45\%) limits for $H_0$
obtained from the analysis of the indicated data sets.
From Ref.~\protect\cite{Gariazzo:2013gua}.}
\end{table}

\begin{table}[t]
\begin{center}
\renewcommand{\arraystretch}{1.2}
\begin{tabular}{clccc}
\multicolumn{2}{c}{data} & $\Neff^{\text{gbf}}$ & $\Neff^{\text{mbf}}\pm1\sigma$ & $2\sigma$ \\
\hline
\multirow{3}{*}{
\begin{varwidth}{4em}
\center no\\SBL\\prior
\end{varwidth}}
	&CMB+$H_0$		&$3.84$	&$3.76^{+0.25}_{-0.23}$	&$3.29 \div 4.26$ \\
	&CMB+$H_0$+BAO		&$3.59$	&$3.71^{+0.23}_{-0.27}$	&$3.17 \div 4.18$ \\
	&CMB+$H_0$+BAO+LGC	&$3.57$	&$3.51^{+0.29}_{-0.29}$	&$3.05 \div 4.01$ \\
\hline
\multirow{4}{*}{
\begin{varwidth}{4em}
\center TH\\SBL\\prior
\end{varwidth}}
	&CMB			&$3.29$	&$3.26^{+0.21}_{-0.10}$	&$3.05 \div 3.67$ \\
	&CMB+$H_0$		&$3.23$	&$3.23^{+0.19}_{-0.12}$	&$3.05 \div 3.66$ \\
	&CMB+$H_0$+BAO		&$3.11$	&$3.23^{+0.15}_{-0.11}$	&$3.05 \div 3.55$ \\
	&CMB+$H_0$+BAO+LGC	&$3.36$	&$3.32^{+0.12}_{-0.09}$	&$3.15 \div 3.57$ \\
\hline
\multirow{4}{*}{
\begin{varwidth}{4em}
\center DW\\SBL\\prior
\end{varwidth}}
	&CMB			&$3.43$	&$3.35^{+0.16}_{-0.15}$	&$3.09 \div 3.73$ \\
	&CMB+$H_0$		&$3.19$	&$3.31^{+0.18}_{-0.13}$	&$3.08 \div 3.70$ \\
	&CMB+$H_0$+BAO		&$3.29$	&$3.30^{+0.13}_{-0.13}$	&$3.08 \div 3.60$ \\
	&CMB+$H_0$+BAO+LGC	&$3.30$	&$3.42^{+0.11}_{-0.11}$	&$3.22 \div 3.67$ \\
\hline
\end{tabular}
\end{center}
\caption[Results on \Neff\ from different dataset]{\label{tab:jhep_neff}
As Tab.~\ref{tab:jhep_H0}, but for $\Neff$.
From Ref.~\protect\cite{Gariazzo:2013gua}.}
\end{table}

\begin{table*}[t]
\begin{center}
\renewcommand{\arraystretch}{1.2}
\resizebox{1\textwidth}{!}{
\begin{tabular}{cl|cccc|ccccc}
\multicolumn{2}{c}{data}
& $m_{s,\text{gbf}}^{\text{eff}}$ & $m_{s,\text{mbf}}^{\text{eff}}$ & $1\sigma$ & $2\sigma$ 
& $m_{s}^{\text{gbf}}$ & $m_{s}^{\text{mbf}}$ & $1\sigma$ & $2\sigma$ \\
\hline
\multirow{3}{*}{
\begin{varwidth}{4em}
\center no\\SBL\\prior
\end{varwidth}}
& CMB+$H_0$		& $0$    & $0$    & $<0.10$	   & $<0.27$          & $0$ & $0$                                           & \begin{tabular}{c}$<0.13$\\$<0.14$\end{tabular}		      & \begin{tabular}{c}$<0.38$\\$<0.44$\end{tabular}                   & \begin{tabular}{c}(TH)\\(DW)\end{tabular} \\
& CMB+$H_0$+BAO		& $0$    & $0$    & $<0.13$	   & $<0.32$          & $0$ & $0$                                           & \begin{tabular}{c}$<0.18$\\$<0.21$\end{tabular}		      & \begin{tabular}{c}$<0.51$\\$<0.65$\end{tabular}                   & \begin{tabular}{c}(TH)\\(DW)\end{tabular} \\
& CMB+$H_0$+BAO+LGC	& $0.41$ & $0.42$ & $0.28\div0.56$ & $0.15 \div 0.70$ & \begin{tabular}{c}$0.67$\\$0.79$\end{tabular} & \begin{tabular}{c}$0.62$\\$0.92$\end{tabular} & \begin{tabular}{c}$0.21\div1.14$\\$0.00\div1.11$\end{tabular} & \begin{tabular}{c}$0.00 \div 2.68$\\$0.00 \div 4.81$\end{tabular} & \begin{tabular}{c}(TH)\\(DW)\end{tabular} \\
\hline
\multirow{4}{*}{
\begin{varwidth}{4em}
\center TH\\SBL\\prior
\end{varwidth}}
& CMB			& $0.45$ & $0.42$ & $0.26 \div 0.67$ & $0.11 \div 0.89$ & $1.30$ & $1.28$ & $1.09 \div 1.36$ & $0.96 \div 1.42$ \\
& CMB+$H_0$		& $0.35$ & $0.38$ & $0.20 \div 0.61$ & $0.05 \div 0.86$ & $1.28$ & $1.28$ & $1.08 \div 1.35$ & $0.95 \div 1.40$ \\
& CMB+$H_0$+BAO		& $0.17$ & $0.37$ & $0.20 \div 0.54$ & $0.08 \div 0.75$ & $1.29$ & $1.27$ & $1.08 \div 1.35$ & $0.95 \div 1.39$ \\
& CMB+$H_0$+BAO+LGC	& $0.47$ & $0.48$ & $0.35 \div 0.60$ & $0.25 \div 0.74$ & $1.12$ & $1.27$ & $1.08 \div 1.35$ & $0.95 \div 1.40$ \\
\hline
\multirow{4}{*}{
\begin{varwidth}{4em}
\center DW\\SBL\\prior
\end{varwidth}}
& CMB			& $0.44$ & $0.36$ & $0.19 \div 0.57$ & $0.06 \div 0.83$ & $1.13$ & $1.28$ & $1.08 \div 1.35$ & $0.96 \div 1.42$ \\
& CMB+$H_0$		& $0.16$ & $0.35$ & $0.16 \div 0.53$ & $0.04 \div 0.77$ & $1.13$ & $1.28$ & $1.07 \div 1.35$ & $0.94 \div 1.39$ \\
& CMB+$H_0$+BAO		& $0.32$ & $0.28$ & $0.16 \div 0.46$ & $0.06 \div 0.64$ & $1.28$ & $1.27$ & $1.07 \div 1.34$ & $0.95 \div 1.39$ \\
& CMB+$H_0$+BAO+LGC	& $0.32$ & $0.45$ & $0.33 \div 0.58$ & $0.22 \div 0.72$ & $1.27$ & $1.28$ & $1.08 \div 1.35$ & $0.95 \div 1.40$ \\
\hline
%\end{tabular}
& SBL \protect\cite{Giunti:2013aea} & & & &
& $1.27$
& $1.27$
& $1.10\div1.36$
& $0.97\div1.42$
\\
\hline
\end{tabular}
}
\end{center}
\caption[Results on \meff{s} and $m_s$ from different dataset]{\label{tab:jhep_meffs}
As Tab.~\ref{tab:jhep_H0}, but for $\meff{s}$.
We give also the corresponding values for $m_{s}$,
see Eqs.~\eqref{eq:THquantities} and \eqref{eq:DWquantities}.
From Ref.~\protect\cite{Gariazzo:2013gua}.}
\end{table*}

\begin{figure}[tp]
\centering
\includegraphics*[page=1, width=\singlefigbig, viewport=14 14 913 913]%
{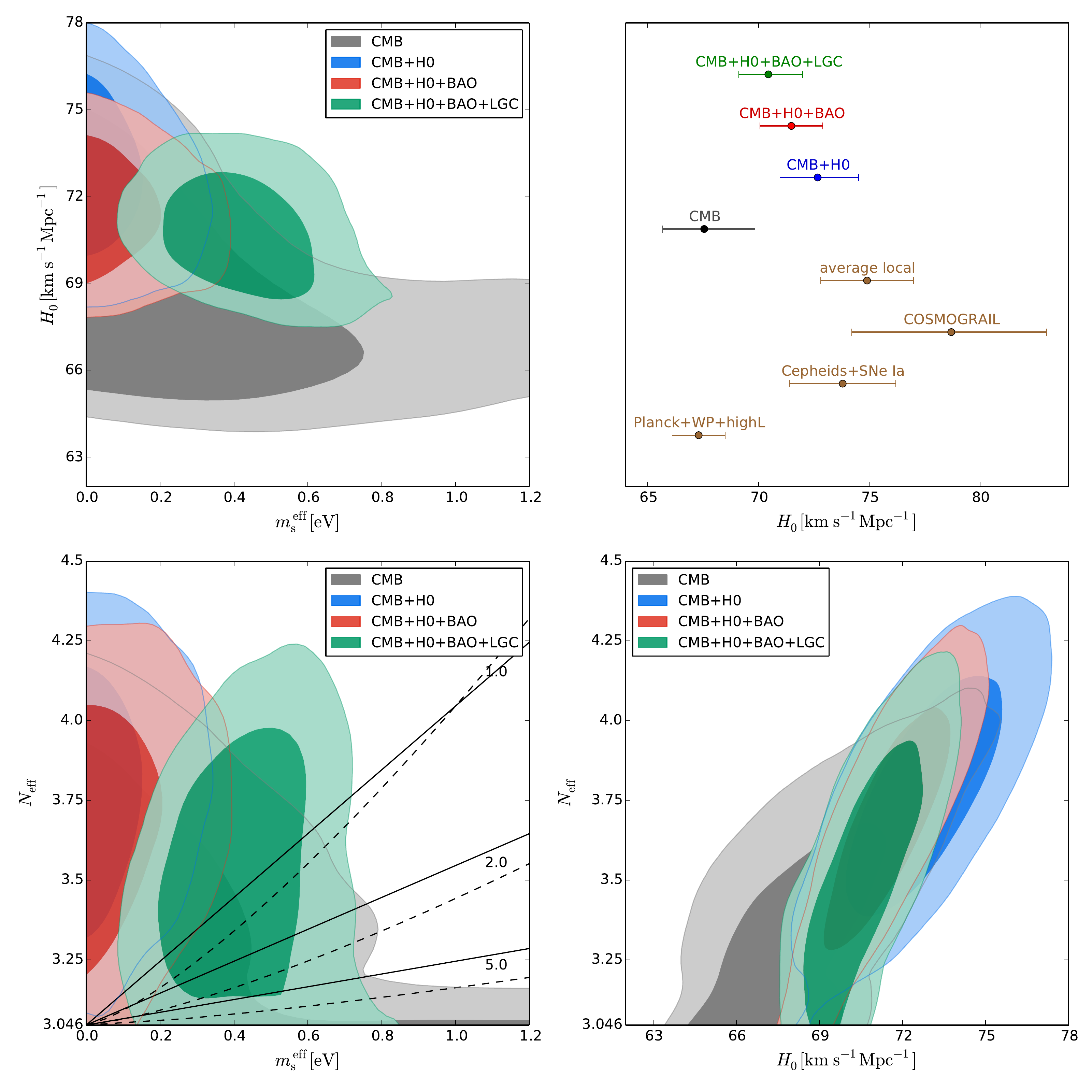}
\caption[Constraints on the \lsn\ from the analysis
of cosmological data]
{ \label{fig:jhep_nosbl}
Results of the analysis of cosmological data alone.
The regions in the 2D plots show,
respectively, the $1\sigma$ and $2\sigma$
marginalized posterior probability regions
obtained from the analysis of the indicated data sets.
The four lower intervals of $H_{0}$ in the upper-right panel
correspond to the measurements from
Planck+WP+highL in the \lcdm\ model \protect\cite{Ade:2013zuv},
Cepheids+SNe Ia \protect\cite{Riess:2011yx},
COSMOGRAIL \protect\cite{Suyu:2012aa},
and a local average %$H_{0}=74.9\pm2.1\Hou$
obtained combining the
two previous measurements
(see Ref.~\protect\cite{Gariazzo:2013gua}).
In the bottom-left panel $m_{s}$ is constant, with the indicated value in eV,
along the dashed lines in the thermal model
and
along the solid lines in the Dodelson-Widrow model.
From Ref.~\protect\cite{Gariazzo:2013gua}.
}
\end{figure}

\begin{figure}[t]
\centering
\includegraphics*[page=5, width=\halfwidth, viewport=403 14 798 394]%
{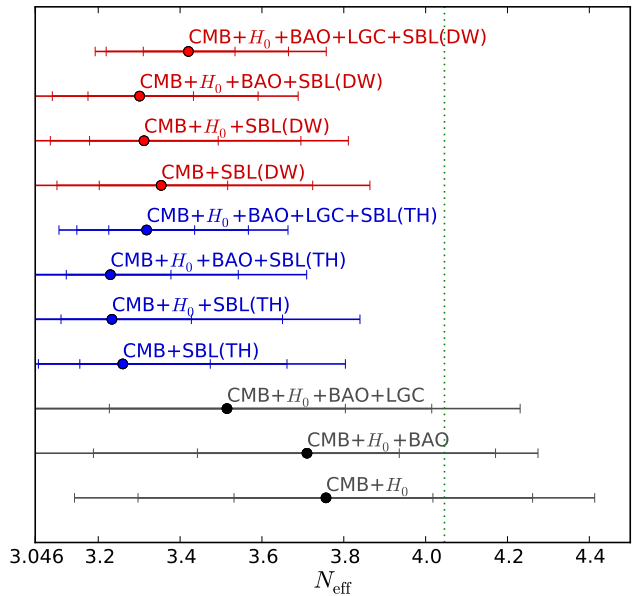}
\caption[Allowed intervals of \Neff\ from different combinations
of cosmological and SBL data]{ \label{fig:jhep_neff}
Comparison of the allowed intervals of $\Neff$
obtained from the fits of
CMB, CMB+$H_0$, CMB+$H_0$+BAO and CMB+$H_0$+BAO+LGC data
without (black) and with the SBL prior in the thermal (blue)
and Dodelson-Widrow (red) models.
The segments in each bar correspond to $1\sigma$, $2\sigma$
and $3\sigma$ probability.
The dotted vertical line corresponds to $\DNeff = 1$.
From Ref.~\protect\cite{Gariazzo:2013gua}.
}
\end{figure}

\begin{figure}[t]
\centering
\includegraphics*[page=5, width=\halfwidth, viewport=2 14 397 394]%
{nucosmo/JHEP_figs.pdf}
\caption[Allowed intervals of \meff{s} from different combinations
of cosmological and SBL data]{ \label{fig:jhep_mseff}
As in Fig.~\ref{fig:jhep_neff}, but for \meff{s}.
From Ref.~\protect\cite{Gariazzo:2013gua}.
}
\end{figure}

\begin{figure}[t]
\centering
\includegraphics*[page=6, width=\halfwidth, viewport=11 13 424 423]%
{nucosmo/JHEP_figs.pdf}
\caption[Allowed intervals of $m_s$ from different combinations
of cosmological and SBL data]{\label{fig:jhep_ms}
As in Fig.~\ref{fig:jhep_neff}, but for $m_s$.
The value indicated with ``SBL'' is obtained from the 3+1 analysis of SBL data
\protect\cite{Giunti:2013aea}.
The out-of-bounds upper limits obtained in the CMB+$H_0$+BAO+LGC analysis are:
$7.4 \, \text{eV}$ ($3\sigma$, TH),
$4.8 \, \text{eV}$ ($2\sigma$, DW),
$17.1 \, \text{eV}$ ($3\sigma$, DW).
From Ref.~\protect\cite{Gariazzo:2013gua}.
}
\end{figure}

\begin{figure}[tp]
\centering
\includegraphics*[page=3, width=\singlefigbig, viewport=14 14 913 913]%
{nucosmo/JHEP_figs.pdf}
\caption[Constraints on a thermal \lsn\ from the analysis of cosmological data]
{ \label{fig:jhep_thermal}
As Fig.~\ref{fig:jhep_nosbl}, but with the inclusion of the SBL prior
for a light sterile neutrino in the thermal model.
From Ref.~\protect\cite{Gariazzo:2013gua}.
}
\end{figure}

\begin{figure}[tp]
\centering
\includegraphics*[page=2, width=\singlefigbig, viewport=14 14 913 913]%
{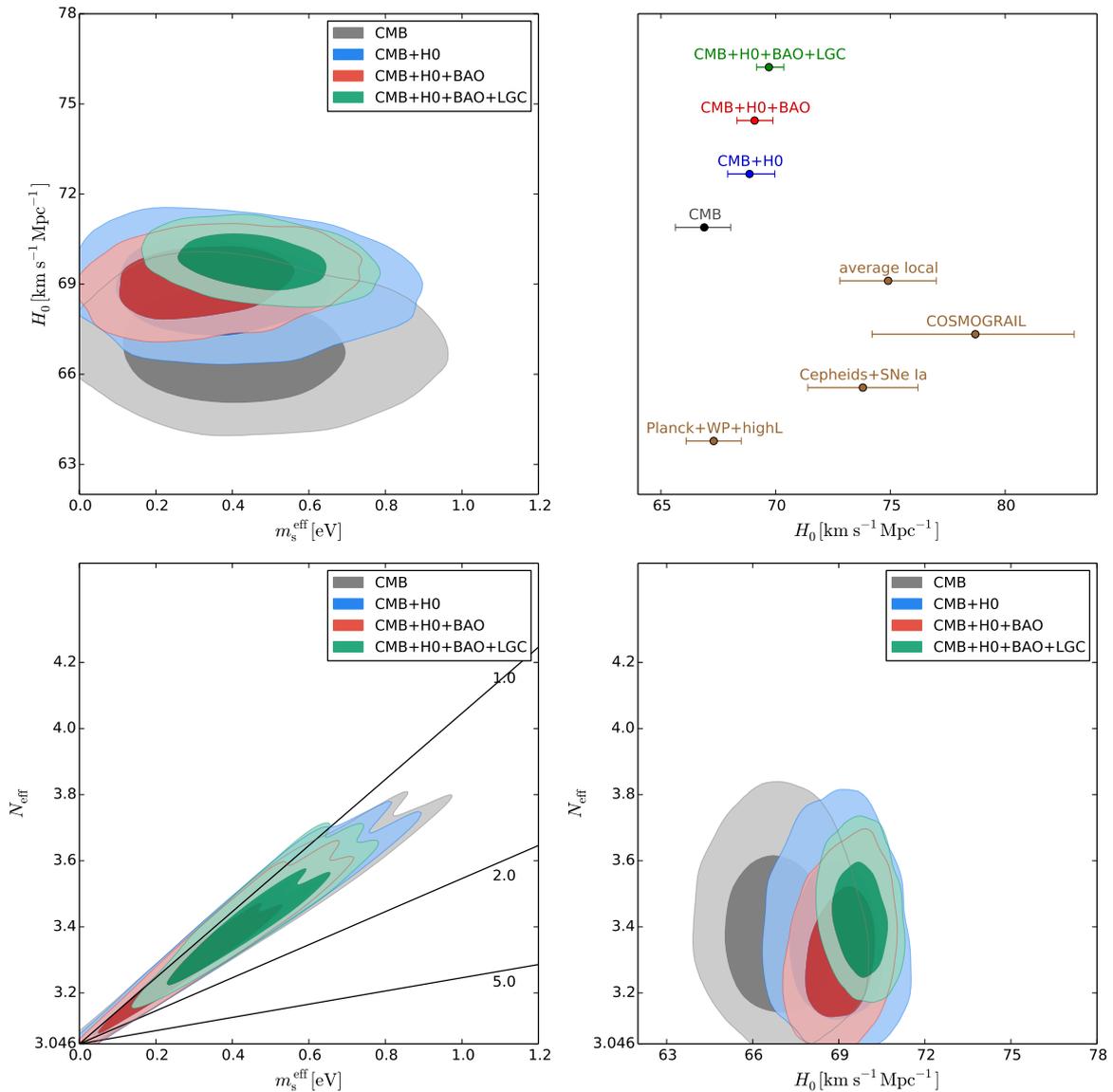}
\caption[Constraints on a DW \lsn\ from the analysis of cosmological data]
{ \label{fig:jhep_DW}
As Fig.~\ref{fig:jhep_nosbl}, but with the inclusion of the SBL prior
for a light sterile neutrino in the Dodelson-Widrow model.
From Ref.~\protect\cite{Gariazzo:2013gua}.}
\end{figure}

\begin{figure}[tp]
\centering
\includegraphics[page=4, width=\singlefigbig, viewport=14 14 913 913]{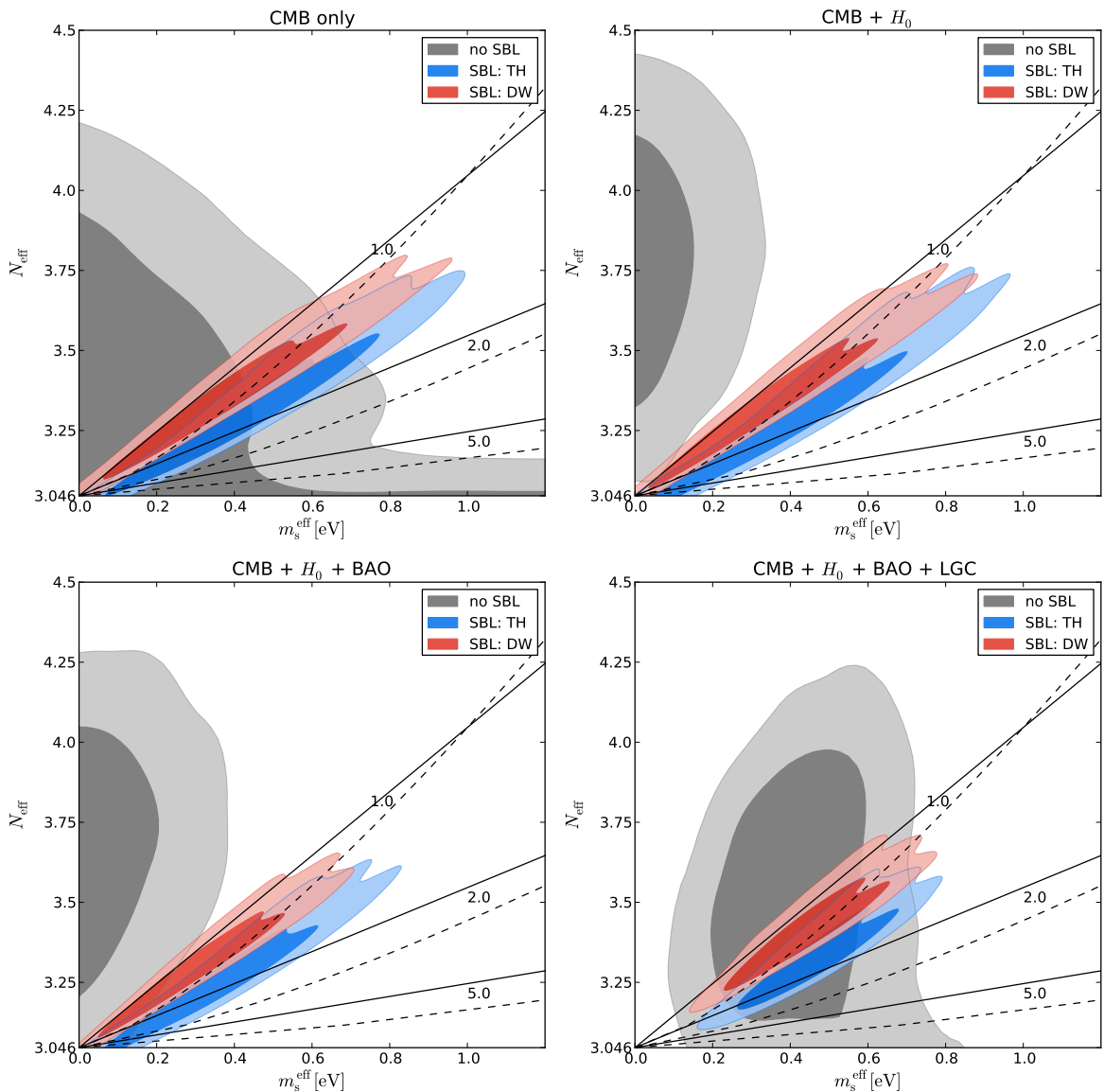}
\caption[Effect of the SBL prior on the results of the analyses of various cosmological data]
{ \label{fig:jhep_meffs}
Illustrations of the effect of the SBL prior on the results of the fits of
CMB, CMB+$H_0$, CMB+$H_0$+BAO and CMB+$H_0$+BAO+LGC data.
The value of $m_{s}$ is constant, equal to the indicated value in eV,
along the dashed (solid) lines in the thermal (Dodelson-Widrow) model.
From Ref.~\protect\cite{Gariazzo:2013gua}.
}
\end{figure}

\subsection{Cosmological Data and Local \texorpdfstring{$H_0$}{H0} Measurements}
\label{ssec:jhep_cosmological}

For our cosmological analysis we used a modified version
of the publicly available software
\texttt{CosmoMC}\footnote{\url{http://cosmologist.info/cosmomc/}}
\cite{Lewis:2002ah},
a Monte Carlo Markov Chain (MCMC) software which computes
the theoretical predictions using
CAMB\footnote{\url{http://camb.info/}} \cite{Lewis:1999bs}.
All the datasets we will use for the analyses have been described
extensively in Chapter~\ref{ch:cosmomeasurements} and
we indicate here only the ones we are going to consider, that are:
\begin{description}

 \item[Planck --]
 The full 2013 Planck data \cite{Ade:2013sjv};
 
 \item[WP --]
 The nine-year large-scale $E$-polarization WMAP data \cite{Bennett:2012zja};
 
 \item[HighL --]
 CMB spectra
 at high multipoles from Atacama Cosmology Telescope (ACT)
 \cite{Das:2013zf} and 
 South Pole Telescope (SPT)
 \cite{Keisler:2011aw,Reichardt:2011yv}.
 We will indicate the Planck+WP+highL dataset with \textbf{CMB};

 \item[BAO --]
 Baryon Acoustic Oscillations (BAO) data from
 the Sloan Digital Sky Survey (SDSS) Data Release 7 (DR7)
 \cite{Abazajian:2008wr,Percival:2009xn,Padmanabhan:2012hf},
 the SDSS Baryon Oscillation Spectroscopic Survey (BOSS) Data Release 9 (DR9)
 \cite{Ahn:2012fh,Anderson:2012sa},
 and 
 the 6dF Galaxy Survey (6dFGS) \cite{Jones:2009yz,Beutler:2011hx};

 \item[LGC --]
 Local Galaxy Cluster data from the Chandra Cluster Cosmology Project
 \cite{Vikhlinin:2008ym,Burenin:2012uy};
 
 \item[$\mathbf{H_0}$ --]
%   A detailed discussion on the $H_0$ tension between local
%   and cosmological measurements can be found in Section~\ref{sec:h0}
%   or in Ref.~\cite{Gariazzo:2013gua}.
%   We use 
  the local determination of the Hubble parameter %as determined
  by the Hubble Space Telescope (HST) observations,
  $
  H_0 = 73.8 \pm 2.4 \Hou
  $ \cite{Riess:2011yx}, used
  as a prior in the cosmological analyses.

\end{description}

\subsection{Results from Cosmology}
Since we are interested in studying the effects on the analyses
of cosmological data
of a sterile neutrino with a mass motivated by SBL oscillation anomalies,
we consider an extension of the standard cosmological model
in which both $\Neff$ and $\meff{s}$
are free parameters to be determined by the data.
The model we adopt is then an extension of the \lcdm\ model
(described in Section~\ref{sec:par_depend})
that includes $\Neff$ and $\meff{s}$, for a total of
eight free parameters.

Figure~\ref{fig:jhep_nosbl}
and
the first parts of
Tabs.~\ref{tab:jhep_H0}, \ref{tab:jhep_neff} and \ref{tab:jhep_meffs}
shows the results for
$H_0$,
$\Neff$ and
$\meff{s}$
obtained from the fits of
CMB, CMB+$H_0$, CMB+$H_0$+BAO and CMB+$H_0$+BAO+LGC data.
In Tab.~\ref{tab:jhep_meffs}
we give also the corresponding results for
$m_{s} \simeq m_{4}$,
which depend on the statistical distribution of sterile neutrinos.
Therefore,
we distinguish the results for
$m_{s}$ obtained in the
thermal (TH) and Dodelson-Widrow (DW) models
using, respectively,
Eqs.~\eqref{eq:THquantities} and \eqref{eq:DWquantities}.
In Figs.~\ref{fig:jhep_neff}, \ref{fig:jhep_mseff} and \ref{fig:jhep_ms}
we compare graphically the allowed ranges of $\Neff$, $\meff{s}$ and $m_{s}$
obtained in the different fits.

From the bottom-left panel in Fig.~\ref{fig:jhep_nosbl},
one can see that the fit of CMB data alone restricts
$\meff{s}$ to small values only for
$\Neff\gtrsim3.2$,
whereas there is a tail of allowed large values of
$\meff{s}$
for smaller $\Neff$.
This is in agreement with Fig.~28-right of Ref.~\cite{Ade:2013zuv},
where the tail at small $\Neff$
has been explained as corresponding to the case
in which the sterile neutrino
behaves as warm dark matter,
because its mass is large and it becomes non-relativistic
well before recombination.
This happens in both the thermal and Dodelson-Widrow models,
as one can infer from Eqs.~\eqref{eq:THquantities} and \eqref{eq:DWquantities}.
The presence of this tail of the posterior distribution of
$\meff{s}$
implies that the posterior distributions of the fitted parameters
depend on the arbitrary upper value chosen for $\meff{s}$
in the \texttt{CosmoMC} runs
(we chose $\meff{s}<5\,\text{eV}$,
whereas the Planck Collaboration chose
$\meff{s}<3\,\text{eV}$).
Hence, we do not present in the tables the numerical results
of the fit of CMB data alone,
which suffer from this arbitrariness.

The addition of the local $H_0$ prior leads to an increase of
$\Neff$
which evicts the
large-$\meff{s}$ and small-$\Neff$
region in which the sterile neutrino
behaves as warm dark matter.
This can be seen from the
CMB+$H_0$
allowed regions in Fig.~\ref{fig:jhep_nosbl},
the corresponding upper limits for
$\meff{s}$
($m_{s}$)
in Figs.~\ref{fig:jhep_mseff} (\ref{fig:jhep_ms}) and in Tab.~\ref{tab:jhep_meffs}.
The further addition of BAO data slightly lowers the
best-fit values and the allowed ranges of
$H_0$
and
$\Neff$
(see Figs.~\ref{fig:jhep_nosbl}, \ref{fig:jhep_neff}
and
Tabs.~\ref{tab:jhep_H0}, \ref{tab:jhep_neff}).
Hence,
the upper limits for
$\meff{s}$
and
$m_{s}$
(see Figs.~\ref{fig:jhep_mseff}, \ref{fig:jhep_ms}
and Tab.~\ref{tab:jhep_meffs})
are slightly larger,
but still rather stringent,
of the order of
$\meff{s} \lesssim 0.3 \, \text{eV}$
and
$m_{s} \lesssim 0.6 \, \text{eV}$
at $2\sigma$.

Comparing the
CMB+$H_0$
and
CMB+$H_0$+BAO
allowed intervals of
$m_{s}$
in Tab.~\ref{tab:jhep_meffs} and Fig.~\ref{fig:jhep_ms}
with those obtained from the analysis of SBL data
in the framework of 3+1 mixing \cite{Giunti:2013aea},
it is clear that there is a tension\footnote{
Possible ways of solving this tension have been discussed
before the Planck 2013 data release in
Refs.~\cite{Giusarma:2011zq,Motohashi:2012wc,Ho:2012br}.
}:
about
$5.0\sigma$,
$4.6\sigma$,
$4.1\sigma$,
$3.5\sigma$,
respectively,
in the
CMB+$H_0$(TH),
CMB+$H_0$(DW),
CMB+$H_0$+BAO(TH)
CMB+$H_0$+BAO(DW)
fits.
The tensions are smaller in the Dodelson-Widrow model
and this could be an indication in favor of this case, if
SBL oscillations will be confirmed by future experiments
(see Refs.~\cite{Cribier:2011fv,Abazajian:2012ys,Bungau:2012ys,
Rubbia:2013ywa,Borexino:2013xxa,Elnimr:2013wfa,Qian:2013ora,Adey:2013pio}).

Let us now consider the inclusion of the LGC data set
in the cosmological fit.
As discussed in Section~\ref{sec:cluster}
and in Ref.~\cite{Wyman:2013lza},
the measured amount of clustering of galaxies
\cite{Burenin:2012uy,Vikhlinin:2008ym}
is smaller than that
obtained by evolving the primordial density fluctuations
with the relatively large matter density at recombination
measured precisely by Planck
\cite{Ade:2013zuv}.
The correlation of a relatively large
matter density and the clustering of galaxies
can be quantified through the approximate relation
$
\sigma_{8}
\propto
\Omega_{m}^{0.563}
$
\cite{Hu:2003pt,Hu:2004kn}
which relates the rms amplitude
of linear fluctuations today at a scale of $8 h^{-1} \Mpc$,
$\sigma_{8}$,
with the present matter density
$\Omega_{m}$.
The value of
$\sigma_{8}$ and the amount of clustering of galaxies
can be lowered by adding 
hot dark matter in the form of sterile neutrinos with eV-scale masses\footnote{
Let us note that there was already a tension between
LGC data and pre-Planck CMB data
and the sterile neutrino solution
was proposed in Refs.~\cite{Burenin:2012uy,Burenin:2013wg}}
to the \lcdm{} cosmological model.
The free-streaming of these sterile neutrinos suppresses the growth of structures
at distances smaller than the free-streaming length,
leading to a suppression of $\sigma_{8}$
with respect to the \lcdm{} approximate relation
$
\sigma_{8}
\propto
\Omega_{m}^{0.563}
$.
In this way,
the relatively large Planck value of $\Omega_{m}$
can be reconciled
with the relatively small amount of local galaxy clustering in the LGC data set.

Hence,
the inclusion of LGC data in the cosmological fits
favors the existence of a sterile neutrino with a mass of the order of that
required by SBL data,
which is at least partially thermalized in the early Universe
\cite{Wyman:2013lza}.
The results of our
CMB+$H_0$+BAO+LGC
fit given in Figs.~\ref{fig:jhep_nosbl}, \ref{fig:jhep_neff}, \ref{fig:jhep_mseff}, \ref{fig:jhep_ms}
and
Tabs.~\ref{tab:jhep_H0}, \ref{tab:jhep_neff}, \ref{tab:jhep_meffs}
confirm this expectation.
In particular,
from the allowed intervals of
$m_{s}$
in Tab.~\ref{tab:jhep_meffs} and Fig.~\ref{fig:jhep_ms}
one can see that the tension between cosmological data and SBL 3+1 oscillations
disappears with the inclusion of LGC data.

In the following Subsection we analyze the cosmological data
using as a prior distribution for $m_s$
the posterior distribution obtained from the analysis of SBL data.
This is perfectly consistent
in the case of
CMB+$H_0$+BAO+LGC
cosmological data.
However,
we present also the results obtained with the
CMB,
CMB+$H_0$ and
CMB+$H_0$+BAO
cosmological data,
in spite of the tension with SBL data discussed above,
because
we think that one cannot dismiss the results of laboratory experiments
on the basis of cosmological observations,
which are indirect probes of the neutrino masses and
whose interpretation has larger uncertainties.

\subsection{Results with the SBL Prior}
\label{ssec:jhep_SBL}

The experimental data that motivate the existence of the \lsn\ and
from which the SBL prior we use here \cite{Giunti:2013aea}
is calculated were presented in the previous Chapter.
Following 
Refs.~\cite{Archidiacono:2012ri,
Archidiacono:2013xxa,Kristiansen:2013mza},
we use the posterior distribution of
$m_s \simeq m_4 \simeq \sqrt{\Delta{m}^2_{41}}$
obtained from the analysis of SBL data
as a prior in the \texttt{CosmoMC} analysis of cosmological data.
The range of $m_{s}$ allowed by the analysis of SBL data \cite{Giunti:2013aea}
is shown in Fig.~\ref{fig:jhep_ms} and Tab.~\ref{tab:jhep_meffs}.
Note that the SBL prior on $m_s$ has different cosmological implications
in the thermal and Dodelson-Widrow models,
because the
$\DNeff$ dependence of the effective mass
$\meff{s}$
is different (see Eqs.~\eqref{eq:THquantities} and \eqref{eq:DWquantities}).

Figure~\ref{fig:jhep_thermal} shows the results of the analysis
of various combinations of datasets
(CMB, CMB+$H_0$, CMB+$H_0$+BAO and CMB+$H_0$+BAO+LGC),
with the SBL prior in the thermal model.
For convenience,
the effect of the SBL prior
on the allowed regions in the
$\meff{s}$--$\Neff$ plane
is illustrated clearly in Fig.~\ref{fig:jhep_meffs},
where each panel shows the change of the allowed regions
due to the inclusion of the SBL prior
in the analysis of the indicated data set.
One can see that in all the four analyses the SBL prior forces the
allowed region to lie near the dashed line which corresponds to
$m_{s} = 1 \, \text{eV}$.
In order to keep $m_{s}$ at the eV scale without increasing too much $\meff{s}$,
which is forbidden by the cosmological data,
$\Neff$
is forced towards low values.

In the case of the CMB+$H_0$+BAO+LGC cosmological data set, after
the addition of the SBL prior the allowed range of
$\meff{s}$
(see Fig.~\ref{fig:jhep_mseff} and Tab.~\ref{tab:jhep_meffs})
is approximately confirmed,
but a lower
$\Neff$ is required
(see Fig.~\ref{fig:jhep_neff} and Tab.~\ref{tab:jhep_neff}),
being $\Neff\lesssim3.7$ with $3\sigma$ probability.
As discussed in Ref.~\cite{Mirizzi:2013gnd},
in the standard cosmological scenario
active-sterile neutrino oscillations
generated by values of the mixing parameters allowed by the fit of SBL data
imply $\DNeff = 1$.
Therefore,
it is likely that the compatibility of
the neutrino oscillation explanation of the SBL anomalies
with cosmological data
requires that active-sterile neutrino oscillations in the early Universe
are somewhat suppressed by a non-standard mechanism,
as, for example,
a large lepton asymmetry
\cite{Hannestad:2012ky,Mirizzi:2012we,Saviano:2013ktj,Hannestad:2013wwj}.

As one can see from
Figs.~\ref{fig:jhep_neff}, \ref{fig:jhep_mseff}, \ref{fig:jhep_DW}
and \ref{fig:jhep_meffs}
and from
Tabs.~\ref{tab:jhep_neff} and \ref{tab:jhep_meffs},
similar conclusions are reached in the Dodelson-Widrow model.
One can note, however, that in this case
slightly larger values of $\Neff$
are allowed with respect to the thermal case,
and there is a slightly better compatibility
of cosmological and SBL data.
This happens because for a given value of $m_{s}$ arising mainly by SBL data
and an upper bound on
$\meff{s}$ given by cosmological data
slightly larger values of
$\DNeff \leq 1$ are allowed by Eq.~\eqref{eq:DWquantities}
in the Dodelson-Widrow model
than by
Eq.~\eqref{eq:THquantities} in the thermal model.

\subsection{Discussion}
\label{ssec:jhep_conclusions}

In this section we have analyzed different cosmological data,
including those of the Planck experiment \cite{Ade:2013sjv,Ade:2013zuv},
taking into account the possible existence of a sterile neutrino
with a mass $m_s$ at the eV scale,
which could have the effect of dark radiation in the early Universe.
We investigated three effects:
1) the contribution of local measurements of the Hubble constant $H_0$;
2) the effect of the measurements of the mass distribution of
local galaxy clusters \cite{Wyman:2013lza};
3) the assumption of a prior distribution for $m_s$
obtained from the analysis of short-baseline oscillation data
in the framework of 3+1 mixing, which requires a sterile neutrino mass
between about 0.9 and 1.5 eV \cite{Giunti:2013aea}.
For the statistical distribution of the sterile neutrinos
we considered the two most studied cases:
the thermal model and the Dodelson-Widrow model \cite{Dodelson:1993je}.

We have shown that the local measurements of the Hubble constant $H_0$
induce an increase of the value of the effective number of
relativistic degrees of freedom
$\Neff$
above the Standard Model value.
This is an indication in favor of the existence of sterile neutrinos
and their contribution to dark radiation.
However,
we obtained that the sterile neutrino mass has a $2\sigma$ upper bound of
about 0.5 eV in the thermal model
and
about 0.6 eV in the Dodelson-Widrow model.
Hence, there is a tension
between cosmological and SBL data.
The Dodelson-Widrow model is slightly more compatible with SBL data
and it may turn out that it is favorite if
SBL oscillations will be confirmed by future experiments
(see Refs.~\cite{Cribier:2011fv,Abazajian:2012ys,Bungau:2012ys,
Rubbia:2013ywa,Borexino:2013xxa,Elnimr:2013wfa,Qian:2013ora,Adey:2013pio}).

The tension between cosmological and SBL data
disappears if we consider also the measurements of
the local galaxy cluster mass distribution,
which favor the existence of sterile neutrinos with eV-scale masses
which can suppress the small-scale clustering of galaxies
through free-streaming \cite{Wyman:2013lza}.
In this case we obtained a cosmologically allowed range
for the sterile neutrino mass
which at $2\sigma$ can be as large as about
2.7 eV in the thermal model
and
4.8 eV in the Dodelson-Widrow model.

In the combined fit of
cosmological and SBL data
the sterile neutrino mass is restricted around 1 eV by the SBL prior
and the cosmological limits on the effective sterile neutrino mass
$\meff{s}$
imply that the contribution of the sterile neutrino
to the effective number of relativistic degrees of freedom
$\Neff$
is likely to be smaller than one.
In this case,
the production of sterile neutrinos in the early Universe
must be somewhat suppressed by a non-standard mechanism,
as, for example,
a large lepton asymmetry
\cite{Hannestad:2012ky,Mirizzi:2012we,Saviano:2013ktj,Hannestad:2013wwj}.
The slightly smaller suppression required by the
Dodelson-Widrow model and the slightly better compatibility
of cosmological and SBL data in this model
may be indications in its favor,
with respect to the thermal model.

%%%%%%%%%%%%%%%%%%%%%%%%%%
% JCAP 1406 (2014) 031   %
%%%%%%%%%%%%%%%%%%%%%%%%%%

\section{Degeneracies between Neutrinos and Tensor Modes}
\label{sec:bicep}
% \cite{Archidiacono:2014apa}

Some months after the 2013 release of Planck data, the publication
of the new data from the BICEP2 experiment \cite{Ade:2014xna}
has indicated a high tensor-to-scalar ratio
corresponding to the existence of primordial tensor perturbations,
that may be significantly correlated with
the neutrino-related parameters.
We want to investigate how the constraints on eV mass sterile neutrinos
are influenced by the BICEP2 claim.
We will demonstrate that eV mass sterile neutrinos
are not significantly constrained by current cosmological data,
given that they contribute with a small amount of
relativistic degrees of freedom \DNeff.
These analyses can be considered a conceptual exercise
and not a new set of physical bounds on the \lsn, since
we know today that the BICEP2 signal did not concern
primordial tensor modes, but it was significantly
contaminated by polarized dust emission
(see Subsection~\ref{ssec:tensor} and 
Refs.~\cite{Adam:2014bub,Ade:2015fwj,Array:2015xqh,Ade:2015tva}).

This Section is structured in this way:
Subsection~\ref{sub:bicep_param} contains a discussion
of the cosmological parameter estimation,
including the cosmological model and the experimental data,
in Subsection \ref{sub:bicep_res} (\ref{sub:bicep_resSBL})
we present the results
of the cosmological (joint) analysis and finally
Subsection \ref{sub:bicep_disc}
contains a thorough discussion of these results.

\subsection{The cosmological analysis}
\label{sub:bicep_param}
As we probed in Section~\ref{sec:jhep} that the TH and the DW scenarios
give very similar results, we restrict our calculations to the
thermal case only.
The setup under investigation here is then a model in which
the neutrino sector is described by 3 massless
or almost massless active species,
as well as one additional sterile species characterized
by a temperature $T_s$.
Since we want to describe only the phenomenology of a TH \lsn,
here we decided to use the physical mass $m_s$ as a free parameter,
instead of the effective mass \meff{s}.

Our cosmological model is a
flat \lcdm{}+$r_{0.002}$+$\nu_s$ model with a total of 
nine parameters
\begin{equation}\label{eq:model}
{\bm \theta} = \{\omega_{\mathrm c},\omega_{\mathrm b},\theta_{\mathrm s},\tau,
\ln(10^{10}A_{s}),n_{s},r_{0.002},m_s,\DNeff\}.
\end{equation}
We recall that
$\omega_\mathrm{c} \equiv \Omega_{\mathrm c} h^2$ and
$\omega_{\mathrm b} \equiv \Omega_{\mathrm b} h^2$
are respectively the present-day CDM and baryon energy densities,
$\theta_{\mathrm s}$ is the angular the sound horizon,
$\tau$ is the optical depth to reionization,
and
$\ln(10^{10}A_{s})$ and $n_s$ denote respectively
the amplitude and spectral index of the initial scalar fluctuations.
The last parameter is $r_{0.002}$ (also indicated with $r$),
the tensor-to-scalar ratio at the pivot scale of $0.002 \mpcinv$.
We assume a flat prior on all of the cosmological parameters,
with the exception of $m_s$.
For the physical mass of the additional sterile neutrino
we shall consider a flat prior only when the SBL data are not included.
To perform the joint analysis of SBL and cosmological data, in turn,
we shall use
the posterior obtained in the analysis of SBL neutrino oscillations
(see Sec.~\ref{sec:sbl})
as a prior on $m_s$.

In this Section we consider SBL and cosmological data.
The latter consist of CMB data, Large Scale Structure (LSS),
Hubble constant $H_0$, $\sigma_8$ measurements from 
the CFHTLenS and the Planck Sunyaev Zel'Dovich (SZ) cluster counts.
We briefly resume here the considered datasets,
that are extensively described in Chapter~\ref{ch:cosmomeasurements}:

\begin{description}
 \item[CMB --]
 The CMB dataset is based on the one adopted in the previous Section.
 We additionally include the B-modes autocorrelation power spectrum
 of the BICEP2 experiment,
 either using all of the nine channels ($20<\ell<340$),
 or only the first five data points ($\ell<200$),
 as in the BICEP2 paper \cite{Ade:2014xna}.

 \item[LSS --]
The information on the matter power spectrum
% at four different redshifts $z=0.22$, $z=0.41$, $z=0.60$ and $z=0.78$
from the WiggleZ Dark Energy Survey \cite{Parkinson:2012vd}.

 \item[$\mathbf{H_0}$ --]
the same measurement we used
in the previous Section, $H_0=73.8\pm2.4\Hou$ \cite{Riess:2011yx}.

 \item[CFHTLenS --]
The weak gravitational lensing signal extracted from
the Canada-France Hawaii Telescope Lensing Survey (CFHTLenS)
\cite{Kilbinger:2012qz,Heymans:2013fya},
that constrains a combination of the total matter density $\Omega_m$
and
the standard deviation of the amplitude of the matter density
fluctuations on a sphere of radius $8h^{-1}\mathrm{Mpc}$, $\sigma_8$. 
This result is included in our analysis as a Gaussian prior
$\sigma_8(\Omega_m/0.27)^{0.46}=0.774\pm0.040$.

 \item[PSZ --]
The number counts of clusters from
the Planck Sunyaev Zel'Dovich catalogue \cite{Ade:2013lmv},
incorporated in our analysis as a Gaussian prior
$\sigma_8(\Omega_m/0.27)^{0.3}=0.782\pm0.010$.
\end{description}

In addition, we consider the SBL neutrino oscillation data
as a prior on the physical mass of the \lsn,
as we did in the previous Section.
Further details on the parameterization of neutrino oscillations and
on the SBL constraints are reported in Chapter~\ref{ch:nu}.

\subsection{Cosmological Results}
\label{sub:bicep_res}

\begin{figure}[t]
\centering
\includegraphics[width=\onethirdwidth]{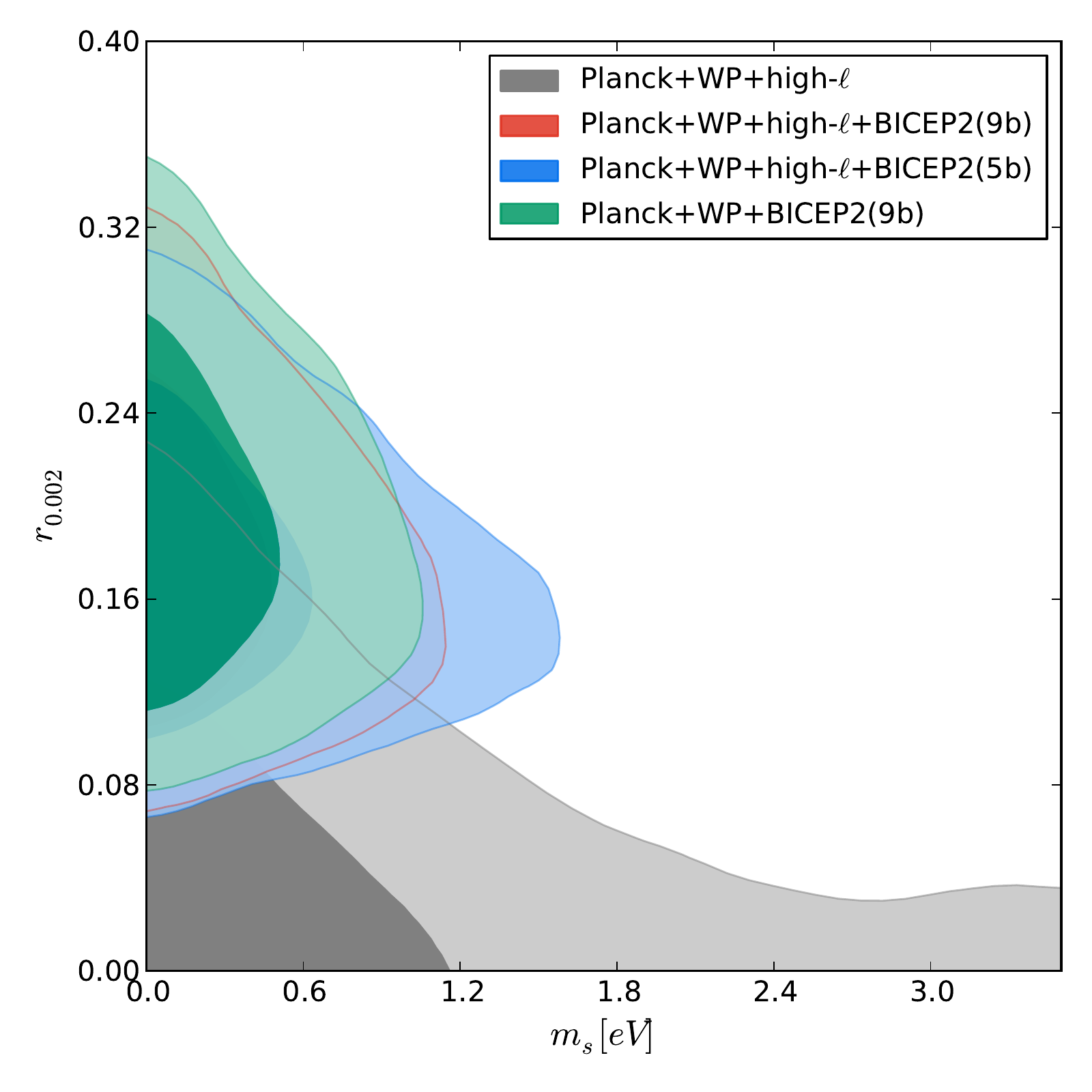}
\includegraphics[width=\onethirdwidth]{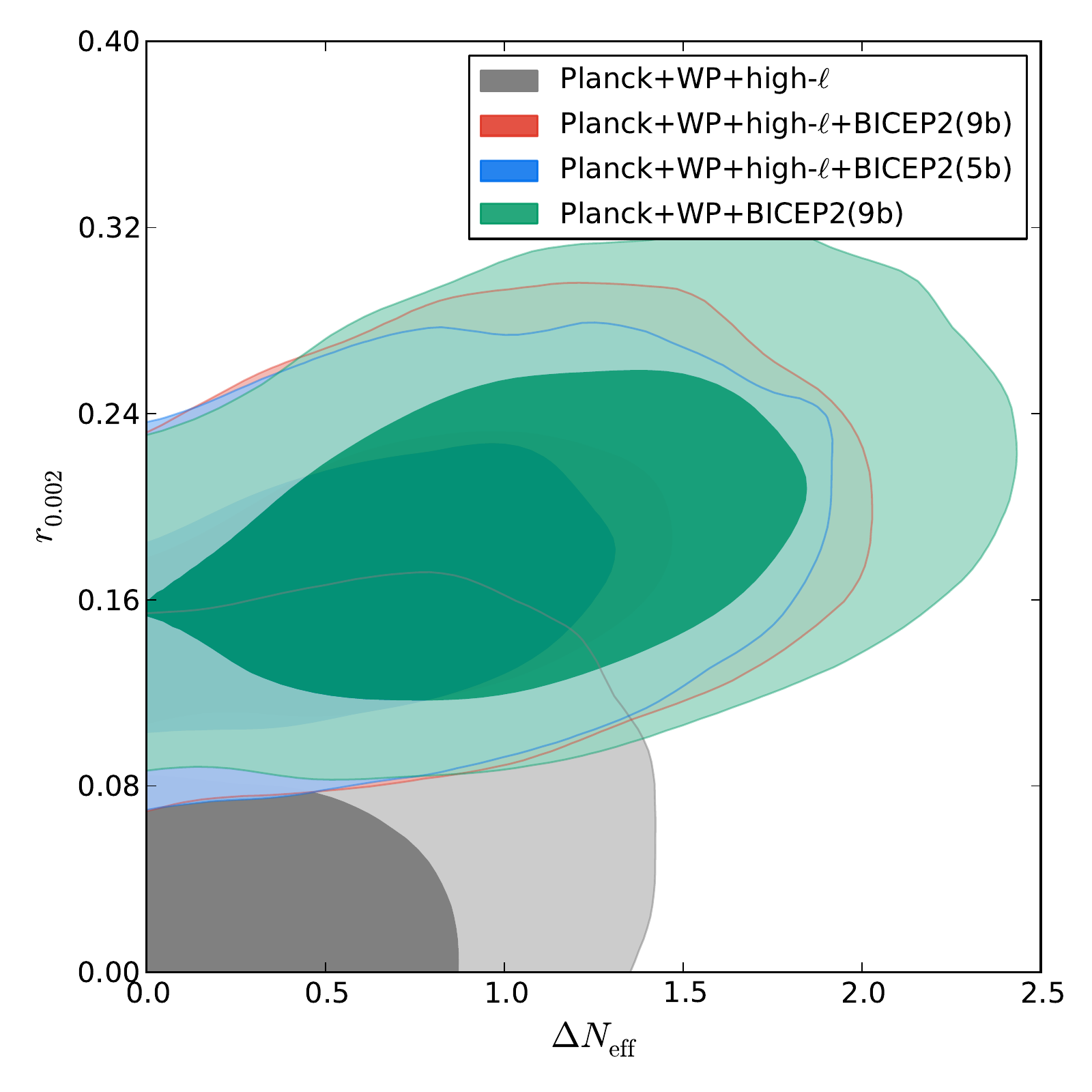}
\includegraphics[width=\onethirdwidth]{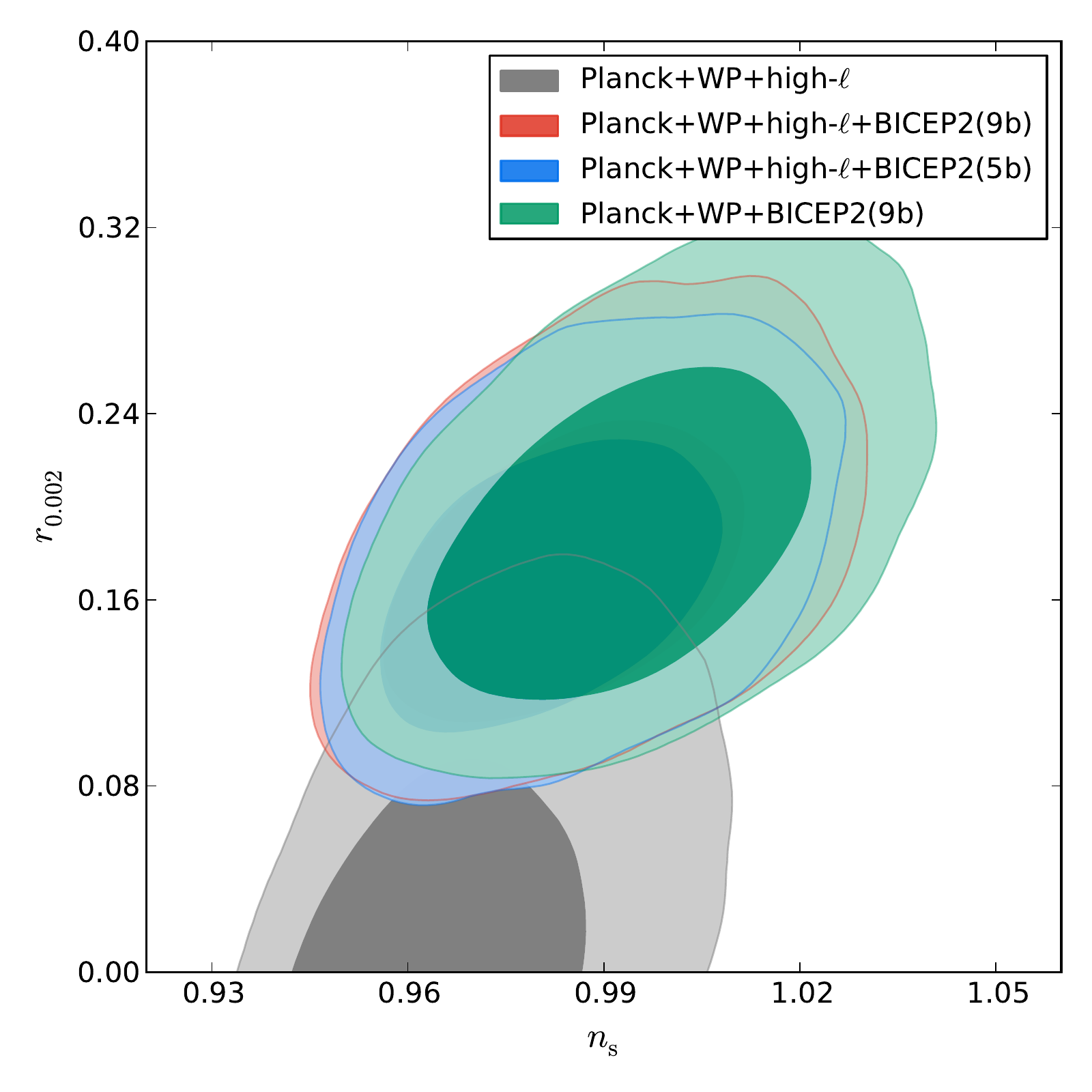} 
\caption{$1\sigma$ and $2\sigma$ marginalized contours
for different combinations of CMB data sets.
From Ref.~\protect\cite{Archidiacono:2014apa}.}
\label{fig:bicep_cmbonly}
\end{figure}

\begin{figure}[tp]
\centering
\includegraphics[width=0.6\textwidth]{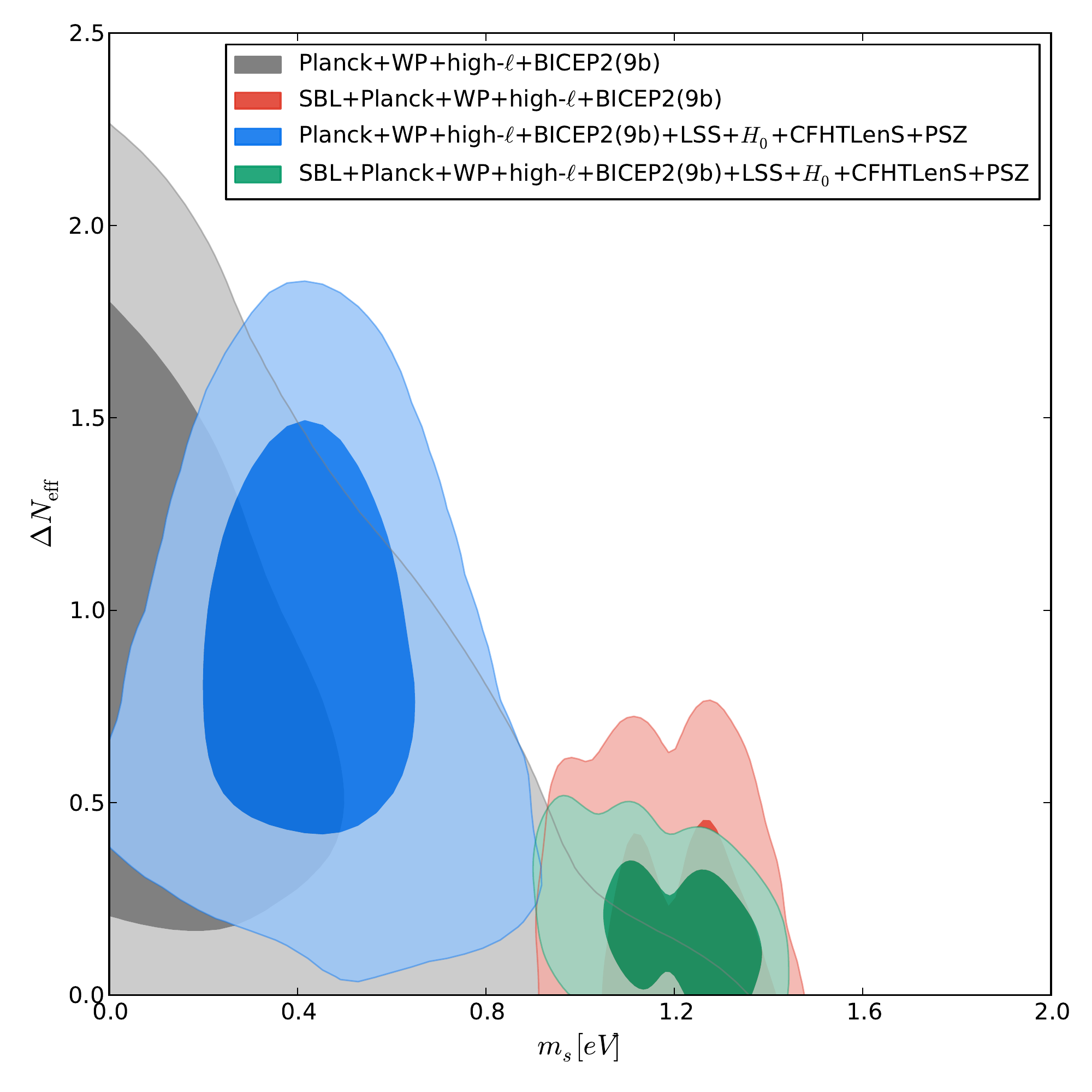} \\
\caption[$1\sigma$ and $2\sigma$ marginalized contours
in the plane $(m_s,\DNeff)$
for different data combinations]
{$1\sigma$ and $2\sigma$ marginalized contours in the plane
$(m_s,\DNeff)$.
The banana shaped regions allowed by cosmology indicate
a sub-eV mass and an excess in $\Neff$,
while the inclusion of SBL data forces the mass to be around 1~eV,
moving the contours towards the warm dark matter limit,
which implies a lower value of $\DNeff$ because
of the strong correlation between the two parameters.
From Ref.~\protect\cite{Archidiacono:2014apa}.}
\label{fig:bicep_ns_ms}
\end{figure}

An interesting question is how the addition of an eventual detection
of an high tensor-to-scalar ratio from inflation
changes the preferred region in $(m_s,\DNeff)$
(see Fig.~\ref{fig:bicep_ns_ms}) space.
Therefore, we first look at CMB data only,
with and without BICEP2 data included.
The result of this analysis can be seen
in Fig.~\ref{fig:bicep_cmbonly} and in Tab.~\ref{tab:bicep_cmbonly}.
As can be seen in Fig.~\ref{fig:bicep_cmbonly},
$m_s$ and $r$ are anti-correlated.
This happens because $r$ adds power on large scales whereas $m_s$
subtracts power on intermediate and small scales.
The inclusion of BICEP2 data therefore tends to strengthen
the bound on $m_s$ in order to keep constant the ratio between
the small and large scales.
Conversely, the addition of the BICEP2 data allows for higher values
of $\Neff$, because $\Neff$ is strongly correlated with $n_s$
and the addition of tensors shifts the allowed values for $n_s$ up.
For the case of CMB data only,
the addition of BICEP2 data therefore strengthens the bound
on $m_s$ slightly while allowing for a much higher $\Neff$.
This is consistent with the analysis presented in
Ref.~\cite{Dvorkin:2014lea}%
% \footnote{Notice that here the notation is different:
% in Ref.~\cite{Dvorkin:2014lea} $m_s$ indicates the effective mass
% of the sterile neutrino, while our $m_s$ is the physical mass.
% A direct comparison of the numerical results is not possible
% due to volume effects in Bayesian marginalization.
% Here we just want to emphasize that,
% concerning the effect due to the inclusion of BICEP2 data,
% both our results and those of Ref.~\cite{Dvorkin:2014lea}
% point towards tighter constraints
% on the additional massive component.}
.

We want now to be more precise.
The fact that the BICEP2 data lead to an enhancement of \Neff\
is due to the correlation between
$\Neff$
and the spectral index $n_s$
of the scalar PPS in Eq.~\eqref{eq:plPPS}.
Keeping fixed the amplitude $A_{s}$ at $k \sim k_{0}$,
which is constrained by the high-precision Planck data,
the scalar contribution to the
large-scale temperature fluctuations with $k \ll k_{0}$
measured by WMAP and Planck
can be decreased by an increase\footnote{
One could think to
alleviate the tension between BICEP2 and WMAP-Planck
by decreasing $n_s$,
if the value of $r$ measured by BICEP2
refers to a wavenumber $k_{1}$
larger than than the wavenumber
$k_{2} = 0.002 \mpcinv$
corresponding to the WMAP and Planck upper bounds
\cite{Ashoorioon:2014nta,Audren:2014cea}.
Since
$
r_{k_{2}}
\simeq
r_{k_{1}}
\left( k_{1} / k_{2} \right)^{n_{s}-1-n_{t}}
$,
where $n_{t}$ is the tensor spectral index,
for $k_{2} < k_{1}$
and
$n_{s}-1-n_{t} < 0$
we have $r_{k_{2}} < r_{k_{1}}$
and the ratio
$r_{k_{2}} / r_{k_{1}}$
decreases by decreasing $n_s$.
However,
one must take into account that WMAP and Planck
did not measure directly the tensor fluctuations as BICEP2,
but measured the temperature fluctuations,
in which the scalar and tensor contributions are indistinguishable.
Hence,
decreasing $n_{s}$ increases the scalar contribution to the
temperature fluctuations measured by
WMAP and Planck at $k_{2} < k_{1}$
and there is less room for a tensor contribution.
Therefore the WMAP and Planck upper bounds on $r_{k_{2}}$
tighten by about the same amount of the decrease of the BICEP2 value of
$r_{k_{2}}$,
maintaining the tension.
}
of the spectral index $n_s$.
In this way,
the WMAP and Planck data leave more space for the tensor contribution
\cite{Knox:1994qj}
and the corresponding bounds on $r$ are relaxed.
However,
the increase of $n_s$ induces an increase of small scale fluctuations
with $k \gg k_{0}$,
which would spoil the fit of high-$\ell$ CMB data
if the increase is not
compensated by an effect beyond the standard cosmological \lcdm\ model.
An increase of $\Neff$ above the standard value
$\Neff^{\text{SM}} = 3.046$ \cite{Mangano:2005cc}
has just the desired effect of decreasing the small scale fluctuations
(see Subsection~\ref{sub:radiationeffects}).
In fact,
from the fit of CMB data without BICEP2 we obtain
$n_{s} = 0.970 {}^{+0.011}_{-0.018}$ ($1\sigma$)
and
$\DNeff < 1.18$ ($2\sigma$),
and adding BICEP2 data we find
$n_{s} = 0.986 {}^{+0.016}_{-0.020}$ ($1\sigma$)
and
$\DNeff = 0.82 {}^{+0.40}_{-0.57}$ ($1\sigma$)
as reported in Tab.~\ref{tab:bicep_cmbonly}.

When the inclusion of the BICEP2 data
is restricted to the first five bins,
the results concerning the basic cosmological parameters
remain unchanged within $1\sigma$,
whereas the bound on the mass becomes slightly weaker
and the constraints on $\DNeff$ are tighter.
If we remove the high multipole CMB data
the bound on the mass remains almost unchanged,
while $\DNeff$ moves towards one additional
fully thermalized sterile neutrino.

\begin{table*}[tp]
\begin{center}
\renewcommand{\arraystretch}{1.2}
\resizebox{1\textwidth}{!}{
\begin{tabular}{|l|c|c|c|c|}
\hline
\hline
 Parameters 
 &Planck+WP+high-$\ell$ &Planck+WP+high-$\ell$&Planck+WP+high-$\ell$
 &Planck+WP \\
 & &+BICEP2(9bins) &+BICEP2(5bins) & +BICEP2(9bins)\\

\hline
$\Omega_{\mathrm b} h^2$ 	
			& $0.02231^{+0.00032}_{-0.00040}\,^{+0.00078}_{-0.00072}$ & $0.02251^{+0.00039}_{-0.00046}\,^{+0.00087}_{-0.00078}$ & $0.02249^{+0.00035}_{-0.00045}\,^{+0.00084}_{-0.00078}$ & $0.02259^{+0.00040}_{-0.00050}\,^{+0.00094}_{-0.00082}$ \\

$\Omega_{\mathrm c} h^2$ 
			& $0.125^{+0.005}_{-0.007}\,^{+0.011}_{-0.010}$           & $0.129^{+0.006}_{-0.007}\,^{+0.013}_{-0.012}$           & $0.128^{+0.005}_{-0.008}\,^{+0.013}_{-0.012}$           & $0.132^{+0.007}_{-0.008}\,^{+0.015}_{-0.014}$           \\

$\theta_{\mathrm s}$ 
			& $1.0404^{+0.0009}_{-0.0008}\,^{+0.0016}_{-0.0017}$      & $1.0399^{+0.0009}_{-0.0009}\,^{+0.0017}_{-0.0017}$      & $1.0401^{+0.0009}_{-0.0009}\,^{+0.0018}_{-0.0017}$      & $1.0395^{+0.0009}_{-0.0009}\,^{+0.0019}_{-0.0018}$      \\

$\tau$ 
			& $0.094^{+0.013}_{-0.016}\,^{+0.031}_{-0.027}$           & $0.097^{+0.013}_{-0.016}\,^{+0.031}_{-0.027}$           & $0.096^{+0.013}_{-0.016}\,^{+0.030}_{-0.029}$           & $0.098^{+0.014}_{-0.017}\,^{+0.031}_{-0.031}$           \\

$n_{\mathrm s}$ 
			& $0.970^{+0.011}_{-0.018}\,^{+0.033}_{-0.027}$           & $0.986^{+0.016}_{-0.020}\,^{+0.035}_{-0.033}$           & $0.983^{+0.014}_{-0.020}\,^{+0.034}_{-0.031}$           & $0.995^{+0.017}_{-0.021}\,^{+0.038}_{-0.036}$           \\

$\log(10^{10} A_s)$ 
			& $3.106^{+0.029}_{-0.036}\,^{+0.068}_{-0.062}$           & $3.120^{+0.030}_{-0.037}\,^{+0.071}_{-0.061}$           & $3.167^{+0.047}_{-0.040}\,^{+0.080}_{-0.089}$           & $3.145^{+0.052}_{-0.046}\,^{+0.090}_{-0.098}$           \\

$r$ 
			& $<0.145$                                                & $0.177^{+0.036}_{-0.050}\,^{+0.093}_{-0.086}$           & $0.172^{+0.035}_{-0.048}\,^{+0.088}_{-0.082}$           & $0.192^{+0.040}_{-0.055}\,^{+0.101}_{-0.092}$           \\
\hline
$\DNeff$ 
			& $<1.18$                                                 & $0.82^{+0.40}_{-0.57};\,<1.66$                           & $0.73^{+0.31}_{-0.59};\,<1.56$                          & $1.08^{+0.49}_{-0.61};\,<2.03$                          \\

$m_s [\mathrm{eV}]$
			& $<2.17$                                                 & $<0.85$                                                 & $<1.15$                                                 & $<0.81$                                                 \\

\hline
\hline
\end{tabular}
}
\vspace{0.3cm}
\caption[Marginalized limits for the cosmological parameters 
from various combinations of CMB data]
{Marginalized $1\sigma$ and $2\sigma$ confidence level limits
for the cosmological parameters, % in various dataset combinations,
given with respect to the mean value.
Upper limits are given at $2\sigma$.
From Ref.~\protect\cite{Archidiacono:2014apa}.}
\label{tab:bicep_cmbonly}
\end{center}
\end{table*}

Having established how the constraints change from CMB data only
we now proceed to study the influence of the auxiliary cosmological data.
From now on, we will consider only the full CMB dataset
that includes the BICEP2 data for all the nine bins.

In Tab.~\ref{tab:bicep_cosmoresults} we report the
mean values and the $1\sigma$ and $2\sigma$ errors
on the cosmological parameters and on the neutrino parameters
in the different combinations of data sets illustrated above,
when the SBL data are not included.

\begin{figure}[tp]
\centering
\includegraphics[width=\singlefigsmall]{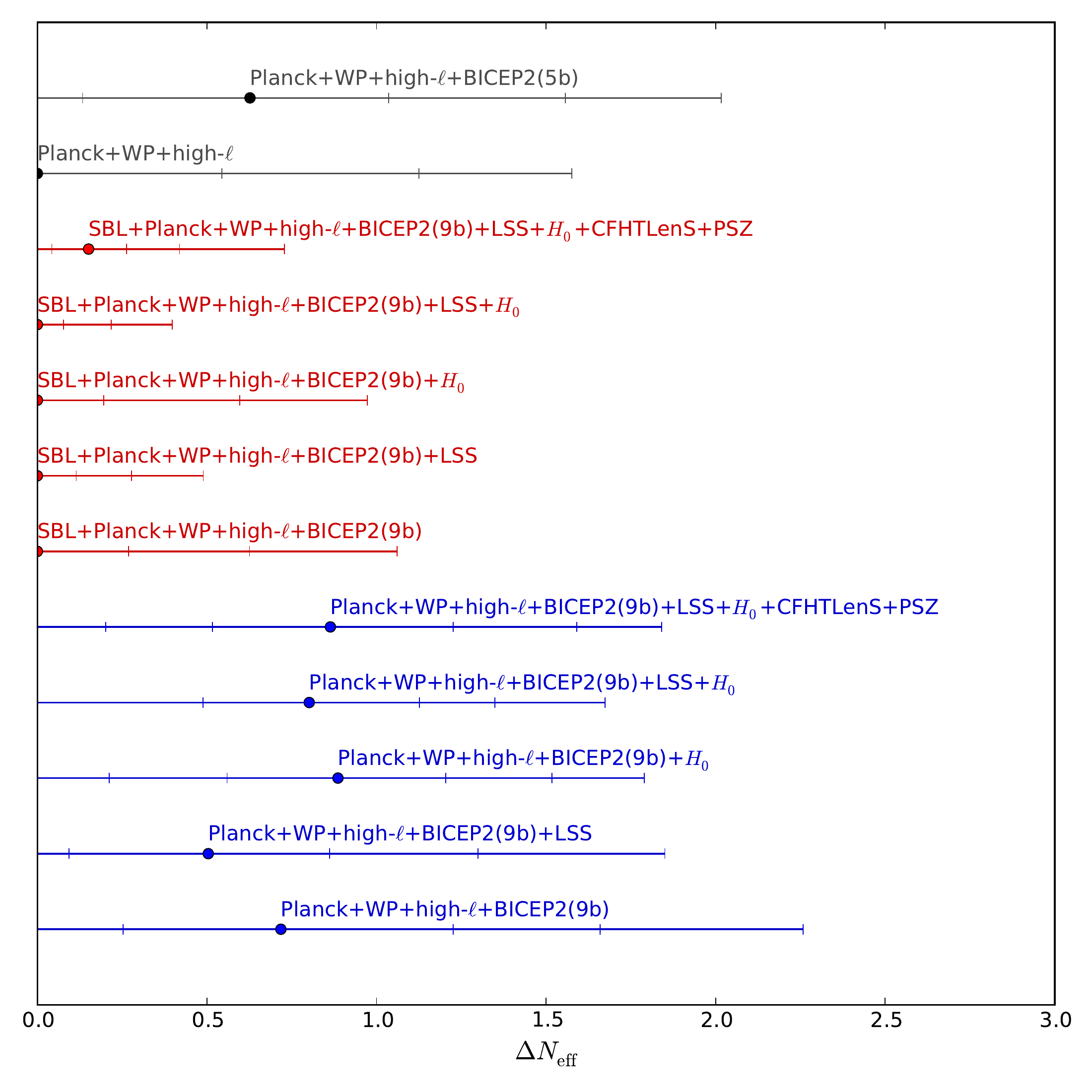} \\
\caption[$1\sigma$, $2\sigma$ and $3\sigma$ confidence level limits
for $\DNeff$, for different dataset combinations]
{$1\sigma$, $2\sigma$ and $3\sigma$ confidence level limits
for $\DNeff$, for different dataset combinations.
The circles indicate the mean value.
From Ref.~\protect\cite{Archidiacono:2014apa}.}
\label{fig:bicep_ns}
\end{figure}

\begin{figure}[tp]
\centering
\includegraphics[width=\singlefigsmall]{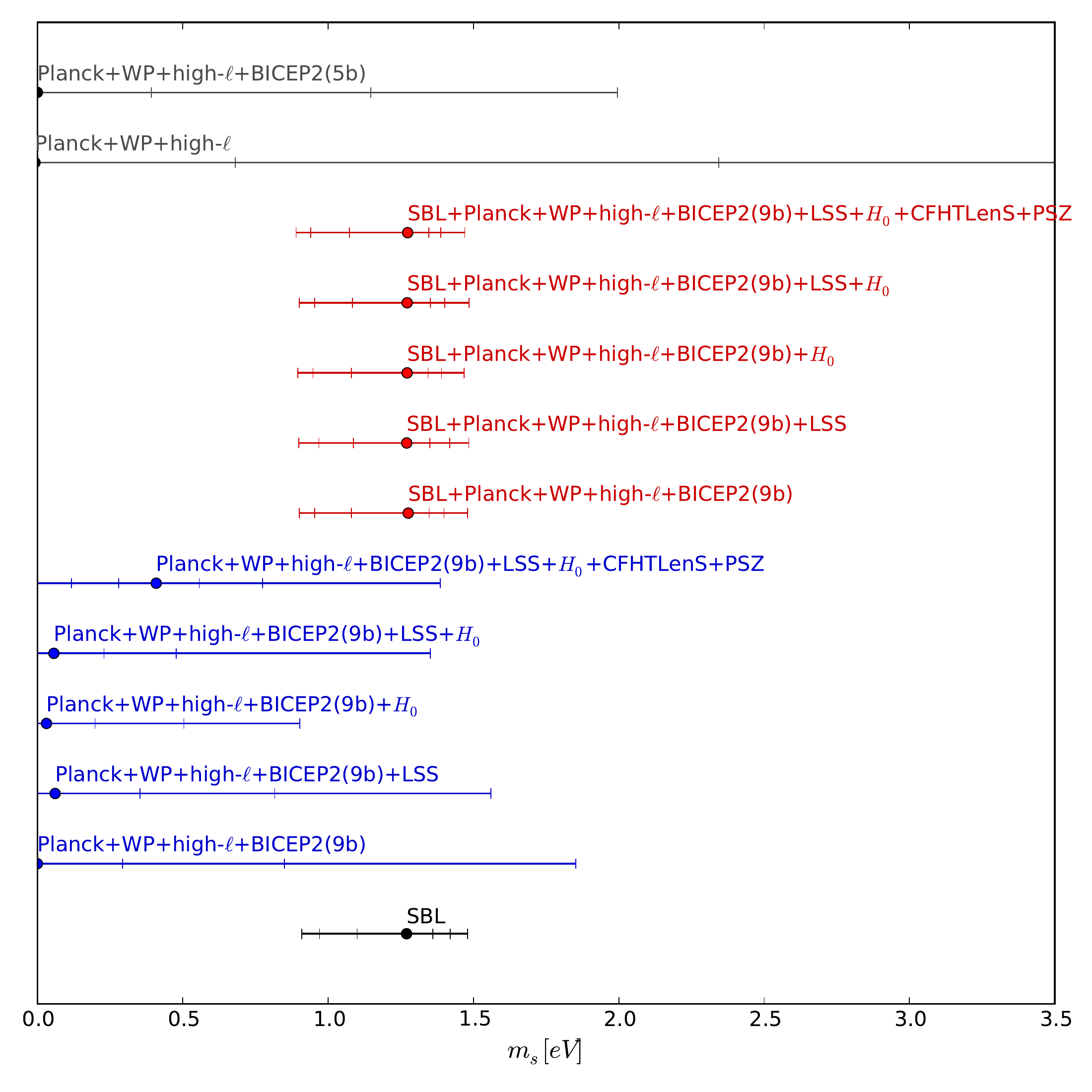} \\
\caption[$1\sigma$, $2\sigma$ and $3\sigma$ confidence level limits
for $m_s$, for different dataset combinations]
{As in Fig.~\ref{fig:bicep_ns}, but for $m_s$.
From Ref.~\protect\cite{Archidiacono:2014apa}.}
\label{fig:bicep_ms}
\end{figure}

As seen above, the Planck CMB data provide a fairly stringent
upper limit on the sterile neutrino mass,
except for very low values of $\Neff$,
corresponding to the warm dark matter limit.
Conversely the preferred value of $\Neff$ is higher than 3,
with 4 only being slightly disfavored.
The inclusion of BICEP2 data pushes the preferred $\Neff$ up,
as it was pointed out by several authors
\cite{Giusarma:2014zza, Zhang:2014dxk, Dvorkin:2014lea}.
However, since $m_s$ and $\Neff$ are anti-correlated
this actually results in a tighter bound
on the sterile neutrino mass from CMB only.

When we include LSS or $H_0$ data the picture remains
qualitatively unchanged although,
since $m_s$ and $H_0$ are anti-correlated,
the addition of the HST $H_0$ data strengthens
the upper bound on the sterile neutrino mass.
In Fig.~\ref{fig:bicep_ms} and Fig.~\ref{fig:bicep_ns}
we can see how the error bars change for $m_s$ and $\DNeff$
respectively, with various dataset combinations.

The picture changes with the inclusion of weak lensing and cluster data,
that leads to an important qualitative change for the preferred range
of $m_s$.
Both these data sets give a preference for a low value of $\sigma_8$.
Given that the amplitude of the fluctuations is fixed on large scales
by the CMB measurements, a low value of $\sigma_8$ can be caused
by a non-zero neutrino mass which specifically reduces the power
on small scales thanks to its free-streaming,
while leaving the large scale power unchanged with respect to the
standard \lcdm\ prediction.
The addition of the CFHTLenS and PSZ data sets yields then
a preferred mass for the sterile neutrino of around 0.5 eV,
with $\DNeff=1$ allowed.

\begin{table*}[tp]
\begin{center}
\renewcommand{\arraystretch}{1.2}
\resizebox{1\textwidth}{!}{
\begin{tabular}{|l|c|c|c|c|c|}
\hline
\hline
            &Planck+WP+high-$\ell$ 	&Planck+WP+high-$\ell$ 	&Planck+WP+high-$\ell$ 	&Planck+WP+high-$\ell$ &Planck+WP+high-$\ell$\\
Parameters  &+BICEP2			&+BICEP2		&+BICEP2		&+BICEP2		&+BICEP2\\
            &         			&+LSS         		&+$H_0$         	&+LSS+$H_0$ 		&+LSS+$H_0$+CFHTLenS+PSZ\\
\hline
$\Omega_{\mathrm b} h^2$ 	
			& $0.02251^{+0.00039}_{-0.00046}\,^{+0.00087}_{-0.00078}$ & $0.02232^{+0.00033}_{-0.00039}\,^{+0.00073}_{-0.00069}$ & $0.02257^{+0.00029}_{-0.00030}\,^{+0.00059}_{-0.00057}$ & $0.02248^{+0.00029}_{-0.00029}\,^{+0.00057}_{-0.00056}$ & $0.02267^{+0.00027}_{-0.00028}\,^{+0.00055}_{-0.00053}$ \\

$\Omega_{\mathrm c} h^2$ 
			& $0.129^{+0.006}_{-0.007}\,^{+0.013}_{-0.012}$           & $0.128^{+0.005}_{-0.006}\,^{+0.011}_{-0.010}$           & $0.130^{+0.006}_{-0.006}\,^{+0.011}_{-0.011}$           & $0.129^{+0.005}_{-0.005}\,^{+0.011}_{-0.011}$           & $0.127^{+0.006}_{-0.006}\,^{+0.011}_{-0.011}$           \\

$\theta_{\mathrm s}$ 
			& $1.0399^{+0.0009}_{-0.0009}\,^{+0.0017}_{-0.0017}$      & $1.0401^{+0.0009}_{-0.0008}\,^{+0.0017}_{-0.0017}$      & $1.0398^{+0.0008}_{-0.0008}\,^{+0.0018}_{-0.0016}$      & $1.0399^{+0.0008}_{-0.0008}\,^{+0.0017}_{-0.0016}$      & $1.0400^{+0.0009}_{-0.0009}\,^{+0.0018}_{-0.0017}$      \\

$\tau$ 
			& $0.097^{+0.013}_{-0.016}\,^{+0.031}_{-0.027}$           & $0.093^{+0.013}_{-0.014}\,^{+0.027}_{-0.027}$           & $0.099^{+0.013}_{-0.015}\,^{+0.029}_{-0.026}$           & $0.095^{+0.013}_{-0.014}\,^{+0.028}_{-0.027}$           & $0.091^{+0.013}_{-0.015}\,^{+0.028}_{-0.027}$           \\

$n_{\mathrm s}$ 
			& $0.986^{+0.016}_{-0.020}\,^{+0.035}_{-0.033}$           & $0.977^{+0.012}_{-0.016}\,^{+0.028}_{-0.027}$           & $0.989^{+0.011}_{-0.011}\,^{+0.021}_{-0.022}$           & $0.985^{+0.011}_{-0.010}\,^{+0.020}_{-0.022}$           & $0.993^{+0.010}_{-0.011}\,^{+0.021}_{-0.021}$           \\

$\log(10^{10} A_s)$ 
			& $3.120^{+0.030}_{-0.037}\,^{+0.071}_{-0.061}$           & $3.182^{+0.042}_{-0.038}\,^{+0.073}_{-0.078}$           & $3.124^{+0.030}_{-0.031}\,^{+0.060}_{-0.058}$           & $3.116^{+0.029}_{-0.030}\,^{+0.060}_{-0.055}$           & $3.124^{+0.031}_{-0.031}\,^{+0.063}_{-0.061}$           \\

$r$ 
			& $0.177^{+0.036}_{-0.050}\,^{+0.093}_{-0.086}$           & $0.168^{+0.034}_{-0.046}\,^{+0.085}_{-0.078}$           & $0.181^{+0.037}_{-0.047}\,^{+0.087}_{-0.081}$           & $0.175^{+0.035}_{-0.045}\,^{+0.083}_{-0.077}$           & $0.206^{+0.041}_{-0.051}\,^{+0.094}_{-0.090}$           \\
\hline
$\DNeff$ 
			& $0.82^{+0.40}_{-0.57};\,<1.66$                          & $0.61^{+0.25}_{-0.52};\,<1.30$                           & $0.88^{+0.32}_{-0.32}\,^{+0.64}_{-0.67}$                & $0.81^{+0.32}_{-0.32};\,<1.35$                          & $0.89^{+0.34}_{-0.37}\,^{+0.70}_{-0.69}$                \\

$m_s [\mathrm{eV}]$
			& $<0.85$                                                 & $<0.82$                                                 & $<0.50$                                                 & $<0.48$                                                 & $0.44^{+0.11}_{-0.16}\,^{+0.33}_{-0.32}$                \\

\hline
\hline
\end{tabular}
}
\vspace{0.3cm}
\caption[Marginalized limits for the cosmological parameters
from various combinations of cosmological data]
{Marginalized $1\sigma$ and $2\sigma$ confidence level limits
for the cosmological parameters, given with respect to the mean value,
from the analyses of cosmological data only.
Upper limits are given at $2\sigma$.
From Ref.~\protect\cite{Archidiacono:2014apa}.}
\label{tab:bicep_cosmoresults}
\end{center}
\end{table*}

\subsection{Results with the SBL prior}
\label{sub:bicep_resSBL}

At this point we can try to understand if the cosmological
and SBL data are really compatible.
When we use cosmological data without weak lensing and cluster data
we find a relatively stringent upper bound on $m_s$.
This is relaxed when $\DNeff$ is low, simply because the suppression
of structure formation scales with the total density in neutrinos
at late times, i.e.\ as $\DNeff^{3/4} m_s$.
However, since CMB data prefers a high $\DNeff$ this possibility
is disfavored, and the conclusion is that CMB and LSS data
require the sterile mass to be low.
The bound can easily be relaxed in models where additional
dark radiation is provided by other particles.
When we add weak lensing and cluster data the sterile mass comes out to be
around 0.5 eV and fully thermalized sterile neutrinos are allowed. 

In Tab.~\ref{tab:bicep_sblresults}
we report the marginalized mean values and the $1\sigma$ and $2\sigma$
errors on the cosmological parameters and on the neutrino parameters
in the different combinations of data sets illustrated above,
when SBL data are included.
As we stated before,
it is easy to see that the anti-correlation between $m_s$ and $\DNeff$,
together with the strong bounds on $m_s$ from the SBL data,
leaves a very small space to a fully thermalized sterile neutrino.
When adding SBL data, the constraints on $m_s$ come only
by the oscillation experiments,
with very small dependence on the cosmological data. 
On the other hand, cosmology provides a strong limit
on $\DNeff$ that is compatible with 0 within $2\sigma$
in all the cases, as we can see in Fig.~\ref{fig:bicep_ns}.
When LSS data are included, the value of $\DNeff$
is even more constrained.
Only when CFHTLenS and PSZ are included there is a little evidence
that $\DNeff>0$ at more than $1\sigma$:
even in this case, however,
a fully thermalized sterile neutrino with $\DNeff=1$
is strongly disfavored.

This tension between cosmological and SBL data,
has been studied also in past works
(see e.g.\ Ref.~\cite{Mirizzi:2013gnd}):
% is not alleviated in the physical case
% (i.e.\ varying $\DNeff$ in the range $[0,1]$):
the mass values preferred by SBL data lay above
the hot dark matter limit and therefore they are disfavored
by cosmology, even if there is only one partially
(or fully, $\DNeff=1$) thermalized sterile neutrino.
Quantitatively speaking, a model with one fully thermalized sterile
neutrino and with a mass fixed at the SBL best-fit %($m_s=1.27$ eV)
has a $\Delta \chi^2 \simeq 18$ compared to the cosmological best-fit model,
if Planck+WP+high-$\ell$ data are considered.
If also BICEP2 data are considered, the value lowers to
$\Delta \chi^2 \simeq 12$:
this is possible since the inclusion of the BICEP2 data strengthens
the limit on $m_s$, but it weakens the limit on $\DNeff$. 

If a partial thermalization is taken into account and $\DNeff$
is free to vary moving towards lower values,
the $\Delta\chi^2$ differences are smaller.
For a $m_s=1.27$ eV neutrino with small $\DNeff$
we have $\Delta \chi^2 \simeq 1$ from Planck+WP+high-$\ell$
and $\Delta \chi^2 \simeq 6$ from Planck+WP+high-$\ell$+BICEP2.

We can conclude that a fully thermalized sterile neutrino
with a mass fixed at the SBL best-fit is less disfavored
by cosmology if the BICEP2 data are included.
On the contrary, if the sterile neutrino is not fully thermalized
the inclusion of BICEP2 data worsens the consistency
of the presence of a 1~eV mass sterile neutrino in cosmology.

\begin{table*}[tp]
\begin{center}
\renewcommand{\arraystretch}{1.2}
\resizebox{1\textwidth}{!}{
\begin{tabular}{|l|c|c|c|c|c|}
\hline
\hline
&SBL+Planck+WP &SBL+Planck+WP &SBL+Planck+WP &SBL+Planck+WP &SBL+Planck+WP\\
 Parameters &+high-$\ell$+BICEP2 &+high-$\ell$+BICEP2 &+high-$\ell$+BICEP2 &+high-$\ell$+BICEP2 &+high-$\ell$+BICEP2\\
                   &         &+LSS         &+$H_0$          &+LSS+$H_0$ &+LSS+$H_0$+CFHTLenS+PSZ\\
\hline

$\Omega_{\mathrm b} h^2$& $0.02214^{+0.00029}_{-0.00029}\,^{+0.00058}_{-0.00058}$ & $0.02200^{+0.00026}_{-0.00025}\,^{+0.00051}_{-0.00052}$ & $0.02230^{+0.00027}_{-0.00027}\,^{+0.00060}_{-0.00054}$ & $0.02214^{+0.00025}_{-0.00025}\,^{+0.00049}_{-0.00051}$ & $0.02236^{+0.00023}_{-0.00023}\,^{+0.00047}_{-0.00047}$ \\

$\Omega_{\mathrm c} h^2$& $0.121^{+0.003}_{-0.004}\,^{+0.008}_{-0.007}$           & $0.121^{+0.002}_{-0.003}\,^{+0.006}_{-0.005}$           & $0.118^{+0.003}_{-0.004}\,^{+0.007}_{-0.006}$           & $0.118^{+0.002}_{-0.002}\,^{+0.005}_{-0.005}$           & $0.117^{+0.002}_{-0.003}\,^{+0.006}_{-0.006}$           \\

$\theta_{\mathrm s}$ & $1.0408^{+0.0008}_{-0.0007}\,^{+0.0015}_{-0.0014}$      & $1.0409^{+0.0006}_{-0.0006}\,^{+0.0012}_{-0.0013}$      & $1.0413^{+0.0007}_{-0.0006}\,^{+0.0013}_{-0.0015}$      & $1.0413^{+0.0006}_{-0.0006}\,^{+0.0012}_{-0.0012}$      & $1.0413^{+0.0006}_{-0.0006}\,^{+0.0013}_{-0.0014}$      \\

$\tau$ & $0.092^{+0.012}_{-0.014}\,^{+0.026}_{-0.025}$           & $0.088^{+0.012}_{-0.014}\,^{+0.027}_{-0.024}$           & $0.094^{+0.012}_{-0.015}\,^{+0.028}_{-0.027}$           & $0.091^{+0.012}_{-0.014}\,^{+0.026}_{-0.024}$           & $0.086^{+0.012}_{-0.014}\,^{+0.026}_{-0.024}$           \\

$n_{\mathrm s}$ & $0.962^{+0.008}_{-0.008}\,^{+0.016}_{-0.015}$           & $0.958^{+0.006}_{-0.006}\,^{+0.013}_{-0.013}$           & $0.967^{+0.007}_{-0.008}\,^{+0.015}_{-0.014}$           & $0.962^{+0.006}_{-0.006}\,^{+0.012}_{-0.012}$           & $0.970^{+0.005}_{-0.005}\,^{+0.011}_{-0.011}$           \\

$\log(10^{10} A_s)$ & $3.213^{+0.031}_{-0.031}\,^{+0.063}_{-0.063}$           & $3.220^{+0.030}_{-0.030}\,^{+0.059}_{-0.059}$           & $3.091^{+0.026}_{-0.030}\,^{+0.057}_{-0.051}$           & $3.085^{+0.025}_{-0.027}\,^{+0.052}_{-0.048}$           & $3.169^{+0.027}_{-0.026}\,^{+0.053}_{-0.052}$           \\

$r$ & $0.160^{+0.034}_{-0.042}\,^{+0.078}_{-0.075}$           & $0.150^{+0.032}_{-0.039}\,^{+0.071}_{-0.067}$           & $0.164^{+0.032}_{-0.043}\,^{+0.079}_{-0.073}$           & $0.158^{+0.032}_{-0.042}\,^{+0.075}_{-0.070}$           & $0.179^{+0.034}_{-0.043}\,^{+0.082}_{-0.076}$           \\
\hline
$\DNeff$ & $<0.63$                                                 & $<0.28$                                                 & $<0.59$                                                 & $<0.22$                                                 & $0.19^{+0.07}_{-0.15};\,<0.42$                          \\

$m_s [\mathrm{eV}]$& $1.21^{+0.14}_{-0.13}\,^{+0.19}_{-0.25}$                & $1.22^{+0.13}_{-0.13}\,^{+0.20}_{-0.25}$                & $1.20^{+0.14}_{-0.12}\,^{+0.19}_{-0.25}$                & $1.21^{+0.14}_{-0.13}\,^{+0.19}_{-0.26}$                & $1.19^{+0.15}_{-0.12}\,^{+0.19}_{-0.25}$                \\

\hline
\hline
\end{tabular}
}
\vspace{0.3cm}
\caption{As in Tab.~\ref{tab:bicep_cosmoresults},
but from the joint analyses of cosmological and SBL data.
From Ref.~\protect\cite{Archidiacono:2014apa}.}
\label{tab:bicep_sblresults}
\end{center}
\end{table*}

\subsection{Discussion}
\label{sub:bicep_disc}

We have performed an analysis of light sterile neutrinos
in the context of both cosmology and
short baseline neutrino oscillation experiments.
Previous analyses have shown that while SBL data points
to the existence of a mainly sterile mass state around 1 eV,
this is not compatible with cosmological data unless
the additional state is somehow prevented from being fully thermalized
in the early Universe~\cite{Archidiacono:2012ri}.

If the BICEP2 data were related to primordial tensor modes,
they would favor
a higher dark radiation content,
but this actually would tighten the cosmological bounds on the mass
of the sterile neutrino, because $m_s$ and $\DNeff$
are strongly anti-correlated.
Cosmological data from the CFHTLenS survey and
the Planck SZ cluster counts actually favor a non-zero mass
of the sterile neutrino, because it alleviates the tension
between the value of $\sigma_8$ inferred from the CMB measurements
in the context of the minimal \lcdm{} model and the lower values indicated
by data CFHTLenS and PSZ data.
The inclusion of these two data sets points towards a sterile neutrino mass
around 0.5 eV,
but with relatively a low $\DNeff$.

The SBL data strongly constrains $m_s$, but not $\DNeff$,
and indicates a \lsn\ mass not much lower than 1 eV.
At the same time the mixing angle is large enough that the additional
state should be almost fully thermalized
\cite{Hannestad:2012ky,Mirizzi:2013gnd,Vincent:2014rja}.
However, this scenario is highly disfavored by cosmological data
(with a $\Delta \chi^2>10$),
which requires $\DNeff$ to be small if the mass is around 1~eV.
Indeed, a model with a mass of 1 eV and a low $\DNeff$ is compatible
with cosmology within roughly $2\sigma$ confidence level.
The conclusion is that light sterile neutrinos as indicated
by SBL data are close to being ruled out by cosmological data,
unless they are somehow prevented
from thermalizing in the early Universe.

A possible way out of this problem is that sterile neutrinos
have new interactions which induce a non-standard matter potential
and block thermalization
\cite{Hannestad:2013ana,Dasgupta:2013zpn,Bringmann:2013vra,
Ko:2014bka,Mirizzi:2012we,Saviano:2013ktj}.
In this case there may be 1 eV sterile neutrinos and an $\Neff$
not much beyond 3, so that the model would be compatible with all existing data.
While this scenario certainly works well and can possibly also
explain some of the astrophysical anomalies related
to cold dark matter, there are without a doubt other possible ways
of making eV sterile neutrino compatible
with both SBL and cosmological data.
For example, some models with low temperature reheating
or non-standard expansion rate of the Universe at the MeV scale
where the new state is thermalized can also prevent thermalization
\cite{Rehagen:2014vna} (see also Section~\ref{sec:nucosmo_conclusions}).
In the next Section we will present a model that involves an invisible decay of
the sterile neutrino, occurring in cosmological time-scales.
This model has the advantage of allowing to reconcile the presence
of a massive sterile neutrino with the CMB data,
provided that $\DNeff=1$ is allowed for massless species.
% Thus, eV mass sterile neutrinos remain an intriguing possibility
% which potentially has wide ranging implications for cosmology. 

%%%%%%%%%%%%%%%%%%%
% arxiv:1404.6160 %
%%%%%%%%%%%%%%%%%%%
\section{Decaying Sterile Neutrino}
\label{sec:lsndecay}
% \cite{Gariazzo:2014pja}

\subsection{Motivations and Theoretical Model}
In the previous Sections we discussed how a light sterile neutrino
could help to reconcile the cosmological and the local
determinations of $H_0$ and
the observed matter fluctuations at small scale
with the value estimated from cosmology.
The mass scale of 1\ev\ that can explain
the SBL neutrino oscillations, however,
is not the same that emerges from the solution of the $\sigma_8$
problem, that requires masses around 0.5\ev.
It turns out that the cosmological and
SBL data on the neutrino mass are compatible only if one assumes
that the \lsn\ with a mass of 1\ev\ is not fully thermalized
in the early Universe
($\DNeff = 0.19 {}^{+0.07}_{-0.15}$ at $1\sigma$).
The case of a fully thermalized sterile neutrino is disfavored by 
$\Delta\chi^2>10$ \cite{Archidiacono:2014apa},
even if the (wrong) BICEP2 results would favor an higher \DNeff.
Similar conclusions have been presented also before
the BICEP2 results
(see e.g.\
Refs.~\cite{Archidiacono:2013xxa,Mirizzi:2013gnd,Gariazzo:2013gua},
which take into account the 2013 Planck data \cite{Ade:2013zuv})
and after the 2015 release of the Planck data \cite{Ade:2015xua},
that strongly disfavors any departures from $\Neff=3.046$.
These results motivated the study of mechanisms which can suppress the
thermalization of sterile neutrinos in the early Universe
due to active-sterile oscillations before neutrino decoupling
\cite{Dolgov:2003sg,Cirelli:2004cz,Chu:2006ua,Hannestad:2012ky}.
Examples are a large lepton asymmetry
\cite{Hannestad:2012ky,Mirizzi:2012we,
Saviano:2013ktj,Hannestad:2013wwj},
an enhanced background potential due to new interactions
in the sterile sector
\cite{Hannestad:2013ana,Dasgupta:2013zpn,Bringmann:2013vra,
Ko:2014bka,Archidiacono:2014nda,
Archidiacono:2015ota,Archidiacono:2015oma},
a larger cosmic expansion rate at the time
of sterile neutrino production
\cite{Rehagen:2014vna},
and
MeV dark matter annihilation
\cite{Ho:2012br}.

\begin{figure}[t]
\centering
\includegraphics[width=\halfwidth,page=1]{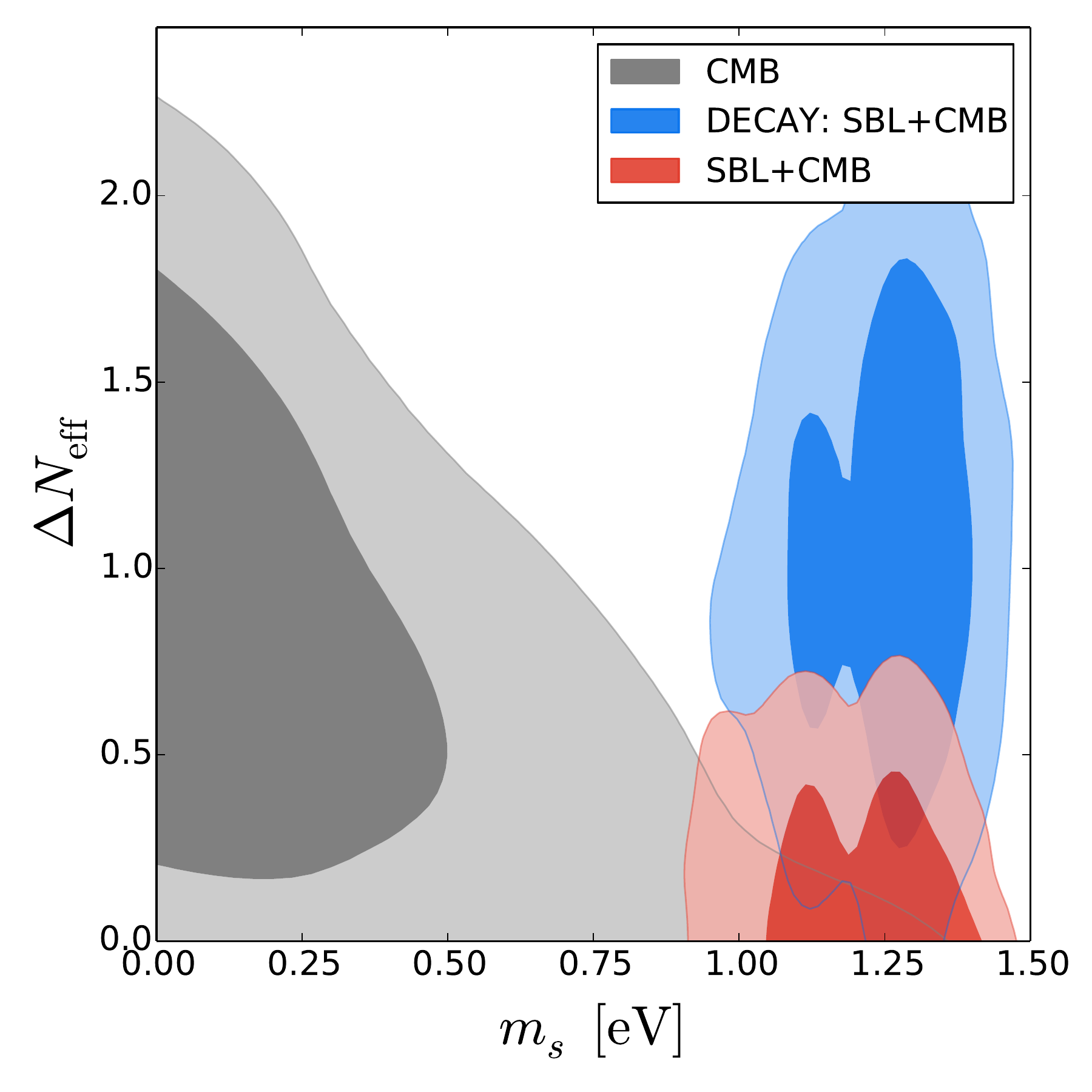}
\includegraphics[width=\halfwidth,page=2]{nucosmo/decay_plots.pdf}
\caption[Marginalized 2D constraints in the ($m_s$--\DNeff) and ($H_0$--\DNeff)
planes from the decaying sterile neutrino model, 
considering CMB data only]{\label{fig:decay_cmb}
$1\sigma$ and $2\sigma$ marginalized allowed regions
obtained with CMB data (Planck+WP+high-$\ell$+BICEP2(9bins);
see Ref.~\cite{Archidiacono:2014apa}),
without and with the inclusion of SBL data.
The gray and red regions are those obtained in
Ref.~\cite{Archidiacono:2014apa} without and with the SBL prior.
The blue regions are obtained by adding the possibility of invisible
decays for a sterile neutrino that explains the SBL oscillations.
}
\end{figure}

In this Section
we propose to solve the problem of the thermalization
of the sterile neutrino with an eV-scale mass
by introducing an invisible decay of the sterile neutrino.
The decay must be invisible in order not
to generate unobserved signals.
We assume that the decay products are very light
(or massless) particles belonging to the sterile sector.
For example, the eV-scale sterile neutrino $\nu_{s}$
could decay into a lighter sterile neutrino $\nu_{s'}$
and a very light invisible (pseudo)scalar boson%
\footnote{
The new invisible light (pseudo)scalar boson
is assumed to interact only with the sterile neutrinos,
without the interactions with the active neutrinos
studied in Refs.~\cite{Beacom:2004yd,Hannestad:2005ex}
and references therein.
}
$\phi$.
The lighter sterile neutrino $\nu_{s'}$ must have very
small mixing with the active neutrinos,
in order to forbid its thermalization in the early Universe and
to preserve the effectiveness of the standard three-neutrino
mixing paradigm
for the explanation of solar and atmospheric neutrino oscillations.
Also the very light invisible boson $\phi$
has a negligible thermal distribution before the decay,
because it belongs to the sterile sector which
may have been in equilibrium at very early times,
but has decoupled from the thermal plasma at a very high temperature.
In this way the densities of all the particles belonging to the sterile sector
have been washed out in the following phase transitions
and heavy particle-antiparticle annihilations
(see, for example, Ref.~\cite{Dolgov:2002wy}).
Another possible decay which does not need the presence
of a light boson is
$\nu_{s} \to \nu_{s'} \bar\nu_{s'} \nu_{s'}$,
which needs an effective four-fermion interaction of sterile neutrinos.

In the invisible decay scenario,
the eV-scale sterile neutrino can be fully thermalized
in the early Universe
through active-sterile oscillations
\cite{Dolgov:2003sg,Cirelli:2004cz,Chu:2006ua,Hannestad:2012ky}
and generates $\DNeff = 1$.
In the first radiation-dominated part of the evolution
of the Universe the mass of the sterile neutrino is not important,
because it is relativistic and it contributes only as radiation.
The mass effect is important in the following matter-dominated
evolution of the Universe,
which leads to the formation of Large Scale Structures (LSS)
and the current matter density.
The sterile neutrinos which decay into invisible relativistic particles
before becoming non-relativistic
do not contribute to the matter budget.
In this way the eV-scale mass of the sterile neutrino
indicated by short-baseline oscillation experiments
becomes compatible with a full thermalization
of the sterile neutrino in the early Universe.

We analyzed the same cosmological data considered
in the previous Section (see Ref.~\cite{Archidiacono:2014apa})
and we
modified the Boltzmann solver \camb\ \cite{Lewis:1999bs}
in order to take into account the invisible decay
of the sterile neutrino.
For simplicity%
\footnote{
A precise calculation requires the solution of the
coupled Boltzmann equations describing the evolution
of the distributions of the sterile neutrino and the decay products.
This is beyond the scope of this calculation.
},
we neglected the energy dependence of the sterile neutrino lifetime
and we considered a sterile neutrino with a Fermi-Dirac distribution 
multiplied by
\begin{equation}
N_{s}(t)
=
\DNeff \, e^{-t/\tau_{s}}
\,,
\label{eq:decay_Ns}
\end{equation}
where $t$ is the cosmic time and
$\tau_{s}$ is the effective lifetime of the sterile neutrino.
We neglect also the energy distributions
of the very light or massless invisible decay products
(which depend on the specific decay model)
and we parameterize their effect with an effective increase
of the amount of radiation by
$\DNeff \left( 1 - e^{-t/\tau_{s}} \right)$.
Following
the analyses of the previous Section,
we take into account the SBL constraint on $m_s$
through a prior given by the posterior of the global analysis of
SBL oscillation data presented in Ref.~\cite{Giunti:2013aea}.

Here we present the same analyses performed in
Ref.~\cite{Gariazzo:2014pja},
but with different results for the complete dataset.
In fact, we improved the numerical calculations and
we fixed an error in the code that affected only the analyses
including the CFHTLenS and PSZ datasets.
Since the most recent cosmological data disfavor $\DNeff=1$,
however, the final conclusions will be the same.

\subsection{Results}

Figure~\ref{fig:decay_cmb}
shows the $1\sigma$ and $2\sigma$ marginalized allowed regions
in the planes
($m_s$--$\DNeff$)
and
($H_0$--$\DNeff$)
obtained by fitting the CMB data
(Planck+WP+high-$\ell$+BICEP2(9bins);
see Ref.~\cite{Archidiacono:2014apa})
with the SBL prior
in a model with free $\DNeff$
and a massive sterile neutrino which decays invisibly.
The corresponding numerical values of the cosmological parameters
are listed in Tab.~\ref{tab:decay_all}.
% From the values of
% $\Delta\chi^2(\text{A})$ and $\Delta\chi^2(\text{B})$
% one can see that the fit of cosmological data is even slightly better than that
% obtained with a stable sterile neutrino without mass constraints (A)
% and much better than that
% obtained with a stable sterile neutrino with the SBL mass prior (B).

\begin{table}
\begin{center}
\renewcommand{\arraystretch}{1.2}
\begin{tabular}{|l|c|c|}
\hline
\hline
Parameters
&
CMB+SBL
&
{
\renewcommand{\arraystretch}{1}
\begin{tabular}{c}
CMB+SBL
\\
+LSS+$H_0$
\\
+CFHTLenS+PSZ
\end{tabular}
}
\\
\hline

$\Omega_{\mathrm b} h^2$   & $0.02276\,^{+0.00043}_{-0.00041}\,^{+0.00084}_{-0.00088}$ & $0.02256\,^{+0.00046}_{-0.00042}\,^{+0.00070}_{-0.00088}$ \\
$\Omega_{\mathrm c} h^2$ & $0.132\,^{+0.007}_{-0.008}\,^{+0.014}_{-0.014}$           & $0.116\,^{+0.003}_{-0.003}\,^{+0.006}_{-0.005}$           \\
$\theta_{\mathrm s}$       & $1.0405\,^{+0.0007}_{-0.0007}\,^{+0.0014}_{-0.0015}$      & $1.0416\,^{+0.0006}_{-0.0006}\,^{+0.0013}_{-0.0012}$      \\
$\tau$                     & $0.101\,^{+0.015}_{-0.016}\,^{+0.034}_{-0.027}$           & $0.080\,^{+0.012}_{-0.012}\,^{+0.024}_{-0.023}$           \\
$n_{\mathrm s}$            & $1.006\,^{+0.018}_{-0.019}\,^{+0.037}_{-0.035}$           & $0.988\,^{+0.011}_{-0.011}\,^{+0.021}_{-0.021}$           \\
$\log(10^{10} A_s)$        & $3.123\,^{+0.045}_{-0.045}\,^{+0.086}_{-0.094}$           & $3.094\,^{+0.033}_{-0.038}\,^{+0.084}_{-0.068}$           \\
$r$                        & $0.193\,^{+0.045}_{-0.053}\,^{+0.111}_{-0.091}$           & $0.202\,^{+0.043}_{-0.048}\,^{+0.099}_{-0.087}$           \\ \hline
$\DNeff$                   & $1.06\,^{+0.46}_{-0.45}\,^{+0.88}_{-0.91}$                & $0.30\,^{+0.16}_{-0.23};\,<{0.69}$                         \\
$m_s [\mathrm{eV}]$        & $1.27\,^{+0.11}_{-0.15}\,^{+0.17}_{-0.23}$                & $1.26\,^{+0.10}_{-0.16}\,^{+0.17}_{-0.27}$                \\
% $\log_{10}(\tau_s)$        & $-4.6\, (2\sigma)$                                        & $<-4.9$; unconstrained                                    \\ \hline
% $\Delta\chi^2 (\text{A})$  & $-11.5$                                                   & $-0.2$		\\
% $\Delta\chi^2 (\text{B})$  & $-12.7$                                                   & $-2.1$		\\
\hline
\end{tabular}
\end{center}
\caption[Marginalized constraints on the cosmological parameters
obtained in the model with a decaying sterile neutrino]
{\label{tab:decay_all}
Marginalized $1\sigma$ and $2\sigma$ confidence level limits
for the cosmological parameters
obtained with the invisible sterile neutrino decays.
% The decay lifetime $\tau_s$ is given in units of the age
% of the Universe $T_0$.
See Fig.~\ref{fig:decay_tau_post} for the constraints
on the decay lifetime $\tau_s$.
% $\Delta\chi^2(\text{A})$ and $\Delta\chi^2(\text{B})$
% give the variation of the cosmological
% $\chi^2$ with respect to the fit of cosmological data without (A) and with (B) the SBL prior
% for the mass of a stable sterile neutrino
% (corresponding respectively to the grey and red regions and lines in the figures).
}
\end{table}

\begin{figure}[t]
\centering
\includegraphics[width=\singlefigland,page=5]{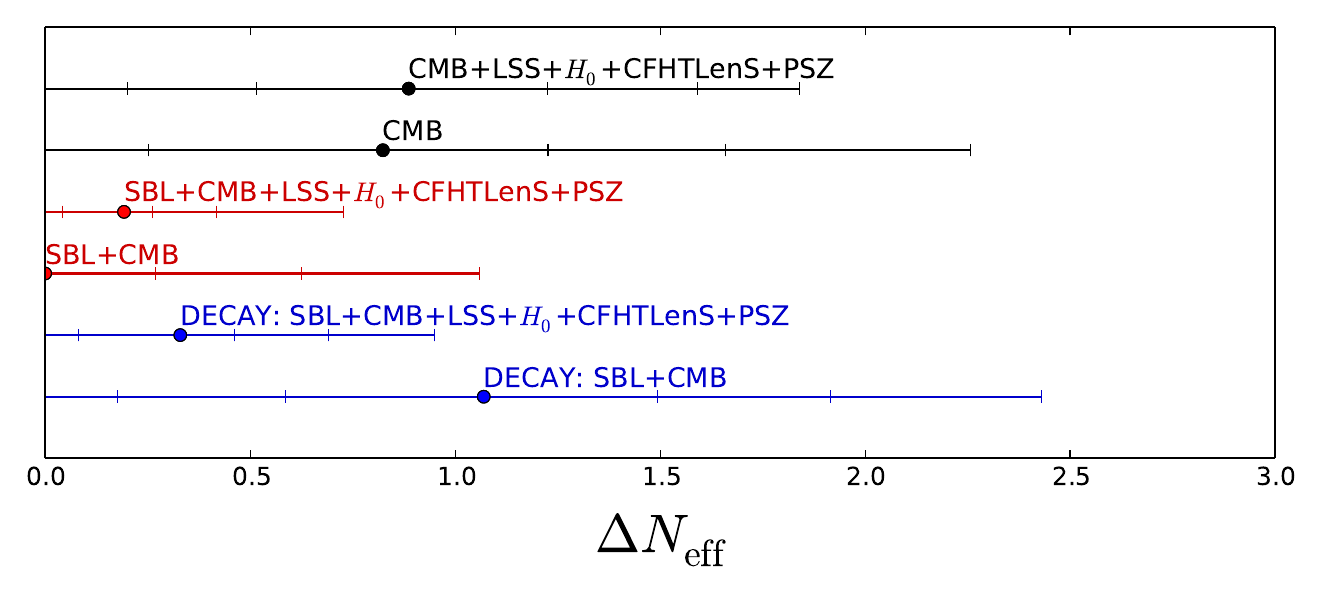}
\includegraphics[width=\singlefigland,page=6]{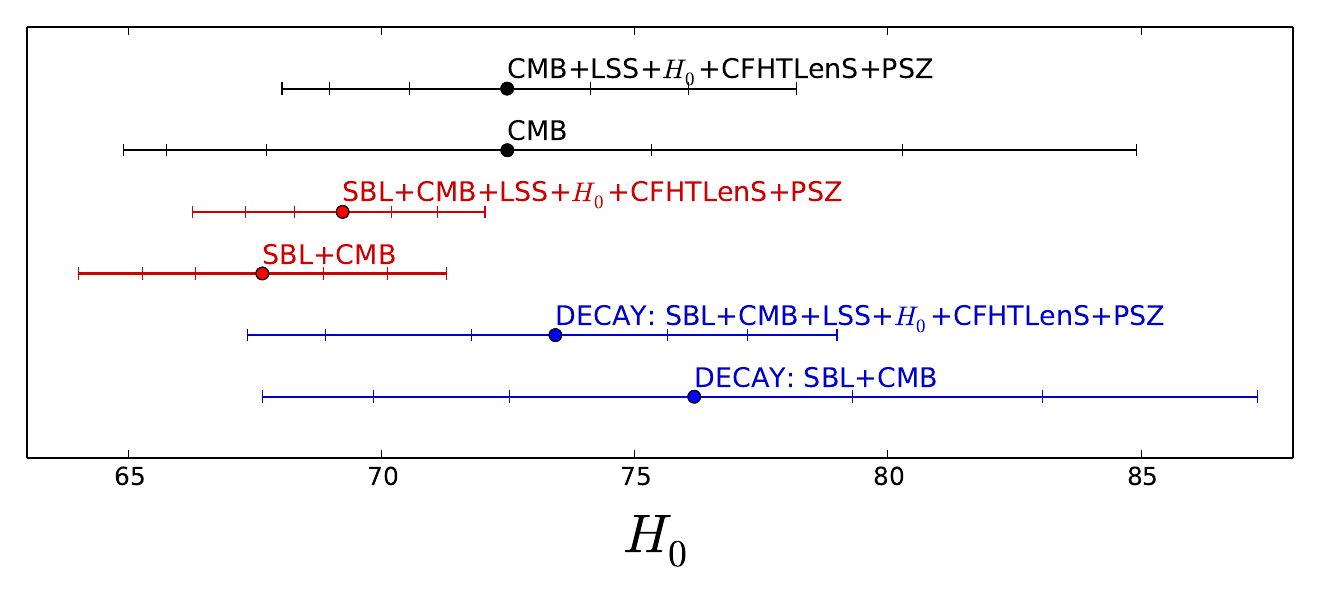}
\caption[Marginalized constraints on \DNeff\ and $H_0$
from the decaying sterile neutrino model]{\label{fig:decay_int}
$1\sigma$, $2\sigma$ and $3\sigma$ marginalized error bars
for $\DNeff$ and $H_0$
obtained in the different fits of the cosmological data considered in
Figs.~\ref{fig:decay_cmb} and \ref{fig:decay_all}.
The circles indicate the marginalized best fit values.
The black and red intervals are taken from
the results in Section~\ref{sec:bicep}.
The blue intervals are obtained by adding the invisible sterile
neutrino decay.
}
\end{figure}

In Fig.~\ref{fig:decay_cmb}
we compared the allowed regions obtained
with the invisible decay of the sterile neutrino
with the corresponding regions
shown in Fig.~\ref{fig:bicep_ns}
for a stable sterile neutrino, without and with the SBL prior.
One can see that the invisible decay of the sterile neutrino allows
$\DNeff = 1$,
which corresponds to the full initial thermalization
of the sterile neutrino,
even if the SBL prior forces the sterile neutrino mass
to assume values around 1.2 eV.
In practice,
the invisible decay of the sterile neutrino
allows us to relax the upper bound of about 0.6 for
$\DNeff$
presented in Tab.~\ref{tab:bicep_sblresults}
with the SBL prior
and bring the allowed range of $\DNeff$
at a level which is similar to that presented in
Tab.~\ref{tab:bicep_cosmoresults}
without the SBL prior
(see also
\cite{Giusarma:2014zza,Zhang:2014dxk,Dvorkin:2014lea,Zhang:2014nta}).
This can also be seen
in the upper panel of Fig.~\ref{fig:decay_int},
which shows the marginalized allowed interval of $\DNeff$.

Figure~\ref{fig:decay_cmb} shows also that by allowing
the sterile neutrino to decay
one can recover a correlation between
$\DNeff$
and
$H_0$
which is similar to that obtained in the analysis
of CMB data without the SBL prior.
Hence,
we obtain that large values of $\DNeff$
are correlated to large values of the Hubble constant $H_0$,
which are in agreement with the local measurements of $H_0$
(see e.g.\ Refs.~\cite{Ade:2013zuv,Gariazzo:2013gua}).

\begin{figure}[t]
\centering
\includegraphics[width=\halfwidth,page=3]{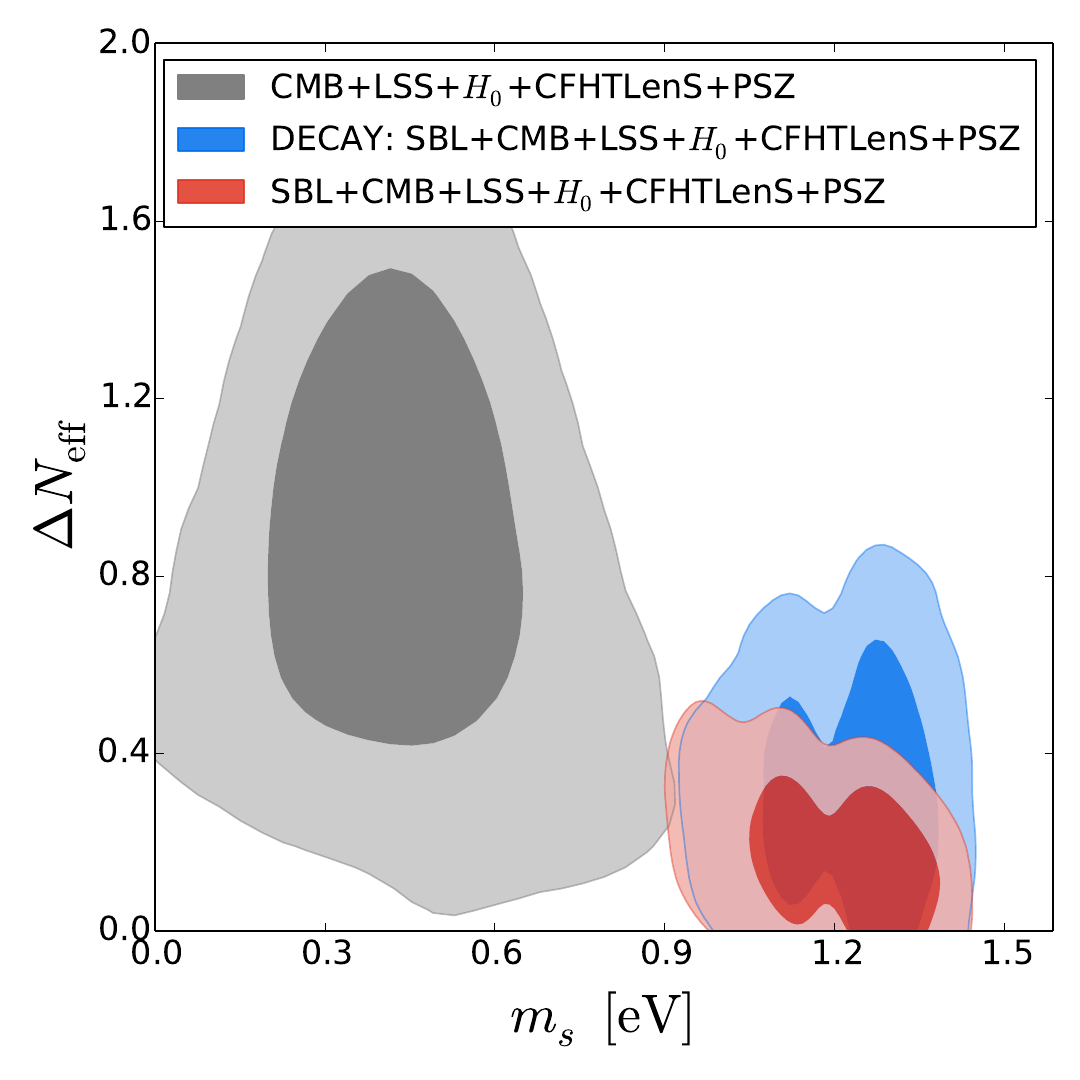}
\includegraphics[width=\halfwidth,page=4]{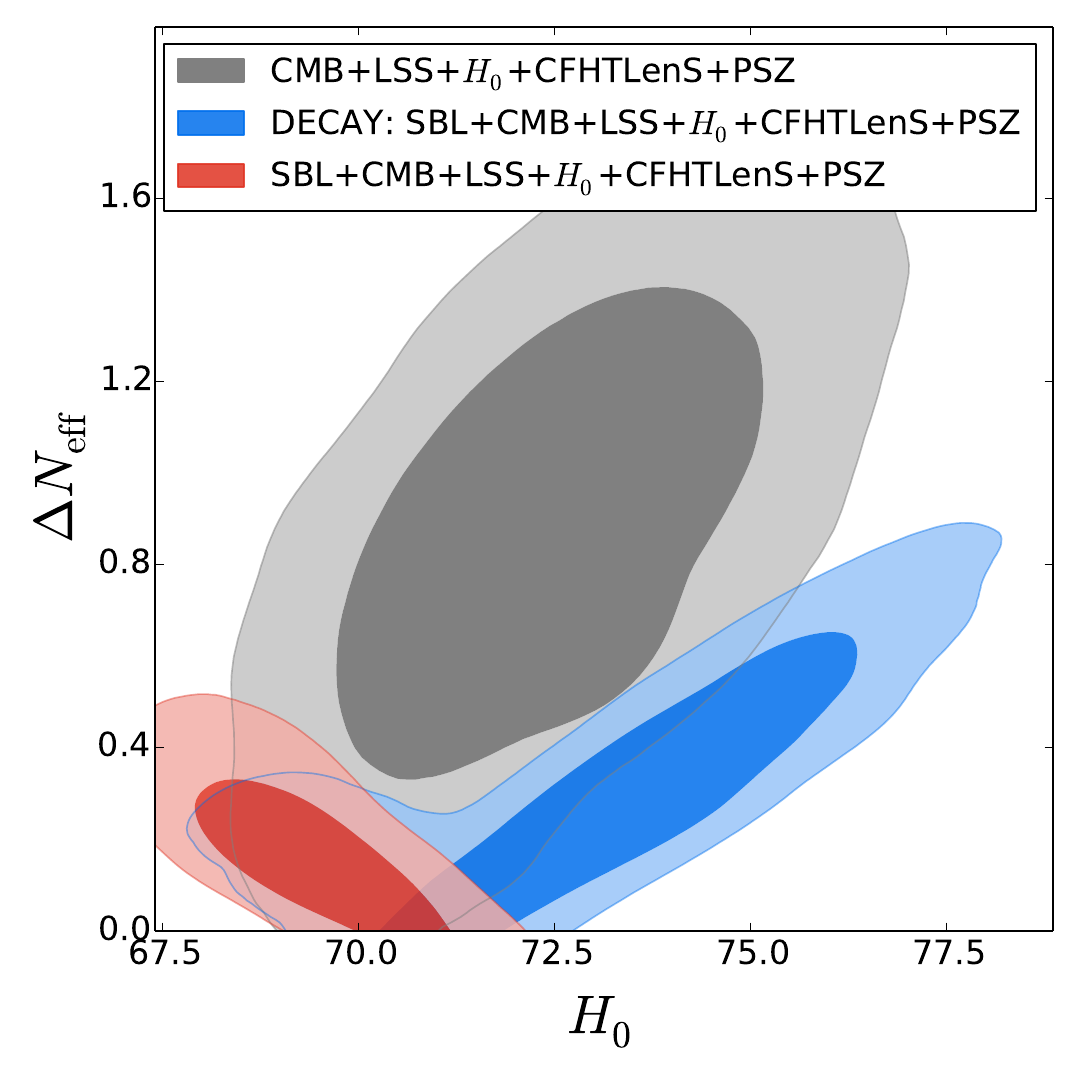}
\caption[Marginalized 2D constraints in the
($m_s$--\DNeff) and ($H_0$--\DNeff)
planes from the decaying sterile neutrino model, 
considering the full dataset]{\label{fig:decay_all}
As in Fig.~\ref{fig:decay_cmb}, but for the complete dataset
(the same CMB data, plus LSS+$H_0$+CFHTLenS+PSZ).
}
\end{figure}

Figure~\ref{fig:decay_all}
shows the $1\sigma$ and $2\sigma$ marginalized allowed regions
corresponding to those of Fig.~\ref{fig:decay_cmb}
and obtained by adding the same cosmological data considered in
the previous Section:
Large Scale Structures (LSS),
local $H_0$ measurements,
cosmic shear (CFHTLenS)
and
Sunyaev-Zel'dovich cluster counts from Planck (PSZ).
One can see that this wide data set allows
more freedom for $\DNeff$,
but the value $\DNeff=1$ is still excluded by the $3\sigma$ limits
(see also Fig.~\ref{fig:decay_int}).
This is a consequence of the fact that the CFHTLenS and PSZ datasets
require that the massive neutrino free-streams at late times
to explain the smaller matter fluctuations
that has been observed in the local Universe.
Clearly, this restricts the possibilities for the neutrino decay.
As we can see in Fig.~\ref{fig:decay_cmb},
$\DNeff$
and
$H_0$
are partially correlated,
indicating relatively large values of $H_0$
for $\DNeff\gtrsim0.5$,
which are in agreement with the local measurements of $H_0$.

\begin{figure}[t]
\centering
\includegraphics[width=\halfwidth,page=7]{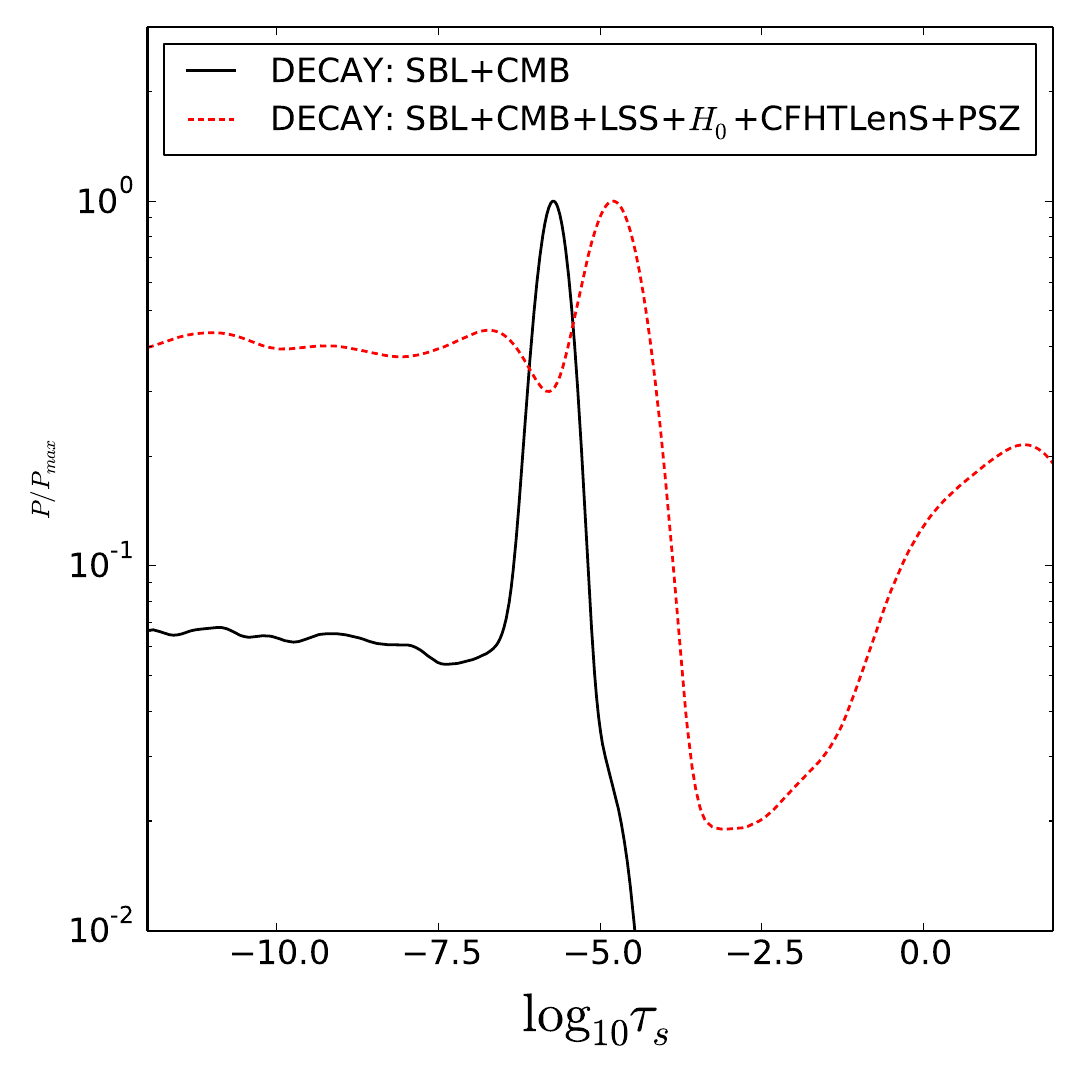}
\caption[Marginalized posterior distribution for $\log_{10}\tau_s$]%
{\label{fig:decay_tau_post}
Marginalized posterior distributions for $\log_{10}\tau_s$.
The decay lifetime $\tau_s$ is given in units of the age
of the Universe $T_0$.
}
\end{figure}

The bounds on the decay lifetime $\tau_s$ are not shown in
Tab.~\ref{tab:decay_all}.
The reason is that the marginalized posterior distribution
of $\tau_s$ is rather complicated and it is not simple to define
a constraint or an upper limit in this case.
The marginalized posterior distributions obtained for $\tau_s$
with the two data combinations are plotted in
Fig.~\ref{fig:decay_tau_post}.
Let we start discussing the one obtained from the CMB data only.
We can see that the curve presents a peak corresponding to 
$\log_{10}\tau_s\simeq-6$.
Since we measure $\tau_s$ in units of the age of the Universe $T_0$,
this means that the most likely value for the decay lifetime is
$\tau_s\simeq10^{-6} T_0$, or approximately $10^4$~years:
it corresponds to a massive sterile neutrino that decays approximately
at the time of its transition to the non-relativistic regime.
As a consequence, its mass has an impact on the Universe evolution 
only for a brief period.
All the values $\log_{10}\tau_s\lesssim-7$ are equally feasible,
since if the sterile neutrino decays when it is completely relativistic
its mass never affects the evolution,
and it gives the same contribution of a massless neutrino.
For this reason, a marginalized constraint on $\tau_s$ would depend on
the lower limit adopted for the prior on $\tau_s$.
As a conclusion, the CMB data requires that the sterile neutrino mass
affects only a short phase of the Universe evolution, approximately
at the time of the sterile neutrino transition to the non-relativistic
regime.

As we already mentioned,
the situation is different if the complete dataset is considered,
because the CFHTLenS and the PSZ data would prefer
a massive sterile neutrino at late times,
in order to have the suppression in the matter fluctuations that would
reconcile the cosmological and the local estimates of $\sigma_8$.
If the sterile neutrino decays in the early Universe,
it cannot free-stream at late times and the
matter fluctuations are not suppressed.
This is a reason for which we see a sort of bimodal distribution
in the posterior of $\tau_s$ obtained from the analysis of the
complete dataset (red curve in Fig.~\ref{fig:decay_tau_post}):
the shape for $\log_{10}\tau_s\lesssim-3$ is similar to the one
obtained from the CMB only dataset, with a small shift towards
higher values of $\tau_s$, but the posterior is enhanced for
$\log_{10}\tau_s\gtrsim-2$ by the phenomenology related to the
CFHTLenS and PSZ datasets.
We can conclude that the bounds for the decaying sterile neutrino
are affected by the tension between the cosmological and local
observations, and a clear result cannot be obtained.

\section{Conclusions and Perspectives}
\label{sec:nucosmo_conclusions}
In conclusion,
we have shown that the cosmological and
the SBL data are compatible only
if the light sterile neutrino is not fully thermalized.
Even if the BICEP2 results about the primordial tensor modes were
correct,
a fully thermalized sterile neutrino
with mass of about 1~eV as indicated by
short-baseline neutrino oscillation data
would not be compatible with cosmology.
Since the mixing parameters obtained from oscillation experiments
would allow a full thermalization of the sterile neutrino
in the early Universe
\cite{Dolgov:2003sg,Cirelli:2004cz,Melchiorri:2008gq,
Hannestad:2012ky,Mirizzi:2013gnd,Hannestad:2015tea},
some new mechanism should be found.

The possibility that the sterile neutrino is not stable and decays
in cosmological times \cite{Gariazzo:2014pja}
is not a good solution for two reasons.
Firstly, as we pointed out in the previous Section,
such a decaying sterile neutrino would not help to solve the tension
between CMB data, that would prefer a rapid decay,
and the local determinations of the matter fluctuations from CFHTLenS
and PSZ, that can be explained only with the free-streaming of 
relic particle that becomes non-relativistic during the evolution
of the structures.
Secondly, even if one neglects the local measurements and considers
only the CMB data,
the decay of the sterile neutrino works well only if $\DNeff$
is allowed to be compatible with 1.
The new analyses of the B-mode polarization data showed that
the signal measured by BICEP2
does not come from the presence of significant primordial tensor modes
\cite{Ade:2015tva}, but mainly from dust emission.
Since the correlation between $r$ and the effective number of
relativistic species was responsible of increasing \Neff\ in the
analyses we presented, this is a point against the robustness
of the solution we proposed.
Moreover, the latest Planck data,
in particular about the small scale polarization,
strongly disfavor $\Neff>3.046$ \cite{Ade:2015xua,Gariazzo:2015rra}.
As a consequence, the decay solution cannot work,
since the decay of the sterile neutrino could explain only
the full thermalization of a massive neutrino
given that $\DNeff=1$ is allowed for massless species.
If $\DNeff=1$ for massless species is disfavored by cosmology,
the decay would not solve the problems and a new solution must be found.

In the past several authors proposed new mechanisms
that can relieve the tension:
among the others, we list
a large lepton asymmetry
\cite{Foot:1995bm,Foot:1996qc,Bell:1998sr,Bell:1998ds,Shi:1999kg,DiBari:1999vg,
DiBari:2001jk,Chu:2006ua,Hannestad:2012ky,Mirizzi:2012we,Saviano:2013ktj,
Hannestad:2013wwj},
new neutrino interactions \cite{Bento:2001xi,Dasgupta:2013zpn,
Hannestad:2013ana,Bringmann:2013vra,Ko:2014bka,Archidiacono:2014nda,
Saviano:2014esa,Mirizzi:2014ama,Tang:2014yla,Chu:2015ipa},
entropy production after neutrino decoupling \cite{Ho:2012br},
a very low reheating temperature \cite{Gelmini:2004ah,Smirnov:2006bu},
time varying dark energy components \cite{Giusarma:2011zq},
a larger cosmic expansion rate at the time of sterile neutrino production
\cite{Rehagen:2014vna}.

In the next Chapters we will present two different mechanisms
that could alleviate the tensions we discussed.
In Chapter~\ref{ch:pps_nu} we discuss the possibility that the effects
on the cosmological observables
due to the presence of a sterile neutrino with mass around 1~eV
are compensated by an additional freedom in the Primordial Power Spectrum (PPS)
of scalar perturbations generated during inflation.
If the PPS can have a shape more complicated than a power-law
(Eq.~\eqref{eq:plPPS}),
a local modification of the initial amplitude of the scalar fluctuations may
cancel the Silk damping effect driven by the high $\Neff$
given by the 3+1 neutrino states (see Section~\ref{sec:nucosmology}).

A completely different possibility that could give an explanation to the $H_0$
and the $\sigma_8$ tensions is discussed in Chapter~\ref{ch:cde}.
We present a model for a phenomenological coupling
between Dark Matter (DM) and Dark Energy (DE).
If there is an energy transfer from DE to DM, the prediction
from the cosmological model gives a smaller $\sigma_8$ and an higher $H_0$,
so that local and CMB estimates for these parameters are reconciled.
This model, however, does not improve the compatibility
between cosmological and SBL data on the presence
of a massive sterile neutrino, whose presence is still disfavored.

%!TeX root=main.tex 
\chapter{Inflationary Freedom and Neutrino Properties}
\label{ch:pps_nu}
\chapterprecis{This Chapter is based on
Refs.~\protect\cite{Gariazzo:2014dla,DiValentino:2016ikp}.}

In this Chapter we discuss how the constraints on the neutrino
properties obtained using the CMB data and several other cosmological
data can be biased by the assumption on the
Primordial Power Spectrum (PPS) of scalar perturbations.
If the PPS presents deviations with respect
to the standard power-law (PL),
as some inflationary models predict,
it is possible to obtain misleading results from the cosmological
analyses.

\section{Motivations for Inflationary Freedom}
We discussed in Chapter~\ref{ch:cosmology} that
Inflation is one of the most successful theories
that explains the ``horizon problem'' and the ``flatness problem''.
Moreover, inflation gives origin
to the primordial density perturbations
that evolved to form the structures we observe today,
that we calculated in Subsection~\ref{sec:initial_conditions}.
The standard inflationary paradigm predicts a simple shape
for the PPS of scalar perturbations:
in this context, the PPS is scale independent and 
it can be described by the power-law expression
in Eq.~\eqref{eq:plPPS}.
Different models that give an inflationary phase
in the early Universe, however, 
can originate more complicated PPS shapes,
with possible features or different behaviors at different scales
(see e.g.~Refs.~\cite{Romano:2014kla,Kitazawa:2014dya}
or the reviews \cite{Martin:2013tda,Chluba:2015bqa}).
It is currently impossible to test directly the physics
at the scale of cosmological inflation and consequently
it is impossible
to check the correctness of the simplest inflationary models.
If the theoretical models are wrong or incomplete,
any cosmological analysis performed assuming
a power-law PPS can lead to biased constraints.
The only possibility we have to test the inflationary predictions
for the PPS is to reconstruct an unknown PPS starting from the
physical observables that we can measure, for example
the CMB spectrum.

If one tries to constrain the PPS
under the assumption of the \lcdm~model
a non-standard behavior can be found.
Firstly, it is necessary
to assume a model for the evolution of the Universe
and to calculate the transfer function.
The physics of the transfer function, introduced in the previous
Chapters, is well understood.
We mentioned also that the CMB anisotropies can be described
very well with a small number of parameters,
that are robustly determined
from the analyses of the latest experimental data
from Planck \cite{Adam:2015rua}.
Few cosmological parameters that are very well known, indeed,
are sufficient to calculate the transfer function.
This can be used to calculate the theoretical predictions
for the CMB spectra using a completely unknown PPS,
and then a comparison with the measured power spectra
allows to put constraints on the unknown PPS.
This process can be deployed using several methods 
that were developed in the past:
for example we can find
regularization methods like
Richardson-Lucy iteration
\cite{Shafieloo:2003gf,Nicholson:2009pi,Hazra:2013ugu,Hazra:2014jwa},
truncated singular value decomposition \cite{Nicholson:2009zj}
and Tikhonov regularization \cite{Hunt:2013bha,Hunt:2015iua},
or methods like the
maximum entropy deconvolution \cite{Goswami:2013uja}
or the
``cosmic inversion'' methods
\cite{Matsumiya:2001xj,Matsumiya:2002tx,Kogo:2003yb,
Kogo:2005qi,Nagata:2008tk}.
In the 2015 release of scientific results,
the Planck collaboration presented a wide discussion 
about inflation and the constraints on the PPS, in Ref.~\cite{Ade:2015lrj}.
All these analyses suggest that the PPS may deviate from the
expected power-law behavior, especially in the region at small wavemodes:
the statistical significances of the deviations are small in some cases,
but it is interesting to note that both 
the CMB power spectra as measured by WMAP \cite{Bennett:2012zja}
and by Planck \cite{Ade:2013sjv, Adam:2015rua} show similar results.
The main source of the difference between the reconstructed PPS
and the power-law is in the region at low multipoles,
where the cosmic variance is larger.
These deviations could be the result of simple statistical fluctuations or 
be the result of a non-standard inflationary mechanism.

The effects that considering a non-standard PPS has
on cosmological parameter estimation have been studied by several authors.
for example, the power-law PPS has been simply modified
with the introduction of
a running in the tilt of the power-law
\cite{Archidiacono:2012gv,Zhao:2012xw,Vazquez:2013dva,
Abazajian:2014tqa},
a running of the running \cite{Cheng:2014bta},
or
a sharp cut-off in the power-law \cite{Abazajian:2014tqa}.
Our main goal is to study how the freedom of the form of the PPS
can affect the existing bounds on different neutrino properties
in the early Universe, such as those on the
sum of the active neutrino masses \summnu,
on the effective number of relativistic species \Neff\
and on the properties of a light sterile neutrino.

Previous analyses of the cosmological data
with a standard power-law PPS have found that
a fully thermalized sterile neutrino is quite disfavored
(see Chapter~\ref{ch:lsn_cosmo} and Refs.~%
\cite{DiValentino:2013qma,Archidiacono:2013xxa,Mirizzi:2013gnd,
Gariazzo:2013gua,Archidiacono:2014apa}).
These results motivated the study of mechanisms which can suppress the
thermalization of sterile neutrinos in the early Universe,
that would be due to active-sterile oscillations before neutrino decoupling
\cite{Dolgov:2003sg,Cirelli:2004cz}.
Examples are a large lepton asymmetry
\cite{Chu:2006ua,Hannestad:2012ky,Mirizzi:2012we,
Saviano:2013ktj,Hannestad:2013wwj},
an enhanced background potential due to new interactions in the sterile sector
\cite{Hannestad:2013ana,Dasgupta:2013zpn,Bringmann:2013vra,
Ko:2014bka,Archidiacono:2014nda,Saviano:2014esa,Mirizzi:2014ama},
a larger cosmic expansion rate at the time of sterile neutrino production
\cite{Rehagen:2014vna},
and
MeV dark matter annihilation
\cite{Ho:2012br}.
We will show in this Chapter that a further possibility consists in the fact
that a free PPS can partially compensate the effects of a light sterile
neutrino on the cosmological observables.

Besides our main objective,
which is to study the robustness of neutrino constraints when
the PPS of scalar perturbations is free to vary,
we are also interested in obtaining information
on the form of the PPS.
With these aims,
we considered a general form of the PPS that allows
the presence of features without forcing a particular shape.
In the literature several model-independent parameterizations
have been used:
for example,
a free PPS can be described with
wavelets \cite{Mukherjee:2003cz,Mukherjee:2003yx,Mukherjee:2003ag,
Mukherjee:2005dc},
principal components \cite{Leach:2005av},
top-hat bins without interpolation \cite{Wang:1998gb},
power-law bins \cite{Hannestad:2000pm,Hazra:2013nca},
linear interpolation
\cite{Wang:2000js,Bridle:2003sa,Hannestad:2003zs,Bridges:2005br,
Spergel:2006hy,Bridges:2006zm,Bridges:2008ta,Vazquez:2013dva},
broken power-law \cite{Hazra:2013nca,Hazra:2014aea},
and
interpolating spline functions
\cite{Sealfon:2005em,Verde:2008zza,Peiris:2009wp,Hlozek:2011pc,
Gauthier:2012aq,dePutter:2014hza,Abazajian:2014tqa,Hu:2014aua}.
We decided to follow part of the prescriptions of the interpolating
spline form presented in Refs.~\cite{Hlozek:2011pc,dePutter:2014hza},
improving the parametrization by using a
``piecewise cubic Hermite interpolating polynomial''
(\pchip),
which is described in Section~\ref{sec:inflfreed_ppsparam}.
This method
allows to avoid the spurious oscillating behavior that can appear
between the nodes of the interpolating splines.

\section{Primordial Power Spectrum Parameterization}
\label{sec:inflfreed_ppsparam}
% JCAP2015... \pchip\ PPS
We adopt a non-parametric description
for the PPS of scalar perturbations:
we describe the function $P_s(k)$
as the interpolation among a series of nodes at fixed wavemodes $k$.
We consider twelve nodes $k_j$ ($j\in[1,12]$)
that cover a wide range of values of $k$:
\begin{align}
k_1     &= 5\e{-6} \mpcinv , \nonumber\\
k_2     &= 10^{-3} \mpcinv , \nonumber\\
k_j     &= k_2 (k_{11}/k_2)^{(j-2)/9} \quad \text{for} \quad j\in[3,10] , \nonumber\\
k_{11}  &= 0.35 \mpcinv , \nonumber\\
k_{12}  &= 10\mpcinv .
\label{eq:nodesspacing}
\end{align}
The most interesting range is located between
$k_2=0.001\mpcinv$ and $k_{11}=0.35\mpcinv$, 
that is approximately the range of wavemodes probed
by CMB experiments. 
In this range we use equally spaced nodes in $\log k$.
Additionally, we consider $k_1=5\e{-6}\mpcinv$ and 
$k_{12}=10\mpcinv$ 
in order to be sure that all the PPS evaluations are
inside the covered range:
we expect that the nodes at these extreme wavemodes
are less constrained by the data.

Having fixed the position of all the nodes,
the free parameters that enter our MCMC analyses
are the values of the PPS at each node, $\psj{j}=P_s(k_j)/P_0$, 
where $P_0$ is the overall normalization.
We use $P_0=2.36\e{-9}$ \cite{Larson:2010gs}
in Section~\ref{sec:inflfreed_sterile} and 
$P_0=2.2\e{-9}$ \cite{Ade:2015xua} in the following
Sections.
Each parameter $\psj{j}$,
whose expected value should be close to 1,
is free to vary in the interval $[0.01,10]$,
on which we adopt a flat prior.

The complete $P_s(k)$ at all $k$ is then described as
the interpolation among the points $\psj{j}$:
\begin{equation}\label{eq:PPS_pchip}
  P_{s}(k)=P_0\times\pchip(k; \psj{1}, \ldots, \psj{12})\,,
\end{equation}
where \pchip\ stands for
``piecewise cubic Hermite interpolating
polynomial''~\cite{Fritsch:1980,Fritsch:1984}.
This function is similar to the natural cubic spline,
but it has the advantage of avoiding the introduction
of spurious oscillations in the interpolation:
this is obtained with a condition
on the first derivative in the nodes,
that is null if there is a change in the monotonicity
of the point series.
If the monotonicity does not change in the node $\psj{j}$,
the derivative is instead fixed using the secants 
between $P_{s,j-1}$, $\psj{j}$ and $\psj{j+1}$.
The price to pay to preserve the original monotonicity
of the nodes series is on the second derivative,
that becomes discontinue in the nodes, differently
from what happens for the natural cubic spline.
A more detailed discussion on the \pchip~PPS description
can be found in Appendix~\ref{ch:app_pchip}.

When presenting our results,
we will compare the constraints obtained in the context of the 
standard \lcdm~model with a standard power-law PPS
and those obtained with the free \pchip~PPS.
In the former case the cosmological model is described
by the six parameters
described in Section~\ref{sec:par_depend}
($\Omega_b h^2$, $\Omega_c h^2$, $\theta$, $\tau$, $A_s$, $n_s$), 
while in the latter case we substitute $A_s$ and $n_s$
with the parameters used to describe the \pchip~PPS,
$\psj{j}$ ($j\in[1,12]$)
and we have a model with 16 free parameters
($\Omega_b h^2$, $\Omega_c h^2$, $\theta$, $\tau$,
$\psj{1},\ldots, \psj{12}$).
These models will be extended to study the properties of neutrinos
or other aspects of the cosmological model.

When comparing the PL and the \pchip\ PPS scenarios,
it is convenient to write the values of the \pchip\ nodes
that correspond to the values of the PL PPS
at the corresponding wavemodes,
given the reference values
$n_s^{\textrm{ref}}$ and $A_s^{\textrm{ref}}$.
These can be converted into reference values
to compare the node \psj{i} with:
\begin{equation}
\label{eq:psj_plpps}
  \psj{i}^{\textrm{ref}}
  \equiv
  \frac{A_s^{\textrm{ref}}}{P_0}
  \,
  \left(\frac{k_i}{k_*}\right)^{n_s^{\textrm{ref}}-1}
  \quad\mbox{ with }
  i\in [1,\ldots,12]
  \,.
\end{equation}

\section{An example: Inflationary Freedom and Light Sterile Neutrinos}
\label{sec:inflfreed_sterile}
% JCAP 2015... results
% \cite{Gariazzo:2014dla}
\subsection{Parameterization and Data}
Before studying separately the degeneracies of the various cosmological 
parameters with the free PPS, we show that
significant variations in the results are allowed if the shape
of the PPS is changed.
We will follow Ref.~\cite{Gariazzo:2014dla},
where it is shown that the constraints on the properties
of a light sterile neutrino change significantly when one
analyzes the same set of cosmological data relaxing the hypothesis
of a power-law PPS for the scalar perturbations.

To do this, we adopt the same parameterization for the light sterile
neutrino and for the cosmological model that we used in
Sections~\ref{sec:bicep} and \ref{sec:lsndecay}:
we use an extended flat \lcdm\ model
to accommodate the presence of a sterile neutrino
and we consider a scenario involving inflationary freedom
in the production of the primordial power spectra.
In the analysis with a power-law (PL) PPS we have then
a cosmological model with a total of
eight parameters:
\begin{equation}\label{eq:inflfreed_lsn_modelbase}
{\bm \theta}
=
\{\omega_\mathrm{c},\omega_\mathrm{b},\theta,\tau,
\ln(10^{10}A_{s}),n_{s},m_s,\DNeff
\}.
\end{equation}
In contrast with previous analyses
(see Chapter~\ref{ch:lsn_cosmo}
and
Refs.~\cite{Gariazzo:2013gua,Archidiacono:2014apa,Gariazzo:2014pja}),
we limit the allowed range of
$\DNeff$ in the interval $0\leq\DNeff\leq1$,
assuming that the additional sterile neutrino cannot contribute
to the relativistic energy density more
than a standard active neutrino.
This is what should happen if sterile neutrinos are produced
in the early Universe by
neutrino oscillations before neutrino decoupling
\cite{Dolgov:2003sg,Cirelli:2004cz}.

We assume a flat prior for all the parameters in
Eq.~\eqref{eq:inflfreed_lsn_modelbase},
except $m_s$,
for which we use a flat prior
for $0\leq m_s/\ev\leq 3$ only in the analyses
which do not take into account the constraints
from short-baseline neutrino oscillation data.
In the analyses which take into account these constraints
we use as prior for $m_s$ the posterior obtained
from the analysis of SBL data presented in
Chapter~\ref{ch:nu}.
As in the previous Chapter,
we neglect the masses of the three light neutrinos
$\nu_{1}$,
$\nu_{2}$,
$\nu_{3}$,
which are assumed to be much smaller than 1 eV.

In order to parameterize the free PPS we follow the prescriptions
presented in Section~\ref{sec:inflfreed_ppsparam} with
$P_0=2.36\e{-9}$ \cite{Larson:2010gs}.
In the \pchip PPS analysis we consider a flat
\lcdm+$\nu_s$ cosmological model with a total of
18 parameters:
\begin{equation}\label{eq:inflfreed_lsn_modelmodified}
{\bm \theta} =
\{\omega_{\mathrm c},\omega_{\mathrm b}, \theta,
\tau, m_s, \DNeff,
P_{s,1}, \ldots, P_{s,12}
\},
\end{equation}
where
$\omega_{\mathrm c}$,
$\omega_{\mathrm b}$,
$\theta$,
$\tau$,
$m_s$ and
$\DNeff$ are the same as in the set \eqref{eq:inflfreed_lsn_modelbase}.

In this Section we use the same datasets as in
Sections~\ref{sec:bicep}, \ref{sec:lsndecay} and
Refs.~\cite{Archidiacono:2014apa,Gariazzo:2014pja},
apart from the controversial BICEP2 data
on the B-mode polarization of the CMB
\cite{Ade:2014xna}
that we neglect.
In the following we will denote the analyses
of all the cosmological data
alone (Planck 2013 + ACT/SPT + WMAP polarization
+ LSS + $H_0$ + PSZ + CFHTLenS, see Subsection~\ref{sub:bicep_param})
as ``\textbf{COSMO}'' 
and those which include also the prior on the sterile neutrino mass
from short-baseline neutrino oscillation as
``\textbf{COSMO+SBL}''.

\begin{table}
\begin{center}
\renewcommand{\arraystretch}{1.2}
\begin{tabular}{|l|c|c|}
\hline
Parameters
&
COSMO
&
COSMO+SBL
\\

\hline

$100\,\omega_{\mathrm b}$ 	
			& $2.263^{+0.026}_{-0.027}\,^{+0.052}_{-0.053}\,^{+0.078}_{-0.080}$        & $2.251^{+0.023}_{-0.025}\,^{+0.049}_{-0.045}\,^{+0.075}_{-0.067}$        \\
                             
$\omega_{\mathrm c}$       
			& $0.120^{+0.004}_{-0.005}\pm0.008\,^{+0.011}_{-0.009}$        & $0.117^{+0.002}_{-0.003}\,^{+0.006}_{-0.005}\,^{+0.010}_{-0.006}$        \\
                             
$\theta$             
			& $1.0412\pm0.0007\pm0.0014\,^{+0.0020}_{-0.0021}$ & $1.0416\pm0.0006\pm0.0012\,^{+0.0018}_{-0.0019}$ \\
                             
$\tau$                       
			& $0.087^{+0.013}_{-0.014}\,^{+0.028}_{-0.026}\,^{+0.045}_{-0.037}$        & $0.087\pm0.013\,^{+0.026}_{-0.025}\,^{+0.040}_{-0.035}$        \\
\hline                       
$\DNeff$                        
			& $0.38^{+0.18}_{-0.33}$; No limit; No limit                               & $0.19^{+0.09}_{-0.12}$; $<0.41$; $<0.60$                               \\
                                                                                    
$m_s [\mathrm{eV}]$                                                                     
			& $0.61^{+0.31}_{-0.42}$; $<2.03$; No limit                                & $1.25^{+0.11}_{-0.16}\,^{+0.17}_{-0.29}\,^{+0.22}_{-0.35}$               \\
\hline                             
$n_{\mathrm s}$                  
			& $0.979^{+0.011}_{-0.010}\pm0.020\,^{+0.030}_{-0.025}$        & $0.969\pm0.005\pm0.011\,^{+0.017}_{-0.016}$        \\
                             
$\log(10^{10} A_s)$          
			& $3.152^{+0.031}_{-0.032}\,^{+0.064}_{-0.058}\,^{+0.094}_{-0.087}$        & $3.178^{+0.024}_{-0.025}\,^{+0.048}_{-0.051}\,^{+0.072}_{-0.075}$        \\
\hline
\end{tabular}
\end{center}
\caption[Constraints on the cosmological parameters
obtained with the power-law PPS]{\label{tab:inflfreed_plPPS} 
Marginalized $1\sigma$, $2\sigma$ and $3\sigma$ confidence level limits for the cosmological parameters
obtained with the power-law parametrization for the PPS.
}
\end{table}

\begin{table}
\begin{center}
\renewcommand{\arraystretch}{1.2}
\begin{tabular}{|l|c|c|}
\hline
Parameters
&
COSMO
&
COSMO+SBL
\\

\hline

$100\,\omega_{\mathrm b}$ 	
			& $2.251^{+0.036}_{-0.036}\,^{+0.073}_{-0.072}\,^{+0.111}_{-0.106}$        & $2.247^{+0.036}_{-0.038}\,^{+0.072}_{-0.078}\,^{+0.111}_{-0.117}$        \\
                        
$\omega_{\mathrm c}$  
			& $0.125^{+0.005}_{-0.004}\,^{+0.007}_{-0.012}\,^{+0.008}_{-0.015}$        & $0.118^{+0.004}_{-0.005}\,^{+0.011}_{-0.007}\,^{+0.016}_{-0.008}$        \\
                        
$\theta$        
			& $1.0406^{+0.0007}_{-0.0008}\,^{+0.0016}_{-0.0014}\,^{+0.0026}_{-0.0019}$ & $1.0413^{+0.0008}_{-0.0007}\,^{+0.0014}_{-0.0016}\,^{+0.0020}_{-0.0024}$ \\
                        
$\tau$                  
			& $0.086^{+0.014}_{-0.015}\,^{+0.031}_{-0.028}\,^{+0.052}_{-0.036}$        & $0.090^{+0.014}_{-0.016}\,^{+0.033}_{-0.029}\,^{+0.051}_{-0.039}$        \\
\hline                                                                                                         
$\DNeff$                                                                                                                              
			& $>0.51$; No limit; No limit                                              & $0.25^{+0.13}_{-0.22}$; $<0.75$; No limit                                \\
                                                                                                                                             
$m_s [\mathrm{eV}]$                                                                                                                              
			& $0.63^{+0.23}_{-0.28}\,^{+1.11}_{-0.59}$; No limit                       & $1.22^{+0.13}_{-0.15}\,^{+0.17}_{-0.28}\,^{+0.24}_{-0.33}$               \\
\hline
$P_{s,1}$
			& $<2.51$; $<7.97$; No limit                                               & $<2.75$; $<8.30$; No limit                                               \\

$P_{s,2}$
			& $1.06^{+0.19}_{-0.22}\,^{+0.44}_{-0.35}\,^{+0.70}_{-0.44}$               & $1.05^{+0.18}_{-0.22}\,^{+0.44}_{-0.35}\,^{+0.75}_{-0.44}$               \\

$P_{s,3}$
			& $0.65^{+0.20}_{-0.19}\,^{+0.38}_{-0.37}\,^{+0.57}_{-0.54}$               & $0.67^{+0.20}_{-0.19}\,^{+0.39}_{-0.36}\,^{+0.61}_{-0.52}$               \\

$P_{s,4}$
			& $1.14^{+0.12}_{-0.11}\,^{+0.23}_{-0.22}\,^{+0.36}_{-0.31}$               & $1.13^{+0.11}_{-0.11}\,^{+0.23}_{-0.21}\,^{+0.34}_{-0.31}$               \\

$P_{s,5}$
			& $0.97^{+0.05}_{-0.06}\,^{+0.11}_{-0.10}\,^{+0.18}_{-0.16}$               & $0.98^{+0.05}_{-0.06}\,^{+0.11}_{-0.10}\,^{+0.17}_{-0.15}$               \\

$P_{s,6}$
			& $0.96\pm0.03\,^{+0.07}_{-0.06}\,^{+0.10}_{-0.08}$                        & $0.98\pm0.03\,^{+0.07}_{-0.06}\,^{+0.11}_{-0.08}$               \\
                                                                                                
$P_{s,7}$                                                                                       
			& $0.94\pm0.03\,^{+0.06}_{-0.05}\,^{+0.10}_{-0.08}$                        & $0.94\pm0.03\pm0.06\,^{+0.10}_{-0.07}$               \\
                                                                                                
$P_{s,8}$                                                                                       
			& $0.93\pm0.03\,^{+0.06}_{-0.05}\,^{+0.10}_{-0.07}$                        & $0.93\pm0.03\pm0.06\,^{+0.10}_{-0.07}$               \\
                                                                                                
$P_{s,9}$                                                                                       
			& $0.93\pm0.03\,^{+0.07}_{-0.06}\,^{+0.11}_{-0.08}$                        & $0.91\pm0.03\,^{+0.07}_{-0.06}\,^{+0.10}_{-0.07}$               \\
                                                                                                
$P_{s,10}$                                                                                      
			& $0.91\pm0.04\pm0.08\,^{+0.12}_{-0.11}$                                   & $0.88^{+0.03}_{-0.04}\,^{+0.08}_{-0.07}\,^{+0.14}_{-0.08}$               \\

$P_{s,11}$
			& $1.13^{+0.17}_{-0.16}\,^{+0.28}_{-0.32}\,^{+0.40}_{-0.39}$               & $1.00^{+0.13}_{-0.17}\,^{+0.35}_{-0.24}\,^{+0.52}_{-0.28}$               \\

$P_{s,12}$
			& $<0.69$; $<1.18$; $<1.55$                                                & $<0.49$; $<1.01$; $<1.33$                                                \\
\hline
\end{tabular}
\end{center}
\caption[Constraints on the cosmological parameters
obtained with the \pchip\ PPS]
{\label{tab:inflfreed_freePPS}
Marginalized $1\sigma$, $2\sigma$ and $3\sigma$ confidence level limits for the cosmological parameters
obtained with the \pchip{} parametrization for the PPS.
From Ref.~\cite{Gariazzo:2014dla}.
}
\end{table}

\subsection{Results}
\label{ssec:results_cosmo}

The results of our COSMO and COSMO+SBL analyses are presented in
Tab.~\ref{tab:inflfreed_plPPS}
for the standard case of a power-law PPS
and
in Tab.~\ref{tab:inflfreed_freePPS}
for the free PPS with the \pchip parameterization.
In the upper part of the tables we listed the common parameters
of the \lcdm\ model,
in the central part we listed the neutrino parameters
$\DNeff$ and $m_s$,
while the lower part concerns the parameters used
to parameterize the PPS:
$n_{\mathrm s}$ and $\log(10^{10} A_s)$ for the power-law PPS
and the $P_{s,j}$ nodes for the \pchip PPS.
We do not discuss here the constraints on the PPS parameters,
that will be presented
in the final Section of this Chapter.
Here we discuss firstly the results relative
to the parameters in the upper part of the Tables~\ref{tab:inflfreed_plPPS}
and
\ref{tab:inflfreed_freePPS}
($\omega_{\mathrm b}$,
$\omega_{\mathrm c}$,
$\theta$ and
$\tau$)
and then the results relative to the parameters
in the central part of the tables,
$\DNeff$ and $m_s$.

The bounds on the parameters of the \lcdm\ model change
slightly when more freedom is admitted for the PPS.
Comparing Tabs.~\ref{tab:inflfreed_plPPS}
and \ref{tab:inflfreed_freePPS},
one can see that
the limits
on the parameters of the \lcdm\ model
are slightly weakened in the \pchip PPS case
and for some parameters there is also a small shift
in the marginalized best-fit value.
In all the cases in which this happens,
the marginalized best-fit values move
inside the $1\sigma$ uncertainties.
% We will discuss in the following Sections that this is a consequence
% of the degeneracies between the various cosmological parameters and the
% PPS nodes.
The freedom of the form of the PPS affects the COSMO results
more than the COSMO+SBL results:
in the former case the $\omega_{\mathrm c}$ and $\theta$
best values change by about $1\sigma$,
while a smaller shift is obtained for $100\,\omega_{\mathrm b}$.
On the other hand, in the COSMO+SBL analysis all the shifts
are much smaller than the $1\sigma$ uncertainties,
since the degeneracies between $m_s$
and the other parameters are less significant,
because the allowed range for $m_s$ is smaller.

\begin{figure}
\centering
\includegraphics[width=\halfwidth]{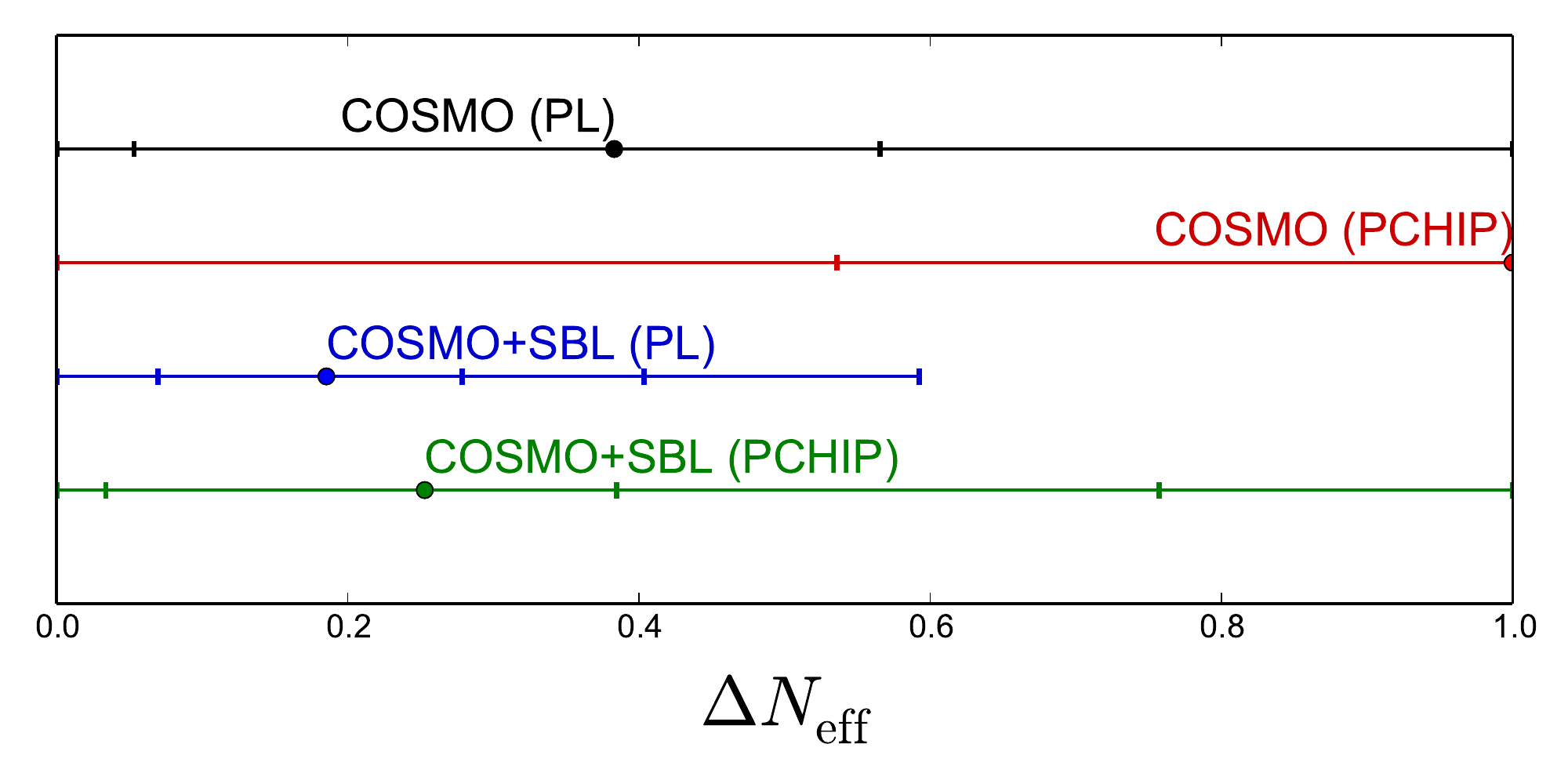}
\includegraphics[width=\halfwidth]{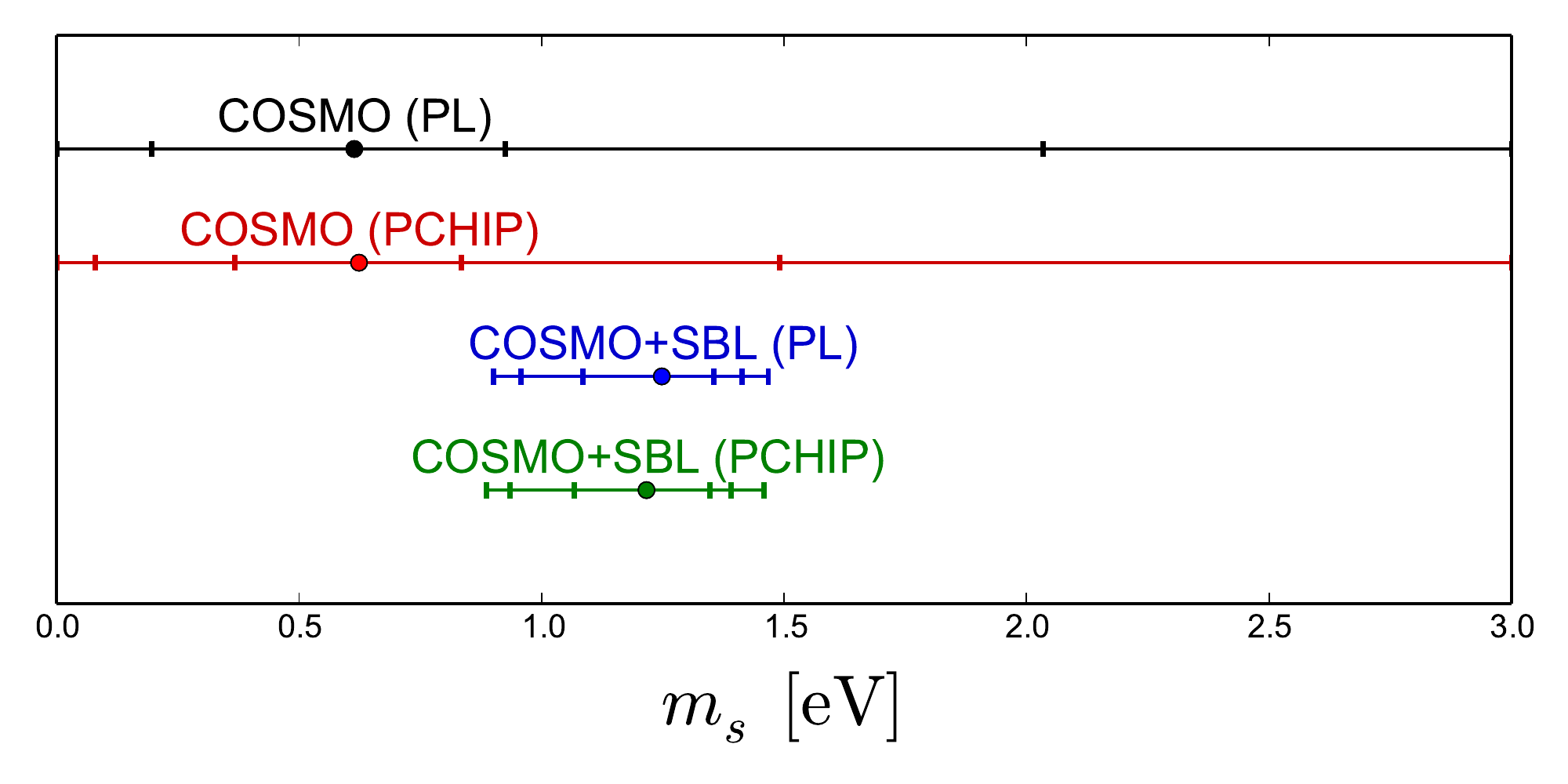}
\caption[Constraints on $\DNeff$ and $m_s$ with different
assumptions on the PPS]
{\label{fig:inflfreed_errorbars}
$1\sigma$, $2\sigma$ and $3\sigma$ marginalized intervals for $\DNeff$ and $m_s$
obtained in the different analyses discussed in the text
(considering $0\leq\DNeff\leq1$ and $0\leq m_s/\ev\leq 3$).
From Ref.~\cite{Gariazzo:2014dla}.
}
\end{figure}

\begin{figure}
\centering
\includegraphics[width=\halfwidth]{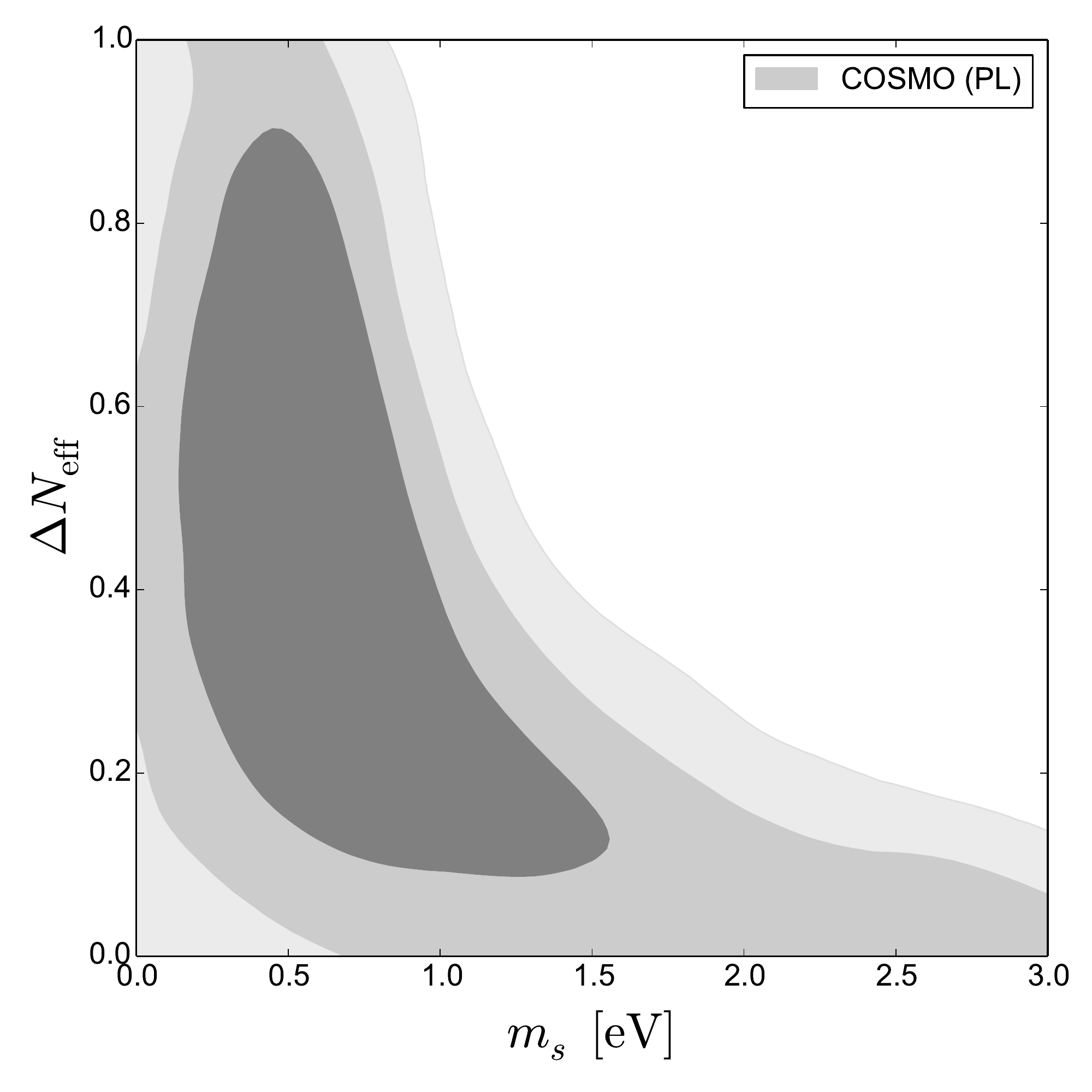}
\includegraphics[width=\halfwidth]{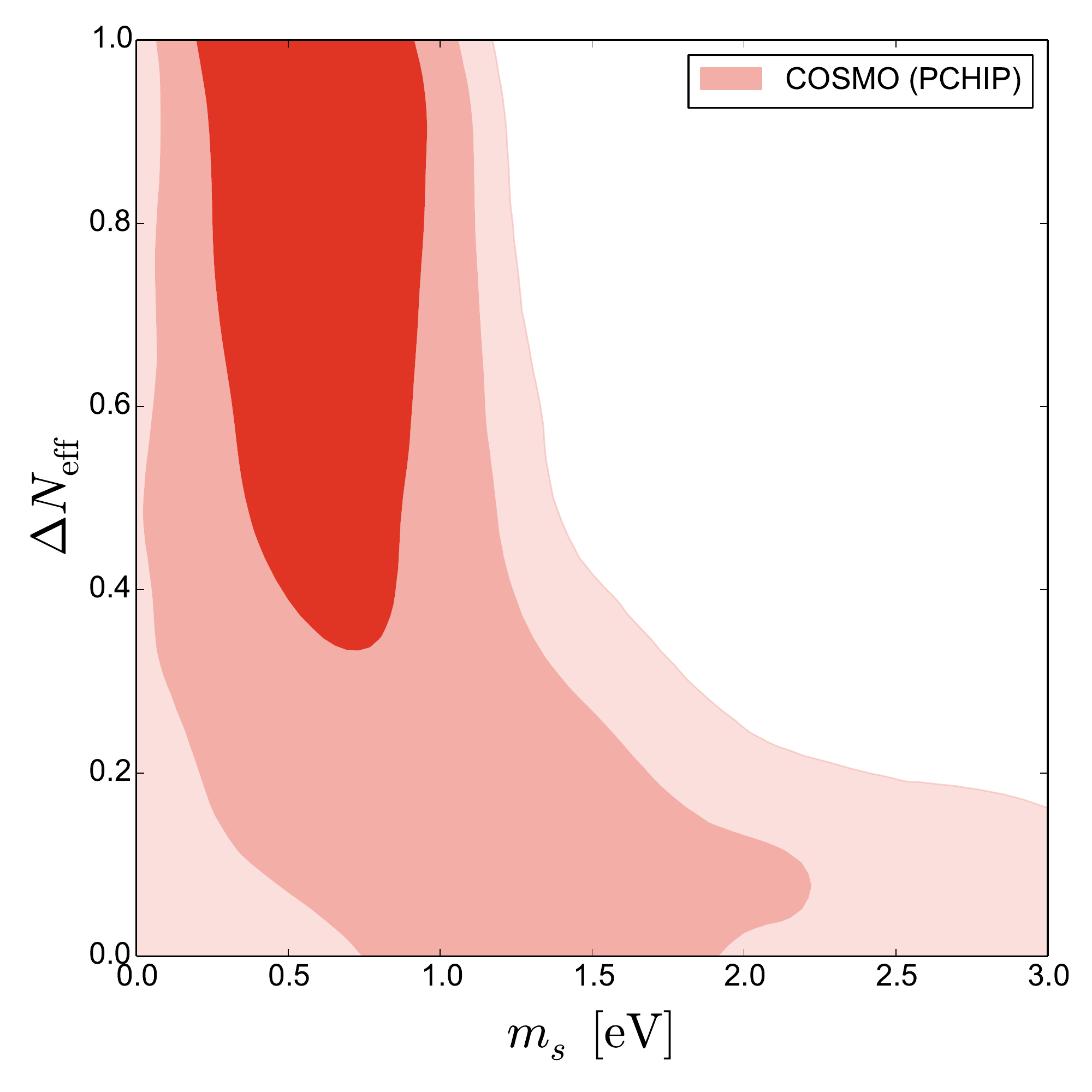}
\caption[Marginalized contours in the ($m_s-\DNeff$) plane
in the fits without the SBL prior]%
{\label{fig:inflfreed_posterior2DnoSBL}
$1\sigma$, $2\sigma$ and $3\sigma$ marginalized contours
in the ($m_s-\DNeff$) plane
in the fits without the SBL prior.
The left and right panels correspond, respectively,
to the standard power-law PPS and the \pchip PPS analyses.
From Ref.~\cite{Gariazzo:2014dla}.
}
\end{figure}

The upper points in
Figure~\ref{fig:inflfreed_errorbars}
show the marginalized $1\sigma$, $2\sigma$ and $3\sigma$
allowed intervals for
$\DNeff$ and $m_s$
that we obtained in the
COSMO(PL)
and
COSMO(PCHIP)
analyses,
without the SBL prior.
Figure~\ref{fig:inflfreed_posterior2DnoSBL} shows the
corresponding marginalized $1\sigma$, $2\sigma$ and $3\sigma$
allowed regions in the $m_s$--$\DNeff$ plane.
We can notice some major changes in the allowed values of both
$\DNeff$ and $m_s$
in the \pchip PPS case with respect to the power-law PPS case.
With a power-law PPS the best-fit value of $\DNeff$ is around 0.4,
whereas with the \pchip PPS it is at $\DNeff=1$,
that is the upper limit for $\DNeff$ assumed in the analysis.
The reason of this behavior is that
the effects of the presence of the additional relativistic energy
in the primordial Universe
can be compensated by
an increase of the \pchip PPS at large $k$
(see Section~\ref{sec:inflfreed_nnu}).
As a result,
the marginalized posterior for $\DNeff$
is increased in the region towards $\DNeff=1$,
together with the higher values
in the \pchip PPS for $k>0.35\mpcinv$.
In the next Sections we will discuss more in details the reasons
that drive to the loosened constraints on the neutrino parameters
when a free PPS is assumed.

Without the SBL constraint on $m_s$,
the different preferences for the value of $\DNeff$
in the power-law and \pchip PPS analyses
correspond to different allowed intervals for $m_s$.
As shown in Fig.~\ref{fig:inflfreed_errorbars},
although in both cases the
best-fit value of $m_{s}$ is near 0.6 eV,
the intermediate preferred region for $\DNeff$ in the power-law PPS analysis
gives for $m_{s}$ an upper limit of about 2 eV at $2\sigma$,
whereas the
large preferred values for $\DNeff$ in the \pchip PPS analysis
gives a tighter upper limit of about 1.5 eV at $2\sigma$,
since the volume of the posterior distribution is shifted
towards lower values of $m_s$.

The SBL prior on the sterile neutrino mass
$m_s$ puts a constraint so strong that in practice
the value of this parameter does not depend
on the inclusion or not of the freedom of the PPS.
In fact, the $m_s$ limits
in Tabs.~\ref{tab:inflfreed_plPPS} and \ref{tab:inflfreed_freePPS}
are similar in the power-law PPS and \pchip PPS analyses.
This can be seen also from the marginalized
allowed intervals of $m_s$ in
Fig.~\ref{fig:inflfreed_errorbars},
comparing the
COSMO+SBL(PL)
and
COSMO+SBL(PCHIP) allowed intervals.

A major difference appears, instead, in the limits for $\DNeff$,
because
the effects of the presence of additional relativistic energy
in the primordial Universe
can be compensated by
an increase in the \pchip PPS at large $k$,
as in the case without the SBL constraint on $m_s$.
As shown in Fig.~\ref{fig:inflfreed_errorbars},
the best-fit and upper limits on $\DNeff$ in the
COSMO+SBL(PL)
and
COSMO+SBL(PCHIP)
are different.
In particular,
in the COSMO+SBL(PCHIP) the $3\sigma$ upper limit on $\DNeff$
allows the presence of a
fully thermalized sterile neutrino compatible
with the SBL constraint on $m_s$.

\begin{figure}
\centering
\includegraphics[width=\halfwidth]{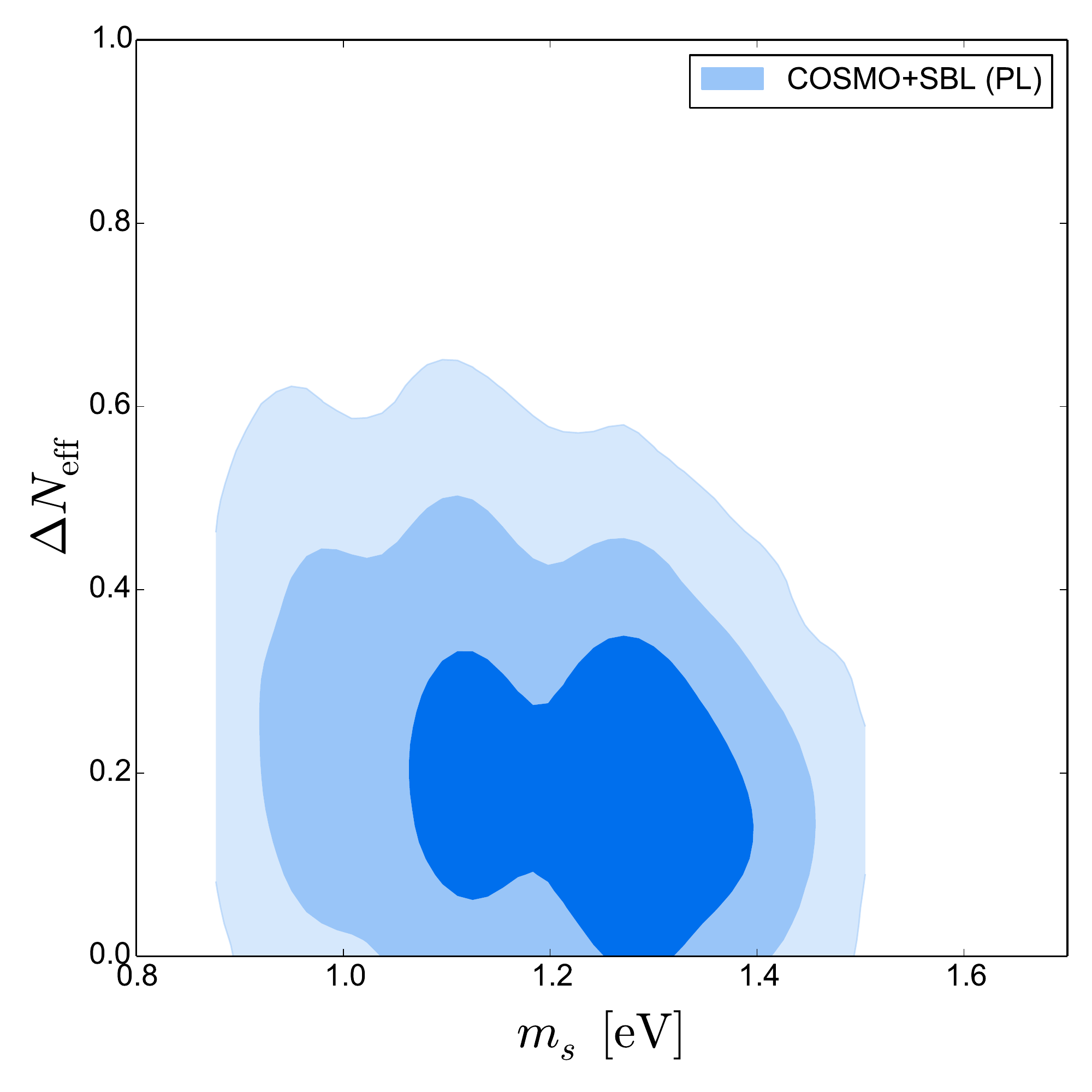}
\includegraphics[width=\halfwidth]{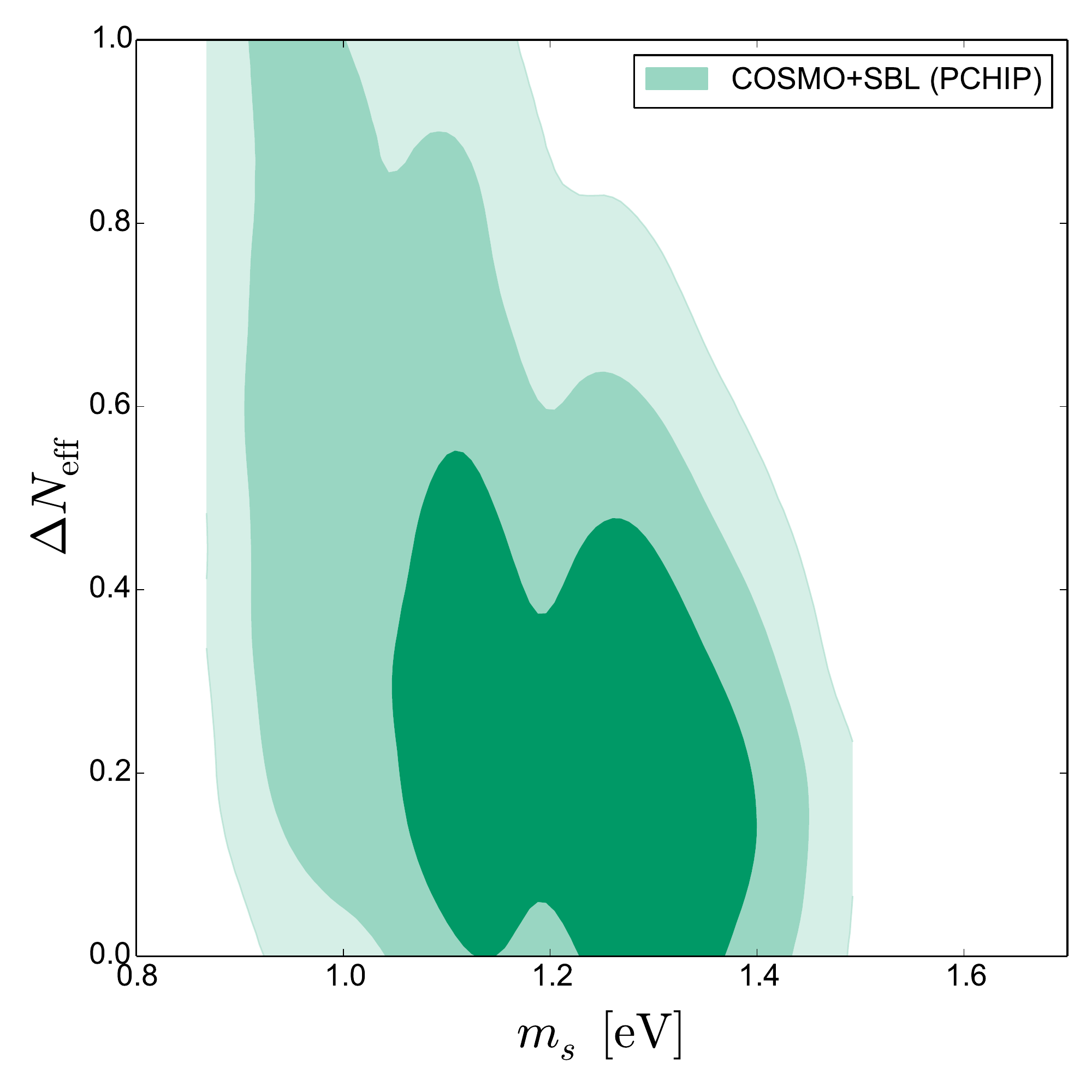}
\caption[Marginalized contours in the ($m_s-\DNeff$) plane
in the fits with the SBL prior]%
{\label{fig:inflfreed_posterior2DSBL}
As in Fig.~\ref{fig:inflfreed_posterior2DnoSBL}, but with
the inclusion of the SBL prior on $m_s$.
From Ref.~\cite{Gariazzo:2014dla}.
}
\end{figure}

Figure~\ref{fig:inflfreed_posterior2DSBL} shows the contour plots of
the marginalized $1\sigma$, $2\sigma$ and $3\sigma$
regions in the $m_s$--$\DNeff$ plane
that we obtained in the
COSMO+SBL(PL)
and
COSMO+SBL(PCHIP)
analyses.
The allowed regions in the left panel
are similar\footnote{
The only difference is that the analysis
in Ref.~\cite{Archidiacono:2014apa}
took into account also the
BICEP2 data
on the B-mode polarization of the CMB
\cite{Ade:2014xna}.
}
to those obtained in 
Ref.~\cite{Archidiacono:2014apa}
with a standard power-law PPS.
One can see that in this case
a fully thermalized sterile neutrino is quite disfavored.
On the other hand,
from the right panel
one can see that in the \pchip PPS analysis
a fully thermalized sterile neutrino with a mass just
below $1\ev$ and with $\DNeff=1$ is even inside the $2\sigma$ region.
This means that a fully thermalized sterile neutrino
can be accommodated in the cosmological model
if the PPS is not forced to be described by a power-law.


At this point we know that the freedom in the inflationary paradigm
can have a significant impact on the constraints derived
from cosmology.
We will study now separately how the \pchip\ PPS
assumption influences the constraints on the base parameters
of the \lcdm\ model (Section~\ref{sec:inflfreed_lcdm})
and on the neutrino properties.
We will consider separately the effective number of
relativistic degrees of freedom in Section~\ref{sec:inflfreed_nnu}
and the sum of the neutrino masses
in Section~\ref{sec:inflfreed_mnu}.
The following results are based on the Planck 2015 data
and they have been presented in Ref.~\cite{DiValentino:2016ikp}.

\section{Base Model and Cosmological Data}
\label{sec:inflfreed_modeldata}
The common underlying model that we will extend
to study various dark radiation properties
is the \lcdm~model already introduced.
From the fundamental cosmological parameters of the \lcdm\ model
we will compute
other derived quantities, such as the Hubble parameter today $H_0$
and the clustering parameter $\sigma_8$,
defined as the mean matter fluctuations
inside a sphere of 8$h^{-1}$~Mpc radius.

We base our following analyses on the latest data released
by the Planck Collaboration \cite{Adam:2015rua},
of which we consider the full temperature power spectrum
at multipoles $2\leq\ell\leq2500$ (\textbf{Planck~TT} hereafter)
and the polarization power spectra in the range
$2\leq\ell\leq29$ (\textbf{lowP}).
We shall also include the polarization data
at $30\leq\ell\leq2500$ (\textbf{TE, EE}) \cite{Aghanim:2015xee}.
Since the polarization spectra at high multipoles
are still under discussion
and some residual systematics
were detected by the Planck Collaboration
\cite{Aghanim:2015xee,Ade:2015xua},
we shall use as baseline dataset the combination
\textbf{Planck~TT+lowP} and
the impact of polarization measurements will be
separately studied in the dataset \textbf{Planck~TT,TE,EE+lowP}.

Additionally,
we will consider the two CMB datasets above in combination
with these additional cosmological measurements
(see Chapter~\ref{ch:cosmomeasurements}):
\begin{description}
  \item[BAO --] Baryon Acoustic Oscillations data as obtained by
    6dFGS \cite{Beutler:2011hx}, 
    by the SDSS Main Galaxy Sample (MGS) \cite{Ross:2014qpa}
    and
    by the BOSS experiment in the DR11 release
    \cite{Anderson:2013zyy};
  \item[MPkW --] the matter power spectrum as measured by the 
    WiggleZ Dark Energy Survey \cite{Parkinson:2012vd};
  \item[lensing --] the reconstruction of the lensing potential
    obtained by the Planck collaboration
    with the CMB trispectrum analysis \cite{Ade:2015zua}.
\end{description}

\section{Constraints in the \texorpdfstring{\lcdm}{LambdaCDM} Model}
\label{sec:inflfreed_lcdm}

\begin{table}[t]
\resizebox{1\textwidth}{!}{
\begin{tabular}{|c||c|c||c|c||c|c||c|c|}
\hline
Parameter	&\multicolumn{2}{c||}{Planck TT+lowP}	&\multicolumn{2}{c||}{Planck TT,TE,EE+lowP}	&\multicolumn{2}{c||}{Planck TT+lowP}	&\multicolumn{2}{c|}{Planck TT,TE,EE+lowP}\\
	&\multicolumn{2}{c||}{}	&\multicolumn{2}{c||}{}	&\multicolumn{2}{c||}{+MPkW}	&\multicolumn{2}{c|}{+MPkW}\\
\hline\hline
$100\Omega_b h^2$                      & $2.222\,^{+0.045}_{-0.043}$    & $2.175\,^{+0.077}_{-0.076}$    & $2.225\,^{+0.032}_{-0.030}$ & $2.215\,^{+0.038}_{-0.037}$    & $2.221\,^{+0.044}_{-0.045}$ & $2.190\,^{+0.072}_{-0.070}$    & $2.223\pm0.031$                & $2.214\,^{+0.035}_{-0.036}$	\\
$\Omega_c h^2$                         & $0.1197\,^{+0.0043}_{-0.0042}$ & $0.1231\,^{+0.0061}_{-0.0059}$ & $0.1198\pm0.0029$           & $0.1209\,^{+0.0035}_{-0.0034}$ & $0.1198\pm0.0039$           & $0.1223\,^{+0.0056}_{-0.0053}$ & $0.1200\,^{+0.0028}_{-0.0027}$ & $0.1210\pm0.0033$   	\\
$100\theta$                            & $1.041\pm0.001$                & $1.040\pm0.001$                & $1.0408\pm0.0006$           & $1.0407\pm0.0006$              & $1.041\pm0.001$             & $1.041\pm0.001$                & $1.0408\pm0.0006$              & $1.0407\pm0.0006$   	\\
$\tau$                                 & $0.078\,^{+0.038}_{-0.036}$    & $0.073\,^{+0.044}_{-0.042}$    & $0.079\pm0.034$             & $0.082\pm0.040$                & $0.075\,^{+0.038}_{-0.039}$ & $0.076\,^{+0.048}_{-0.046}$    & $0.076\,^{+0.034}_{-0.033}$    & $0.083\,^{+0.038}_{-0.037}$      	\\
$n_S$                                  & $0.966\pm0.012$                & --                             & $0.964\pm0.010$             & --                             & $0.965\pm0.011$             & --                             & $0.964\pm0.009$                & --	\\
$\ln[10^{10}A_s]$                      & $3.089\,^{+0.072}_{-0.069}$    & --                             & $3.094\pm0.066$             & --                             & $3.084\,^{+0.073}_{-0.074}$ & --                             & $3.087\,^{+0.066}_{-0.065}$    & --	\\
$H_0\,\mathrm{[km\,s^{-1}\,Mpc^{-1}]}$ & $67.3\,^{+1.9}_{-1.8}$         & $65.7\pm2.7$                   & $67.3\pm1.3$                & $66.8\pm1.5$                   & $67.3\,^{+1.7}_{-1.8}$      & $66.1\pm2.5$                   & $67.2\pm1.2$                   & $66.7\,^{+1.5}_{-1.4}$           	\\
$\sigma_8$                             & $0.83\pm0.03$                  & $0.87\pm0.06$                  & $0.83\pm0.03$               & $0.88\,^{+0.05}_{-0.06}$       & $0.83\pm0.03$               & $0.84\,^{+0.04}_{-0.03}$       & $0.83\pm0.03$                  & $0.83\pm0.03$         	\\
\hline
$\psj{1}$  & $\equiv 1.365$ & $<7.93$                  & $\equiv 1.397$ & $<7.69$                  & $\equiv 1.371$ & $<7.90$                  & $\equiv 1.388$ & $<7.68$         	\\
$\psj{2}$  & $\equiv 1.140$ & $1.15\,^{+0.38}_{-0.35}$ & $\equiv 1.155$ & $1.14\,^{+0.39}_{-0.36}$ & $\equiv 1.139$ & $1.14\,^{+0.39}_{-0.36}$ & $\equiv 1.147$ & $1.14\,^{+0.38}_{-0.36}$         	\\
$\psj{3}$  & $\equiv 1.115$ & $0.73\,^{+0.39}_{-0.37}$ & $\equiv 1.128$ & $0.71\,^{+0.38}_{-0.35}$ & $\equiv 1.113$ & $0.73\,^{+0.39}_{-0.38}$ & $\equiv 1.120$ & $0.72\,^{+0.38}_{-0.37}$         	\\
$\psj{4}$  & $\equiv 1.091$ & $1.19\,^{+0.26}_{-0.25}$ & $\equiv 1.102$ & $1.22\,^{+0.23}_{-0.22}$ & $\equiv 1.088$ & $1.19\pm0.25$            & $\equiv 1.094$ & $1.22\pm0.22$         	\\
$\psj{5}$  & $\equiv 1.067$ & $1.07\pm0.11$            & $\equiv 1.076$ & $1.08\,^{+0.11}_{-0.10}$ & $\equiv 1.063$ & $1.07\,^{+0.12}_{-0.11}$ & $\equiv 1.069$ & $1.08\pm0.10$         	\\
$\psj{6}$  & $\equiv 1.043$ & $1.06\,^{+0.09}_{-0.08}$ & $\equiv 1.051$ & $1.07\,^{+0.08}_{-0.08}$ & $\equiv 1.040$ & $1.06\pm0.09$            & $\equiv 1.044$ & $1.07\,^{+0.08}_{-0.07}$         	\\
$\psj{7}$  & $\equiv 1.021$ & $1.04\,^{+0.09}_{-0.08}$ & $\equiv 1.027$ & $1.04\pm0.08$            & $\equiv 1.016$ & $1.03\pm0.09$            & $\equiv 1.020$ & $1.04\,^{+0.08}_{-0.07}$         	\\
$\psj{8}$  & $\equiv 0.998$ & $0.99\,^{+0.09}_{-0.08}$ & $\equiv 1.003$ & $1.01\pm0.08$            & $\equiv 0.993$ & $1.00\pm0.09$            & $\equiv 0.996$ & $1.01\,^{+0.08}_{-0.07}$         	\\
$\psj{9}$  & $\equiv 0.976$ & $0.97\,^{+0.09}_{-0.08}$ & $\equiv 0.980$ & $0.99\,^{+0.08}_{-0.07}$ & $\equiv 0.971$ & $0.98\pm0.09$            & $\equiv 0.973$ & $0.99\,^{+0.08}_{-0.07}$         	\\
$\psj{10}$ & $\equiv 0.955$ & $0.97\,^{+0.10}_{-0.09}$ & $\equiv 0.957$ & $0.98\pm0.09$            & $\equiv 0.949$ & $0.95\pm0.09$            & $\equiv 0.951$ & $0.96\pm0.08$         	\\
$\psj{11}$ & $\equiv 0.934$ & $<4.03$                  & $\equiv 0.935$ & $2.44\,^{+2.00}_{-2.37}$ & $\equiv 0.928$ & $0.82\,^{+0.45}_{-0.38}$ & $\equiv 0.929$ & $0.81\,^{+0.45}_{-0.38}$         	\\
$\psj{12}$ & $\equiv 0.833$ & nb                       & $\equiv 0.829$ & nb                       & $\equiv 0.825$ & $<3.93$                  & $\equiv 0.823$ & $<3.44$         	\\
\hline
\end{tabular}}
\caption[Constraints on the parameters of the \lcdm\ model,
comparing the PL and \pchip\ PPS results]
{\label{tab:inflfreed_lcdm}
Constraints on the cosmological parameters from 
the Planck TT+lowP and 
Planck TT,TE,EE+lowP datasets,
and also in combination with the matter power spectrum
shape measurements from WiggleZ (MPkW),
in the \lcdm\ model (\emph{nb} stands for \emph{no bound}).
For each combination,
we report the limits obtained for the two parameterizations
of the primordial power spectrum, namely the power-law model
(first column)
and the polynomial expansion (second column of each data combination).
Limits are at 95\% CL around the mean value
of the posterior distribution.
For each dataset, in the case of the power-law model,
the values of \psj{i} are computed according to
Eq.~\eqref{eq:psj_plpps}.
From Ref.~\cite{DiValentino:2016ikp}.
}
\end{table}

In this Section we shall only consider a limited number
of data combinations,
mostly focusing on the variations driven by the inclusion of the
\pchip PPS in the analyses.
We add to the Planck~TT+lowP measurements only the datasets
that can improve the constraints on the \pchip PPS
at small scales, which are the Planck polarization measurements
at high-$\ell$ and
the MPkW constraints on the matter power spectrum.

\begin{figure}
\centering
\includegraphics[width=\singlefigsmall]{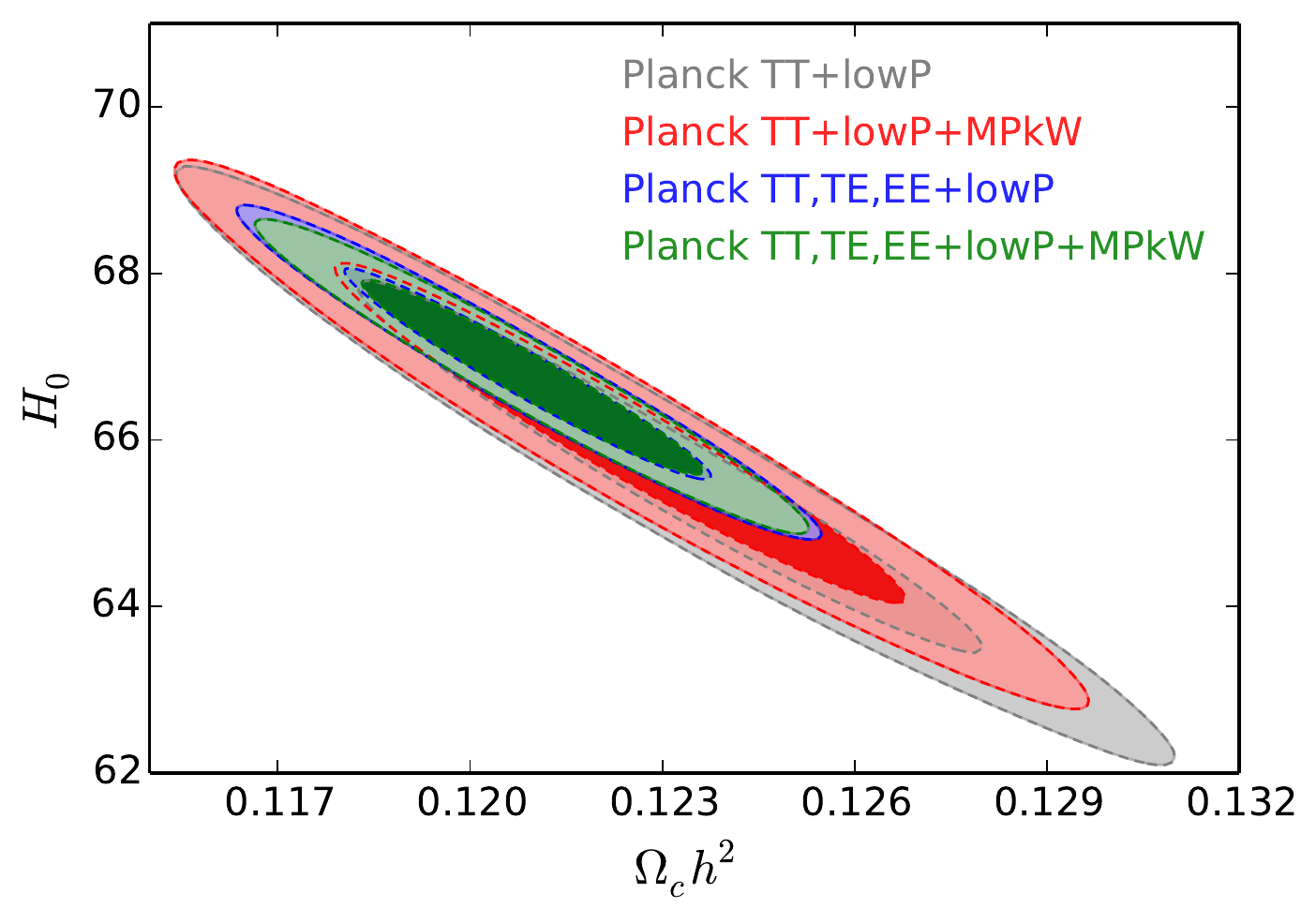}
\caption[2D constraints in the ($\Omega_c h^2$, $H_0$) plane,
obtained in the \lcdm\ model with a \pchip\ PPS]%
{\label{fig:inflfreed_lcdm-omegach2_H0}
2D constraints at 68\% and 95\% CL in the ($\Omega_c h^2$,
  $H_0$) plane, obtained in the
 \lcdm\ model considering the \pchip PPS description,
 for different data combinations.
 From Ref.~\cite{DiValentino:2016ikp}.}
\end{figure}

The results we obtain for the \lcdm\ model are reported in 
Tab.~\ref{tab:inflfreed_lcdm}.
For each dataset, we list the constraints on the different
parameters obtained using both the standard power-law PPS and 
the model independent approach (\pchip) for the PPS.
In the
absence of high multipole polarization or large scale structure data,
the errors are generically enlarged for all the parameters:
those showing a larger difference between their values in the
\pchip PPS case
and in the power-law PPS case
are $\Omega_b h^2$, $\Omega_c h^2$, $H_0$ and $\sigma_8$,
with deviations of the order of 1$\sigma$
in the \pchip PPS case with respect to the power-law PPS case. 
This is a consequence of the numerous degeneracies, and we illustrate
an example in  Fig.~\ref{fig:inflfreed_lcdm-omegach2_H0},
which depicts the constraints in the
($\Omega_c h^2$, $H_0$) plane for different data combinations, in the
\lcdm\ model, assuming the \pchip PPS description.
Simultaneous variations of the parameters can produce effects 
on the CMB spectrum that can be compensated by the freedom in the PPS.

The differences between the \pchip and
the power-law PPS parameterizations
are much smaller
for the ``Planck TT,TE,EE+lowP+MPkW'' dataset,
and the two descriptions of the PPS
give bounds for the \lcdm\ parameters that are in full agreement.

\begin{figure}
\centering
\includegraphics[width=\singlefigland,page=1]{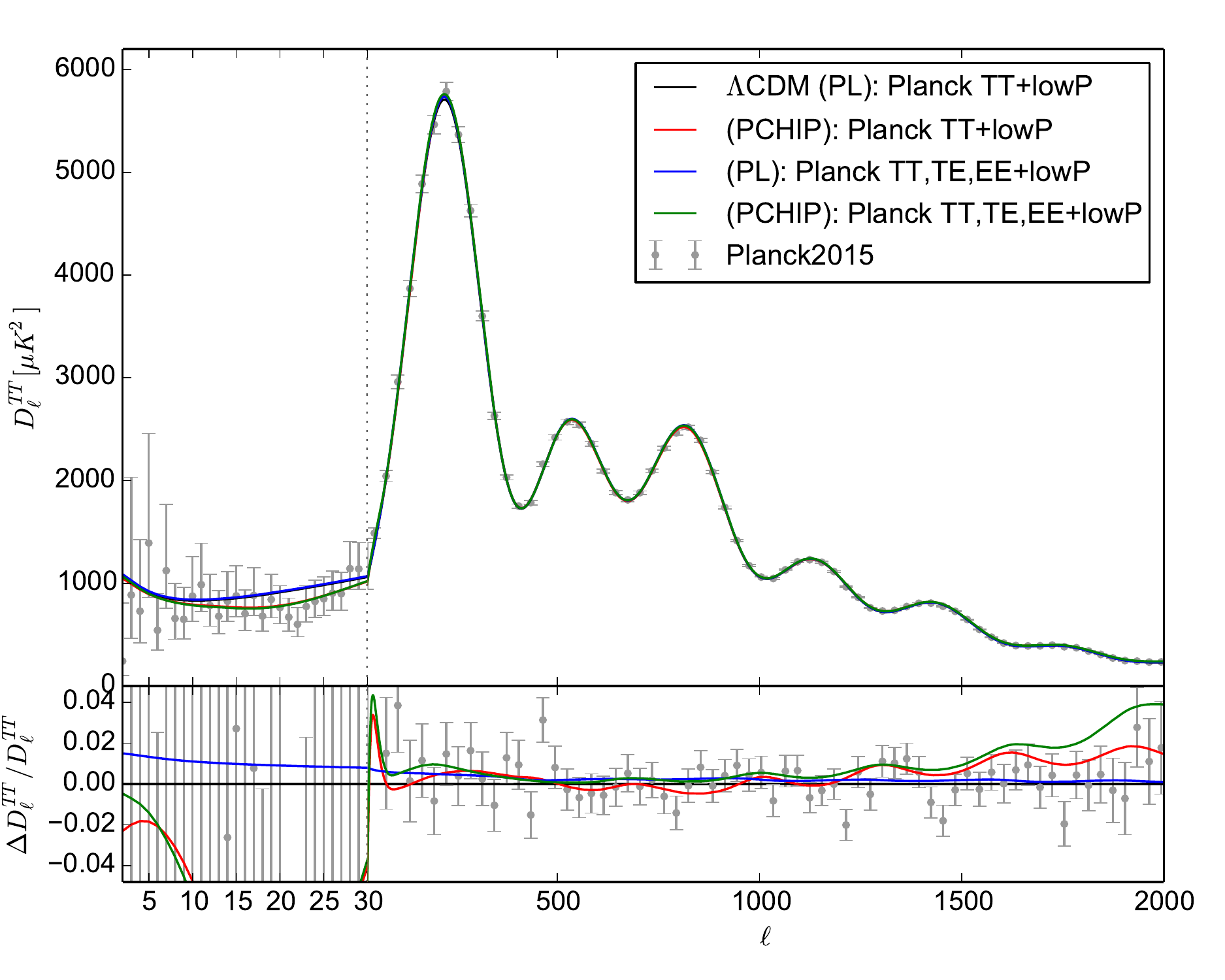}
\includegraphics[width=\singlefigland,page=2]{inflfreed/darkrad/cl_lcdm.pdf}
\caption[Comparison of the measured CMB spectra with the
theoretical predictions of a \lcdm\ model with a power-law or a
\pchip\ PPS]
{\label{fig:inflfreed_cl_lcdm}
 Comparison of the Planck 2015 data \cite{Adam:2015rua}
 with the TT and TE spectra obtained using the marginalized
 best-fit values from the analyses of 
 Planck~TT+lowP (black) and 
 Planck~TT,TE,EE+lowP (blue) in the \lcdm\ model with the power-law (PL) PPS,
 and from the analyses of 
 Planck~TT+lowP (red) and 
 Planck~TT,TE,EE+lowP (green) in the \lcdm\ model with the \pchip\ PPS.
 The adopted values for each spectrum are reported in
 Tab.~\ref{tab:inflfreed_lcdm}.
 We plot the $D_\ell=\ell(\ell+1)\,C_\ell/(2\pi)$ spectra
 and the relative (absolute for the TE spectra) difference
 between each spectrum and the one obtained
 in the \lcdm\ (power-law PPS) model from the Planck~TT+lowP data (black line).
 From Ref.~\cite{DiValentino:2016ikp}.}
\end{figure}

The addition of the high multipole polarization spectra
has a profound
impact in our analyses.
Figure~\ref{fig:inflfreed_cl_lcdm}
depicts the CMB spectra measured by Planck
\cite{Adam:2015rua},
together with the theoretical spectra obtained from the best-fit
values arising from our analyses.
More concretely, we use the marginalized best-fit values reported in 
Tab.~\ref{tab:inflfreed_lcdm}
for the \lcdm\ model with a power-law PPS obtained
from the analyses of the
Planck~TT+lowP (in black) and Planck~TT,TE,EE+lowP (in blue)
datasets,
plus the best-fit values in the \lcdm\ model with a \pchip\ PPS,
from the Planck~TT+lowP (red) and Planck~TT,TE,EE+lowP (green)
datasets.
We plot the $D_\ell=\ell(\ell+1)\,C_\ell/(2\pi)$ spectra
of the TT and TE anisotropies as well as the relative
(absolute for the TE spectra) difference
between each spectrum and the one obtained
from the Planck~TT+lowP data in the \lcdm\ model
with the power-law PPS.
Notice that in the case of TT, the best-fit
spectra are in good agreement with the observational data, even if
there are variations among the \lcdm\ parameters, 
as they can be compensated by the freedom in the PPS.
However, for the TE cross-correlation spectrum, such a
compensation is no longer possible
and the existing degeneracies are broken.
Consequently, the inclusion of the TE spectrum in the analyses
has a strong impact on the derived bounds.
In particular, in the region between $600\leq\ell\leq1200$
in the TE cross-correlation spectra
(see the lower panel of Fig.~\ref{fig:inflfreed_cl_lcdm})
it is possible to notice that the line representing
the results obtained in the \pchip\ PPS approach
without polarization data deviates significantly
from the observational data points.
The addition of high multipole polarization data results
in a good agreement with the predictions obtained
using the power-law PPS.

The bounds on the nodes of the \pchip PPS parameterization
are also reported in Tab.~\ref{tab:inflfreed_lcdm}.
The most significant deviations from the power-law PPS appear at the
extreme wavemodes.
At small $k$, the deviations appear because
the \pchip\ PPS can
reproduce the fluctuations in the CMB temperature spectrum
(see the red and green curves in the upper panel of
Fig.~\ref{fig:inflfreed_cl_lcdm}),
while at high $k$ the data have smaller precision
and therefore the \pchip\ nodes 
are less constrained.
We will describe the bounds on the \pchip nodes
and on the form of the 
reconstructed PPS in Sec.~\ref{sec:inflfreed_ppsconstr},
underlying the common aspects
and the differences that appear
in the various extensions of the \lcdm\ model.

\section{Massless Neutrinos}
\label{sec:inflfreed_nnu}

\subsection{Parameterization}
We already said that massless species account as radiation
during all the evolution of the Universe.
The contribution of the relativistic particles
to the total energy density can be written using the effective
number of degrees of freedom \neff, as in Eq.~\eqref{eq:neff}.
The standard value is $\neff=3.046$ \cite{Mangano:2005cc}
for the three active neutrino standard scenario. 
Deviations of $\neff$ from its standard value may indicate
that the thermal history of the active neutrino 
is different from what we expect,
or that additional relativistic particles are present
in the Universe, as 
additional sterile neutrinos or thermal axions
(see Chapter~\ref{ch:ther_ax} for this last possibility).

We recall that a non-standard value of $\neff$
affects the Big Bang Nucleosynthesis era,
and also the matter-radiation equality.
A shift in the matter-radiation equality would cause a change
in the expansion rate at decoupling, affecting the sound horizon
and the angular scale of the peaks of the CMB spectrum,
as well as in the contribution of the
\emph{early Integrated Sachs Wolfe (ISW) effect}
(see Section~\ref{sub:radiationeffects}).
To avoid such a shift and its consequences,
it is possible to change simultaneously the energy densities
of matter and dark energy,
in order to keep fixed all the relevant scales in the Universe.
In this case, the CMB spectrum is affected only
by an increased Silk damping at small scales
(see Fig.~\ref{fig:neff_cl}). 

Considering the \lcdm\ + \neff\ model,
we will now present the constraints on the effective number of
relativistic species obtained
both in the power-law and the \pchip\ PPS scenarios.

\subsection{Results}
\begin{figure}
\centering
\includegraphics[width=\singlefigland]{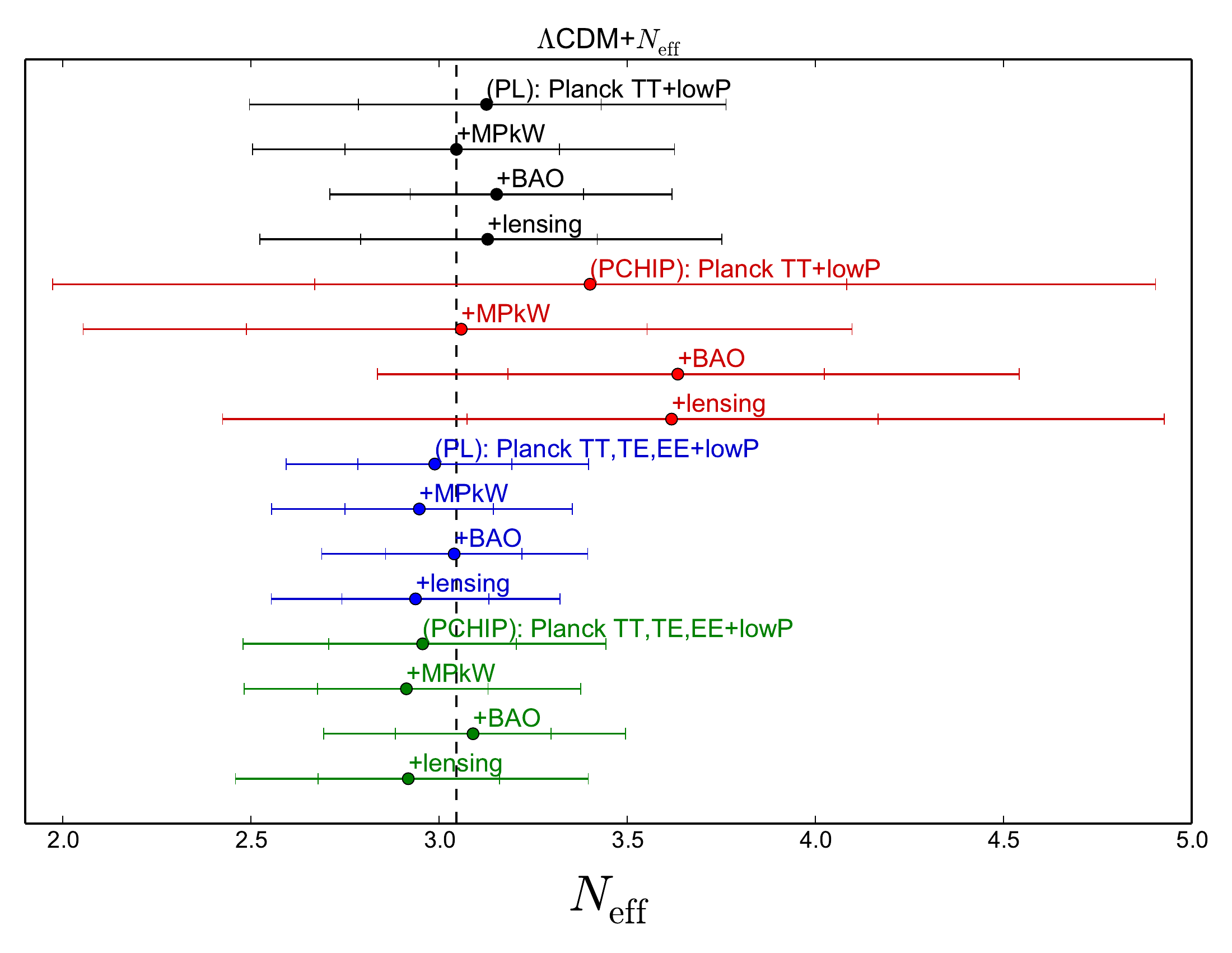}
\caption[Constraints on \neff\ obtained in the \lcdm\ + \neff\
model with a power-law or \pchip\ PPS]
{68\% and 95\% CL constraints on $\neff$, obtained in the
 \lcdm\ + \neff\ model.
 Different colors indicate
 Planck~TT+lowP with PL PPS (black),
 Planck~TT+lowP with PCHIP PPS (red),
 Planck~TT,TE,EE+lowP with PL PPS (blue) and
 Planck~TT,TE,EE+lowP with PCHIP PPS (green).
 For each color we plot 4 different datasets:
 from top to bottom, we have
 CMB only, CMB+MPkW, CMB+BAO and CMB+lensing.
 From Ref.~\cite{DiValentino:2016ikp}.}
\label{fig:inflfreed_nnu_bars}
\end{figure}

The constraints on $\neff$ are summarized in
Fig.~\ref{fig:inflfreed_nnu_bars},
where we plot the 68\% and 95\% CL constraints on $\neff$ obtained
with different datasets and PPS combinations
for the \lcdm\ + \neff\ model.

\begin{figure}%sideways
\centering
\includegraphics[width=\columnwidth,page=1]{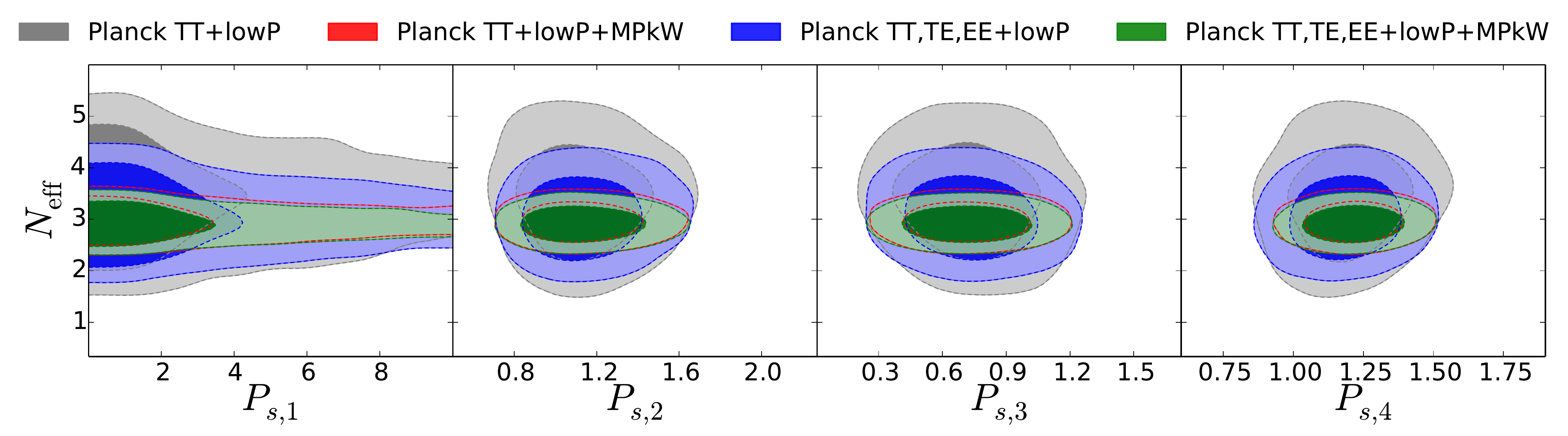}
\includegraphics[width=\columnwidth,page=2]{inflfreed/darkrad/nnu_corr.pdf}
\includegraphics[width=\columnwidth,page=3]{inflfreed/darkrad/nnu_corr.pdf}
\caption[Constraints in the ($\neff$, \psj{j})
planes, obtained in the \lcdm\ + \neff\ model]
{68\% and 95\% CL constraints in the ($\neff$, \psj{j})
planes, obtained in the \lcdm\ + \neff\ model.
We show the results for 
Planck~TT+lowP (gray),
Planck~TT+lowP+MPkW (red),
Planck~TT,TE,EE+lowP (blue) and
Planck~TT,TE,EE+lowP+MPkW (green).
Adapted from Ref.~\cite{DiValentino:2016ikp}.
}
\label{fig:inflfreed_nnu_corr}
\end{figure}%sideways

\begin{table}[p]
\resizebox{1\textwidth}{!}{
\begin{tabular}{|c||c| c||c|c||c|c||c|c|}
\hline
Parameter	
	&\multicolumn{2}{c||}{Planck TT+lowP}
	&\multicolumn{2}{c||}{Planck TT+lowP+MPkW}
	&\multicolumn{2}{c||}{Planck TT+lowP+BAO}
	&\multicolumn{2}{c| }{Planck TT+lowP+lensing}\\
\hline\hline
$\Omega_b h^2$                         & $2.230\,^{+0.075}_{-0.071}$    & $2.189\,^{+0.107}_{-0.105}$    & $2.221\,^{+0.066}_{-0.063}$    & $2.186\,^{+0.081}_{-0.082}$    & $2.233\pm0.047$                & $2.205\,^{+0.060}_{-0.057}$    & $2.232\,^{+0.074}_{-0.069}$    & $2.198\,^{+0.093}_{-0.091}$\\
$\Omega_c h^2$                         & $0.1205\,^{+0.0081}_{-0.0077}$ & $0.1272\,^{+0.0189}_{-0.0182}$ & $0.1198\,^{+0.0077}_{-0.0073}$ & $0.1226\,^{+0.0148}_{-0.0141}$ & $0.1207\,^{+0.0077}_{-0.0074}$ & $0.1294\,^{+0.0153}_{-0.0146}$ & $0.1195\,^{+0.0079}_{-0.0073}$ & $0.1287\,^{+0.0169}_{-0.0161}$   \\
$100\theta$                            & $1.041\pm0.001$                & $1.040\pm0.002$                & $1.041\pm0.001$                & $1.041\pm0.002$                & $1.041\pm0.001$                & $1.0400\,^{+0.0015}_{-0.0014}$ & $1.041\pm0.001$                & $1.0401\,^{+0.0017}_{-0.0015}$   \\
$\tau$                                 & $0.080\,^{+0.044}_{-0.042}$    & $0.076\,^{+0.050}_{-0.047}$    & $0.075\,^{+0.040}_{-0.039}$    & $0.075\,^{+0.048}_{-0.043}$    & $0.082\,^{+0.035}_{-0.036}$    & $0.079\,^{+0.046}_{-0.041}$    & $0.069\,^{+0.040}_{-0.038}$    & $0.066\,^{+0.042}_{-0.038}$      \\
$\neff$                                & $3.13\,^{+0.64}_{-0.63}$       & $3.40\,^{+1.50}_{-1.43}$       & $3.05\,^{+0.58}_{-0.54}$       & $3.06\,^{+1.04}_{-1.00}$       & $3.15\,^{+0.47}_{-0.44}$       & $3.63\,^{+0.91}_{-0.80}$       & $3.13\,^{+0.62}_{-0.61}$       & $3.62\,^{+1.31}_{-1.19}$         \\
$n_S$                                  & $0.969\,^{+0.032}_{-0.030}$    & --                             & $0.965\,^{+0.027}_{-0.026}$    & --                             & $0.971\,^{+0.018}_{-0.017}$    & --                             & $0.971\,^{+0.030}_{-0.028}$    & --                               \\
$\ln[10^{10}A_s]$                      & $3.096\,^{+0.095}_{-0.089}$    & --                             & $3.083\,^{+0.085}_{-0.084}$    & --                             & $3.100\,^{+0.074}_{-0.075}$    & --                             & $3.070\,^{+0.085}_{-0.079}$    & --                               \\
$H_0\,\mathrm{[km\,s^{-1}\,Mpc^{-1}]}$ & $68.0\,^{+5.7}_{-5.6}$         & $68.2\,^{+11.4}_{-11.1}$       & $67.3\,^{+4.8}_{-4.6}$         & $66.0\,^{+7.4}_{-7.2}$         & $68.3\,^{+3.0}_{-2.9}$         & $70.2\,^{+4.6}_{-4.2}$         & $68.5\,^{+5.6}_{-5.3}$         & $70.2\,^{+9.4}_{-8.8}$           \\
$\sigma_8$                             & $0.83\,^{+0.05}_{-0.04}$       & $0.88\,^{+0.10}_{-0.09}$       & $0.83\pm0.04$                  & $0.84\pm0.06$                  & $0.84\pm0.04$                  & $0.90\pm0.08$                  & $0.82\pm0.04$                  & $0.88\pm0.08$         \\ \hline
$\psj{1}$                              & $\equiv 1.337$                 & $<7.96$                        & $\equiv 1.369$                 & $<7.97$                        & $\equiv 1.318$                 & $<8.06$                        & $\equiv 1.279$                 & $<7.87$                          \\
$\psj{2}$                              & $\equiv 1.135$                 & $1.14\,^{+0.40}_{-0.37}$       & $\equiv 1.138$                 & $1.14\,^{+0.39}_{-0.36}$       & $\equiv 1.130$                 & $1.14\,^{+0.41}_{-0.38}$       & $\equiv 1.097$                 & $1.14\,^{+0.39}_{-0.37}$         \\
$\psj{3}$                              & $\equiv 1.112$                 & $0.73\,^{+0.41}_{-0.38}$       & $\equiv 1.112$                 & $0.73\,^{+0.40}_{-0.37}$       & $\equiv 1.109$                 & $0.72\,^{+0.41}_{-0.38}$       & $\equiv 1.076$                 & $0.70\,^{+0.39}_{-0.37}$         \\
$\psj{4}$                              & $\equiv 1.090$                 & $1.20\,^{+0.27}_{-0.25}$       & $\equiv 1.087$                 & $1.19\pm0.25$                  & $\equiv 1.088$                 & $1.20\,^{+0.27}_{-0.26}$       & $\equiv 1.056$                 & $1.18\,^{+0.26}_{-0.25}$         \\
$\psj{5}$                              & $\equiv 1.068$                 & $1.07\,^{+0.13}_{-0.12}$       & $\equiv 1.062$                 & $1.07\pm0.11$                  & $\equiv 1.068$                 & $1.06\pm0.12$                  & $\equiv 1.036$                 & $1.04\pm0.10$         \\
$\psj{6}$                              & $\equiv 1.047$                 & $1.06\,^{+0.10}_{-0.09}$       & $\equiv 1.038$                 & $1.06\,^{+0.09}_{-0.08}$       & $\equiv 1.048$                 & $1.06\pm0.09$                  & $\equiv 1.017$                 & $1.03\,^{+0.07}_{-0.06}$         \\
$\psj{7}$                              & $\equiv 1.026$                 & $1.05\,^{+0.10}_{-0.09}$       & $\equiv 1.015$                 & $1.03\,^{+0.09}_{-0.08}$       & $\equiv 1.028$                 & $1.05\,^{+0.09}_{-0.08}$       & $\equiv 0.998$                 & $1.02\,^{+0.08}_{-0.07}$         \\
$\psj{8}$                              & $\equiv 1.005$                 & $1.00\,^{+0.11}_{-0.10}$       & $\equiv 0.992$                 & $1.00\,^{+0.10}_{-0.09}$       & $\equiv 1.009$                 & $1.02\pm0.09$                  & $\equiv 0.979$                 & $0.99\pm0.09$         \\
$\psj{9}$                              & $\equiv 0.985$                 & $1.00\,^{+0.14}_{-0.13}$       & $\equiv 0.970$                 & $0.97\,^{+0.11}_{-0.10}$       & $\equiv 0.990$                 & $1.02\pm0.09$                  & $\equiv 0.961$                 & $0.99\,^{+0.12}_{-0.11}$         \\
$\psj{10}$                             & $\equiv 0.965$                 & $1.01\,^{+0.20}_{-0.19}$       & $\equiv 0.948$                 & $0.95\,^{+0.15}_{-0.14}$       & $\equiv 0.972$                 & $1.05\pm0.12$                  & $\equiv 0.943$                 & $1.02\pm0.17$         \\
$\psj{11}$                             & $\equiv 0.946$                 & $<3.78$                        & $\equiv 0.927$                 & $0.85\,^{+0.58}_{-0.45}$       & $\equiv 0.954$                 & $<3.83$                        & $\equiv 0.925$                 & $<3.55$                          \\
$\psj{12}$                             & $\equiv 0.853$                 & nb                             & $\equiv 0.824$                 & $<4.24$                        & $\equiv 0.865$                 & nb                             & $\equiv 0.840$                 & nb                               \\
\hline
\end{tabular}}
\caption[Constraints on cosmological parameters 
in the \lcdm\ + \neff\ model,
without CMB polarization at high multipoles]
{Constraints on cosmological parameters from 
the Planck TT+lowP dataset
alone and in combination with the matter power spectrum
shape measurements from WiggleZ (MPkW),
the BAO data and the lensing constraints from Planck,
in the \lcdm\ + \neff\ model (\emph{nb} stands for \emph{no bound}).
For each combination,
we report the limits obtained for the two parameterizations
of the primordial power spectrum, namely the power-law model
(first column)
and the polynomial expansion (second column of each pair).
Limits are at 95\% CL around the mean value
of the posterior distribution.
For each dataset, in the case of power-law model,
the values of \psj{i} are computed according to
Eq.~\eqref{eq:psj_plpps}.
From Ref.~\protect\cite{DiValentino:2016ikp}.
}\label{tab:inflfreed_nnu}
\end{table}

\begin{table}[p]
\resizebox{1\textwidth}{!}{
\begin{tabular}{|c||c| c||c|c||c|c||c|c|}
\hline
Parameter	&\multicolumn{2}{c||}{Planck TT,TE,EE+lowP}	&\multicolumn{2}{c||}{Planck TT,TE,EE+lowP}	&\multicolumn{2}{c||}{Planck TT,TE,EE+lowP}	&\multicolumn{2}{c|}{Planck TT,TE,EE+lowP}\\
	&\multicolumn{2}{c||}{}	&\multicolumn{2}{c||}{+MPkW}	&\multicolumn{2}{c||}{+BAO}	&\multicolumn{2}{c|}{+lensing}\\
\hline\hline
$100\Omega_b h^2$                      & $2.220\pm0.048$                & $2.206\,^{+0.054}_{-0.055}$    & $2.214\,^{+0.047}_{-0.046}$    & $2.203\pm0.049$                & $2.229\pm0.038$                & $2.226\,^{+0.041}_{-0.040}$    & $2.216\,^{+0.045}_{-0.046}$    & $2.204\,^{+0.055}_{-0.053}$\\
$\Omega_c h^2$                         & $0.1191\,^{+0.0062}_{-0.0061}$ & $0.1197\,^{+0.0072}_{-0.0071}$ & $0.1186\,^{+0.0062}_{-0.0061}$ & $0.1191\,^{+0.0070}_{-0.0067}$ & $0.1192\,^{+0.0060}_{-0.0059}$ & $0.1203\,^{+0.0067}_{-0.0068}$ & $0.1178\,^{+0.0058}_{-0.0057}$ & $0.1184\,^{+0.0069}_{-0.0067}$   \\
$100\theta$                            & $1.0409\pm0.0009$              & $1.0408\,^{+0.0010}_{-0.0009}$ & $1.0409\pm0.0009$              & $1.0409\pm0.0009$              & $1.0409\,^{+0.0009}_{-0.0008}$ & $1.0407\pm0.0009$              & $1.0410\,^{+0.0009}_{-0.0008}$ & $1.0410\,^{+0.0010}_{-0.0009}$   \\
$\tau$                                 & $0.077\pm0.035$                & $0.081\,^{+0.040}_{-0.039}$    & $0.073\,^{+0.036}_{-0.035}$    & $0.080\,^{+0.039}_{-0.037}$    & $0.082\pm0.032$                & $0.087\pm0.040$                & $0.060\pm0.028$                & $0.064\,^{+0.034}_{-0.032}$      \\
$\neff$                                & $2.99\,^{+0.41}_{-0.39}$       & $2.96\,^{+0.49}_{-0.48}$       & $2.95\,^{+0.41}_{-0.39}$       & $2.91\,^{+0.46}_{-0.43}$       & $3.04\pm0.35$                  & $3.09\pm0.40$                  & $2.94\pm0.38$                  & $2.92\,^{+0.48}_{-0.46}$         \\
$n_S$                                  & $0.962\pm0.019$                & --                             & $0.960\pm0.019$                & --                             & $0.966\pm0.015$                & --                             & $0.961\,^{+0.019}_{-0.018}$    & --                               \\
$\ln[10^{10}A_s]$                      & $3.088\pm0.074$                & --                             & $3.078\,^{+0.075}_{-0.072}$    & --                             & $3.098\,^{+0.067}_{-0.069}$    & --                             & $3.049\,^{+0.058}_{-0.056}$    & --                               \\
$H_0\,\mathrm{[km\,s^{-1}\,Mpc^{-1}]}$ & $66.8\,^{+3.2}_{-3.1}$         & $66.1\,^{+3.9}_{-3.8}$         & $66.5\pm3.1$                   & $65.8\,^{+3.6}_{-3.4}$         & $67.5\pm2.4$                   & $67.6\,^{+2.6}_{-2.5}$         & $66.7\pm3.0$                   & $66.2\,^{+3.9}_{-3.7}$           \\
$\sigma_8$                             & $0.83\,^{+0.04}_{-0.03}$       & $0.87\pm0.07$                  & $0.82\,^{+0.04}_{-0.03}$       & $0.83\pm0.04$                  & $0.83\pm0.03$                  & $0.88\,^{+0.06}_{-0.08}$       & $0.81\,^{+0.03}_{-0.02}$       & $0.86\pm0.06$         \\ \hline
$\psj{1}$                              & $\equiv 1.415$                 & $<7.62$                        & $\equiv 1.427$                 & $<7.79$                        & $\equiv 1.377$                 & $<7.27$                        & $\equiv 1.373$                 & $<8.15$                          \\
$\psj{2}$                              & $\equiv 1.157$                 & $1.14\,^{+0.38}_{-0.35}$       & $\equiv 1.154$                 & $1.14\,^{+0.38}_{-0.35}$       & $\equiv 1.150$                 & $1.14\,^{+0.38}_{-0.36}$       & $\equiv 1.117$                 & $1.14\,^{+0.38}_{-0.35}$         \\
$\psj{3}$                              & $\equiv 1.128$                 & $0.72\,^{+0.37}_{-0.34}$       & $\equiv 1.125$                 & $0.72\,^{+0.37}_{-0.35}$       & $\equiv 1.125$                 & $0.73\,^{+0.38}_{-0.37}$       & $\equiv 1.089$                 & $0.68\,^{+0.36}_{-0.34}$         \\
$\psj{4}$                              & $\equiv 1.101$                 & $1.22\pm0.22$                  & $\equiv 1.096$                 & $1.22\pm0.22$                  & $\equiv 1.100$                 & $1.23\,^{+0.22}_{-0.21}$       & $\equiv 1.062$                 & $1.20\pm0.21$         \\
$\psj{5}$                              & $\equiv 1.074$                 & $1.08\pm0.10$                  & $\equiv 1.068$                 & $1.08\,^{+0.10}_{-0.09}$       & $\equiv 1.076$                 & $1.09\,^{+0.11}_{-0.10}$       & $\equiv 1.035$                 & $1.05\,^{+0.09}_{-0.08}$         \\
$\psj{6}$                              & $\equiv 1.048$                 & $1.06\pm0.08$                  & $\equiv 1.040$                 & $1.06\,^{+0.08}_{-0.07}$       & $\equiv 1.053$                 & $1.07\,^{+0.09}_{-0.08}$       & $\equiv 1.009$                 & $1.03\,^{+0.07}_{-0.06}$         \\
$\psj{7}$                              & $\equiv 1.022$                 & $1.04\pm0.08$                  & $\equiv 1.013$                 & $1.04\,^{+0.08}_{-0.07}$       & $\equiv 1.030$                 & $1.05\,^{+0.09}_{-0.08}$       & $\equiv 0.984$                 & $1.00\pm0.06$         \\
$\psj{8}$                              & $\equiv 0.997$                 & $1.00\,^{+0.09}_{-0.08}$       & $\equiv 0.987$                 & $1.00\,^{+0.08}_{-0.07}$       & $\equiv 1.007$                 & $1.02\,^{+0.09}_{-0.08}$       & $\equiv 0.959$                 & $0.97\pm0.07$         \\
$\psj{9}$                              & $\equiv 0.973$                 & $0.98\,^{+0.09}_{-0.08}$       & $\equiv 0.962$                 & $0.98\,^{+0.09}_{-0.08}$       & $\equiv 0.985$                 & $1.00\,^{+0.09}_{-0.08}$       & $\equiv 0.935$                 & $0.95\pm0.07$         \\
$\psj{10}$                             & $\equiv 0.949$                 & $0.97\,^{+0.11}_{-0.10}$       & $\equiv 0.937$                 & $0.94\pm0.10$                  & $\equiv 0.964$                 & $1.00\,^{+0.11}_{-0.09}$       & $\equiv 0.912$                 & $0.94\,^{+0.10}_{-0.09}$         \\
$\psj{11}$                             & $\equiv 0.926$                 & $<4.30$                        & $\equiv 0.913$                 & $0.77\,^{+0.42}_{-0.37}$       & $\equiv 0.943$                 & $2.60\,^{+2.01}_{-2.52}$       & $\equiv 0.889$                 & $2.57\,^{+1.96}_{-2.17}$         \\
$\psj{12}$                             & $\equiv 0.815$                 & nb                             & $\equiv 0.799$                 & $<3.32$                        & $\equiv 0.841$                 & nb                             & $\equiv 0.780$                 & nb                               \\
\hline
\end{tabular}}
\caption[As Tab.~\ref{tab:inflfreed_nnu},
but using the full CMB data]{
As Tab.~\ref{tab:inflfreed_nnu},
but using the Planck~TT,TE,EE+lowP dataset.
From Ref.~\protect\cite{DiValentino:2016ikp}.
}\label{tab:inflfreed_nnu_pol}
\end{table}

The introduction of $\neff$ as a free parameter does not change
significantly the results for the \lcdm\ parameters
if a power-law PPS is considered.
However, once the freedom in the PPS is introduced,
a strong degeneracy between the \pchip\ 
nodes \psj{j} and \neff\ appears.
Even if the constraints on \neff\ are loosened
for the \pchip\ PPS case, all the dataset combinations
give constraints on \neff\ that are compatible
with the standard value 3.046 at 95\% CL, 
as we can notice from Fig.~\ref{fig:inflfreed_nnu_bars}.
The mild preference for $\neff>3.046$ arises mainly
as a volume effect in the Bayesian analysis,
since the \pchip\ PPS parameters can be tuned to reproduce
the observed CMB temperature spectrum
for a wide range of \neff\ values.
As expected, the degeneracy between the nodes \psj{j} and \neff\ 
shows up at high wavemodes,
where the Silk damping effect is dominant,
see Fig.~\ref{fig:inflfreed_nnu_corr}.
As a consequence of this correlation,
the values preferred for the nodes \psj{6} to \psj{10}
are slightly larger than the  best-fit values
of the power-law PPS at the same wavemodes.

\begin{figure}
\centering
\includegraphics[width=\singlefigsmall]{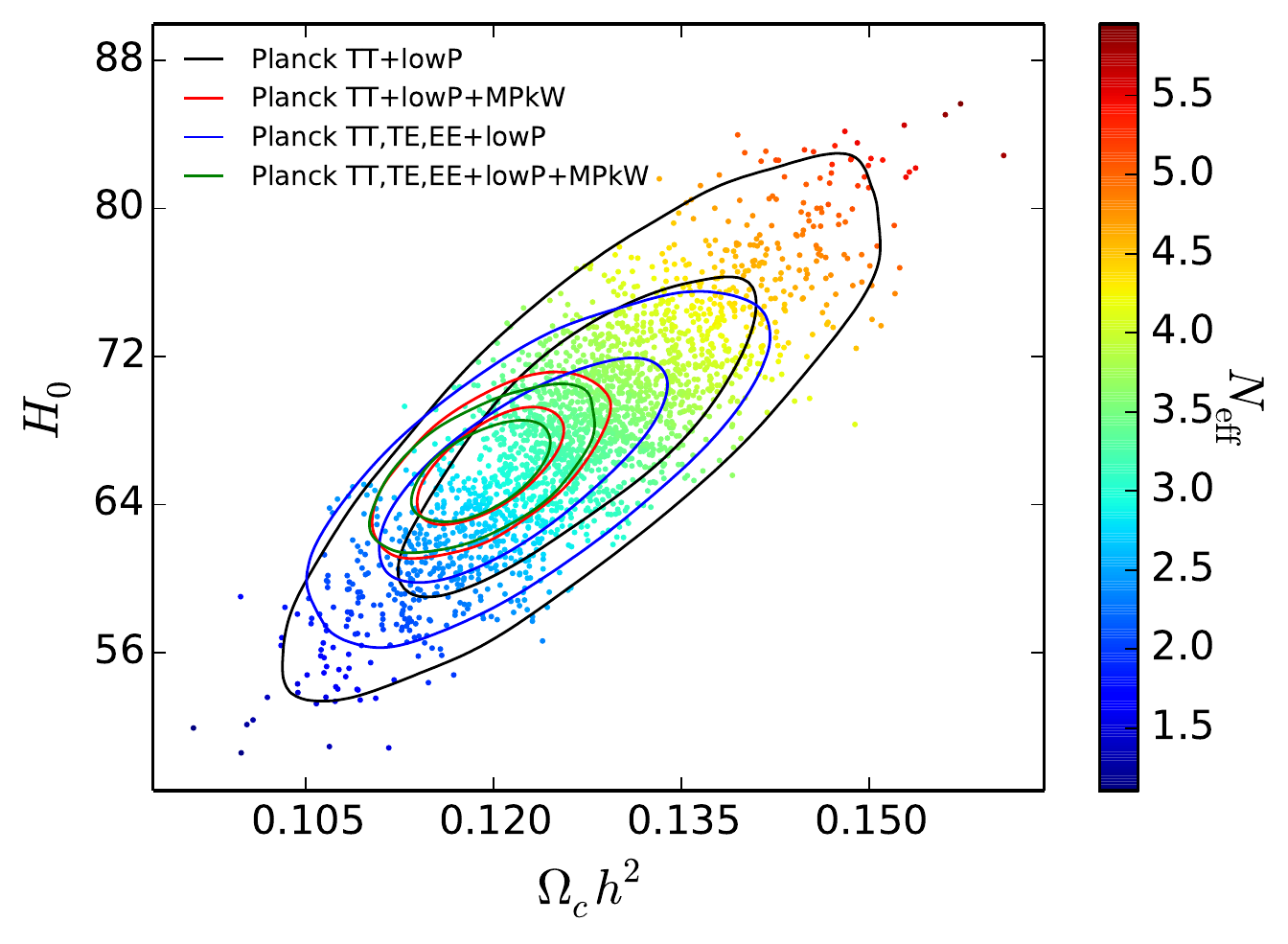}
\caption[Constraints in the 
($\Omega_c h^2$, $H_0$) plane obtained in the
\lcdm\ + \neff\ model with a \pchip PPS]
{2D constraints at 68\% and 95\% CL in the 
($\Omega_c h^2$, $H_0$) plane obtained in the
\lcdm\ + \neff\ model with a \pchip PPS,
for different data combinations.
The coloured points are obtained in the same model,
from the Planck~TT+lowP
analysis, and show the correlation with \neff.
From Ref.~\cite{DiValentino:2016ikp}.
}
\label{fig:inflfreed_lcdm_nnu-omegach2_H0}
\end{figure}

The cosmological limits for a number of parameters change
as a consequence of the various degeneracies with \neff.
For example, to compensate the shift of the matter-radiation
equality redshift due to the increased radiation energy density,
the CDM energy density $\Omega_c h^2$ mean value
is slightly shifted and
its constraints are weakened.
At the same time, the uncertainty on the Hubble parameter $H_0$
is considerably relaxed,
because $H_0$ must be also changed accordingly. 

It is interesting to note that the introduction of \neff\ 
as a free parameter induces a change in the degeneracy
between $\Omega_c h^2$ and $H_0$.
This effect can be noticed by comparing
Fig.~\ref{fig:inflfreed_lcdm-omegach2_H0},
obtained in the \lcdm\ model,
and Fig.~\ref{fig:inflfreed_lcdm_nnu-omegach2_H0},
obtained in the \lcdm\ + \neff\ model.
The reason for which this degeneracy changes is related to
the fact that \neff\ and $\Omega_c h^2$ control
the matter-radiation equality redshift.
If \neff\ is freely varying,
larger values of this parameter will require
a larger matter content $\Omega_c h^2$
to leave unchanged the equality era,
and the $H_0$ parameter will move toward larger values.
On the other hand, if \neff\ is fixed to its standard value
and $\Omega_c h^2$ is increased,
in order to keep unchanged the matter-radiation equality era,
a lower value of $H_0$ would be required to compensate the effect.

The results obtained with the inclusion of
the full CMB polarization data are shown in
Tab.~\ref{tab:inflfreed_nnu_pol}.
The introduction of the polarization data helps in improving
the constraints in the models with a \pchip\ PPS,
since the effects of increasing \neff\ and changing the PPS are
different for the temperature-temperature,
the temperature-polarization and the polarization-polarization
correlation spectra,
as previously discussed in the context of the \lcdm\ model.
When the degeneracies are broken, the preferred value of \neff\ is
very close to the standard value 3.046.
Apparently, the Planck polarization data seem to  prefer 
a value of \neff\ slightly smaller than 3.046
for all the datasets except those including the BAO data,
but the effect is not statistically significant
(see the blue and green points in Fig.~\ref{fig:inflfreed_nnu_bars}).

As the bounds for \neff\ are compatible with 3.046,
the \lcdm\ + \neff\ model gives results that are very close
to those obtained in the simple \lcdm\ model,
but with slightly larger parameter uncertainties,
in particular for $H_0$ and $\Omega_c h^2$.

\section{Massive Neutrinos}
\label{sec:inflfreed_mnu}
\subsection{Parameterization}
Neutrinos oscillations have robustly established the existence
of neutrino masses (see Chapter~\ref{ch:nu}).
However, neutrino mixing data only provide information
on the squared mass differences and not on the absolute scale
of neutrino masses.
Cosmology provides an independent tool to test it,
as massive neutrinos leave a non negligible imprint
in different cosmological observables
\cite{Reid:2009nq,Hamann:2010pw,dePutter:2012sh,Giusarma:2012ph,
Zhao:2012xw,Hou:2012xq,Archidiacono:2013lva,Giusarma:2013pmn,
Riemer-Sorensen:2013jsa,Hu:2014qma,Giusarma:2014zza,
DiValentino:2015sam}.
We recall that
the primary effect of varying the neutrino mass scale
on the CMB temperature spectrum is related
to the early ISW effect (see Subsection~\ref{sub:masseffects}).
The neutrino transition from the relativistic
to the non-relativistic regime affects the decay
of the gravitational potentials at the decoupling period,
producing an enhancement of the small-scale perturbations,
especially near the first acoustic peak.

The baseline scenario we analyze here 
is an extension of the \lcdm\ model where we assume
three active massive neutrino species with degenerate masses.
As we did in the previous Section,
we will study the \lcdm\ + \mnu\ model to test the robustness
of the constraints on the neutrino mass scale under
the assumption of a free PPS.

\subsection{Results}

\begin{figure}
\centering
\includegraphics[width=\singlefigland,page=1]{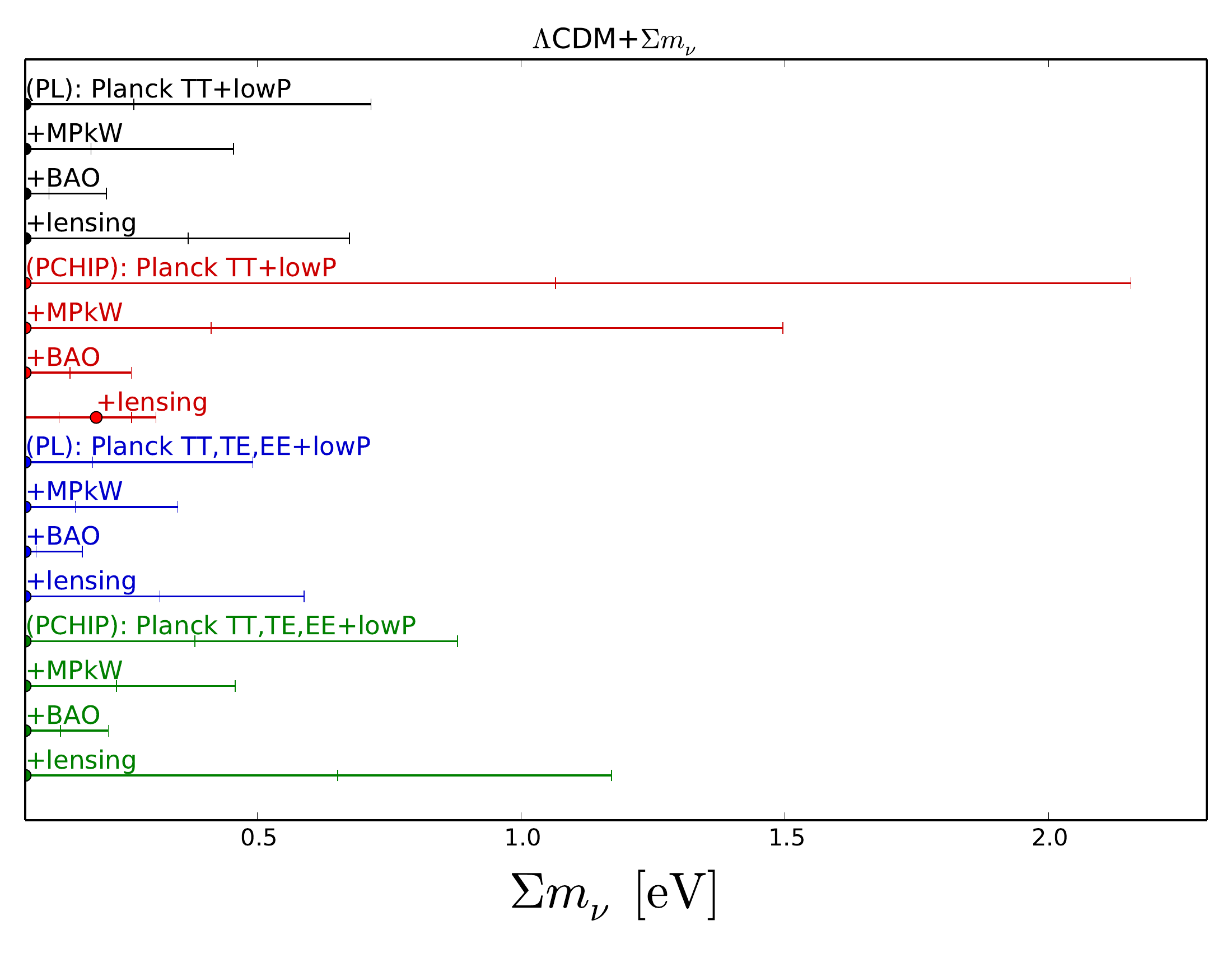}
\caption[As Fig.~\ref{fig:inflfreed_nnu_bars} but for the
\lcdm\ + $\mnu$ case]{
As Fig.~\ref{fig:inflfreed_nnu_bars}
but for the \lcdm\ + $\mnu$ case.
From Ref.~\cite{DiValentino:2016ikp}.}
\label{fig:inflfreed_mnu_bars}
\end{figure}

The 68\% and 95\% CL bounds on \mnu\ obtained
with different dataset and PPS combinations are summarized
in Fig.~\ref{fig:inflfreed_mnu_bars}.
We shall discuss these results in detail below.

\begin{table}[p]
\resizebox{1\textwidth}{!}{
\begin{tabular}{|c||c|c||c|c||c|c||c|c|}
\hline
Parameter	&\multicolumn{2}{c||}{Planck TT+lowP}	
&\multicolumn{2}{c||}{Planck TT+lowP+MPkW}	
&\multicolumn{2}{c||}{Planck TT+lowP+BAO} 
&\multicolumn{2}{c|}{Planck TT+lowP+lensing}\\
\hline\hline
$\Omega_b h^2$    & $2.214_{-0.052}^{+0.054}$    & $2.127_{-0.100}^{+0.097}$    & $2.217_{-0.047}^{+0.045}$    & $2.161_{-0.103}^{+0.089}$    & $2.228_{-0.039}^{+0.041}$    & $2.219_{-0.061}^{+0.063}$    & $2.210_{-0.057}^{+0.048}$    & $2.136_{-0.079}^{+0.081}$\\
$\Omega_c h^2$    & $0.1202_{-0.0047}^{+0.0044}$ & $0.1253_{-0.0064}^{+0.0067}$ & $0.1200_{-0.0039}^{+0.0043}$ & $0.1239_{-0.0067}^{+0.0072}$ & $0.1188_{-0.0029}^{+0.0028}$ & $0.1186_{-0.0033}^{+0.0032}$ & $0.1197_{-0.0042}^{+0.0041}$ & $0.1243_{-0.0059}^{+0.0060}$\\
$100\theta$       & $1.0407_{-0.0011}^{+0.0010}$ & $1.0398\pm0.0014$            & $1.041\pm0.001$              & $1.0401_{-0.0015}^{+0.0013}$ & $1.0410_{-0.0008}^{+0.0009}$ & $1.0410\pm0.0008$            & $1.0408_{-0.0009}^{+0.0010}$ & $1.0399_{-0.0011}^{+0.0012}$\\
$\tau$            & $0.080\pm0.038$              & $0.075_{-0.044}^{+0.048}$    & $0.077_{-0.037}^{+0.038}$    & $0.077_{-0.043}^{+0.050}$    & $0.082_{-0.037}^{+0.038}$    & $0.093_{-0.047}^{+0.050}$    & $0.072_{-0.032}^{+0.034}$    & $ 0.071_{-0.037}^{+0.040}$\\
$\mnu [eV]$       & $<0.75$                      & $<2.16$                      & $<0.46$                      & $<1.15$                      & $<0.22$                      & $<0.26$                      & $<0.63$                      & $<1.64$\\
$n_S$             & $0.964_{-0.013}^{+0.014}$    & --                           & $0964\pm0.012$               & --                           & $0.968\pm0.009$              & --                           & $0.963\pm0.014$              & --	\\
$\ln[10^{10}A_s]$ & $3.095_{-0.073}^{+0.074}$    & --                           & $3.089_{-0.070}^{+0.074}$    & --                           & $3.096\pm0.073$              & --                           & $3.077_{-0.059}^{+0.061}$    & --	\\
$H_0 [\Hou]$      & $65.5_{-4.4}^{+5.9}$         & $58.4_{-10.4}^{+8.8}$        & $66.3_{-3.8}^{+3.2}$         & $62.4_{-10.9}^{+6.3}$        & $67.6\pm1.3$                 & $67.1_{-1.4}^{+1.3}$         & $65.2_{-3.8}^{+3.5}$         & $58.7_{-6.8}^{+7.1}$	\\
$\sigma_8$        & $0.79_{-0.08}^{+0.11}$       & $0.72_{-0.20}^{+0.18}$       & $0.81_{-0.07}^{+0.06}$       & $0.77_{-0.19}^{+0.10}$       & $0.83\pm0.04$                & $0.87_{-0.07}^{+0.07}$       & $0.77_{-0.06}^{+0.05}$       & $0.71_{-0.14}^{+0.14}$\\ \hline
$\psj{1}$         & $\equiv1.399$                & $<8.23$                      & $\equiv1.390$                & $<7.81$                      & $\equiv1.349$                & $<8.08$                      & $\equiv1.386$                & $<7.74$	\\
$\psj{2}$         & $\equiv1.156$                & $1.20_{-0.36}^{+0.40}$       & $\equiv1.149$                & $1.17_{-0.36}^{+0.38}$       & $\equiv1.139$                & $1.11_{-0.34}^{+0.38}$       & $\equiv1.140$                & $1.21_{-0.36}^{+0.37}$	\\
$\psj{3}$         & $\equiv1.129$                & $0.74_{-0.37}^{+0.38}$       & $\equiv1.122$                & $0.74_{-0.37}^{+0.38}$       & $\equiv1.116$                & $0.77_{-0.40}^{+0.41}$       & $\equiv1.113$                & $ 0.73\pm0.39$	\\
$\psj{4}$         & $\equiv1.103$                & $1.22_{-0.26}^{+0.28}$       & $\equiv1.096$                & $1.20\pm0.26$                & $\equiv1.093$                & $1.21\pm0.26$                & $\equiv1.086$                & $ 1.23\pm0.26$	\\
$\psj{5}$         & $\equiv1.077$                & $1.13_{-0.15}^{+0.17}$       & $\equiv1.071$                & $1.09\pm0.13$                & $\equiv1.070$                & $1.08_{-0.12}^{+0.13}$       & $\equiv1.060$                & $1.11\pm0.12$\\
$\psj{6}$         & $\equiv1.052$                & $1.109_{-0.097}^{+0.104}$    & $\equiv1.046$                & $1.080_{-0.087}^{+0.090}$    & $\equiv1.048$                & $1.077_{-0.100}^{+0.104}$    & $\equiv1.035$                & $1.076_{-0.073}^{+0.075}$	\\
$\psj{7}$         & $\equiv1.028$                & $1.049_{-0.087}^{+0.093}$    & $\equiv1.022$                & $1.044_{-0.085}^{+0.091}$    & $\equiv1.026$                & $1.054_{-0.093}^{+0.100}$    & $\equiv1.010$                & $1.034_{-0.064}^{+0.069}$	\\
$\psj{8}$         & $\equiv1.004$                & $0.998_{-0.085}^{+0.096}$    & $\equiv0.998$                & $1.002_{-0.089}^{+0.098}$    & $\equiv1.005$                & $1.026_{-0.100}^{+0.105}$    & $\equiv0.986$                & $0.988_{-0.069}^{+0.076}$	\\
$\psj{9}$         & $\equiv0.981$                & $0.973_{-0.084}^{+0.097}$    & $\equiv0.975$                & $0.977_{-0.089}^{+0.098}$    & $\equiv0.984$                & $1.011_{-0.097}^{+0.102}$    & $\equiv0.963$                & $0.966_{-0.069}^{+0.077}$	\\
$\psj{10}$        & $\equiv0.958$                & $0.966_{-0.095}^{+0.098}$    & $\equiv0.953$                & $0.956_{-0.089}^{+0.097}$    & $\equiv0.964$                & $1.005_{-0.096}^{+0.106}$    & $\equiv0.940$                & $0.968_{-0.077}^{+0.085}$\\
$\psj{11}$        & $\equiv0.936$                & $2.03_{-2.02}^{+1.91}$       & $\equiv0.930$                & $0.97_{-0.75}^{+1.77}$       & $\equiv0.944$                & $2.74_{-2.69}^{+2.07}$       & $\equiv0.918$                & $2.74_{-2.15}^{+1.53}$	\\
$\psj{12}$        & $\equiv0.830$                & nb                           & $\equiv0.825$                & $<3.89$                      & $\equiv0.848$                & nb                           & $\equiv0.811$                & nb	\\
\hline
\end{tabular}}
\caption[As Tab.~\ref{tab:inflfreed_nnu},
but for the \lcdm\ + $\mnu$ model]
{\label{tab:inflfreed_mnu}
As Tab.~\ref{tab:inflfreed_nnu},
but for the \lcdm\ + $\mnu$ model.
From Ref.~\cite{DiValentino:2016ikp}.}
\end{table}

\begin{table}[p]
\resizebox{1\textwidth}{!}{
\begin{tabular}{|c||c|c||c|c||c|c||c|c|}
\hline
Parameter	&\multicolumn{2}{c||}{Planck TT,TE,EE+lowP}	&\multicolumn{2}{c||}{Planck TT,TE,EE+lowP}	&\multicolumn{2}{c||}{Planck TT,TE,EE+lowP}	&\multicolumn{2}{c|}{Planck TT,TE,EE+lowP}\\
	&\multicolumn{2}{c||}{}	&\multicolumn{2}{c||}{+MPkW}	&\multicolumn{2}{c||}{+BAO}	&\multicolumn{2}{c|}{+lensing}\\
\hline\hline
$\Omega_b h^2$    & $2.221_{-0.034}^{+0.032}$    & $2.2080_{-0.040}^{+0.039}$   & $2.223_{-0.027}^{+0.028}$    & $2.209_{-0.038}^{+0.037}$    & $2.223\pm0.027$           & $2.226\pm0.033$              & $2.215\pm0.033$           & $2.203\pm0.041$\\
$\Omega_c h^2$    & $0.1200_{-0.0030}^{+0.0031}$ & $0.1212_{-0.0034}^{+0.0035}$ & $0.1199_{-0.0027}^{+0.0028}$ & $0.1212_{-0.0033}^{+0.0035}$ & $0.1192\pm0.0023$         & $0.1191_{-0.0025}^{+0.0024}$ & $0.1101\pm0.0030$         & $0.1207_{-0.0035}^{+0.0033}$\\
$100\theta$       & $1.0407\pm0.0007$            & $1.0405\pm-0.0007$           & $1.0407\pm0.0006$            & $1.0406\pm-0.0007$           & $1.0408\pm0.0006$         & $1.0408\pm-0.0006$           & $1.0406\pm0.0007$         & $1.0405\pm-0.0007$	\\
$\tau$            & $0.081_{-0.034}^{+0.033}$    & $0.085_{-0.040}^{+0.042}$    & $0.080\pm0.034$              & $0.088\pm0.037$              & $0.083_{-0.032}^{+0.033}$ & $0.088_{-0.040}^{+0.045}$    & $0.076_{-0.032}^{+0.033}$ & $ 0.082\pm0.035$\\
$\mnu [eV]$       & $<0.50$                      & $<0.88$                      & $<0.35$                      & $<0.46$                      & $<0.18$                   & $<0.22$                      & $<0.63$                   & $<1.17$\\
$n_S$             & $0.97\pm0.01$                & --                           & $0.964\pm0.009$              & --                           & $0.966\pm0.008$           & --                           & $0.963\pm0.009$           & --	\\
$\ln[10^{10}A_s]$ & $3.098_{-0.065}^{+0.064}$    & --                           & $3.095_{-0.066}^{+0.065}$    & --                           & $3.100_{-0.064}^{+0.063}$ & --                           & $3.086_{-0.061}^{+0.063}$ & --	\\
$H_0 [\Hou]$      & $66.3_{-3.8}^{+2.9}$         & $64.3_{-5.0}^{+3.9}$         & $66.7_{-2.7}^{+2.3}$         & $64.4_{-3.1}^{+2.1}$         & $67.5_{-1.2}^{+1.1}$      & $67.1_{-1.2}^{+1.3}$         & $65.0_{-3.8}^{+3.3}$      & $62.8_{-5.6}^{+5.1}$	\\
$\sigma_8$        & $0.81_{-0.08}^{+0.06}$       & $0.82_{-0.14}^{+0.11}$       & $0.82_{-0.06}^{+0.05}$       & $0.81_{-0.06}^{+0.05}$       & $0.83\pm0.03$             & $0.87_{-0.08}^{+0.07}$       & $0.78_{-0.06}^{+0.05}$    & $0.71_{-0.13}^{+0.12}$\\ \hline
$\psj{1}$         & $\equiv1.405$                & $<7.52$                      & $\equiv1.399$                & $<7.43$                      & $\equiv1.380$             & $<7.59$                      & $\equiv1.399$             & $<7.91$	\\
$\psj{2}$         & $\equiv1.160$                & $1.16_{-0.35}^{+0.37}$       & $\equiv1.156$                & $1.15_{-0.36}^{+0.40}$       & $\equiv1.153$             & $1.13_{-0.36}^{+0.39}$       & $\equiv1.150$             & $1.18_{-0.36}^{+0.38}$	\\
$\psj{3}$         & $\equiv1.133$                & $0.73_{-0.36}^{+0.39}$       & $\equiv1.129$                & $0.73_{-0.38}^{+0.39}$       & $\equiv1.127$             & $0.73_{-0.37}^{+0.39}$       & $\equiv1.123$             & $ 0.73_{-0.35}^{+0.37}$	\\
$\psj{4}$         & $\equiv1.107$                & $1.24_{-0.22}^{+0.23}$       & $\equiv1.103$                & $1.23\pm0.23$                & $\equiv1.103$             & $1.23_{-0.22}^{+0.23}$       & $\equiv1.096$             & $ 1.24\pm0.23$	\\
$\psj{5}$         & $\equiv1.081$                & $1.10\pm0.11$                & $\equiv1.077$                & $1.10\pm0.10$                & $\equiv1.079$             & $1.09_{-0.10}^{+0.11}$       & $\equiv1.070$             & $1.09\pm0.11$\\
$\psj{6}$         & $\equiv1.056$                & $1.073_{-0.085}^{+0.091}$    & $\equiv1.052$                & $1.079_{-0.073}^{+0.078}$    & $\equiv1.055$             & $1.069_{-0.085}^{+0.093}$    & $\equiv1.044$             & $1.065_{-0.072}^{+0.076}$	\\
$\psj{7}$         & $\equiv1.031$                & $1.050_{-0.087}^{+0.086}$    & $\equiv1.028$                & $1.055_{-0.072}^{+0.077}$    & $\equiv1.032$             & $1.046_{-0.083}^{+0.092}$    & $\equiv1.019$             & $1.039_{-0.068}^{+0.069}$	\\
$\psj{8}$         & $\equiv1.007$                & $1.016\pm0.084$              & $\equiv1.004$                & $1.021_{-0.073}^{+0.077}$    & $\equiv1.009$             & $1.019_{-0.088}^{+0.089}$    & $\equiv0.995$             & $1.007_{-0.072}^{+0.070}$	\\
$\psj{9}$         & $\equiv0.984$                & $0.996_{-0.081}^{+0.082}$    & $\equiv0.981$                & $0.998_{-0.071}^{+0.075}$    & $\equiv0.987$             & $1.003_{-0.079}^{+0.087}$    & $\equiv0.972$             & $0.988_{-0.070}^{+0.068}$	\\
$\psj{10}$        & $\equiv0.961$                & $1.00_{-0.08}^{+0.09}$       & $\equiv0.958$                & $0.97_{-0.08}^{+0.9}$        & $\equiv0.966$             & $1.00_{-0.09}^{+0.10}$       & $\equiv0.948$             & $0.98_{-0.07}^{+0.08}$\\
$\psj{11}$        & $\equiv0.938$                & $2.77_{-2.63}^{+1.88}$       & $\equiv0.936$                & $0.82_{-0.45}^{+0.56}$       & $\equiv0.944$             & $2.79_{-2.72}^{+2.02}$       & $\equiv0.926$             & $3.015_{-2.14}^{+1.51}$	\\
$\psj{12}$        & $\equiv0.831$                & nb                           & $\equiv0.830$                & $<3.20$                      & $\equiv0.843$             & nb                           & $\equiv0.818$             & nb	\\
\hline
\end{tabular}}
\caption[As Tab.~\ref{tab:inflfreed_nnu_pol},
but for the \lcdm\ + $\mnu$ model]
{\label{tab:inflfreed_mnu_pol}
As Tab.~\ref{tab:inflfreed_nnu_pol},
but for the \lcdm\ + $\mnu$ model.
From Ref.~\cite{DiValentino:2016ikp}.}
\end{table}

Table~\ref{tab:inflfreed_mnu} depicts the 95\% CL
constraints on the sum of the three
active neutrino masses arising from Planck TT+lowP
CMB measurements plus other external datasets.
Notice that for all the data combinations
the bounds on neutrino masses are weaker
when considering the \pchip PPS with respect
to the power-law PPS case.
This loosened bounds are due to the degeneracy between \mnu\ and
the nodes $\psj{5}$ and $\psj{6}$,
that correspond to the wavenumbers where
the contribution of the early ISW effect is located.
Therefore, the change induced on these angular scales
by a larger neutrino mass could be compensated
by increasing $\psj{5}$ and $\psj{6}$.   

% \begin{figure}%sideways
% \centering
% \includegraphics[width=\columnwidth]{inflfreed/darkrad/mnu_corr1.pdf}
% \includegraphics[width=\columnwidth]{inflfreed/darkrad/mnu_corr2.pdf}
% \caption{As Fig.~\ref{fig:nnu_corr} but for
% the \lcdm\ + $\mnu$ case.}
% \label{fig:mnu_corr}
% \end{figure}%sideways

% The addition of the matter power spectrum measurements, MPkW, leads to an upper bound on  \mnu\ of $1.1497$~eV at 95\% CL in the~\pchip parameterization,  which is twice the value obtained when considering the power-law PPS with the same dataset.
The most stringent constraints on the sum of the three active
neutrino masses are obtained when the BAO data are considered.
In particular, we have $\mnu<0.26$~eV ($\mnu<0.22$~eV) at 95\% CL
when considering the \pchip\ (power-law) PPS parameterization.
This is the consequence of the fact that the BAO data strongly
constrains the energy densities of the massive species,
so that the degeneracy between \mnu\ and the PPS is broken.
% Finally, the combination of Planck TT+lowP data with the Planck CMB lensing measurements provide a bound on neutrino masses of $\mnu<1.636$~eV at 95\% CL in the~\pchip case.

\begin{figure}[t]
\centering
\includegraphics[width=\halfwidth]{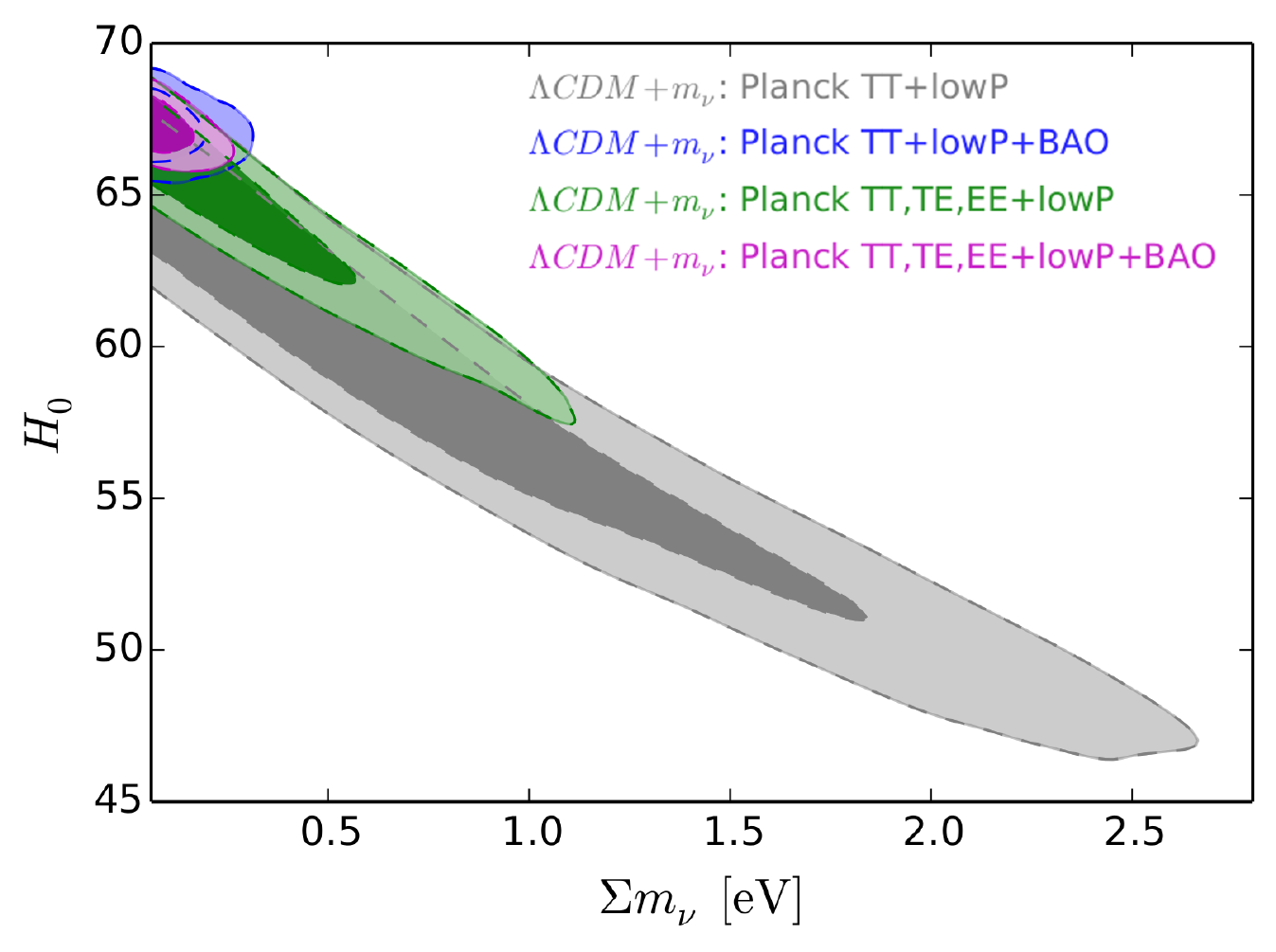}
\includegraphics[width=\halfwidth]{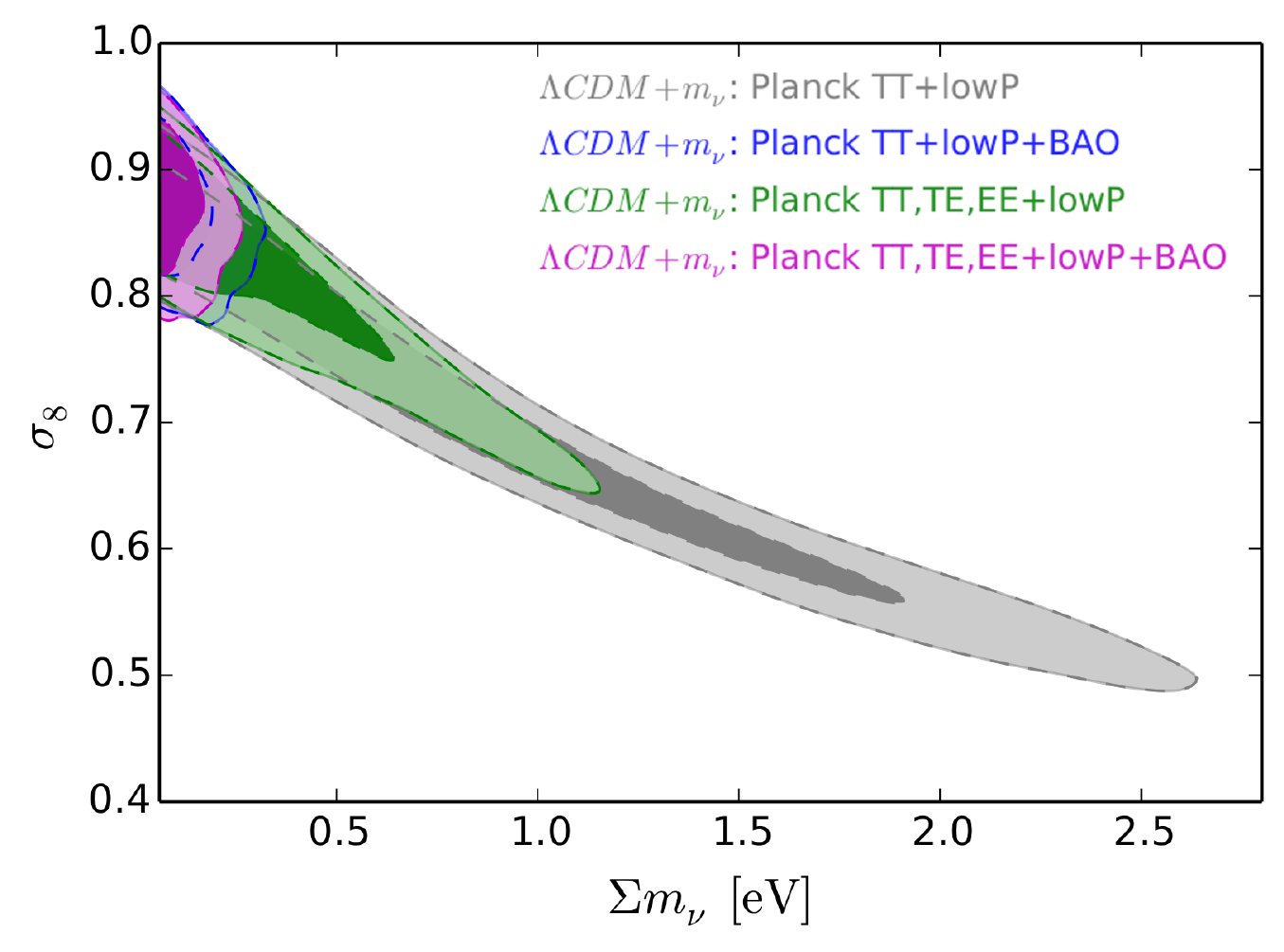}
 \caption[Allowed regions in the
 ($\mnu$, $H_0$) and ($\mnu$, $\sigma_8$) planes,
 obtained in the \lcdm\ + \mnu\ model
 within the \pchip PPS parameterization]
 {$68\%$ and $95\%$~CL allowed regions in the
 ($\mnu$, $H_0$) plane (left panel) and
 in the ($\mnu$, $\sigma_8$) plane (right panel),
 obtained in the \lcdm\ + \mnu\ model
 within the \pchip PPS parameterization.
 From Ref.~\cite{DiValentino:2016ikp}.}
\label{fig:inflfreed_deg_mnu}
\end{figure}

It can be noticed that using the \pchip PPS parameterization
there is a shift not only for the preferred value of \mnu,
but also for other cosmological parameters, such as
the Hubble constant and the clustering parameter $\sigma_8$.
This occurs because there exists a strong degeneracy between
the neutrino mass and the Hubble constant,
as shown in the left panel of Fig.~\ref{fig:inflfreed_deg_mnu}
and between \mnu\ and $\sigma_8$, as shown in the right
panel of Fig.~\ref{fig:inflfreed_deg_mnu}.
In particular, considering CMB data only,
a higher value of $\mnu$ will alter
the angular diameter distance to the last scattering surface,
change that can be compensated with a smaller value
of the Hubble constant $H_0$.
The mean values of the clustering parameter $\sigma_8$
are also displaced by $\sim2\sigma$ (except for the BAO case)
toward lower values in the~\pchip PPS approach
with respect to those obtained using the power-law PPS,
since the free-streaming of a heavier neutrino 
causes a larger suppression of the perturbations at small scales
(see Ref.~\citelesg{}).
The fact that the larger allowed range for \mnu\ causes
a shift in the mean values of $H_0$ and $\sigma_8$ 
is a simple consequence of volume effects that arise
during the Bayesian marginalization.

Table~\ref{tab:inflfreed_mnu_pol} presents the constraints
on the cosmological parameters from the Planck TT,TE,EE+lowP data
alone and in combination with the MPkW, BAO and lensing measurements.
If one considers the high-$\ell$ polarization measurements,
the bounds on the sum of the neutrino masses are larger
when using the \pchip parameterization with respect
to the ones obtained with the power-law approach.
However, these bounds are more stringent than those
obtained using the Planck TT+lowP data only.
As we discussed in the previous Sections,
the reason for this improvement is due to the fact that
the inclusion of the polarization measurements removes
many of the degeneracies among the parameters,
but in particular between \mnu\ and the PPS.
The constraints on \mnu\ from all the data combinations and the
PPS parameterizations are plotted in
Fig.~\ref{fig:inflfreed_mnu_bars}.
Also when the full CMB polarization spectra are included
the data combination that gives the most stringent constraints
is the one involving BAO datasets,
since it provides a 95\% CL upper bound $\mnu<0.22$~eV
in the \pchip PPS case and $\mnu<0.18$~eV in the power-law PPS case.

\section{Constraints on the Primordial Power Spectrum}
\label{sec:inflfreed_ppsconstr}
% XI di arxiv:1601.xxxxx

From the MCMC analyses presented in the previous sections we obtained 
constraints on the nodes used to parameterize the \pchip~PPS.
Using these information, we can
obtain a reconstruction of the spectrum shape for the different extensions
of the \lcdm~model.
Since the form of the reconstructed PPS
is similar for the different models,
we discuss now the common features of the \pchip~PPS
as obtained for the 
\lcdm~model.

\begin{figure}
  \centering
  \includegraphics[width=\singlefigland,page=1]{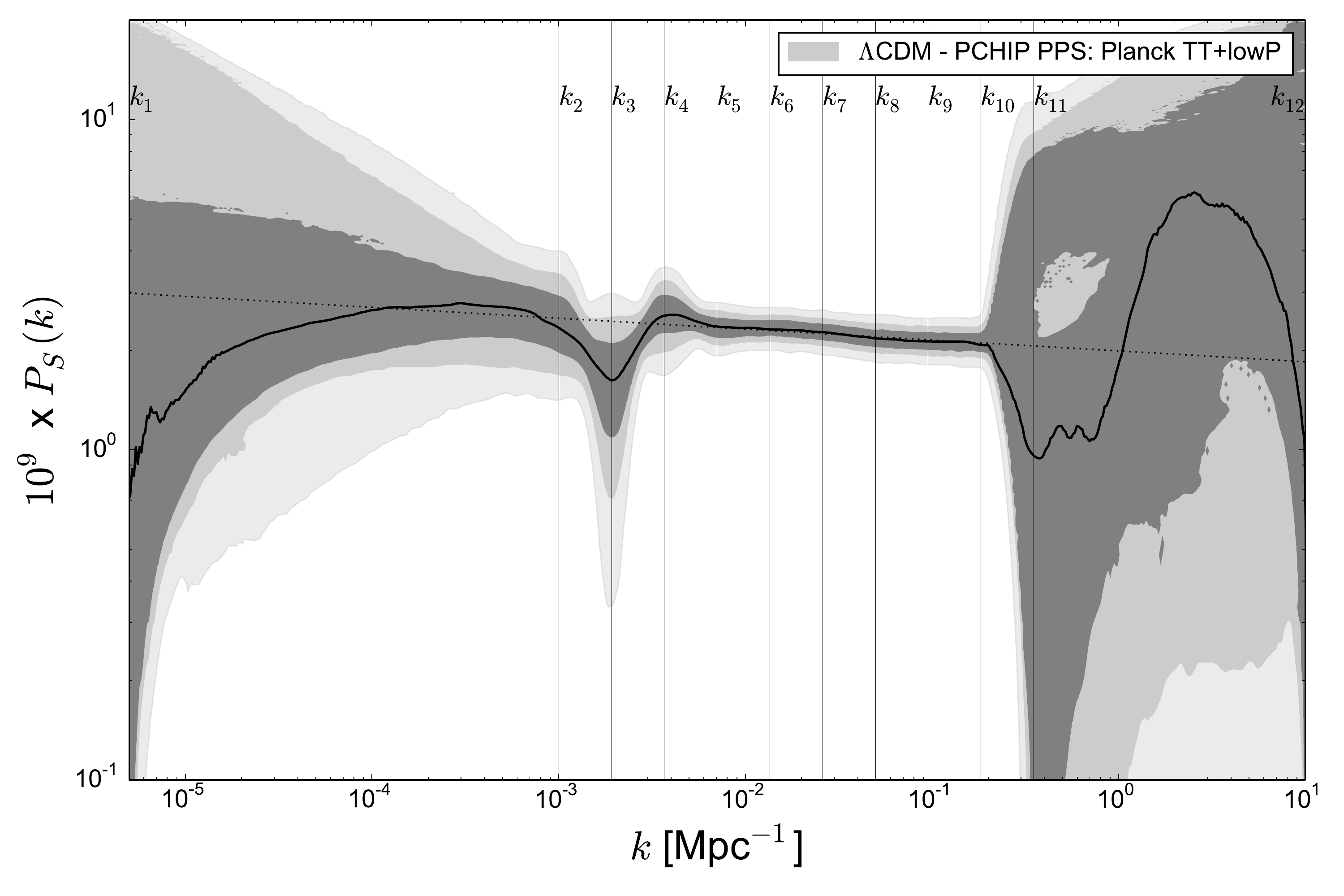}
  \caption[Reconstruction of 
    \pchip~PPS obtained in the \lcdm~model,
    with the ``Planck~TT+lowP'' dataset]
    {\label{fig:inflfreed_pps_planck} 
    Reconstruction of the marginalized best fit
    \pchip~PPS (solid line) with 68\%, 95\% and 99\% confidence bands
    as obtained in the \lcdm~model,
    with the ``Planck~TT+lowP'' dataset.
    The dotted line represents the power-law PPS
    corresponding to the Planck
    best fit \cite{Ade:2015xua}.
    From Ref.~\cite{DiValentino:2016ikp}.}
\end{figure}

\begin{figure}
  \centering
  \includegraphics[width=\singlefigland,page=3]{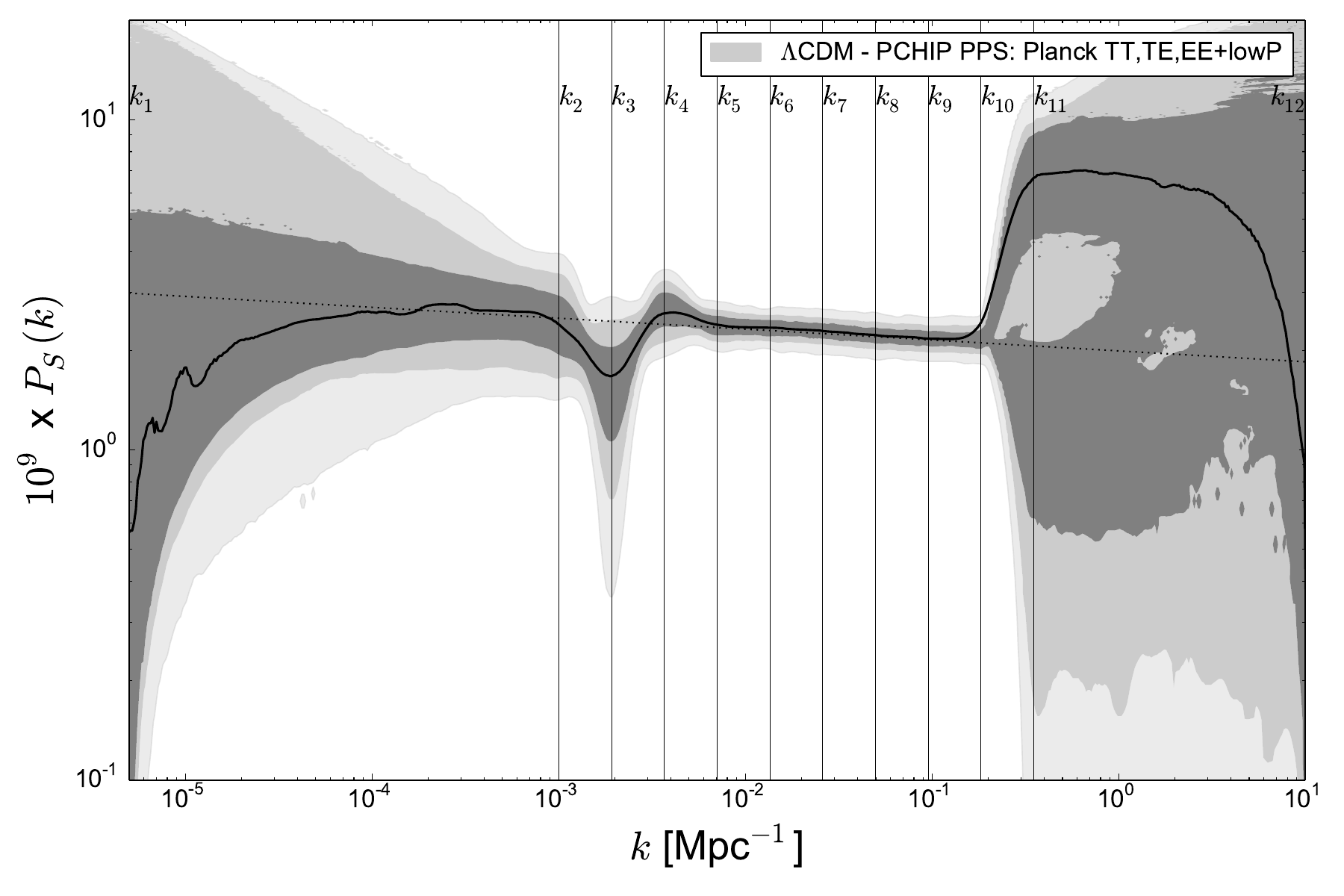}
  \caption[As in Fig.~\ref{fig:inflfreed_pps_planck}, but with
    the ``Planck~TT,TE,EE+lowP'' dataset]
    {\label{fig:inflfreed_pps_planck_pol} 
    As in Fig.~\ref{fig:inflfreed_pps_planck}, but with
    the ``Planck~TT,TE,EE+lowP'' dataset.
    From Ref.~\cite{DiValentino:2016ikp}.
  }
\end{figure}

\begin{figure}
  \centering
  \includegraphics[width=\singlefigland,page=4]{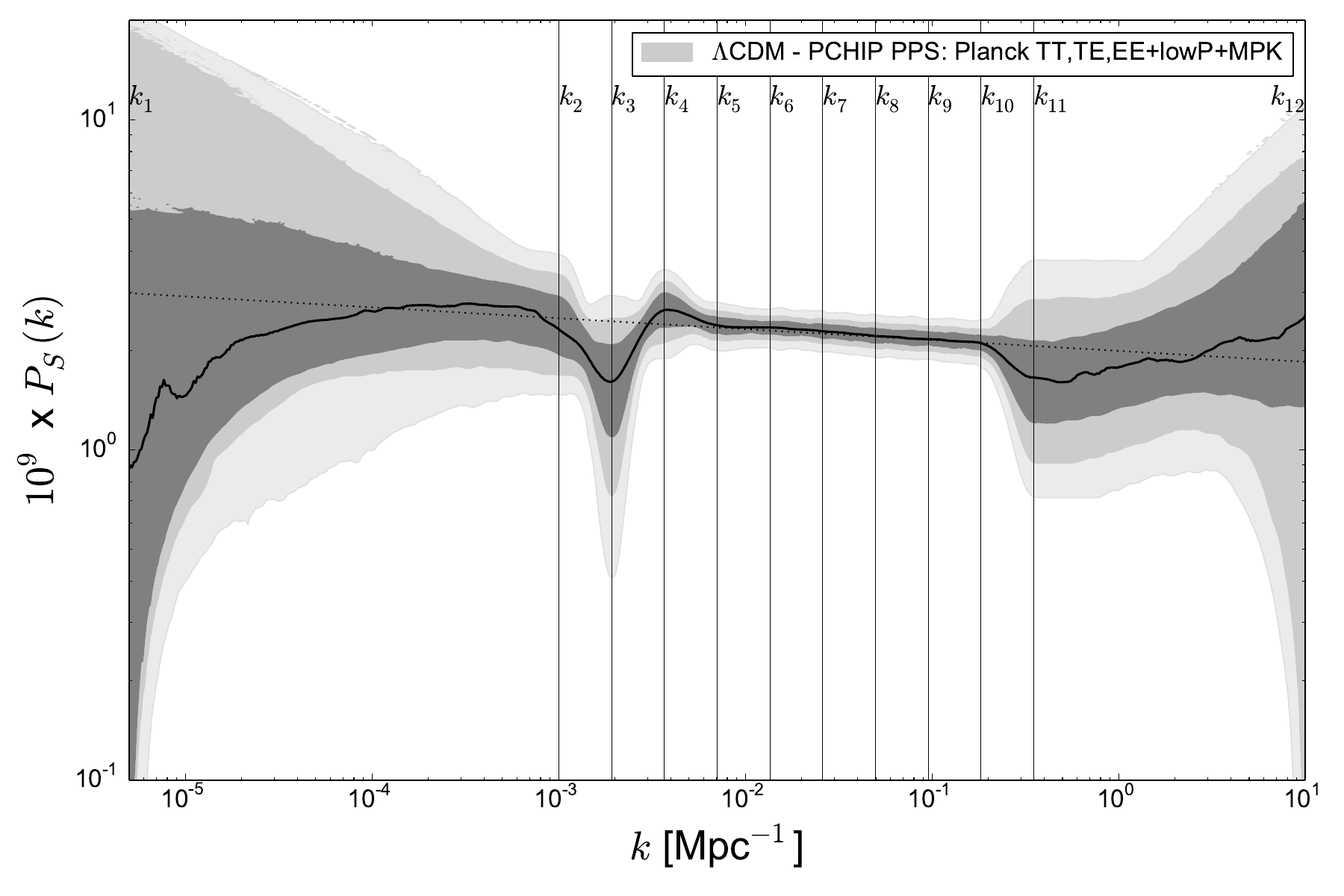}
  \caption[As in Fig.~\ref{fig:inflfreed_pps_planck}, but with
    the ``Planck~TT,TE,EE+lowP+MPkW'' dataset]
    {\label{fig:inflfreed_pps_planck_pol_mpkw} 
    As in Fig.~\ref{fig:inflfreed_pps_planck}, but with
    the ``Planck~TT,TE,EE+lowP+MPkW'' dataset.
    From Ref.~\cite{DiValentino:2016ikp}.}
\end{figure}

\begin{figure}
\centering
\includegraphics[width=\singlefigland]{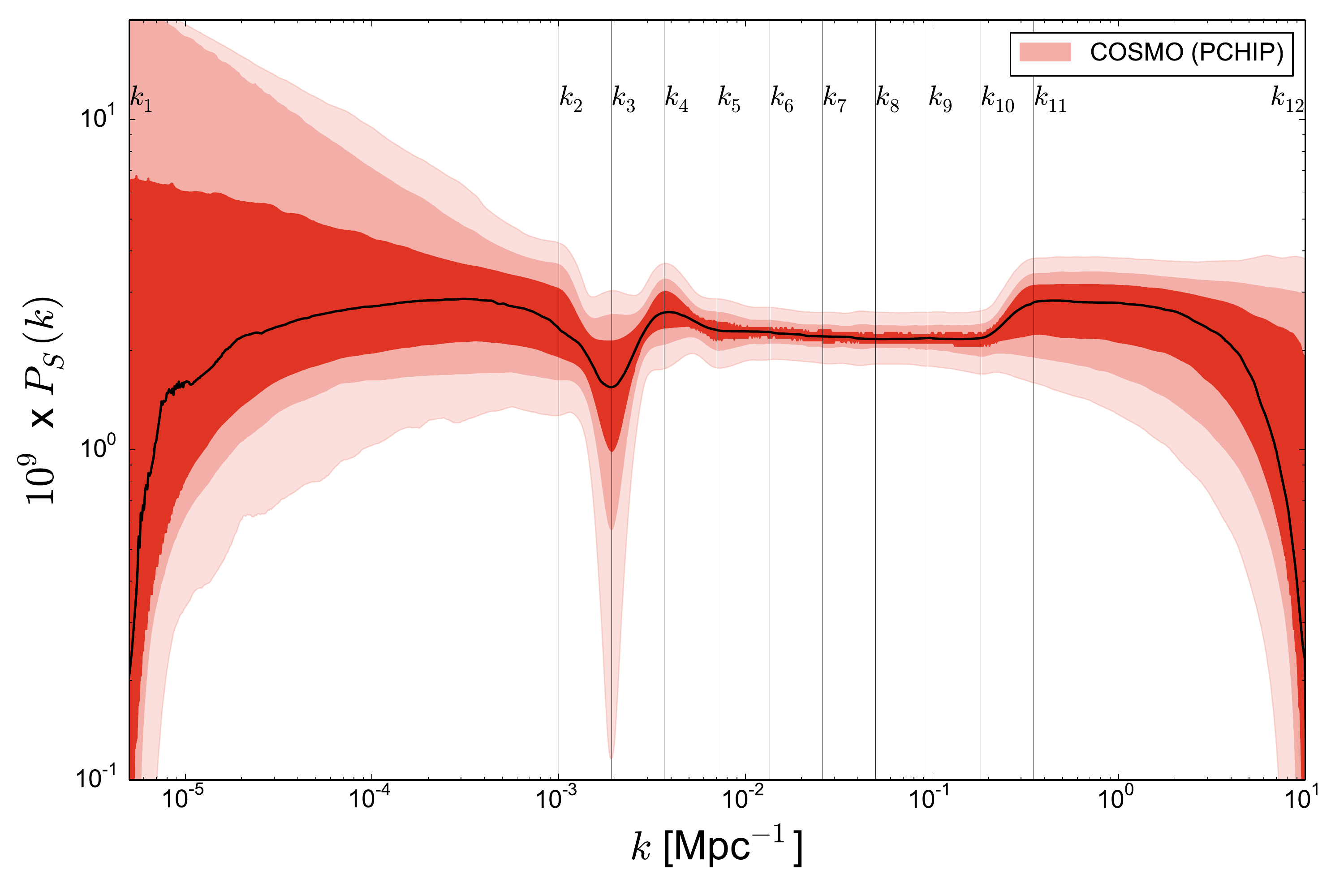}
\caption[Allowed
$1\sigma$,
$2\sigma$ and
$3\sigma$
bands of the
\pchip PPS obtained in the analyses
without (COSMO)
% and with (COSMO+SBL)
the SBL prior]{\label{fig:inflfreed_bands}
Allowed
$1\sigma$,
$2\sigma$ and
$3\sigma$
bands of the
\pchip PPS obtained in the analyses
without (COSMO)
% and with (COSMO+SBL)
the SBL prior.
The bands have been obtained by marginalizing
the posterior distribution for each value of the wavenumber $k$
in a fine grid.
The black curves correspond to the maximum
of the posterior distribution
for each value of $k$.
From Ref.~\cite{Gariazzo:2014dla}.
}
\end{figure}

We show the results for the 
dataset combinations that give the most interesting results
for the PPS, as for the bounds on the parameters that
we reported in Tab.~\ref{tab:inflfreed_lcdm}:
Planck~TT+lowP (Fig.~\ref{fig:inflfreed_pps_planck}),
Planck~TT,TE,EE+lowP (Fig.~\ref{fig:inflfreed_pps_planck_pol}) and
Planck~TT,TE,EE+lowP+MPkW
(Fig.~\ref{fig:inflfreed_pps_planck_pol_mpkw}).
Additionally, we show in Figure~\ref{fig:inflfreed_bands}
the results obtained in Ref.~\cite{Gariazzo:2014dla}
from the analyses of the former Planck 2013 spectra, together with
the WMAP polarization and the ACT/SPT data at high multipoles.
The plotted bands correspond to the constraints reported for the
COSMO analysis in Tab.~\ref{tab:inflfreed_freePPS}.
In each of these figures we show
the marginalized best-fitting reconstruction
of the \pchip~PPS (solid line),
the uncertainty bands at 68\%, 95\% and 99\% CL
at different gray-scales
and the best-fitting power-law PPS (dotted line)
as obtained by the Planck collaboration
for the \lcdm~model~\cite{Ade:2015xua}, as a comparison.
The bands are obtained marginalizing over all the values of the PPS
separately for each bin in $k$.

Notice that the nodes $\psj{1}$ and $\psj{12}$
are badly constrained, due to the fact that these nodes are selected
to cover a wide range of wavemodes for computational reasons,
but there are no available data to constrain them directly.
Also the node $\psj{11}$ is not very well constrained
by the Planck temperature
data, as it is possible to see in
Fig.~\ref{fig:inflfreed_pps_planck}.
The bounds on $\psj{11}$ and $\psj{12}$
can be improved with the inclusion of the
high-multipole polarization data (TE,EE),
for which the reconstructed PPS is presented in
Fig.~\ref{fig:inflfreed_pps_planck_pol}:
the improvement is
particularly significant for $\psj{11}$.
The inclusion of the MPkW data allows
to notably improve the constraints
on the last two nodes of the \pchip~PPS parameterization,
see Fig.~\ref{fig:inflfreed_pps_planck_pol_mpkw}.
The impact of the polarization on the nodes
at high $k$ is smaller than
the one of the matter power spectrum data,
since the MPkW dataset provides
stronger constraints on the smallest angular scales.
The situation is slightly different
for the PPS reconstruction presented in
Fig.~\ref{fig:inflfreed_bands}, for which
the tight constraints for \psj{11} and \psj{12}
arise from the CMB data at high multipoles, provided
by the ACT and SPT experiments (see Sec.~\ref{sec:cmb}).

The bounds on the nodes at small wavemodes ($\psj{1}$ to $\psj{4}$)
are almost insensitive to the inclusion of additional datasets
or to the change in the underlying cosmological model,
with only small variations well inside the 1$\sigma$ range
between the different results.
The error bars on the nodes are larger in this part of the spectrum,
since it corresponds to low multipoles of the CMB power spectra,
where the cosmic variance is larger.
In this part of the PPS we have the most evident deviations from the
simple power-law PPS.
The features are described by the node $\psj{3}$, 
for which the value corresponding to the power-law PPS is
approximately $2\sigma$ away from the reconstructed result,
and by the node $\psj{4}$, 
which is mildly discrepant with the power-law value
($1\sigma$ level).
These nodes describe the behavior of the CMB temperature spectrum
at low-$\ell$, where the observations of the Planck
and WMAP experiments
show a lack of power at $\ell\simeq20$ and an excess of power
at $\ell\simeq40$.
The detection of these features is in agreement with several previous
studies
\cite{Shafieloo:2003gf,Nicholson:2009pi,Hazra:2013ugu,Hazra:2014jwa,
Nicholson:2009zj,Hunt:2013bha,Hunt:2015iua,
Goswami:2013uja,Matsumiya:2001xj,Matsumiya:2002tx,
Kogo:2003yb,Kogo:2005qi,Nagata:2008tk,Ade:2015lrj,
Gariazzo:2014dla,DiValentino:2015zta}.
Since this behavior of the CMB spectrum at low multipoles
has been reported by analyses of both Planck and WMAP data,
it is unlikely that it is the consequence of some instrumental
systematics.
It is possible that this feature is simply the result of
a large statistical fluctuation in a region of the spectrum
where cosmic variance is very large.
On the other hand, the lack of power at a precise scale
can be the signal of some non-standard
inflationary mechanism that produced a non standard
spectrum for the initial scalar perturbations.
Future investigations will possibly clarify these properties of the PPS.

The central part of the reconstructed PPS, from
\psj{5} to \psj{10}, is very well constrained by the data.
In this range of wavemodes, no deviations from the power-law PPS
are visible, thus confirming the validity of the assumption that
the PPS is almost scale-invariant for a wide range of wavemodes.
This is also the region where the PPS shape is more sensitive
to the changes in the
\lcdm~model caused by its extensions.
As we can see from the results presented in previous sections,
the constraints on the nodes \psj{5} to \psj{10}
are different for each extension
of the \lcdm~model, in agreement with
the results obtained for $\ln[10^{10}A_s]$
and $n_s$ when considering the power-law PPS.
In the various tables, when presenting
the results on the power-law PPS,
we listed the values of the \pchip nodes that would correspond 
to the best-fitting $A_s$ and $n_s$,
to simplify the comparison with the \pchip PPS constraints.
These values are calculated using Eq.~\eqref{eq:psj_plpps}.
In the range between $k\simeq 0.007$ and $k\simeq0.2$,
the constraints in the \pchip nodes
correspond, for most of the cases,
to the values expected by the power-law PPS
analyses, within their allowed 1$\sigma$ range.
There are a few exceptions: for example,
in the \lcdm\ + \neff\ model and with the Planck~TT+lowP+BAO dataset,
the node \psj{10} deviates from the expected value corresponding
to the power-law PPS
by more than 1$\sigma$ (see Tab.~\ref{tab:inflfreed_nnu}).
This is a consequence of the large correlation and
the large variability range
that this dataset allows for \neff.
The inclusion of polarization data at high-$\ell$,
limiting the range for \neff, 
does not allow for these deviations from the power-law PPS.

% \begin{figure}
%   \centering
%   \includegraphics[width=\singlefigland,page=2]
%   {inflfreed/darkrad/lcdm_sPPS.pdf}
%   \caption{\label{fig:pps_planck_mpkw} 
%     As in Fig.~\ref{fig:pps_planck}, but for
%     the ``Planck~TT+lowP+MPkW'' dataset.}
% \end{figure}

\section{Discussion and Conclusions}
The description of the cosmological model
may require a non-standard
Primordial Power Spectrum (PPS) of scalar perturbations generated
during the inflationary phase at the beginning of the Universe.
Several analyses have considered the possible deviations
from the PPS power-law exploiting both the WMAP and the Planck
data measurements of the CMB temperature
power
spectrum~\cite{Shafieloo:2003gf,Nicholson:2009pi,Hazra:2013ugu,Hazra:2014jwa,
Nicholson:2009zj,Hunt:2013bha,Hunt:2015iua,
Goswami:2013uja,Matsumiya:2001xj,Matsumiya:2002tx,
Kogo:2003yb,Kogo:2005qi,Nagata:2008tk,Ade:2015lrj,
Gariazzo:2014dla,DiValentino:2015zta}.
Even if the significance of these deviations is small,
it leaves some freedom for the PPS assumed form. 
Here we test the robustness of the cosmological bounds
on several cosmological parameters when the PPS is allowed
to have a model-independent shape, that we describe using
a \pchip\ function to interpolate a series of twelve nodes \psj{j}.
Our results show that the constraints can significantly
change if one considers only the temperature spectrum
of the CMB in the data analyses,
since the free PPS form can be changed to compensate for the variations
in the cosmological parameters.
These degeneracies are broken by the inclusion
of the polarization spectra measured
by Planck.
In particular, we show that they are removed due to the inclusion
of the temperature-polarization cross-correlation
spectrum.
For this reason, we stress here the importance of including several 
datasets in the analyses, since they are
crucial for solving the possible degeneracies
between the PPS generated during inflation and the parameters that govern 
the subsequent evolution, in order to avoid misleading results.

We have explored the impact of a non-canonical PPS in several different
extensions of the \lcdm\ model, varying the effective number of relativistic
species and the masses of the active and the light sterile neutrinos.

Using the 2013 Planck data, we found that
the freedom of the form of the PPS does not affect significantly the
fitted values of the parameters in the \lcdm\ model,
while the results concerning the existence
of a sterile neutrino in the early Universe
can change drastically.
If we do not impose any prior on
the sterile neutrino mass $m_s$ from the results
of short-baseline oscillation experiments
(see Chapter~\ref{ch:nu}),
a larger value for the sterile neutrino contribution
$\DNeff$
to the effective number of relativistic degrees of freedom
before photon decoupling
is preferred in the \pchip PPS parameterization
with respect to the standard power-law parameterization.
The marginalized best fit of $\DNeff$ is moved towards one,
which corresponds to a fully thermalized sterile neutrino.
This shift corresponds to a tightening
of the cosmological preferred values for $m_s$.

In the analysis with a prior on $m_s$ obtained from the fit of
short-baseline oscillation experiments
\cite{Giunti:2013aea},
the freedom of the \pchip PPS affects only the bound on $\DNeff$,
because the allowed range of $m_s$
is strongly constrained by the SBL prior.
We found that a free form of the PPS allows
the existence in the early Universe of
a fully thermalized sterile neutrino with a mass of about 1 eV
\cite{Dolgov:2003sg,Cirelli:2004cz}.
This possibility is quite disfavored by
the analysis of cosmological data
with a power-law PPS
\cite{DiValentino:2013qma,Archidiacono:2013xxa,Mirizzi:2013gnd,
Gariazzo:2013gua,Archidiacono:2014apa}.
Hence,
the freedom of the PPS may allow us to reconcile
the cosmological data with
short-baseline neutrino oscillations
without the need of an additional mechanism which suppresses the
thermalization of the sterile neutrino
\cite{Chu:2006ua,Hannestad:2012ky,Mirizzi:2012we,
Saviano:2013ktj,Hannestad:2013wwj,Hannestad:2013ana,
Dasgupta:2013zpn,Bringmann:2013vra,Ko:2014bka,
Archidiacono:2014nda,Saviano:2014esa,Mirizzi:2014ama,
Rehagen:2014vna,Ho:2012br}.
The updated analyses that include the full temperature and polarization
data released in 2015 by Planck, however,
forbid this reconciliation,
since the CMB polarization at high multipoles breaks the degeneracies
between \neff\ and the \pchip\ nodes.

We studied then the degeneracies between the PPS shape and the different
cosmological parameters, separately.
We considered the most recent CMB data from the 2015 release of the Planck
collaboration.

Concerning the effective number of degrees of freedom \neff,
we find that the results are in good agreement
with the standard value of 3.046,
if one assumes the standard power-law PPS.
Increasing \neff\ has the main effect of increasing the Silk damping
of the CMB spectrum
at small scales and therefore it is easy change the PPS shape
at that scales to compensate the increased damping.
This results in a strong degeneracy
between the relevant \pchip\ PPS nodes and \neff.
As a consequence of volume effects in the Bayesian analyses,
the constraints on \neff\ are significantly loosened.
For some data combinations we obtain $\neff\simeq4.8$ allowed at 95\% CL.
However, the \neff\ effects can not be compensated by the \pchip\ nodes
in the polarization spectra,
in particular in the case of the TE cross-correlation.
This is the reason for which the inclusion of CMB polarization measurements
in the analyses allows to break the degeneracies and to restore
the \neff\ bounds very close to 3.046 for all the data combinations, with 
$\neff>3.5$ excluded at more than 95\% CL for all the datasets.

In the minimal three active massive neutrinos scenario,
the constraints on \mnu\ from the free PPS scenario
are relaxed with respect to the PPS power-law ones.
This is due to the degeneracy between \mnu\ and
the nodes $\psj{5}$ and $\psj{6}$,
that correspond to the scales at which the early Integrated Sachs-Wolfe
effect contributes to the CMB spectrum.
Also in this case these degeneracies are broken by the inclusion
of additional datasets, as the CMB polarization at high multipoles
and the BAO measurements.
The tightest limits we find is $\mnu<0.18$~eV (0.22~eV) at 95\% CL
from the combination of Planck~TT,TE,EE+lowP+BAO data,
when considering a power-law (\pchip) PPS.

Even if we presented only the results in the
\lcdm\ + \neff\ and \lcdm\ + \mnu\ models, similar constraints would be obtained
if the neutrino parameters were varied together.
The degeneracies with the PPS, in fact, are related to different scales.
The results in the 
\lcdm\ + \neff\ + \mnu\ and
\lcdm\ + \neff\ + \mnu\ +\meff{s} models are reported
in Ref.~\cite{DiValentino:2016ikp}.

From the MCMC analyses we have also the opportunity to reconstruct
and study the shape of the PPS.
We find that the reconstructed spectrum is perfectly described by a
power-law in the region between
$k\simeq0.007\mpcinv$ and $k\simeq0.2\mpcinv$,
but there are indications that a small dip (at $k\simeq0.002\mpcinv$)
and a statistically less relevant bump (at $k\simeq0.0035\mpcinv$)
appear at large scales.
These features are found both considering the WMAP and the Planck
CMB spectra.
If confirmed by future surveys, they will indicate that
the simplest inflationary model is not complete and some
new physical mechanism during inflation introduces
a scale dependency in the PPS.

In summary, we have shown that dangerous degeneracies among
the parameters of the \lcdm\ model (and its possible extensions)
and the PPS shape arise when considering CMB temperature power spectrum
measurements only.
Fortunately, these degeneracies disappear with the inclusion
of the CMB polarization data at high multipoles.
This is due to the fact that all these cosmological parameters influence
the TT, TE and EE spectra in different ways.
This confirms the robustness of both the \lcdm\ model and
the simplest inflationary models,
that predict a power-law PPS that successfully explains
the observations at small scales.
The large scale fluctuations of the CMB spectrum, however,
seem to point towards
something new in the scenarios that describe inflation.
It must be clarified whether these features  are indicating
a more complicated inflationary mechanism 
or they are instead simple statistical fluctuations of the CMB temperature
anisotropies.

%!TeX root=main.tex 
\chapter{Thermal Axion Properties}
\label{ch:ther_ax}
\chapterprecis{This Chapter is based on
Refs.~\protect\cite{DiValentino:2015zta,DiValentino:2016ikp}.}

In the previous Chapters we discussed mainly the properties
of active and sterile neutrinos.
Among the possible candidates of hot dark matter, however, 
other particles can be listed.
In this Chapter we present the case of the thermal axions,
which are introduced in Section~\ref{sec:ax_intro}.
In the following Section~\ref{sec:ax_method}
we present the cosmological model and the data
that we consider in our analyses.
The results are presented in Sections~\ref{sec:ax_res_ma}
for the axion mass alone, 
and in Section~\ref{sec:ax_mnu}
for the joint constraints on the active neutrino and thermal axion masses.

\section{Introduction}
\label{sec:ax_intro}
% PRD91

The axion field is the solution proposed by Peccei and
Quinn~\cite{Peccei:1977hh,Peccei:1977ur,Weinberg:1977ma,
Wilczek:1977pj} 
to solve the strong CP problem in Quantum ChromoDynamics,
by adding a new global Peccei-Quinn symmetry $U(1)_{PQ}$ that,
when spontaneously broken at an energy scale $f_a$,
generates a Pseudo-Nambu-Goldstone boson, the axion particle. 
Non-thermal axions, as those produced by the misalignment mechanism, 
while being a negligible hot dark matter candidate,
may constitute a fraction or the total cold dark matter
component of the Universe.
We do not explore such a possibility here. 
Thermal axions \cite{Turner:1986tb,Chang:1993gm,Masso:2002np}, instead,
affect the cosmological observables in a very
similar way to that induced by the presence of neutrino masses and/or
extra sterile neutrino species. 
Massive thermal axions as hot relics affect large scale structures,
since they only cluster at scales larger than
their free-streaming scale when they become non-relativistic,
suppressing therefore structure formation at small scales. 
Concerning the Cosmic Microwave Background (CMB) physics, 
an axion mass leads to a signature in the CMB photon
temperature anisotropies via the early integrated Sachs-Wolfe effect.
In addition, extra light species as thermal axions
contribute to the dark radiation content of the Universe, or, in
other words, lead to an increase of the effective number
of relativistic degrees of freedom $\Neff$, defined 
in Eq.~\eqref{eq:neff}.
The extra contribution
to $\Neff$ arising from thermal axions can modify both the
CMB anisotropies (via Silk damping) and
the primordial abundances of light elements predicted by Big Bang
Nucleosynthesis. 
The former cosmological signatures of thermal axions
have been extensively
exploited in the literature to derive bounds on the thermal axion
mass, see
Refs.~\cite{Melchiorri:2007cd,Hannestad:2007dd,
Hannestad:2010yi,Archidiacono:2013cha,Giusarma:2014zza}.

The most relevant process for the axion thermalization purpose
is the interaction with the pion \cite{Chang:1993gm}:
\begin{equation}
\label{eq:axpion}
 \pi + \pi \rightarrow \pi+a.
\end{equation}
Assuming this process for the interaction, 
the axion coupling constant $f_a$ can be related
to the axion mass by the following relation
\cite{Agashe:2014kda}:
\begin{equation}
\label{eq:axionmass}
m_a
=
\frac{f_\pi m_\pi}{f_a} \frac{\sqrt{R}}{1 + R}
=
0.6\ \mathrm{eV}\ \frac{10^7\, \mathrm{GeV}}{f_a}~,
\end{equation}
where $m_\pi=135$~MeV is the pion mass,
$R=0.56$ is the up-to-down quark masses ratio,
and
$f_\pi = 92$~MeV is the pion decay constant.
To consider other values of $R$ in the range $0.35-0.60$
\cite{Agashe:2014kda} does not affect in a significant way
this relationship \cite{Archidiacono:2015mda}.

Axions decouple from the primordial plasma
at a temperature $T_D$,
when the thermally averaged interaction rate $\Gamma(T)$
of the interaction \eqref{eq:axpion},
falls below the expansion rate of the Universe $H(T)$.
This decoupling process is known as
the freeze out condition for a thermal relic, and is given by:
\begin{equation}
\Gamma (T_D) = H (T_D)~,
\label{eq:decouplinga}
\end{equation}
where \cite{Agashe:2014kda}
\begin{equation}
\Gamma = \frac{3}{1024\pi^5}\frac{1}{f_a^2f_{\pi}^2}C_{a\pi}^2 I_a~.
\end{equation}
In this formula the axion-pion coupling constant is
$C_{a\pi} = (1-R)/[3(1+R)]$ \cite{Chang:1993gm}.
The integral $I_a$ can be expressed in the following way
\cite{Chang:1993gm}:
\be
I_a 
=
n_a^{-1}T^8\int dx_1dx_2\frac{x_1^2x_2^2}{y_1y_2}
f(y_1)f(y_2) 
% \times
\int^{1}_{-1}
d\omega\frac{(s-m_{\pi}^2)^3(5s-2m_{\pi}^2)}{s^2T^4}~,
\ee
with $n_a=(\zeta_{3}/\pi^2) T^3$ the number density for axions
in thermal equilibrium and the function $f(y)=1/(e^y-1)$
the pion thermal distribution.
Moreover, we have three different kinematical variables,
$x_i=|\vec{p}_i|/T$,  $y_i=E_i/T$ ($i=1,2$) and
$s=2(m_{\pi}^2+T^2(y_1y_2-x_1x_2\omega))$. 

\begin{figure}[!t]
\centering
\includegraphics[width=\singlefigbig]{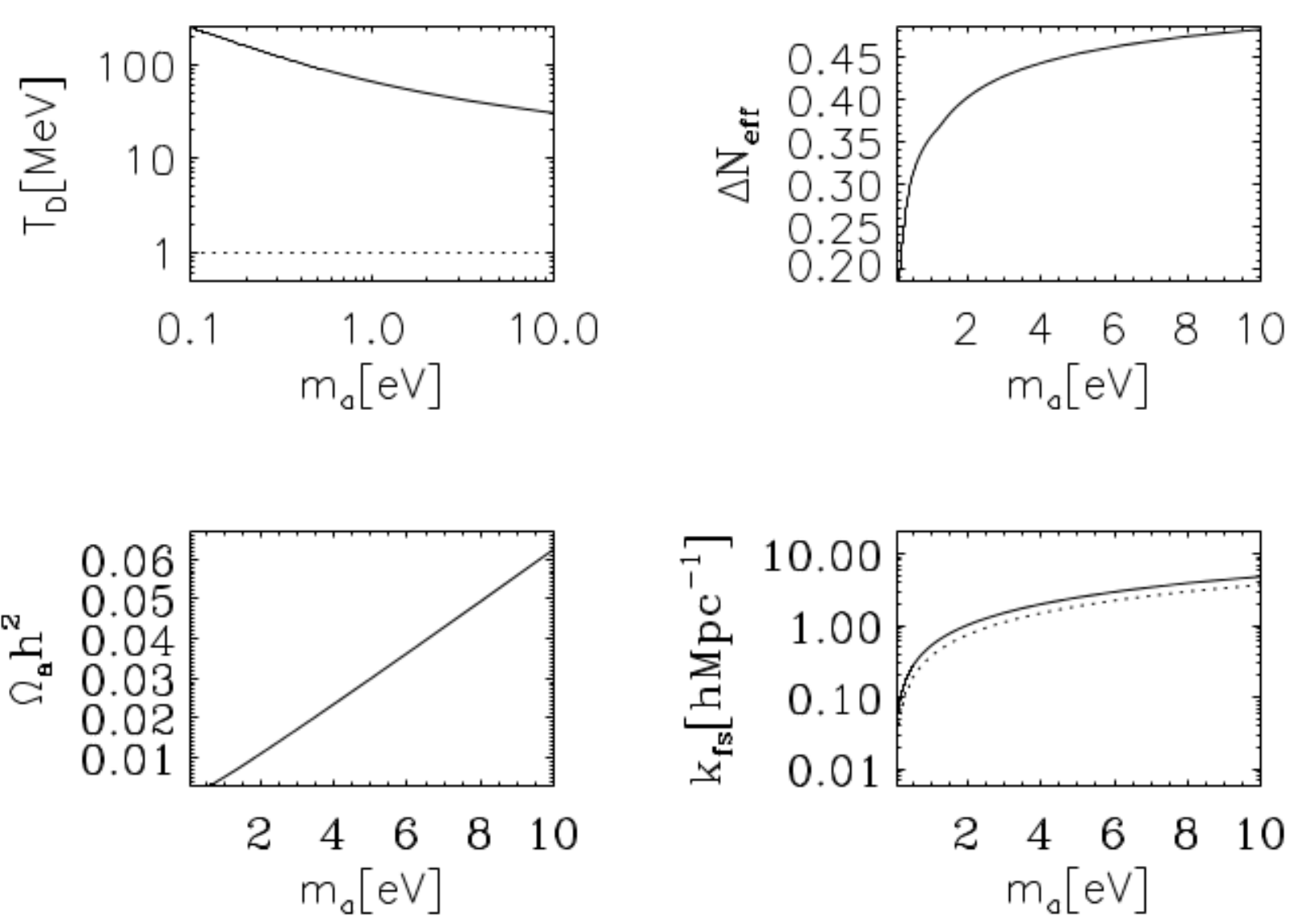}
\caption[Cosmological quantities related to the thermal axion,
as a function of the axion mass]%
{The upper left panel shows the temperature of decoupling
as a function of the axion mass (solid curve),
as well as the Big Bang Nucleosynthesis temperature,
$T_{\textrm{BBN}}\simeq 1$~MeV (dashed curve).
The upper right panel shows the axion contribution
to the extra dark radiation content of the Universe,
while the bottom right plot depicts the free-streaming scale
of an axion (solid curve) or a neutrino (dashed curve)
versus the axion/neutrino mass, in~eV.
The bottom left panel shows the current axion mass-energy density
as a function of the axion mass.
From~Ref.~\protect\cite{DiValentino:2015zta}.}
\label{fig:ax_91_maref}
\end{figure}

The freeze out equation above, Eq.~\eqref{eq:decouplinga},
can be numerically solved \cite{Hannestad:2005df},
obtaining the axion decoupling temperature
$T_D$ as a function of the axion mass $m_a$.
The upper left panel of Fig.~\ref{fig:ax_91_maref} shows
the axion decoupling temperature as a function of the axion mass,
in~eV.
Notice that the higher is the axion mass,
the lower is the temperature of decoupling.
Afterwards it is possible to obtain the present axion number density,
that is related to the current photon density $n_\gamma$
by the following equation \cite{Hannestad:2005df}: 
\begin{equation}
n_a
=
\frac{g_{\star S}(T_0)}{g_{\star S}(T_D)}
\times \frac{n_\gamma}{2}~, 
\label{eq:numberdens}
\end{equation}
where $g_{\star S}$ is the number of \emph{entropic} degrees of
freedom and $g_{\star S}(T_0) = 3.91$. 

The contribution of the relic axion
to the total mass-energy density of the Universe is given
by the product of the axion mass times the axion number density.
The quantity $\Omega_a h^2$ at the present epoch is depicted
in the bottom left panel of Fig.~\ref{fig:ax_91_maref}.
Notice that
a $1$~eV axion will give rise to $\Omega_a h^2\simeq 0.005$ today,
while a neutrino of the same mass will contribute
to the total mass-energy density of the Universe with
$\Omega_\nu h^2\simeq 0.01$.
Notice however that $\Omega_a h^2$ represents the contribution
from relic, thermal axion states only.
Non-thermal processes, as the misalignment production,
could also produce a non-thermal axion population
which we do not consider here.
See Ref.~\cite{DiValentino:2015wba}
for the most recent cosmological constraints on such scenario.

The thermal axion can have, as massive neutrinos,
the transition between the relativistic
to the non relativistic regime.
When the thermal axion is still a relativistic particle
it increases the effective number of relativistic
degrees of freedom $\Neff$,
enhancing the amount of radiation in the Universe.
The contribution to \Neff\ from the thermal axion is given by
\cite{Hannestad:2005df}:
\begin{equation}
\label{eq:ax_dneff}
\DNeff
=
\frac{4}{7}\left(\frac{3}{2}\frac{n_a}{n_\nu}\right)^{4/3},
\end{equation}
where $n_a$ is given by Eq.~\eqref{eq:numberdens} and $n_\nu$ refers
to the present neutrino plus antineutrino number density per flavor.
The upper right panel of Fig.~\ref{fig:ax_91_maref} shows
the axion contribution to the radiation content
of the Universe as a function of the axion mass.
Notice that the extra dark radiation arising from a $1$~eV axion
is still compatible (at $95\%$~CL) with the most recent measurements
of $\Neff$ from the Planck mission~\cite{Ade:2015xua}.

The last crucial cosmological quantity
is the axion free streaming scale,
i.e.\ the wavenumber $k\fs{}$ below which
the axion density perturbations will contribute
to clustering once the axion is a non-relativistic particle.
This scale is illustrated in Fig.~\ref{fig:ax_91_maref}
(solid line),
in the bottom right panel, together with that corresponding
to a neutrino of the same mass (dashed line).
Notice that they cover the same scales for our choice of priors
for $m_a$ and $\sum m_\nu$ and therefore one can expect
a large correlation between these two quantities in measurements
of galaxy clustering.

Several papers in the literature provide bounds
on the thermal axion mass, see for example
Refs.~\cite{Melchiorri:2007cd,Hannestad:2007dd,Hannestad:2010yi,
Archidiacono:2013cha,Giusarma:2014zza,DiValentino:2014zna,
DiValentino:2015wba}.
Here we present the results obtained in
Ref.~\cite{DiValentino:2015zta,DiValentino:2016ikp},
studying the constraints on the thermal axion mass,
and testing their robustness against the assumption
of a free Primordial Power Spectrum (PPS) of
scalar perturbations, as we did for the neutrino properties
in the previous Chapter.
In Section \ref{sec:ax_mnu} we also take into account the fact
that thermal axions and massive neutrinos affect
the cosmological observables in a very similar way,
and we will consider the sum of the neutrino masses and
the axion mass free to vary at the same time.

\section{Method}
\label{sec:ax_method}

\subsection{Cosmological model}
\label{sub:ax_model}

The thermal axion can be parameterized through its 
coupling constant $f_a$ or through its mass $m_a$.
Even if they are equivalent (see Eq.~\eqref{eq:axionmass}),
for our purposes
it is more convenient to use the axion mass $m_a$.
All the other cosmological quantities can be derived
as a function of the axion mass $m_a$,
as we showed in the previous Section and in Fig.~\ref{fig:ax_91_maref}.

The baseline scenario we consider here is the $\Lambda$CDM model,
extended to include the thermal axion.
We also adopt the \pchip\ PPS prescriptions presented in
Section~\ref{sec:inflfreed_ppsparam}.
When considering the \pchip\ PPS,
for the numerical analyses 
we use the following set of parameters:
\begin{equation}\label{parameterPPS}
\{\omega_b,\omega_c, \theta, \tau, m_a, P_{s,1}, \ldots, P_{s,12}\}~,
\end{equation}
where the cosmological parameters are the same presented
in Section~\ref{sec:inflfreed_modeldata},
with the only exception of
$m_a$.
The $P_{s,1}, \ldots, P_{s,12}$ nodes describe the \pchip PPS
(see Section~\ref{sec:inflfreed_ppsparam}).
We shall consider a scenario in which massive neutrinos
are also present,
to explore the expected degeneracy between
the sum of the neutrino masses and
the thermal axion mass~\cite{Giusarma:2014zza},
in Section~\ref{sec:ax_mnu}. 

In order to compare the results obtained with the \pchip PPS
to the results obtained with the usual power-law PPS model,
we describe the latter case with the following set of parameters:
\begin{equation}\label{parameterPL}
\{\omega_b,\omega_c, \theta, \tau, m_a, n_s, \log[10^{10}A_{s}]\}~,
\end{equation}
where $n_s$ and $A_{s}$ are the spectral index and
the amplitude of the scalar power-law PPS
written in Eq.~\eqref{eq:plPPS}
and the other parameters are the same ones described above.

As we discussed extensively the constraints on the reconstructed PPS
in Section~\ref{sec:inflfreed_ppsconstr},
in this Chapter
we will not focus on the constraints obtained for the nodes \psj{i},
since they are very similar to those already presented.

\subsection{Cosmological measurements}
\label{sub:ax_data}

Our baseline data set consists of CMB measurements.
We will adopt the same datasets presented in
Section~\ref{sec:inflfreed_modeldata}, and we will use as baseline
datasets the combinations \textbf{Planck~TT+lowP} and
\textbf{Planck~TT,TE,EE+lowP}.
Additionally,
we will consider these two CMB datasets in combination
with the
\textbf{BAO}, \textbf{MPkW} and \textbf{lensing} datasets.

We will also stress the role that the thermal axion can have
in solving the tension between local and cosmological determinations
of $\sigma_8$.
In this case we will indicate with \textbf{CMB}
the combination of the
temperature data from the 2013 release of the Planck satellite
\cite{Ade:2013sjv,Ade:2013kta},
the WMAP 9-year polarization measurements \cite{Bennett:2012zja}
and
the high multipole data from the
South Pole Telescope (SPT) \cite{Reichardt:2011yv} and
the Atacama Cosmology Telescope (ACT) \cite{Das:2013zf} experiments.
\textbf{HST} indicates
a gaussian prior on the Hubble constant
$H_0=70.6\pm3.3\Hou$~\cite{Efstathiou:2013via}.
We will present the results obtained studying the
weak lensing measurements from CFHTLenS (\textbf{CFHT})
\cite{Heymans:2013fya},
described in Section~\ref{sec:shear},
and
on the cluster normalization condition
as measured by
the Planck Sunyaev-Zel'dovich (\textbf{PSZ})
2013 catalogue~\cite{Ade:2013lmv},
obtained
using both the assumption of a fixed mass bias and a free mass bias
(see Section~\ref{sec:cluster}).

\begin{figure}[!t]
\begin{center}
\includegraphics[width=\singlefigland]{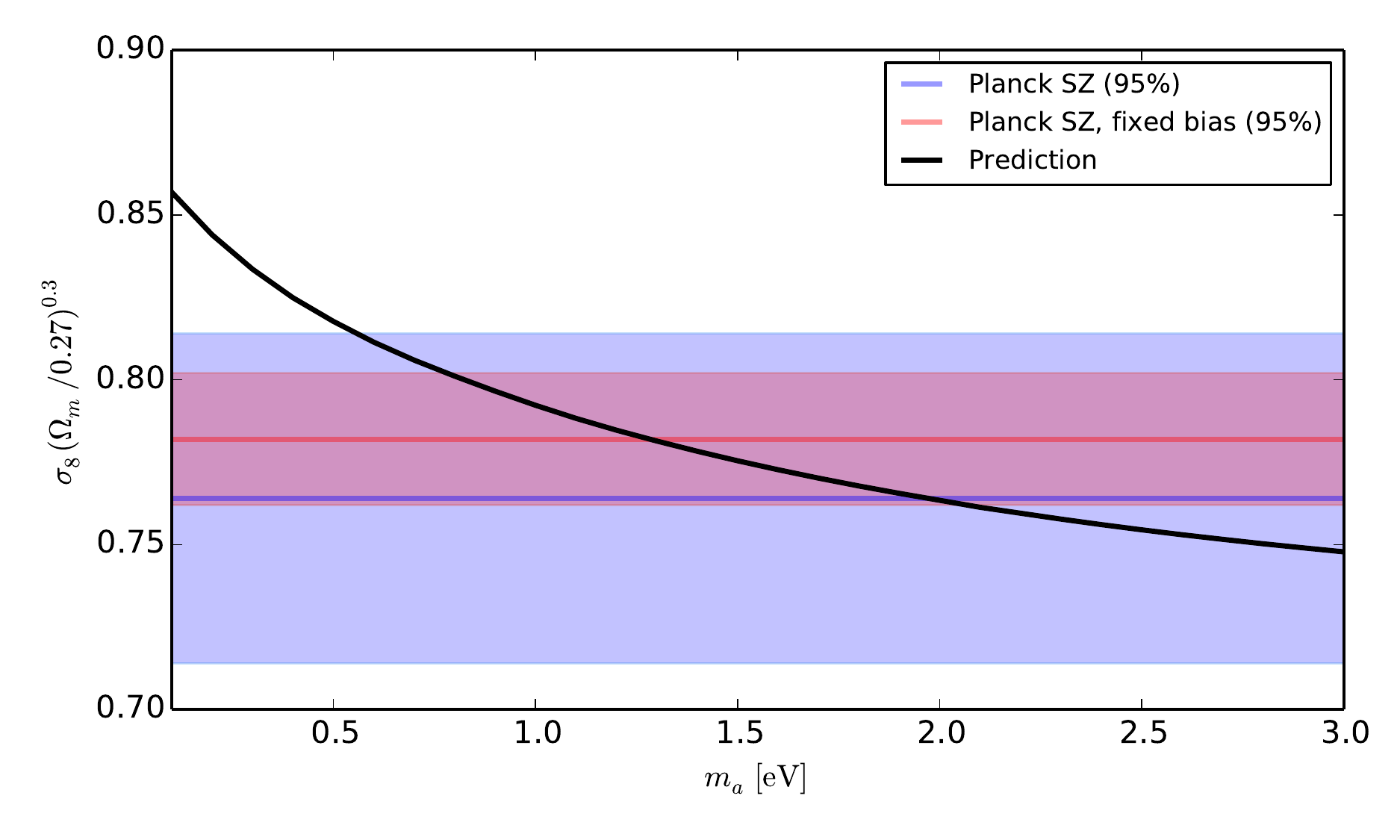}
\end{center}
 \caption[Cluster normalization condition as a function of
 the thermal axion mass]%
 {Cluster normalization condition,
 $\sigma_8 (\Omega_m/0.27)^{0.3}$,
 as a function of the thermal axion mass.
 We also show the current PSZ measurements~\protect\cite{Ade:2013lmv}
 with their associated $95\%$~CL uncertainties.
 From~Ref.~\protect\cite{DiValentino:2015zta}.}
\label{fig:ax_91_sigma8}
\end{figure}

Figure \ref{fig:ax_91_sigma8} illustrates the prediction
for the cluster normalization condition,
$\sigma_8 (\Omega_m/0.27)^{0.3}$,
as a function of the thermal axion mass.
We also show the PSZ measurements~\cite{Ade:2013lmv}
with their associated
$95\%$~CL uncertainties,
including those in which the cluster mass bias parameter is fixed.
Notice that the normalization condition decreases
as the axion mass increases,
as a consequence of the free-streaming nature of the axion:
the larger is the axion mass,
the larger is the reduction in the matter power spectra,
as it happens for massive neutrinos.

\begin{figure}[!t]
\begin{center}
\includegraphics[width=\singlefigland]{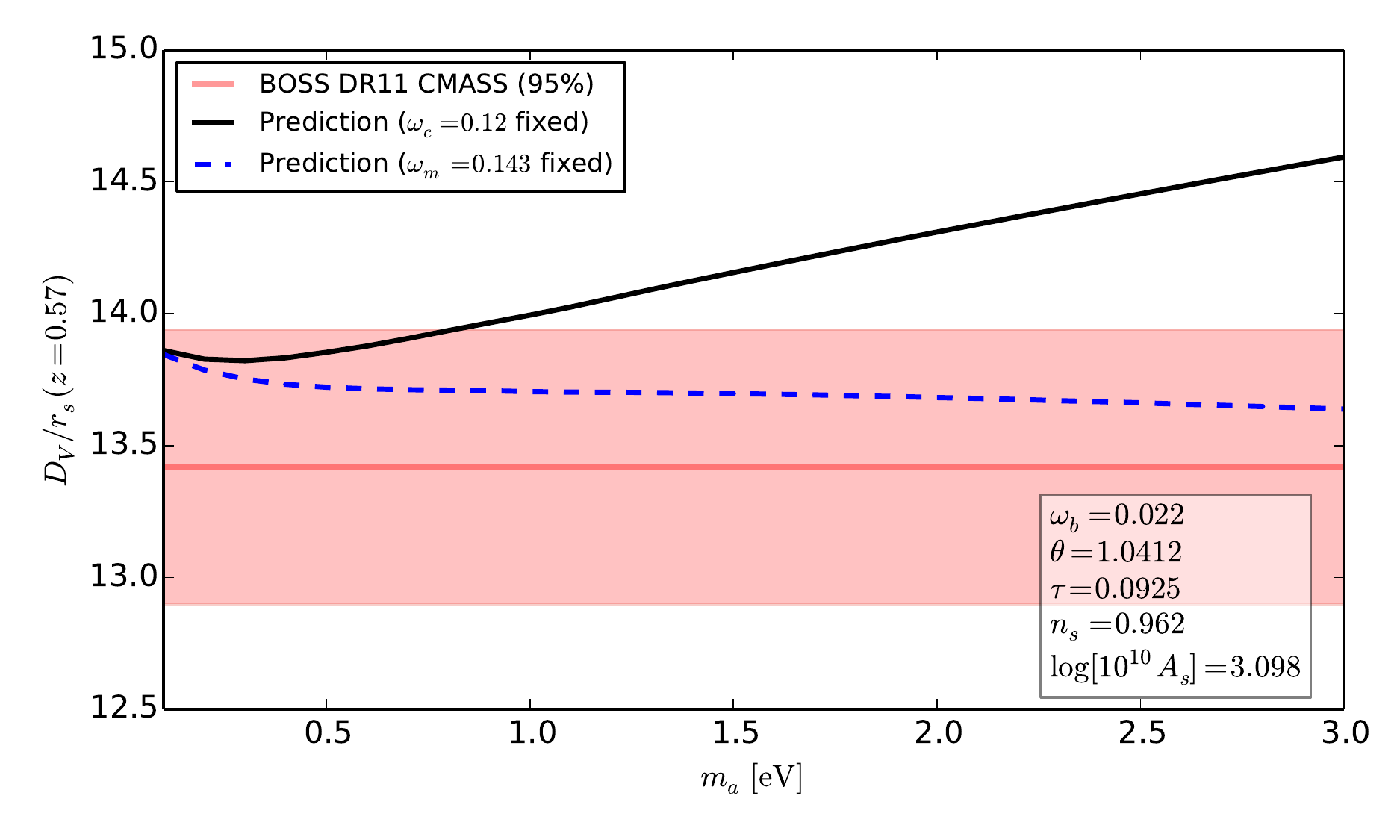}
\end{center}
\caption[The spherically averaged BAO distance
as a function of the axion mass, fixing the CDM or the total matter
energy density]%
{The solid black line depicts
the spherically averaged BAO distance $D_V(z)$
as a function of the axion mass at a redshift of $z=0.57$,
after keeping fixed all the remaining cosmological parameters,
included the cold dark matter energy density.
The dashed blue line depicts the equivalent obtained keeping fixed
the total matter mass-energy density.
The bands show the measurement from the BOSS experiment (DR11)
\protect\cite{Anderson:2013zyy}
with its associated $95\%$~CL error.
From Ref.~\protect\cite{DiValentino:2015zta}.}
\label{fig:ax_91_dvboss}
\end{figure}

Concerning the BAO constraints,
we want to point out an interesting effect that affects the results
that we will present.
Figure \ref{fig:ax_91_dvboss} illustrates
the spherically averaged BAO distance,
$D_V(z) \propto D^2_A(z)/H(z)$ at a redshift of $z=0.57$
as a function of the axion mass,
as well as the measurement from the BOSS experiment
with $95\%$~CL error bars~\cite{Anderson:2013zyy}.
Notice that, from background measurements only,
there exists a strong degeneracy between
the CDM and the axion mass-energy densities.
The solid black line in Fig.~\ref{fig:ax_91_dvboss}
shows the spherically averaged BAO distance
if all the cosmological parameters are fixed, including $\omega_c$.
The spherically averaged BAO distance deviates strongly 
from the $\Lambda$CDM prediction.
However, if $\omega_c$ is varied while $m_a$ is changed,
in order to keep the total matter mass-energy density constant
the spherically averaged BAO distance approaches
its expected value in a $\Lambda$CDM cosmology
(see the dotted blue line in Fig.~\ref{fig:ax_91_dvboss}).

\begin{figure}[!t]
\includegraphics[width=\halfwidth]{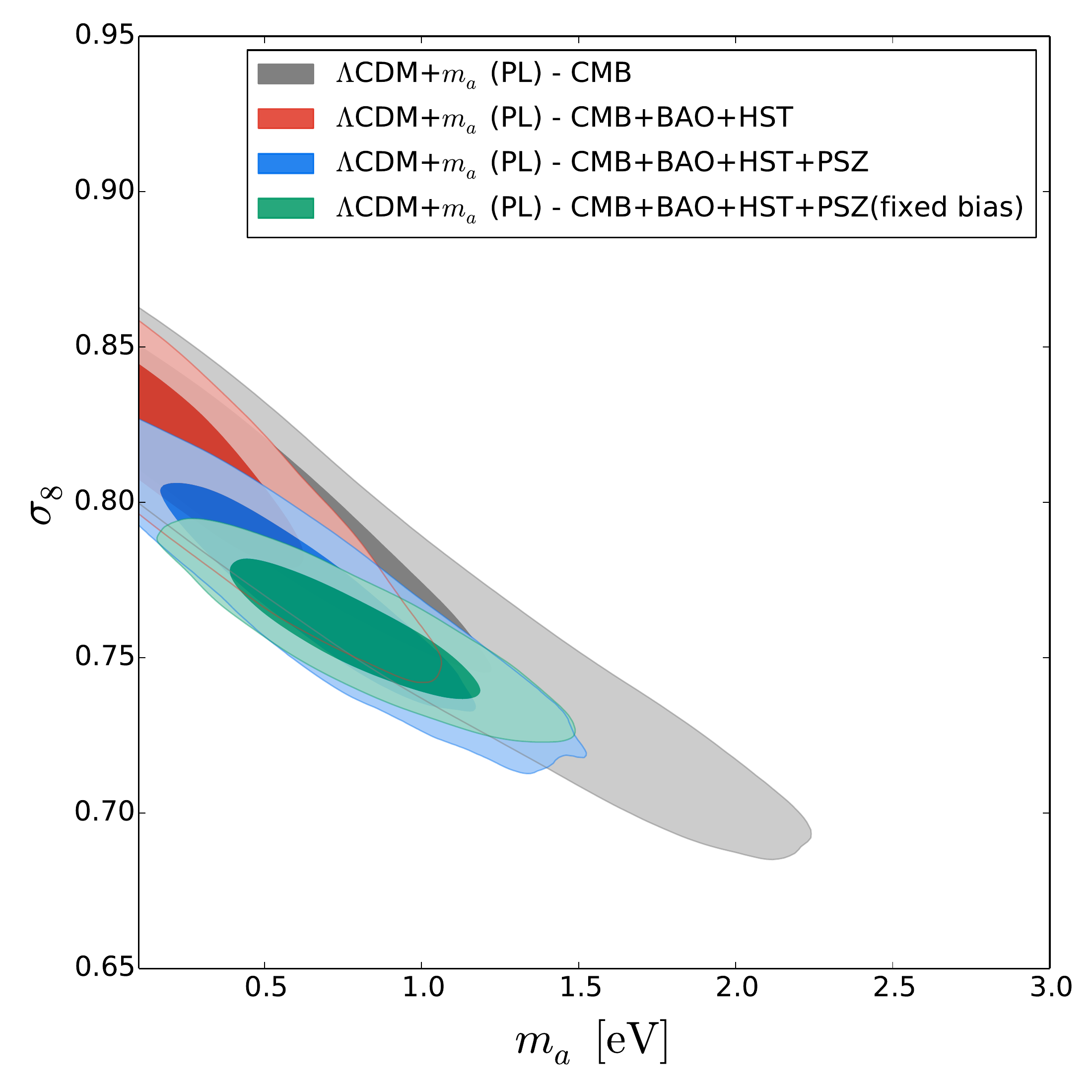}
\includegraphics[width=\halfwidth]{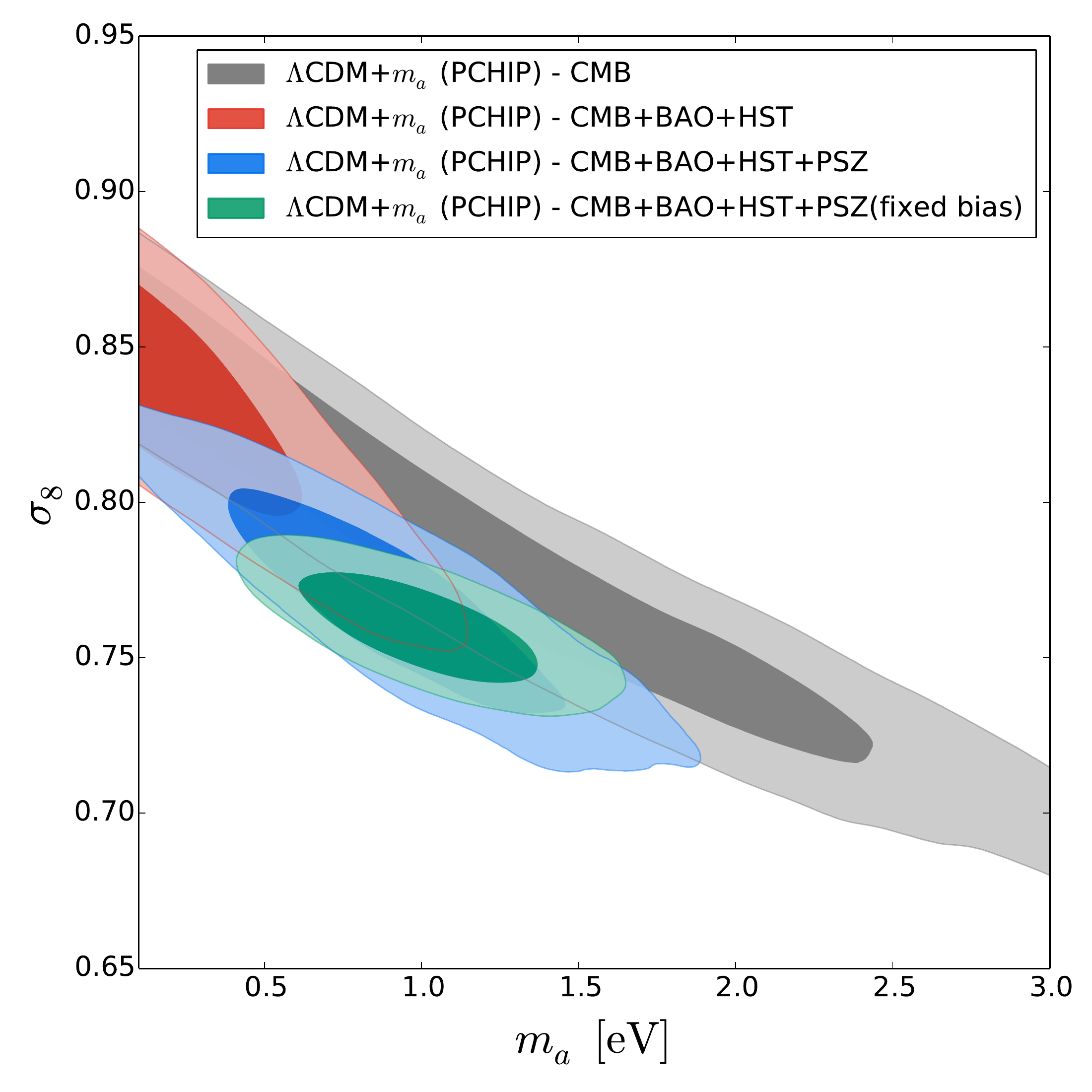}
\caption[Marginalized 2D constraints in the ($m_a$, $\sigma_8$)
plane from the \lcdm\ + $m_a$ model,
with a power-law or a \pchip\ PPS]
{\label{fig:ax_91_ma_pchip}
$68\%$ and $95\%$~CL allowed
regions in the ($m_a$, $\sigma_8$) plane for different possible
data combinations,
when a power-law (left panel) or
a \pchip{} (right panel) PPS is assumed.
From Ref.~\cite{DiValentino:2015zta}.}
\end{figure}

\section{Constraints on the Thermal Axion Mass}
\label{sec:ax_res_ma}

\begin{figure}
\centering
\includegraphics[width=\singlefigland]{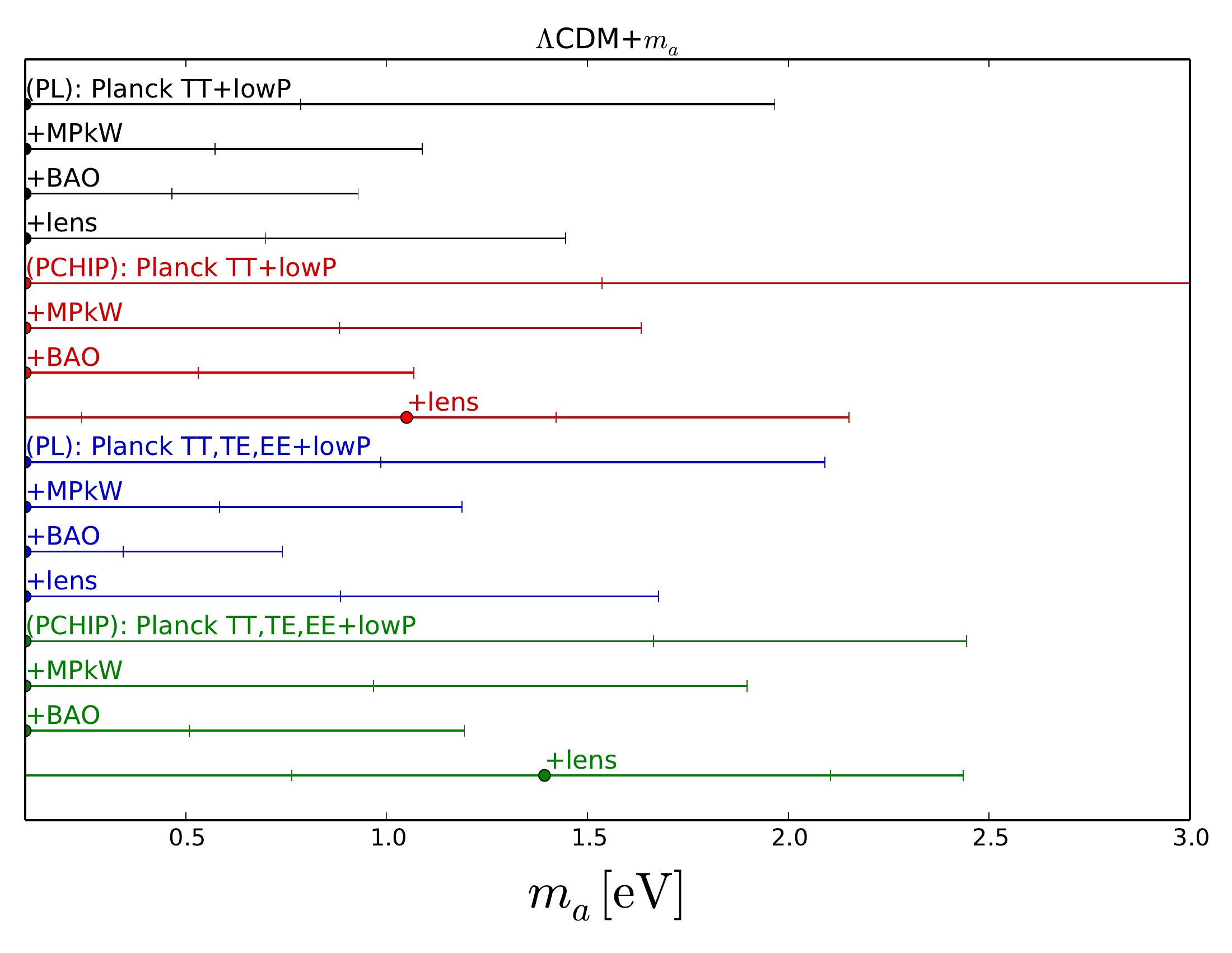}
\caption[As Fig.~\ref{fig:inflfreed_nnu_bars} but in the context of the
 \lcdm\ + $m_a$ model, focusing on the thermal axion mass $m_a$ parameter]
 {As Fig.~\ref{fig:inflfreed_nnu_bars} but in the context of the
 \lcdm\ + $m_a$ model, focusing on the thermal axion mass $m_a$ parameter.
 From Ref.~\cite{DiValentino:2016ikp}.}
 \label{fig:ax_ma_bars}
\end{figure}

Tables \ref{tab:ax_ma} and \ref{tab:ax_ma_pol} summarize our results
for the extended \lcdm\ + $m_a$ scenario, comparing,
for each dataset considered here,
the constraints arising in the power-law PPS scheme to the bounds obtained 
in the \pchip PPS formalism.
We can observe that the bounds on the axion mass are relaxed in the \pchip PPS
scenario, as illustrated in Fig.~\ref{fig:ax_ma_bars}
and in Tabs.~\ref{tab:ax_ma}
and \ref{tab:ax_ma_pol}.
This effect is related to the relaxed bound we have on $\neff$
when letting it free to vary in an extended \lcdm\ + $\neff$ scenario
that we discussed in Section~\ref{sec:inflfreed_nnu}.
From the results presented in  Tab.~\ref{tab:inflfreed_nnu},
we found $\neff=3.40 _{-1.43}^{+1.50}$ at $95\%$~CL
for the \pchip PPS parameterization, implying that the \pchip formalism
favors extra dark radiation, and therefore a higher axion mass is allowed.
As a consequence, we find that the axion mass is totally unconstrained
using the Planck~TT+lowP data in the \pchip PPS approach,
with respect to the bound $m_a<1.97$~eV at $95\%$~CL
we have for the standard power-law case.

However, when considering the Planck~TT,TE,EE+lowP dataset
for the \lcdm\ + $\neff$ model,
we find $\neff=2.99 _{-0.39}^{+0.41}$ at $95\%$~CL
for the power-law PPS, and $\neff=2.96_{-0.48}^{+0.49}$
for the \pchip PPS approach (see Tab.~\ref{tab:inflfreed_nnu_pol}),
perfectly in agreement with the standard value $\neff=3.046$.
First of all, this implies that the axion mass constraints arising
from high-$\ell$ polarization data are slightly weaker than those obtained
with Planck~TT+lowP dataset in the power-law PPS formalism.
In fact, the weakening of these bounds is driven by the fact
that $\neff$ is forced to be greater than standard value,
as discussed more in detail in the next Section.
Secondly, this means that the \pchip parametrization of the PPS
no longer favors an extra dark radiation component,
and the effective neutrino number is perfectly in agreement
with the bounds obtained by the Planck collaboration.
Therefore, these tighter values in the \pchip approach
will lead to stronger constraints on the thermal axion mass from CMB data only,
finding $m_a<2.44$~eV with the \pchip PPS,
mildly larger than the bound $m_a<2.09$~eV obtained within the power-law PPS,
both at $95\%$~CL.

\begin{table}[p]
\resizebox{1\textwidth}{!}{
\renewcommand{\arraystretch}{1.2}
\begin{tabular}{|c||c|c||c|c||c|c||c|c|}
\hline
Parameter	&\multicolumn{2}{c||}{Planck TT+lowP}	&\multicolumn{2}{c||}{Planck TT+lowP+MPkW}	&\multicolumn{2}{c||}{Planck TT+lowP+BAO} &\multicolumn{2}{c|}{Planck TT+lowP+lensing}\\
\hline\hline
$100\Omega_b h^2$
&$2.245_{-0.046}^{+0.048}$	&$2.178_{-0.079}^{+0.080}$	
&$2.240_{-0.043}^{+0.045}$	&$2.191_{-0.071}^{+0.074}$	
&$2.248_{-0.040}^{+0.043}$	&$2.224_{-0.062}^{+0.065}$
&$2.245_{-0.047}^{+0.046}$	&$2.182_{-0.077}^{+0.082}$\\

$\Omega_c h^2$	&$0.1229_{-0.0047}^{+0.0049}$	&$0.1267_{-0.0061}^{+0.0062}$	
&$0.1234_{-0.0043}^{+0.0045}$	&$0.1262_{-0.0056}^{+0.0058}$	
&$0.1219_{-0.0028}^{+0.0027}$	&$0.1222\pm0.0032$
&$0.1222_{-0.0044}^{+0.0043}$	&$0.1253_{-0.0059}^{+0.0058}$\\

$100\theta$	
&$1.041\pm0.001$	
&$1.0399\pm0.0011$	
&$1.0405_{-0.0010}^{+0.0009}$	
&$1.0401_{-0.0011}^{+0.0010}$	
&$1.0407\pm0.0008$	
&$1.0406_{-0.0009}^{+0.0008}$
&$1.0406\pm0.0009$	
&$1.0401\pm0.0010$	\\

$\tau$	&$0.088_{-0.038}^{+0.039}$		&$0.074_{-0.043}^{+0.047}$	
&$0.084_{-0.039}^{+0.040}$	&$0.076_{-0.043}^{+0.049}$	
&$0.090\pm0.038$	&$0.091_{-0.043}^{+0.046}$
&$0.078\pm0.034$		&$0.062_{-0.037}^{+0.038}$\\

$m_a\,[\mathrm{eV}]$	&$<1.97$	& nb	
&$<1.09$	&$<1.63$	
&$<0.93$	&$<1.07$
&$<1.45$	&$<2.15$\\

$n_S$	&$0.974_{-0.015}^{+0.014}$	&--	
&$0.974\pm0.012$	&--	
&$0.978\pm0.010$	&--
&$0.977_{-0.013}^{+0.012}$	&--	\\

$\ln[10^{10}A_s]$	&$3.119_{-0.074}^{+0.075}$	&--	
&$3.112\pm0.077$	&--	
&$3.121_{-0.075}^{+0.076}$	&--
&$3.096_{-0.061}^{+0.062}$	&--	\\

$H_0 [\mathrm{Km \, s^{-1} \, Mpc^{-1}}]$	&$67.9_{-2.8}^{+2.4}$	&$65.2\pm3.4$	
&$68.1_{-2.3}^{+2.0}$	&$66.3_{-3.1}^{+2.9}$	
&$68.8\pm 1.1$	&$68.4\pm 1.3$
&$68.4_{-2.5}^{+2.2}$	&$66.0\pm3.0$	\\

$\sigma_8$	&$0.799_{-0.086}^{+0.063}$	&$0.800_{-0.097}^{+0.099}$	
&$0.812_{-0.050}^{+0.045}$	&$0.801_{-0.070}^{+0.066}$	
&$0.817_{-0.049}^{+0.044}$	&$0.859_{-0.081}^{+0.078}$
&$0.794_{-0.059}^{+0.046}$	&$0.804_{-0.085}^{+0.076}$\\

\hline
$\psj{1}$	&$\equiv1.307$	&$<7.36$			&$\equiv1.297$	&$<8.0$				&$\equiv1.262$	&$<7.93$			&$\equiv1.242$	&$<7.95$	\\

$\psj{2}$	&$\equiv1.138$	&$1.18_{-0.37}^{+0.40}$		&$\equiv1.131$	&$1.16_{-0.37}^{+0.41}$		&$\equiv1.123$	&$1.12_{-0.36}^{+0.39}$		&$\equiv1.100$	&$1.18_{-0.37}^{+0.40}$	\\

$\psj{3}$	&$\equiv1.119$	&$0.71_{-0.37}^{+0.39}$		&$\equiv1.112$	&$0.72_{-0.40}^{+0.41}$		&$\equiv1.107$	&$0.76_{-0.39}^{+0.41}$		&$\equiv1.083$	&$0.68\pm 0.37$	\\

$\psj{4}$	&$\equiv1.101$	&$1.20\pm0.27$			&$\equiv1.093$	&$1.20_{-0.26}^{+0.27}$		&$\equiv1.091$	&$1.22_{-0.26}^{+0.27}$		&$\equiv1.067$	&$1.19\pm0.26$	\\

$\psj{5}$	&$\equiv1.082$	&$1.09\pm0.12$			&$\equiv1.075$	&$1.08_{-0.11}^{+0.12}$		&$\equiv1.076$	&$1.08_{-0.12}^{+0.13}$		&$\equiv1.051$	&$1.057_{-0.098}^{+0.099}$\\

$\psj{6}$	&$\equiv1.064$	&$1.070_{-0.089}^{+0.093}$	&$\equiv1.057$	&$1.071_{-0.083}^{+0.093}$	&$\equiv1.061$	&$1.072_{-0.095}^{+0.097}$	&$\equiv1.036$	&$1.043_{-0.066}^{+0.064}$	\\

$\psj{7}$	&$\equiv1.046$	&$1.047_{-0.081}^{+0.090}$	&$\equiv1.039$	&$1.042_{-0.086}^{+0.091}$	&$\equiv1.045$	&$1.056_{-0.087}^{+0.095}$	&$\equiv1.020$	&$1.011_{-0.059}^{+0.064}$	\\

$\psj{8}$	&$\equiv1.029$	&$1.003_{-0.089}^{+0.093}$	&$\equiv1.021$	&$1.007_{-0.091}^{+0.097}$	&$\equiv1.031$	&$1.028_{-0.093}^{+0.097}$	&$\equiv1.005$	&$0.974_{-0.066}^{+0.072}$	\\

$\psj{9}$	&$\equiv1.011$	&$0.988_{-0.087}^{+0.092}$	&$\equiv1.004$	&$0.991_{-0.090}^{+0.097}$	&$\equiv1.016$	&$1.021_{-0.091}^{+0.095}$	&$\equiv0.990$	&$0.964_{-0.072}^{+0.073}$	\\

$\psj{10}$	&$\equiv0.994$	&$1.00_{-0.09}^{+0.10}$		&$\equiv0.987$	&$0.987_{-0.095}^{+0.099}$	&$\equiv1.001$	&$1.03_{-0.10}^{+0.11}$		&$\equiv0.975$	&$0.986_{-0.082}^{+0.084}$\\

$\psj{11}$	&$\equiv0.978$	&$<3.69$			&$\equiv0.971$	&$0.90_{-0.56}^{+0.75}$		&$\equiv0.987$	&$2.6_{-2.5}^{+1.9}$		&$\equiv0.961$	&$2.5_{-1.7}^{+1.5}$	\\

$\psj{12}$	&$\equiv0.896$	&nb				&$\equiv0.890$	&$<3.41$			&$\equiv0.917$	&nb				&$\equiv0.890$	&nb	\\

\hline
\end{tabular}
}
\caption[Constraints on cosmological parameters 
in the \lcdm\ + $m_a$ model,
without CMB polarization at high multipoles]
{Constraints on cosmological parameters from 
the Planck TT+lowP dataset
alone and in combination with the matter power spectrum
shape measurements from WiggleZ (MPkW),
the BAO data and the lensing constraints from Planck,
in the \lcdm\ + $m_a$ model (\emph{nb} stands for \emph{no bound}).
For each combination,
we report the limits obtained for the two parameterizations
of the primordial power spectrum, namely the power-law model
(first column)
and the polynomial expansion (second column of each pair).
Limits are at 95\% CL around the mean value
of the posterior distribution.
For each dataset, in the case of power-law model,
the values of \psj{i} are computed according to
Eq.~\eqref{eq:psj_plpps}.
From Ref.~\cite{DiValentino:2016ikp}.
}\label{tab:ax_ma}
\end{table}

\begin{table}[p]
\resizebox{1\textwidth}{!}{
\renewcommand{\arraystretch}{1.2}
\begin{tabular}{|c||c| c||c|c||c|c||c|c|}
\hline
Parameter	&\multicolumn{2}{c||}{Planck TT,TE,EE+lowP}	&\multicolumn{2}{c||}{Planck TT,TE,EE+lowP}	&\multicolumn{2}{c||}{Planck TT,TE,EE+lowP}	&\multicolumn{2}{c|}{Planck TT,TE,EE+lowP}\\
	&\multicolumn{2}{c||}{}	&\multicolumn{2}{c||}{+MPkW}	&\multicolumn{2}{c||}{+BAO}	&\multicolumn{2}{c|}{+lensing}\\
\hline\hline
$100\Omega_b h^2$	&$2.248\pm0.032$	&$2.241_{-0.038}^{+0.039}$	
&$2.245_{-0.031}^{+0.030}$	&$2.236_{-0.038}^{+0.037}$	
&$2.250_{-0.029}^{+0.030}$	&$2.248_{-0.036}^{+0.038}$
&$2.248_{-0.030}^{+0.033}$	&$2.242_{-0.038}^{+0.039}$	\\

$\Omega_c h^2$	&$0.1232_{-0.0036}^{+0.0034}$	&$0.1233_{-0.0043}^{+0.0041}$	
&$0.1236_{-0.0033}^{+0.0032}$	&$0.1241_{-0.0040}^{+0.0037}$	
&$0.1224_{-0.0024}^{+0.0023}$	&$0.1223\pm0.0029$
&$0.1231_{-0.0033}^{+0.0032}$	&$0.1224_{-0.0043}^{+0.0039}$	\\

$100\theta$	
&$1.0403\pm0.0007$	
&$1.0402_{-0.0006}^{+0.0007}$	
&$1.0403\pm0.0007$	
&$1.0402\pm0.0007$	
&$1.0406\pm0.0006$	
&$1.0405\pm0.0006$
&$1.0404\pm0.0007$	
&$1.0403\pm0.0006$	\\

$\tau$	&$0.090_{-0.034}^{+0.033}$		&$0.090_{-0.042}^{+0.043}$	
&$0.087\pm0.034$	&$0.091\pm 0.039$	
&$0.092\pm0.034$	&$0.093_{-0.042}^{+0.043}$
&$0.075\pm0.028$		&$0.071_{-0.032}^{+0.034}$	\\

$m_a\,[\mathrm{eV}]$	&$<2.09$	& $<2.44$	
&$<1.19$	&$<1.90$	
&$<0.74$	&$<1.19$
&$<1.68$	&$<2.44$\\

$n_S$	&$0.972_{-0.012}^{+0.011}$	&--	
&$0.9734\pm0.0098$	&--	
&$0.9754_{-0.0089}^{+0.0092}$	&--
&$0.974_{-0.011}^{+0.010}$	&--	\\

$\ln[10^{10}A_s]$	&$3.125_{-0.067}^{+0.065}$	&--	
&$3.119_{-0.068}^{+0.067}$	&--	
&$3.125\pm0.067$	&--
&$3.092\pm0.053$	&--	\\

$H_0 [\mathrm{Km \, s^{-1} \, Mpc^{-1}}]$	&$67.6_{-2.2}^{+1.9}$	&$66.8\pm2.2$	
&$67.9_{-1.8}^{+1.6}$	&$67.3_{-2.1}^{+2.0}$	
&$68.6\pm 1.0$	&$68.5\pm 1.1$
&$67.9_{-2.0}^{+1.9}$	&$66.9_{-1.9}^{+2.1}$	\\

$\sigma_8$	&$0.798_{-0.090}^{+0.067}$	&$0.806_{-0.10}^{+0.11}$	
&$0.815_{-0.054}^{+0.045}$	&$0.801_{-0.078}^{+0.068}$	
&$0.827_{-0.039}^{+0.037}$	&$0.871_{-0.084}^{+0.072}$
&$0.788_{-0.066}^{+0.051}$	&$0.790_{-0.085}^{+0.092}$	\\

\hline
$\psj{1}$	&$\equiv1.339$	&$<7.74$			&$\equiv1.319$	&$<7.85$			&$\equiv1.302$	&$<7.71$			&$\equiv1.272$	&$<7.74$	\\

$\psj{2}$	&$\equiv1.154$	&$1.15_{-0.36}^{+0.39}$		&$\equiv1.143$	&$1.14_{-0.36}^{+0.40}$		&$\equiv1.141$	&$1.12_{-0.36}^{+0.39}$		&$\equiv1.108$	&$1.18_{-0.37}^{+0.40}$	\\

$\psj{3}$	&$\equiv1.133$	&$0.72_{-0.37}^{+0.40}$		&$\equiv1.123$	&$0.74_{-0.37}^{+0.38}$		&$\equiv1.122$	&$0.74_{-0.38}^{+0.40}$		&$\equiv1.090$	&$0.68_{-0.34}^{+0.37}$	\\

$\psj{4}$	&$\equiv1.113$	&$1.25\pm0.24$			&$\equiv1.103$	&$1.24\pm 0.23$			&$\equiv1.104$	&$1.24\pm 0.23$			&$\equiv1.071$	&$1.23_{-0.22}^{+0.23}$	\\

$\psj{5}$	&$\equiv1.093$	&$1.11_{-0.11}^{+0.12}$		&$\equiv1.084$	&$1.11_{-0.10}^{+0.11}$		&$\equiv1.086$	&$1.10_{-0.11}^{+0.12}$		&$\equiv1.053$	&$1.071_{-0.088}^{+0.092}$\\

$\psj{6}$	&$\equiv1.073$	&$1.089_{-0.091}^{+0.098}$	&$\equiv1.065$	&$1.087_{-0.081}^{+0.083}$	&$\equiv1.069$	&$1.077_{-0.088}^{+0.096}$	&$\equiv1.036$	&$1.013_{-0.059}^{+0.064}$	\\

$\psj{7}$	&$\equiv1.054$	&$1.058_{-0.087}^{+0.090}$	&$\equiv1.047$	&$1.061_{-0.077}^{+0.079}$	&$\equiv1.052$	&$1.056_{-0.087}^{+0.094}$	&$\equiv1.018$	&$1.013_{-0.059}^{+0.064}$	\\

$\psj{8}$	&$\equiv1.035$	&$1.035_{-0.085}^{+0.091}$	&$\equiv1.029$	&$1.037_{-0.079}^{+0.080}$	&$\equiv1.035$	&$1.036_{-0.085}^{+0.093}$	&$\equiv1.001$	&$0.995_{-0.060}^{+0.066}$	\\

$\psj{9}$	&$\equiv1.016$	&$1.020_{-0.083}^{+0.088}$	&$\equiv1.011$	&$1.020_{-0.078}^{+0.080}$	&$\equiv1.018$	&$1.027_{-0.089}^{+0.090}$	&$\equiv0.984$	&$0.982_{-0.061}^{+0.067}$	\\

$\psj{10}$	&$\equiv0.998$	&$1.03_{-0.09}^{+0.10}$		&$\equiv0.993$	&$1.009_{-0.085}^{+0.088}$	&$\equiv1.002$	&$1.04\pm0.10$			&$\equiv0.968$	&$0.998_{-0.071}^{+0.079}$\\

$\psj{11}$	&$\equiv0.980$	&$2.8_{-2.4}^{+1.6}$		&$\equiv0.976$	&$0.94_{-0.8}^{+1.0}$		&$\equiv0.985$	&$2.9_{-2.6}^{+1.8}$		&$\equiv0.952$	&$3.1_{-1.7}^{+1.4}$	\\

$\psj{12}$	&$\equiv0.892$	& $<8.89$			&$\equiv0.891$	&$<3.06$			&$\equiv0.906$	&$<8.66$			&$\equiv0.872$	&nb	\\

\hline
\end{tabular}
}
\caption[As Tab.~\ref{tab:ax_ma},
but using the full CMB data]{
As Tab.~\ref{tab:ax_ma},
but using the Planck~TT,TE,EE+lowP dataset.
From Ref.~\protect\cite{DiValentino:2016ikp}.
}\label{tab:ax_ma_pol}
\end{table}

\subsection{Thermal Axions and Small Scales Perturbations}
\label{ssec:ax_sigma8}

In parallel to what we did for the massive sterile neutrino in the 
previous Chapter, we present here some results obtained
when the constraints on the small scales matter perturbations
are included in the cosmological analyses involving the thermal axions.
The data considered here are not the most recent ones.
The constraints from CFHTLenS and the Planck SZ cluster counts
obtained in the most recent analyses, taking into account
a large number of possible astrophysical systematics, tend to show a smaller tension
with the CMB data (see Sections \ref{sec:cluster} and \ref{sec:shear}).
Therefore, a thermal axion would not be needed to reconcile the two sets of data.
It is however possible that future experiments will be able
to distinguish the various systematics
and to improve the measurements.
In the case that the tension will appear again, explanations as the one
we provide here will be necessary.

When the CFHT bounds on the $\sigma_8$--$\Omega_m$ relationship
are considered in addition to the CMB constraints,
the bounds on the thermal axion mass become weaker. 
The reason is related to the lower $\sigma_8$ values preferred
by weak lensing measurements, 
values that can be achieved by allowing for higher axion masses. 
The larger is the axion mass,
the larger is the reduction of the matter power
spectrum at small (i.e.\ cluster) scales, 
leading consequently to a smaller value
of the clustering parameter $\sigma_8$.  

If we instead consider the PSZ data set with fixed cluster mass
bias, together with the CMB, BAO and HST measurements, a non-zero
value of the thermal axion mass of $\sim 1$~eV ($\sim 0.80$~eV) is
favored at $\sim4\sigma$ ($\sim3\sigma$) level, when considering the
\pchip (standard power-law) PPS approach \cite{DiValentino:2015zta}~%
\footnote{A similar effect when considering PSZ data for constraining either
thermal axion or neutrino masses has also been found in
Refs.~\cite{Hamann:2013iba,Wyman:2013lza,Gariazzo:2013gua,
Giusarma:2014zza,Dvorkin:2014lea,Archidiacono:2014apa}.}.
However, these results must be regarded as an illustration of what could be 
achieved with future cluster mass calibrations, as the Planck collaboration
has recently shown in their analyses of the 2015 Planck cluster
catalogue~\cite{Ade:2015fva}.
When more realistic approaches for the cluster mass bias are used,
the errors on the so-called cluster normalization condition are larger,
and consequently the preference for a non-zero axion mass 
of $1$~eV is only mild in the \pchip PPS case, while in the case
of a standard power-law PPS such an evidence completely disappears.

The left (right) panel of Fig.~\ref{fig:ax_91_ma_pchip}
shows the $68\%$ and $95\%$~CL
allowed regions in the ($m_a$, $\sigma_8$) plane
in the power-law (\pchip) PPS scenario.
The lower values of the $\sigma_8$ clustering parameter preferred by PSZ data
are translated into a preference for non-zero thermal axion masses.
Larger values of $m_a$ will enhance the matter power spectrum suppression
at scales below the axion free-streaming scale,
leading to smaller values of the $\sigma_8$ clustering parameter,
as preferred by PSZ measurements.
The evidence for non-zero axion masses is more significant when
the cluster mass bias is fixed in the PSZ data analyses.
These results are the analogous of what we found for the sterile neutrino
in Chapter~\ref{ch:lsn_cosmo}.

\subsection{Planck TT+lowP}
The most stringent constraints on the axion properties are obtained
with the most recent CMB data, released in 2015 by the Planck
collaboration \cite{Adam:2015rua}, that we are going to consider now.

Table \ref{tab:ax_ma} shows our results at $95\%$~CL arising
from the Planck~TT+lowP data alone and in combination with the MPkW,
BAO and lensing measurements, for an extended \lcdm\ + $m_a$ scenario,
in the context of the two PPS parameterizations explored here. 

As we discussed before, the first thing to note is that bounds
on the axion mass are largely relaxed when considering the \pchip PPS
with respect to the ones obtained in the power-law PPS, in the case
of the CMB measurements only.
The Planck~TT+lowP dataset cannot constrain the axion mass
in the \pchip approach.
However, when adding the matter power spectrum measurements
via the MPkW dataset, the upper limit on the axion mass is reduced
by a half in the power-law approach: we have the limit $m_a<1.09$~eV
at $95\%$~CL, that becomes $m_a<1.63$~eV at $95\%$~CL
in the \pchip parametrization.

The most stringent bounds arise when using the BAO data,
since they are directly sensitive to the free-streaming nature
of the thermal axion.
While the MPkW measurements are also sensitive
to this small scale structure suppression, BAO measurements
are able to constrain better the cold dark matter density $\Omega_c h^2$,
strongly correlated with $m_a$.
The lower is the thermal axion mass,
the lower is the amount of hot dark matter and consequently the lower must be
the cold dark matter component, and viceversa. 
We find $m_a<0.93$~eV at $95\%$~CL in the standard case,
and a slightly weaker constraint in the \pchip case, $m_a<1.07$~eV at $95\%$~CL,
both obtained using the Planck~TT+lowP+BAO dataset.

Finally, when considering the lensing dataset, we obtain $m_a<1.45$~eV
at $95\%$~CL in the power-law PPS case,
that is slightly relaxed in the \pchip PPS, $m_a<2.15$~eV at $95\%$~CL.
For this combination of datasets, a mild preference appears
for an axion mass different from zero:
$m_a=1.05_{-0.81}^{+0.37}$ at $68\%$~CL,
only when considering the \pchip approach, as depicted in
Fig.~\ref{fig:ax_ma_bars}.
This is probably due to the existing tension between the
Planck data on the lensing reconstruction from the CMB trispectrum
and the lensing effect observed in the CMB spectrum,
see e.g.\ Refs.~\cite{Ade:2015xua,DiValentino:2015ola}.

The weakening of the axion mass constraints in most of the data combinations 
obtained in the \pchip PPS scheme is responsible for the shift at more
than 1$\sigma$ of the cold dark matter mass-energy density,
due to the existing degeneracy between $m_a$ and $\Omega_c h^2$.
Interestingly, this effect has also an impact on the Hubble constant,
leading to a shift of about 2$\sigma$ towards lower values 
of the mean value of $H_0$ due to parameter degeneracies, as previously
discussed in Chapter~\ref{ch:pps_nu}.
Furthermore, a shift in the optical depth towards a lower mean value is
also present when analyzing the \pchip PPS scenario.
One can explain this shift via the existing degeneracies
between $\tau$ and $H_0$ and between $\tau$ and $\Omega_c h^2$.
Once BAO measurements are included in the data analyses,
the degeneracies are largely removed and there is no significant shift
in the values of the $\Omega_c h^2$, $H_0$ and $\tau$ parameters
within the \pchip PPS approach, when comparing to their mean values
in the power-law PPS.  

\begin{figure*}
\centering
\includegraphics[width=\textwidth]{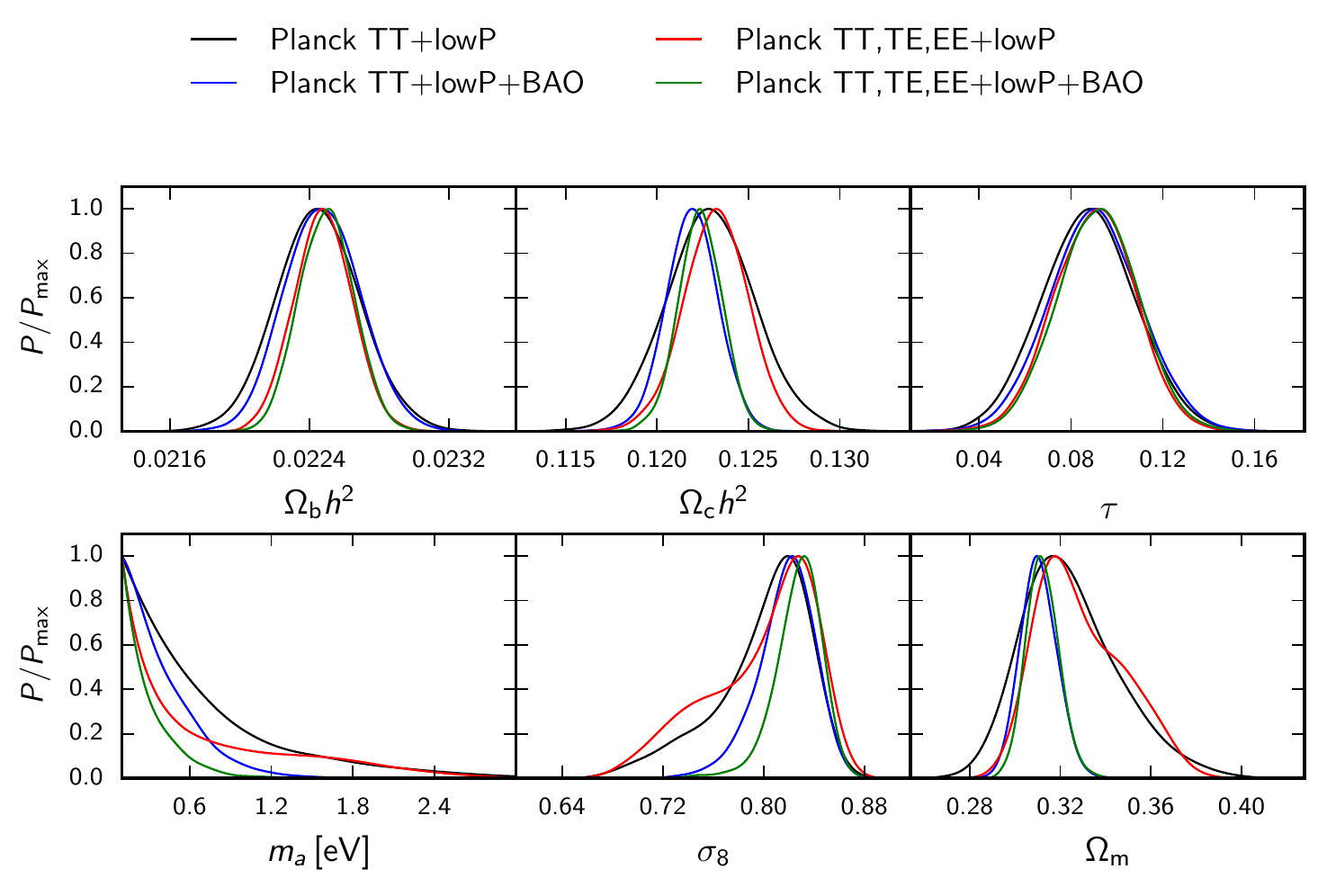}
\caption[Marginalized 1D posterior distributions in the 
\lcdm\ + $m_a$ model.]
{One-dimensional posterior probability for the most relevant
cosmological parameters for the combination of datasets labeled in the figure,
for the power-law approach in the \lcdm\ + $m_a$ scenario.
From Ref.~\cite{DiValentino:2016ikp}.}\label{fig:ax_1d_ma}
\end{figure*}

\subsection{Planck TT,TE,EE+lowP}
Table \ref{tab:ax_ma_pol} shows our results at $95\%$~CL
from the Planck~TT,TE,EE+lowP data alone and in combination with MPkW,
BAO and lensing measurements, for an extended \lcdm\ + $m_a$ scenario,
comparing the power-law PPS and the \pchip PPS bounds. 

In general, the constraints arising from high-$\ell$ polarization measurements
are slightly weaker than those obtained previously.
The weakening of the axion mass is driven by the preference of
Planck~TT,TE,EE+lowP for a lower value of $\neff$, as pointed out before.
As shown in Fig.~\ref{fig:ax_91_maref}, the additional contribution to $\neff$
due to thermal axions is a steep function of the axion mass,
at least for low thermal axion masses (i.e.\ below $\sim1$~eV).
The lower value of $\neff$ preferred by small-scale polarization dramatically
sharpens the posterior of $m_a$ at low mass (see Fig.~\ref{fig:ax_1d_ma}).
At higher masses, however, the axion contribution 
to \neff\ depends weakly on $m_a$:
as a consequence,
the posterior distribution flattens at high $m_a$ and overlaps
with the one resulting from Planck TT+lowP,
since CMB polarization does not help in improving the constraints on $\Omega_m$.
We can in fact notice the presence of a bump in the posterior distributions
of $\Omega_m$ and $\sigma_8$ for Planck~TT,TE,EE+lowP.
The mismatch in the values of $\Omega_m$ preferred by low and high thermal
axion masses leads to a worsening of the constraints on $m_a$ with respect
to the Planck TT+lowP scenario, since the volume of the posterior distribution
is now mainly distributed at higher masses.
When the BAO data are considered, we get the tightest bounds on $m_a$.
In addition, the bump in both the $\Omega_m$ and $\sigma_8$ distributions 
disappears completely, due to the higher constraining power
of the BAO data.
As a result, the tail of the $m_a$ distribution is excluded
when the BAO measurements are considered, and the constraints
do not suffer of the problem related to the volume effects discussed above.

Furthermore, the thermal axion mass bounds are
relaxed within the \pchip PPS formalism.
In particular, concerning the CMB measurements only,
$m_a<2.44$~eV at $95\%$~CL in the \pchip approach,
compared to the bound $m_a<2.09$~eV at $95\%$~CL
in the standard power-law PPS description.
When adding the matter power spectrum measurements (MPkW),
we find upper limits on the axion mass that are $m_a<1.19$~eV at $95\%$~CL
in the power-law PPS and $m_a<1.90$~eV at $95\%$~CL
in the \pchip parametrization.
When considering the lensing dataset,
we obtain $m_a<1.68$~eV at $95\%$~CL in the power-law PPS case,
that is relaxed when using the \pchip PPS, that gives $m_a<2.44$~eV at $95\%$~CL.
A mild preference for an axion mass different from zero appears from
this particular data combination ($m_a=1.39_{-0.63}^{+0.71}$ at $68\%$~CL)
only when considering the \pchip PPS approach, see Fig.~\ref{fig:ax_ma_bars}.  

It is important to note that, when the CMB polarization at
high multipoles is included,
the shifts induced in the mean value of the optical depth and
in the abundance of the cold dark matter disappear.

\section{Thermal Axions and massive neutrinos}\label{sec:ax_mnu}

In this Section we show the bounds in a scenario that includes both
massive neutrinos and the thermal axion relics.
In principle, it should be possible to distinguish between these
two relic populations because thermal axions increase the amount
of radiation expected in the standard model, where the neutrino contribution
is fixed to $\neff=3.046$,
modifying the $\neff$ value through the Eq.~\eqref{eq:ax_dneff}.
In addition,
axions are expected to be colder and have a larger mass than neutrinos. 
First of all it is important to note that the thermal axion mass bounds
are unchanged in the extended \lcdm\ + $m_a$ + $\mnu$ model with respect
to the \lcdm\ + $m_a$ scenario.
After comparing among the results shown in
Tabs.~\ref{tab:ax_ma} (\ref{tab:ax_ma_pol}) and
\ref{tab:ax_mamnu} (\ref{tab:ax_mamnu_pol})
for the Planck TT+lowP (Planck TT,TE,EE+lowP) dataset baseline
we can notice that the axion mass constraints are almost identical.
In other words, massive neutrinos do not affect the upper limits obtained
for the thermal axion mass.
On the other hand, the presence of thermal axions
tightens the neutrino mass bounds,
presented in Sec.~\ref{sec:inflfreed_mnu},
as both the thermal relics behave
as hot dark matter with a free-streaming nature.

\begin{table}[p]
\resizebox{1\textwidth}{!}{
\renewcommand{\arraystretch}{1.2}
\begin{tabular}{|c||c| c||c|c||c|c||c|c|}
\hline
Parameter	&\multicolumn{2}{c||}{Planck TT+lowP}	&\multicolumn{2}{c||}{Planck TT+lowP+MPkW}	&\multicolumn{2}{c||}{Planck TT+lowP+BAO} &\multicolumn{2}{c|}{Planck TT+lowP+lensing}\\
\hline\hline
$100\Omega_b h^2$	&$2.237_{-0.055}^{+0.051}$	&$2.134_{-0.093}^{+0.098}$	
&$2.237\pm 0.046$	&$2.170_{-0.10}^{+0.09}$	
&$2.248_{-0.042}^{+0.044}$	&$2.226_{-0.065}^{+0.070}$
&$2.236\pm 0.051$	&$2.150_{-0.082}^{+0.087}$	\\

$\Omega_c h^2$	&$0.1234\pm 0.0048$	&$0.1288_{-0.0069}^{+0.0068}$	
&$0.1235_{-0.0042}^{+0.0045}$	&$0.1279_{-0.0066}^{+0.0075}$	
&$0.1217_{-0.0032}^{+0.0030}$	&$0.1220_{-0.0037}^{+0.0033}$
&$0.1230_{-0.0046}^{+0.0049}$	&$0.1278\pm 0.0064$\\

$100\theta$	
&$1.040\pm 0.001$	
&$1.0393_{-0.0014}^{+0.0013}$	
&$1.0404_{-0.0010}^{+0.0009}$	
&$1.0397_{-0.0014}^{+0.0013}$	
&$1.0407\pm0.0009$	
&$1.0406_{-0.0008}^{+0.0009}$
&$1.040\pm0.001$	
&$1.0395\pm0.0012$	\\

$\tau$	&$0.090_{-0.039}^{+0.040}$		&$0.075_{-0.042}^{+0.046}$	
&$0.087_{-0.037}^{+0.039}$	&$0.076_{-0.046}^{+0.048}$	
&$0.092\pm0.038$	&$0.092_{-0.047}^{+0.048}$
&$0.085_{-0.035}^{+0.037}$		&$0.071_{-0.037}^{+0.040}$\\

$\mnu\,[\mathrm{eV}]$	&$<0.62$	& $<2.20$
&$<0.40$	&$<1.24$	
&$<0.20$	&$<0.21$
&$<0.57$	&$<1.42$\\

$m_a\,[\mathrm{eV}]$	&$<1.91$	& nb	
&$<1.03$	&$<1.65$	
&$<0.94$	&$<1.03$
&$<1.39$	&$<2.13$\\

$n_S$	&$0.973_{-0.016}^{+0.015}$	&--	
&$0.974\pm0.012$	&--	
&$0.978_{-0.011}^{+0.010}$	&--
&$0.974_{-0.014}^{+0.013}$	&--\\

$\ln[10^{10}A_s]$	&$3.123_{-0.075}^{+0.076}$	&--	
&$3.116_{-0.073}^{+0.076}$	&--	
&$3.123\pm 0.074$	&--
&$3.111_{-0.064}^{+0.068}$	&--	\\

$H_0 [\mathrm{Km \, s^{-1} \, Mpc^{-1}}]$	&$66.5_{-5.1}^{+4.2}$	&$59_{-10}^{+9}$	
&$67.3_{-3.4}^{+3.0}$	&$63.1_{-9.4}^{+6.3}$	
&$68.7\pm 1.3$	&$68.4\pm 1.4$
&$66.6_{-4.7}^{+4.1}$	&$60\pm7$\\

$\sigma_8$	&$0.78_{-0.11}^{+0.09}$	&$0.68_{-0.20}^{+0.18}$	
&$0.798_{-0.066}^{+0.060}$	&$0.75_{-0.15}^{+0.11}$	
&$0.814_{-0.052}^{+0.048}$	&$0.858_{-0.081}^{+0.078}$
&$0.771_{-0.074}^{+0.064}$	&$0.70\pm 0.13$\\

\hline
$\psj{1}$	&$\equiv1.324$	&$<7.66$			&$\equiv1.303$	&$<7.70$			&$\equiv1.264$	&$<7.75$			&$\equiv1.296$	&$<7.61$	\\

$\psj{2}$	&$\equiv1.147$	&$1.23_{-0.37}^{+0.40}$		&$\equiv1.135$	&$1.20_{-0.38}^{+0.40}$		&$\equiv1.125$	&$1.12_{-0.37}^{+0.40}$		&$\equiv1.129$	&$1.23_{-0.38}^{+0.39}$	\\

$\psj{3}$	&$\equiv1.128$	&$0.73_{-0.37}^{+0.40}$		&$\equiv1.116$	&$0.73_{-0.38}^{+0.41}$		&$\equiv1.109$	&$0.76\pm 0.43$			&$\equiv1.110$	&$0.71_{-0.36}^{+0.40}$\\

$\psj{4}$	&$\equiv1.108$	&$1.23_{-0.27}^{+0.28}$		&$\equiv1.097$	&$1.21_{-0.26}^{+0.27}$		&$\equiv1.093$	&$1.23_{-0.26}^{+0.27}$		&$\equiv1.092$	&$1.23_{-0.27}^{+0.28}$\\

$\psj{5}$	&$\equiv1.089$	&$1.15_{-0.16}^{+0.18}$		&$\equiv1.079$	&$1.10_{-0.12}^{+0.14}$		&$\equiv1.078$	&$1.08_{-0.12}^{+0.13}$		&$\equiv1.073$	&$1.11_{-0.12}^{+0.13}$	\\

$\psj{6}$	&$\equiv1.070$	&$1.11\pm 0.11$			&$\equiv1.061$	&$1.083_{-0.090}^{+0.093}$	&$\equiv1.063$	&$1.07_{-0.09}^{+0.10}$		&$\equiv1.055$	&$1.078_{-0.073}^{+0.076}$	\\

$\psj{7}$	&$\equiv1.051$	&$1.057_{-0.082}^{+0.090}$	&$\equiv1.043$	&$1.047_{-0.086}^{+0.088}$	&$\equiv1.048$	&$1.058_{-0.095}^{+0.098}$	&$\equiv1.038$	&$1.039_{-0.065}^{+0.069}$\\

$\psj{8}$	&$\equiv1.033$	&$1.007_{-0.088}^{+0.091}$	&$\equiv1.025$	&$1.007_{-0.091}^{+0.095}$	&$\equiv1.033$	&$1.03\pm 0.10$			&$\equiv1.020$	&$0.995_{-0.069}^{+0.078}$\\

$\psj{9}$	&$\equiv1.015$	&$0.987_{-0.086}^{+0.090}$	&$\equiv1.008$	&$0.990_{-0.091}^{+0.095}$	&$\equiv1.018$	&$1.02\pm 0.10$			&$\equiv1.003$	&$0.981_{-0.074}^{+0.078}$	\\

$\psj{10}$	&$\equiv0.997$	&$1.00\pm 0.10$			&$\equiv0.991$	&$0.991_{-0.096}^{+0.097}$	&$\equiv1.003$	&$1.04_{-0.10}^{+0.11}$		&$\equiv0.986$	&$1.004_{-0.081}^{+0.088}$\\

$\psj{11}$	&$\equiv0.980$	&$<3.72$			&$\equiv0.975$	&$1.1_{-0.8}^{+1.3}$		&$\equiv0.989$	&$2.6_{-2.5}^{+1.9}$		&$\equiv0.970$	&$2.8_{-1.9}^{+1.5}$	\\

$\psj{12}$	&$\equiv0.895$	&nb				&$\equiv0.893$	&$<3.14$			&$\equiv0.919$	&nb				&$\equiv0.889$	&nb\\

\hline
\end{tabular}
}
\caption[As Tab.~\ref{tab:ax_ma},
but for the \lcdm\ + $m_a$ + \mnu\ model]
{\label{tab:ax_mamnu}
As Tab.~\ref{tab:ax_ma},
but for the \lcdm\ + $m_a$ + \mnu\ model.
From Ref.~\protect\cite{DiValentino:2016ikp}.}
\end{table}

\begin{table}[p]
\resizebox{1\textwidth}{!}{
\renewcommand{\arraystretch}{1.2}
\begin{tabular}{|c||c| c||c|c||c|c||c|c|}
\hline
Parameter	&\multicolumn{2}{c||}{Planck TT,TE,EE+lowP}	&\multicolumn{2}{c||}{Planck TT,TE,EE+lowP}	&\multicolumn{2}{c||}{Planck TT,TE,EE+lowP}	&\multicolumn{2}{c|}{Planck TT,TE,EE+lowP}\\
	&\multicolumn{2}{c||}{}	&\multicolumn{2}{c||}{+MPkW}	&\multicolumn{2}{c||}{+BAO}	&\multicolumn{2}{c|}{+lensing}\\
\hline\hline
$100\Omega_b h^2$
&$2.244_{-0.035}^{+0.034}$
&$2.237\pm 0.040$	
&$2.242_{-0.031}^{+0.032}$	
&$2.233_{-0.036}^{+0.037}$	
&$2.250_{-0.030}^{+0.031}$	
&$2.248_{-0.036}^{+0.038}$
&$2.242_{-0.037}^{+0.033}$	
&$2.234_{-0.040}^{+0.041}$\\

$\Omega_c h^2$	&$0.1235_{-0.0036}^{+0.0034}$	&$0.1235_{-0.0043}^{+0.0040}$	
&$0.1236\pm 0.0033$	&$0.1243_{-0.0039}^{+0.0037}$	
&$0.1223\pm0.0023$	&$0.1222_{-0.0030}^{+0.0028}$
&$0.1235\pm 0.0034$	&$0.1230_{-0.0042}^{+0.0040}$	\\

$100\theta$	
&$1.0402\pm0.0007$	
&$1.0401\pm0.0007$	
&$1.0403\pm0.0007$	
&$1.0402\pm0.0007$	
&$1.0405\pm0.0006$	
&$1.0405\pm0.0006$
&$1.0403\pm0.0007$	
&$1.0401\pm0.0007$\\

$\tau$	&$0.093_{-0.036}^{+0.035}$		&$0.090\pm0.043$	
&$0.088_{-0.037}^{+0.034}$	&$0.094_{-0.038}^{+0.040}$	
&$0.092_{-0.035}^{+0.034}$	&$0.093_{-0.041}^{+0.042}$
&$0.083_{-0.032}^{+0.033}$		&$0.084\pm 0.037$\\

$\mnu\,[\mathrm{eV}]$	&$<0.44$	& $<0.72$	
&$<0.32$	&$<0.37$	
&$<0.16$	&$<0.18$
&$<0.53$	&$<0.94$\\

$m_a\,[\mathrm{eV}]$	&$<2.05$	& $<2.37$	
&$<1.12$	&$<1.79$	
&$<0.76$	&$<1.18$
&$<1.66$	&$<2.33$\\

$n_S$	&$0.972_{-0.012}^{+0.011}$	&--	
&$0.973\pm0.010$	&--	
&$0.9754_{-0.0089}^{+0.0093}$	&--
&$0.972\pm 0.011$	&--	\\

$\ln[10^{10}A_s]$	&$3.130_{-0.070}^{+0.068}$	&--	
&$3.120_{-0.071}^{+0.067}$	&--	
&$3.126_{-0.068}^{+0.066}$	&--
&$3.109_{-0.061}^{+0.063}$	&--\\

$H_0 [\mathrm{Km \, s^{-1} \, Mpc^{-1}}]$	&$66.7_{-3.6}^{+3.1}$	&$65.3_{-4.7}^{+4.0}$	
&$67.4_{-2.7}^{+2.3}$	&$66.6_{-3.0}^{+2.8}$	
&$68.6\pm 1.1$	&$68.4_{-1.2}^{+1.1}$
&$66.4_{-3.7}^{+3.3}$	&$63.9_{-5.1}^{+4.7}$	\\

$\sigma_8$	&$0.781_{-0.094}^{+0.081}$	&$0.78_{-0.12}^{+0.13}$	
&$0.806_{-0.062}^{+0.054}$	&$0.791_{-0.081}^{+0.075}$	
&$0.827_{-0.042}^{+0.039}$	&$0.871_{-0.081}^{+0.073}$
&$0.767_{-0.072}^{+0.066}$	&$0.73_{-0.11}^{+0.12}$	\\

\hline
$\psj{1}$	&$\equiv1.346$	&$<7.75$			&$\equiv1.320$	&$<7.57$			&$\equiv1.299$	&$<7.70$			&$\equiv1.318$	&$<7.26$	\\

$\psj{2}$	&$\equiv1.160$	&$1.17_{-0.37}^{+0.39}$		&$\equiv1.144$	&$1.15_{-0.36}^{+0.39}$		&$\equiv1.140$	&$1.13_{-0.36}^{+0.38}$		&$\equiv1.136$	&$1.19_{-0.36}^{+0.39}$\\

$\psj{3}$	&$\equiv1.139$	&$0.73_{-0.38}^{+0.40}$		&$\equiv1.124$	&$0.74_{-0.38}^{+0.40}$		&$\equiv1.122$	&$0.73_{-0.37}^{+0.38}$		&$\equiv1.115$	&$0.71_{-0.36}^{+0.39}$\\

$\psj{4}$	&$\equiv1.119$	&$1.26_{-0.23}^{+0.24}$		&$\equiv1.105$	&$1.24_{-0.23}^{+0.24}$		&$\equiv1.104$	&$1.24_{-0.22}^{+0.23}$		&$\equiv1.095$	&$1.26\pm 0.24$\\

$\psj{5}$	&$\equiv1.098$	&$1.11_{-0.11}^{+0.12}$		&$\equiv1.085$	&$1.11_{-0.10}^{+0.11}$		&$\equiv1.087$	&$1.10\pm 0.11$			&$\equiv1.076$	&$1.11_{-0.10}^{+0.11}$\\

$\psj{6}$	&$\equiv1.079$	&$1.089_{-0.093}^{+0.097}$	&$\equiv1.066$	&$1.093_{-0.084}^{+0.086}$	&$\equiv1.069$	&$1.076_{-0.086}^{+0.091}$	&$\equiv1.056$	&$1.077_{-0.076}^{+0.079}$\\

$\psj{7}$	&$\equiv1.059$	&$1.060_{-0.087}^{+0.093}$	&$\equiv1.048$	&$1.068_{-0.076}^{+0.085}$	&$\equiv1.052$	&$1.056_{-0.083}^{+0.089}$	&$\equiv1.037$	&$1.042_{-0.071}^{+0.073}$\\

$\psj{8}$	&$\equiv1.040$	&$1.036_{-0.086}^{+0.092}$	&$\equiv1.030$	&$1.042_{-0.081}^{+0.084}$	&$\equiv1.036$	&$1.036_{-0.083}^{+0.090}$	&$\equiv1.018$	&$1.021_{-0.071}^{+0.074}$\\

$\psj{9}$	&$\equiv1.021$	&$1.022_{-0.087}^{+0.088}$	&$\equiv1.012$	&$1.026_{-0.081}^{+0.083}$	&$\equiv1.019$	&$1.027\pm 0.087$		&$\equiv1.000$	&$1.007_{-0.071}^{+0.074}$	\\

$\psj{10}$	&$\equiv1.003$	&$1.03_{-0.09}^{+0.10}$		&$\equiv0.994$	&$1.014_{-0.085}^{+0.091}$	&$\equiv1.003$	&$1.04\pm0.10$			&$\equiv0.982$	&$1.022_{-0.080}^{+0.082}$	\\

$\psj{11}$	&$\equiv0.985$	&$3.0_{-2.6}^{+1.5}$		&$\equiv0.977$	&$0.94_{-0.7}^{+1.1}$		&$\equiv0.987$	&$3.0_{-2.6}^{+1.8}$		&$\equiv0.964$	&$3.3\pm 1.3$	\\

$\psj{12}$	&$\equiv0.896$	& $<8.61$			&$\equiv0.892$	&$<2.99$			&$\equiv0.909$	&$<8.53$			&$\equiv0.878$	&nb	\\

\hline
\end{tabular}
}
\caption[As Tab.~\ref{tab:ax_mamnu},
but from the Planck~TT,TE,EE+lowP dataset]
{\label{tab:ax_mamnu_pol}
As Tab.~\ref{tab:ax_mamnu},
but for the Planck TT, TE, EE+lowP dataset.
From Ref.~\protect\cite{DiValentino:2016ikp}.}
\end{table}

\subsection{Results with Planck~TT+lowP}

\begin{figure}
\centering
\includegraphics[width=\singlefigland, page=1]{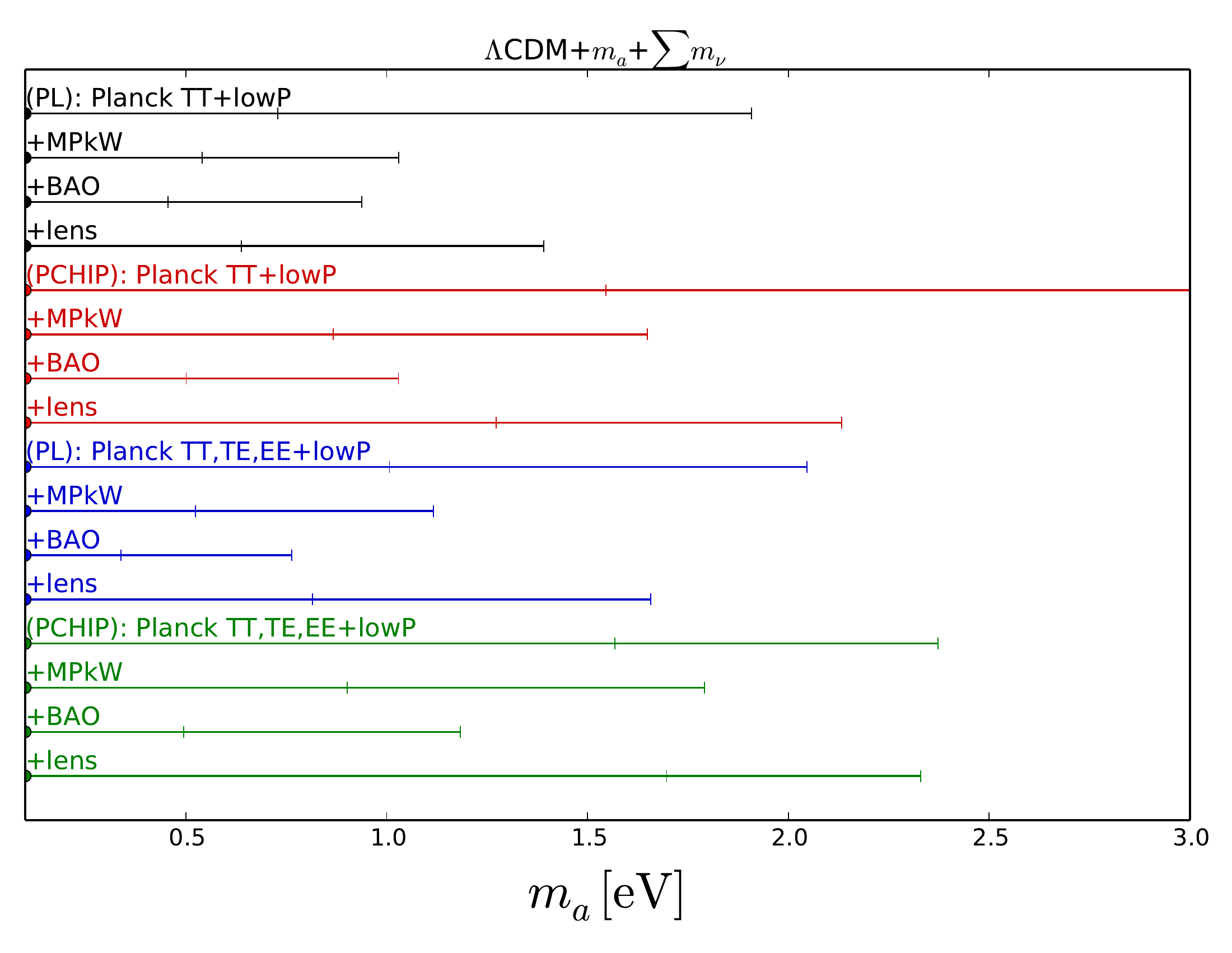}
\caption[As Fig.~\ref{fig:inflfreed_nnu_bars} but in the context of a 
 \lcdm\ + $m_a$ + $\mnu$ model, focusing on the $m_a$ parameter]
 {As Fig.~\ref{fig:inflfreed_nnu_bars} but in the context of a 
 \lcdm\ + $m_a$ + $\mnu$ model, focusing on the $m_a$ parameter.
 From Ref.~\cite{DiValentino:2016ikp}.}
 \label{fig:ax_mamnu_ma}
\end{figure}

\begin{figure}
\centering
\includegraphics[width=\singlefigland, page=2]{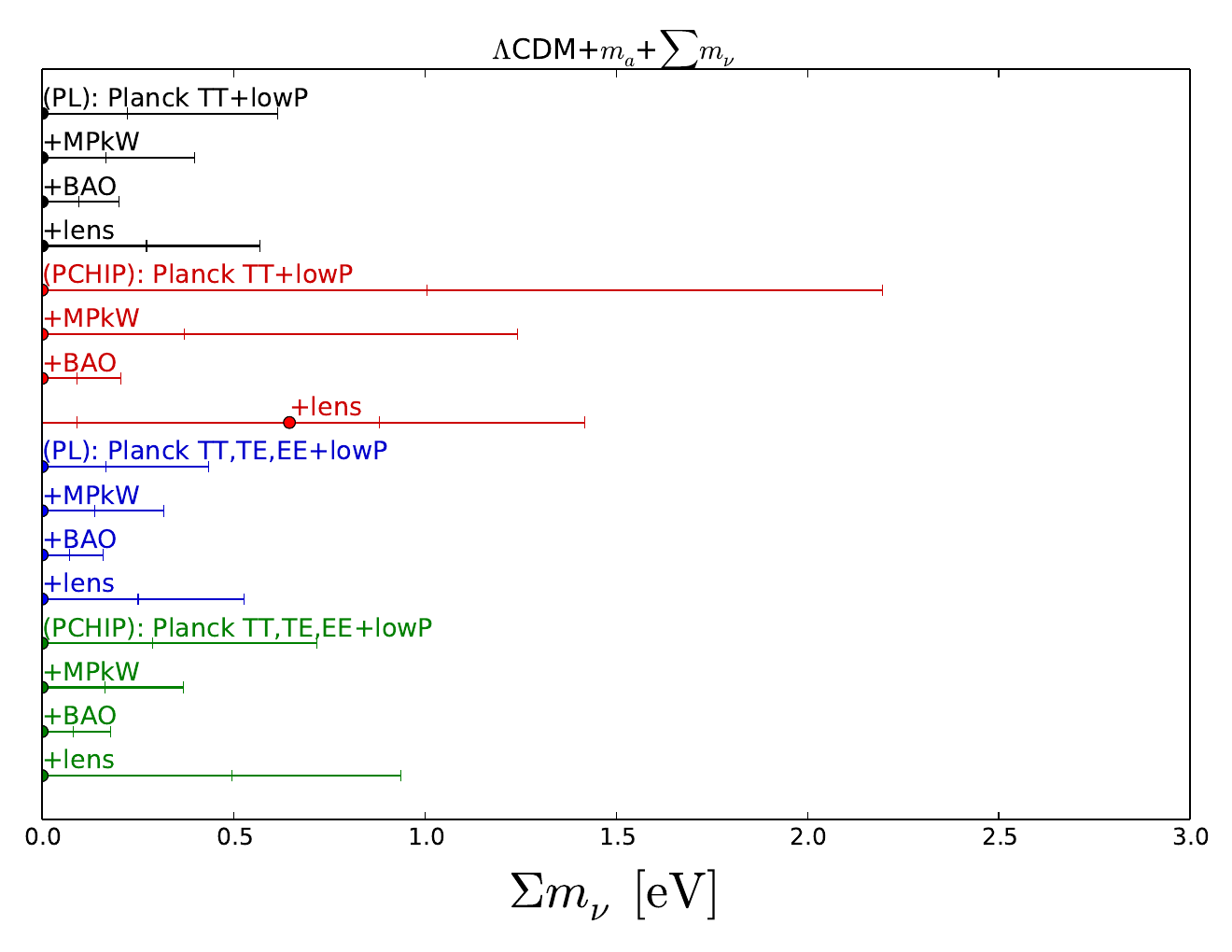}
\caption[As Fig.~\ref{fig:ax_mamnu_ma},
but for the $\mnu$ parameter]
{As Fig.~\ref{fig:ax_mamnu_ma},
but for the $\mnu$ parameter.
From Ref.~\cite{DiValentino:2016ikp}.}\label{fig:ax_mamnu_mnu}
\end{figure}

Table \ref{tab:ax_mamnu} presents our results at $95\%$~CL
from the Planck~TT+lowP data alone and in combination
with the MPkW, BAO and lensing measurements,
for an extended \lcdm\ + $m_a$ + $\mnu$ scenario,
in the two PPS parameterizations exploited here.
As discussed before for the \lcdm\ + $m_a$ model,
for any combination of datasets the bounds on the axion mass are relaxed
when considering the \pchip PPS with respect to the power-law PPS ones
(see Fig.~\ref{fig:ax_mamnu_ma}).
In addition, in this case, we can also notice a weakening
of the total neutrino mass constraints when using the \pchip approach
(see Fig.~\ref{fig:ax_mamnu_mnu}).
The only exception appears when considering the BAO measurements,
since they are directly sensitive to the free-streaming nature
of the two relic particles. 

Concerning the CMB measurements only, the Planck~TT+lowP data are not able
to constrain the axion mass in the \pchip approach, providing $\mnu<2.20$~eV
at $95\%$~CL versus the $\mnu<0.62$~eV at $95\%$~CL limit obtained
for the power-law approach.
When adding the matter power spectrum measurements (MPkW), both
the upper limits on the axion mass and on the neutrino masses
are reduced by about a half in the canonical power-law PPS scenario,
and become $m_a<1.65$~eV at $95\%$~CL and $\mnu<1.24$~eV at $95\%$~CL
in the \pchip parameterization.
As in the previous sections, the most stringent bounds arise
when using the BAO data in both parameterizations:
we have $m_a<1.03$~eV at $95\%$~CL and $\mnu<0.21$~eV at $95\%$~CL
in the \pchip case.
Finally, when considering the lensing dataset with the \pchip PPS,
we obtain $m_a<2.13$~eV at $95\%$~CL and $\mnu<1.42$~eV at $95\%$~CL. 

As in the \lcdm\ + $m_a$ model,
we find a shift of about 2$\sigma$ towards lower values of the mean value
of the Hubble constant in the \pchip parameterization, except when BAO data
are included in the analyses.
However, in this case, the value of $H_0$ is strongly degenerate
with the total neutrino mass, as explained in the previous Chapter. 

In addition, in this extended scenario with massive neutrinos and
within the \pchip approach, we have a shift of about 2$\sigma$ toward
higher values in the mean value of the cold dark matter density.
These shifts are larger than those reported in the \lcdm\ + $m_a$ scenario,
as now we have one extra degeneracy (between $\mnu$ and $\Omega_c h^2$).
A shift in the optical depth $\tau$ is also present in this case,
due to the degeneracy between $\Omega_c h^2$ and $\tau$.
Once BAO measurements are considered, the shifts in the mean values
of the parameters are not significant. 

\subsection{Results with Planck TT,TE,EE+lowP}

Table \ref{tab:ax_mamnu_pol} shows our results at $95\%$~CL
from the Planck~TT,TE,EE+lowP data alone and in combination
with MPkW, BAO and lensing measurements,
for an extended \lcdm\ + $m_a$ + $\mnu$ scenario,
comparing the power-law PPS and the \pchip PPS bounds.
As noticed above in the Planck~TT+lowP baseline results,
the bounds on the axion mass and on the total neutrino mass are relaxed
when considering the \pchip PPS with respect to the power-law PPS ones. 

The axion (neutrino) mass constraints are, in general,
slightly weaker (stronger) than those obtained using only
the temperature power spectrum at small angular scales,
for the reasons explained above.
In particular, focusing on the CMB measurements alone,
the Planck~TT,TE,EE+lowP data provides upper limits
on the thermal axion mass and on the neutrino masses
in the \pchip approach of $m_a<2.37$~eV at $95\%$~CL and
$\mnu<0.72$~eV at $95\%$~CL, respectively.

Furthermore, when adding the matter power spectrum measurements (MPkW)
we find $m_a<1.79$~eV at $95\%$~CL and $\mnu<0.37$~eV at $95\%$~CL
in the \pchip parametrization.
This last constraint on the neutrino masses is about half 
the bound obtained with the Planck~TT,TE,EE+lowP dataset.
The most stringent bounds on both the axion mass and
on the total neutrino mass arise, as usual, from the addition of BAO data.
We find $m_a<1.18$~eV at $95\%$~CL and $\mnu<0.18$~eV at $95\%$~CL
in the \pchip PPS, respectively.
Finally, when considering the lensing dataset within the \pchip PPS
we obtain $m_a<2.33$~eV at $95\%$~CL and $\mnu<0.94$~eV at $95\%$~CL. 

The mean values of the optical depth or of the cold dark matter density
do not suffer from the shifts detailed in the absence
of high multipole polarization data.
There is a (mild) shift,
caused by the degeneracy between \mnu\ and $H_0$,
toward lower values
in the Hubble constant case
within the \pchip approach,
which gets accentuated when including the lensing likelihood.
As expected, the shift in the mean value of the clustering parameter $\sigma_8$
is larger than in previous cases, due to the presence of
two hot dark matter species reducing the small-scale matter fluctuations.

\section{Conclusions}
After discussing the cosmological
properties of active and sterile neutrinos,
that may contribute as hot or warm dark matter components
depending on their mass,
we studied in this Chapter another possible candidate
for hot dark matter: the thermal axion.
The contribution of the thermal axion to the cosmological quantities
can be described using the scale $f_a$ at which
the $U(1)_{PQ}$ symmetry
is spontaneously broken, or equivalently its mass $m_a$.
Since the thermal axion contributes as a relavitistic particle
in the early Universe and
as a massive component in the late Universe,
its effects are similar to those of the massive neutrinos and
a degeneracy exist between the two particles.
In particular,
the thermal axion free-streaming can explain the discrepancy
between local measurements and cosmological estimates
of the clustering parameter $\sigma_8$.

The most recent analyses from the Planck collaboration put
strong constraints on the additional relativistic particles in the early Universe
and no deviations from $\neff=3.046$ are shown.
Our analyses in Chapters~\ref{ch:lsn_cosmo} and \ref{ch:pps_nu}
confirmed these results, and the most stringent bound from CMB data only
we have found is $\neff=2.99\,^{+0.41}_{-0.39}$,
obtained from the Planck~TT,TE,EE+lowP dataset.
The corresponding bound at 68\% CL is
$\neff=2.99\pm0.20$ with a power-law PPS
($\neff=2.96\pm0.25$ with a \pchip\ PPS),
for which the 68\% CL upper constraint is $\neff\simeq3.2$.
The minimum contribution for a thermal axion
is $\DNeff\simeq0.2$, corresponding to
the minimum value $m_a\simeq0.1\ev$
allowed to perform the calculations for the thermal axion.
For smaller axion masses, in fact,
the decoupling temperature is above the QCD scale and
there are no particles that can efficiently interact with the axion
and allow its decoupling
\cite{Masso:2002np,Graf:2010tv,Salvio:2013iaa},
that jumps to very high temperatures.
In this case, the upper value on \neff\ allowed
by the CMB data at 68\% CL
is smaller than the minimum value of \neff\ that is possible
when a thermal axion is included, that would be $\neff\simeq3.25$.
As a consequence, the presence of a thermal axion
is excluded at 68\% CL by CMB data.

Concerning the robustness of the bounds on the axion mass against
changes in the assumptions on the power spectrum
of initial curvature perturbations,
we can notice that the axion mass bounds are largely relaxed when a
free PPS is assumed.
When including the small scale CMB polarization we find
a further weakening of the axion mass constraints.
This is due to the fact that polarization constrains
significantly the contribution of the axion to \neff,
but this depends weakly on $m_a$, if it is large.
As a consequence, the posterior distribution
is smaller at small $m_a$, but is
unchanged for large $m_a$.
The reduced volume of the posterior distribution
for small axion masses is then translated into a broadening
of the marginalized constraints towards higher values for $m_a$.
The strongest bound we find on the thermal axion mass within
the \pchip approach is $m_a<1.07$~eV at 95\% CL when considering
the Planck TT+lowP+BAO data combination.
In the standard power-law scenario, the most stringent bound is
$m_a<0.74$~eV at 95\% CL, obtained with the further inclusion of the
polarization at high multipoles 
(Planck TT,TE,EE+lowP+BAO).

When we vary also the massive neutrino mass to test
the degeneracy with the thermal axion mass,
we find that the constraints on the total neutrino mass
are tighter than those obtained without thermal axions
(see Section~\ref{sec:inflfreed_mnu}),
while the bounds on the thermal axion mass are unchanged.
The strongest bounds we find for the thermal axion mass
and the total neutrino mass in the \pchip approach are
$m_a<1.03$~eV at 95\% CL and $\mnu<0.18$~eV at 95\% CL,
when considering
the Planck TT+lowP+BAO and Planck TT,TE,EE+lowP+BAO
dataset combinations, respectively.
In the power-law PPS scenario the strongest bounds are
$m_a<0.76$~eV at 95\% CL and $\mnu<0.16$~eV at 95\% CL,
obtained both for the Planck TT,TE,EE+lowP+BAO dataset.

Finally,
from the analyses we performed in this Chapter
it is possible to obtain constraints on the PPS shape.
These results are not discussed in this Chapter,
however, since they are very similar
to those presented in Chapter~\ref{ch:pps_nu}.

%!TeX root=main.tex 
\chapter{Inflationary Freedom and Primordial non-Gaussianities}
\label{ch:ng}
\chapterprecis{This Chapter is based on
Ref.~\protect\cite{Gariazzo:2015qea}.}

% % % % % \begin{abstract}
The simplest models of inflation predict small non-Gaussianities
and a featureless power spectrum.
As we discussed in Chapter~\ref{ch:pps_nu},
however, a large number of well-motivated theoretical
scenarios of inflation predict features in the power-spectrum.
Hints of these feature have been observed in the CMB temperature
spectrum at small multipoles.
The scenarios that give origin to the features in the PPS
could also generate large non-Gaussianities.
We adopt the \pchip\ parameterization presented in
Section~\ref{sec:inflfreed_ppsparam}
to study, in a model-independent manner,
how
the constraints from future large scale structures (LSS) surveys
on the local non-Gaussianity parameter $\fnl$
change
if the assumption of a power-law (PL)
spectrum of initial perturbations is relaxed.
% If the analyses are restricted to the large scale-dependent bias
% induced in the linear matter power spectrum by non-Gaussianities,
% the errors on the $\fnl$ parameter could be increased
% by $60\%$  when exploiting data from the future DESI survey,
% if dealing with only one possible dark matter tracer. 
% In the same context, a nontrivial bias
% $|\delta \fnl| \sim 2.5$ could be induced
% if future data are fitted to the wrong primordial power spectrum.
% Combining all the possible DESI objects slightly ameliorates
% the problem, as the forecasted errors on $\fnl$
% would be degraded by $40\%$ when relaxing the assumptions
% concerning the primordial power spectrum shape. 
% Also the shift on the non-Gaussianity parameter is reduced
% in this case, $|\delta \fnl| \sim 1.6$. 
% The addition of Cosmic Microwave Background priors ensure
% robust future $\fnl$ bounds, as the forecasted errors
% obtained including these measurements are almost independent
% on the primordial power spectrum features,
% and $|\delta \fnl| \sim 0.2$,
% close to the standard single-field slow-roll paradigm prediction.
% % % % % \end{abstract}

\section{Introduction}
\label{sec:ng_intro}
We already discussed the fact that inflation has been
introduced to explain the flatness problem,
the horizon problem and
the generation of the primordial perturbations
seeding the evolution of our current
Universe~\cite{Guth:1980zm,Linde:1981mu,Starobinsky:1982ee,
Hawking:1982cz,Albrecht:1982wi,Mukhanov:1990me,Mukhanov:1981xt,
Lucchin:1984yf,Lyth:1998xn,Bassett:2005xm,Baumann:2008bn}.
The inflationary theories, however, could be confirmed as responsible
for the Universe we observe today only if 
a signal of primordial gravitational waves would be detected.
The different theories, nevertheless, may give different predictions
for the power spectrum of the initial curvature perturbations
$\mcp_\mcr(k)$.
As we discussed in Subsection~\ref{ssec:inCurvPert},
the Primordial Power Spectrum (PPS)
is usually assumed to be featureless,
described by a simple power-law $\mcp_\mcr(k)\propto k^{n_s-1}$
(see Eq.~\eqref{eq:plPPS}),
with $n_s$ the scalar spectral index.
This might not be the correct case, and a vast number of models proposed
in the past predict a non-standard PPS
(see e.g.\ the review \cite{Chluba:2015bqa}).
That is the case of slow-roll induced by phase transitions 
in the early
Universe~\cite{Adams:1997de,Hunt:2004vt,Hotchkiss:2009pj}, 
by some inflationary
potentials~\cite{Starobinsky:1992ts,Leach:2001zf,Gong:2005jr,
Adams:2001vc,Chen:2006xjb,Chen:2008wn,Lerner:2008ad,
Dvorkin:2009ne,Adshead:2011jq,Hodges:1989dw,Leach:2000yw,Joy:2007na, 
Jain:2007au,Bean:2008na,Ashoorioon:2006wc,Ashoorioon:2008qr,
Saito:2008em,Achucarro:2010da,Goswami:2010qu,Brax:2011si,
Arroja:2011yu,Liu:2011cw,Romano:2014kla,Kitazawa:2014dya},
by resonant particle 
production~\cite{Chung:1999ve,Mathews:2004vu,Romano:2008rr,
Barnaby:2009mc,Barnaby:2010sq},
variation in the sound speed of adiabatic 
modes~\cite{Achucarro:2012fd,Palma:2014hra} or 
by trans-Planckian 
physics~\cite{Brandenberger:2000wr,Danielsson:2002kx,Schalm:2004xg,
Greene:2005aj,Easther:2005yr}.
All the non-standard scenarios, of which this list is just a small fraction,
as well as other non-canonical 
schemes~\cite{Burgess:2002ub,Piao:2003zm,Powell:2006yg,
Nicholson:2007by,Lasenby:2003ur,Ribeiro:2012ar,Dvali:2003us,
Langlois:2004px}, 
could lead to a PPS which may notably differ
from the simple power-law parameterization. 

Most of the inflationary models we listed above predict also 
deviations from the pure Gaussian initial conditions.
Non-Gaussianities are usually described by a single parameter,
$\fnl$.  
In the matter-dominated Universe,
the gauge-invariant Bardeen potential on large scales
can be parametrized 
as~\cite{Salopek:1990jq,Gangui:1993tt,Verde:1999ij,Komatsu:2001rj} 
\begin{equation}
\Phi_{\mathrm{NG}}
=
\Phi+\fnl\left(\Phi^2-\langle \Phi^2 \rangle\right)\,,
\label{eq:fnl}
\end{equation}
where $\Phi$ is a Gaussian random field.
The non-Gaussianity parameter $\fnl$ is often considered
to be a constant,
yielding non-Gaussianities of the \textit{local} type. 

Traditionally, the standard observable to constrain
non-Gaussianities is the Cosmic Microwave Background (CMB),
through the three point correlation function, or \emph{bispectrum}. 
As the odd power correlation functions vanish for the case
of Gaussian random variables, the bispectrum provides
the lowest order statistic to test any departure from Gaussianity. 
The bispectrum is much richer than the power spectrum,
as it depends on both the scale and the shape
of the power spectra of primordial perturbations. 
The current bound from the complete Planck mission for
the local non-Gaussianity parameter is
$\fnl=0.8\pm 5.0$ ($68\%$~CL)~\cite{Ade:2015ava}. 

The large scale structures (LSS) of the Universe provide
an independent tool to test primordial non-Gaussianites,
as shown in the pioneer works of Refs.~\cite{Dalal:2007cu}
and~\cite{Matarrese:2008nc}. 
Dark matter halos will be affected by the presence 
of non-Gaussianities, and a scale-dependent bias will characterize
the non-Gaussian signal at large 
scales~\cite{Slosar:2008hx,Afshordi:2008ru,Carbone:2008iz,
Grossi:2009an,Desjacques:2008vf,Pillepich:2008ka,Alvarez:2014vva}. 
The strongest bounds on primordial non-Gaussianities obtained 
using exclusively 
LSS data are those obtained
from the DR8 photometric data,
see Ref.~\cite{Leistedt:2014zqa},
which exploits 800000 quasars and
finds $-49 < \fnl\ <31$
(see also Ref.~\cite{Agarwal:2013qta}). 
While current LSS constraints are highly penalized
by the systematic uncertainties,
it has been shown by a number of authors that upcoming future LSS
surveys will reach 
$\sigma(\fnl)<1$~\cite{
Alvarez:2014vva,dePutter:2014lna,Dore:2014cca,Byun:2014cea,
Raccanelli:2014awa,Yamauchi:2014ioa,Camera:2014bwa,Ferramacho:2014pua,
Ferraro:2014jba,Fedeli:2010ud,Carbone:2010sb,Giannantonio:2011ya}.
 
Despite the fact that the simplest models of inflation
(i.e.\ single field, slow-rolling with a canonical kinetic term)
predict small non-Gaussianities,
there are some theoretical scenarios in which large non-Gaussianities
could be generated,
see e.g.\ Ref.~\cite{Bartolo:2004if} and references therein. 
The same deviations from the standard slow-roll inflation
that give rise to non-Gaussianities could also be
a potential source for other features in the 
PPS~\cite{Hotchkiss:2009pj},
which are absent in the simplest models of inflation. 
For example, both a non-canonical PPS and large non-Gaussianities
can be generated simultaneously in scenarios involving
particle production during inflation~\cite{Barnaby:2010sq}. 
These two phenomena could also appear together in single field
models with non-standard inflationary 
potentials~\cite{Chen:2006xjb,Chen:2008wn,Adshead:2011jq,
Saito:2008em,Goswami:2010qu,Arroja:2011yu}, 
as well as in multi-field inflationary models~\cite{Achucarro:2010da}
and Brane Inflation~\cite{Bean:2008na}.
Finally, preheating 
scenarios~\cite{Chambers:2007se,Bond:2009xx}
are other examples of models that give rise to both a
non-standard PPS and non-Gaussianities. 

As nature could have chosen other inflationary scenarios rather
than the single field slow-roll paradigm,
it is interesting to explore
how the forecasts for LSS surveys
concerning future measurements of $\fnl$ are affected
when the assumption of a standard PPS is relaxed,
possibly adopting a model-independent description of the PPS.
This has never been done before while forecasting errors
on the $\fnl$ parameter and it is a mandatory calculation,
because models which will produce non-Gaussianities will likely
give rise to a non-standard PPS as well. 
Even if non-Gaussianities and distortions of the PPS are expected to be governed
by the same fundamental physics,
the underlying inflationary mechanism is unknown a priori. 
A conservative and general approach is therefore to treat these
two physical effects as independent and
to be determined simultaneously.
In this Chapter we adopt this strategy.

Following Ref.~\cite{Gariazzo:2015qea} on which this Chapter is based,
we structure the discussion in this way:
we start describing the parameterization of the PSS used here
in Sec.~\ref{sec:ng_pps},
we describe the scale-dependent halo bias
in the matter power spectrum in Subsection~\ref{sec:ng_nghb}, 
while Subsection~\ref{sec:ng_method} is about
the methodology followed for our calculations as well as
the specifications of the future LSS survey
illustrated here.
We present our results in Subsec.~\ref{sec:ng_results} and
we draw our conclusions in Sec.~\ref{sec:ng_concl}. 

\section{Primordial power spectrum}
\label{sec:ng_pps}
% The simplest models of inflation predict a power-law form for the PPS
% of scalar and tensor perturbations. 
We discussed in Chapter~\ref{ch:pps_nu} that,
in principle, a non-standard shape for the PPS
(see Ref.~\cite{Chluba:2015bqa} and references therein),
can be generated by many inflationary models
(see e.g.\ Ref.~\cite{Martin:2013tda} for some compilation)
that goes beyond the simplest one.
A power-law PPS of scalar and tensor perturbations is the simplest
possibility, but it may not be the correct one.
In order to explore the robustness of future forecasted errors
from LSS surveys on the local non-Gaussianity parameter $\fnl$, 
we assume a non-parametric form for the PPS,
following the prescriptions 
reported in Section~\ref{sec:inflfreed_ppsparam}.
This is one of a number of possible methods explored in the 
literature~\cite{Hunt:2013bha,Hazra:2014jwa,Mukherjee:2003cz, Shafieloo:2006hs,
Leach:2005av,Wang:1998gb,Bridle:2003sa,Hannestad:2003zs,Bridges:2008ta,
Verde:2008zza,Ichiki:2009zz,Hu:2014aua,Vazquez:2012ux,Nicholson:2009zj,
Hunt:2015iua,Goswami:2013uja,Matsumiya:2001xj,Matsumiya:2002tx,Kogo:2003yb,
Kogo:2005qi,Nagata:2008tk,Ade:2015lrj,Hazra:2013nca,Aslanyan:2014mqa,
Shafieloo:2003gf,TocchiniValentini:2004ht,Paykari:2014cna,Kogo:2004vt,
Nicholson:2009pi,Hamann:2009bz,Gauthier:2012aq,Hazra:2013ugu,dePutter:2014hza,
Iqbal:2015tta}.

In brief, 
we describe the PPS of the scalar perturbations as a function that
interpolates the PPS values in a series of nodes at fixed position. 
The function we exploit to interpolate is named 
\emph{piecewise cubic Hermite interpolating polynomial}, 
the \pchip\ algorithm~\cite{Fritsch:1980},
described in details in Appendix~\ref{ch:app_pchip}.
The nodes we use to interpolate the PPS are twelve, located
at the values of $k$ listed in Eq.~\eqref{eq:nodesspacing}.
The nodes are equally spaced (in logarithmic scale) in the range $(k_2, k_{11})$,
that has been shown to be well constrained by current cosmological
data~\cite{dePutter:2014hza}.
The extreme nodes in $k_1$ and $k_{12}$ are fixed to allow
for a non-constant behavior of the PPS outside the well-constrained range.
The \pchip\ PPS is given by Eq.~\eqref{eq:PPS_pchip}
and we parameterize the value of the PPS in the nodes with \psj{j}.

\section{Forecasts}
\label{sec:ng_future}
\subsection{Non-Gaussian halo bias} 
\label{sec:ng_nghb}
Non-Gaussianities as introduced in Eq.~\eqref{eq:fnl} induce
a scale-dependent bias that affects the matter power spectrum at large scales.
This scale-dependent bias reads
as~\cite{Dalal:2007cu,Slosar:2008hx}
\begin{equation}
\delta_g
=
b\, \delta_{\mathrm{dm}}
\quad
\mbox{where}
\quad b=b_{\mathrm{G}}+\Delta b\,,
\label{eq:deltaNG}
\end{equation}
where
$\delta_g (\delta_{\mathrm{dm}})$ are the galaxy (dark matter) overdensities,
$b_{\mathrm{G}}$ is the Gaussian bias
and $\Delta b$ reads as
\begin{equation}
\Delta b
=
3 \fnl(1-b_{\mathrm{G}})
\delta_{\mathrm c}
\frac{H_0^2\Omega_{\mathrm m}}{k^2 T(k) D(a)}\,,
\label{eq:bfnl}
\end{equation}
where $T(k)$ is the linear transfer function.
The growth factor $D(a)$ is defined as
$\delta_{\mathrm{dm}}(a)/\delta_{\mathrm{dm}}(a=1)$
and $\delta_c$ refers
to the critical linear overdensity for spherical collapse~\cite{Kitayama:1996pk}.
The power spectrum with the inclusion of non-Gaussianities is obtained using 
\begin{equation}
P_{\textrm{ng}}
=
P \left(b_G+\Delta b+f\mu_k^2\right)^2\,,
\label{eq:Pstb}
\end{equation}
where $\mu_k$ is the cosine of the angle between the line of sight and
the wave vector $k$
and $f$ is defined as $d\ln\delta_{\mathrm{dm}}/d\ln a$.
$P$ is the dark matter power spectrum, whose $k$
dependence is driven either by Eq.~\eqref{eq:PPS_pchip}
or by the standard power-law PPS in Eq.~\eqref{eq:plPPS}
(given the amplitude $A_s$ and the slope $n_s$).

\begin{figure}
\begin{center}
\includegraphics[width=0.9\textwidth]{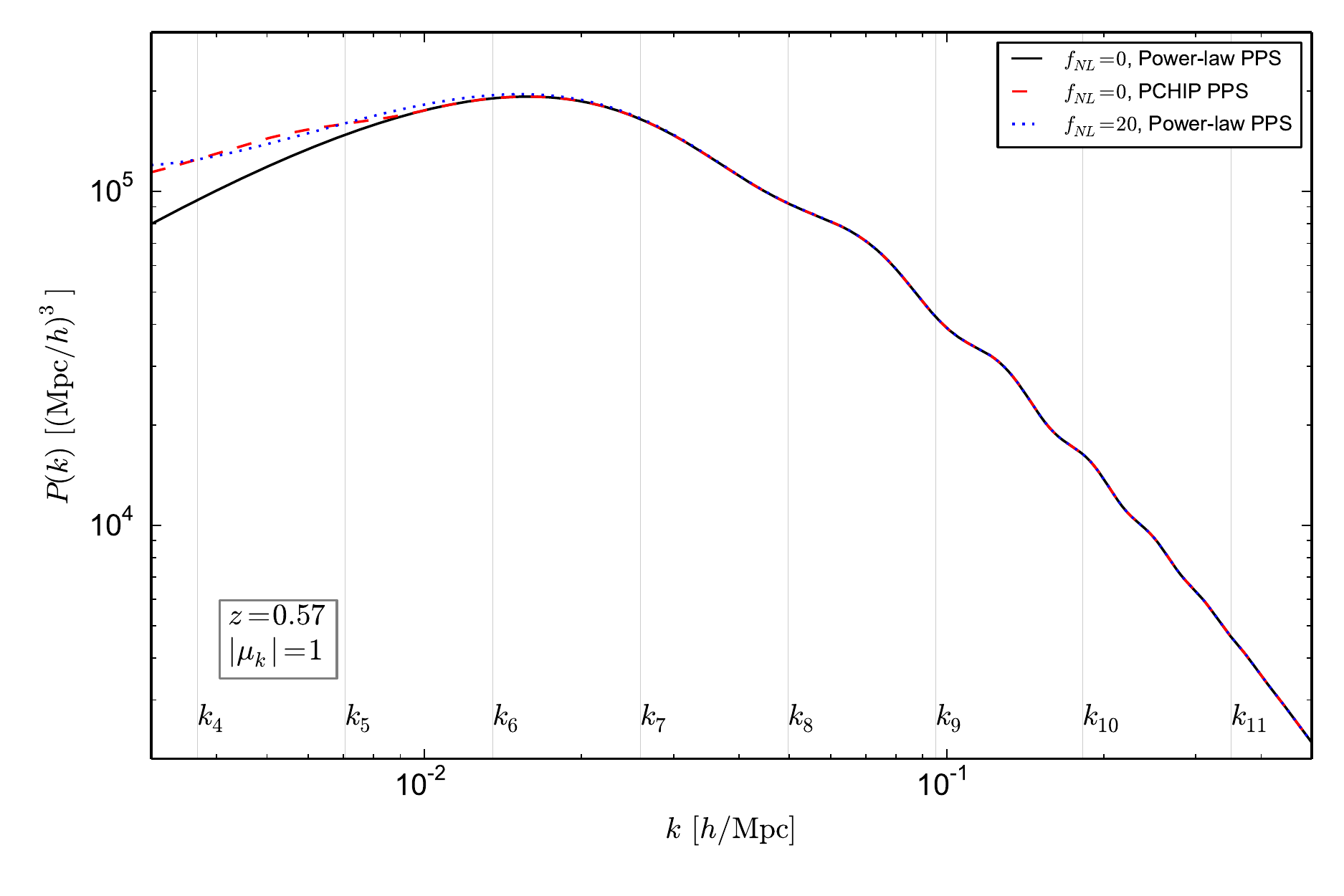}

\includegraphics[width=0.9\textwidth]{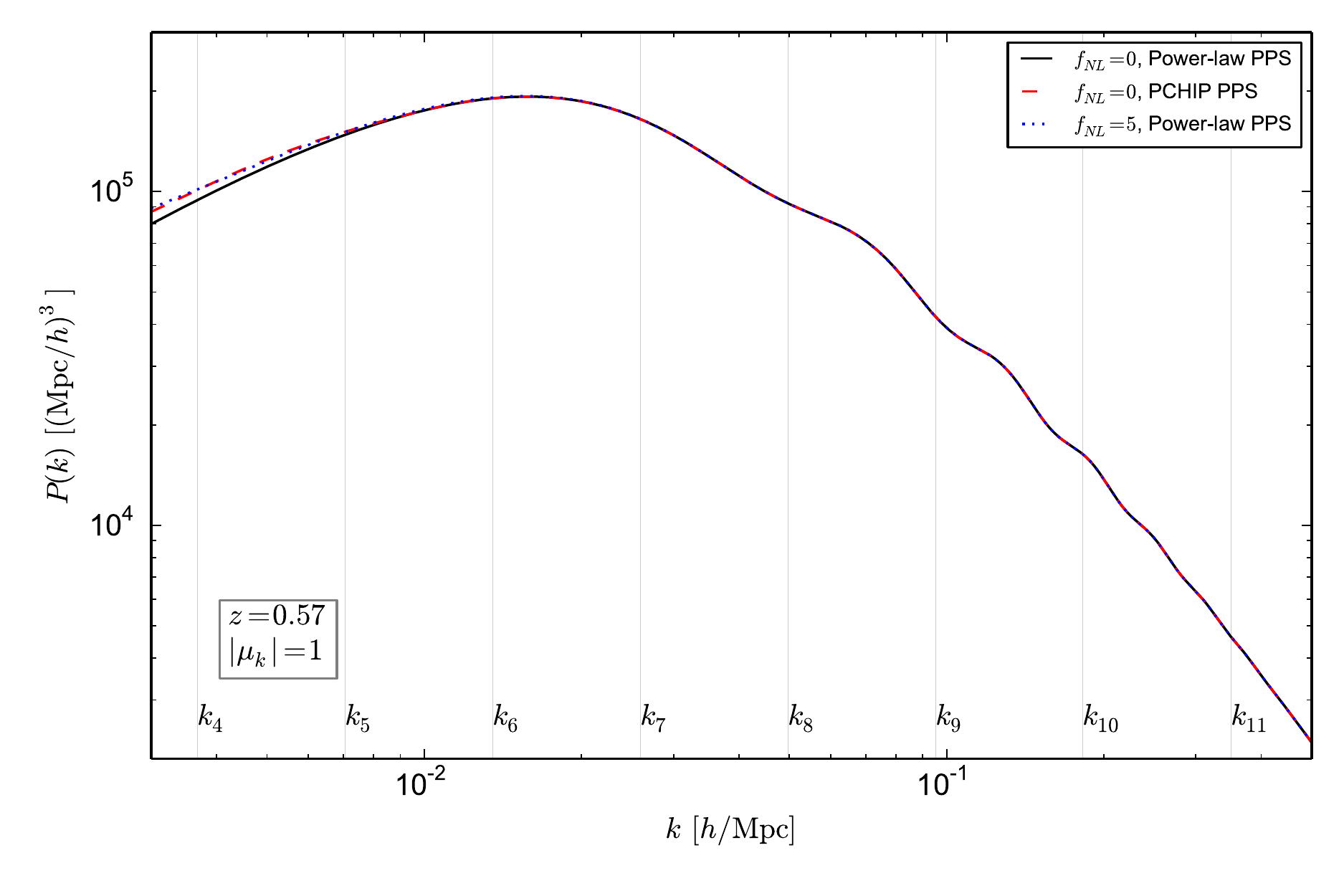}
\end{center}
\caption[Comparison of the galaxy power spectra obtained with the
power-law PPS and the \pchip\ PPS for different values of \fnl]
{
The top panel depicts the galaxy power spectrum obtained with the power-law PPS case,
for $\fnl=0$ (black solid curve) and $\fnl=20$ (blue dotted curve),
together with one obtained with a \pchip\ PPS (red dashed lines) for $\fnl=0$.  
The values of the \pchip\ PPS nodes are chosen accordingly
to match the predictions of the $\fnl=20$ case. 
The bottom panel shows the same for $\fnl=5$,
with appropriate changes of the values of the \pchip\ PPS nodes. 
The labels $k_i$ for $i=4,\ldots,11$ indicate the $k$ position of the five
nodes considered in our analysis ($i=5,\ldots,9$),
plus the nodes $k_{4}$, $k_{10}$, $k_{11}$ that lie outside the $k$ range probed by the DESI experiment. 
The galaxy power spectra are obtained for $z=0.57$,
$|\mu_k|=1$ and assuming a constant Gaussian bias $b_G$.
From Ref.~\cite{Gariazzo:2015qea}.}
\label{fig:ng_fig1}
\end{figure}

In the top panel of Fig.~\ref{fig:ng_fig1}
we plot the galaxy power spectrum in absence of non-Gaussianities
(i.e.\ for $\fnl=0$) and for $\fnl=20$. 
The red dashed line indicates that using a \pchip\ PPS with $\fnl=0$
it is possible to reproduce the galaxy power spectrum obtained
with a standard power-law PPS and $\fnl\neq 0$. 
The $P_{s,j}$ values needed to obtain such an effect were taken
within their $95\%$~CL allowed regions~\cite{Gariazzo:2014dla}.
This shows that large degeneracies between the $P_{s,j}$ nodes
and the $\fnl$ parameter may appear.
The large value $\fnl=20$,
albeit allowed by the current LSS limits on local non-Gaussianities,
is much larger than the expected errors from the upcoming galaxy surveys
(see e.g.\ Refs.~\cite{Byun:2014cea,Sartoris:2015aga}).
Therefore, we also illustrate in the bottom panel of Fig.~\ref{fig:ng_fig1}
the equivalent plot for $\fnl=5$.
In this case,
the values for the PPS nodes $P_{s,j}$ required to match the predictions
obtained with the PL PPS
lie within their $68\%$~CL allowed regions~\cite{Gariazzo:2014dla}.
We can notice that the degeneracies are still present:
we therefore expect that the forecasted errors on $\fnl$
are largely affected by the uncertainties on the precise PPS shape.

\subsection{Methodology} 
\label{sec:ng_method}

We focus here on the future spectroscopic galaxy survey DESI
(Dark Energy Spectroscopic Instrument) experiment~\cite{Levi:2013gra}.
Although multi-band, full-sky imaging surveys have been shown
to be the optimal setups to constrain non-Gaussianities
via LSS measurements~\cite{Alvarez:2014vva,dePutter:2014lna},
the purpose of the current analysis is to explore the degeneracies with
the PPS parameterization rather than to optimize the $\fnl$ sensitivity. 
For this reason, we restrict ourselves to the DESI galaxy redshift survey, but
similar results would be obtained with the results of the ESA
Euclid instrument~\cite{Scaramella:2015rra}.

In order to compute the expected errors on the local non-Gaussianity parameter,
we follow here the usual Fisher matrix approach.
The elements of the Fisher matrix,
as long as the posterior distribution for the parameters
can be approximated by a Gaussian function,
are~\cite{Tegmark:1996bz,Jungman:1995bz,Fisher:1935bi}
\begin{equation}
\label{eq:fish}
F_{\alpha\beta}
=
\frac{1}{2}{\mathrm{Tr}}\left[C^{-1}C_{,\alpha}C^{-1}C_{,\beta}\right]~,
\end{equation}
where $C=S+N$ is the total covariance.
The covariance matrix contains both the signal $S$ and the noise $N$ terms,
and $C_{,\alpha}$ indicates its derivatives with respect
to the cosmological parameter $p_\alpha$ in the context
of the underlying fiducial model.
The $68\%$~CL marginalized errors on a given parameter $p_\alpha$ are
$\sigma(p_\alpha)=\sqrt{({F}^{-1})_{\alpha\alpha}}$,
where ${F}^{-1}$ is the inverse of the Fisher matrix.
In order to highlight the differences in the errors on the $\fnl$ parameter
arising from different PPS choices,
we only consider information concerning non-Gaussianities from LSS data, 
and we neglect the information that could be added
from the measurements of the CMB bispectrum.  

Our LSS Fisher matrix reads as~\cite{Seo:2003pu}
\begin{eqnarray}  
F^{\mathrm LSS}_{\alpha \beta}
&=&
\int_{\vec{k}_{\mathrm{min}}} ^ {\vec{k}_{\mathrm{max}}}
\frac{\partial \ln P_{\mathrm{ng}}(\vec{k})}{\partial p_\alpha}
\frac{\partial \ln P_{\mathrm{ng}}(\vec{k})}{\partial p_\beta}
V_\eff(\vec{k})
\frac{d\vec{k}}{2(2 \pi)^3}  
\label{eq:Fij} \\
&=&
\int_{-1}^{1} \int_{k_{\mathrm{min}}}^{k_{\mathrm{max}}}
\frac{\partial \ln P_{\mathrm{ng}}(k,\mu_k)}{\partial p_\alpha}
\frac{\partial \ln P_{\mathrm{ng}}(k,\mu_k)}{\partial p_\beta}
V_{\mathrm{eff}}(k,\mu_k)
\nonumber\\
& &
\frac{2\pi k^2 dk d\mu_k}{2(2 \pi)^3}\,,
\nonumber 
\end{eqnarray}
where $V_\eff$ is the effective volume of the survey.
It is calculated as
\begin{eqnarray}
V_{\mathrm{eff}}(k,\mu_k)
&=&
\left[
  \frac{{n}P_{\mathrm{ng}}(k,\mu_k)}{{n}P_{\mathrm{ng}}(k,\mu_k)+1}
\right ]^2
V_{\textrm{survey}}\,,
\label{eq:Veff} 
\end{eqnarray}
where $P_{\mathrm{ng}}$ is the power spectrum calculated with the inclusion
of non-Gaussianities (see Eq.~\eqref{eq:Pstb})
and $n$ refers to the galaxy number density per redshift bin.
We assume $k_{\mathrm{max}}=0.1 h$/Mpc
and we choose $k_{\mathrm{min}}=2\pi/V^{1/3}$,
where $V$ represents the volume of the redshift bin.
The DESI survey is expected to cover 14000 deg$^2$ of the sky
in the range $0.15<z<1.85$,
divided in redshift bins of width $\Delta z=0.1$. 
We follow Ref.~\cite{Font-Ribera:2013rwa} for the number densities $n(z)$
and biases $b_G(z)$ associated to the three types of DESI tracers:
Luminous Red Galaxies (LRGs),
Emission Line Galaxies (ELGs) and
high-redshift quasars (QSOs). 
We include the redshift dependence of the (fiducial) bias $b_G$
in Eq.~\eqref{eq:Pstb} as follows:
$b_{G}(z) D(z) = 0.84, 1.7, 1.2$ for ELG, LRG and QSO's respectively,
where $D(z)$ is the growth factor as a function of the redshift,
as in Eq.~\eqref{eq:bfnl}. 
Since we want to combine the three different Fishers matrices
from the three DESI tracers (LRGs, ELGs and QSOs),
we adopt the multi-tracer formalism developed in Ref.~\cite{Abramo:2013awa}.
In the work, the authors present a generic expression
for the Fisher information matrix of surveys with any number of tracers. 
The multi-tracer technique provides constraints that can surpass
those set by cosmic variance, since the possible tracers of LSS
can present differences in their clustering.

We remind that the observed size of an object or of a feature
at a redshift $z$ is obtained in terms of the redshift and
the angular quantities $\Delta z$ and $\Delta \theta$.
These two quantities are related to the comoving distances $r_\parallel$
and $r_\perp$, along and across the line of sight respectively,
through the angular diameter distance $D_A(z)$ and the Hubble rate $H(z)$.
The same applies to the Fourier transform associated variables,
$k_\parallel$ and $k_\perp$ for the dual coordinates of
$r_\parallel$ and $r_\perp$.
Therefore, when reconstructing the measurements of galaxy redshifts and
positions in some reference cosmological model which differs
from a given fiducial cosmology,
one has to take into account the geometrical effects~\cite{Seo:2003pu}:
\begin{equation}
P_{obs}(k_\parallel^{ref}, k_\perp^{ref})
=
\frac{D_A(z)|_{ref}^2}{D_A(z)^2}\, 
\frac{H(z)}{H(z)|_{ref}}\, 
P_{fid}(k_\parallel, k_\perp)\,,
\end{equation}
where the \textit{ref} sub/superscript denote quantities in the
reference cosmological model%
\footnote{
$k_\parallel
=
k_\parallel^{ref}\,
D_A(z)|_{ref}/D_A(z)$
and
$k_\perp
=
k_\perp^{ref} \,
H(z)/H(z)|_{ref}$.}.
We properly take into account these effects in our Fisher matrix forecasts
when taking numerical derivatives of the galaxy power spectrum
with respect to the cosmological parameters at given values
of $|\mathbf{k}|$ and $\mu_k$,
that are the equivalent of $k_\parallel$ and $k_\perp$.

In addition to the Fisher matrix forecasts,
we will also compute the expected shift in the $\fnl$ parameter
if the \psj{j} parameters (with $j=5,\ldots,9$) describing the \pchip\ PPS
are incorrectly set to values different from their fiducial ones.  
For that purpose,
we use the method developed by the authors of Ref.~\cite{Heavens:2007ka}. 
This is the main idea:
if the future DESI data are fitted assuming a cosmological model
with fixed values of \psj{j},
corresponding to fix both $n_s$ and $A_s$ to their best-fit values,
the model is characterized by $n^\prime= 5$ parameters
$\mcm'=\{\Omega_b h^2$,  $\Omega_c h^2$, $h$,  $\fnl$, $w\}$.
If the true underlying cosmology is a model with different values of the \psj{j}
and it is characterized by $n=10$ parameters
$\mcm=\{\Omega_b h^2$,  $\Omega_c h^2$, $h$,  $\fnl$, $w$, $\psj{j}\}$
(with $j=5,\ldots,9$),
the values inferred for the $n^\prime=5$ parameters will be shifted
from their true values to compensate for the fact
that the model used to fit the data is wrong.
Under the assumption of a Gaussian likelihood,
the shifts in the $n^\prime$ parameters are~\cite{Heavens:2007ka}
\begin{eqnarray}
\delta\theta'_\alpha
=
-(F'^{-1})_{\alpha\beta}G_{\beta\zeta}\delta\psi_\zeta \qquad
& &
\alpha,\beta=1\ldots n',
\nonumber\\
& &
\zeta=n'+1\ldots n \label{eq:offset}\,,
\end{eqnarray}
where $F^\prime$ is the Fisher matrix for the model with $n'$ parameters
(with fixed \psj{j})
and
$G$ denotes the Fisher matrix for the $n$ parameters model
(including the previous $n'$ parameters and the \pchip\ parameters). 

In the following, unless otherwise stated,
we adopt the best-fit values from the recent Planck release~\cite{Ade:2015xua},
which corresponds to $A_s=2.2\e{-9}$ and $n_s=0.965$
at the pivot scale $k_{pivot}=0.05$ for the standard power-law PPS.
When we consider the \pchip\ parameterization,
the best-fit values of the nodes we considered in the numerical analyses are:
$P_{s,5}=1.07099$,
$P_{s,6}=1.04687$,
$P_{s,7}=1.02329$,
$P_{s,8}=1.00024$ and
$P_{s,9}=0.97771$.
These values are obtained calculating the value of the best-fit PL PPS
at the positions of the nodes $k_5$ to $k_9$
using Eq.~\eqref{eq:psj_plpps},
given the Planck 2015 best-fit values for $A_s$ and $n_s$.
The nodes \psj{j}
corresponding to $j<5$ and $j>9$
are outside the range of wavemodes that
the DESI survey is expected to cover,
considering the values of 
$k_{\mathrm{max}}$ and $k_{\mathrm{min}}$ that we adopt here.

\subsection{Results} 
\label{sec:ng_results}

We present now the results obtained from
our Fisher matrix calculations,
for the two fiducial cosmologies explored here:
one in which the PPS is described by the standard power-law form,
and a second one where we assume a free PPS,
described by the \pchip\ parameterization.
The parameters describing the model with a PL PPS are
the baryon and cold dark matter energy densities
$\Omega_b h^2$ and $\Omega_c h^2$,
the reduced Hubble parameter $h$,
the scalar spectral index $n_s$,
the amplitude of the PPS $A_s$
and the equation of state of the dark energy component $w$. 
The \pchip\ PPS case is also described by
$\Omega_b h^2$, $\Omega_c h^2$, $h$, $w$,
plus five nodes $\psj{j}$ with $j\in 5,\ldots,9$.
Non-Gaussianities of the local type are included
in both the fiducial cosmologies via the $\fnl$ parameter.
All the results described below, unless otherwise stated,
refer to the analysis of the three DESI tracers (ELGs, LRGs and QSOs).
This means that they have been obtained exploiting exclusively
the scale-dependent biases imprinted in the power spectra 
of these three types of tracers. 

\begin{table*}[t]
\begin{center}
\begin{tabular}{c|c|c c c c }
\hline
&fiducial & LRG  & ELG & QSO& All   \\
\hline
$\Omega_b h^2$ & $ 0.02267	$ & $ 4.78\e{-3}$ & $4.86\e{-3} $ & $ 5.11\e{-3} $ & $ 2.38\e{-3}$ \\
$\Omega_c h^2$ & $0.1131$ & $ 1.75\e{-2}$ & $1.65\e{-2} $ & $1.51\e{-2} $ & $ 7.70\e{-3}$ \\
$h	$ & $	0.705$ & $ 5.02\e{-2}$ & $ 5.01\e{-2}$ & $ 4.69\e{-2}$ & $ 2.42\e{-2}$ \\
$n_s	$ & $	0.96	$  & $ 5.68\e{-2} $ & $4.28\e{-2} $ & $ 4.12\e{-2}$ & $1.96\e{-2} $ \\
$A_s	$ & $	2.2\e{-9}$ & $0.341 $ & $0.331 $ & $ 0.302$ & $ 0.156$ \\
$\fnl$ & $20$ & $ 19.9$ & $10.1$ & $8.56 $ & $ 4.79$ \\
$w	$ & $	-1	$ & $	5.38\e{-2}  $  & $4.09\e{-2} $ & $ 6.18\e{-2} $ & $2.36\e{-2} $ \\
\hline
\end{tabular}
\end{center}
\caption[Marginalized 1$\sigma$ constraints in the model
with a PL PPS, assuming $\fnl=20$]
{Marginalized 1$\sigma$ constraints on the parameters associated
to the PL PPS assuming a fiducial value $\fnl=20$.
The error on the amplitude of the power spectrum is evaluated
on $A_s/(2.2\cdot10^{-9})$.
From Ref.~\cite{Gariazzo:2015qea}.
}
\label{tab:ng_sk1}
\end{table*}

\begin{table*}[t]
\begin{center}
\begin{tabular}{c|c|c c c c}
\hline
&fiducial & LRG & ELG & QSO& All \\
\hline
$\Omega_b h^2$ & $ 0.02267$ & $7.85\e{-3}$ & $ 3.65\e{-3}$ & $4.70\e{-3} $& $2.30\e{-3}  $\\ 
$\Omega_c h^2$ & $0.1131$ & $2.30\e{-2}$ & $1.11\e{-2} $ & $ 1.41\e{-2} $& $6.36\e{-3}  $\\ 
$h	$ & $	0.705$ & $7.67\e{-2}$ & $3.59\e{-2} $ & $4.62\e{-2}  $& $ 2.12\e{-2}$\\ 
$P_{s,5}	$ & $	1.07099$ & $ 0.340$ & $0.169$ & $ 0.212$& $ 0.111$\\ 
$P_{s,6}	$ & $	1.04687$ & $ 0.419$ & $ 0.198$ & $0.254$& $0.119$\\ 
$P_{s,7}	$ & $1.02329$ & $ 0.451$ & $0.216$ & $ 0.276$& $ 0.125$\\ 
$P_{s,8}	$ & $	1.00024$ & $ 0.479$ & $0.229 $ & $0.293$& $0.132$\\ 
$P_{s,9}	$ & $	0.97771$ & $ 0.482$ & $0.234$ & $ 0.298$& $ 0.134 $\\ 
$\fnl$ & $20$ & $32.2$ & $13.3 $ & $ 12.6$& $ 6.43$\\ 
$w	$ &$ 	-1$ & $4.03\e{-2}$ & $2.80\e{-2} $ & $4.45\e{-2} $& $ 2.45\e{-2}$\\ 
\hline
\end{tabular}
\end{center}
\caption[Marginalized 1$\sigma$ constraints in the model
with a \pchip\ PPS,
assuming $\fnl=20$]
{Marginalized 1$\sigma$ constraints on the parameters associated
to the non-standard PPS assuming $\fnl=20$.
From Ref.~\cite{Gariazzo:2015qea}.}
\label{tab:ng_nsk1}
\end{table*}

\begin{table*}[t]
\begin{center}
\begin{tabular}{c|c|c c c c }
\hline
&fiducial & LRG  & ELG & QSO& All   \\
\hline
$\Omega_b h^2$ & $ 0.02267	$ & $2.67\e{-4} $ & $2.63\e{-4} $ & $ 2.66\e{-4}$ & $2.59\e{-4}$ \\
$\Omega_c h^2$ & $0.1131$ & $ 1.64\e{-3}$ & $1.44\e{-3} $ & $1.52\e{-3} $ & $ 1.24\e{-3}$ \\
$h$ & $0.705$ & $ 6.66 \e{-3}$ & $5.24 \e{-3} $ & $5.86 \e{-3} $ & $ 4.12 \e{-3}$\\
$n_s	$ & $	0.96	$  & $6.72\e{-2} $ & $ 6.41\e{-2}$ & $6.53\e{-2} $ & $5.84\e{-3}$ \\
$A_s	$ & $	2.2\e{-9}	$ & $3.87\e{-2} $ & $ 3.28\e{-2} $ & $3.51\e{-2} $ & $2.71\e{-2} $ \\
$\fnl$ & $20$ & $ 17.4$ & $9.14 $ & $ 7.58$ & $ 4.56$ \\
$w	$ & $	-1	$ & $	 4.51\e{-2} $ &  $ 3.36\e{-2}$ & $ 5.44\e{-2}$ & $ 2.17\e{-2} $ \\
\hline
\end{tabular}
\end{center}
\caption[As Tab.~\ref{tab:ng_sk1} but including CMB priors]
{As Tab.~\ref{tab:ng_sk1} but including CMB priors.
From Ref.~\cite{Gariazzo:2015qea}.}
\label{tab:ng_sk1cmb}
\end{table*}

\begin{table*}[t]
\begin{center}
\begin{tabular}{c|c|c c c c c}
\hline
&fiducial & LRG  & ELG& QSO& all \\
\hline
$\Omega_b h^2$ & $ 0.02267$ & $3.92\e{-4}$ & $ 3.79\e{-4}$ & $3.87\e{-4} $& $3.74\e{-4} $\\ 
$\Omega_c h^2$ & $0.1131$ & $1.36\e{-3}$ & $1.10\e{-3} $ & $1.18\e{-3} $& $ 1.04\e{-3} $\\ 
$h	$ & $	0.705$ & $ 4.13\e{-3}$ & $3.14\e{-3} $ & $3.62\e{-3}  $& $2.93\e{-3}  $\\ 
$P_{s,5}	$ & $	1.07099$ & $2.98\e{-2}$ & $2.69\e{-2} $ & $2.77\e{-2}  $& $ 2.60\e{-2}$\\ 
$P_{s,6}	$ & $	1.04687$ & $2.89\e{-2}$ & $2.10\e{-2}  $ & $2.32\e{-2}  $& $ 1.99\e{-2}$\\ 
$P_{s,7}	$ & $1.02329$ & $2.00\e{-2}$ & $1.73\e{-2}  $ & $1.84\e{-2}  $& $1.69\e{-2} $\\ 
$P_{s,8}	$ & $	1.00024$ & $1.92\e{-2}$ & $1.76\e{-2}  $ & $ 1.86\e{-2} $& $ 1.73\e{-2}$\\ 
$P_{s,9}	$ & $	0.97771$ & $2.59\e{-2}$ & $2.31\e{-2}  $ & $ 2.42\e{-2} $& $ 2.22\e{-2}$\\ 
$\fnl$ & $20$ & $13.0$ & $ 6.85$ & $ 5.64$& $4.75 $\\ 
$w	$ &$ 	-1$ & $3.24\e{-2}$ & $ 2.46\e{-2}$ & $4.0\e{-2} $& $2.28\e{-2}  $\\ 
\hline
\end{tabular}
\end{center}
\caption[As Tab.~\ref{tab:ng_nsk1} but including CMB priors]
{As Tab.~\ref{tab:ng_nsk1} but including CMB priors.
From Ref.~\cite{Gariazzo:2015qea}.}
\label{tab:ng_nsk1cmb}
\end{table*}

%%%%%%%%%%%%%%%%%%%%%%%%%%%%%%%%%%

\begin{table*}[t]
\begin{center}
\begin{tabular}{c|c|c c c c }
\hline
&fiducial & LRG  & ELG & QSO& All   \\
\hline
$\Omega_b h^2$ & $ 0.02267	$ & $4.78\e{-3} $ & $ 5.17\e{-3} $ & $ 5.18\e{-3}$ & $ 2.45\e{-3}$ \\
$\Omega_c h^2$ & $0.1131$ & $1.73\e{-2} $ & $1.73\e{-2}  $ & $1.52\e{-2}  $ & $ 7.88\e{-3}$ \\
$h	$ & $	0.705$ & $ 5.0\e{-2}$ & $ 5.29\e{-2}$ & $ 4.75\e{-2}$ & $ 2.48\e{-2}$ \\
$n_s	$ & $	0.96	$  & $ 5.59\e{-2} $ & $ 4.40\e{-2} $ & $ 4.11\e{-2} $ & $2.0\e{-2} $ \\
$A_s	$ & $	2.2\e{-9}$ & $0.339 $ & $ 0.347$ & $ 0.305$ & $ 0.160$ \\
$\fnl$ & $5$ & $18.9 $ & $9.32 $ & $ 7.83$ & $ 4.45$ \\
$w	$ & $	-1$ & $	5.38\e{-2} $  & $ 4.13\e{-2}$ & $ 6.19\e{-2}$ & $2.38\e{-2} $ \\
\hline
\end{tabular}
\end{center}
\caption[Marginalized 1-$\sigma$ constraints in the model
with a PL PPS, assuming $\fnl=5$]
{Marginalized 1-$\sigma$ constraints on  the parameters associated
to the PL PPS assuming a fiducial value $\fnl=5$.
The error on the amplitude of the power spectrum is evaluated
on $A_s/(2.2\cdot10^{-9})$.
From Ref.~\cite{Gariazzo:2015qea}.}
\label{tab:ng_sk5}
\end{table*}

\begin{table*}[t]
\begin{center}
\begin{tabular}{c|c|c c c c}
\hline
&fiducial & LRG  & ELG& QSO& All \\
\hline
$\Omega_b h^2$ & $ 0.02267$ & $7.72\e{-3}$ & $ 3.61\e{-3}$ & $4.61\e{-3} $& $2.31\e{-3}  $\\ 
$\Omega_c h^2$ & $0.1131$ & $2.28\e{-2}$ & $1.09\e{-2} $ & $ 1.38\e{-2} $& $6.37\e{-3}  $\\ 
$h	$ & $	0.705$ & $7.56\e{-2}$ & $3.54\e{-2} $ & $4.52\e{-2}  $& $ 2.13\e{-2}$\\ 
$P_{s,5}	$ & $	1.07099$ & $ 0.342$ & $0.169$ & $ 0.215$& $ 0.113$\\ 
$P_{s,6}	$ & $	1.04687$ & $ 0.415$ & $ 0.196$ & $0.251$& $0.120$\\ 
$P_{s,7}	$ & $1.02329$ & $ 0.445$ & $0.212$ & $ 0.270$& $ 0.126$\\ 
$P_{s,8}	$ & $	1.00024$ & $ 0.472$ & $0.225 $ & $0.287$& $0.133$\\ 
$P_{s,9}	$ & $	0.97771$ & $ 0.476$ & $0.230$ & $ 0.292$& $ 0.135 $\\ 
$\fnl$ & $5$ & $29.3$ & $11.9$ & $ 10.7$& $ 5.97$\\ 
$w	$ &$ 	-1$ & $4.02\e{-2}$ & $2.79\e{-2} $ & $4.45\e{-2} $& $ 2.44\e{-2}$\\ 
\hline
\end{tabular}
\end{center}
\caption[Marginalized 1$\sigma$ constraints in the model
with a \pchip\ PPS, assuming $\fnl=5$]
{Marginalized 1$\sigma$ constraints on the parameters associated
to the non-standard PPS assuming $\fnl=5$.
From Ref.~\cite{Gariazzo:2015qea}.}
\label{tab:ng_nsk5}
\end{table*}

\begin{table*}[t]
\begin{center}
\begin{tabular}{c|c|c c c c }
\hline
&fiducial & LRG & ELG & QSO& All   \\
\hline
$\Omega_b h^2$ & $ 0.02267	$ & $2.67\e{-4} $ & $ 2.63\e{-4}$ & $ 2.67\e{-4}$ & $2.59\e{-4}$ \\
$\Omega_c h^2$ & $0.1131$ & $1.64\e{-3} $ & $1.43\e{-3} $ & $1.52\e{-3} $ & $ 1.24\e{-3}$ \\
$h$ & $0.705$ & $ 6.66 \e{-3} $ & $ 5.23 \e{-3}$ & $  5.85 \e{-3}$ & $ 4.11 \e{-3}$\\
$n_s	$ & $	0.96	$  & $ 6.71\e{-3}$ & $6.40\e{-3} $ & $6.53\e{-3}  $ & $5.84\e{-3}$ \\
$A_s	$ & $	2.2\e{-9}	$ & $ 3.87\e{-2} $ & $ 3.27\e{-2}$ & $3.51\e{-2} $ & $2.70\e{-2} $ \\
$\fnl$ & $5$ & $ 16.8$ & $8.56 $ & $ 7.12$ & $ 4.27$ \\
$w	$ & $	-1	$ & $	4.50\e{-2} $ &  $3.36\e{-2} $ & $ 5.43\e{-2} $ & $ 2.17\e{-2} $ \\
\hline
\end{tabular}
\end{center}
\caption[As Tab.~\ref{tab:ng_sk5} but including CMB priors]
{As Tab.~\ref{tab:ng_sk5} but including CMB priors.
From Ref.~\cite{Gariazzo:2015qea}.}
\label{tab:ng_sk5cmb}
\end{table*}

\begin{table*}[t]
\begin{center}
\begin{tabular}{c|c|c c c c c}
\hline
&fiducial & LRG  & ELG& QSO& all \\
\hline
$\Omega_b h^2$ & $ 0.02267$ & $3.92\e{-4}$ & $ 3.79\e{-4}$ & $3.86\e{-4} $& $3.75\e{-4} $\\ 
$\Omega_c h^2$ & $0.1131$ & $1.36\e{-3}$ & $1.10\e{-3} $ & $1.18\e{-3} $& $ 1.04\e{-3} $\\ 
$h	$ & $	0.705$ & $ 4.10\e{-3}$ & $3.13\e{-3} $ & $3.59\e{-3}  $& $2.92\e{-3}  $\\ 
$P_{s,5}	$ & $	1.07099$ & $2.98\e{-2}$ & $2.68\e{-2} $ & $2.77\e{-2}  $& $ 2.60\e{-2}$\\ 
$P_{s,6}	$ & $	1.04687$ & $2.89\e{-2}$ & $2.11\e{-2}  $ & $2.33\e{-2}  $& $ 2.0\e{-2}$\\ 
$P_{s,7}	$ & $1.02329$ & $2.00\e{-2}$ & $1.73\e{-2}  $ & $1.84\e{-2}  $& $1.69\e{-2} $\\ 
$P_{s,8}	$ & $	1.00024$ & $1.92\e{-2}$ & $1.76\e{-2}  $ & $ 1.86\e{-2} $& $ 1.73\e{-2}$\\ 
$P_{s,9}	$ & $	0.97771$ & $2.50\e{-2}$ & $2.31\e{-2}  $ & $ 2.43\e{-2} $& $ 2.22\e{-2}$\\ 
$\fnl$ & $5$ & $12.4$ & $ 6.42$ & $ 5.23$& $4.46 $\\ 
$w	$ &$ 	-1$ & $3.23\e{-2}$ & $ 2.46\e{-2}$ & $3.99\e{-2} $& $2.27\e{-2}  $\\ 
\hline
\end{tabular}
\end{center}
\caption[As Tab.~\ref{tab:ng_nsk5} but including CMB priors]
{As Tab.~\ref{tab:ng_nsk5} but including CMB priors.
From Ref.~\cite{Gariazzo:2015qea}.}
\label{tab:ng_nsk5cmb}
\end{table*}

Table~\ref{tab:ng_sk1} (\ref{tab:ng_nsk1}) shows
the $1\sigma$ marginalized errors for the case of a standard (\pchip) PPS,
for a fiducial value $\fnl=20$ for each of the DESI tracers
and from the combination of all of them,
obtained using the multi-tracer technique.
Even if such a value of the $\fnl$ parameter is larger
than the expected sensitivity from future probes,
it is still allowed by current LSS bounds
on primordial non-Gaussianities.  
Notice that, for the standard PL PPS,
the expected error on $\fnl$ is
$19.9$, $10.1$ and $8.56$ for
LRGs, ELGs and QSOs respectively,
while in the case of the \pchip\ parameterization, we obtain
$\sigma(\fnl)=32.2$, $13.3$ and $12.6$ respectively.
Therefore, the error on \fnl\ is much larger
when a \pchip\ PPS is assumed, up to the $60\%$ level.
The constraints on the remaining cosmological parameters
are barely affected by the different assumption on the PPS.
In some cases,
their error is even smaller than in the standard power-law scenario.
This is indeed the case of the equation of state parameter $w$,
or of $\Omega_b h^2$ and $\Omega_c h^2$.
The errors on the latter two parameters are smaller than in the PL PPS approach
only when exploiting either ELGs or QSOs tracers.
The combination of the data from the three tracers exploiting the multi-tracer
technique alleviates the problem with the error on $\fnl$.
In fact, the value of $\sigma(\fnl)$ increases only of about $40\%$
when relaxing the assumption of a PL PPS, rather than of $60\%$ as obtained
with the separate tracers.

\begin{figure*}
\begin{center}
\begin{tabular}{c c}
 \includegraphics[width=0.35\textwidth]{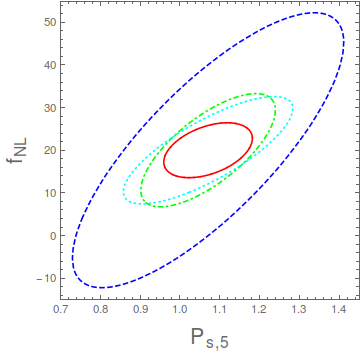}
&\includegraphics[width=0.35\textwidth]{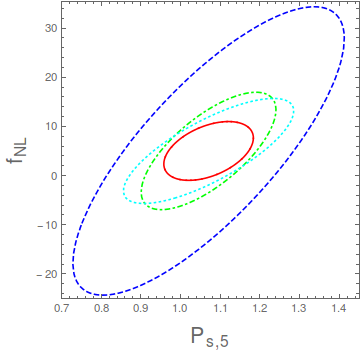}\\
 \includegraphics[width=0.35\textwidth]{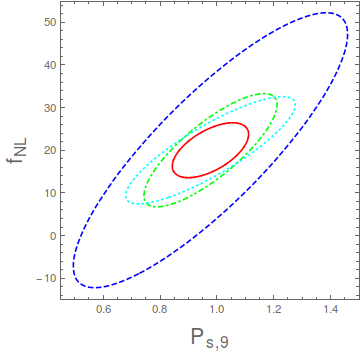}
&\includegraphics[width=0.35\textwidth]{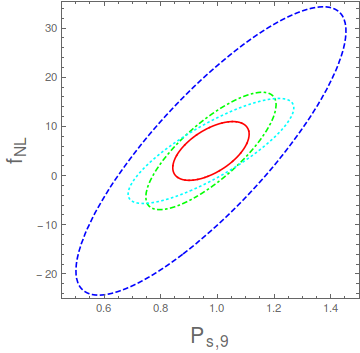}\\
\end{tabular}
\end{center}
\caption[Degeneracies between \fnl\ and two of the \pchip\ PPS
parameters, \psj{5} and \psj{9}]
{The upper left (right) panel shows the degeneracy between
\fnl\ and \psj{5},
for a fiducial cosmology with $\fnl=20$ ($\fnl=5$),
assuming $k_{\mathrm{max}}=0.1h$/Mpc.
We show the $1\sigma$ marginalized contours associated to the
LRGs (in dashed blue lines),
ELGs (in dot-dashed green lines),
QSOs (in dotted cyan lines) and
multi-tracer (in solid red) Fisher matrix analyses.
The bottom panels shows the analogous but 
for the degeneracy between \fnl\ and \psj{9}.
From Ref.~\cite{Gariazzo:2015qea}.}
\label{fig:ng_deg}
\end{figure*}

This generic increase in the error on $\fnl$ arises from
the large degeneracies between the non-Gaussianity parameter
and the \psj{j} nodes,
which is reduced when combining the tracers.
The top and bottom panels of Fig.~\ref{fig:ng_deg}
illustrate the large degeneracies between the $\fnl$ parameter
and two of the \pchip\ PPS nodes, $\psj{5}$ and $\psj{9}$,
for the fiducial value $\fnl=20$.  
We only show the degeneracies with two nodes,
but they are similar to the ones with the remaining nodes.

The problem of the degeneracy could be solved in two ways,
either exploiting smaller scales in the observed galaxy or quasar power spectra,
or using CMB priors.
In practice,
going to the mildly non-linear regime would require new additional \psj{j} nodes,
with the consequence that new degeneracies between these additional \psj{j} nodes
and the non-Gaussianity parameter $\fnl$ would appear.
Indeed,
we have numerically checked that such a possibility does not solve the problem.
Furthermore,
a non-linear description of the matter power spectrum
would depend on additional parameters,
enlarging the number of degeneracies.
In contrast,
the CMB priors on the PPS parameters, as well as
on the dark matter and baryon mass-energy densities,
help enormously in solving the problem of the large degeneracies between
the PPS parameterization and non-Gaussianities. 
Tables~\ref{tab:ng_sk1cmb} and \ref{tab:ng_nsk1cmb} show the equivalent
of Tables~\ref{tab:ng_sk1} and \ref{tab:ng_nsk1} with the inclusion of
CMB priors from the Planck mission 2013 data~\cite{Ade:2013zuv}. 
Notice that the impact of the Planck priors is largely more significant
in the \pchip\ parameterization case: 
the $\fnl$ errors arising from the three different dark matter tracers
when the CMB information is included are smaller in the \pchip\ PSS description 
than in the PL PSS approach. 
When the multi-tracer technique is applied,
the overall errors after considering Planck 2013 CMB constraints
are very similar, regardless on the PPS description and
close to $\sigma(\fnl)\simeq 5$. 

Table~\ref{tab:ng_sk5} (\ref{tab:ng_nsk5}) shows
the $1\sigma$ marginalized errors for the case of a PL (\pchip) PPS,
for another possible fiducial value of the non-Gaussianity parameter,
$\fnl=5$.
Again, the errors are obtained from each of the DESI tracers,
as well as from the multi-tracer technique that combines all of them.
As in the case of $\fnl=20$,
the error on the non-Gaussianity parameter increases
when the PPS parameterization is changed,
reaching in some cases a $60\%$ increment.
The results are very similar to those obtained and illustrated
before for the larger non-Gaussianities.
The errors on the other cosmological parameters remain unaffected
by the choice of the PPS parameterization.
The dark energy equation of state parameter is extracted
with a smaller error in the \pchip\ PPS case,
and also $\Omega_b h^2$ and $\Omega_c h^2$ are determined
with a smaller error in that case from the analyses of the ELGs and QSOs tracers. 
The multi-tracer technique provides a reduction on the $\fnl$ error
that is similar to the one obtained in the previous case with $\fnl=20$. 
The top and bottom right panels of Fig.~\ref{fig:ng_deg}
illustrate the large degeneracies between the non-Gaussianity parameter $\fnl$
and the nodes \psj{5} and \psj{9}, for the fiducial value $\fnl=5$.
We can notice that the degeneracy pattern appears to be independent
of the value of $\fnl$. 
The addition of the CMB priors reduces the errors
on all the cosmological parameters, including $\fnl$,
to the same values in both PPS parameterizations,
as shown in Tabs.~\ref{tab:ng_sk5cmb} and \ref{tab:ng_nsk5cmb}.

We now perform an additional forecast.
We focus on the shift induced in the local non-Gaussianity parameter $\fnl$,
which we set to zero in the two cosmologies $\mcm$ and $\mcm'$.
For the purpose of this analysis,
in the case of the $\mcm'$ cosmology
we fix all the nodes \psj{j} to their best-fit values
according to the Planck 2013 results for the PL PPS
(see Eq.~\eqref{eq:psj_plpps}).
A shift in $\fnl$ is expected to compensate for the fact that
the \pchip\ nodes \psj{j} are additional parameters in $\mcm$,
while they are not considered as free parameters in the $\mcm'$ analysis.
If we displace the \psj{j} parameters (with $j=5,\ldots,9$) from their fixed
fiducial values in $\mcm'$,
we are adding them as additional parameters in the cosmological model,
so that they must be determined by the data.
Referring to the notations of Eq.~\eqref{eq:offset} and
using a shift $\delta\psi_{\psj{j}}=0.1$,
which is smaller than the $1\sigma$ expected errors
(see Tabs.~\ref{tab:ng_nsk1} and \ref{tab:ng_nsk5}),
we obtain that the corresponding shift in the
$\fnl$ parameter is $\delta \theta_{\fnl}\simeq 2.5$,
regardless of the exploited dark matter tracer.
This is a quite large displacement of the local non-Gaussianity parameter
which will induce a non-negligible bias
in the reconstruction of the inflationary mechanism.
While the remaining cosmological parameters are also
slightly displaced with respect to their fiducial values,
their shifts will not induce a misinterpretation
of the underlying true cosmology.
The shift of the non-Gaussianity parameter $\delta \theta_{\fnl}$
could be a potential problem when extracting the (true) value
of the $\fnl$ parameter not only from the DESI survey,
but also for other future experiments
with improved sensitivities to non-Gaussianities,
such as SPHEREx~\cite{Dore:2014cca}.
The combination of all the three possible DESI tracers
leads to a smaller shift in the $\fnl$ parameter
($\delta \theta_{\fnl}\simeq 1.6$).
If CMB priors are applied the shift is considerably reduced to
$\delta \theta_{\fnl} \simeq 0.2$,
which is close to the expectations for non-Gaussianities
in the most economical inflationary models,
i.e.\ within single field slow-roll 
inflation~\cite{Bartolo:2004if,Maldacena:2002vr}.

\section{Conclusions} 
\label{sec:ng_concl}
While in the simplest inflationary models the primordial power spectrum (PPS)
of the curvature perturbations $\mcp_\mcr(k)$ 
can be described by a simple power-law without features,
there exists a large number of well-motivated inflationary scenarios
that could give rise to a non-standard PPS. 
The majority of these models will also generate non-Gaussianities.
The Large Scale Structures (LSS) of the Universe provide,
together with the CMB bispectrum,
a tool to test primordial non-Gaussianities. 

In the literature, it is possible to find several works devoted to forecast
the expectations from upcoming galaxy surveys,
such as the Dark Energy Spectroscopic Instrument (DESI) experiment. 
The forecasted errors and bounds on the non-Gaussianity local parameter
$\fnl$
are usually derived under the assumption of a power-law PPS. 
We relax this assumption and we compute the sensitivity to $\fnl$
expected from the DESI experiment.
To do this we assume that the precise shape of the PPS and
the non-Gaussianity parameter need to be extracted simultaneously.
If the analysis is restricted to LSS data,
the standard errors computed assuming a featureless PPS are enlarged
by $60\%$ when using the \pchip\ PPS parameterization
and treating each of the possible dark matter tracers individually.

Another potential problem in future galaxy surveys could be induced
by the possibly wrong assumption of a featureless PSS,
if nature could have chosen a more complicated inflationary mechanism
that results in a non-trivial PPS. 
If future data will be fitted using the wrong PPS cosmology,
a shift $|\delta \theta_{\fnl}|\simeq 2.5$ would be inferred
(for $k_{\mathrm{max}}=0.1 h$/Mpc) even if the true cosmology has $\fnl=0$.
The former two problems may be alleviated using the multi-tracer technique.
After combining all the DESI possible tracers,
when compared to the value obtained with the PL PPS parameterization,
the forecasted errors on $\fnl$ will be degraded by $40\%$
and the resulting shift will be reduced to $|\delta \theta_{\fnl}|\simeq 1.6$.
The addition of CMB priors from the Planck 2013 data
on the PPS parameters and on the energy densities of dark matter and baryons
leads to an error on $\fnl$ which is independent
of the PPS parameterization used in the analysis. 
After considering CMB priors,
the value of the shift $|\delta \theta_{\fnl}|$ is reduced to $0.2$,
which is of the order of standard predictions for
single-field slow-roll inflation~\cite{Bartolo:2004if,Maldacena:2002vr}.

%!TeX root=main.tex 
\chapter{Coupling between Dark Matter and Dark Energy}
\label{ch:cde}
\chapterprecis{This Chapter is based on
Ref.~\protect\cite{Murgia:2016ccp}.}

% \abstract{
% In this work we consider a phenomenological non-gravitational coupling between Dark Energy (DE) and Dark Matter (DM)
% and we study how it is constrained by Cosmic Microwave Background (CMB) data and other cosmological observations,
% such as SuperNovae (SN), Baryon Acoustic Oscillations (BAO), Redshift Space Distortions (RSD) and gravitational lensing.
% We parameterize the interaction in the dark sector with an energy transfer term $Q=\xi H\rhode$
% and we consider both positive and negative values for the dimensionless coupling strength $\xi$,
% thus studying the possibility of DM decay in DE, that we name MOD1, but also the opposite choice,
% corresponding to DE decaying into DM (MOD2).
% We find that MOD1 predicts low values for $H_0$ and high values for $\sigma_8$,
% increasing the tension between CMB and local determinations of these parameters:
% this arises from the higher amount of DM in the Early Universe, leading to a stronger clustering during the evolution.
% On the contrary, MOD2 allows a full reconciliation between high- and low-redshift observations.
% The inclusion of SN and RSD data allows to put strong constraints on the effective equation of state of 
% interacting dark energy, corresponding to a small preference for non-zero coupling.
% }
In the previous Chapters we presented the constraints coming
from cosmology on different dark radiation candidates, 
as neutrinos and thermal axions.
Among the different aims of the analyses,
there was the need
to find a possible explanation to the small tensions that appear
in the context of the \lcdm\ model between local measurements
and cosmological estimates of the Hubble parameter $H_0$ and
of the small scales matter fluctuations $\sigma_8$.

In this Chapter we show a new possibility that allows
to solve these tensions, not involving new particles, but 
considering a new
interaction between the dark components of the present Universe:
dark energy and dark matter.
A specific theoretical model for the interaction
would require a particle physics model
that explains the nature of dark matter and dark energy.
Several proposed scenarios exist, but we
do not have a well established model:
therefore we will consider only a phenomenological 
parameterization for the interaction.

\section{Introduction}
\label{sec:intro}
The results obtained analyzing the recent data
of the Planck collaboration \cite{Adam:2015rua} show us that
only up to the 5\% of the total energy density of the Universe today
is provided by baryon matter,
while the remaining 95\% comes from currently unknown constituents,
divided in two different classes,
being radiation negligible today.
The 26\% of the total energy density comes
from some matter component that feels gravity, 
but does not interact with photons and
is then named Dark Matter (DM).
The remaining 69\% comes from a diffuse fluid that is responsible
of the accelerated expansion we observe in the recent history
of the Universe.
The fluid that provides this kind of energy density
is named Dark Energy (DE),
behaving differently from any other massive component.
The leading candidate for DE is the cosmological constant $\Lambda$ 
that represents the vacuum energy in the equations
of General Relativity:
it is described by the equation of state (EoS)
$p_\Lambda=\wla\rhode$, 
where $\wla=-1$ and $p_\Lambda$, $\rhode$ are the pressure
and the energy density of DE, respectively.
Further details on DM and the cosmological constant are discussed
in Chapter~\ref{ch:cosmology}.

It is difficult to understand the value
of the cosmological constant in terms of fundamental physics, 
since it is well below the vacuum energy which can be obtained
in the context of quantum field theory,
in the Standard Model of Particle Physics. 
This problem is usually referred to as the
``cosmological constant problem''.
Beside this, there is another problem related to the cosmological
constant, that is called ``coincidence problem'':
it appears unnatural that matter and DE, today, contribute
to the total energy density with approximately the same amount.
Possible solutions to these problems are related to the nature of DE.
One possibility is that the DE energy density is not provided
by the cosmological constant, but by some dynamical mechanism:
for example, it is possible to obtain
the same EoS with $\wla\simeq-1$
by means of a dynamic scalar field, $\phi(t)$,
that is rolling down a potential $V(\phi)$.
This mechanism is similar to the one we presented
in Subsection~\ref{ssec:inCurvPert} for single-field inflation.
In fact, a $\Lambda$-dominated Universe expands exponentially,
in analogy with the behavior that appears during an inflationary phase.
With such a dynamical mechanism, the ``cosmological constant''
and ``coincidence'' problems are partially solved, 
since the smallness of the vacuum energy and the relative amount
of DM and DE energy densities come from a dynamical condition
and not from a fine tuning of the parameters.

Cosmology gives us an evidence that DE and DM exist through
the determination of their energy densities, 
but it does not give us the characteristics they have:
until particle physics experiments will not give
suitable candidates to account for DM and DE,
we will not have any information on their characteristics.
In particular, any type of non-gravitational interaction
involving DE or DM is only constrained 
by astrophysical observations,
with upper bounds on the interaction strength.
In this light, it is interesting to extend the \lcdm~model
to study the effects of a new non-gravitational interaction
in the dark sector, 
involving DE and DM (see e.g.\ 
Refs.~\cite{Friedman:1991dj,Gradwohl:1992ue,Wetterich:1994bg,
Amendola:1999er,Amendola:2003wa,Pettorino:2008ez,
Valiviita:2008iv,He:2008tn,Gavela:2009cy,He:2009mz,
Abdalla:2014cla,Salvatelli:2014zta} 
and the review \cite{Bolotin:2013jpa}).
The new interaction can be phenomenologically introduced
in cosmology in different ways,
see e.g.\ 
Refs.~\cite{He:2008tn,Gavela:2009cy,He:2009mz,Abdalla:2014cla,
He:2008si,Jackson:2009mz,Salvatelli:2013wra,
Costa:2013sva,Gavela:2010tm,He:2010im}
and Ref.~\cite{Koyama:2009gd} for a classification.
We will parameterize it through a new term
in the stress-energy tensor that enters the Einstein equations. 
In the coupled scenario, the DE and DM components of the
stress-energy tensor $T^{\mu\nu}$ are no longer separately conserved:
\begin{subequations}
\begin{eqnarray}
 \nabla_\mu T^{\mu\nu}_{dm}
 & = &
 Q u^\nu_{dm}/a\label{eq:stressenergyTDM}\,,\\
 \nabla_\mu T^{\mu\nu}_{\mathrm{DE}}
 & = &
 -Q u^\nu_{dm}/a\label{eq:stressenergyTDE}\,,
\end{eqnarray}
\end{subequations}
where the coefficient $Q$ encodes the interaction rate, 
$u^\nu_{dm}$ is the dark matter four-velocity and 
$a$ is the time-dependent scale factor of the Universe
\cite{He:2008tn,Gavela:2009cy,He:2009mz,Abdalla:2014cla,He:2008si,
Jackson:2009mz,Salvatelli:2013wra, Costa:2013sva,Gavela:2010tm,He:2010im}.
The introduction of the coupling term in
Eqs.~\eqref{eq:stressenergyTDM} and \eqref{eq:stressenergyTDE} leads
to the following conservation equations for the energy densities
of DM and DE:
\begin{subequations}\label{eq:dmdecou}
  \begin{eqnarray}
 \dot{\rho}_{dm}+3\hub\rhodm
 &=&
 +Q \;, \label{eq:dmdecou_dm}\\
 \dot{\rhode}+3\hub(1+\wla)\rhode
 &=&
 -Q \,,  \label{eq:dmdecou_de}
  \end{eqnarray}
\end{subequations}
where 
$\rho_{\mathrm{DM}(\Lambda)}$ is the energy density for DM (DE),
$\wla$ gives the EoS $p_\Lambda=\wla \rhode$ for DE,
$\hub = \dot{a}/a$ is the Hubble parameter.
With the introduction of the coupling term,
the energy densities of the dark components
are not individually conserved, 
because there exists an energy flux between them:
if $Q$ is positive the energy flux is
from DE to DM and DE decays into DM, while 
if $Q$ is negative the energy flux has the opposite direction and
DM decays into DE.

Several interaction models has been proposed in the literature, 
for example 
Refs.~\cite{Amendola:1999er,Amendola:2006dg,Valiviita:2008iv,
He:2008si,CalderaCabral:2009ja,Jackson:2009mz,Pavan:2011xn},
where the role of Coupled Dark Energy (CDE in the following)
is played by a scalar field.
In our work, instead of focusing on the theoretical framework
that gives origin to a CDE scenario,
we use a phenomenological approach and we study a CDE model with
\cite{He:2008tn,Gavela:2009cy,He:2009mz,Abdalla:2014cla,He:2008si,
Jackson:2009mz,Salvatelli:2013wra, Costa:2013sva,Gavela:2010tm,He:2010im}
\be\label{eq:Qcoupling}
Q=\xi \hub \rhode\,, 
\ee
where $\xi$ is the dimensionless coupling parameter:
in this way the coupling is spatially-independent
and
the time dependency of the interaction rate
is governed by the Hubble parameter 
$\hub=\dot a/a$
\cite{Zimdahl:2001ar,Salvatelli:2013wra, Costa:2013sva}.
Standard cosmology corresponds to $\xi=0$.

In the following we will test this CDE model against
cosmological observables and derive bounds
on the relevant model parameters,
that in our approach are $\wla$ and $\xi$.
We will also discuss whether the ensuing results help
in alleviating the tension on the determination
of $H_0$ and $\sigma_8$ which arises
from high and low redshift cosmological observables.

The outline of this Chapter is the following:
in Section~\ref{sec:cde_param} we describe our parameterization
for the \lcdm~model,
its extension that include a coupling between DE and DM,
and the cosmological data we used.
In Section~\ref{sec:cde_results} we present and discuss the results.
In Section~\ref{sec:cde_sterilenuDM} we study the possibility
that DM is composed of one interacting fraction
and one stable fraction, represented by a sterile neutrino.
Finally, we summarize our conclusions in Section~\ref{sec:cde_disc}.

\section{Method}
\label{sec:cde_param}
% In this section we present the parameterization
% we used to perform our cosmological analysis.
% In Subsection~\ref{ssec:lcdm} we show the baseline decoupled model, that is the standard \lcdm~model,
% while in
% Subsection~\ref{ssec:cde} we present the modifications required to study the coupled model 
% and the equations for the perturbations and the background evolutions.

\subsection{Parameterization}
\label{ssec:cde_param}
Our baseline model is the well studied and confirmed \lcdm~model, 
already adopted in the previous Chapters
and described in Section~\ref{sec:par_depend}.
% where $\Lambda$ indicates the cosmological constant and CDM is for Cold Dark Matter.
In this Chapter we use the following set of parameters:
\be
{\bm \theta}
=
\{
\Omega_\mathrm{c}h^2,
\Omega_\mathrm{b}h^2,
\theta,
\tau,
\ln(10^{10}A_{s}),
n_{s},
\wla,
\xi
\},
\ee
where we have
the present baryon density $\Omega_bh^2$, 
the present CDM density $\Omega_ch^2$,
the ratio of the sound horizon
to the angular diameter distance at decoupling $\theta$,
the optical depth at reionization $\tau$, 
the amplitude $A_s$ and
the spectral index $n_s$ 
of the primordial power spectrum of scalar perturbations.
The parameters $\wla$ and $\xi$ are used for the CDE models,
while they are fixed to $\xi=0$ and $\wla=-1$ in the \lcdm\ model.

In the first part of our analysis we do not consider
the effects of varying the parameters
that describe the neutrino sector:
the sum of the neutrino masses $\sum m_\nu$, 
that we fix to the minimal value allowed
by the neutrino oscillations, $\sum m_\nu=0.06$~eV 
for two almost massless and one massive neutrino, 
and the effective number of relativistic species $\neff$, 
that we fix to the standard value $\neff^{\mathrm{sm}}=3.046$
\cite{Mangano:2005cc} obtained for the three active neutrinos.
In Section~\ref{sec:cde_sterilenuDM}, instead,
we will study the constraints on an additional light sterile neutrino
using the same parameterization adopted in Section~\ref{sec:jhep}.

% For the \lcdm~parameters we adopt flat priors in the ranges listed in Tab.~\ref{tab:priorslcdm}.

% \begin{table}[t]
% \begin{center}
% \renewcommand{\arraystretch}{1.2}
% \begin{tabular}{|c|c|c|}
%   \hline
%   Parameter		& Prior		\\	\hline	
%   $\Omega_bh^2$ 	& [0.005, 0.1]	\\
%   $\Omega_ch^2$ 	& [0.001, 0.5]	\\
%   $100\theta$ 		& [0.5, 10]	\\ 	
%   $\tau$  		& [0.01, 0.8]	\\ 
%   $\logA$ 	& [2.7, 4]	\\ 	
%   $n_s$ 		& [0.9, 1.1]	\\ 	\hline
%   $\sum m_{\nu}$	& 0.06 eV	\\
%   $N_{\nu}$		& 3.046 \cite{Mangano:2005cc}	\\ 	\hline
%   $H_0$~[\Hou]
% 			& [20,100]	\\	\hline
% \end{tabular}
% \caption[Priors on the cosmological parameters in the \lcdm\
% and Coupled Dark Energy models]
% {The priors on the cosmological parameters that we use for the analysis of the \lcdm~and the CDE models.
% The first six priors are for the standard \lcdm~free parameters:
% all the priors are flat in the listed intervals.
% The total neutrino mass and effective number are kept fixed.
% We limit the derived parameter $H_0$ to exclude models that predict extreme and very unlikely values for the Hubble rate today.
% From Ref.~\cite{Murgia:2016ccp}.}
% \label{tab:priorslcdm}
% \end{center}
% \end{table}

% \subsection{Coupling between DE and DM}
% \label{ssec:cde}
We introduce a phenomenological coupling between
the dark components in the Universe,
parameterized through a coupling term $Q$,
written in Eq.~\eqref{eq:Qcoupling}.
After introducing the coupling,
Eqs.~\eqref{eq:dmdecou_dm} and \eqref{eq:dmdecou_de} can be derived 
from the time component
of the stress-energy momentum conservation equation.
The decoupled Equations (with $Q=0$) correspond to $\xi=0$.
Using the coupling term in Eq.~\eqref{eq:Qcoupling}, 
it is possible to solve Eqs.~\eqref{eq:dmdecou_dm}
and \eqref{eq:dmdecou_de} and to write explicitly 
the background equations for the energy densities of DM and DE 
\cite{Izquierdo:2010qy,Gavela:2009cy,Costa:2013sva}:
\begin{subequations}
\label{eq:coupl_bg}
  \begin{eqnarray}
  \rhodm
  &=&
  \rhodm^0\,a^{-3} + {\rhode^0 a^{-3} \Bigg[\frac{\xi}{3\wla+\xi}
  \big(1-a^{-3\wla-\xi}\big)\Bigg]}\label{eq:coupl_bgDM}\,,\\
  \rhode
  &=&
  \rhode^0\, a^{-3(\wla+1)-\xi}\,,\label{eq:coupl_bgDE}
  \end{eqnarray}
\end{subequations}
where $\rho_{i}^0$ is the energy density of the species $i$ today.
We emphasize that $\xi<0$ correspond
to an energy flux from DM to DE, with DM decaying into DE, 
whereas $\xi>0$ correspond
to an energy flux from DE to DM, with DE decaying into DM.
In the following we will refer to the former case
as Model 1 (MOD1) and to the latter case
as Model 2 (MOD2) 
for sake of brevity.
From Eq.~\eqref{eq:coupl_bgDE} we can see that DE obeys
an effective EoS given by $\wlae=\wla+\xi/3$:
this allows to write Eq.~\eqref{eq:coupl_bgDE} in the usual form
$\rhode = \rhode^0\, a^{-3(\wlae+1)}$.

In the presence of the coupling term in Eq.~\eqref{eq:Qcoupling}, 
the interaction model does not suffer gravitational instabilities 
if $\wla \neq -1$~\cite{Valiviita:2008iv,He:2008si}:
for this reason we will consider a constant
$\wla \neq -1$ when $\xi\neq0$.
Early time instabilities can however rise up also
when $\wla\neq-1$ if the coupling is strong \cite{Gavela:2009cy}:
in particular the instability is not present
if $\xi$ and $\wla+1$ have opposite sign,
but they can be present if the two quantities have the same sign.
We will consider only constant values $\wla>-1$ for MOD1,
for which $\xi<0$, and
constant values $\wla<-1$ for MOD2,
for which $\xi>0$,
in order to avoid the instabilities.
It is worthwhile to note that in the latter case 
the DM energy density can assume negative values in the past 
for particular combinations of $\wla$ and $\xi$
(Eq.~\eqref{eq:coupl_bgDM}),
while the DE energy density is always positive
(Eq.~\eqref{eq:coupl_bgDE}).
To avoid unphysical values of $\rhodm$,
we must therefore impose $\xi \lesssim -\wla$:
this is automatic for $\xi<0$ (MOD1)
unless $\wla$ assumes positive values, 
but this do not occur since the accelerated expansion of the Universe
at late times requires $\wla<-1/3$.
For MOD2, instead, we impose the prior $0\leq\xi\leq0.5$,
but we will find that the largest values of $\xi$ in this interval
are disfavored by our analyses.

From Eq.~\eqref{eq:coupl_bgDE} we note that $\rhode$
increases with the scale factor if~$\wla < -1 -{\xi}/{3}$:
in this region DE has an effective phantom behavior,
that is the unbounded increase of $\rhode$ in future times. 
The effective phantom behavior occurs in both
the models MOD1 and MOD2.
Even when $\wla > -1$ and $\xi<0$ (MOD1)
the phantom regime can be present since when $a$ increases
$\rhode$ can be increased by the energy transfer from DM to DE,,
instead of following the decreasing behavior driven by $\wla > -1$.
This effective behavior, however,
has the advantage of being free from the instabilities 
that can occur for a true phantom dark energy
\cite{Huey:2004qv,Das:2005yj}.

Looking at Eqs.~\eqref{eq:coupl_bgDM} and \eqref{eq:coupl_bgDE},
we notice that 
it is difficult to disentangle the effects of the DE EoS parameter $w$
and the coupling $\xi$ by only studying the background evolution.
We must include the perturbation evolution equations,
which are also affected by the additional coupling.
To obtain the new equations for the linear perturbation in DM and DE 
one has to perform the calculations in the perturbed space time,
following the method we presented in Section~\ref{sec:pertUniv}.
As a result, the coupled perturbation equations
in the synchronous gauge can be obtained \cite{Costa:2013sva}:
\begin{subequations}
\label{eq:pertcou}
\begin{eqnarray}
\dot{\delta}_{dm}
& = &
-\left(kv_{dm}+\frac{\dot{h}}{2}\right)
+ {\xi \hub \frac{\rhode}{\rhodm}
(\delta_\Lambda-\delta_{dm})}
\,;\label{eq:pertcouDMd}\\ 
\dot{v}_{dm}
& = &
-\hub v_{dm}\left(1+{\xi\frac{\rhode}{\rhodm}}\right)
\,;\label{eq:pertcouDMv}\\ 
\dot{\delta}_{\Lambda}
& = &
-(1+\wla)\left(kv_\Lambda+\frac{\dot{h}}{2}\right)
-3\hub(1-\wla)\cdot
\left(\delta_\Lambda \hub(3(1+\wla)+{\xi})\frac{v_\Lambda}{k}\right)
\,;\label{eq:pertcouDEd}\\ 
\dot{v}_{\Lambda}
& = &
-2\hub
\left(1+{\frac{\xi}{1+\wla}}\right)
v_\Lambda+k\frac{\delta_\Lambda}{1+\wla}
\,;\label{eq:pertcouDEv}
\end{eqnarray}
\end{subequations}
where $h=6\phi$ is the synchronous gauge metric perturbation 
and the DM peculiar velocity ${v}_{dm}$
is fixed to zero using the gauge freedom.
Moreover, the DE sound speed is fixed: $c_{s,\Lambda}=1$.
The uncoupled equations for $\delta_{dm}$ and $v_{dm}$
have been presented in Eqs.~\eqref{eq:pertDMFourDelta} and
\eqref{eq:pertDMFourV} in the conformal Newtonian gauge.
They can be recovered using $\xi=0$ and
changing appropriately the gauge.
We adopt the adiabatic initial conditions 
(see Section~\ref{sec:initial_conditions})
for the CDE component 
\cite{Valiviita:2008iv,He:2008si,Gavela:2010tm}
as for all the other cosmological constituents \cite{Ma:1995ey}.

\begin{figure}[t]
  \centering
  \includegraphics[page=1,width=\singlefigland]{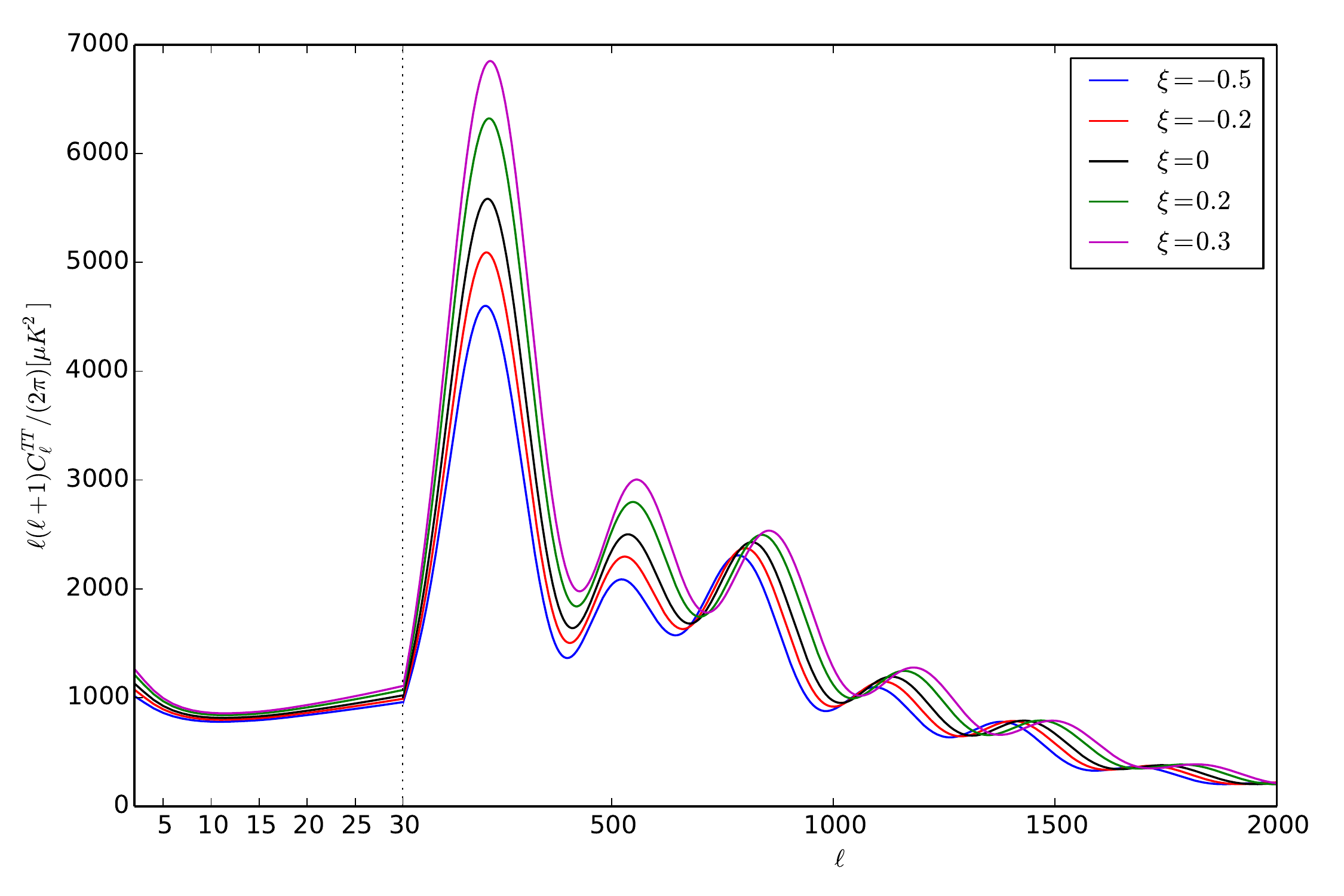}
  \includegraphics[page=2,width=\singlefigland]{cde/cls.pdf}
  \caption[Dependence of the CMB spectrum
  on the coupling strength and the Dark Matter energy density today]
  {
  Dependence of the CMB spectrum on two
  cosmological parameters:
  the coupling strength $\xi$ (upper panel) and
  the DM energy density today $\Omega_c h^2$
  (lower panel).
  All the other parameters are kept fixed.
  The black curve is the same in the different panels.
  From Ref.~\cite{Murgia:2016ccp}.
  }
  \label{fig:cls}
\end{figure}

The effects of the additional coupling are visible in different ways
on the cosmological observables.
Since we expect a strong degeneracy between the coupling parameter
$\xi$ and the DM density today $\Omega_c h^2$, 
due to the conversion of DM into DE (or vice versa)
that reduce (increase) the DM abundance at different times,
we briefly list the effects that the dark coupling has on cosmology
when we consider $\Omega_c h^2$ fixed.
The DM density today will be degenerate with the variations
in the coupling strength $\xi$, 
since $\xi$ impacts the CMB spectra through
the corresponding DM energy density 
at the matter-radiation equality epoch or
at the CMB decoupling, that is higher (smaller) 
if the coupling parameter is negative (positive).
When $\Omega_c h^2$ is fixed, 
the presence of the coupling provides a shift in the position
and a change in the envelope of the CMB peaks, 
due mainly to the different background evolution
and to the different DM density in the early Universe,
and a change in the low-$\ell$ spectrum,
due to a different contribution
to the integrated Sachs-Wolfe (ISW) effect
\cite{He:2010im, Salvatelli:2013wra}. 
The upper panel of Fig.~\ref{fig:cls} shows
the dependence of the CMB spectrum on $\xi$.
The DE EoS parameter $\wla$, in turn,
has an impact mainly on the low-$\ell$ part of the spectrum 
and on the position of the acoustic peaks,
leaving their envelope almost unchanged: %, 
% as it is possible to see in the lower panel of Fig.~\ref{fig:cls}:
this gives the opportunity of breaking the degeneracy 
arising from the background evolution Equations
\eqref{eq:coupl_bgDM} and \eqref{eq:coupl_bgDE}
when studying the CMB spectrum in a wide range of multipoles.

The DM abundance, instead, is relevant for the matter-radiation
equality and for the expansion rate at the time of CMB decoupling, 
that influences the comoving sound horizon and consequently
the angular scale of the peaks: 
it is difficult to distinguish
the impact of the DM energy density and the coupling strength
from CMB data alone, 
as it is possible to see comparing the panels of Fig.~\ref{fig:cls}.
The degeneracy with the DM density
can be studied with additional data 
on the gravitational lensing and
on the clustering, since the coupling introduces
a non-standard time-dependency of the DM density. 
The fact of having different amounts of DM at different epochs leads
to different evolution histories of the small scale fluctuations
under the effect of gravity.
If DM decays into DE, for example, there is much more DM
in the early Universe, leading to a stronger clustering
and to an anticipated nonlinear regime for the evolution
of the perturbations.

For our cosmological analyses
we implemented all the relevant equations 
into the numerical Boltzmann solver \camb\cite{Lewis:1999bs}
and we modified the Markov Chain Monte Carlo (MCMC)
code \cosmomc\cite{Lewis:2002ah} 
in order to include $\xi$ as an additional parameter.
We then use \cosmomc to obtain the cosmological constraints and 
we compare the results obtained in the standard \lcdm~model
with those obtained considering the CDE scenarios, MOD1 and MOD2.
We restrict $\xi$ and $\wla$ to the intervals
in Tab.~\ref{tab:priorscde} for the reasons explained above
and we consider flat priors in these ranges for our MCMC analyses.

Finally, we want to underline the connections of
the parameters used in the Equations
presented above with the parameters that appear in the \lcdm\ 
model and in the \camb/ \cosmomc software
that we use for the analyses.
The DM energy density $\rhodm$ is proportional to the parameter
$\Omega_ch^2$, since
$\Omega_c=\rhodm/\rho_c\propto\rhodm/h^2$
(see Section~\ref{sec:feq}):
the physical energy density of DM today is then proportional to
$\Omega_ch^2$.
On the contrary, $\Omega_c$ depends on Hubble parameter today.
This observation will be useful when we will discuss the results
for the CDE models.
On the contrary, we will present the results for
$\Omega_\Lambda
=
1-\Omega_k-(\Omega_\gamma+\Omega_b+\Omega_c+\Omega_\nu)$,
where we always consider $\Omega_k=0$ (flat Universe).
$\Omega_\Lambda$ is a derived parameter in our analyses,
and it is not proportional to the physical energy density $\rhode$,
but it depends on the Hubble parameter today
($\Omega_\Lambda\propto\rhode/h^2$).

\begin{table}[t]
\begin{center}
\renewcommand{\arraystretch}{1.2}
\begin{tabular}{|c|c|c|c|}
  \hline
  	& \multicolumn{3}{c|}{Prior}		\\ \hline
  Parameter	& \lcdm	& MOD1		& MOD2 	\\ \hline
  $\wla$  	&-1	& [-0.999, -0.1]& [-2.5, -1.001]\\
  $\xi$ 	& 0	& [-1, 0]	& [0, 0.5]	\\ \hline
  &no interaction& DM decays into DE & DE decays into DM \\ \hline
\end{tabular}
\caption[Priors on the Dark Energy parameters in the \lcdm\
and Coupled Dark Energy models]
{The priors on parameters for the coupling scenario,
the coupling parameter $\xi$ and the DE EoS parameter $\wla$,
that we use for the analyses of the different models.
All the priors are flat in the listed intervals.
From Ref.~\cite{Murgia:2016ccp}.}
\label{tab:priorscde}
\end{center}
\end{table}

\subsection{Cosmological Data}
\label{ssec:cde_data}
We base our analyses on the Cosmic Microwave Background (CMB) data
(see Section~\ref{sec:cmb})
from the 2015 Planck release \cite{Adam:2015rua}, 
in particular we consider as our minimal data combination
the full temperature autocorrelation spectrum
in the range $2\leq\ell\leq2500$ (denoted as \textbf{PlanckTT}) plus the
low-$\ell$ Planck polarization spectra in the range $2 \leq\ell\leq 29$
(denoted as \textbf{lowP}) \cite{Aghanim:2015xee}.
Additionally, we consider and add separately 
the high-$\ell$ Planck polarization spectra in the range
$30\leq\ell< 2500$ (hereafter \textbf{highP}) \cite{Aghanim:2015xee}.

Since the coupling between DE and DM introduces
a time-dependency in the background evolution of DE and DM 
(see Eqs.~\eqref{eq:coupl_bgDM} and \eqref{eq:coupl_bgDE}), 
it is important to test our theoretical models using data
at many different redshift with respect to the CMB measurements.
In particular, in MOD1 we expect a higher amount of DM
in the early Universe than in the \lcdm{} model, 
with stronger gravitational effects in the initial phases of the evolution.
On the opposite side, in the MOD2 the amount of DM is smaller
in the early Universe 
and the gravitational clustering is reduced until enough DE is decayed into DM.
For these reasons, it is important to consider observations
at various redshift to constrain the CDE models,
as they can distinguish the different evolution histories.

One of the most important probes of the expansion
and of the existence of DE
are the Supernovae (SNe) of type Ia.
We consider the luminosity distances of SN Ia 
from the SNLS and SDSS catalogs as re-analyzed in the joint analysis
\cite{Betoule:2014frx} (\textbf{JLA} hereafter),
introduced in Section~\ref{sec:sn}.

Another interesting probe of the Universe evolution comes
from the Redshift Space Distortions
(RSD, see Subsection~\ref{ssec:rsd}), 
namely distortions of the shape of galaxy clusters in the redshift space
due to peculiar motions of the single objects
along the line of sight.
We include also the Baryon Acoustic Oscillations (BAO)
data as determined by
6dFGS \cite{Beutler:2011hx}, 
SDSS-MGS \cite{Ross:2014qpa} and
BOSS DR11 \cite{Anderson:2013zyy},
together with the RSD determinations from BOSS DR11 
\cite{Samushia:2013yga}.
We will refer to the combination of these measurements as to the
\textbf{BAO/RSD} dataset.

The amount of DM affects also the strength of the gravitational lensing.
We include information on the power spectrum
of the lensing potential reconstructed by Planck 
from the trispectrum detection \cite{Ade:2015zua} (hereafter \textbf{lens}).
We do not consider weak lensing determinations obtained
from the cosmic shear measurements of the CFHTLenS survey
\cite{Heymans:2012gg} for the reasons explained in Section~\ref{sec:shear}.
We also do not consider the other local determinations of $\sigma_8$ 
from local measurements (see Section~\ref{sec:cluster})
for the same reasons, nor
any constraints on the Hubble parameter $H_0$,
the expansion rate of the Universe today,
due to the tensions that exist between local determinations and CMB estimates
also for this observable (see Section~\ref{sec:h0}).
It is important, however, to discuss and possibly solve the small tensions 
that currently are present between the CMB observations
and the local measurements, 
and new physics beyond the standard cosmological model
can help in this direction.
As we will show in the next Section,
the CDE model can reconcile local and cosmological measurements 
for both $H_0$ and $\sigma_8$.

In our analyses we will explore different combinations of the listed dataset: 
our starting point will be the CMB-only dataset \textbf{PlanckTT+lowP}, 
then we will add one of the other datasets at a time (highP, lens,
JLA, BAO/RSD) and 
finally we will consider a combination involving all the dataset,
``PlanckTT+lowP + highP + lens + JLA + BAO/RSD'', 
that we will indicate with \textbf{ALL} for sake of brevity.
For each of these data combinations we will test
the three cosmological models (\lcdm, MOD1, MOD2) to study 
how the constraints change.

\section{Results}
\label{sec:cde_results}

\begin{table}[t]                                                                           
\begin{center}        
\renewcommand{\arraystretch}{1.2}                                                                  
\begin{tabular}{|c|c|c|c|}                                                              
\hline
Parameter &	\lcdm	& MOD1	& MOD2\\
\hline
$100\Omega_bh^2$
			& $2.222\,^{+0.047}_{-0.043}$    & $2.216\,^{+0.046}_{-0.045}$    & $2.226\,^{+0.047}_{-0.046}$   
\\
$\Omega_ch^2$
			& $0.120\,^{+0.004}_{-0.004}$    & $0.069\,^{+0.053}_{-0.065}$    & $0.133\,^{+0.019}_{-0.016}$   
\\
$100\theta$
			& $1.0409\,^{+0.0009}_{-0.0009}$ & $1.0441\,^{+0.0052}_{-0.0040}$ & $1.0402\,^{+0.0013}_{-0.0013}$
\\
$\tau$
			& $0.078\,^{+0.039}_{-0.037}$    & $0.077\,^{+0.039}_{-0.038}$    & $0.077\,^{+0.039}_{-0.038}$   
\\
$n_s$
			& $0.965\,^{+0.012}_{-0.012}$    & $0.964\,^{+0.013}_{-0.012}$    & $0.966\,^{+0.013}_{-0.012}$   
\\
$\logA$
			& $3.089\,^{+0.074}_{-0.072}$    & $3.088\,^{+0.073}_{-0.073}$    & $3.087\,^{+0.073}_{-0.074}$   
\\ 	\hline
$\xi$
			& $0$                            & $(-0.789,0]$                      & $[0,0.269)$   
\\
$\wla$
			& $-1$                           & $[-1,-0.703)$                      & $-1.543\,^{+0.524}_{-0.447}$  
\\ 	\hline
$H_0$~[\Hou\ ]
			& $67.28\,^{+1.92}_{-1.89}$      & $67.91\,^{+7.44}_{-7.87}$      & $>68.32$   
\\
$\sigma_8$
			& $0.830\,^{+0.029}_{-0.028}$    & $1.464\,^{+1.948}_{-1.037}$    & $0.898\,^{+0.163}_{-0.160}$   
\\ 	\hline
\end{tabular}
\caption[Marginalized constraints on the cosmological parameters
from the CMB data set]
{Marginalized limits at 2$\sigma$ for various parameters
considered in our analyses, 
obtained with the ``PlanckTT+lowP'' dataset
for the three different models (\lcdm, MOD1 and MOD2).
When an interval denoted with parenthesis is given,
it refers to the  2$\sigma$ C.L.\ range starting from the
prior extreme, listed in Tab.~\ref{tab:priorscde}.
$H_0$ is limited to the range $[20,100]$.
From Ref.~\cite{Murgia:2016ccp}.
}
\label{tab:cmb}
\end{center}
\end{table}

\begin{table}[t]                                                                           
\begin{center}        
\renewcommand{\arraystretch}{1.2}                                                                  
\begin{tabular}{|c|c|c|c|}                                                              
\hline
Parameter &	\lcdm	& MOD1	& MOD2\\
\hline
$100\Omega_bh^2$
			& $2.229\,^{+0.028}_{-0.028}$    & $2.228\,^{+0.030}_{-0.030}$    & $2.227\,^{+0.031}_{-0.030}$    
\\
$\Omega_ch^2$
			& $0.119\,^{+0.002}_{-0.002}$    & $0.091\,^{+0.029}_{-0.033}$    & $0.135\,^{+0.014}_{-0.014}$    
\\
$100\theta$
			& $1.0409\,^{+0.0006}_{-0.0006}$ & $1.0426\,^{+0.0022}_{-0.0019}$ & $1.0400\,^{+0.0010}_{-0.0010}$ 
\\
$\tau$
			& $0.062\,^{+0.025}_{-0.025}$    & $0.063\,^{+0.027}_{-0.026}$    & $0.059\,^{+0.028}_{-0.027}$    
\\
$n_s$
			& $0.966\,^{+0.008}_{-0.008}$    & $0.966\,^{+0.009}_{-0.009}$    & $0.966\,^{+0.009}_{-0.009}$    
\\
$\logA$
			& $3.055\,^{+0.045}_{-0.046}$    & $3.058\,^{+0.049}_{-0.049}$    & $3.050\,^{+0.050}_{-0.051}$    
\\ 	\hline
$\xi$
			& $0$                            & $(-0.463,0]$                      & $[0,0.300)$    
\\                                                   
$\wla$                                          
			& $-1$                           & $[-1,-0.829)$                      & $(-1.129,-1]$   
\\ 	\hline
$H_0$~[\Hou\ ]
			& $67.72\,^{+1.01}_{-0.97}$      & $67.57\,^{+1.81}_{-1.79}$      & $67.83\,^{+1.90}_{-1.75}$      
\\
$\sigma_8$
			& $0.812\,^{+0.017}_{-0.017}$    & $0.994\,^{+0.294}_{-0.219}$    & $0.749\,^{+0.069}_{-0.063}$    
\\ 	\hline
\end{tabular}
\caption[Marginalized constraints on the cosmological parameters
from the complete data set]
{The same as in Tab.~\ref{tab:cmb}, for the results
obtained with the ``ALL'' dataset.
From Ref.~\cite{Murgia:2016ccp}.
}
\label{tab:all}
\end{center}
\end{table}

In this section we present the result obtained
in the cosmological analyses.
We compare
the three different models (\lcdm, MOD1, MOD2)
and the constraints provided by the different datasets.
We list in Tables \ref{tab:cmb} (``CMB only'')
and \ref{tab:all} (``ALL'' dataset) 
the 2$\sigma$ constraints for the parameters we considered.
Most of the standard \lcdm~parameters are not sensitive
to the coupling in the dark sector and
the ensuing results are unchanged when moving from
the \lcdm~model to the MOD1 and MOD2 scenarios: 
the baryon density today $\Omega_b h^2$,
the optical depth at reionization $\tau$,
the tilt $n_s$ and the amplitude $\logA$
of the power spectrum of scalar perturbations.
Their determination is therefore robust
against modified expansion histories induced
by the new DM/DE coupling.

Slightly larger variations occur for the ratio
of the sound horizon to the angular diameter distance at decoupling, $\theta$,
but even in this case the differences between the various models
are well inside the mutual 2$\sigma$ limits.
Interestingly, the addition of the external data in the ``ALL'' dataset
reduces the uncertainties on various parameters,
but requires a shift towards lower values 
for the optical depth at reionization $\tau$ and for the amplitude
of the scalar perturbations power spectrum $\logA$.
These parameters suffer of a mild tension
in the recent Planck results, as discussed in \cite{Ade:2015xua},
since the analyses that consider the low-$\ell$ temperature spectrum
point towards higher values of $\tau$ with respect
to the results obtained from the polarization spectra only. 
If one considers the lensing information and the BAO measurements
together with the temperature spectrum,
the results are in good agreement with the indications
in favor of a small $\tau$ coming from the Planck polarization spectra.
As the CMB observations constrain the combination $A_s e^{-2\tau}$,
a smaller $\tau$ reflects in a smaller $A_s$.

\begin{figure}
  \centering
  \includegraphics[page=1,width=\singlefigsmall]{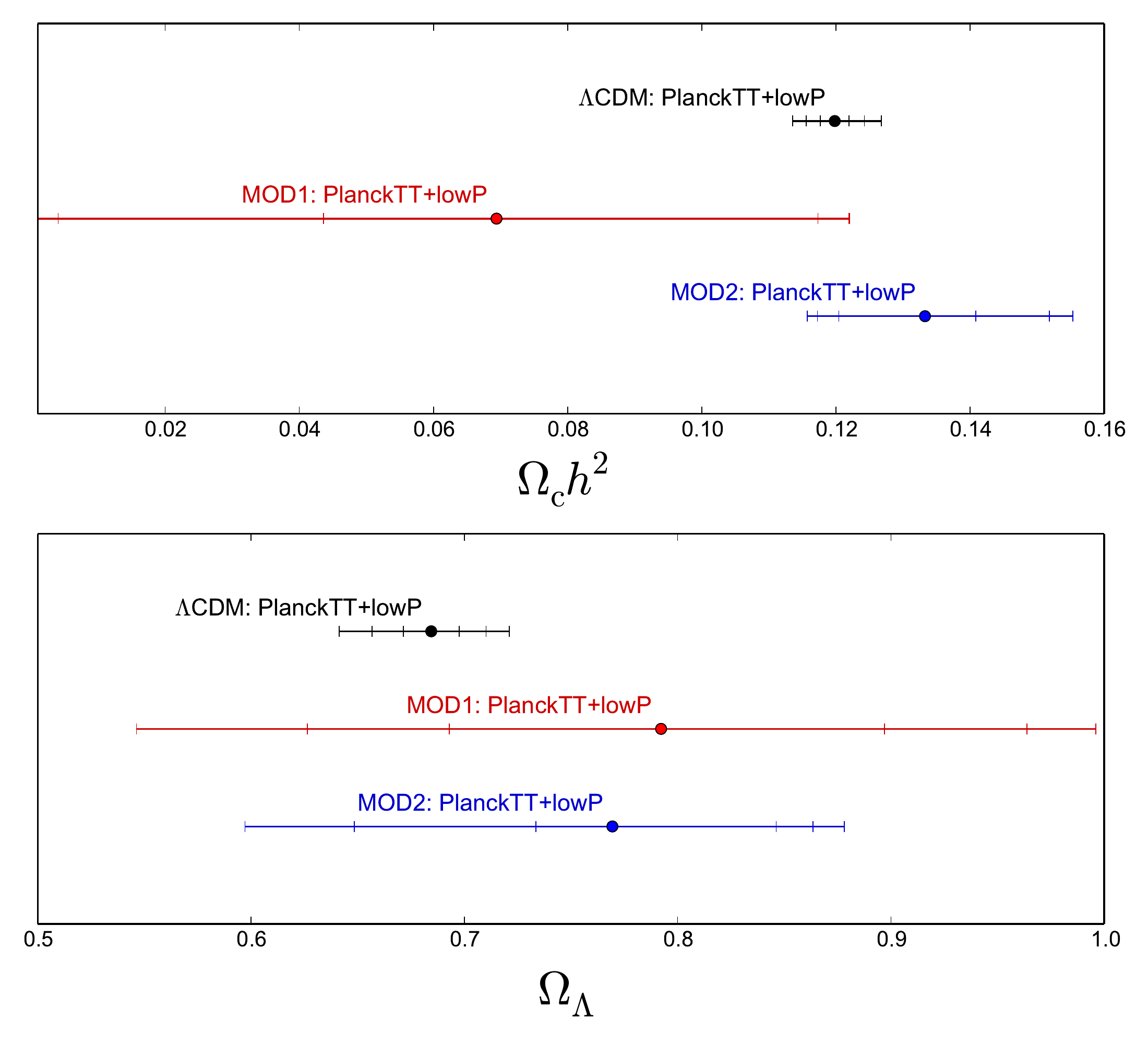}
  \caption[Marginalized limits from the ``PlanckTT+lowP'' dataset
  for $\Omega_ch^2$ and $\Omega_\Lambda$,
  considering the \lcdm, MOD1 and MOD2 scenarios]{
  Marginalized 1, 2 and 3$\sigma$ limits from the ``PlanckTT+lowP'' dataset
  for $\Omega_ch^2$ and $\Omega_\Lambda$,
  for the three different cosmological models: \lcdm, MOD1 and MOD2.
  MOD1 predicts a smaller amount of DM today with respect to \lcdm,
  as one would expect in a model in which the energy flux is from DM to DE; 
  on the other hand, MOD2 predicts more DM today compared to \lcdm,
  since in that model the energy flux is opposite,
  i.e.~DE decays into DM.
  From Ref.~\cite{Murgia:2016ccp}.}
  \label{fig:omega1D}
\end{figure}

As we would expect, 
there is a strong correlation between the coupling parameter $\xi$
and the current DM energy density $\Omega_ch^2$.
For $\xi<0$ (MOD1), the bigger is the interaction,
the smaller is the DM abundance today,
i.e.~more DM decayed into DE during the evolution.
Conversely, in $\xi>0$ (MOD2) a larger current DM abundance is predicted.
Since CMB data mainly constrain the DM abundance in the early Universe, 
the best fit values for $\Omega_ch^2$ can be very different
in the \lcdm, MOD1 or MOD2 cases,
as it is possible to see from Tabs.~\ref{tab:cmb} and \ref{tab:all}
and the upper panel in Fig.~\ref{fig:omega1D}, 
where the 1, 2 and 3$\sigma$ limits for $\Omega_ch^2$
in the different models are shown.
Given a flat Universe, this reflects also in different values
for the DE energy density today in the different models
(see the 1, 2 and 3$\sigma$ limits for $\Omega_\Lambda$
in the lower panel in Fig.~\ref{fig:omega1D}).

\begin{figure}
  \centering
  \includegraphics[page=2,width=\singlefigsmall]{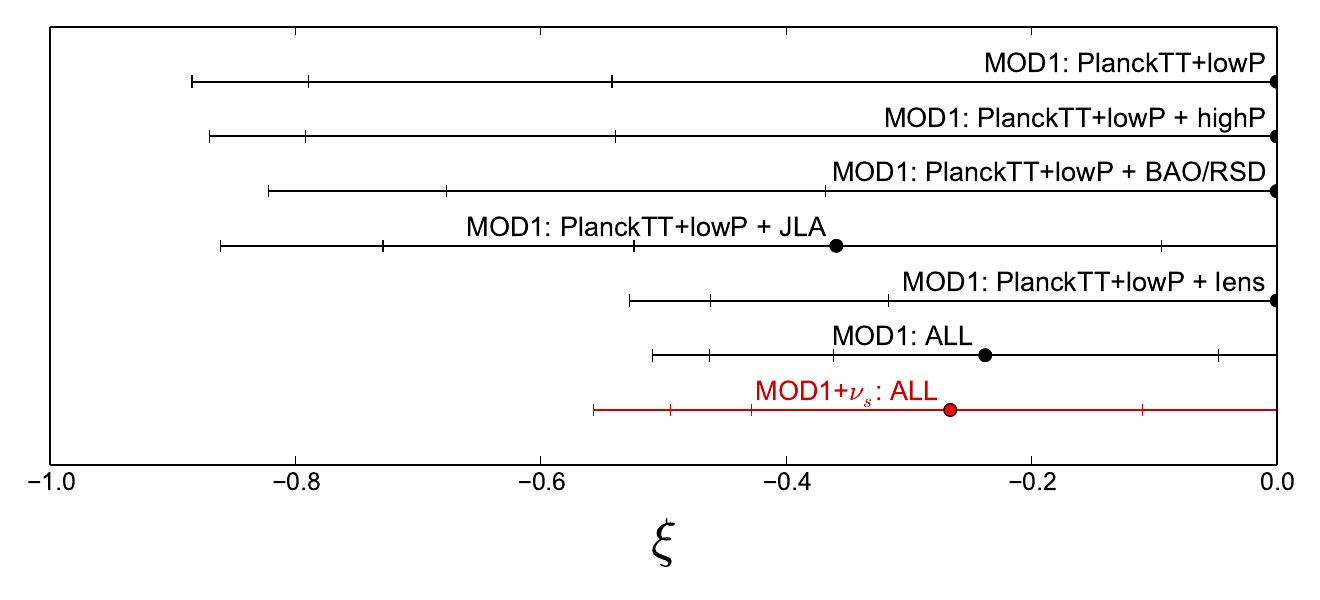}
  \includegraphics[page=3,width=\singlefigsmall]{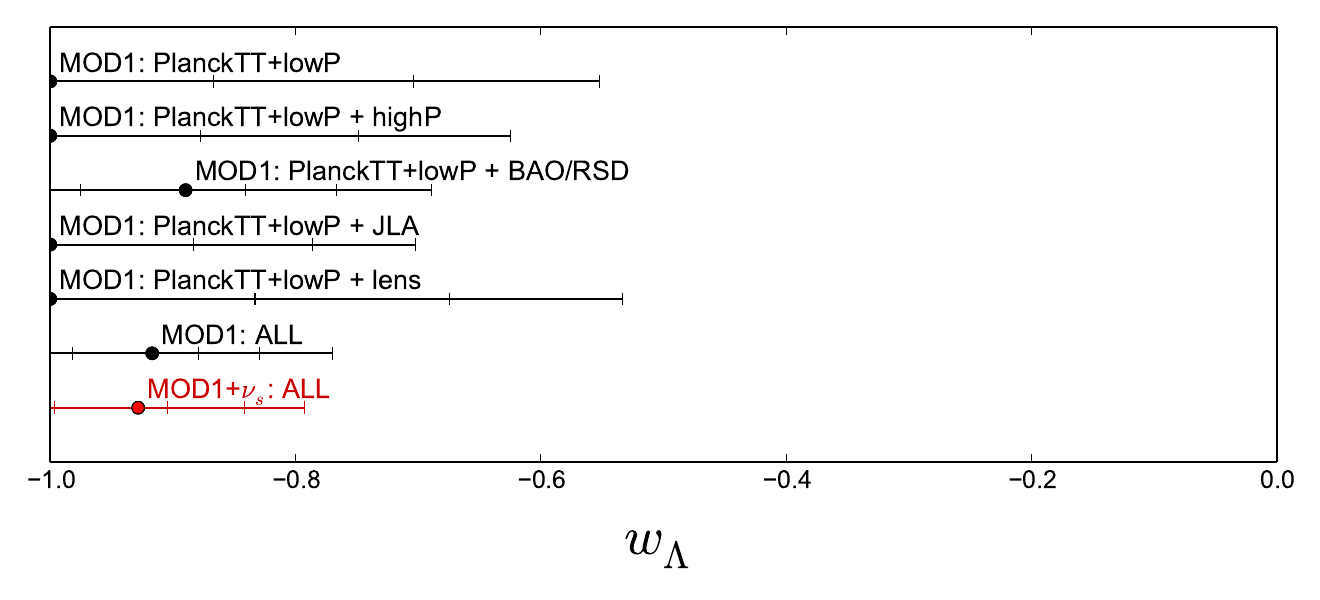}
  \caption[Marginalized limits for $\xi$ and $\wla$,
  considering the MOD1 scenario]{
  Marginalized 1, 2 and 3$\sigma$ limits on $\xi$ and $\wla$ in the MOD1, for different datasets.
  When the error bars are not visible, they coincide with the limit in the prior, as listed in Tab.~\ref{tab:priorscde}.
  The red point is for the MOD1\nus~model, discussed in Section~\ref{sec:cde_sterilenuDM}.
  From Ref.~\cite{Murgia:2016ccp}.
  }
  \label{fig:mod1_csi_w}
\end{figure}
\begin{figure}
  \centering
  \includegraphics[page=4,width=\singlefigsmall]{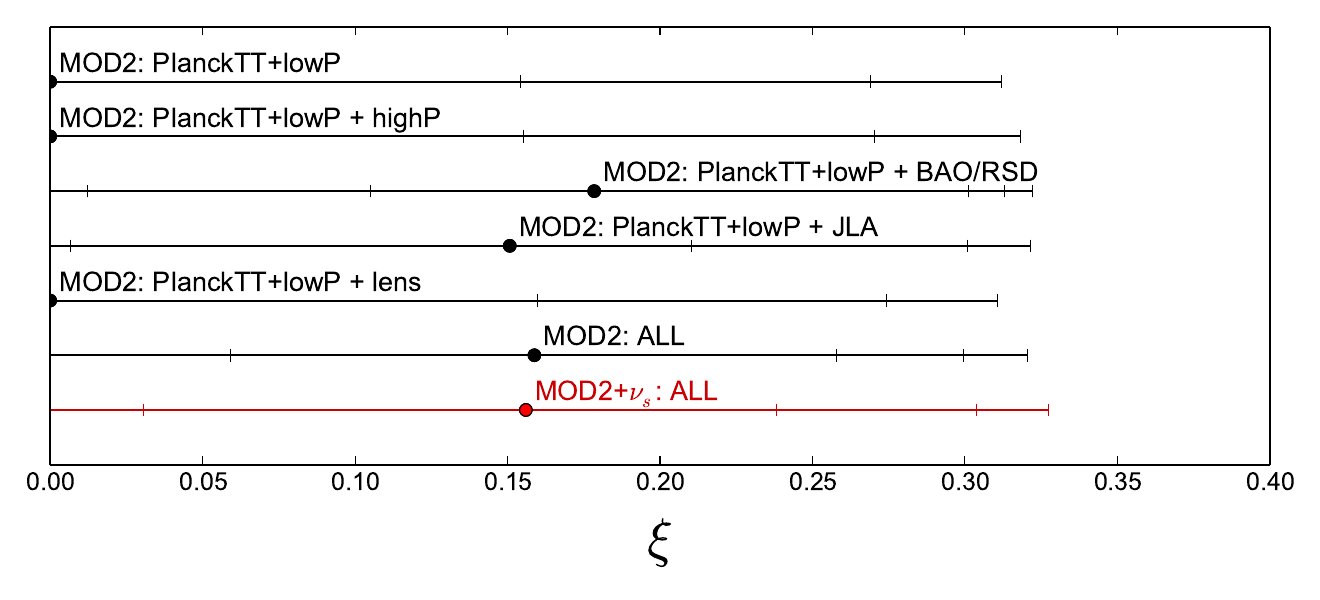}
  \includegraphics[page=5,width=\singlefigsmall]{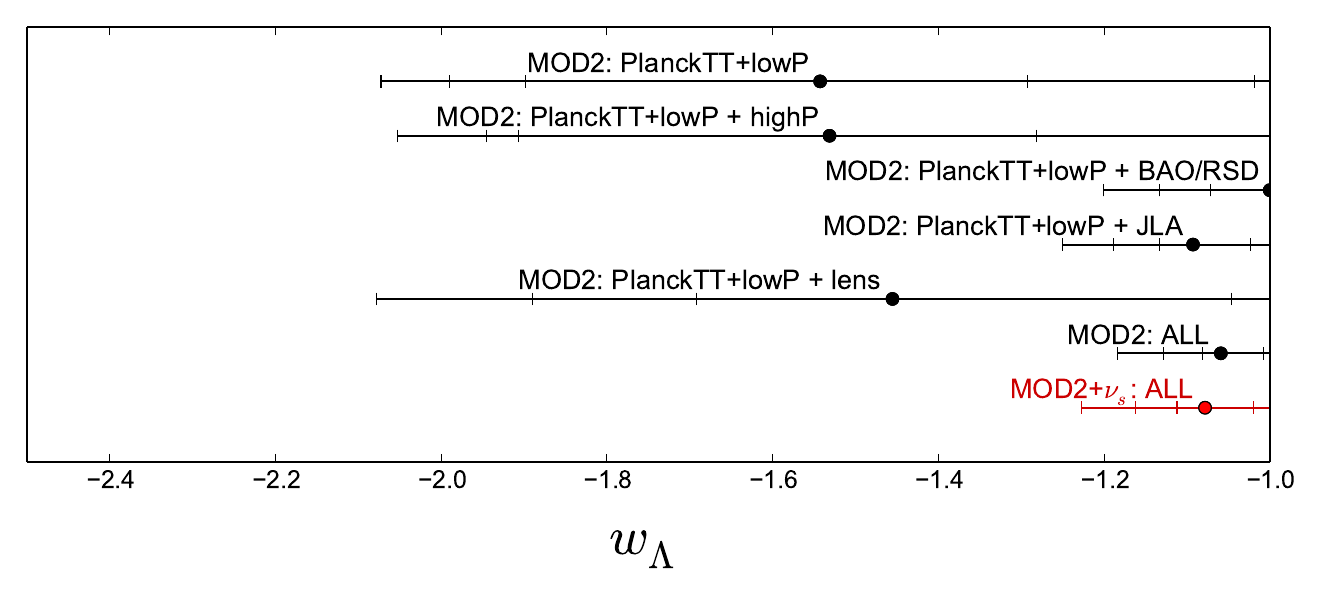}
  \caption[Marginalized limits for $\xi$ and $\wla$,
  considering the MOD2 scenario]{
  Marginalized 1, 2 and 3$\sigma$ limits from $\xi$ and $\wla$
  in the MOD2, for different datasets.
  When the error bars are not visible,
  they coincide with the limit in the prior, as listed in
  Tab.~\ref{tab:priorscde}.
  The red point is for the MOD2\nus~model,
  discussed in Section~\ref{sec:cde_sterilenuDM}.
  From Ref.~\cite{Murgia:2016ccp}.
  }
  \label{fig:mod2_csi_w}
\end{figure}

\begin{figure}
  \centering
  \includegraphics[page=2,width=\halfwidth]{cde/csi_w.pdf}
  \includegraphics[page=1,width=\halfwidth]{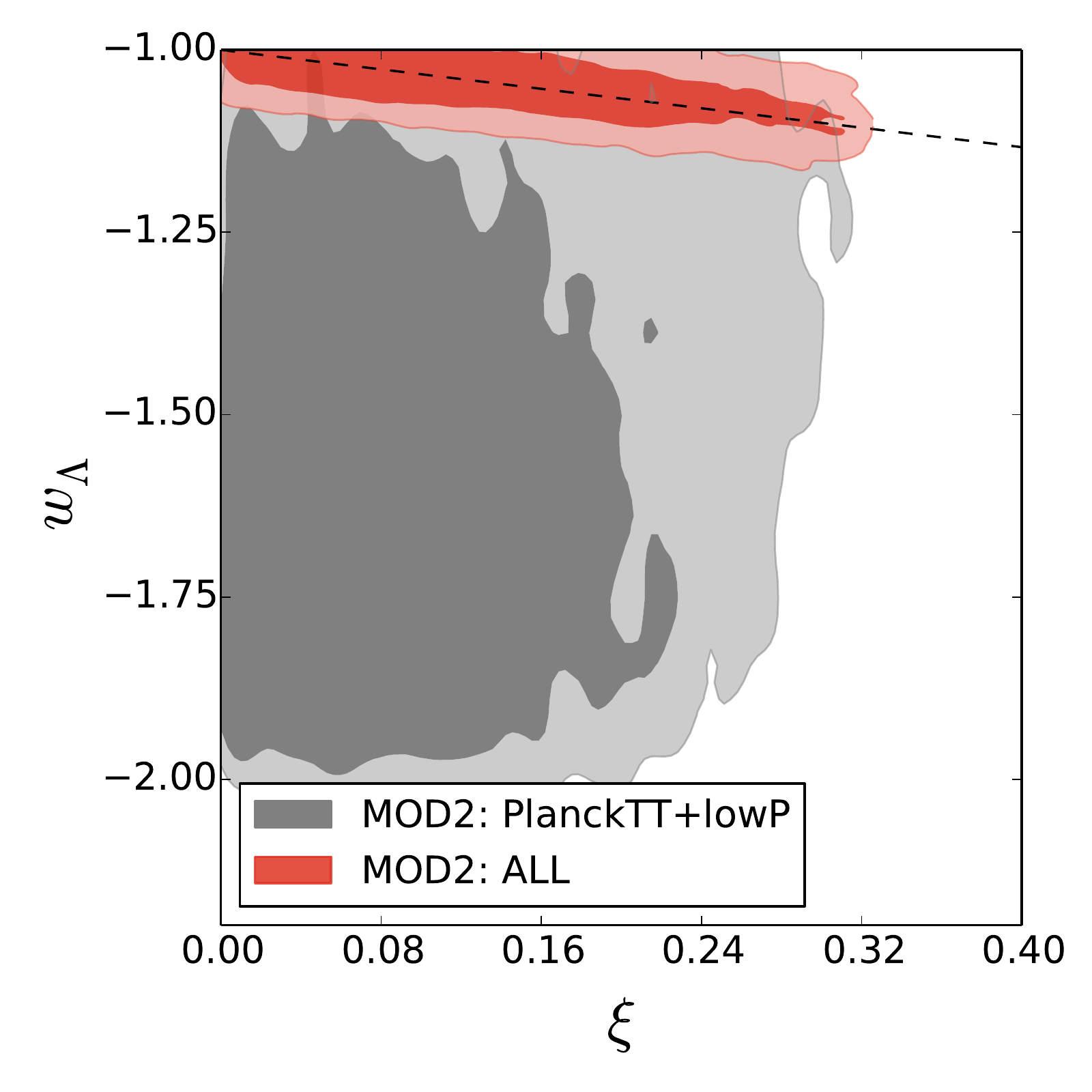}
  \caption[Marginalized constraints in the ($\xi$, $\wla$) plane
  for the MOD1 and MOD2 scenarios]
  {
  Marginalized 1 and 2$\sigma$ allowed regions in the ($\xi$, $\wla$)
  plane in the MOD1 (left) and MOD2 (right) scenarios, 
  for different datasets.
  Points in the regions below the dashed lines
  (representing $\wlae=\wla+\xi/3=-1$) 
  correspond to an increasing energy density for DE in the future.
  From Ref.~\cite{Murgia:2016ccp}.
  }
  \label{fig:csi_w}
\end{figure}

Figs.~\ref{fig:mod1_csi_w} and \ref{fig:mod2_csi_w}
show the 1, 2 and 3$\sigma$ limits
on $\xi$ (upper panels) and $\wla$ (lower panels) obtained
with different datasets, 
for both the CDE models MOD1 (Fig.~\ref{fig:mod1_csi_w})
and MOD2 (Fig.~\ref{fig:mod2_csi_w}). 
The constraints are almost insensitive to the addition
of the CMB polarization at high multipoles (``highP'').
The lensing information, instead, leads to
stronger constraints for $\xi$ in MOD1:
as expected, this comes from the bounds
on the DM abundance during the expansion history
that are provided by the lensing detection.
Both in MOD1 and MOD2, the addition of the JLA and BAO/RSD dataset leads
to stronger bounds on the DE EoS $\wla$, 
that is constrained towards -1.
Actually the cosmological data constrains the effective DE EoS
parameter $\wlae=\wla+\xi/3$
that drives the background evolution in Eq.~\eqref{eq:coupl_bgDE}.
As we can see in Fig.~\ref{fig:csi_w}, for both MOD1 (left panel)
and MOD2 (right panel) the marginalized regions in the 
($\xi$, $\wla$) plane are well constrained around the $\wlae=-1$ (dashed) line,
thus indicating a preference for a DE energy density
that is effectively constant over time.

From Tab.~\ref{tab:cmb} we can also see how the CMB data
only gives poor constraints on both
the derived quantities $H_0$ and $\sigma_8$.
For the Hubble parameter, this is due to the strong correlation
between $H_0$ and the DE EoS parameter:
as we can see in Eq.~\eqref{eq:coupl_bgDE},
when $\wla<-1$ the DE density today is larger
for larger values of $|\wla|$.
Since the Universe is DE-dominated at late times,
the total energy density $\rho_{\mathrm{tot}}$ increases with $\rhode$
and consequently the Hubble rate today
$H\propto\sqrt{\rho_{\mathrm{tot}}}$ is larger.
When $\wla>-1$, instead, the situation is opposite and values
for $H_0$ lower than the CMB predictions can be found.
The CMB alone, however, is not a good way to constrain the DE EoS:
with the introduction of additional data, in particular the BAO/RSD
and JLA datasets, the constraints on $\wla$ are much stronger, 
especially in MOD2, and consequently the allowed regions
for $H_0$ are better identified.

\begin{figure}
  \centering
  \includegraphics[page=1,width=\halfwidth]{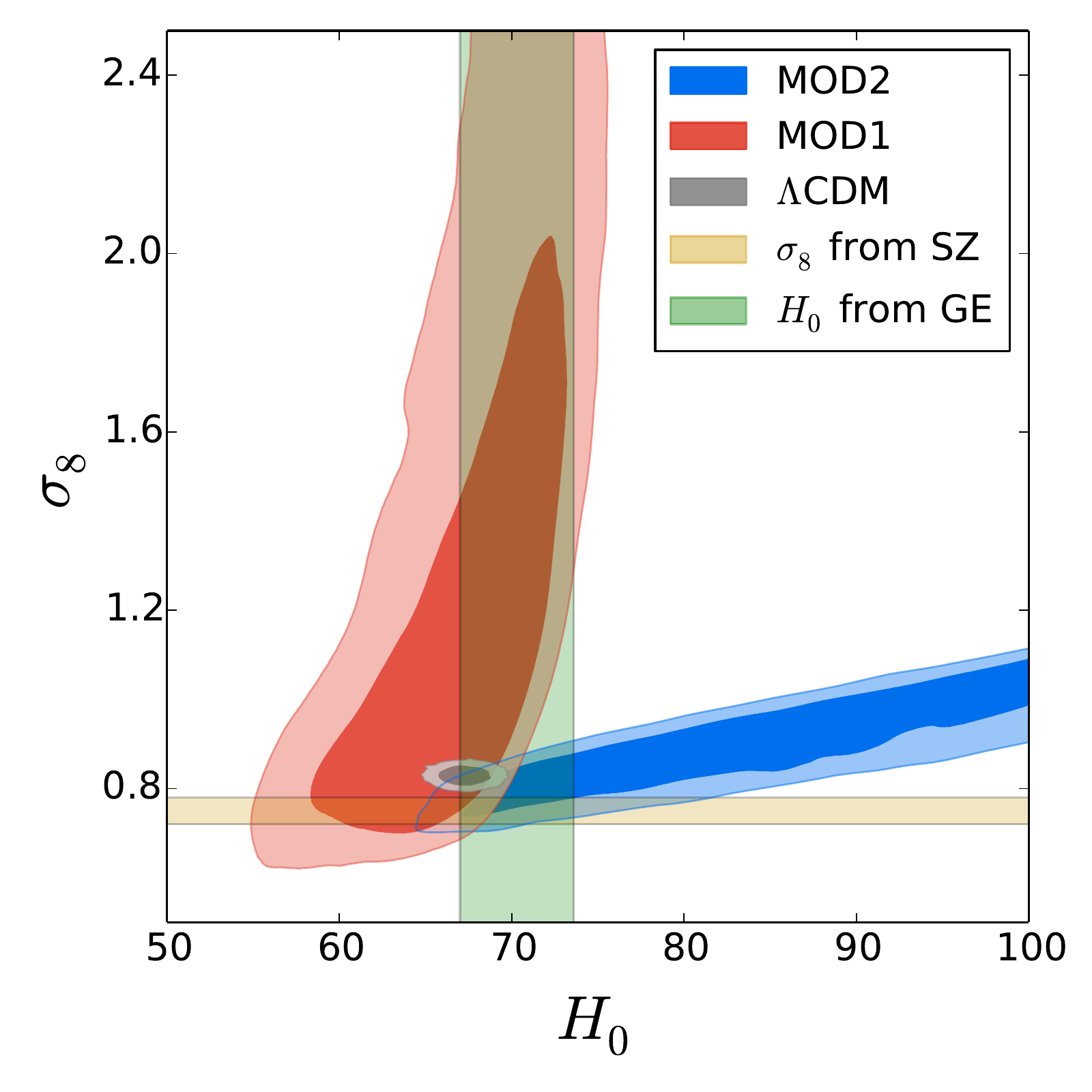}
  \includegraphics[page=2,width=\halfwidth]{cde/sigma8_h0.pdf}
  \caption[Marginalized constraints in the ($\sigma_8$, $H_0$) plane
  for the \lcdm, MOD1 and MOD2 scenarios]{
  Marginalized 1 and 2$\sigma$ allowed regions
  in the ($\sigma_8$, $H_0$) plane for different models:
  \lcdm~(gray), MOD1 (red) and MOD2 (blue).
  The left panel correspond to the CMB only dataset ``PlanckTT+lowP'', 
  while the panel on the right correspond
  to the full combination considered here (``ALL'').
  The green band is $H_0=70.6\pm3.3\Hou$ \cite{Efstathiou:2013via} (GE), 
  while the dark yellow band is $\sigma_8=0.75\pm0.03$ \cite{Ade:2013lmv} (SZ).
  From Ref.~\cite{Murgia:2016ccp}.
  }
  \label{fig:sigma8_h0}
\end{figure}

It is interesting to note that MOD1 predicts
a value for $\sigma_8$ significantly larger than the \lcdm~prediction
(see both Tab.~\ref{tab:cmb} and Tab.~\ref{tab:all}):
since MOD1 predicts a larger amount of DM in the early Universe,
there is more clustering in the primordial Universe, 
that results in an earlier transition to the nonlinear evolution
and hence to unavoidably larger values for $\sigma_8$
with respect to the \lcdm~predictions.
Even if the $\sigma_8$ values as determined
by local measurements are underestimates of the true value,
as the CMB determinations within the \lcdm\ model seems to suggest,
this can be a strong argument against a CDE parameterization through MOD1.
On the contrary, in MOD2 the DM abundance is bigger
in the late Universe with respect to the earlier epochs:
the nonlinear evolution is entered later
during the Universe evolution and $\sigma_8$ does not increase significantly, 
because at late times the DE is dominant and prevents clustering.
A hint for late-time appearance of DM was found also
in the recent study \cite{Pigozzo:2015swa},
thus giving another point in favor of MOD2.

In Fig.~\ref{fig:sigma8_h0} we summarize the results
on $H_0$ and $\sigma_8$
in the three models (\lcdm, MOD1, MOD2) for both the CMB only (left panel)
and the ``ALL'' (right panel) datasets.
As reference, we plot
two bands representing the local determinations
of $\sigma_8=0.75\pm0.03$ from Planck \cite{Ade:2013lmv},
obtained leaving the mass bias free to vary,
and $H_0=70.6\pm3.3$ \cite{Efstathiou:2013via} as a comparison.
Both the plots display that in MOD1 it is impossible
to obtain high $H_0$ values with low $\sigma_8$ values
and the uncertainty on $\sigma_8$ reflects also in an uncertainty
on $H_0$ \cite{Odderskov:2015fba}.
On the contrary, in MOD2 $H_0$ can assume larger values
without implying very large values for $\sigma_8$.
This is due to the opposite correlation of $\sigma_8$
with the coupling parameter $\xi$:
whereas in MOD1 a larger $\sigma_8$ arises from a larger interaction rate,
MOD2 shows an opposite behavior, namely lower values 
of $\sigma_8$ correspond to a stronger coupling in the dark sector
and possibly to high values of $H_0$.
In this sense, MOD2 should be preferred over MOD1, since
in this context the cited tensions regarding $\sigma_8$ and $H_0$ 
can be solved.

\section{Sterile neutrinos as stable DM component}
\label{sec:cde_sterilenuDM}
Up to now we did not consider the possibility
that the total amount of DM energy density
is provided by two or more different species,
with only one of them coupled to DE.
In this situation, the DM is composed by a stable and an interacting fraction, 
with the consequence that only part of the DM can
feed (or be fed by) DE during the Universe evolution.
A model with an interacting DM component combined with a stable one
was studied for example in Ref.~\cite{Berezhiani:2015yta},
where the authors report a preference
for the existence of two separate components.

Among the most investigated DM candidates,
sterile neutrinos have been widely studied in the past
(see e.g.\ Chapters \ref{ch:nu}, \ref{ch:lsn_cosmo} and \ref{ch:pps_nu}).
We present here a comparison of the bounds obtained
for the sterile neutrino properties
when the underlying cosmological model is changed from the \lcdm~model
to the CDE scenarios MOD1 and MOD2,
to test the possibility that the additional neutrino represents
the stable DM fraction.

To include the additional neutrino in the cosmological analysis
we use the parameterization presented in
Ref.~\cite{Ade:2013zuv} and adopted in Section~\ref{sec:jhep}.
The additional neutrino acts as a relativistic component
in the early Universe and gives a contribution
to the effective number of relativistic species \neff\ 
that is $\Delta\neff=\neff-\neff^{\mathrm{sm}}$
and it can be obtained from Eq.~\eqref{eq:dneff}.
In the late Universe, when the sterile neutrino becomes non-relativistic,
its mass becomes important and it behaves as a massive component.
Since we will not study the compatibility of the cosmological constraints
with SBL neutrino oscillations, in this case
it is more convenient to use the effective mass $\meff{s}$
(see Eq.~\eqref{eq:meffs}) instead of the physical mass $m_s$.
The effective mass is more convenient than $m_s$ also because
we are particularly interested in the degeneracy between
$\Omega_s h^2\propto\meff{s}$
and the DM energy density $\Omega_ch^2$.
For both $\neff$ and $\meff{s}$ we adopt flat priors in the intervals
listed in Tab.~\ref{tab:priorssterilenu}.

\begin{table}[t]
\begin{center}
\renewcommand{\arraystretch}{1.2}
\begin{tabular}{|c|c|c|}
  \hline
  	& \multicolumn{2}{c|}{Prior}	\\ \hline
  Parameter	& \lcdm	& $\nu_s$	\\ \hline
  $\meff{s}$  	& 0	& [0,15]	\\
  $\neff$ 	& 3.046	& [3.046, 6]	\\ \hline
\end{tabular}
\caption[Priors on the neutrino parameters]
{The priors on the neutrino parameters $\neff$ and $\meff{s}$,
flat in the listed intervals.
From Ref.~\cite{Murgia:2016ccp}.}
\label{tab:priorssterilenu}
\end{center}
\end{table}

We study the constraints on the sterile neutrino properties using
only the full data combination ``ALL'',
that gives the strongest constraints on the CDE models.
We compare the results obtained in the \lcdm\nus, MOD1\nus~and MOD2\nus~models
in Tab.~\ref{tab:lsn_all} for all the relevant parameters.
The inclusion of an additional neutrino do not change significantly
the constraints on the \lcdm~parameters,
with the only exception of $\Omega_ch^2$.
For the baryon energy density there is a small shift of less than 1$\sigma$, 
while the errors on $\tau$, $n_s$ and $\logA$ are slightly increased,
but these changes are independent on the CDE model.

\begin{figure}
  \centering
  \includegraphics[page=1,width=\halfwidth]{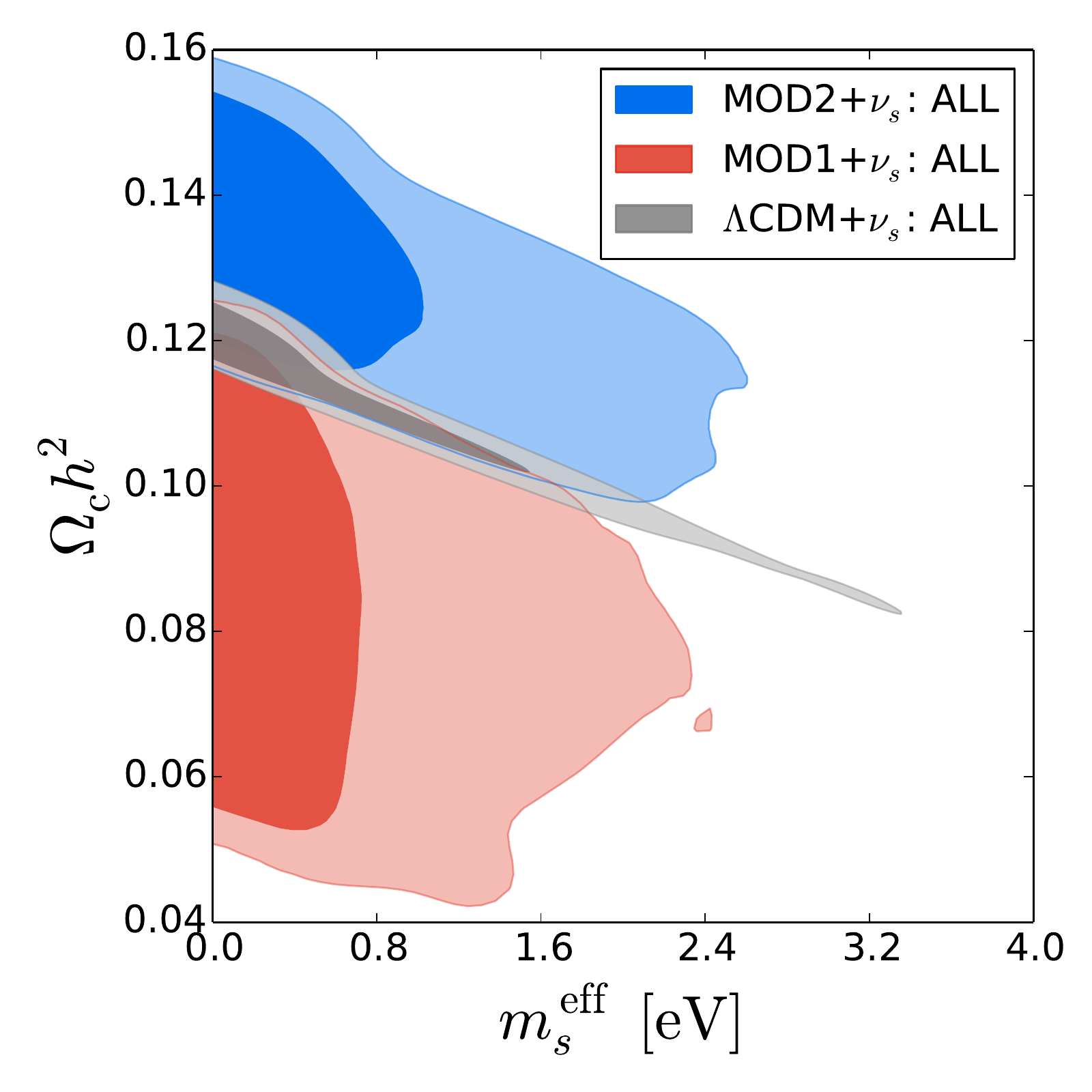}
  \caption[Marginalized constraints in the ($\Omega_ch^2$, $\meff{s}$) plane
  for the \lcdm, MOD1 and MOD2 scenarios with the addition of
  a sterile neutrino]{
  Marginalized 1 and 2$\sigma$ allowed regions in the
  ($\Omega_ch^2$, $\meff{s}$) plane for different models:
  \lcdm\nus~(gray), MOD1\nus~(red) and MOD2\nus~(blue),
  obtained with the full data combination considered here (``ALL'').
  From Ref.~\cite{Murgia:2016ccp}.
  }
  \label{fig:lsn_omc_meff}
\end{figure}

As expected, the quantity that varies most is the CDM energy density
$\Omega_ch^2$, that is lower and more uncertain in all the models.
This is due to the fact that the sterile neutrino acts
as a massive component in the late Universe
and it contributes to the total amount of matter with
$\Omega_s h^2\propto\meff{s}$:
a degeneracy with DM exists.
The degeneracy is shown in Fig.~\ref{fig:lsn_omc_meff},
where it is clear that a higher DM energy density
corresponds to a lower $\meff{s}$, for all the models.
The differences in $\Omega_ch^2$ between the CDE\nus,
MOD1\nus~and MOD2\nus\ models, however,
are the same we discussed without the sterile neutrino.

Constraints on the parameters \neff~and \meff{s} are almost the same
in the different models, with only very small differences:
this means that the properties of the sterile neutrino as DM
are robust against the introduction of the new interaction.
In parallel, also the constraints on the coupling parameter $\xi$
and on the DE EoS parameter $\wla$ are almost insensitive
to the presence of the additional neutrino.
The 1, 2 and 3$\sigma$ limits on $\xi$ and $\wla$ 
are plotted in red in Figures~\ref{fig:mod1_csi_w}
and \ref{fig:mod2_csi_w} for MOD1 and MOD2 respectively:
the ``ALL'' dataset, independently of the $\nu_s$ presence,
gives a 1$\sigma$ preference for a non-zero interaction in the dark sector.

\begin{figure}
  \centering
  \includegraphics[page=1,width=\halfwidth]{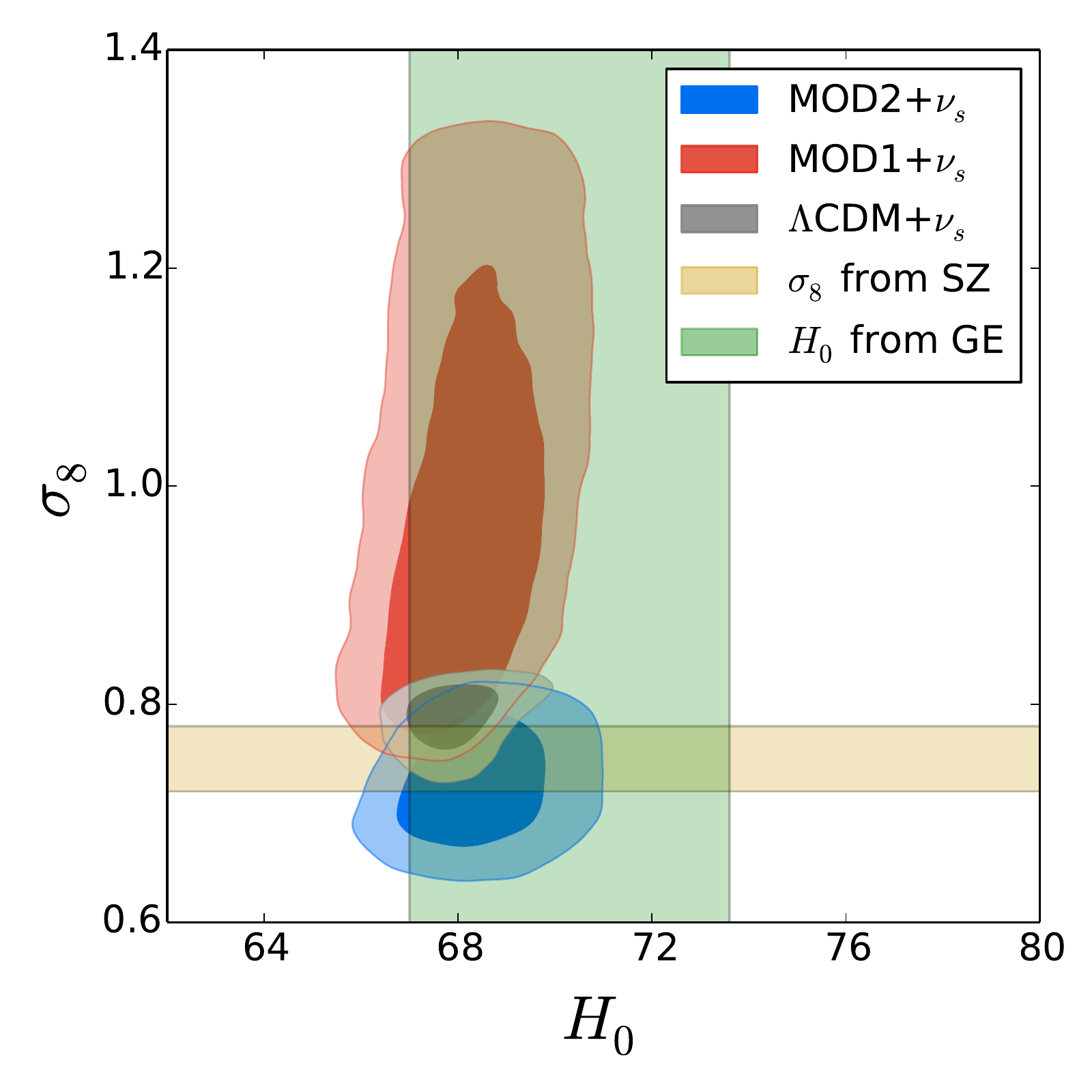}
  \caption[Marginalized constraints in the ($\sigma_8$, $H_0$) plane
  for the \lcdm, MOD1 and MOD2 scenarios with the addition of
  a sterile neutrino]{
  Marginalized 1 and 2$\sigma$ allowed regions in the
  ($\sigma_8$, $H_0$) plane for different models:
  \lcdm\nus~(gray), MOD1\nus~(red) and MOD2\nus~(blue),
  obtained with the full data combination considered here (``ALL'').
  The green band is $H_0=70.6\pm3.3\Hou$ \cite{Efstathiou:2013via} (GE), 
  while the dark yellow band is $\sigma_8=0.75\pm0.03$ \cite{Ade:2013lmv} (SZ).
  From Ref.~\cite{Murgia:2016ccp}.
  }
  \label{fig:lsn_h0_sigma8}
\end{figure}

The presence of an additional component that acts
as a relativistic particle in the early Universe
and a non-relativistic one in the late Universe
gives a suppression in the clustering, due to the free-streaming effect,
and an increase of the Hubble parameter, due to the necessity
of increasing both the DM and DE energy densities in the Universe to avoid
a shift of the matter-radiation equality and of the coincidence time.
As a consequence, the inclusion of the sterile neutrino shifts
the predictions for $H_0$ towards slightly higher values
and lowers those for $\sigma_8$.
In Fig.~\ref{fig:lsn_h0_sigma8} we show the equivalent
of Fig.~\ref{fig:sigma8_h0} for the models with the additional neutrino.
Apart for the fact that the regions are slightly wider,
there are no significant variations with respect
to the right panel of Fig.~\ref{fig:sigma8_h0}.
As a consequence of the lowering of $\sigma_8$, however,
models with the sterile neutrino show a higher compatibility
with the low-$\sigma_8$ measurements as, for example,
the Planck cluster counts (SZ, yellow band in the plots).

\begin{table}[t]                                                                           
\begin{center}        
\renewcommand{\arraystretch}{1.2}                                                                  
\begin{tabular}{|c|c|c|c|}                                                              
\hline
Parameter &	\lcdm	& MOD1	& MOD2\\
\hline
$100\Omega_bh^2$
			& $2.237\,^{+0.034}_{-0.031}$    & $2.237\,^{+0.036}_{-0.032}$    & $2.236\,^{+0.035}_{-0.032}$    
\\
$\Omega_ch^2$
			& $0.113\,^{+0.014}_{-0.019}$    & $0.083\,^{+0.034}_{-0.033}$    & $0.129\,^{+0.024}_{-0.025}$    
\\
$100\theta$
			& $1.0408\,^{+0.0006}_{-0.0007}$ & $1.0426\,^{+0.0022}_{-0.0020}$ & $1.0400\,^{+0.0010}_{-0.0011}$ 
\\
$\tau$
			& $0.063\,^{+0.032}_{-0.033}$    & $0.064\,^{+0.034}_{-0.035}$    & $0.060\,^{+0.034}_{-0.035}$    
\\
$n_s$
			& $0.969\,^{+0.012}_{-0.011}$    & $0.968\,^{+0.013}_{-0.012}$    & $0.968\,^{+0.012}_{-0.012}$    
\\
$\logA$
			& $3.059\,^{+0.066}_{-0.067}$    & $3.061\,^{+0.068}_{-0.070}$    & $3.054\,^{+0.070}_{-0.069}$    
\\ 	\hline
$\xi$
			& $0$                            & $(-0.494,0]$                      & $[0,0.304)$    
\\                      
$\wla$                  
			& $-1$                           & $[-1,-0.841)$                      & $(-1.162,-1]$   
\\ 	\hline
$\meff{s}$~[eV]
			& $<2.1$                         & $<1.9$                         & $<2.2$                         
\\                      
$\neff$                 
			& $<3.34$                        & $<3.38$                        & $<3.35$                        
\\ 	\hline
$H_0$~[\Hou]
			& $67.91\,^{+1.33}_{-1.26}$      & $68.23\,^{+2.21}_{-2.00}$      & $68.43\,^{+2.16}_{-2.07}$      
\\
$\sigma_8$
			& $0.789\,^{+0.039}_{-0.045}$    & $0.988\,^{+0.300}_{-0.229}$    & $0.727\,^{+0.073}_{-0.072}$    
\\ 	\hline
\end{tabular}
\caption[Marginalized constraints on the cosmological parameters
from the CMB data set, including a sterile neutrino]
{Marginalized limits at 2$\sigma$ for various parameters considered
in our analyses, 
obtained with the ``ALL'' dataset for the three different models
(\lcdm\nus, MOD1\nus~and MOD2\nus).
When an interval denoted with parenthesis is given,
it refers to the  2$\sigma$ C.L.\ range starting from the
prior extreme.
These are listed in Tabs.~\ref{tab:priorscde}
and \ref{tab:priorssterilenu}.
From Ref.~\cite{Murgia:2016ccp}.
}
\label{tab:lsn_all}
\end{center}
\end{table}

\section{Conclusions}
\label{sec:cde_disc}
The largest part of the energy density
of our Universe is represented by a dark sector,
formed by dark matter and dark energy.
Both these components are known only for they gravitational effects,
but we still ignore if they can be explained in the context of fundamental physics:
while many candidates of DM have been proposed,
the true nature of DE is completely unknown from this point of view.
Apart for gravity, we ignore how DM and DE
interact with the other particles.
The existence of a non-gravitational coupling involving DE or DM
cannot be excluded:
this additional interaction would have an impact on cosmology
and it can be tested, in principle,
studying the various cosmological observables.
A coupling with standard matter is disfavored by observations both for DE and DM,
but it is possible that the interaction does not involve baryons
nor photons or other particles in the standard model.
We studied the possibility that DM and DE are coupled
to each other in a non-gravitational way.
We introduced a phenomenological interaction rate $Q=\xi H\rhode$ 
\cite{Zimdahl:2001ar,Salvatelli:2013wra, Costa:2013sva},
where the dimensionless parameter $\xi$ encodes the coupling strength:
for our choice, positive $\xi$ values correspond to DM decaying in DE,
while a negative $\xi$ gives a scenario with DE decaying in DM.

We test the coupled model using several cosmological data:
CMB data and gravitational lensing reconstructions from the 2015 Planck release,
SuperNovae distance calibrations, BAO and RSD measured by several experiments.
All these measurements have the aim to constrain the evolution
of the Universe at different redshifts
and to test the gravitational interaction at different epochs.
The time-dependency of DE and DM energy densities
is indeed modified by the introduction of the coupling 
that influences both the background and the perturbations evolution.

In the context of an extended \lcdm~model, we obtained constraints
on the coupling parameter $\xi$ and on the DE Equation of State (EoS) $\wla$.
We base our analysis on the Planck observations for CMB temperature
and polarization \cite{Adam:2015rua,Aghanim:2015xee},
but we obtain the strongest constraints from the inclusion
of additional information at several different redshifts.
The introduction of Supernovae data from the joint analysis
of Ref.~\cite{Betoule:2014frx}
strongly constrains the effective DE EoS
$\wlae=\wla+\xi/3$ to be -1,
while the BAO/RSD
\cite{Beutler:2011hx,Ross:2014qpa,Anderson:2013zyy,Samushia:2013yga}
data gives a mild preference for a non-zero coupling, both for MOD1 and MOD2.

If we consider the predicted values of the Hubble parameter $H_0$
and of $\sigma_8$, however,
we note that the phenomenology of MOD1, that was more studied in the past
(see e.g.~Refs.~\cite{Costa:2013sva, Salvatelli:2013wra}),
increases the tension with the low-redshift measurements
of $H_0$ \cite{Riess:2011yx, Efstathiou:2013via}
and the local determinations of $\sigma_8$
\cite{Ade:2013lmv,Ade:2015fva,Reichardt:2012yj,
Vikhlinin:2008ym,Bohringer:2014ooa, Heymans:2012gg}.
The reason is that in MOD1 a higher amount of DM
in the early Universe is required to have some residual DM today.
This higher DM amount increases the clustering effect and drives
the evolution to nonlinear scales earlier.
In MOD2, on the contrary, $\sigma_8$ is smaller than in the \lcdm~model
and CMB estimates can be reconciled with low-redshift probes.

We studied also the possible presence of a sterile neutrino
\cite{Merle:2013gea,Gariazzo:2015rra,Bergstrom:2014fqa, Merle:2013wta,
Gariazzo:2014dla,Gariazzo:2014pja,Gariazzo:2013gua,Archidiacono:2014apa}
as an additional and stable dark matter component.
In this case we find that the sterile neutrino parameters
are completely insensitive to
the parameters of the CDE model and the constraints are practically
the same for the \lcdm\nus, the MOD1\nus~and the MOD2\nus~models.

In conclusion, a coupled DM/DE cosmology is a viable option, 
compatible with a large host of cosmological data.
Moreover, a model where DE decays into DM during the evolutionary history
of the Universe can help solving the small tensions that currently
exist between different high- and low-redshift observations
in the context of the \lcdm~model, therefore providing
an interesting new opportunity
of investigation for models of the dark sectors of the Universe.

% \input{pps.tex}
% \input{darkrad.tex}

% \part{Summary and Conclusions}
\chapter{Summary and Conclusions}
%!TeX root=main.tex 
\label{ch:conclusions}
Our knowledge of the Universe is rather robust.
Most of the predictions of the theoretical model based on
the theory of General Relativity proposed by A.~Einstein have been
experimentally confirmed.
The last, exciting probe of General Relativity is
the recent first detection
of the gravitational waves by the LIGO/VIRGO collaboration
\cite{Abbott:2016blz}.
Modern cosmology is based on the models derived from
the Einstein's theory, that are tested using the
numerous experimental data collected in several different observations.
The strongest tools to study the models of the Universe evolution
are the observations of the
Cosmic Microwave Background (CMB),
that is the relic photon radiation
emitted in the early Universe.
This Thesis is devoted to study several aspects of the
cosmological evolution using mainly the CMB results obtained
by the Planck experiment.
We considered different extensions of the standard \lcdm\ model:
we included additional particles (neutrinos, axions),
we assumed non-standard inflationary scenarios
and
we introduced
an additional coupling between dark matter and dark energy.

The first results that we reported concern the cosmological constraints
on the light sterile neutrino.
We found that the CMB data disfavor the presence of
an additional massive neutrino,
if it is thermalized with the active neutrinos.
A light neutrino relic is favored, instead, by the local measurements
of $\sigma_8$ and $H_0$
(see Chapter~\ref{ch:lsn_cosmo}), because
the free-streaming nature of the neutrino 
allows to reduce the amount of matter fluctuations at small
scales even if $H_0$ is simultaneously increased, as a consequence of
the correlation with the presence of additional relativistic
particles.
The mass required to reconcile the $H_0$ and $\sigma_8$ tensions,
however, is smaller than the one required by SBL neutrino oscillations.

As a consequence of the anticorrelation
between the sterile neutrino mass $m_s$
and its contribution \DNeff\ 
to the effective number of relativistic species \Neff\ 
(see Chapter~\ref{ch:lsn_cosmo}),
the presence of a neutrino with 1~eV mass is allowed only
if its contribution to \neff\ is much smaller
than the one from each active neutrino.
The strongest constraints on \DNeff\ can be obtained considering
the 2015 data on the CMB anisotropies by the Planck collaboration,
from which
it is possible to obtain $\DNeff\lesssim0.4$ at 95\% C.L.,
with small variations due to the inclusion of different datasets
(see Section~\ref{sec:inflfreed_nnu}).
This confirms the problem of the missing thermalization of the
sterile neutrino.
Previous studies \cite{Hannestad:2012ky,Saviano:2013ktj,Vincent:2014rja}
have shown that the mixing parameters derived
from the SBL analyses are large enough to allow the sterile neutrino
to be in equilibrium with the active neutrinos.
Since this does not happen, some new physical mechanism should operate.
Some of the possibilities include:
a large lepton asymmetry
\cite{Foot:1995bm,Foot:1996qc,Bell:1998sr,Bell:1998ds,Shi:1999kg,
DiBari:1999vg,DiBari:2001jk,Chu:2006ua,Hannestad:2012ky,
Mirizzi:2012we,Saviano:2013ktj,Hannestad:2013wwj},
new neutrino interactions \cite{Bento:2001xi,Dasgupta:2013zpn,
Hannestad:2013ana,Bringmann:2013vra,Ko:2014bka,Archidiacono:2014nda,
Saviano:2014esa,Mirizzi:2014ama,Tang:2014yla,Chu:2015ipa},
entropy production after neutrino decoupling \cite{Ho:2012br},
very low reheating temperatures \cite{Gelmini:2004ah,Smirnov:2006bu},
time varying dark energy components \cite{Giusarma:2011zq},
a larger cosmic expansion rate at the time
of sterile neutrino production \cite{Rehagen:2014vna}.

The mechanism of the sterile neutrino decay proposed originally
in Ref.~\cite{Gariazzo:2014pja} has been studied
in Section~\ref{sec:lsndecay}.
We showed that the decay of the light sterile neutrino may help
in solving the incomplete thermalization problem only if the CMB
data would allow $\DNeff=1$ for massless species.
In this case, indeed, if the fully thermalized sterile neutrino decays 
into massless species when it is still relativistic,
its mass is not relevant for the evolution, but
the amount of radiation is given by $\neff\simeq4$.
In the decay scenario we found a tension
between the measurements at low-redshift and the CMB:
if the local determinations of $\sigma_8$ would favor
the presence of a massive neutrino in the late-times evolution,
in order to suppress
the matter fluctuations through the free-streaming effect,
the CMB data strongly prefer a rapid decay of the sterile neutrino.
These requirements are clearly incompatible.
For this reason and since $\neff\simeq4$ is strongly disfavored by
the current data,
the sterile neutrino decay scenario is not a viable solution
to reconcile the presence of the
light sterile neutrino in cosmology.

Another possibility that we proposed is to assume a scenario
that we denoted as ``Inflationary Freedom''.
With this name we indicated the possibility that the Primordial
Power Spectrum (PPS) of scalar perturbations generated during
inflation can be more complicated than a simple power-law,
as the simplest inflationary models predict.
Since the final CMB spectrum is the convolution 
of the scalar PPS and
of the transfer function, robustly calculated from the theory
discussed in Chapter~\ref{ch:cmbr},
changes in the transfer function can be
compensated by variations in the PPS.
Previous analyses of the WMAP and Planck (2013, 2015) CMB spectra
showed that there are indications 
for deviations from the power-law shape of the PPS,
especially at large scales
\cite{Shafieloo:2003gf,Nicholson:2009pi,Hazra:2013ugu,Hazra:2014jwa,
Nicholson:2009zj,Hunt:2013bha,Hunt:2015iua,
Goswami:2013uja,Matsumiya:2001xj,Matsumiya:2002tx,
Kogo:2003yb,Kogo:2005qi,Nagata:2008tk,Ade:2015lrj,
Gariazzo:2014dla,DiValentino:2015zta,DiValentino:2016ikp}.

We used a model independent parameterization for the free PPS
and we showed that strong degeneracies between the PPS parameters
and the neutrino parameters exist, in particular at small scales.
The effective number of relativistic species is degenerate with the
PPS because the presence of additional radiation leads to an enhanced
Silk damping effect (see Section~\ref{sec:nucosmology}),
that can be compensated with an
enhancement of the PPS at the relevant scales.
One of the main effects of the neutrino mass in cosmology
is to alter the contribution of the early ISW effect
(see Section~\ref{sec:nucosmology}).
Also in this case
a variation of the PPS at the scales corresponding to the early ISW
contribution can partially compensate the effects of increasing
the neutrino masses.
The result is that the bounds on \neff\ and on the neutrino mass scale
are significantly relaxed, if only the temperature spectrum of CMB
anisotropies is considered.
Since the impact of the cosmological parameters and of the PPS
are different in the temperature and polarization spectra, however,
the degeneracy between the PPS and the neutrino parameters
can be broken with the inclusion of the TE and EE spectra
at high multipoles measured by the Planck collaboration.
If the results obtained without the CMB polarization data
in the context of ``Inflationary Freedom''
would allow the presence of a fully thermalized sterile neutrino
(see Section~\ref{sec:inflfreed_sterile}),
this is no more true when the polarization data are included
(see Sections~\ref{sec:inflfreed_nnu}
and~\ref{sec:inflfreed_mnu}).
% The strongest bounds
% we find are $\summnu<0.18\,(0.22)$~eV at 95\% C.L.\ 
% from the combination of Planck~TT,TE,EE+lowP+BAO data,
% when considering a power-law (free) PPS,
% while the results on \neff\ 
% based on the Planck~TT,TE,EE+lowP dataset are
% almost equivalent and we find
% $2.5\lesssim\neff\lesssim3.5$ at 95\% C.L., almost independent
% of the inclusion of low-redshift datasets.

In Chapter~\ref{ch:ther_ax} we studied a different candidate for
hot dark matter: the thermal axion.
Axions are pseudo-Nambu-Goldstone bosons generated by
the spontaneous breaking of the global
Peccei-Quinn symmetry $U(1)_{PQ}$, introduced to solve the strong
CP problem in Quantum ChromoDynamics.
The new symmetry is spontaneously broken at the scale $f_a$,
to which the thermal axion mass is connected
by Eq.~\eqref{eq:axionmass}.
Since it is relativistic in the early Universe, contributing to \neff,
and non-relativistic at late times, the thermal axion
contribution to the cosmological evolution is similar
to that of a massive neutrino.
Thanks to its free-streaming properties, a thermal axion can
reduce the matter fluctuations at small scales and
help to reconcile the $\sigma_8$ tension.
Also in this case, however, we found that
the full CMB data from the 2015 release of Planck disfavor
the presence of the additional thermal axion in cosmology.
The constraint comes in particular from the fact that a thermal axion
gives a minimum contribution $\DNeff\simeq0.2$ to the amount
of radiation in the early Universe,
but this is outside the limits at 68\% C.L. obtained from the
full Planck dataset.

Despite the fact that the presence of the axion is disfavored
by the CMB data at 68\% C.L.,
the significance of this result is not high
and the presence of a thermal axion is still allowed
by the at 95\% C.L.\ constraints.
For this reason,
we studied the bounds on the axion mass in the context of a power-law
and of a free PPS.
% The strongest bound we find on the thermal axion mass within
% the free PPS approach is $m_a<1.07$~eV at 95\% C.L.\ 
% (Planck~TT+lowP+BAO).
% In the standard power-law scenario, the most stringent bound is
% $m_a<0.74$~eV at 95\% CL, obtained with the further inclusion of the
% polarization at high multipoles 
% (Planck TT,TE,EE+lowP+BAO).
When we varied also the neutrino masses to test
the degeneracy with the thermal axion mass,
we found that the constraints on the total neutrino mass
are tighter than those obtained without thermal axions,
while the bounds on the thermal axion mass are unchanged.
% The strongest bounds we find for the thermal axion mass
% and the total neutrino mass in the free PPS approach are
% $m_a<1.03$~eV at 95\% C.L.\ and $\mnu<0.18$~eV at 95\% C.L.,
% when considering
% the Planck TT+lowP+BAO and Planck TT,TE,EE+lowP+BAO
% dataset combinations, respectively.
% In the power-law PPS scenario the strongest bounds are
% $m_a<0.76$~eV at 95\% C.L.\ and $\mnu<0.16$~eV at 95\% C.L.,
% obtained both for the Planck TT,TE,EE+lowP+BAO dataset.
In both cases we find only upper limits on the axion
and neutrino masses,
unless the Planck SZ cluster counts data are
included in the analyses.
In the latter case we found the only evidence
for a non-null axion mass
(see Subsection~\ref{ssec:ax_sigma8}).
As we discussed in Chapter~\ref{ch:cosmomeasurements}, however,
the local determinations of $\sigma_8$ may suffer the presence
of unaccounted systematics.
If these present results will be confirmed in future experiments,
the evidence of a non-zero axion mass will be strengthened.
As of today, anyhow, the evidence for $m_a>0$
must be treated with caution, because the CMB results 
that disfavor the presence of massive thermal axions are more robust
than the local determinations of $\sigma_8$.

The majority of the models 
that generate a non-standard PPS
also generate primordial non-Gaussianities,
that can be studied using
the Large Scale Structures (LSS) of the Universe and
the CMB bispectrum.
In Chapter~\ref{ch:ng} we studied how
the expectations for
the Dark Energy Spectroscopic Instrument (DESI) experiment,
an upcoming galaxy survey,
change when
the hypothesis of a power-law PPS is relaxed.
To do this, we assumed that the precise shape of the PPS and
the non-Gaussianity parameter $\fnl$ need to be extracted simultaneously
from the data.
We considered three different DESI tracers of the matter distribution at
various redshifts:
luminous red galaxies,
emission line galaxies and
high-redshift quasars.
If the analysis is restricted to LSS data,
the standard errors computed assuming a power-law PPS are enlarged
by $60\%$ when using the free PPS parameterization
and treating each of the possible dark matter tracers individually.
The problem is then that determining the PPS and $\fnl$ simultaneously
may cause a degrading of the obtained constraints.
Another problem could be induced in this way:
if nature have chosen a more complicated inflationary mechanism
that results in a non-trivial PPS,
all the analyses performed under the possibly wrong assumption
of a power-law PPS may give biased results,
as a consequence of the degeneracy
between the PPS and the non-Gaussianities.
This degeneracy may be reduced using
the multi-tracer technique,
or combining the DESI tracers
with the CMB priors on the PPS parameters.
The addition of CMB priors
on the PPS parameters and on the energy densities
of dark matter and baryons
leads to an error on $\fnl$ which is independent
of the PPS parameterization used in the analysis. 

In the context of the \lcdm\ model,
it is possible to obtain predictions on values of
the Hubble parameter $H_0$
and of the clustering parameter $\sigma_8$ today
from the analyses of CMB data.
These predictions are in tension with the low-redshift measurements
of $H_0$ \cite{Riess:2011yx, Efstathiou:2013via}
and the local determinations of $\sigma_8$
\cite{Ade:2013lmv,Ade:2015fva,Reichardt:2012yj,
Vikhlinin:2008ym,Bohringer:2014ooa, Heymans:2012gg}.
These tensions may be alleviated by the presence of a massive neutrino,
that can reduce the perturbations at small scales thanks to its
free-streaming properties that influence the Universe evolution,
but this is not the only possibility.
In Chapter~\ref{ch:cde} we proposed a solution
that involves the introduction of
a phenomenological non-gravitational coupling
between dark matter and dark energy.
Dark matter and dark energy are known only for
their gravitational effects,
but we still ignore if they can be explained in the context
of fundamental physics.
The existence of a non-gravitational coupling involving DE or DM
cannot be excluded:
this additional interaction would have an impact on cosmology
and it can be tested
studying the various cosmological observables.
We introduced a phenomenological interaction rate that
describes the energy transfer from dark matter to dark energy.
For our choice, a positive coupling (MOD1 for sake of brevity)
corresponds to DM decaying in DE,
while a negative coupling (MOD2)
gives a scenario with DE decaying in DM.
We tested the coupled model using several cosmological data
at different redshifts,
since the time-dependency of the DE and DM energy densities
is modified by the introduction of the coupling,
that influences both the background and the perturbations evolution.
In the context of an extended \lcdm~model, we obtained constraints
on the coupling parameter $\xi$ and
on the dark energy equation of state parameter $\wla$.
The introduction of Supernovae data from the joint analysis
of Ref.~\cite{Betoule:2014frx}
strongly constrains the effective DE EoS parameter
$\wlae=\wla+\xi/3$ (see Subsection \ref{ssec:cde_param})
to be $-1$,
while the BAO/RSD
\cite{Beutler:2011hx,Ross:2014qpa,Anderson:2013zyy,Samushia:2013yga}
data gives a preference for a non-zero coupling,
both for MOD1 and MOD2.
We noticed that the phenomenology of MOD1, that was more studied in the past
(see e.g.\ Refs.~\cite{Costa:2013sva, Salvatelli:2013wra}),
increases the tension with the low-redshift measurements
of $H_0$ \cite{Riess:2011yx, Efstathiou:2013via}
and of $\sigma_8$
\cite{Ade:2013lmv,Ade:2015fva,Reichardt:2012yj,
Vikhlinin:2008ym,Bohringer:2014ooa, Heymans:2012gg}.
The reason is that in MOD1 a higher amount of DM
in the early Universe is required to have some residual DM today,
and
this higher DM amount increases the clustering effect and
accelerates the nonlinear evolution.
In MOD2, on the contrary,
$\sigma_8$ is smaller than in the \lcdm~model
and CMB estimates can be reconciled with low-redshift probes.

In conclusion, the \lcdm\ model is extremely robust
and most of the currently available cosmological data disfavor
(or strongly constrain) any deviation from the simplest description
of the Universe.
Despite this, some small tensions are present.
It is still not clear if they are the consequence of
unaccounted systematics, incomplete or approximated calculations,
or unaccounted astrophysical effects in the analyses
of low-redshift observations.
Maybe they are just hints that some new physics exists.
Some new mechanism during inflation may be responsible
of the features at large scales
observed in the power spectrum of initial scalar fluctuations.
A coupled DM/DE cosmology is a viable option
to solve the tensions that exists
between different high- and low-redshift observations
of $H_0$ and $\sigma_8$,
if DE decays into DM during the evolution of the Universe.
This is also an interesting new opportunity
of investigation for models of the dark sector of the Universe.
This solution, however, does not help to solve the problem
of the thermalization of the sterile neutrino.
Future neutrino oscillation experiments will confirm
if the sterile neutrino with mass around 1~eV exists.
If its existence will be proved,
we will have to understand the reasons
for which neutrino oscillations, that would allow its thermalization,
are suppressed
in the hot and dense primordial plasma.

\part{Appendix}
\appendix
%!TeX root=main.tex 
\chapter{\texttt{PCHIP} Parametrization of the Primordial Power Spectrum}
\label{ch:app_pchip}
\chapterprecis{
This Chapter appears as Appendix~A in Ref.~\protect\cite{Gariazzo:2014dla}.}

In this work we parameterized the PPS with a
``piecewise cubic Hermite interpolating polynomial''
(\pchip)
\cite{Fritsch:1980,Fritsch:1984}.
We decided to adopt this interpolating function in order to avoid
spurious oscillations of the interpolating function between the nodes
which is often obtained in spline interpolations.
This problem occurs because a natural cubic spline requires the values of the function, the first and the second derivatives to be continuous in the nodes \cite{NR}.

The \pchip function, instead, is constructed in order to preserve the
shape of the set of points to be interpolated.
This is achieved with
a modification of the ``monotone piecewise cubic interpolation''
\cite{Fritsch:1980}
which can accommodate non-monotone functions
and preserves the local monotonicity.

Let us consider a function with known values $y_{j}$
in N nodes $x_{j}$,
with $j=1,\ldots,N$.
A piecewise cubic interpolation is performed with
$N-1$ cubic functions between the nodes.
The determination of these
$N-1$ cubic functions requires the determination of
$4(N-1)$ coefficients.
Besides the
$2(N-1)$ constraints obtained by requiring that the initial and final
point of each cubic function match the known values of the original function
in the corresponding nodes,
one needs a prescription for the other $2(N-1)$ necessary constraints.
In the case of a natural cubic spline interpolation one gets
$2(N-2)$ constraints by requiring the continuity of the first and second derivatives in the nodes
and the remaining two constraints
are obtained by requiring that the second derivatives in the first and last nodes vanish.
The drawback of this method is that the interpolating curve
is determined by a set of linear equations without any local control.
In fact,
all the interpolating curve is affected by the change of a single point.

\begin{figure*}
\centering
\includegraphics[width=0.8\textwidth]{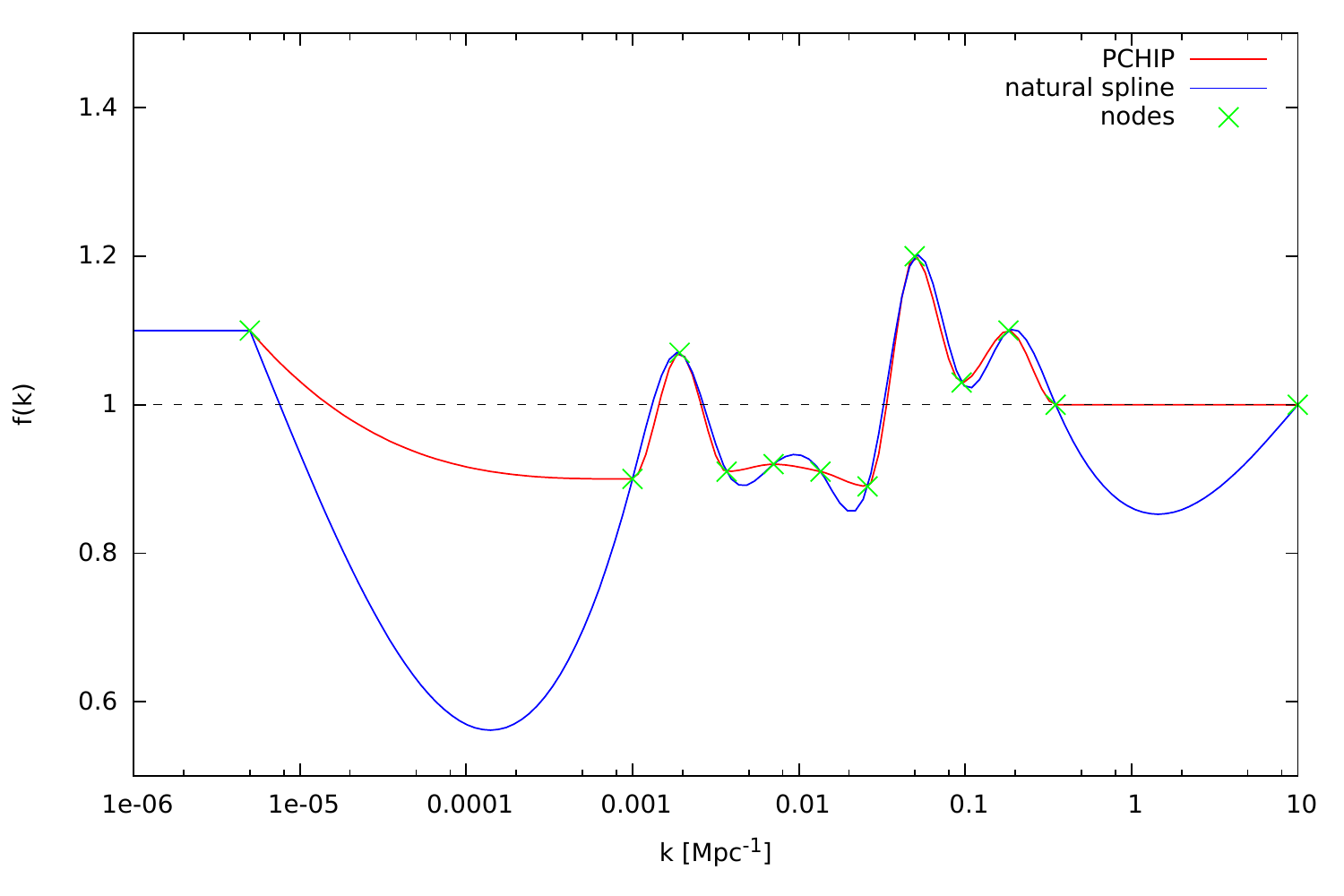}
\caption[Difference between the \pchip\ and the natural cubic spline
interpolating functions]
{\label{fig:interpolation}
Illustration of the difference between the \pchip (red line)
and the natural spline (blue line) interpolations
$f(\log{k}; y_{1}, \ldots, y_{12})$
of a function with known values
$y_{1}, \ldots, y_{12}$
in 12 nodes (green crosses) at the values of $k$ in 
Eq.~\eqref{eq:nodesspacing}.
The values $y_{1}, \ldots, y_{12}$ in the nodes are
1.1,
0.9,
1.07,
0.91,
0.92,
0.91,
0.89,
1.2,
1.03,
1.1,
1.0,
1.0.
}
\end{figure*}

Local control of the interpolating curve can be achieved by
relaxing the requirement of continuity of the second derivatives
in the nodes
and using the resulting freedom to adjust the first derivatives
with a local prescription.
In order to see how it can be done, it is convenient to write the cubic interpolating polynomial
between the nodes $x_{j}$ and $x_{j+1}$ in the Hermite form
\begin{equation}
f(x; y_{1}, \ldots, y_{N})
=
\frac{\left( h_{j} + 2 t \right) \left( h_{j} - t \right)^2}{h_{j}^3} y_{j}
+
\frac{\left( 3 h_{j} - 2 t \right) t^2}{h_{j}^3} y_{j+1}
+
\frac{\left( h_{j} - t \right)^2 t}{h_{j}^2} d_{j}
+
\frac{t^2 \left( h_{j} - t \right)}{h_{j}^2} d_{j+1}
,
\label{a01}
\end{equation}
where
$t = x-x_{j}$
and
$h_{j} = x_{j+1}-x_{j}$.
Here
$d_{j}$ and $d_{j+1}$
are the values of the derivatives in the two nodes.
In the \pchip method the derivatives are chosen in order to preserve the local
monotonicity of the interpolated points.
This is done by considering the relative differences
\begin{equation}
\delta_{j}
=
\frac{y_{j+1} - y_{j}}{x_{j+1}-x_{j}}
.
\label{a02}
\end{equation}
The \pchip  prescription is:
\begin{itemize}

\item
If $\delta_{j-1}$ and $\delta_{j}$ have opposite signs,
then $x_{j}$ is a discrete local minimum or maximum
and
$d_{j}=0$.

\item
If $\delta_{j-1}$ and $\delta_{j}$ have the same sign,
then
$d_{j}$
is determined by the weighted harmonic mean
\begin{equation}
\frac{w_{1}+w_{2}}{d_{j}}
=
\frac{w_{1}}{\delta_{j-1}}
+
\frac{w_{2}}{\delta_{j}}
,
\label{a03}
\end{equation}
with
$w_{1}=2h_{j}+h_{j-1}$
and
$w_{2}=h_{j}+2h_{j-1}$.

\item
The derivatives in the first and last nodes
are determined by a shape-preserving prescription
based on a quadratic fit of three points.
For $d_{1}$ we consider the three points
$(x_{1},y_{1})$,
$(x_{2},y_{2})$,
$(x_{3},y_{3})$.
The derivative in $x_{1}$ of the parabola which passes through these three points is
given by
\begin{equation}
d(h_{1}, h_{2}, \delta_{1}, \delta_{2})
=
\frac{\left( 2 h_{1} + h_{2} \right) \delta_{1} - h_{1} \delta_{2}}{h_{1} + h_{2}}
\,.
\label{a04}
\end{equation}
The shape-preserving prescription for $d_{1}$ is:

\begin{itemize}

\item
If the signs of $d(h_{1}, h_{2}, \delta_{1}, \delta_{2})$
and $\delta_{1}$ are different,
then
$d_{1} = 0$.

\item
If the signs of $\delta_{1}$ and $\delta_{2}$ are different
and
$|d(h_{1}, h_{2}, \delta_{1}, \delta_{2})| > 3 |\delta_{1}|$,
then
$d_{1} = 3 \delta_{1}$.

\item
Else $d_{1} = d(h_{1}, h_{2}, \delta_{1}, \delta_{2})$.

\end{itemize}

For $d_{N}$ one must replace $1 \to N-1$ and $2 \to N-2$.

\end{itemize}

We fit the power spectrum
$P_{s}(k)$
with Eq.~\eqref{eq:PPS_pchip},
in which the function
$\pchip(k; P_{s,1}, \ldots, P_{s,12})$
is calculated with the \pchip prescription
in the logarithmic scale of $k$:
\begin{equation}
\pchip(k; P_{s,1}, \ldots, P_{s,12})
=
f(\log{k}; P_{s,1}, \ldots, P_{s,12})
\,.
\label{a11}
\end{equation}

A comparison between the natural cubic spline and the \pchip interpolations
of the PPS is presented in Fig.~\ref{fig:interpolation}.
We choose the same nodes positions that we used for the PPS parametrization in our cosmological analysis
and we choose the values of the function in the nodes
in order to show the difference between
the natural cubic spline and the \pchip interpolations.
One can see that
the \pchip interpolation can reproduce the shape of the points without adding
the spurious features between the points that are clearly visible
in the natural cubic spline interpolation.

\backmatter
% \cleardoublepage
% \phantomsection 
% \bibliography{all}{}
% \bibliographystyle{utcaps}
\providecommand{\href}[2]{#2}\begingroup\raggedright\endgroup

\cleardoublepage
% \phantomsection
% \addcontentsline{toc}{chapter}{List of Figures}
\listoffigures

\cleardoublepage
% \phantomsection 
% \addcontentsline{toc}{chapter}{List of Tables}
\listoftables

\end{document}